\documentclass[letterpaper,12pt]{article}
\usepackage{tabularx} 
\usepackage{amsmath,bm}  
\usepackage{float}
\usepackage{indentfirst}
\usepackage{graphicx} 
\usepackage[margin=1in,letterpaper]{geometry} 
\usepackage{caption}
\usepackage{cite} 
\usepackage[colorlinks,linkcolor=magenta,anchorcolor=cyan,citecolor=blue]{hyperref} 
\usepackage{tensor}
\hypersetup{
	colorlinks=true,       
	linkcolor=blue,        
	citecolor=blue,        
	filecolor=magenta,     
	urlcolor=blue
}

\def\be{\begin{equation}}
\def\ee{\end{equation}}
\def\ba{\begin{eqnarray}}
\def\ea{\end{eqnarray}}

              \def\.{\cdot}

\begin{document}
\begin{center}

	\vspace{10pt}
	\large{\bf{Second Law of Black Holes Thermodynamics in Non-Quasi-Static Processes}}
	
	\vspace{15pt}
	Xin-Yang Wang $^\text{a}$, Jie Jiang $^\text{a}$
	
	\vspace{15pt}
	\small{\it $^\text{a}$ Faculty of Arts and Sciences, Beijing Normal University, Zhuhai 519087, China}
	\vspace{30pt}
\end{center}
\begin{abstract}
	The Bekenstein-Hawking entropy satisfies the generalized second law of black hole thermodynamics for arbitrary thermodynamic evolution within Einstein-Maxwell theory. In contrast, the black hole entropy that satisfies the second law in low-energy effective modified gravity theories related to quantum gravity is derived solely by considering linear perturbations within quasi-static processes. Since astrophysical processes are typically non-quasi-static, deriving a rigorous expression for entropy within the quasi-static framework is not feasible. By treating the effective quantum corrections as first-order perturbations, the black hole entropy in Einstein-Maxwell gravity with generalized quadratic corrections for arbitrary dynamical processes is derived. This entropy is distinct from the Iyer-Wald and Dong-Wald entropies and is applicable to non-quasi-static processes. Furthermore, it is shown that the black hole entropy is consistent with the generalized covariant entropy boundary. This work establishes a framework for examining the generalized second law of black hole thermodynamics in non-static processes within modified gravity theories.
\end{abstract}
	\vfill {\footnotesize ~\\ $^\text{a}$ xinyangwang@bnu.edu.cn \\ Corresponding author: \\ $^\text{a}$ jiejiang@mail.bnu.edu.cn}
\newpage

\section{Introduction}

Entropy is a fundamental physical quantity that plays a central role in both classical and statistical thermodynamics. In classical thermodynamics, entropy is defined as a macroscopic quantity that characterizes the state of a thermodynamic system and is essential to the formulation of the first law of thermodynamics. Moreover, the intrinsic tendency of entropy to increase during irreversible processes forms a foundational basis for validating and interpreting whether a thermodynamic system satisfies the second law of thermodynamics. In statistical thermodynamics, entropy quantifies the disorder, randomness, and uncertainty associated with the microscopic states of a thermodynamic system. The distinct roles of entropy in classical and statistical thermodynamics provide a fundamental framework for studying thermodynamic systems, establishing a crucial link between macroscopic thermodynamic phenomena and underlying microscopic states.

Black holes, as predicted by General Relativity (GR), represent a distinctive class of spacetime structures. Rigorous theoretical investigations have shown that black holes possess well-defined temperature and entropy, with the entropy in GR specifically referred to as the Bekenstein-Hawking (BH) entropy \cite{Hawking:1971tu, Bekenstein:1973ur, Hawking:1974sw}. The introduction of temperature and entropy suggests that black holes should not be viewed merely as unique spacetime structures, but also as thermodynamic systems. As a result, the four laws of black hole thermodynamics have been formulated within the GR framework, demonstrating a remarkable parallel with the four fundamental laws of classical thermodynamics \cite{Bekenstein:1972tm, Bardeen:1973gs}. However, the applicability of these laws in gravitational theories extending beyond GR remains uncertain, requiring further rigorous investigation. The first law of black hole thermodynamics has been rigorously formulated in any diffeomorphism invariant gravitational theory, where the black hole entropy, commonly referred to as the Wald entropy, is defined as \cite{Wald:1993nt, Iyer:1994ys}
\begin{equation}
	\begin{split}
		S_{\text{W}} = \int_B d^{D-2} x \sqrt{h} s_{\text{W}}\,,
	\end{split}
\end{equation}
where $s_{\text{W}}$ denotes the density of the Wald entropy, which is expressed as 
\begin{equation}
	\begin{split}
		s_{\text{W}} = - 2 \pi \frac{\partial \mathcal{L}}{\partial R_{abcd}} \hat{\boldsymbol{\epsilon}}_{ab} \hat{\boldsymbol{\epsilon}}_{cd}\,.
	\end{split}
\end{equation}
In the definition of the Wald entropy, $B$ represents a cross-section of the event horizon, $h$ denotes the determinant of the induced metric on $B$, $\mathcal{L}$ is the Lagrangian of the gravitational theory, $R_{abcd}$ corresponds to the Riemann curvature tensor, and $\hat{\boldsymbol{\epsilon}}_{ab}$ is the binormal to the cross-section. When the gravitational theory reduces to GR, the Wald entropy simplifies to the BH entropy. While the Wald entropy accurately characterizes the black hole entropy in a specific thermodynamic equilibrium state, it may not fully account for entropy variations during the dynamical evolution of black holes. This limitation arises because, in the context of thermodynamic evolution, additional undetermined contributions extend beyond the Wald entropy, explicitly describing entropy variations throughout the dynamical process. Consequently, the entropy density of black holes during dynamical evolution should generally be expressed as
\begin{equation}
	\begin{split}
		s_{\text{bh}} = s_{\text{W}} + s_{\text{dyna}}\,,
	\end{split}
\end{equation}
where $s_{\text{dyna}}$ represents the entropy density associated with variations in black hole entropy during thermodynamic evolution. This implies that the Wald entropy alone is insufficient for a complete characterization of black hole entropy, as the inclusion of $s_{\text{dyna}}$ provides the necessary flexibility for a more accurate description of entropy changes throughout the dynamical process. Therefore, determining the explicit form of $s_{\text{dyna}}$ is crucial for achieving a comprehensive and precise characterization of black hole entropy.

The entropy of black holes should strictly satisfy the generalized second law of black hole thermodynamics throughout the entire thermodynamic evolution. The entropy density $s_{\text{dyna}}$, arising from the dynamical evolution of black holes, should evolve in accordance with the second law. In quasi-static evolution, where the process is treated as a perturbation of a static black hole spacetime background, research on the generalized second law primarily focuses on formulating an expression for black hole entropy that consistently satisfies the linearized second law of thermodynamics. Black hole entropy satisfying the linearized second law in higher-curvature gravity theories has been extensively explored, with its expression characterized by the Dong-Wald entropy formula \cite{Wall:2015raa}. Additionally, formulations of black hole entropy satisfying the linearized second law have been established in gravitational theories involving non-minimal interactions between gravity and matter fields, such as electromagnetic and scalar fields \cite{Wang:2020svl, Wang:2021zyt, Wang:2022stg, Wang:2022xop}. A recent advancement introduced a novel formulation of black hole entropy, employing the Noether charge method under the linear-order perturbation to describe the entropy of dynamical black holes \cite{Hollands:2024vbe}. However, most astrophysical processes, including gravitational collapse, black hole accretion, and binary black hole mergers, are inherently non-quasi-static, rendering perturbative methods inadequate for investigating the second law of black hole thermodynamics in these contexts. Therefore, developing methodologies to verify the second law in fully dynamical regimes is essential. In Einstein-Maxwell gravitational theory, the BH entropy strictly adheres to the second law of black hole thermodynamics in non-quasi-static processes. However, in other gravitational theories, a rigorous formulation of black hole entropy that satisfies the second law in such contexts remains undeveloped. Consequently, deriving a general formulation of black hole entropy that satisfies the second law in non-quasi-static processes is crucial for accurately characterizing black hole evolution in finite dynamical regimes.

The limitations of contemporary quantum gravity theories hinder the full quantization of gravity, posing significant challenges to understanding its fundamental properties and interactions with matter fields at the quantum scale. In the absence of a complete quantum gravity theory, low-energy effective field theory provides a useful approximation for exploring quantum gravitational effects. Within this framework, gravitational self-interactions and interactions with matter fields are incorporated into the effective gravitational Lagrangian through quantum correction terms. These corrections, characterized by coupling constants treated as small parameters, encode signatures of underlying quantum gravitational effects and serve as perturbative modifications to the classical gravitational theory. It is important to emphasize that quantum corrections are intrinsic to the effective theory, naturally emerging from the low-energy approximation of quantum gravity. This approach forms a solid foundation for investigating the second law of black hole thermodynamics in Einstein-Maxwell gravity with generalized quadratic corrections. The Lagrangian of this gravitational theory is given by
\begin{equation}\label{totallagrangian}
	\boldsymbol{\mathcal{L}} =\boldsymbol{\mathcal{L}}_{\text{EM}} + \boldsymbol{\mathcal{L}}_{\text{int}} + \boldsymbol{\mathcal{L}}_{\text{mt}}\,.
\end{equation}
The first component on the right-hand side of Eq. (\ref{totallagrangian}) is the Lagrangian of the original Einstein-Maxwell gravity, which can be given as 
\begin{equation}
	\boldsymbol{\mathcal{L}}_{\text{EM}} = \frac{1}{16 \pi} \left(R - F_{ab}F^{ab} \right) \boldsymbol{\epsilon}\,.
\end{equation}
The second component in Eq. (\ref{totallagrangian}) encompasses all quantum correction terms in the gravitational theory and can be further decomposed into two distinct components as follows
\begin{equation}\label{correctterm}
	\boldsymbol{\mathcal{L}}_{\text{int}} = \boldsymbol{\mathcal{L}}_{\text{gem}} + \boldsymbol{\mathcal{L}}_{\text{grav}}\,.
\end{equation}
The first term, $\boldsymbol{\mathcal{L}}_{\text{gem}}$, describes the non-minimal coupling interactions between the gravitational and the electromagnetic fields, as well as the self-interactions of the electromagnetic field. The second term, $\boldsymbol{\mathcal{L}}_{\text{grav}}$, exclusively accounts for the self-interactions of the gravitational field. The explicit expressions of these two components are given as
\begin{equation}\label{deltalgem}
	\begin{split}
		\boldsymbol{\mathcal{L}}_{\text{gem}} = & \frac{1}{16 \pi} \left(a_1 R F_{ab} F^{ab} + a_2 R_{ab} F^{ac} F^{b}_{\ c} + a_3 R_{abcd} F^{ab} F^{cd} \right. \\
		&\left. + a_4 F_{ab} F^{ab} F_{cd} F^{cd} + a_5 F_{ab} F^{bc} F_{cd} F^{da}\right) \boldsymbol{\epsilon}
	\end{split}
\end{equation}
and 
\begin{equation}\label{deltalgrav}
	\boldsymbol{\mathcal{L}}_{\text{grav}} = \frac{1}{16 \pi} \left(b_1 R^2 + b_2 R_{ab} R^{ab} + b_3 R_{abcd} R^{abcd} \right) \boldsymbol{\epsilon}\,.
\end{equation}
The third component in Eq. (\ref{totallagrangian}) corresponds to the Lagrangian of additional matter fields within the gravitational theory. In the explicit formulation of the gravitational Lagrangian, $R$ represents the Ricci scalar, $R_{ab}$ is the Ricci tensor, and $F_{ab} = 2 \nabla_{[a} A_{b]}$ refers to the electromagnetic field strength tensor, where $A_a$ is the electromagnetic vector potential. The tensor $\boldsymbol{\epsilon}$ serves as the spacetime volume element, constructed within the framework of the gravitational theory.

All coupling constants in the quantum corrections are treated as small parameters within the framework of the low-energy effective approximation. In the Einstein-Maxwell gravitational theory with generalized quadratic corrections, these constants appear exclusively as linear terms in the Lagrangian, with no higher-order contributions. When quantum corrections are considered as perturbative modifications to the original gravitational theory, these correction terms can be systematically interpreted as linear perturbations within the Einstein-Maxwell framework. Importantly, these perturbations are intrinsic to the fundamental structure of the gravitational theory and are independent of any subsequent analysis related to the second law of black hole thermodynamics.

The second law of black hole thermodynamics in Einstein-Maxwell gravity with generalized quadratic corrections, as defined by the Lagrangian in Eq. (\ref{deltalgrav}), has been analyzed in the quasi-static regime \cite{Zhu:2023xbn}. The black hole entropy satisfying the second law is identified as the Wald entropy. This result underscores the limitations of Einstein-Maxwell gravity with generalized quadratic corrections in accounting for the entropy density $s_{\text{dyna}}$ associated with quasi-static black hole evolution. To derive an explicit expression for $s_{\text{dyna}}$, it is essential to extend the analysis of the second law beyond quasi-static processes to non-quasi-static regimes. Since the Lagrangian includes only linear-order quantum correction terms, the second law can be analyzed independently for each component of the Lagrangian. This approach enables an effective investigation of the second law in Einstein-Maxwell gravity with generalized quadratic corrections, focusing specifically on the following Lagrangian as
\begin{equation}\label{usedlagrangian}
	\boldsymbol{\mathcal{L}} =\boldsymbol{\mathcal{L}}_{\text{EM}} + \boldsymbol{\mathcal{L}}_{\text{gem}} + \boldsymbol{\mathcal{L}}_{\text{mt}}\,.
\end{equation}
From the Lagrangian in Eq. (\ref{usedlagrangian}), the equation of motion corresponding to the gravitational component can be formally written as
\begin{equation}\label{equationofmotion}
	E_{ab} = 8 \pi T_{ab}\,.
\end{equation}
The left-hand side of the equation of motion in Eq. (\ref{equationofmotion}) is expressed as
\begin{equation}\label{exphab}
	E_{ab} = G_{ab} + \frac{1}{2} \sum_{i = 1}^{3} a_i H_{ab}^{(i)}\,,
\end{equation}
where $G_{ab} = R_{ab} - 1 / 2 R g_{ab}$ is the Einstein tensor, and $H_{ab}^{(i)}$ includes terms that are proportional to the coupling constant $a_i$ in the equations of motion. The right-hand side of Eq. (\ref{equationofmotion}) represents the stress-energy tensor, which can be expressed as 
\begin{equation}\label{totalstressenergy}
	\begin{split}
		T_{ab} = T^{\text{em}}_{ab} + T^{\text{mt}}_{ab}\,,
	\end{split}
\end{equation}
where
\begin{equation}\label{stentensorem}
	\begin{split}
		T^{\text{em}}_{ab} = \frac{1}{4 \pi} \left(F_{ac} F_{b}^{\ c} - \frac{1}{4} g_{ab} F_{cd} F^{cd} \right)\,.
	\end{split}
\end{equation}
is the stress-energy tensor of the electromagnetic field, and $T^{\text{mt}}_{ab}$ denotes the stress-energy tensor of the matter fields in spacetime.

\section{The second law of black hole thermodynamics}

Before analyzing the second law of black hole thermodynamics, it is essential to construct the spacetime geometry and adopt a coordinate system that is appropriately suited to the spacetime. A null hypersurface $L$ is selected within the spacetime, with the null vector $k^a = \left(\partial / \partial u \right)^a$ serving as its tangent vector, where $u$ is an affine parameter. A compact $(D-2)$-dimensional cross-section $B (u)$ on $L$ is treated as a spatial hypersurface. The hypersurface $L$ is bounded by $B (0)$ and $B (1)$, corresponding to $u = 0$ and $u = 1$, respectively. Another null vector, $l^a = \left(\partial / \partial z \right)^a$, is defined on $L$, where $z$ is the affine parameter along $l^a$. Using $u$ and the coordinates $x = \left(x^1\,, \cdots\,, x^{D-2} \right)$ on each cross-section, the coordinates $\left(u\,, x \right)$ can be established on $L$. To facilitate the expression of black hole entropy and simplify subsequent calculations, Gaussian null coordinates $\left\{u\,, z\,, x \right\}$ are introduced to cover the vicinity of $L$. In these coordinates, the line element of the spacetime can be written as
\begin{equation}\label{gaussiannullcoordinates}
	\begin{split}
		ds^2 = 2 \left(d z + z^2 \alpha d u + z \beta_i d x^i \right) + \gamma_{ij} d x^i d x^j\,,
	\end{split}
\end{equation}
where the coefficients $\alpha$, $\beta_i$, and $\gamma_{ij}$ are functions of $\left(u\,, z\,, x \right)$. The line element on $L$ corresponds specifically to the condition $z = 0$. The indices $i\,, j\,, \cdots$ refer to the coordinates of the cross-section $B (u)$. The density of the Wald entropy can be expressed as 
\begin{equation}\label{waldentropydensity}
	\begin{split}
		s_{\text{W}} = & 1 + a_1 \left(- 2 F_{uz} F_{uz} - 4 \gamma^{ij} F_{ui} F_{zj} + \gamma^{ij} \gamma^{kl} F_{ik} F_{jl} \right) - a_2 \left(F_{uz} F_{uz} + \gamma^{ij} F_{ui} F_{zj} \right)\\
		& - a_3 \left(2 F_{uz} F_{uz} \right)\,.
	\end{split}
\end{equation}
A simplification convention is introduced in Eq. (\ref{waldentropydensity}), where tensor indices contracted with the null vector $k^a$ are represented by the affine parameter $u$, and those contracted with the null vector $l^a$ are denoted by the parameter $z$. This convention is employed to streamline the expressions in the subsequent analysis.

The second law of black hole thermodynamics asserts that the entropy of a black hole should not decrease during adiabatic thermodynamic evolution. To rigorously assess the validity of this law in Einstein-Maxwell gravity with generalized quadratic corrections, the evolution of black hole entropy throughout the thermodynamic process should be examined. The generalized expansion, which quantifies the change in entropy per unit area of the cross-section of the null hypersurface during thermodynamic evolution, is expressed as
\begin{equation}\label{defgeneralexpansion}
	\begin{split}
		\Theta \left(u\,, x \right) = \frac{1}{\sqrt{\gamma \left(u\,, x \right)}} \partial_u \left[\sqrt{\gamma \left(u\,, x \right)} s_{bh} \left(u\,, x \right) \right]\,,
	\end{split}
\end{equation}
where $\partial_u$ denotes shorthand for $k^a \nabla_a$. During the transition of the black hole between thermodynamic equilibrium states, the second law of black hole thermodynamics dictates that the entropy should converge to a well-defined value once a new equilibrium state is reached. This implies that the generalized expansion during dynamical evolution satisfy the condition
\begin{equation}\label{partialutheta}
	\begin{split}
		\partial_u \Theta (u\,, x) \le 0\,.
	\end{split}
\end{equation}
From the equation of motion in Eq. (\ref{equationofmotion}), the derivative of the generalized expansion with respect to the affine parameter $u$ is given by 
\begin{equation}\label{rayequation}
	\begin{split}
		\partial_u \Theta \left(u\,, x \right) = - 8 \pi \mathcal{T} \left(u\,, x \right) + \mathcal{F} \left(u\,, x \right)\,,
	\end{split}
\end{equation} 
where 
\begin{equation}
	\begin{split}
		\mathcal{T} \left(u\,, x \right) = T^{\text{mt}}_{uu}
	\end{split}
\end{equation}
and
\begin{equation}
	\begin{split}
		\mathcal{F} \left(u\,, x \right) = E_{uu} - 8 \pi T^{\text{em}}_{uu} + \partial_u \Theta \left(u\,, x \right)\,.
	\end{split}
\end{equation}
Eq. (\ref{rayequation}) represents the Raychaudhuri equation in Gaussian null coordinates for the spacetime derived from Einstein-Maxwell gravity with generalized quadratic corrections. The stress-energy tensor in Eq. (\ref{totalstressenergy}) is required to satisfy the null energy condition, i.e., $T_{uu} = T^{\text{em}}_{uu} + T^{\text{mt}}_{uu} \ge 0$. To analyze the evolution of the generalized expansion during the thermodynamic process and assess its compliance with the criteria outlined in Eq. (\ref{partialutheta}), the explicit form of the function $\mathcal{F} \left(u\,, x \right)$ should be derived and examined to ensure that it satisfies the condition $\mathcal{F} \left(u\,, x \right) \le 0$.

In Einstein-Maxwell gravity with generalized quadratic corrections, all quantum correction terms are treated as linear-order perturbations to the original Einstein-Maxwell theory within the low-energy effective approximation. As a result, the coupling constants associated with these corrections are considered small quantities, and the coupling constants $a_i$ (with $i = 1\,, 2\,, 3$) are conveniently represented by the symbol $a$. To verify the condition $\mathcal{F} \left(u\,, x \right) \le 0$ under the constraint of the linear-order quantum corrections, the black hole entropy $s_{\text{bh}}$ is reformulated as $s_{\text{bh}} = s_{\text{bh}}^{\text{(nc)}} + s_{\text{bh}}^{\text{(cc)}}$, where $s_{\text{bh}}^{\text{(nc)}} = 1$ denotes the component independent of $a$, and $s_{\text{bh}}^{\text{(cc)}} = s_{\text{W}} - 1 + s_{\text{dyna}}$ includes $a$. The generalized expansion is decomposed as $\Theta \left(u\,, x \right) = \Theta^{\text{(nc)}} \left(u\,, x \right) + \Theta^{\text{(cc)}} \left(u\,, x \right)$. Based on the expression of the Ricci tensor and the stress-energy tensor of the electromagnetic field in Eq. (\ref{stentensorem}), the function $\mathcal{F} \left(u\,, x \right)$ can be further expressed as
\begin{equation}\label{formalexpfunctionf}
	\begin{split}
		\mathcal{F} \left(u\,, x \right) = - K^{ij} K_{ij} - 2 F_{ui} F_{u}^{\ i} + \frac{1}{2} \sum_{i = 1}^{3} a H_{uu}^{(i)} + \partial_u \Theta^{\text{(cc)}} \left(u\,, x \right)\,,
	\end{split}
\end{equation}
where $K_{ij}$ denotes the extrinsic curvature of the cross-section, and the first term is derived from the function $\Theta^{\text{(nc)}} (u\,, x)$. 

Furthermore, to examine whether each term in the function $\mathcal{F} \left(u\,, x \right)$ satisfies the constraints imposed by the linear-order quantum corrections in the gravitational theory, the tensor $X_{a_1 \cdots a_n}$ is introduced to represent any tensor appearing in the function. The Frobenius norm of the tensor on $B (u)$ is defined as
\begin{equation}\label{assumeorderk}
	\begin{split}
		\left\|X \right\| = \sqrt{\int_{B(u)} d^{D-2} x \sqrt{\gamma} X_{a_1 \cdots a_n} X^{a_1 \cdots a_n}}\,,
	\end{split}
\end{equation}
where $\sqrt{\gamma}$ is an abbreviation for $\sqrt{\gamma(u, x)}$. Building upon this, an arbitrary tensor $Y_{a_1 \cdots a_n}$ is introduced. By applying the Cauchy-Schwarz inequality, the contraction between $X_{a_1 \cdots a_n}$ and $Y_{a_1 \cdots a_n}$ over $B (u)$ satisfies
\begin{equation}\label{orderofxy}
	\begin{split}
		\left|\int_{B(u)} d^{D-2} x \sqrt{\gamma} X_{a_1 \cdots a_n} Y^{a_1 \cdots a_n} \right| \le \left\|X \right\| \left\|Y \right\| = \mathcal{O} \left(\left\|X \right\| \right)\,.
	\end{split}
\end{equation}
Since the equation of motion involves covariant derivative operators $D_a$ acting on tensors on $B (u)$, the behavior of the tensor $D_{a_1} X_{a_2 \cdots a_n}$ should be further considered. Following an approach analogous to that employed in Eq. (\ref{orderofxy}), the contraction between $D_{a_1} X_{a_2 \cdots a_n}$ and $Y^{a_1 \cdots a_n}$ over $B (u)$ can be expressed as
\begin{equation}\label{orderxdk}
	\begin{split}
		& \left| \int_{B(u)} d^{D-2} x \sqrt{\gamma} Y^{a_1 \cdots a_n} \left(D_{a_1} X_{a_2 \cdots a_n} \right) \right| = \left| \int_{B(u)} d^{D-2} x \sqrt{\gamma} \left(D_{a_1} Y^{a_1 \cdots a_n} \right) X_{a_2 \cdots a_n} \right|\\
		& \le \left\|D Y \right\| \left\|X \right\| = \mathcal{O} \left(\left\|X \right\| \right)\,.
	\end{split}
\end{equation}
The order on the right-hand side of inequalities (\ref{orderofxy}) and (\ref{orderxdk}) is primarily determined by the order of $\left\|X \right\|$, as tensor $Y^{a_1 \cdots a_n}$ is arbitrary and typically of order $\mathcal{O} \left(1 \right)$.

By employing the explicit forms of $H^{(i)}_{uu}\,, i = 1\,, 2\,, 3$ and the definition of the generalized expansion in Eq. (\ref{defgeneralexpansion}), the integral of $\mathcal{F} \left(u\,, x \right)$ over $B(u)$ can be expressed as 
\begin{equation}\label{finalexpressionoff}
	\begin{split}
		& \int_{B(u)} d^{D-2} x \sqrt{\gamma} \mathcal{F} \left(u\,, x \right)\\
		= & - \left\|K \right\|^2 - 2 \left\|F \right\|^2 + \partial_u^2 \int_{B(u)} d^{D-2} x \left[\sqrt{\gamma} \left(- 8 a_1 \gamma^{ij} F_{ui} F_{zj} - 2 a_2 \gamma^{ij} F_{ui} F_{zj} + s_{\text{dyna}} \right) \right]\\
		& + a_i \mathcal{O} \left(\left\|K \right\|^2\,, \left\|F \right\|^2\,, \left\|K \right\| \left\|F \right\| \right)\,.
	\end{split}
\end{equation}
From the norm definition in Eq. (\ref{assumeorderk}), $\left\|K \right\|$ and $\left\|F \right\|$ denote the norms of $K_{ij}$ and $F_{ui}$, respectively. During the derivation of Eq. (\ref{finalexpressionoff}), the symbols $s_1$ and $s_2$ represent the powers of the coupling constant $a$, with values $s_1, s_2 \in [0\,, 1]$ due to the linear-order constraint of the quantum corrections. The orders of the integrals of the first two components in the function $\mathcal{F} \left(u, x \right)$ over $B (u)$ in Eq. (\ref{formalexpfunctionf}) are related to $a$ by $\mathcal{O} \left(\left\|K \right\|^2 \right) = \mathcal{O} \left(a^{2 s_1} \right)$ and $\mathcal{O} \left(\left\|F \right\|^2 \right) = \mathcal{O} \left(a^{2 s_2} \right)$. Based on the analysis of Eqs. (\ref{orderofxy}) and (\ref{orderxdk}), it follows that any term in the latter two components of $\mathcal{F} \left(u, x \right)$ involving $K_{ij}$, $D_k K_{ij}$, $F_{ui}$, or $D_{j} F_{ui}$ is at least of order $\mathcal{O} \left(a^{s_1+1}, a^{s_2+1} \right)$. As a result, the contributions from the first two components of $\mathcal{F} \left(u, x \right)$ are significantly larger than those from the latter two components. This dominance implies that the leading behavior of $\mathcal{F} \left(u, x \right)$ is primarily determined by the first two components. Therefore, any term in the third or fourth components of $\mathcal{F} \left(u\,, x \right)$ involving $K_{ij}$, $D_k K_{ij}$, $F_{ui}$, or $D_{j} F_{ui}$ can be grouped into the third component in Eq. (\ref{finalexpressionoff}). Since these terms have comparatively lower weights than the first two components and do not influence the final expression of $\mathcal{F}(u, x)$ required for analyzing the second law of thermodynamics, the explicit expressions of these terms are omitted from Eq. (\ref{finalexpressionoff}) for brevity. Additionally, the relative magnitudes of $s_1$ and $s_2$ do not affect the validity of $\mathcal{F} \left(u\,, x\right) < 0$, as both $\left\|K \right\|^2$ and $\left\|F \right\|^2$ are negative in Eq. (\ref{finalexpressionoff}). To ensure that the condition $\mathcal{F} \left(u\,, x\right) < 0$ is consistently satisfied, the second component in Eq. (\ref{finalexpressionoff}) should be eliminated, as its sign cannot be definitively determined. Consequently, when $\mathcal{F} \left(u\,, x\right) < 0$ holds, the entropy density $s_{\text{dyna}}$ can be explicitly determined as
\begin{equation}\label{expsdyna}
	\begin{split}
		s_{\text{dyna}} = 8 a_1 \gamma^{ij} F_{ui} F_{zj} + 2 a_2 \gamma^{ij} F_{ui} F_{zj}\,.
	\end{split}
\end{equation}
By combining the results from Eqs. (\ref{waldentropydensity}) and (\ref{expsdyna}), the expression for black hole entropy, which rigorously satisfies the second law of thermodynamics during non-quasi-static processes within the Einstein-Maxwell gravitational theory with generalized quadratic corrections, has been fully derived.

\section{The generalized covariant entropy bound}

In a $D$-dimensional spacetime governed by a gravitational theory, a connected $(D-2)$-dimensional spatial hypersurface $B(u)$ can be identified, with its boundary area denoted as $A(B(u))$. The hypersurface $L$ is defined as a null hypersurface within the spacetime, generated by one of the four null congruences emanating from $B(u)$. When the expansion of the congruence is non-positive at every point on $L$, the null hypersurface is referred to as a light sheet. The quantity $S_{L}$ represents the entropy flux passing through the null hypersurface $L$.

For weak gravitational systems or strong gravitational systems described by GR, the variation of entropy on a spatial hypersurface due to the entropy flux $S_{L}$ is expected to satisfy the inequality $S_{L} \le A (B) / 4$, as articulated by the holographic principle. This inequality, known as the Bousso bound, establishes a fundamental upper limit on entropy, corresponding to the maximum entropy of a black hole. Furthermore, the Bousso bound has been extended to more general scenarios where the light sheet $L$ originates from one spatial hypersurface $B(u)$ and terminates at another hypersurface $B(u^\prime)$. However, the Bousso bound does not apply to modified gravity theories or other strong gravitational systems. To address this limitation, the generalized covariant entropy bound has been introduced to constrain entropy variations induced by entropy flux. In this framework, the upper limit of the inequality is replaced by the black hole entropy $S_{\text{bh}} \left(B(u)\right)$, as defined in modified gravity theories.

According to the two fundamental assumptions outlined in Ref. \cite{Matsuda:2020yvl} and the Raychaudhuri equation in Eq. (\ref{rayequation}), the entropy flux in Einstein-Maxwell gravity with generalized quadratic corrections satisfies the inequality
\begin{equation}\label{inequalityslwithfunctionf}
	\begin{split}
		S_L \le & S_{\text{bh}} \left(B(0)\right) - S_{\text{bh}} \left(B(1) \right)\\
		& + \frac{1}{4} \int_{0}^{1} d u \int_{0}^{u} d \tilde{u} \int_{B(u)} d^{D-2} x \sqrt{\gamma} \mathcal{F} \left(\tilde{u}\,, x \right)\,.
	\end{split}
\end{equation}
Based on Eqs. (\ref{finalexpressionoff}) and (\ref{expsdyna}), the integral corresponding to the third term in Eq. (\ref{inequalityslwithfunctionf}) is negative and can therefore be disregarded without compromising the validity of the inequality. By omitting this integral, the resulting simplified inequality aligns with the generalized covariant entropy bound.

\section{Connection with the perturbation method}

Perturbations induced by matter fields in spacetime are commonly employed to examine the linearized formulation of the second law of black hole thermodynamics in quasi-static processes. Assuming a regular bifurcation surface in spacetime, the Boost Symmetry approach has been utilized to derive expressions for black hole entropy that satisfy the second law in the linear-order approximation of perturbations \cite{Wall:2015raa}. For an arbitrary tensor $X_{a_1 \cdots a_m b_1 \cdots b_n}$, dimensional analysis establishes a relationship between the tensor contracted with $m$ null vectors $k^a$ and $n$ null vectors $l^a$, and the affine parameter $u$, which can be expressed as
\begin{equation}\label{boostsymmetry}
	\begin{split}
		k^{a_1} \cdots k^{a_m} l^{b_1} \cdots l^{b_n} X_{a_1 \cdots a_m b_1 \cdots b_n} = X_{u_1 \cdots u_m z_1 \cdots z_n} \propto \frac{1}{u^{m - n}}\,.
	\end{split}
\end{equation}
The relationship in Eq. (\ref{boostsymmetry}) indicates that the quantity $X_{u_1 \cdots u_m z_1 \cdots z_n}$ diverges near the bifurcation surface (as $u \to 0$) when $m > n$. This divergence implies that the regularity condition of the bifurcation surface precludes the presence of such a quantity in the background spacetime. From the perspective of perturbation theory, quantities present in the background spacetime are classified as zeroth-order terms in the perturbative expansion, while $X_{u_1 \cdots u_m z_1 \cdots z_n}$ for $m > n$ is considered a first-order quantity. In summary, Boost Symmetry can be succinctly stated as follows: any tensor with more contractions involving $k^a$ than $l^a$ is a first-order perturbative quantity, whereas any tensor with an equal or greater number of contractions involving $l^a$ compared to $k^a$ is a background or zeroth-order perturbative quantity. Based on the concept of Boost Symmetry, the entropy density $s_{\text{dyna}}$ is simplified as
\begin{equation}\label{sdyna}
\begin{split}
s_{\text{dyna}} \simeq 0,,
\end{split}
\end{equation}
where the symbol $\simeq$ denotes equality to linear-order in the perturbative approximation. This result implies that, in accordance with the second law, the black hole entropy reduces to the Wald entropy within the perturbative framework.

\section{Discussions and Conclusions}

In Einstein-Maxwell gravity with generalized quadratic corrections, the correction terms are interpreted as first-order quantum corrections derived from the low-energy effective theory of quantum gravity. These corrections impose a universal constraint, requiring that all physical quantities include only linear-order quantum corrections, with higher-order corrections considered negligible. Since the Wald entropy typically satisfies the first law of black hole thermodynamics, the black hole entropy density, under this constraint, is expressed as the Wald entropy density with an additional correction term, $s_{\text{dyna}}$, which accounts for the dynamic aspects of black hole thermodynamic evolution. This corrected formulation of the Wald entropy ensures compliance with the second law of black hole thermodynamics.

Since astrophysical phenomena are inherently non-quasi-static processes, verifying the validity of the second law of black hole thermodynamics requires that the black hole spacetime asymptotically approaches a stable state after the dynamical evolution, without imposing additional constraints on the spacetime background. When generalized quadratic corrections are treated as perturbations to the Einstein-Maxwell theory, the corrected entropy density $s_{\text{dyna}}$ is derived in Eq. (\ref{expsdyna}) within the framework of effective gravitational theory. As shown in Eq. (\ref{finalexpressionoff}), $s_{\text{dyna}}$ emerges solely during the thermodynamic evolution of the black hole. Once the black hole returns to thermodynamic equilibrium, the black hole entropy reduces to the Wald entropy. Therefore, the Wald entropy, incorporating $s_{\text{dyna}}$, satisfies both the first and second laws of black hole thermodynamics. In contrast to the Iyer-Wald entropy and Dong-Wald entropies, this formulation can describe black hole entropy in non-quasi-static processes. This work establishes a novel framework for exploring the generalized second law of black hole thermodynamics in non-static processes within the context of modified gravity.

The derived expression for the function $\mathcal{F} \left(u,, x \right)$ inherently validates the generalized covariant entropy bound. This bound asserts that, for any thermodynamic system (whether or not it involves a black hole) enclosed within a spatial hypersurface, the entropy on the boundary of the hypersurface should satisfy the second law of thermodynamics during finite dynamical evolution driven by the entropy flux $S_L$. The entropy density $s_{\text{dyna}}$, derived from a non-quasi-static perspective, naturally satisfies the conditions imposed by this entropy bound. This result further reinforces the validity of the derived black hole entropy expression in non-quasi-static scenarios, confirming its applicability to finite thermodynamic evolution processes.

From a perturbative perspective, the black hole entropy that satisfies the linearized second law during quasi-static processes is identified as the Wald entropy using the Boost Symmetry method \cite{Wang:2021zyt}. Within this framework, the entropy density $s_{\text{dyna}}$ does not contribute to the black hole entropy. This implies that the entropy derived through the non-quasi-static method, which satisfies the second law of black hole thermodynamics, provides a comprehensive formulation of black hole entropy associated with thermodynamic evolution, while also encompassing the results from the perturbative approach.

\section*{Acknowledgement}
X.-Y. W. is supported by the National Natural Science Foundation of China with Grant No. 12105015, the Guangdong Basic and Applied Basic Research Foundation with Grant No. 2023A1515012737, and the Talents Introduction Foundation of Beijing Normal University with Grant No. 111032109. J. J. is supported by the National Natural Science Foundation of China with Grant No. 12205014, the Guangdong Basic and Applied Research Foundation with Grant No. 2021A1515110913, and the Talents Introduction Foundation of Beijing Normal University with Grant No. 310432102.

\appendix

\section*{Appendices}

\section*{Appendix A: Preparation for deriving the expression of the function $\mathcal{F} \left(u\,, x \right)$}

We aim to investigate the second law of black hole thermodynamics in non-quasi-static processes within the Einstein-Maxwell gravitational theory with generalized quadratic corrections and to derive a general expression for black hole entropy that satisfies the second law during this thermodynamic evolution. Considering the gravitational theory that includes only first-order perturbations from quantum corrections, the Lagrangian of the gravitational theory, as employed in the study of the second law of black hole thermodynamics within this framework, can be expressed as
\begin{equation}\label{apusedlagrangian}
	\boldsymbol{\mathcal{L}} =\boldsymbol{\mathcal{L}}_{\text{EM}} + \boldsymbol{\mathcal{L}}_{\text{gem}} + \boldsymbol{\mathcal{L}}_{\text{mt}}\,,
\end{equation}
where 
\begin{equation}
	\boldsymbol{\mathcal{L}}_{\text{EM}} = \frac{1}{16 \pi} \left(R - F_{ab}F^{ab} \right) \boldsymbol{\epsilon}
\end{equation}
and 
\begin{equation}\label{apdeltalgem}
	\begin{split}
		\boldsymbol{\mathcal{L}}_{\text{gem}} = & \frac{1}{16 \pi} \left(a_1 R F_{ab} F^{ab} + a_2 R_{ab} F^{ac} F^{b}_{\ c} + a_3 R_{abcd} F^{ab} F^{cd} \right. \\
		&\left. + a_4 F_{ab} F^{ab} F_{cd} F^{cd} + a_5 F_{ab} F^{bc} F_{cd} F^{da}\right) \boldsymbol{\epsilon}\,.
	\end{split}
\end{equation}
The third component, $\boldsymbol{\mathcal{L}}_{\text{mt}}$, in Eq. (\ref{apusedlagrangian}) corresponds to the Lagrangian of additional matter fields within the gravitational theory. From the Lagrangian in Eq. (\ref{apusedlagrangian}), the equation of motion for the gravitational theory can be expressed as
\begin{equation}\label{apequationofmotion}
	E_{ab} = 8 \pi T_{ab}\,.
\end{equation}
The left-hand side of the equation of motion in Eq. (\ref{apequationofmotion}) is
\begin{equation}\label{apexpresseab}
	E_{ab} = G_{ab} + \frac{1}{2} \sum_{i = 1}^{3} a_i H_{ab}^{(i)}\,,
\end{equation}
where $G_{ab}$ is the Einstein tensor, which can be expanded as $G_{ab} = R_{ab} - 1 / 2 R g_{ab}$. 
The explicit expressions for the three quantities $H^{(i)}_{ab}$, where $i = 1\,, 2\,, 3$, are obtained as
\begin{equation}\label{hkki}
	\begin{split}
		H_{ab}^{(1)} = & 2 R_{ab} F_{cd} F^{cd} - 4 R F_{a}^{\ c} F_{bc} - g_{ab} R F_{cd} F^{cd} - 2 F^{cd} \nabla_a \nabla_b F_{cd} \\
		& - 4 \nabla_a  F^{cd} \nabla_b F_{cd} - 2 F^{cd} \nabla_b \nabla_a F_{cd} + 4 g_{ab} F^{cd} \nabla_e \nabla^e F_{cd} \\
		& + 4 g_{ab} \nabla_e F_{cd} \nabla^e F^{cd}\,, \\
		H_{ab}^{(2)} = & \nabla_a F_b^{\ c} \nabla_d F_c^{\ d} - 2 R_{cd} F_{a}^{\ c} F_{b}^{\ d} -  g_{ab} F^{cd} R_{de} F_{c}^{\ e} +  \nabla_b F_a^{\ c} \nabla_d F_c^{\ d} \\
		& + F^{cd} \nabla_d \nabla_a F_{bc} +  F_b^{\ c} \nabla_d \nabla_a F_c^{\ d} +  F^{cd} \nabla_d \nabla_b F_{ac} + F_a^{\ c} \nabla_d \nabla_b F_c^{\ d} \\
		& + F_b^{\ c} \nabla_d \nabla^d F_{ac} +  F_a^{\ c} \nabla_d \nabla^d F_{bc} +  g_{ab} F^{cd} \nabla_d \nabla_e F_{c}^{\ e} +  \nabla_b F_{cd} \nabla^d F^{\ c}_a \\
		& +  \nabla_a F_{cd} \nabla^d F_b^{\ c} -  g_{ab} \nabla_c F^{cd} \nabla_e F_d^{\ e} +  g_{ab} F^{cd} \nabla_e \nabla_d F_c^{\ e} + g_{ab} \nabla_d F_{ce} \nabla^e F^{cd} \\
		& + 2 \nabla_d F_{bc} \nabla^d F_a^{\ c}\,, \\
		H_{ab}^{(3)} = & 2 F_{b}^{\ c} \nabla_c \nabla_d F_{a}^{\ d} - R_{acde} F_{b}^{\ c} F^{de} - F_{a}^{\ c} F^{de} R_{bcde} - g_{ab} R_{cdef} F^{cd} F^{ef} \\
		& + 2 F_a^{\ c} \nabla_c \nabla_d F_b^{\ d}+ 4 \nabla_c F_a^{\ c} \nabla_d F_{b}^{\ d} + 2 F_b^{\ c} \nabla_d \nabla_c F_a^{\ d} + 2 F_a^{\ c} \nabla_d \nabla_c F_b^{\ d} \\
		& + 4 \nabla_c F_{bd} \nabla^d F_{a}^{\ c}\,.
	\end{split}
\end{equation}
The right-hand side of Eq. (\ref{apequationofmotion}) represents the stress-energy tensor, which can be expressed as 
\begin{equation}\label{aptotalstressenergy}
	\begin{split}
		T_{ab} = T^{\text{em}}_{ab} + T^{\text{mt}}_{ab}\,,
	\end{split}
\end{equation}
where
\begin{equation}\label{apstentensorem}
	\begin{split}
		T^{\text{em}}_{ab} = \frac{1}{4 \pi} \left(F_{ac} F_{b}^{\ c} - \frac{1}{4} g_{ab} F_{cd} F^{cd} \right)\,.
	\end{split}
\end{equation}
is the stress-energy tensor of the electromagnetic field, and $T^{\text{mt}}_{ab}$ denotes the stress-energy tensor of the matter fields in spacetime.

The primary objective is to derive a general expression of black hole entropy that adheres to the second law of black hole thermodynamics within the context of Einstein-Maxwell gravity with generalized quadratic corrections. Before addressing the second law of black hole thermodynamics, it is necessary to first consider the $D$-dimensional spacetime relevant to the gravitational theory. In this spacetime, an appropriate null hypersurface $L$ is chosen, with its tangent vector defined as $k^a = \left(\partial / \partial u \right)^a$, where $u$ serves as an affine parameter on $L$. An additional null vector $l^a = \left(\partial / \partial z \right)^a$ is introduced, with $z$ labeling the hypersurfaces generated by $l^a$. The $(D-2)$-dimensional compact cross-section of $L$, denoted as $B(u)$, represents a spatial hypersurface. The null hypersurface $L$ is bounded by $B(0)$ and $B(1)$. The spatial coordinates $x = \left(x^1, \ldots, x^{D-2} \right)$ are introduced on $B(u)$. Furthermore, the Gaussian null coordinates $\left\{u, z, x \right\}$ provide a natural framework to describe the spacetime in the vicinity of $L$. The line element of the spacetime in the coordinates is expressed as
\begin{equation}\label{apgaussiannullcoordinates}
	ds^2 \left(\lambda \right) = 2 \left(dz + z^2 \alpha du + z \beta_i dx^i \right) du + \gamma_{ij} dx^i dx^j\,,
\end{equation}
where the coefficients $\alpha$, $\beta_i$, and $\gamma_{ij}$ are all functions of $\left\{u, z, x \right\}$, and the indices $i\,, j\,, k\,, \cdots $ correspond to the spatial components of tensors. From the expression of the line element of the Gaussian null coordinates in Eq. (\ref{apgaussiannullcoordinates}), the inverse form of the metric can be obtained as 
\begin{equation}\label{apinversemetric}
	\begin{split}
		g^{\mu \nu} & = \left(
		\begin{array}{ccc}
		0 & 1 & 0 \\
		1 & z^2 \left(\beta^2 - 2 \alpha \right) & - z \beta^j \\
		0 & - z \beta^i & \gamma^{ij} \\
		\end{array}
		\right)\,.
	\end{split}
\end{equation}

In a state of thermodynamic equilibrium, the entropy of black hole should satisfy the first law of black hole thermodynamics. For a general diffeomorphism-invariant gravitational theory, the entropy of black hole that adheres to the first law of black hole thermodynamics can generally be formulated as Wald entropy. The definition of the Wald entropy can be expressed as 
\begin{equation}
	\begin{split}
		S_{\text{W}} = \int_B d^{D-2} x \sqrt{h} s_{\text{W}}\,,
	\end{split}
\end{equation}
where $s_{\text{W}}$ denotes the density of the Wald entropy, which is expressed as 
\begin{equation}
	\begin{split}
		s_{\text{W}} = - 2 \pi \frac{\partial \mathcal{L}}{\partial R_{abcd}} \hat{\boldsymbol{\epsilon}}_{ab} \hat{\boldsymbol{\epsilon}}_{cd}\,.
	\end{split}
\end{equation}
Based on the definition, for Einstein-Maxwell gravity with generalized quadratic corrections, the density of the Wald entropy in the Gaussian null coordinate can be obtained as 
\begin{equation}\label{apwaldentropydensity}
	\begin{split}
		s_{\text{W}} = & 1 + a_1 \left(- 2 F_{uz} F_{uz} - 4 \gamma^{ij} F_{ui} F_{zj} + \gamma^{ij} \gamma^{kl} F_{ik} F_{jl} \right) - a_2 \left(F_{uz} F_{uz} + \gamma^{ij} F_{ui} F_{zj} \right)\\
		& - a_3 \left(2 F_{uz} F_{uz} \right)\,,
	\end{split}
\end{equation}
where the index $u$ represents the contraction of tensor indices with the null vector $k^a$, and the index $z$ denotes the contraction of tensor indices with the null vector $l^a$. However, the Wald entropy alone is insufficient to fully characterize the entropy of a black hole undergoing dynamical thermodynamic evolution. This limitation implies that Wald entropy does not satisfy the requirements of the second law of black hole thermodynamics. To address this, additional terms accounting for the entropy of dynamical black holes are incorporated into the expression of the Wald entropy. The total entropy of dynamically evolving black holes can be formally expressed as
\begin{equation}\label{aptotalentropydensity}
	\begin{split}
		s_{\text{bh}} = s_{\text{W}} + s_{\text{dyna}}\,,
	\end{split}
\end{equation}
where $s_{\text{dyna}}$ corresponds to the entropy of black holes during thermodynamic evolution. When the total entropy density $s_{\text{bh}}$ of the black hole satisfies the second law of black hole thermodynamics, deriving the precise expression for the black hole entropy in accordance with this law requires determining the explicit form of $s_{\text{dyna}}$. 

The variation of black hole entropy along the null hypersurface $L$ during a thermodynamic evolution process is characterized by the generalized expansion, which is mathematically defined as
\begin{equation}\label{apdefgeneralexpansion}
	\begin{split}
		\Theta \left(u\,, x \right) = \frac{1}{\sqrt{\gamma \left(u\,, x \right)}} \partial_u \left[\sqrt{\gamma \left(u\,, x \right)} s_{\text{bh}} \left(u\,, x \right) \right]\,,
	\end{split}
\end{equation}
where $\gamma(u\,, x)$ is the determinant of the induced metric on the cross-section $B(u)$. The generalized expansion quantifies the entropy change per unit area on the cross-section of the null hypersurface $L$ during the thermodynamic process. When investigating the second law of black hole thermodynamics in the Einstein-Maxwell gravitational theory with generalized quadratic corrections, it is essential to consider a thermodynamic evolution process in which the black hole transitions between thermodynamic equilibrium states. During this process, the black hole entropy should continuously increase, as required by the second law of black hole thermodynamics. At the conclusion of the evolution, the entropy of the black hole converges to a definite value upon reaching a new thermodynamic equilibrium state. Consequently, this imposes a requirement on the generalized expansion, which characterizes the variation of black hole entropy along $L$, to satisfy specific conditions
\begin{equation}\label{appartialuthetale0}
	\begin{split}
		\partial_u \Theta \left(u\,, x \right) \le 0\,.
	\end{split}
\end{equation}
throughout the thermodynamic evolution process. The evolution of the generalized expansion along the null hypersurface $L$ is governed by the Raychaudhuri equation in the spacetime corresponding to the Einstein-Maxwell gravity with generalized quadratic corrections. In the Gaussian null coordinate system, this equation is expressed as
\begin{equation}\label{aprayequation}
	\begin{split}
		\partial_u \Theta \left(u\,, x \right) = - 8 \pi T^{\text{mt}}_{uu} + \mathcal{F} \left(u\,, x \right)\,,
	\end{split}
\end{equation} 
where
\begin{equation}\label{aporiginalfunctionF}
	\begin{split}
		\mathcal{F} \left(u\,, x \right) = E_{uu} - 8 \pi T^{\text{em}}_{uu} + \partial_u \Theta \left(u\,, x \right)\,.
	\end{split}
\end{equation}
In the Raychaudhuri equation, $T^{\text{em}}_{uu}$ and $T^{\text{mt}}_{uu}$ represent the stress-energy tensor of the electromagnetic fields and the stress-energy tensor of the matter fields in the spacetime, contracted with the tangent vector $k^a$, respectively.

The first term, $E_{uu}$, in Eq. (\ref{aporiginalfunctionF}) originates from the contraction of the left-hand side of the equation of motion in Eq. (\ref{apexpresseab}) with the tangent vector $k^a$, and can be expressed as
\begin{equation}\label{apexphab}
	E_{uu} = R_{uu} + \frac{1}{2} \sum_{i = 1}^{3} a_i H_{uu}^{(i)}\,,
\end{equation} 
where $R_{uu}$ is the Ricci tensor contracting with the vector $k^a$, and the expressions of $H_{uu}^{(i)}$ $\left(i = 1\,, 2\,, 3\right)$ are given as 
\begin{equation}\label{ohkk1}
	\begin{split}
		H_{uu}^{(1)} = & 2 k^a k^b R_{ab} F_{cd} F^{cd} - 4 R k^a k^b F_{a}^{\ c} F_{bc} - 2 k^a k^b \nabla_a \nabla_b \left(F^{cd} F_{cd} \right)\,,
	\end{split}
\end{equation}
\begin{equation}\label{ohkk2}
	\begin{split}
		H_{uu}^{(2)} = & k^a k^b \nabla_a F_{b}^{\ c} \nabla_d F_{c}^{\ d} - 2 k^a k^b F_{a}^{\ c} F_b^{\ d} R_{cd} + k^a k^b \nabla_b F_{a}^{\ c} \nabla_d F_{c}^{\ d} \\
		& + k^a k^b F^{cd} \nabla_d \nabla_a F_{bc} + k^a k^b F_{b}^{\ c} \nabla_d \nabla_a F_{c}^{\ d} + k^a k^b F^{cd} \nabla_d \nabla_b F_{ac} \\
		& + k^a k^b F_{a}^{\ c} \nabla_d \nabla_b F_{c}^{\ d} + k^a k^b F_{b}^{\ c} \nabla_d \nabla^d F_{ac} + k^a k^b F_{a}^{\ c} \nabla_d \nabla^d F_{bc}\\
		& + k^a k^b \nabla_b F_{cd} \nabla^d F_{a}^{\ c} + k^a k^b \nabla_a F_{cd} \nabla^d F_{b}^{\ c} + 2 k^a k^b \nabla_d F_{bc} \nabla^d F_{a}^{\ c}\,,
	\end{split}
\end{equation}
\begin{equation}\label{ohkk3}
	\begin{split}
		H_{uu}^{(3)} = & 2 k^a k^b F_{b}^{\ c} \nabla_c \nabla_d F_{a}^{\ d} - k^a k^b R_{acde} F_{b}^{\ c} F^{de} - k^a k^b R_{bcde} F_{a}^{\ c} F^{de} \\
		& + 2 k^a k^b F_{a}^{\ c} \nabla_c \nabla_d F_{b}^{\ d} + 4 k^a k^b \nabla_c F_{a}^{\ c} \nabla_d F_{b}^{\ d} + 2 k^a k^b F_{b}^{\ c} \nabla_d \nabla_c F_{a}^{\ d}\\
		& + 2 k^a k^b F_{a} ^{\ c} \nabla_d \nabla_c F_{b}^{\ d} + 4 k^a k^b \nabla_c F_{bd} \nabla^d F_{a}^{\ c}\,.
	\end{split}
\end{equation}
The stress-energy tensors of the matter and electromagnetic fields in the spacetime are assumed to satisfy the null energy condition. This implies that the total stress-energy tensor should satisfy the condition $T_{uu} = T^{\text{em}}_{uu} + T^{\text{mt}}_{uu} \ge 0$. Therefore, based on Eqs. (\ref{aprayequation}) and (\ref{aporiginalfunctionF}), if the condition in Eq. (\ref{appartialuthetale0}) is satisfied, the function $\mathcal{F} \left(u\,, x \right)$ should be derived and further satisfy the condition $\mathcal{F} \left(u\,, x \right) \le 0$. To derive the explicit expression for the function $\mathcal{F} \left(u\,, x \right) \le 0$ that fulfills this inequality, the specific expression of the Ricci tensor $R_{uu}$ in the Gaussian null coordinates should be utilized. Furthermore, as the three terms $H^{(i)}_{uu}$ $\left(i = 1\,, 2\,, 3 \right)$ incorporate covariant derivative operators, these operators should be further expanded during the computation of the explicit expression of the function $\mathcal{F} \left(u\,, x \right)$. The expressions for the Christoffel symbols should be used during the expansion of the covariant derivative operators, while the Riemann curvature tensor is also required when the order of two covariant derivative operators is interchanged. Therefore, it is essential to determine the component forms of the Christoffel symbols, the Riemann curvature tensor, and the Ricci tensor in the Gaussian null coordinate system for calculating the expression for $\mathcal{F} \left(u\,, x \right)$.

According to the definition of the Christoffel symbol
\begin{equation}
	\Gamma^{\sigma}_{\ \mu \nu} = \frac{1}{2} g^{\sigma \rho} \left(\partial_\nu g_{\rho \mu} + \partial_\mu g_{\nu \rho} - \partial_\rho g_{\mu \nu} \right)\,,
\end{equation}
while using the expression of the line element in Eq. (\ref{apgaussiannullcoordinates}) and the the inverse form of the metric in Eq. (\ref{apinversemetric}), all components of the Christoffel symbol in the Gaussian null coordinate system are given as
\begin{equation}\label{allcomponentchrissym}
	\begin{split}
		\Gamma^{u}_{\ uu} = & - z^2 \partial_z \alpha - 2 z \alpha\,,\\
		\Gamma^{u}_{\ u z} = & \Gamma^{u}_{\ z u} = 0\,,\\
		\Gamma^{u}_{\ u i} = & \Gamma^{u}_{\ i u} = - \frac{1}{2} \beta_i - \frac{1}{2} z \partial_z \beta_i\,,\\
		\Gamma^{u}_{\ z z} = & 0\,,\\
		\Gamma^{u}_{\ z i} = & \Gamma^{u}_{\ i z} = 0\,,\\
		\Gamma^{u}_{\ i j} = & \Gamma^{u}_{\ j i} = - \frac{1}{2} \partial_z \gamma_{ij}\,,\\
		\Gamma^{z}_{\ z z} = & 0\,,\\
		\Gamma^{z}_{\ z u} = & \Gamma^{z}_{\ u z} = z^2 \partial_z \alpha + 2 z \alpha - \frac{1}{2} z \beta^i \beta_i - \frac{1}{2} z^2 \beta^i \partial_z \beta_i\,,\\
		\Gamma^{z}_{\ z i} = & \Gamma^{z}_{\ i z} = \frac{1}{2} \beta_i + \frac{1}{2} z \partial_z \beta_i - \frac{1}{2} \left(z \beta^j \right) \partial_z \gamma_{i j}\,,\\
		\Gamma^{z}_{\ u u} = & z^2 \partial_u \alpha - z^2 \left(\beta^2 - 2 \alpha \right) \left(z^2 \partial_z \alpha + 2 z \alpha \right) - z \beta^i \left(z \partial_u \beta_i - z^2 \partial_i \alpha \right)\,,\\
		\Gamma^{z}_{\ u i} = & \Gamma^{z}_{\ i u} = z^2 \partial_i \alpha - \frac{1}{2} z^2 \left(\beta^2 - 2 \alpha \right) \left(\beta_i + z \partial_z \beta_i \right) - \frac{1}{2} z \beta^j \left(z \partial_i \beta_j + \partial_u \gamma_{ij} - z \partial_j \beta_i \right)\,,\\
		\Gamma^{z}_{\ i j} = & \frac{1}{2} \left(z \partial_j \beta_i + z \partial_i \beta_j - \partial_u \gamma_{ij} \right) - \frac{1}{2} z^2 \left(\beta^2 - 2 \alpha \right) \left(\partial_z \gamma_{i j} \right)\\
		& - \frac{1}{2} \left(z \beta^k \right) \left(\partial_j \gamma_{k i} + \partial_i \gamma_{j k} - \partial_k \gamma_{i j} \right)\,,\\
		\Gamma^{i}_{\ u u} = & \left(z \beta^i \right) \left(z^2 \partial_z \alpha + 2 z \alpha \right) + \gamma^{i j} \left(z \partial_u \beta_j - z^2 \partial_j \alpha \right)\,,\\
		\Gamma^{i}_{\ u z} = & \Gamma^{i}_{\ z u} = \frac{1}{2} \gamma^{i j} \left(\beta_j + z \partial_z \beta_j \right)\,,\\
		\Gamma^{i}_{\ z z} = & 0\,,\\
		\Gamma^{i}_{\ u j} = & \Gamma^{i}_{\ j u} = \frac{1}{2} \left(z \beta^i \right) \left(\beta_j + z \partial_z \beta_j \right) + \frac{1}{2} \gamma^{i k} \left(z \partial_j \beta_k + \partial_u \gamma_{j k} - z \partial_k \beta_j \right)\,,\\
		\Gamma^{i}_{\ z j} = & \Gamma^{i}_{\ j z} = \frac{1}{2} \gamma^{i k} \partial_z \gamma_{j k}\,,\\
		\Gamma^{i}_{\ j k} = & \Gamma^{i}_{\ k j} = \frac{1}{2} \left(z \beta^i \right) \partial_z \gamma_{j k} + \hat{\Gamma}^{i}_{\ j k}\,.
	\end{split}
\end{equation}
In analyzing the second law of black hole thermodynamics and deriving an expression for black hole entropy consistent with the second law, the primary focus is on the evolutionary properties of the cross-section on the null hypersurface $L$. Therefore, the derivation of the entropy expression requires restricting the analysis to properties confined to the hypersurface $L$. This restriction implies that the evaluation of physical quantities within the spacetime framework is necessary only on the hypersurface $L$, specifically under the condition $z = 0$ in Gaussian null coordinates. Based on the explicit expressions of all components of the Christoffel symbols provided in Eq. (\ref{allcomponentchrissym}), the non-zero components of the Christoffel symbols under the condition $z = 0$ in Gaussian null coordinates can be expressed as
\begin{equation}\label{allcomponentchrissymz0}
	\begin{split}
		\Gamma^{u}_{\ u i} = & \Gamma^{u}_{\ i u} = - \frac{1}{2} \beta_i\,, \quad \Gamma^{u}_{\ i j} = \Gamma^{u}_{\ j i} = - \frac{1}{2} \partial_z \gamma_{ij}\,, \quad \Gamma^{z}_{\ z i} = \Gamma^{z}_{\ i z} = \frac{1}{2} \beta_i\,,\\
		\Gamma^{z}_{\ i j} = & - \frac{1}{2} \partial_u \gamma_{ij}\,, \quad \Gamma^{i}_{\ u z} = \frac{1}{2} \gamma^{i j} \beta_j\,,\quad \Gamma^{i}_{\ u j} = \Gamma^{i}_{\ j u} = \frac{1}{2} \gamma^{i k} \left(\partial_u \gamma_{j k} \right)\,,\\
		\Gamma^{i}_{\ z j} = & \Gamma^{i}_{\ j z} = \frac{1}{2} \gamma^{i k} \partial_z \gamma_{j k}\,, \quad \Gamma^{i}_{\ j k} = \Gamma^{i}_{\ k j} = \hat{\Gamma}^{i}_{\ j k}\,.
	\end{split}
\end{equation}
On the other hand, all non-zero components of the metric and its inverted version can be directly expressed as 
\begin{equation}\label{nonzerocomponentmetric}
	\begin{split}
		& g_{uz} = g_{zu} = 1\,, \qquad g_{ij} = \gamma_{ij}\,,\\
		& g^{uz} = g^{zu} = 1\,, \qquad g^{ij} = \gamma^{ij}\,.
	\end{split}
\end{equation}
According to the definition of the Riemann tensor
\begin{equation}
	R_{\mu \nu \sigma}^{\ \ \ \ \rho} = \partial_\nu \Gamma^{\rho}_{\ \mu \sigma} - \partial_\mu \Gamma^{\rho}_{\ \nu \sigma} + \Gamma^{\lambda}_{\ \sigma \mu} \Gamma^{\rho}_{\ \nu \lambda} - \Gamma^{\lambda}_{\ \sigma \nu} \Gamma^{\rho}_{\ \mu \lambda}
\end{equation}
and utilizing the expression of the components of the Christoffel symbol in Eq. (\ref{allcomponentchrissym}), all independent components of the Riemann curvature tensor with $z = 0$ are derived as
\begin{equation}\label{allcomponentsriemanntensor}
	\begin{split}
		R_{uzuz} = & - 2 \alpha + \frac{1}{4} \beta^2\,,\\
		R_{uzui} = & \frac{1}{2} \partial_u \beta_i + \frac{1}{4} \beta^j \partial_u \gamma_{ij}\,,\\
		R_{uzij} = & \frac{1}{2} \gamma_{jk} \left(\partial_z \gamma^{kl} \right) \left(\partial_u \gamma_{il} \right) + \frac{1}{2} \partial_i \beta_j - \frac{1}{2} \partial_j \beta_i - \frac{1}{2} \gamma_{jk} \left(\partial_u \gamma^{kl} \right) \partial_z \gamma_{il}\\
		& + \frac{1}{4} \gamma^{lm} \left(\partial_u \gamma_{im} \right) \partial_z \gamma_{lj} - \frac{1}{4} \gamma^{lm} \left(\partial_z \gamma_{im} \right) \left(\partial_u \gamma_{lj} \right)\,,\\
		R_{uiuj} = & - \frac{1}{2} \gamma_{j k} \left(\partial_u \gamma^{kl} \right) \left(\partial_u \gamma_{il} \right) - \frac{1}{2} \partial_u^2 \gamma_{i j} - \frac{1}{4} \gamma^{lm} \left(\partial_u \gamma_{im} \right) \left(\partial_u \gamma_{lj} \right)\,,\\
		R_{uizj} = & \frac{1}{2} \gamma_{j k} \partial_i \beta^k - \frac{1}{2} \gamma_{j k} \left(\partial_u \gamma^{kl} \right) \partial_z \gamma_{il} - \frac{1}{2} \partial_u \partial_z \gamma_{i j} + \frac{1}{2} \gamma_{j k} \beta^l \hat{\Gamma}^{k}_{\ i l} - \frac{1}{4} \beta_i \beta_j \\
		& - \frac{1}{4} \gamma^{l m} \left(\partial_z \gamma_{i m} \right) \partial_u \gamma_{l j}\,,\\
		R_{uijk} = & \frac{1}{2} \gamma_{k l} \left(\partial_i \gamma^{l m} \right) \left(\partial_u \gamma_{j m} \right) + \frac{1}{2} \partial_i \partial_u \gamma_{j k} - \gamma_{k l} \partial_u \hat{\Gamma}^{l}_{\ i j} - \frac{1}{4} \beta_j \partial_u \gamma_{i k}\\
		& + \frac{1}{2} \hat{\Gamma}^{l}_{\ i m} \gamma_{k l} \gamma^{m n} \partial_u \gamma_{j n} + \frac{1}{4} \beta_k \partial_u \gamma_{ji} - \frac{1}{2} \hat{\Gamma}^{m}_{\ j i} \partial_u \gamma_{m k}\,,\\
		R_{zuzi} = & - \frac{1}{2}\partial_z \beta_i + \frac{1}{4} \beta^j \partial_z \gamma_{ij}\,,\\
		R_{zizj} = & - \frac{1}{2} \gamma_{j k} \left(\partial_z \gamma^{k m} \right) \partial_z \gamma_{i m} - \frac{1}{2} \partial_z^2 \gamma_{i j} - \frac{1}{4} \gamma^{l m} \left(\partial_z \gamma_{i m} \right) \partial_z \gamma_{l j}\,,\\
		R_{zijk} = & \frac{1}{2} \gamma_{k l} \left(\partial_i \gamma^{l m} \right) \partial_z \gamma_{j m} + \frac{1}{2} \partial_i \partial_z \gamma_{j k} - \gamma_{k l} \partial_z \hat{\Gamma}^{l}_{\ i j} + \frac{1}{4} \beta_j \partial_z \gamma_{i k}\\
		& + \frac{1}{2} \gamma_{kl} \gamma^{mn} \hat{\Gamma}^{l}_{\ im} \partial_z \gamma_{jn} - \frac{1}{2} \hat{\Gamma}^{m}_{\ j i} \partial_z \gamma_{m k}\,,\\
		R_{ijkl} = & \gamma_{l m} \partial_j \hat{\Gamma}^{m}_{\ i k} - \gamma_{l m} \partial_i \hat{\Gamma}^{m}_{\ j k} - \frac{1}{4} \left(\partial_z \gamma_{k i} \right) \partial_u \gamma_{j l} - \frac{1}{4} \left(\partial_u \gamma_{k i} \right) \partial_z \gamma_{j l} + \gamma_{l m} \hat{\Gamma}^{n}_{\ k i} \hat{\Gamma}^{m}_{\ j n} \\
		& + \frac{1}{4} \left(\partial_z \gamma_{k j} \right) \partial_u \gamma_{i l} + \frac{1}{4} \left(\partial_u \gamma_{k j} \right) \partial_z \gamma_{i l} - \gamma_{l m} \hat{\Gamma}^{n}_{\ k j} \hat{\Gamma}^{m}_{\ i n}\,.
	\end{split}
\end{equation}
Furthermore, all independent components of Ricci scalar with $z = 0$ can be given as 
\begin{equation}\label{allcomponentsricciscalar}
	\begin{split}
		R_{uu} = & - \frac{1}{2} \left(\partial_u \gamma^{ij} \right) \left(\partial_u \gamma_{ij} \right) - \frac{1}{2} \gamma^{ij} \partial_u^2 \gamma_{ij} - \frac{1}{4} \gamma^{ij} \gamma^{kl} \left(\partial_u \gamma_{ik} \right) \left(\partial_u \gamma_{jl} \right)\,,\\
		R_{uz} = & 2 \alpha - \frac{1}{4} \beta^2 + \frac{1}{2} \partial_i \beta^i - \frac{1}{2} \left(\partial_u \gamma^{ij} \right) \partial_z \gamma_{ij} - \frac{1}{2} \gamma^{i j} \partial_u \partial_z \gamma_{i j} + \frac{1}{2} \beta^j \hat{\Gamma}^{i}_{\ ij} - \frac{1}{4} \beta_i \beta^i \\
		& - \frac{1}{4} \gamma^{ij} \gamma^{kl} \left(\partial_z \gamma_{ik} \right) \partial_u \gamma_{jl}\,,\\
		R_{ui} = & - \frac{1}{2} \partial_u \beta_i - \frac{1}{4} \beta^j \partial_u \gamma_{ij} + \frac{1}{2} \left(\partial_j \gamma^{jk} \right) \left(\partial_u \gamma_{ik} \right) + \frac{1}{2} \gamma^{j k} \partial_j \partial_u \gamma_{i k} - \partial_u \hat{\Gamma}^{j}_{\ j i}\\
		& - \frac{1}{4} \gamma^{j k} \beta_i \partial_u \gamma_{j k} + \frac{1}{2} \hat{\Gamma}^{j}_{\ jk} \gamma^{kl} \partial_u \gamma_{il} + \frac{1}{4} \beta^{j} \partial_u \gamma_{ij} - \frac{1}{2} \gamma^{j k} \hat{\Gamma}^{l}_{\ i j} \partial_u \gamma_{lk}\,,\\
		R_{zz} = & - \frac{1}{2} \left(\partial_z \gamma^{ij} \right) \partial_z \gamma_{ij} - \frac{1}{2} \gamma^{ij} \partial_z^2 \gamma_{ij} - \frac{1}{4} \gamma^{ij} \gamma^{kl} \left(\partial_z \gamma_{il} \right) \partial_z \gamma_{kj}\,,\\
		R_{zi} = & \frac{1}{2}\partial_z \beta_i - \frac{1}{4} \beta^j \partial_z \gamma_{ij} + \frac{1}{2} \left(\partial_j \gamma^{jm} \right) \partial_z \gamma_{im} + \frac{1}{2} \gamma^{jk} \partial_j \partial_z \gamma_{ik} - \partial_z \hat{\Gamma}^{j}_{\ ji} + \frac{1}{4} \gamma^{jk} \beta_i \partial_z \gamma_{jk}\\
		& + \frac{1}{2} \gamma^{mn} \hat{\Gamma}^{j}_{\ jm} \partial_z \gamma_{in} - \frac{1}{2} \gamma^{jk} \hat{\Gamma}^{m}_{\ ij} \partial_z \gamma_{mk}\,,\\
		R_{ij} = & \gamma_{j k} \partial_i \beta^k - \gamma_{j k} \left(\partial_u \gamma^{kl} \right) \partial_z \gamma_{il} - \partial_u \partial_z \gamma_{i j} + \gamma_{j k} \beta^l \hat{\Gamma}^{k}_{\ i l} - \frac{1}{2} \beta_i \beta_j \\
		& - \frac{1}{2} \gamma^{l m} \left(\partial_z \gamma_{i m} \right) \partial_u \gamma_{l j} + \partial_k \hat{\Gamma}^{k}_{\ i j} - \partial_i \hat{\Gamma}^{k}_{\ k j} - \frac{1}{4} \gamma^{k l} \left(\partial_z \gamma_{j i} \right) \partial_u \gamma_{k l}\\
		& - \frac{1}{4} \gamma^{k l} \left(\partial_u \gamma_{j i} \right) \partial_z \gamma_{k l} + \hat{\Gamma}^{l}_{\ j i} \hat{\Gamma}^{k}_{\ kl} + \frac{1}{4} \gamma^{k l} \left(\partial_z \gamma_{j k} \right) \partial_u \gamma_{i l} + \frac{1}{4} \gamma^{k l} \left(\partial_u \gamma_{j k} \right) \partial_z \gamma_{i l} \\
		& - \hat{\Gamma}^{l}_{\ j k} \hat{\Gamma}^{k}_{\ il}\,.
	\end{split}
\end{equation}

\section*{Appendix B: Calculation of the expression for the function $\mathcal{F} \left(u\,, x \right)$}

Since all quantum corrections in Einstein-Maxwell gravity with generalized quadratic corrections are treated as linear-order perturbations to the original Einstein-Maxwell gravity within the low energy effective approximation, the coupling constants associated with these corrections are considered small quantities. To verify whether the function $\mathcal{F}(u, x)$ satisfies the condition $\mathcal{F}(u, x) \le 0$ under the linear-order constraints of the quantum corrections, the entropy density $s_{\text{bh}}$ should be further reformulated according to its relationship with the coupling constants. Based on the expression of the density of black hole entropy in Eq. (\ref{apwaldentropydensity}), the black hole density can be rewritten as
\begin{equation}
	\begin{split}
		s_{\text{bh}} = s^{\text{(nc)}}_{\text{bh}} + s^{\text{(cc)}}_{\text{bh}}\,,
	\end{split}
\end{equation}
where $s^{\text{(nc)}}_{\text{bh}} = 1$ represents the black hole entropy independent of the coupling constant, and $s^{\text{(cc)}}_{\text{bh}} = s_{\text{W}} - 1 + s_{\text{dyna}}$ denotes the contribution to the black hole entropy density arising from the coupling constants. Furthermore, according to the definition of the generalized expansion in Eq. (\ref{apdefgeneralexpansion}), the generalized expansion $\Theta \left(u\,, x \right)$ through the re-expressed black hole density can be expressed as
\begin{equation}
	\Theta \left(u\,, x \right) = \Theta^{\text{(nc)}} \left(u\,, x \right) + \Theta^{\text{(cc)}} \left(u\,, x \right)\,.
\end{equation}
Substituting Eqs. (\ref{apstentensorem}) and (\ref{apexphab}) into Eq. (\ref{aporiginalfunctionF}), and utilizing the expression for $R_{uu}$ and the Wald entropy density, the components of the function $\mathcal{F}(u, x)$ that are independent of the coupling constants can be systematically evaluated. The expression for $\mathcal{F}(u, x)$ can be further formulated as
\begin{equation}\label{apformalexpfunctionf}
	\begin{split}
		\mathcal{F} \left(u\,, x \right) = - K^{ij} K_{ij} - 2 F_{ui} F_{u}^{\ i} + \frac{1}{2} \sum_{i = 1}^{3} a_i H_{uu}^{(i)} + \partial_u \Theta^{\text{(cc)}} \left(u\,, x \right)\,,
	\end{split}
\end{equation}
where the first term is derived from $\Theta^{\text{(nc)}} \left(u\,, x \right)$. As elaborated in the main text, deriving the expression for the function $\mathcal{F} \left(u, x \right)$ under the constraints of linear-order quantum corrections in gravitational theory requires the application of a selection criterion. This criterion specifies that contributions from terms containing quantities such as $K_{ij}$, $D_{k} K_{ij}$, $F_{ui}$, or $D_{j} F_{ui}$, when appearing in the third or fourth components of the function $\mathcal{F} \left(u, x \right)$ in Eq. (\ref{apformalexpfunctionf}), are negligible and can be disregarded. Consequently, this criterion facilitates the derivation of the expression of the function $\mathcal{F} \left(u, x \right)$ under the linear-order constraints of the quantum corrections in the following calculation. Moreover, a convenient convention is introduced to streamline the computational process. As indicated in Eq. (\ref{allcomponentchrissym}), the Christoffel symbols with two subscripts $u$, denoted as $\Gamma^{a}_{\ u u}$, are uniformly zero for $a = \{u\,, z\,, i \}$. This implies that any term in the expression of $H_{uu}^{(i)}\,, i = 1\,, 2\,, 3$ containing $\Gamma^{a}_{\ u u}$ is automatically zero. Using an arbitrary tensor $X_a$, the mathematical formulation of this convention is given as
\begin{equation}\label{kkgammaxeq0}
	\begin{split}
		\Gamma^{a}_{\ u u} X_a = 0\,.
	\end{split}
\end{equation}
The result from Eq. (\ref{kkgammaxeq0}) will be directly applied in the derivation of the expression for the function $\mathcal{F} \left(u,, x \right)$.

The computation of the function $\mathcal{F} \left(u,, x \right)$ proceeds in accordance with the established criteria and conventions. Before deriving the expression for the function, it is essential to emphasize that, to obtain the expression for the black hole entropy that satisfies the second law of black hole thermodynamics within the Einstein-Maxwell gravitational theory with generalized quadratic corrections, it is necessary to verify that the integral of the function $\mathcal{F} \left(u,, x \right)$ over the cross-section $B(u)$ satisfies the condition 
\begin{equation}
	\begin{split}
		\int_{B(u)} d^2 x \sqrt{\gamma(u\,, x)} \mathcal{F} \left(u\,, x \right) \le 0\,.
	\end{split}
\end{equation}
In deriving the expression for the function $\mathcal{F} \left(u,, x \right)$ under the constraints of linear-order quantum corrections, the integral sign and induced metric are initially disregarded for computational convenience and to simplify the expression. Once the representation of the function $\mathcal{F} \left(u,, x \right)$ is obtained, the integral sign and induced metric are subsequently reintroduced. Moreover, according to the selection criterion proposed in the above discussion, the terms containing $K_{ij}$, $D_{k} K_{ij}$, $F_{ui}$, or $D_{j} F_{ui}$ in the third and fourth components of the function $\mathcal{F} \left(u\,, x \right)$ contribute less significantly than the first two components. Consequently, these terms have minimal impact on the expression of $\mathcal{F} \left(u\,, x \right)$ under the constraints of linear-order quantum corrections and do not substantially affect the derivation of the black hole entropy expression that satisfies the second law of thermodynamics. Therefore, for the sake of simplification in the subsequent calculations, we will directly neglect these less influential terms.

The process initiates with the evaluation of the three components of the function $\mathcal{F} \left(u\,, x \right)$, denoted as $H^{(i)}_{uu}, \, i = 1, 2, 3$. The calculation starts by deriving the expression for $H_{uu}^{(1)}$ under the constraints of linear-order quantum corrections. Specifically, the third term on the right-hand side of Eq. (\ref{ohkk1}) for $H_{uu}^{(1)}$ is evaluated as
\begin{equation}
	\begin{split}
		- 2 k^a k^b \nabla_a \nabla_b \left(F^{cd} F_{cd} \right) = - 4 k^a k^b \left(\nabla_a \nabla_b F^{cd} \right) F_{cd} - 4 k^a k^b \left(\nabla_b F^{cd} \right) \left(\nabla_a F_{cd} \right)\,.
	\end{split}
\end{equation}
Therefore, the expression of $H_{uu}^{(1)}$ can be further written as 
\begin{equation}\label{rhkk1}
	\begin{split}
		H_{uu}^{(1)} = & - 4 k^a k^b \left(\nabla_a \nabla_b F^{cd} \right) F_{cd} - 4 k^a k^b \left(\nabla_b F^{cd} \right) \left(\nabla_a F_{cd} \right) + 2 k^a k^b R_{ab} F_{cd} F^{cd}\\
		& - 4 R k^a k^b F_{a}^{\ c} F_{bc}\,.
	\end{split}
\end{equation}
In the four terms on the right-hand side of Eq. (\ref{rhkk1}), each term contains repeated indices associated with the spacetime metric. These indices can be expanded using the explicit expressions of the metric in the Gaussian null coordinate system. Following this expansion, the first term on the right-hand side of Eq. (\ref{rhkk1}) can be written as
\begin{equation}\label{huu1firsttermwithmetric}
	\begin{split}
		- 4 k^a k^b \left(\nabla_a \nabla_b F^{cd} \right) F_{cd} = - 4 k^a k^b g^{ce} g^{df} \left(\nabla_a \nabla_b F_{ef} \right) F_{cd}\,.
	\end{split}
\end{equation}
The covariant derivative operators in the expression should be further expanded to include ordinary derivatives and Christoffel symbols. For any tensor $X^{b_1 \cdots b_k}_{\qquad c_1 \cdots c_l}$, the covariant derivative acting on the tensor can be expressed as
\begin{equation}\label{defcovariantderive}
	\begin{split}
		\nabla_a X^{b_1 \cdots b_k}_{\qquad c_1 \cdots c_l} = \partial_a X^{b_1 \cdots b_k}_{\qquad c_1 \cdots c_l} + \sum_{i} \Gamma^{b_{i}}_{\ a d} X^{b_1 \cdots d \cdots b_k}_{\qquad \quad c_1 \cdots c_l} - \sum_{j} \Gamma^{d}_{\ a c_j} X^{b_1 \cdots b_k}_{\qquad c_1 \cdots d \cdots c_l}\,.
	\end{split}
\end{equation}
Based on the expanded expression of the covariant derivation in Eq. (\ref{defcovariantderive}), the expression in Eq. (\ref{huu1firsttermwithmetric}) can be further elaborated as 
\begin{equation}\label{rhkk1first}
	\begin{split}
		& - 4 k^a k^b g^{ce} g^{df} \left(\nabla_a \nabla_b F_{ef} \right) F_{cd}\\
		= & - 4 k^a k^b g^{ce} g^{df} \left(\partial_a \partial_b F_{ef} \right) F_{cd} + 4 k^a k^b g^{ce} g^{df} \Gamma^{g}_{\ ab} \left(\partial_g F_{ef} \right) F_{cd}\\
		& + 4 k^a k^b g^{ce} g^{df} \Gamma^{g}_{\ ae} \left(\partial_b F_{gf} \right) F_{cd} + 4 k^a k^b g^{ce} g^{df} \Gamma^{g}_{\ af} \left(\partial_b F_{eg} \right) F_{cd}\\
		& + 4 k^a k^b g^{ce} g^{df} \left(\partial_a \Gamma^{g}_{\ be} \right) F_{gf} F_{cd} - 4 k^a k^b g^{ce} g^{df} \Gamma^{h}_{\ ab} \Gamma^{g}_{\ he} F_{gf} F_{cd}\\
		& - 4 k^a k^b g^{ce} g^{df} \Gamma^{h}_{\ ae} \Gamma^{g}_{\ bh} F_{gf} F_{cd} + 4 k^a k^b g^{ce} g^{df} \Gamma^{g}_{\ ah} \Gamma^{h}_{\ be} F_{gf} F_{cd}\\
		& + 4 k^a k^b g^{ce} g^{df} \Gamma^{g}_{\ be} \left(\partial_a F_{gf} \right) F_{cd} - 4 k^a k^b g^{ce} g^{df} \Gamma^{g}_{\ be} \Gamma^{h}_{\ ag} F_{hf} F_{cd}\\
		& - 4 k^a k^b g^{ce} g^{df} \Gamma^{g}_{\ be} \Gamma^{h}_{\ af} F_{gh} F_{cd} + 4 k^a k^b g^{ce} g^{df} \left(\partial_a \Gamma^{g}_{\ bf} \right) F_{eg} F_{cd}\\
		& - 4 k^a k^b g^{ce} g^{df} \Gamma^{h}_{\ ab} \Gamma^{g}_{\ hf} F_{eg} F_{cd} - 4 k^a k^b g^{ce} g^{df} \Gamma^{h}_{\ af} \Gamma^{g}_{\ bh} F_{eg} F_{cd}\\
		& + 4 k^a k^b g^{ce} g^{df} \Gamma^{g}_{\ ah} \Gamma^{h}_{\ bf} F_{eg} F_{cd} + 4 k^a k^b g^{ce} g^{df} \Gamma^{g}_{\ bf} \left(\partial_a F_{eg} \right) F_{cd}\\
		& - 4 k^a k^b g^{ce} g^{df} \Gamma^{g}_{\ bf} \Gamma^{h}_{\ ae} F_{hg} F_{cd} - 4 k^a k^b g^{ce} g^{df} \Gamma^{g}_{\ bf} \Gamma^{h}_{\ ag} F_{eh} F_{cd}\,. 
	\end{split}
\end{equation}
Since each term in the results of the above equation involves two formal components of the metric, the explicit expression for the metric in the Gaussian null coordinate system should be substituted to derive the specific expression of the equation. As previously emphasized, the derivation of the black hole entropy expression consistent with the second law of thermodynamics requires focusing exclusively on relevant physical quantities on the null hypersurface $L$ corresponding to the condition $z = 0$ in the Gaussian null coordinate system. Therefore, to streamline subsequent calculations and simplify equation presentation, we will directly use the expressions for the metric and its inverse from Eq. (\ref{nonzerocomponentmetric}), the Christoffel symbols from Eqs. (\ref{allcomponentchrissym}) and (\ref{allcomponentchrissymz0}), the Riemann curvature tensor from Eq. (\ref{allcomponentsriemanntensor}), and the Ricci tensor from Eq. (\ref{allcomponentsricciscalar}). These components will be applied to derive the expressions for each term in the third and fourth components of the function $\mathcal{F} \left(u\,, x \right)$. In the following analysis, non-essential substitution steps will be omitted for clarity, and only the final results will be presented.

The first term of Eq. (\ref{rhkk1first}) is 
\begin{equation}
	\begin{split}
		& - 4 k^a k^b g^{ce} g^{df} \left(\partial_a \partial_b F_{ef} \right) F_{cd}\\
		= & - 4 \left(\partial_u \partial_u F_{zu} \right) F_{uz} - 4 \left(\partial_u \partial_u F_{uz} \right) F_{zu} - 4 \gamma^{ij} \left(\partial_u \partial_u F_{uj} \right) F_{zi}\\
		& - 4 \gamma^{ij} \left(\partial_u \partial_u F_{ju} \right) F_{iz} - 4 \gamma^{ij} \gamma^{kl} \left(\partial_u \partial_u F_{jl} \right) F_{ik}\,.
	\end{split}
\end{equation}
Therefore, utilizing the antisymmetric property of the electromagnetic field tensor, the final result of the first term in Eq. (\ref{rhkk1first}) is obtained as 
\begin{equation}
	\begin{split}
		& - 4 k^a k^b \left(\nabla_a \nabla_b F^{cd} \right) F_{cd}\\
		= & 8 \left(\partial_u \partial_u F_{uz} \right) F_{uz} - 8 \gamma^{ij} \left(\partial_u \partial_u F_{ui} \right) F_{zj} - 4 \gamma^{ij} \gamma^{kl} \left(\partial_u \partial_u F_{il} \right) F_{jk}\,.
	\end{split}
\end{equation}
We will use the antisymmetry property of the electromagnetic tensor directly in the subsequent calculations without providing further explicit clarification. 

The second term of Eq. (\ref{rhkk1first}) is
\begin{equation}
	\begin{split}
		& 4 k^a k^b g^{ce} g^{df} \Gamma^{g}_{\ ab} \left(\partial_g F_{ef} \right) F_{cd} = 4 g^{ce} g^{df} \Gamma^{g}_{\ uu} \left(\partial_g F_{ef} \right) F_{cd} = 0\,.
	\end{split}
\end{equation}

The third term of Eq. (\ref{rhkk1first}) is
\begin{equation}
	\begin{split}
		& 4 k^a k^b g^{ce} g^{df} \Gamma^{g}_{\ ae} \left(\partial_b F_{gf} \right) F_{cd} = 4 g^{ce} g^{df} \Gamma^{g}_{\ ue} \left(\partial_u F_{gf} \right) F_{cd}\\
		= & 4 \Gamma^{g}_{\ uz} \left(\partial_u F_{gu} \right) F_{uz} + 4 \gamma^{ij} \Gamma^{g}_{\ uj} \left(\partial_u F_{gu} \right) F_{iz} + 4 \gamma^{ij} \gamma^{kl} \Gamma^{g}_{\ uj} \left(\partial_u F_{gl} \right) F_{ik}\,.
	\end{split}
\end{equation}
The repeated index $g$ should be further expanded by the metric as
\begin{equation}\label{rhkk1firstthird}
	\begin{split}
		& 4 \Gamma^{g}_{\ uz} \left(\partial_u F_{gu} \right) F_{uz} + 4 \gamma^{ij} \Gamma^{g}_{\ uj} \left(\partial_u F_{gu} \right) F_{iz} + 4 \gamma^{ij} \gamma^{kl} \Gamma^{g}_{\ uj} \left(\partial_u F_{gl} \right) F_{ik}\\
		= & 4 \Gamma^{u}_{\ uz} \left(\partial_u F_{uu} \right) F_{uz} + 4 \Gamma^{z}_{\ uz} \left(\partial_u F_{zu} \right) F_{uz} + 4 \Gamma^{i}_{\ uz} \left(\partial_u F_{iu} \right) F_{uz}\\
		& + 4 \gamma^{ij} \Gamma^{u}_{\ uj} \left(\partial_u F_{uu} \right) F_{iz} + 4 \gamma^{ij} \Gamma^{z}_{\ uj} \left(\partial_u F_{zu} \right) F_{iz} + 4 \gamma^{ij} \Gamma^{k}_{\ uj} \left(\partial_u F_{ku} \right) F_{iz}\\
		& + 4 \gamma^{ij} \gamma^{kl} \Gamma^{u}_{\ uj} \left(\partial_u F_{ul} \right) F_{ik} + 4 \gamma^{ij} \gamma^{kl} \Gamma^{z}_{\ uj} \left(\partial_u F_{zl} \right) F_{ik} + 4 \gamma^{ij} \gamma^{kl} \Gamma^{m}_{\ uj} \left(\partial_u F_{ml} \right) F_{ik}\\
		= & 4 \Gamma^{z}_{\ uz} \left(\partial_u F_{zu} \right) F_{uz} + 4 \Gamma^{i}_{\ uz} \left(\partial_u F_{iu} \right) F_{uz} + 4 \gamma^{ij} \Gamma^{z}_{\ uj} \left(\partial_u F_{zu} \right) F_{iz}\\
		& + 4 \gamma^{ij} \Gamma^{k}_{\ uj} \left(\partial_u F_{ku} \right) F_{iz} + 4 \gamma^{ij} \gamma^{kl} \Gamma^{u}_{\ uj} \left(\partial_u F_{ul} \right) F_{ik} + 4 \gamma^{ij} \gamma^{kl} \Gamma^{z}_{\ uj} \left(\partial_u F_{zl} \right) F_{ik}\\
		& + 4 \gamma^{ij} \gamma^{kl} \Gamma^{m}_{\ uj} \left(\partial_u F_{ml} \right) F_{ik}\,.
	\end{split}
\end{equation}
Therefore, the third term of Eq. (\ref{rhkk1first}) is obtained as 
\begin{equation}
	\begin{split}
		& 4 k^a k^b g^{ce} g^{df} \Gamma^{g}_{\ ae} \left(\partial_b F_{gf} \right) F_{cd}\\
		= & 2 \gamma^{ij} \beta_j \left(\partial_u F_{iu} \right) F_{uz} + 2 \gamma^{ij} \gamma^{kl} \left(\partial_u \gamma_{jl} \right) \left(\partial_u F_{ku} \right) F_{iz} - 2 \gamma^{ij} \gamma^{kl} \beta_j \left(\partial_u F_{ul} \right) F_{ik}\\
		& + 2 \gamma^{ij} \gamma^{kl} \gamma^{mn} \left(\partial_u \gamma_{jn} \right) \left(\partial_u F_{ml} \right) F_{ik}\\
		= & 2 \gamma^{ij} \beta_j \left(\partial_u F_{iu} \right) F_{uz} + 4 \gamma^{ij} \gamma^{kl} K_{jl} \left(\partial_u F_{ku} \right) F_{iz} - 2 \gamma^{ij} \gamma^{kl} \beta_j \left(\partial_u F_{ul} \right) F_{ik}\\
		& + 4 \gamma^{ij} \gamma^{kl} \gamma^{mn} K_{jn} \left(\partial_u F_{ml} \right) F_{ik}\\
		= & 2 \gamma^{ij} \beta_j \left(\partial_u F_{iu} \right) F_{uz} - 2 \gamma^{ij} \gamma^{kl} \beta_j \left(\partial_u F_{ul} \right) F_{ik}\,.
	\end{split}
\end{equation}

The fourth term of Eq. (\ref{rhkk1first}) is
\begin{equation}
	\begin{split}
		& 4 k^a k^b g^{ce} g^{df} \Gamma^{g}_{\ af} \left(\partial_b F_{eg} \right) F_{cd} = 4 g^{ce} g^{df} \Gamma^{g}_{\ uf} \left(\partial_u F_{eg} \right) F_{cd}\\
		= & 4 \Gamma^{g}_{\ uz} \left(\partial_u F_{ug} \right) F_{zu} + 4 \gamma^{ij} \Gamma^{g}_{\ uj} \left(\partial_u F_{ug} \right) F_{zi} + 4 \gamma^{ij} \gamma^{kl} \Gamma^{g}_{\ ul} \left(\partial_u F_{jg} \right) F_{ik}\,.
	\end{split}
\end{equation}
The index $g$ should be further expanded.
\begin{equation}\label{rhkk1firstfourth}
	\begin{split}
		&  4 \Gamma^{g}_{\ uz} \left(\partial_u F_{ug} \right) F_{zu} + 4 \gamma^{ij} \Gamma^{g}_{\ uj} \left(\partial_u F_{ug} \right) F_{zi} + 4 \gamma^{ij} \gamma^{kl} \Gamma^{g}_{\ ul} \left(\partial_u F_{jg} \right) F_{ik}\\
		= & 4 \Gamma^{u}_{\ uz} \left(\partial_u F_{uu} \right) F_{zu} + 4 \Gamma^{z}_{\ uz} \left(\partial_u F_{uz} \right) F_{zu} + 4 \Gamma^{i}_{\ uz} \left(\partial_u F_{ui} \right) F_{zu}\\
		& + 4 \gamma^{ij} \Gamma^{u}_{\ uj} \left(\partial_u F_{uu} \right) F_{zi} + 4 \gamma^{ij} \Gamma^{z}_{\ uj} \left(\partial_u F_{uz} \right) F_{zi} + 4 \gamma^{ij} \Gamma^{k}_{\ uj} \left(\partial_u F_{uk} \right) F_{zi}\\
		& + 4 \gamma^{ij} \gamma^{kl} \Gamma^{u}_{\ ul} \left(\partial_u F_{ju} \right) F_{ik} + 4 \gamma^{ij} \gamma^{kl} \Gamma^{z}_{\ ul} \left(\partial_u F_{jz} \right) F_{ik} + 4 \gamma^{ij} \gamma^{kl} \Gamma^{m}_{\ ul} \left(\partial_u F_{jm} \right) F_{ik}\\
		= & 4 \Gamma^{z}_{\ uz} \left(\partial_u F_{uz} \right) F_{zu} + 4 \Gamma^{i}_{\ uz} \left(\partial_u F_{ui} \right) F_{zu} + 4 \gamma^{ij} \Gamma^{z}_{\ uj} \left(\partial_u F_{uz} \right) F_{zi}\\
		& + 4 \gamma^{ij} \Gamma^{k}_{\ uj} \left(\partial_u F_{uk} \right) F_{zi} + 4 \gamma^{ij} \gamma^{kl} \Gamma^{u}_{\ ul} \left(\partial_u F_{ju} \right) F_{ik} + 4 \gamma^{ij} \gamma^{kl} \Gamma^{z}_{\ ul} \left(\partial_u F_{jz} \right) F_{ik}\\
		& + 4 \gamma^{ij} \gamma^{kl} \Gamma^{m}_{\ ul} \left(\partial_u F_{jm} \right) F_{ik}\,.
	\end{split}
\end{equation}
Therefore, the fourth term of Eq. (\ref{rhkk1first}) is obtained as
\begin{equation}
	\begin{split}
		& 4 k^a k^b g^{ce} g^{df} \Gamma^{g}_{\ af} \left(\partial_b F_{eg} \right) F_{cd}\\
		= & 2 \gamma^{ij} \beta_j \left(\partial_u F_{ui} \right) F_{zu} + 2 \gamma^{ij} \gamma^{kl} \left(\partial_u \gamma_{jl} \right) \left(\partial_u F_{uk} \right) F_{zi} - 2 \gamma^{ij} \gamma^{kl} \beta_l \left(\partial_u F_{ju} \right) F_{ik}\\
		& + 2 \gamma^{ij} \gamma^{kl} \gamma^{mn} \left(\partial_u \gamma_{ln} \right) \left(\partial_u F_{jm} \right) F_{ik}\\
		= & 2 \gamma^{ij} \beta_j \left(\partial_u F_{ui} \right) F_{zu} + 4 \gamma^{ij} \gamma^{kl} K_{jl} \left(\partial_u F_{uk} \right) F_{zi} - 2 \gamma^{ij} \gamma^{kl} \beta_l \left(\partial_u F_{ju} \right) F_{ik}\\
		& + 4 \gamma^{ij} \gamma^{kl} \gamma^{mn} K_{ln} \left(\partial_u F_{jm} \right) F_{ik}\\
		= & 2 \gamma^{ij} \beta_j \left(\partial_u F_{ui} \right) F_{zu} - 2 \gamma^{ij} \gamma^{kl} \beta_l \left(\partial_u F_{ju} \right) F_{ik}\,.
	\end{split}
\end{equation}

The fifth term of Eq. (\ref{rhkk1first}) is
\begin{equation}
	\begin{split}
		& 4 k^a k^b g^{ce} g^{df} \left(\partial_a \Gamma^{g}_{\ be} \right) F_{gf} F_{cd} = 4 g^{ce} g^{df} \left(\partial_u \Gamma^{g}_{\ ue} \right) F_{gf} F_{cd}\\
		= & 4 \left(\partial_u \Gamma^{g}_{\ uz} \right) F_{gu} F_{uz} + 4 \gamma^{ij} \left(\partial_u \Gamma^{g}_{\ uj} \right) F_{gu} F_{iz} + 4 \gamma^{ij} \gamma^{kl} \left(\partial_u \Gamma^{g}_{\ uj} \right) F_{gl} F_{ik}\,.
	\end{split}
\end{equation}
The repeated index $g$ should be further expanded through metric as 
\begin{equation}\label{rhkk1firstfifth}
	\begin{split}
		& 4 \left(\partial_u \Gamma^{g}_{\ uz} \right) F_{gu} F_{uz} + 4 \gamma^{ij} \left(\partial_u \Gamma^{g}_{\ uj} \right) F_{gu} F_{iz} + 4 \gamma^{ij} \gamma^{kl} \left(\partial_u \Gamma^{g}_{\ uj} \right) F_{gl} F_{ik}\\
		= & 4 \left(\partial_u \Gamma^{u}_{\ uz} \right) F_{uu} F_{uz} + 4 \left(\partial_u \Gamma^{z}_{\ uz} \right) F_{zu} F_{uz} + 4 \left(\partial_u \Gamma^{i}_{\ uz} \right) F_{iu} F_{uz}\\
		& + 4 \gamma^{ij} \left(\partial_u \Gamma^{u}_{\ uj} \right) F_{uu} F_{iz} + 4 \gamma^{ij} \left(\partial_u \Gamma^{z}_{\ uj} \right) F_{zu} F_{iz} + 4 \gamma^{ij} \left(\partial_u \Gamma^{k}_{\ uj} \right) F_{ku} F_{iz}\\
		& + 4 \gamma^{ij} \gamma^{kl} \left(\partial_u \Gamma^{u}_{\ uj} \right) F_{ul} F_{ik} + 4 \gamma^{ij} \gamma^{kl} \left(\partial_u \Gamma^{z}_{\ uj} \right) F_{zl} F_{ik} + 4 \gamma^{ij} \gamma^{kl} \left(\partial_u \Gamma^{m}_{\ uj} \right) F_{ml} F_{ik}\\
		= & 4 \left(\partial_u \Gamma^{z}_{\ uz} \right) F_{zu} F_{uz} + 4 \gamma^{ij} \left(\partial_u \Gamma^{z}_{\ uj} \right) F_{zu} F_{iz} + 4 \gamma^{ij} \gamma^{kl} \left(\partial_u \Gamma^{z}_{\ uj} \right) F_{zl} F_{ik}\\
		& + 4 \gamma^{ij} \gamma^{kl} \left(\partial_u \Gamma^{m}_{\ uj} \right) F_{ml} F_{ik}\,.
	\end{split}
\end{equation}
The first term of Eq. (\ref{rhkk1firstfifth}) is 
\begin{equation}
	\begin{split}
		& 4 \left(\partial_u \Gamma^{z}_{\ uz} \right) F_{zu} F_{uz}\\
		= & 4 \partial_u \left(z^2 \partial_z \alpha + 2 z \alpha - \frac{1}{2} z \beta^i \beta_i - \frac{1}{2} z^2 \beta^i \partial_z \beta_i \right) F_{zu} F_{uz}\\
		= & 4 z \partial_u \left(z \partial_z \alpha + 2 \alpha - \frac{1}{2} \beta^i \beta_i - \frac{1}{2} z \beta^i \partial_z \beta_i \right) F_{zu} F_{uz}\\
		= & 0\,.
	\end{split}
\end{equation}
The second term of Eq. (\ref{rhkk1firstfifth}) is
\begin{equation}
	\begin{split}
		& 4 \gamma^{ij} \left(\partial_u \Gamma^{z}_{\ uj} \right) F_{zu} F_{iz}\\
		= & 4 \gamma^{ij} \partial_u \left[z^2 \partial_j \alpha - \frac{1}{2} z^2 \left(\beta^2 - 2 \alpha \right) \left(\beta_j + z \partial_z \beta_j \right) - \frac{1}{2} z \beta^k \left(z \partial_j \beta_k + \partial_u \gamma_{jk} - z \partial_k \beta_j \right) \right]\\
		& \times F_{zu} F_{iz}\\
		= & 4 z \gamma^{ij} \partial_u \left[z \partial_j \alpha - \frac{1}{2} z \left(\beta^2 - 2 \alpha \right) \left(\beta_j + z \partial_z \beta_j \right) - \frac{1}{2} \beta^k \left(z \partial_j \beta_k + \partial_u \gamma_{jk} - z \partial_k \beta_j \right) \right]\\
		& \times F_{zu} F_{iz}\\
		= & 0\,.
	\end{split}
\end{equation}
The third term of Eq. (\ref{rhkk1firstfifth}) is
\begin{equation}
	\begin{split}
		& 4 \gamma^{ij} \gamma^{kl} \left(\partial_u \Gamma^{z}_{\ uj} \right) F_{zl} F_{ik}\\
		= & 4 \gamma^{ij} \gamma^{kl} \partial_u \left[z^2 \partial_j \alpha - \frac{1}{2} z^2 \left(\beta^2 - 2 \alpha \right) \left(\beta_j + z \partial_z \beta_j \right) - \frac{1}{2} z \beta^m \left(z \partial_j \beta_m + \partial_u \gamma_{jm} - z \partial_m \beta_j \right) \right]\\
		& \times F_{zl} F_{ik}\\
		= & 4 z \gamma^{ij} \gamma^{kl} \partial_u \left[z \partial_j \alpha - \frac{1}{2} z \left(\beta^2 - 2 \alpha \right) \left(\beta_j + z \partial_z \beta_j \right) - \frac{1}{2} \beta^m \left(z \partial_j \beta_m + \partial_u \gamma_{jm} - z \partial_m \beta_j \right) \right]\\
		& \times F_{zl} F_{ik}\\
		= & 0\,.
	\end{split}
\end{equation}
The fourth term of Eq. (\ref{rhkk1firstfifth}) is
\begin{equation}
	\begin{split}
		& 4 \gamma^{ij} \gamma^{kl} \left(\partial_u \Gamma^{m}_{\ uj} \right) F_{ml} F_{ik}\\
		= & 4 \gamma^{ij} \gamma^{kl} \partial_u \left[\frac{1}{2} \left(z \beta^m \right) \left(\beta_j + z \partial_z \beta_j \right) + \frac{1}{2} \gamma^{mn} \left(z \partial_j \beta_n + \partial_u \gamma_{jn} - z \partial_n \beta_j \right) \right] F_{ml} F_{ik}\\
		= & 4 \gamma^{ij} \gamma^{kl} \left[\frac{1}{2} \left(z \partial_u \beta^m \right) \left(\beta_j + z \partial_z \beta_j \right) + \frac{1}{2} \left(z \beta^m \right) \left(\partial_u \beta_j + z \partial_u \partial_z \beta_j \right)\right.\\
		& \left. + \frac{1}{2} \left(\partial_u \gamma^{mn} \right) \left(z \partial_j \beta_n + \partial_u \gamma_{jn} - z \partial_n \beta_j \right)\right.\\
		& \left. + \frac{1}{2} \gamma^{mn} \left(z \partial_u \partial_j \beta_n + \partial_u \partial_u \gamma_{jn} - z \partial_u \partial_n \beta_j \right) \right] F_{ml} F_{ik}\\
		= & 4 \gamma^{ij} \gamma^{kl} \left[\frac{1}{2} \left(\partial_u \gamma^{mn} \right) \left(\partial_u \gamma_{jn} \right) \right] F_{ml} F_{ik} + 4 \gamma^{ij} \gamma^{kl} \left(\frac{1}{2} \gamma^{mn} \partial_u \partial_u \gamma_{jn} \right) F_{ml} F_{ik}\\
		= & 2 \gamma^{ij} \gamma^{kl} \left(\partial_u \gamma^{mn} \right) \left(\partial_u \gamma_{jn} \right) F_{ml} F_{ik} + 2 \gamma^{ij} \gamma^{kl} \gamma^{mn} \left(\partial_u \partial_u \gamma_{jn} \right) F_{ml} F_{ik}\,.
	\end{split}
\end{equation}
Therefore, the fifth term of Eq. (\ref{rhkk1first}) is obtained as 
\begin{equation}
	\begin{split}
		& 4 k^a k^b g^{ce} g^{df} \left(\partial_a \Gamma^{g}_{\ be} \right) F_{gf} F_{cd}\\
		= & 2 \gamma^{ij} \gamma^{kl} \left(\partial_u \gamma^{mn} \right) \left(\partial_u \gamma_{jn} \right) F_{ml} F_{ik} + 2 \gamma^{ij} \gamma^{kl} \gamma^{mn} \left(\partial_u \partial_u \gamma_{jn} \right) F_{ml} F_{ik}\\
		= & 8 \gamma^{ij} \gamma^{kl} K^{mn} K_{jn} F_{ml} F_{ik} + 2 \gamma^{ij} \gamma^{kl} \gamma^{mn} \left(\partial_u \partial_u \gamma_{jn} \right) F_{ml} F_{ik}\\
		= & 2 \gamma^{ij} \gamma^{kl} \gamma^{mn} \left(\partial_u \partial_u \gamma_{jn} \right) F_{ml} F_{ik}\,.
	\end{split}
\end{equation}

The sixth term of Eq. (\ref{rhkk1first}) is
\begin{equation}
	\begin{split}
		- 4 k^a k^b g^{ce} g^{df} \Gamma^{h}_{\ ab} \Gamma^{g}_{\ he} F_{gf} F_{cd} = - 4 g^{ce} g^{df} \Gamma^{h}_{\ uu} \Gamma^{g}_{\ he} F_{gf} F_{cd} = 0\,.
	\end{split}
\end{equation}

The seventh term of Eq. (\ref{rhkk1first}) is
\begin{equation}
	\begin{split}
		& - 4 k^a k^b g^{ce} g^{df} \Gamma^{h}_{\ ae} \Gamma^{g}_{\ bh} F_{gf} F_{cd} = - 4 g^{ce} g^{df} \Gamma^{h}_{\ ue} \Gamma^{g}_{\ uh} F_{gf} F_{cd}\\
		= & - 4 \Gamma^{h}_{\ uz} \Gamma^{g}_{\ uh} F_{gu} F_{uz} - 4 \gamma^{ij} \Gamma^{h}_{\ uj} \Gamma^{g}_{\ uh} F_{gu} F_{iz} - 4 \gamma^{ij} \gamma^{kl} \Gamma^{h}_{\ uj} \Gamma^{g}_{\ uh} F_{gl} F_{ik}\,.
	\end{split}
\end{equation}
The repeated index $g$ should be further expanded as 
\begin{equation}\label{rhkk1firstseventh}
	\begin{split}
		& - 4 \Gamma^{h}_{\ uz} \Gamma^{g}_{\ uh} F_{gu} F_{uz} - 4 \gamma^{ij} \Gamma^{h}_{\ uj} \Gamma^{g}_{\ uh} F_{gu} F_{iz} - 4 \gamma^{ij} \gamma^{kl} \Gamma^{h}_{\ uj} \Gamma^{g}_{\ uh} F_{gl} F_{ik}\\
		= & - 4 \Gamma^{h}_{\ uz} \Gamma^{u}_{\ uh} F_{uu} F_{uz} - 4 \Gamma^{h}_{\ uz} \Gamma^{z}_{\ uh} F_{zu} F_{uz} - 4 \Gamma^{h}_{\ uz} \Gamma^{i}_{\ uh} F_{iu} F_{uz}\\
		& - 4 \gamma^{ij} \Gamma^{h}_{\ uj} \Gamma^{u}_{\ uh} F_{uu} F_{iz} - 4 \gamma^{ij} \Gamma^{h}_{\ uj} \Gamma^{z}_{\ uh} F_{zu} F_{iz} - 4 \gamma^{ij} \Gamma^{h}_{\ uj} \Gamma^{k}_{\ uh} F_{ku} F_{iz}\\
		& - 4 \gamma^{ij} \gamma^{kl} \Gamma^{h}_{\ uj} \Gamma^{u}_{\ uh} F_{ul} F_{ik} - 4 \gamma^{ij} \gamma^{kl} \Gamma^{h}_{\ uj} \Gamma^{z}_{\ uh} F_{zl} F_{ik} - 4 \gamma^{ij} \gamma^{kl} \Gamma^{h}_{\ uj} \Gamma^{m}_{\ uh} F_{ml} F_{ik}\\
		= & - 4 \Gamma^{h}_{\ uz} \Gamma^{z}_{\ uh} F_{zu} F_{uz} - 4 \gamma^{ij} \Gamma^{h}_{\ uj} \Gamma^{z}_{\ uh} F_{zu} F_{iz} - 4 \gamma^{ij} \gamma^{kl} \Gamma^{h}_{\ uj} \Gamma^{z}_{\ uh} F_{zl} F_{ik}\\
		& - 4 \gamma^{ij} \gamma^{kl} \Gamma^{h}_{\ uj} \Gamma^{m}_{\ uh} F_{ml} F_{ik}\,.
	\end{split}
\end{equation}
The repeated index $h$ also needs to be expanded using the metric. The first term of Eq. (\ref{rhkk1firstseventh}) is 
\begin{equation}
	\begin{split}
		& - 4 \Gamma^{h}_{\ uz} \Gamma^{z}_{\ uh} F_{zu} F_{uz}\\
		= & - 4 \Gamma^{u}_{\ uz} \Gamma^{z}_{\ uu} F_{zu} F_{uz} - 4 \Gamma^{z}_{\ uz} \Gamma^{z}_{\ uz} F_{zu} F_{uz} - 4 \Gamma^{i}_{\ uz} \Gamma^{z}_{\ ui} F_{zu} F_{uz}\\
		= & 0\,.
	\end{split}
\end{equation}
The second term of Eq. (\ref{rhkk1firstseventh}) is
\begin{equation}
	\begin{split}
		& - 4 \gamma^{ij} \Gamma^{h}_{\ uj} \Gamma^{z}_{\ uh} F_{zu} F_{iz}\\
		= & - 4 \gamma^{ij} \Gamma^{u}_{\ uj} \Gamma^{z}_{\ uu} F_{zu} F_{iz} - 4 \gamma^{ij} \Gamma^{z}_{\ uj} \Gamma^{z}_{\ uz} F_{zu} F_{iz} - 4 \gamma^{ij} \Gamma^{k}_{\ uj} \Gamma^{z}_{\ uk} F_{zu} F_{iz}\\
		= & 0\,.
	\end{split}
\end{equation}
The third term of Eq. (\ref{rhkk1firstseventh}) is
\begin{equation}
	\begin{split}
		& - 4 \gamma^{ij} \gamma^{kl} \Gamma^{h}_{\ uj} \Gamma^{z}_{\ uh} F_{zl} F_{ik}\\
		= & - 4 \gamma^{ij} \gamma^{kl} \Gamma^{u}_{\ uj} \Gamma^{z}_{\ uu} F_{zl} F_{ik} - 4 \gamma^{ij} \gamma^{kl} \Gamma^{z}_{\ uj} \Gamma^{z}_{\ uz} F_{zl} F_{ik} - 4 \gamma^{ij} \gamma^{kl} \Gamma^{m}_{\ uj} \Gamma^{z}_{\ um} F_{zl} F_{ik}\\
		= & 0\,.
	\end{split}
\end{equation}
The fourth term of Eq. (\ref{rhkk1firstseventh}) is
\begin{equation}
	\begin{split}
		& - 4 \gamma^{ij} \gamma^{kl} \Gamma^{h}_{\ uj} \Gamma^{m}_{\ uh} F_{ml} F_{ik}\\
		= & - 4 \gamma^{ij} \gamma^{kl} \Gamma^{u}_{\ uj} \Gamma^{m}_{\ uu} F_{ml} F_{ik} - 4 \gamma^{ij} \gamma^{kl} \Gamma^{z}_{\ uj} \Gamma^{m}_{\ uz} F_{ml} F_{ik} - 4 \gamma^{ij} \gamma^{kl} \Gamma^{n}_{\ uj} \Gamma^{m}_{\ un} F_{ml} F_{ik}\\
		= & - 4 \gamma^{ij} \gamma^{kl} \left(\frac{1}{2} \gamma^{no} \partial_u \gamma_{jo} \right) \left(\frac{1}{2} \gamma^{mp} \partial_u \gamma_{np} \right) F_{ml} F_{ik}\\
		= & - \gamma^{ij} \gamma^{kl} \gamma^{no} \gamma^{mp} \left(\partial_u \gamma_{jo} \right) \left(\partial_u \gamma_{np} \right) F_{ml} F_{ik}\,.
	\end{split}
\end{equation}
Therefore, the seventh term of Eq. (\ref{rhkk1first}) is obtained as
\begin{equation}
	\begin{split}
		& - 4 k^a k^b g^{ce} g^{df} \Gamma^{h}_{\ ae} \Gamma^{g}_{\ bh} F_{gf} F_{cd} = - \gamma^{ij} \gamma^{kl} \gamma^{no} \gamma^{mp} \left(\partial_u \gamma_{jo} \right) \left(\partial_u \gamma_{np} \right) F_{ml} F_{ik}\\
		= & - 4 \gamma^{ij} \gamma^{kl} \gamma^{no} \gamma^{mp} K_{jo} K_{np} F_{ml} F_{ik}\\
		= & 0\,.
	\end{split}
\end{equation}

The eighth term of Eq. (\ref{rhkk1first}) is
\begin{equation}
	\begin{split}
		& 4 k^a k^b g^{ce} g^{df} \Gamma^{g}_{\ ah} \Gamma^{h}_{\ be} F_{gf} F_{cd} = 4 g^{ce} g^{df} \Gamma^{g}_{\ uh} \Gamma^{h}_{\ ue} F_{gf} F_{cd}\\
		= & 4 \Gamma^{g}_{\ uh} \Gamma^{h}_{\ uz} F_{gu} F_{uz} + 4 \gamma^{ij} \Gamma^{g}_{\ uh} \Gamma^{h}_{\ uj} F_{gu} F_{iz} + 4 \gamma^{ij} \gamma^{kl} \Gamma^{g}_{\ uh} \Gamma^{h}_{\ uj} F_{gl} F_{ik}\,.
	\end{split}
\end{equation}
The repeated index $g$ should be further expanded as
\begin{equation}\label{rhkk1firsteighth}
	\begin{split}
		& 4 \Gamma^{g}_{\ uh} \Gamma^{h}_{\ uz} F_{gu} F_{uz} + 4 \gamma^{ij} \Gamma^{g}_{\ uh} \Gamma^{h}_{\ uj} F_{gu} F_{iz} + 4 \gamma^{ij} \gamma^{kl} \Gamma^{g}_{\ uh} \Gamma^{h}_{\ uj} F_{gl} F_{ik}\\
		= & 4 \Gamma^{u}_{\ uh} \Gamma^{h}_{\ uz} F_{uu} F_{uz} + 4 \Gamma^{z}_{\ uh} \Gamma^{h}_{\ uz} F_{zu} F_{uz} + 4 \Gamma^{i}_{\ uh} \Gamma^{h}_{\ uz} F_{iu} F_{uz}\\
		& + 4 \gamma^{ij} \Gamma^{u}_{\ uh} \Gamma^{h}_{\ uj} F_{uu} F_{iz} + 4 \gamma^{ij} \Gamma^{z}_{\ uh} \Gamma^{h}_{\ uj} F_{zu} F_{iz} + 4 \gamma^{ij} \Gamma^{k}_{\ uh} \Gamma^{h}_{\ uj} F_{ku} F_{iz}\\
		& + 4 \gamma^{ij} \gamma^{kl} \Gamma^{u}_{\ uh} \Gamma^{h}_{\ uj} F_{ul} F_{ik} + 4 \gamma^{ij} \gamma^{kl} \Gamma^{z}_{\ uh} \Gamma^{h}_{\ uj} F_{zl} F_{ik} + 4 \gamma^{ij} \gamma^{kl} \Gamma^{m}_{\ uh} \Gamma^{h}_{\ uj} F_{ml} F_{ik}\\
		= & 4 \Gamma^{z}_{\ uh} \Gamma^{h}_{\ uz} F_{zu} F_{uz} + 4 \gamma^{ij} \Gamma^{z}_{\ uh} \Gamma^{h}_{\ uj} F_{zu} F_{iz} + 4 \gamma^{ij} \gamma^{kl} \Gamma^{z}_{\ uh} \Gamma^{h}_{\ uj} F_{zl} F_{ik}\\
		& + 4 \gamma^{ij} \gamma^{kl} \Gamma^{m}_{\ uh} \Gamma^{h}_{\ uj} F_{ml} F_{ik}\,.
	\end{split}
\end{equation}
The repeated index $h$ should be further expanded. The first term in Eq. (\ref{rhkk1firsteighth}) is 
\begin{equation}
	\begin{split}
		& 4 \Gamma^{z}_{\ uh} \Gamma^{h}_{\ uz} F_{zu} F_{uz}\\
		= & 4 \Gamma^{z}_{\ uu} \Gamma^{u}_{\ uz} F_{zu} F_{uz} + 4 \Gamma^{z}_{\ uz} \Gamma^{z}_{\ uz} F_{zu} F_{uz} + 4 \Gamma^{z}_{\ ui} \Gamma^{i}_{\ uz} F_{zu} F_{uz}\\
		= & 0\,.
	\end{split}
\end{equation}
The second term of Eq. (\ref{rhkk1firsteighth}) is
\begin{equation}
	\begin{split}
		& 4 \gamma^{ij} \Gamma^{z}_{\ uh} \Gamma^{h}_{\ uj} F_{zu} F_{iz}\\
		= & 4 \gamma^{ij} \Gamma^{z}_{\ uu} \Gamma^{u}_{\ uj} F_{zu} F_{iz} + 4 \gamma^{ij} \Gamma^{z}_{\ uz} \Gamma^{z}_{\ uj} F_{zu} F_{iz} + 4 \gamma^{ij} \Gamma^{z}_{\ uk} \Gamma^{k}_{\ uj} F_{zu} F_{iz}\\
		= & 0\,.
	\end{split}
\end{equation}
The third term of Eq. (\ref{rhkk1firsteighth}) is
\begin{equation}
	\begin{split}
		& 4 \gamma^{ij} \gamma^{kl} \Gamma^{z}_{\ uh} \Gamma^{h}_{\ uj} F_{zl} F_{ik}\\
		= & 4 \gamma^{ij} \gamma^{kl} \Gamma^{z}_{\ uu} \Gamma^{u}_{\ uj} F_{zl} F_{ik} + 4 \gamma^{ij} \gamma^{kl} \Gamma^{z}_{\ uz} \Gamma^{z}_{\ uj} F_{zl} F_{ik} + 4 \gamma^{ij} \gamma^{kl} \Gamma^{z}_{\ um} \Gamma^{m}_{\ uj} F_{zl} F_{ik}\\
		= & 0\,.
	\end{split}
\end{equation}
The fourth term of Eq. (\ref{rhkk1firsteighth}) is
\begin{equation}
	\begin{split}
		& 4 \gamma^{ij} \gamma^{kl} \Gamma^{m}_{\ uh} \Gamma^{h}_{\ uj} F_{ml} F_{ik}\\
		= & 4 \gamma^{ij} \gamma^{kl} \Gamma^{m}_{\ uu} \Gamma^{u}_{\ uj} F_{ml} F_{ik} + 4 \gamma^{ij} \gamma^{kl} \Gamma^{m}_{\ uz} \Gamma^{z}_{\ uj} F_{ml} F_{ik} + 4 \gamma^{ij} \gamma^{kl} \Gamma^{m}_{\ un} \Gamma^{n}_{\ uj} F_{ml} F_{ik}\\
		= & 4 \gamma^{ij} \gamma^{kl} \left(\frac{1}{2} \gamma^{mo} \partial_u \gamma_{no} \right) \left(\frac{1}{2} \gamma^{np} \partial_u \gamma_{jp} \right) F_{ml} F_{ik}\\
		= & \gamma^{ij} \gamma^{kl} \gamma^{mo} \gamma^{np} \left(\partial_u \gamma_{no} \right) \left(\partial_u \gamma_{jp} \right) F_{ml} F_{ik}\,.
	\end{split}
\end{equation}
Therefore, the eighth term of Eq. (\ref{rhkk1first}) is obtained as
\begin{equation}
	\begin{split}
		& 4 k^a k^b g^{ce} g^{df} \Gamma^{g}_{\ ah} \Gamma^{h}_{\ be} F_{gf} F_{cd} = \gamma^{ij} \gamma^{kl} \gamma^{mo} \gamma^{np} \left(\partial_u \gamma_{no} \right) \left(\partial_u \gamma_{jp} \right) F_{ml} F_{ik}\\
		= & 4 \gamma^{ij} \gamma^{kl} \gamma^{mo} \gamma^{np} K_{no} K_{jp} F_{ml} F_{ik}\\
		= & 0\,.
	\end{split}
\end{equation}

The ninth term of Eq. (\ref{rhkk1first}) is
\begin{equation}
	\begin{split}
		& 4 k^a k^b g^{ce} g^{df} \Gamma^{g}_{\ be} \left(\partial_a F_{gf} \right) F_{cd} = 4 g^{ce} g^{df} \Gamma^{g}_{\ ue} \left(\partial_u F_{gf} \right) F_{cd}\\
		= & 4 \Gamma^{g}_{\ uz} \left(\partial_u F_{gu} \right) F_{uz} + 4 \gamma^{ij} \Gamma^{g}_{\ uj} \left(\partial_u F_{gu} \right) F_{iz} + 4 \gamma^{ij} \gamma^{kl} \Gamma^{g}_{\ uj} \left(\partial_u F_{gl} \right) F_{ik}\,.
	\end{split}
\end{equation}
The index $g$ should be further expanded as 
\begin{equation}\label{rhkk1firstninth}
	\begin{split}
		& 4 \Gamma^{g}_{\ uz} \left(\partial_u F_{gu} \right) F_{uz} + 4 \gamma^{ij} \Gamma^{g}_{\ uj} \left(\partial_u F_{gu} \right) F_{iz} + 4 \gamma^{ij} \gamma^{kl} \Gamma^{g}_{\ uj} \left(\partial_u F_{gl} \right) F_{ik}\\
		= & 4 \Gamma^{u}_{\ uz} \left(\partial_u F_{uu} \right) F_{uz} + 4 \Gamma^{z}_{\ uz} \left(\partial_u F_{zu} \right) F_{uz} + 4 \Gamma^{i}_{\ uz} \left(\partial_u F_{iu} \right) F_{uz}\\
		& + 4 \gamma^{ij} \Gamma^{u}_{\ uj} \left(\partial_u F_{uu} \right) F_{iz} + 4 \gamma^{ij} \Gamma^{z}_{\ uj} \left(\partial_u F_{zu} \right) F_{iz} + 4 \gamma^{ij} \Gamma^{k}_{\ uj} \left(\partial_u F_{ku} \right) F_{iz}\\
		& + 4 \gamma^{ij} \gamma^{kl} \Gamma^{u}_{\ uj} \left(\partial_u F_{ul} \right) F_{ik} + 4 \gamma^{ij} \gamma^{kl} \Gamma^{z}_{\ uj} \left(\partial_u F_{zl} \right) F_{ik} + 4 \gamma^{ij} \gamma^{kl} \Gamma^{m}_{\ uj} \left(\partial_u F_{ml} \right) F_{ik}\\
		= & 4 \Gamma^{z}_{\ uz} \left(\partial_u F_{zu} \right) F_{uz} + 4 \Gamma^{i}_{\ uz} \left(\partial_u F_{iu} \right) F_{uz} + 4 \gamma^{ij} \Gamma^{z}_{\ uj} \left(\partial_u F_{zu} \right) F_{iz}\\
		& + 4 \gamma^{ij} \Gamma^{k}_{\ uj} \left(\partial_u F_{ku} \right) F_{iz} + 4 \gamma^{ij} \gamma^{kl} \Gamma^{u}_{\ uj} \left(\partial_u F_{ul} \right) F_{ik} + 4 \gamma^{ij} \gamma^{kl} \Gamma^{z}_{\ uj} \left(\partial_u F_{zl} \right) F_{ik}\\
		& + 4 \gamma^{ij} \gamma^{kl} \Gamma^{m}_{\ uj} \left(\partial_u F_{ml} \right) F_{ik}\,.
	\end{split}
\end{equation}
Therefore, the ninth term of Eq. (\ref{rhkk1first}) is obtained as
\begin{equation}
	\begin{split}
		& 4 k^a k^b g^{ce} g^{df} \Gamma^{g}_{\ be} \left(\partial_a F_{gf} \right) F_{cd}\\
		= & 2 \gamma^{ij} \beta_j \left(\partial_u F_{iu} \right) F_{uz} + 2 \gamma^{ij} \gamma^{kl} \left(\partial_u \gamma_{jl} \right) \left(\partial_u F_{ku} \right) F_{iz} - 2 \gamma^{ij} \gamma^{kl} \beta_j \left(\partial_u F_{ul} \right) F_{ik}\\
		& + 2 \gamma^{ij} \gamma^{kl} \gamma^{mn} \left(\partial_u \gamma_{jn} \right) \left(\partial_u F_{ml} \right) F_{ik}\\
		= & 2 \gamma^{ij} \beta_j \left(\partial_u F_{iu} \right) F_{uz} + 4 \gamma^{ij} \gamma^{kl} K_{jl} \left(\partial_u F_{ku} \right) F_{iz} - 2 \gamma^{ij} \gamma^{kl} \beta_j \left(\partial_u F_{ul} \right) F_{ik}\\
		& + 2 \gamma^{ij} \gamma^{kl} \gamma^{mn} K_{jn} \left(\partial_u F_{ml} \right) F_{ik}\\
		= & 2 \gamma^{ij} \beta_j \left(\partial_u F_{iu} \right) F_{uz} - 2 \gamma^{ij} \gamma^{kl} \beta_j \left(\partial_u F_{ul} \right) F_{ik}\,.
	\end{split}
\end{equation}

The tenth term of Eq. (\ref{rhkk1first}) is
\begin{equation}
	\begin{split}
		& - 4 k^a k^b g^{ce} g^{df} \Gamma^{g}_{\ be} \Gamma^{h}_{\ ag} F_{hf} F_{cd} = - 4 g^{ce} g^{df} \Gamma^{g}_{\ ue} \Gamma^{h}_{\ ug} F_{hf} F_{cd}\\
		= & - 4 \Gamma^{g}_{\ uz} \Gamma^{h}_{\ ug} F_{hu} F_{uz} - 4 \gamma^{ij} \Gamma^{g}_{\ uj} \Gamma^{h}_{\ ug} F_{hu} F_{iz} - 4 \gamma^{ij} \gamma^{kl} \Gamma^{g}_{\ uj} \Gamma^{h}_{\ ug} F_{hl} F_{ik}\,.
	\end{split}
\end{equation}
The index $g$ should be further expanded.
\begin{equation}\label{rhkk1firsttenth}
	\begin{split}
		& - 4 \Gamma^{g}_{\ uz} \Gamma^{h}_{\ ug} F_{hu} F_{uz} - 4 \gamma^{ij} \Gamma^{g}_{\ uj} \Gamma^{h}_{\ ug} F_{hu} F_{iz} - 4 \gamma^{ij} \gamma^{kl} \Gamma^{g}_{\ uj} \Gamma^{h}_{\ ug} F_{hl} F_{ik}\\
		= & - 4 \Gamma^{u}_{\ uz} \Gamma^{h}_{\ uu} F_{hu} F_{uz} - 4 \Gamma^{z}_{\ uz} \Gamma^{h}_{\ uz} F_{hu} F_{uz} - 4 \Gamma^{i}_{\ uz} \Gamma^{h}_{\ ui} F_{hu} F_{uz}\\
		& - 4 \gamma^{ij} \Gamma^{u}_{\ uj} \Gamma^{h}_{\ uu} F_{hu} F_{iz} - 4 \gamma^{ij} \Gamma^{z}_{\ uj} \Gamma^{h}_{\ uz} F_{hu} F_{iz} - 4 \gamma^{ij} \Gamma^{k}_{\ uj} \Gamma^{h}_{\ uk} F_{hu} F_{iz}\\
		& - 4 \gamma^{ij} \gamma^{kl} \Gamma^{u}_{\ uj} \Gamma^{h}_{\ uu} F_{hl} F_{ik} - 4 \gamma^{ij} \gamma^{kl} \Gamma^{z}_{\ uj} \Gamma^{h}_{\ uz} F_{hl} F_{ik} - 4 \gamma^{ij} \gamma^{kl} \Gamma^{m}_{\ uj} \Gamma^{h}_{\ um} F_{hl} F_{ik}\\
		= & - 4 \Gamma^{z}_{\ uz} \Gamma^{h}_{\ uz} F_{hu} F_{uz} - 4 \Gamma^{i}_{\ uz} \Gamma^{h}_{\ ui} F_{hu} F_{uz} - 4 \gamma^{ij} \Gamma^{z}_{\ uj} \Gamma^{h}_{\ uz} F_{hu} F_{iz}\\
		& - 4 \gamma^{ij} \Gamma^{k}_{\ uj} \Gamma^{h}_{\ uk} F_{hu} F_{iz}  - 4 \gamma^{ij} \gamma^{kl} \Gamma^{z}_{\ uj} \Gamma^{h}_{\ uz} F_{hl} F_{ik} - 4 \gamma^{ij} \gamma^{kl} \Gamma^{m}_{\ uj} \Gamma^{h}_{\ um} F_{hl} F_{ik}\,.
	\end{split}
\end{equation}
The repeated index $h$ should be further expanded using the metric. The first term of Eq. (\ref{rhkk1firsttenth}) is
\begin{equation}
	\begin{split}
		- 4 \Gamma^{z}_{\ uz} \Gamma^{h}_{\ uz} F_{hu} F_{uz} = 0\,.
	\end{split}
\end{equation}
The second term of Eq. (\ref{rhkk1firsttenth}) is
\begin{equation}
	\begin{split}
		& - 4 \Gamma^{i}_{\ uz} \Gamma^{h}_{\ ui} F_{hu} F_{uz}\\
		= & - 4 \Gamma^{i}_{\ uz} \Gamma^{u}_{\ ui} F_{uu} F_{uz} - 4 \Gamma^{i}_{\ uz} \Gamma^{z}_{\ ui} F_{zu} F_{uz} - 4 \Gamma^{i}_{\ uz} \Gamma^{j}_{\ ui} F_{ju} F_{uz}\\
		= & 0\,.
	\end{split}
\end{equation}
The third term of Eq. (\ref{rhkk1firsttenth}) is
\begin{equation}
	\begin{split}
		& - 4 \gamma^{ij} \Gamma^{z}_{\ uj} \Gamma^{h}_{\ uz} F_{hu} F_{iz}\\
		= & - 4 \gamma^{ij} \Gamma^{z}_{\ uj} \Gamma^{u}_{\ uz} F_{uu} F_{iz} - 4 \gamma^{ij} \Gamma^{z}_{\ uj} \Gamma^{z}_{\ uz} F_{zu} F_{iz} - 4 \gamma^{ij} \Gamma^{z}_{\ uj} \Gamma^{k}_{\ uz} F_{ku} F_{iz}\\
		= & 0\,.
	\end{split}
\end{equation}
The fourth term of Eq. (\ref{rhkk1firsttenth}) is
\begin{equation}
	\begin{split}
		& - 4 \gamma^{ij} \Gamma^{k}_{\ uj} \Gamma^{h}_{\ uk} F_{hu} F_{iz}\\
		= & - 4 \gamma^{ij} \Gamma^{k}_{\ uj} \Gamma^{u}_{\ uk} F_{uu} F_{iz} - 4 \gamma^{ij} \Gamma^{k}_{\ uj} \Gamma^{z}_{\ uk} F_{zu} F_{iz} - 4 \gamma^{ij} \Gamma^{k}_{\ uj} \Gamma^{l}_{\ uk} F_{lu} F_{iz}\\
		= & 0\,.
	\end{split}
\end{equation}
The fifth term of Eq. (\ref{rhkk1firsttenth}) is
\begin{equation}
	\begin{split}
		- 4 \gamma^{ij} \gamma^{kl} \Gamma^{z}_{\ uj} \Gamma^{h}_{\ uz} F_{hl} F_{ik} = 0\,.
	\end{split}
\end{equation}
The sixth term of Eq. (\ref{rhkk1firsttenth}) is
\begin{equation}
	\begin{split}
		& - 4 \gamma^{ij} \gamma^{kl} \Gamma^{m}_{\ uj} \Gamma^{h}_{\ um} F_{hl} F_{ik}\\
		= & - 4 \gamma^{ij} \gamma^{kl} \Gamma^{m}_{\ uj} \Gamma^{u}_{\ um} F_{ul} F_{ik} - 4 \gamma^{ij} \gamma^{kl} \Gamma^{m}_{\ uj} \Gamma^{z}_{\ um} F_{zl} F_{ik} - 4 \gamma^{ij} \gamma^{kl} \Gamma^{m}_{\ uj} \Gamma^{n}_{\ um} F_{nl} F_{ik}\\
		= & - \gamma^{ij} \gamma^{kl} \gamma^{mo} \gamma^{np} \left(\partial_u \gamma_{jo} \right) \left(\partial_u \gamma_{mp} \right) F_{nl} F_{ik}\,.
	\end{split}
\end{equation}
Therefore, the tenth term of Eq. (\ref{rhkk1first}) is obtained as
\begin{equation}
	\begin{split}
		& - 4 k^a k^b g^{ce} g^{df} \Gamma^{g}_{\ be} \Gamma^{h}_{\ ag} F_{hf} F_{cd} = - \gamma^{ij} \gamma^{kl} \gamma^{mo} \gamma^{np} \left(\partial_u \gamma_{jo} \right) \left(\partial_u \gamma_{mp} \right) F_{nl} F_{ik}\\
		= & - 4 \gamma^{ij} \gamma^{kl} \gamma^{mo} \gamma^{np} K_{jo} K_{mp} F_{nl} F_{ik}\\
		= & 0\,.
	\end{split}
\end{equation}

The eleventh term of Eq. (\ref{rhkk1first}) is
\begin{equation}
	\begin{split}
		& - 4 k^a k^b g^{ce} g^{df} \Gamma^{g}_{\ be} \Gamma^{h}_{\ af} F_{gh} F_{cd} = - 4 g^{ce} g^{df} \Gamma^{g}_{\ ue} \Gamma^{h}_{\ uf} F_{gh} F_{cd}\\
		= & - 4 \Gamma^{g}_{\ uz} \Gamma^{h}_{\ uu} F_{gh} F_{uz} - 4 \Gamma^{g}_{\ uu} \Gamma^{h}_{\ uz} F_{gh} F_{zu} - 4 \gamma^{ij} \Gamma^{g}_{\ uu} \Gamma^{h}_{\ uj} F_{gh} F_{zi}\\
		& - 4 \gamma^{ij} \Gamma^{g}_{\ uj} \Gamma^{h}_{\ uu} F_{gh} F_{iz} - 4 \gamma^{ij} \gamma^{kl} \Gamma^{g}_{\ uj} \Gamma^{h}_{\ ul} F_{gh} F_{ik}\\
		= & - 4 \gamma^{ij} \gamma^{kl} \Gamma^{g}_{\ uj} \Gamma^{h}_{\ ul} F_{gh} F_{ik}\,.
	\end{split}
\end{equation}
The repeated index $g$ should be further expanded as 
\begin{equation}\label{rhkk1firsteleventh}
	\begin{split}
		& - 4 \gamma^{ij} \gamma^{kl} \Gamma^{g}_{\ uj} \Gamma^{h}_{\ ul} F_{gh} F_{ik}\\
		= & - 4 \gamma^{ij} \gamma^{kl} \Gamma^{u}_{\ uj} \Gamma^{h}_{\ ul} F_{uh} F_{ik} - 4 \gamma^{ij} \gamma^{kl} \Gamma^{z}_{\ uj} \Gamma^{h}_{\ ul} F_{zh} F_{ik} - 4 \gamma^{ij} \gamma^{kl} \Gamma^{m}_{\ uj} \Gamma^{h}_{\ ul} F_{mh} F_{ik}\,.
	\end{split}
\end{equation}
The repeated index $h$ should be further expand. The first term of Eq. (\ref{rhkk1firsteleventh}) is 
\begin{equation}
	\begin{split}
		& - 4 \gamma^{ij} \gamma^{kl} \Gamma^{u}_{\ uj} \Gamma^{h}_{\ ul} F_{uh} F_{ik}\\
		= & - 4 \gamma^{ij} \gamma^{kl} \Gamma^{u}_{\ uj} \Gamma^{u}_{\ ul} F_{uu} F_{ik} - 4 \gamma^{ij} \gamma^{kl} \Gamma^{u}_{\ uj} \Gamma^{z}_{\ ul} F_{uz} F_{ik} - 4 \gamma^{ij} \gamma^{kl} \Gamma^{u}_{\ uj} \Gamma^{m}_{\ ul} F_{um} F_{ik}\\
		= & 0\,.
	\end{split}
\end{equation}
The second term of Eq. (\ref{rhkk1firsteleventh}) is
\begin{equation}
	\begin{split}
		- 4 \gamma^{ij} \gamma^{kl} \Gamma^{z}_{\ uj} \Gamma^{h}_{\ ul} F_{zh} F_{ik} = 0\,.
	\end{split}
\end{equation}
The third term of Eq. (\ref{rhkk1firsteleventh}) is
\begin{equation}
	\begin{split}
		& - 4 \gamma^{ij} \gamma^{kl} \Gamma^{m}_{\ uj} \Gamma^{h}_{\ ul} F_{mh} F_{ik}\\
		= & - 4 \gamma^{ij} \gamma^{kl} \Gamma^{m}_{\ uj} \Gamma^{u}_{\ ul} F_{mu} F_{ik} - 4 \gamma^{ij} \gamma^{kl} \Gamma^{m}_{\ uj} \Gamma^{z}_{\ ul} F_{mz} F_{ik} - 4 \gamma^{ij} \gamma^{kl} \Gamma^{m}_{\ uj} \Gamma^{n}_{\ ul} F_{mn} F_{ik}\\
		= & - \gamma^{ij} \gamma^{kl} \gamma^{mo} \gamma^{np} \left(\partial_u \gamma_{jo} \right) \left(\partial_u \gamma_{lp} \right) F_{mn} F_{ik}\,.
	\end{split}
\end{equation}
Therefore, the eleventh term of Eq. (\ref{rhkk1first}) is obtained as
\begin{equation}
	\begin{split}
		& - 4 k^a k^b g^{ce} g^{df} \Gamma^{g}_{\ be} \Gamma^{h}_{\ af} F_{gh} F_{cd} = - \gamma^{ij} \gamma^{kl} \gamma^{mo} \gamma^{np} \left(\partial_u \gamma_{jo} \right) \left(\partial_u \gamma_{lp} \right) F_{mn} F_{ik}\\
		= & - 4 \gamma^{ij} \gamma^{kl} \gamma^{mo} \gamma^{np} K_{jo} K_{lp} F_{mn} F_{ik}\\
		= & 0\,.
	\end{split}
\end{equation}

The twelfth term of Eq. (\ref{rhkk1first}) is
\begin{equation}
	\begin{split}
		& 4 k^a k^b g^{ce} g^{df} \left(\partial_a \Gamma^{g}_{\ bf} \right) F_{eg} F_{cd} = 4 g^{ce} g^{df} \left(\partial_u \Gamma^{g}_{\ uf} \right) F_{eg} F_{cd}\\
		= & 4 \left(\partial_u \Gamma^{g}_{\ uu} \right) F_{zg} F_{uz} + 4 \left(\partial_u \Gamma^{g}_{\ uz} \right) F_{ug} F_{zu} + 4 \gamma^{ij} \left(\partial_u \Gamma^{g}_{\ uj} \right) F_{ug} F_{zi}\\
		& + 4 \gamma^{ij} \left(\partial_u \Gamma^{g}_{\ uu} \right) F_{jg} F_{iz} + 4 \gamma^{ij} \gamma^{kl} \left(\partial_u \Gamma^{g}_{\ ul} \right) F_{jg} F_{ik}\,.
	\end{split}
\end{equation}
The repeated index $g$ should be further expanded as 
\begin{equation}\label{rhkk1firsttwelfth}
	\begin{split}
		& 4 \left(\partial_u \Gamma^{g}_{\ uu} \right) F_{zg} F_{uz} + 4 \left(\partial_u \Gamma^{g}_{\ uz} \right) F_{ug} F_{zu} + 4 \gamma^{ij} \left(\partial_u \Gamma^{g}_{\ uj} \right) F_{ug} F_{zi}\\
		& + 4 \gamma^{ij} \left(\partial_u \Gamma^{g}_{\ uu} \right) F_{jg} F_{iz} + 4 \gamma^{ij} \gamma^{kl} \left(\partial_u \Gamma^{g}_{\ ul} \right) F_{jg} F_{ik}\\
		= & 4 \left(\partial_u \Gamma^{u}_{\ uu} \right) F_{zu} F_{uz} + 4 \left(\partial_u \Gamma^{z}_{\ uu} \right) F_{zz} F_{uz} + 4 \left(\partial_u \Gamma^{i}_{\ uu} \right) F_{zi} F_{uz}\\
		& + 4 \left(\partial_u \Gamma^{u}_{\ uz} \right) F_{uu} F_{zu} + 4 \left(\partial_u \Gamma^{z}_{\ uz} \right) F_{uz} F_{zu} + 4 \left(\partial_u \Gamma^{i}_{\ uz} \right) F_{ui} F_{zu}\\
		& + 4 \gamma^{ij} \left(\partial_u \Gamma^{u}_{\ uj} \right) F_{uu} F_{zi} + 4 \gamma^{ij} \left(\partial_u \Gamma^{z}_{\ uj} \right) F_{uz} F_{zi} + 4 \gamma^{ij} \left(\partial_u \Gamma^{k}_{\ uj} \right) F_{uk} F_{zi}\\
		& + 4 \gamma^{ij} \left(\partial_u \Gamma^{u}_{\ uu} \right) F_{ju} F_{iz} + 4 \gamma^{ij} \left(\partial_u \Gamma^{z}_{\ uu} \right) F_{jz} F_{iz} + 4 \gamma^{ij} \left(\partial_u \Gamma^{k}_{\ uu} \right) F_{jk} F_{iz}\\
		& + 4 \gamma^{ij} \gamma^{kl} \left(\partial_u \Gamma^{u}_{\ ul} \right) F_{ju} F_{ik} + 4 \gamma^{ij} \gamma^{kl} \left(\partial_u \Gamma^{z}_{\ ul} \right) F_{jz} F_{ik} + 4 \gamma^{ij} \gamma^{kl} \left(\partial_u \Gamma^{m}_{\ ul} \right) F_{jm} F_{ik}\\
		= & 4 \left(\partial_u \Gamma^{u}_{\ uu} \right) F_{zu} F_{uz} + 4 \left(\partial_u \Gamma^{i}_{\ uu} \right) F_{zi} F_{uz} + 4 \left(\partial_u \Gamma^{z}_{\ uz} \right) F_{uz} F_{zu}\\
		& + 4 \gamma^{ij} \left(\partial_u \Gamma^{z}_{\ uj} \right) F_{uz} F_{zi} + 4 \gamma^{ij} \left(\partial_u \Gamma^{z}_{\ uu} \right) F_{jz} F_{iz} + 4 \gamma^{ij} \left(\partial_u \Gamma^{k}_{\ uu} \right) F_{jk} F_{iz}\\
		& + 4 \gamma^{ij} \gamma^{kl} \left(\partial_u \Gamma^{z}_{\ ul} \right) F_{jz} F_{ik} + 4 \gamma^{ij} \gamma^{kl} \left(\partial_u \Gamma^{m}_{\ ul} \right) F_{jm} F_{ik}\,.
	\end{split}
\end{equation}
The first term of Eq. (\ref{rhkk1firsttwelfth}) is 
\begin{equation}
	\begin{split}
		& 4 \left(\partial_u \Gamma^{u}_{\ uu} \right) F_{zu} F_{uz}\\
		= & 4 \partial_u \left(- z^2 \partial_z \alpha - 2 z \alpha \right) F_{zu} F_{uz}\\
		= & 4 z \partial_u \left(- z \partial_z \alpha - 2 \alpha \right) F_{zu} F_{uz}\\
		= & 0\,.
	\end{split}
\end{equation}
The second term of Eq. (\ref{rhkk1firsttwelfth}) is 
\begin{equation}
	\begin{split}
		& 4 \left(\partial_u \Gamma^{i}_{\ uu} \right) F_{zi} F_{uz}\\
		= & 4 \partial_u \left[\left(z \beta^i \right) \left(z^2 \partial_z \alpha + 2 z \alpha \right) + \gamma^{ij} \left(z \partial_u \beta_j - z^2 \partial_j \alpha \right) \right] F_{zi} F_{uz}\\
		= & 4 z \partial_u \left[\beta^i \left(z^2 \partial_z \alpha + 2 z \alpha \right) + \gamma^{ij} \left(\partial_u \beta_j - z \partial_j \alpha \right) \right] F_{zi} F_{uz}\\
		= & 0\,.
	\end{split}
\end{equation}
The third term of Eq. (\ref{rhkk1firsttwelfth}) is 
\begin{equation}
	\begin{split}
		& 4 \left(\partial_u \Gamma^{z}_{\ uz} \right) F_{uz} F_{zu}\\
		= & 4 \partial_u \left(z^2 \partial_z \alpha + 2 z \alpha - \frac{1}{2} z \beta^i \beta_i - \frac{1}{2} z^2 \beta^i \partial_z \beta_i \right) F_{uz} F_{zu}\\
		= & 4 z \partial_u \left(z \partial_z \alpha + 2 \alpha - \frac{1}{2} \beta^i \beta_i - \frac{1}{2} z \beta^i \partial_z \beta_i \right) F_{uz} F_{zu}\\
		= & 0\,.
	\end{split}
\end{equation}
The fourth term of Eq. (\ref{rhkk1firsttwelfth}) is 
\begin{equation}
	\begin{split}
		& 4 \gamma^{ij} \left(\partial_u \Gamma^{z}_{\ uj} \right) F_{uz} F_{zi}\\
		= & 4 \gamma^{ij} \partial_u \left[z^2 \partial_j \alpha - \frac{1}{2} z^2 \left(\beta^2 - 2 \alpha \right) \left(\beta_j + z \partial_z \beta_j \right) - \frac{1}{2} z \beta^k \left(z \partial_j \beta_k + \partial_u \gamma_{jk} - z \partial_k \beta_j \right) \right]\\
		& \times F_{uz} F_{zi}\\
		= & 4 z \gamma^{ij} \partial_u \left[z \partial_j \alpha - \frac{1}{2} z \left(\beta^2 - 2 \alpha \right) \left(\beta_j + z \partial_z \beta_j \right) - \frac{1}{2} \beta^k \left(z \partial_j \beta_k + \partial_u \gamma_{jk} - z \partial_k \beta_j \right) \right]\\
		& \times F_{uz} F_{zi}\\
		= & 0\,.
	\end{split}
\end{equation}
The fifth term of Eq. (\ref{rhkk1firsttwelfth}) is 
\begin{equation}
	\begin{split}
		& 4 \gamma^{ij} \left(\partial_u \Gamma^{z}_{\ uu} \right) F_{jz} F_{iz}\\
		= & 4 \gamma^{ij} \partial_u \left[z^2 \partial_u \alpha - z^2 \left(\beta^2 - 2 \alpha \right) \left(z^2 \partial_z \alpha + 2 z \alpha \right) - z \beta^k \left(z \partial_u \beta_k - z^2 \partial_k \alpha \right) \right] F_{jz} F_{iz}\\
		= & 4 z \gamma^{ij} \partial_u \left[z \partial_u \alpha - z \left(\beta^2 - 2 \alpha \right) \left(z^2 \partial_z \alpha + 2 z \alpha \right) - \beta^k \left(z \partial_u \beta_k - z^2 \partial_k \alpha \right) \right] F_{jz} F_{iz}\\
		= & 0\,.
	\end{split}
\end{equation}
The sixth term of Eq. (\ref{rhkk1firsttwelfth}) is 
\begin{equation}
	\begin{split}
		& 4 \gamma^{ij} \left(\partial_u \Gamma^{k}_{\ uu} \right) F_{jk} F_{iz}\\
		= & 4 \gamma^{ij} \partial_u \left[\left(z \beta^k \right) \left(z^2 \partial_z \alpha + 2 z \alpha \right) + \gamma^{kl} \left(z \partial_u \beta_l - z^2 \partial_l \alpha \right) \right] F_{jk} F_{iz}\\
		= & 4 z \gamma^{ij} \partial_u \left[\beta^k \left(z^2 \partial_z \alpha + 2 z \alpha \right) + \gamma^{kl} \left(\partial_u \beta_l - z \partial_l \alpha \right) \right] F_{jk} F_{iz}\\
		= & 0\,.
	\end{split}
\end{equation}
The seventh term of Eq. (\ref{rhkk1firsttwelfth}) is 
\begin{equation}
	\begin{split}
		& 4 \gamma^{ij} \gamma^{kl} \left(\partial_u \Gamma^{z}_{\ ul} \right) F_{jz} F_{ik}\\
		= & 4 \gamma^{ij} \gamma^{kl}\\
		& \times \partial_u \left[z^2 \partial_i \alpha - \frac{1}{2} z^2 \left(\beta^2 - 2 \alpha \right) \left(\beta_i + z \partial_z \beta_i \right) - \frac{1}{2} z \beta^j \left(z \partial_i \beta_j + \partial_u \gamma_{ij} - z \partial_j \beta_i \right) \right] F_{jz} F_{ik}\\
		= & 4 z \gamma^{ij} \gamma^{kl}\\
		& \times \partial_u \left[z \partial_i \alpha - \frac{1}{2} z \left(\beta^2 - 2 \alpha \right) \left(\beta_i + z \partial_z \beta_i \right) - \frac{1}{2} \beta^j \left(z \partial_i \beta_j + \partial_u \gamma_{ij} - z \partial_j \beta_i \right) \right] F_{jz} F_{ik}\\
		= & 0\,.
	\end{split}
\end{equation}
The eighth term of Eq. (\ref{rhkk1firsttwelfth}) is
\begin{equation}
	\begin{split}
		& 4 \gamma^{ij} \gamma^{kl} \left(\partial_u \Gamma^{m}_{\ ul} \right) F_{jm} F_{ik}\\
		= & 4 \gamma^{ij} \gamma^{kl} \partial_u \left[\frac{1}{2} \left(z \beta^m \right) \left(\beta_l + z \partial_z \beta_l \right) + \frac{1}{2} \gamma^{mn} \left(z \partial_l \beta_n + \partial_u \gamma_{ln} - z \partial_n \beta_l \right) \right] F_{jm} F_{ik}\\
		= & 4 \gamma^{ij} \gamma^{kl} \left[\frac{1}{2} \left(z \partial_u \beta^m \right) \left(\beta_l + z \partial_z \beta_l \right) + \frac{1}{2} \left(z \beta^m \right) \left(\partial_u \beta_l + z \partial_u \partial_z \beta_l \right)\right.\\
		& \left. + \frac{1}{2} \left(\partial_u \gamma^{mn} \right) \left(z \partial_l \beta_n + \partial_u \gamma_{ln} - z \partial_n \beta_l \right) + \frac{1}{2} \gamma^{mn} \left(z \partial_u \partial_l \beta_n + \partial_u \partial_u \gamma_{ln} - z \partial_u \partial_n \beta_l \right) \right]\\
		& \times F_{jm} F_{ik}\\
		= & 4 \gamma^{ij} \gamma^{kl} \left[\frac{1}{2} \left(\partial_u \gamma^{mn} \right) \left(\partial_u \gamma_{ln} \right) \right] F_{jm} F_{ik} + 4 \gamma^{ij} \gamma^{kl} \left(\frac{1}{2} \gamma^{mn} \partial_u \partial_u \gamma_{ln} \right) F_{jm} F_{ik}\\
		= & 2 \gamma^{ij} \gamma^{kl} \left(\partial_u \gamma^{mn} \right) \left(\partial_u \gamma_{ln} \right) F_{jm} F_{ik} + 2 \gamma^{ij} \gamma^{kl} \gamma^{mn} \left(\partial_u \partial_u \gamma_{ln} \right) F_{jm} F_{ik}\,.
	\end{split}
\end{equation}
Therefore, the twelfth term of Eq. (\ref{rhkk1first}) is obtained as
\begin{equation}
	\begin{split}
		& 4 k^a k^b g^{ce} g^{df} \left(\partial_a \Gamma^{g}_{\ bf} \right) F_{eg} F_{cd}\\
		= & 2 \gamma^{ij} \gamma^{kl} \left(\partial_u \gamma^{mn} \right) \left(\partial_u \gamma_{ln} \right) F_{jm} F_{ik} + 2 \gamma^{ij} \gamma^{kl} \gamma^{mn} \left(\partial_u \partial_u \gamma_{ln} \right) F_{jm} F_{ik}\\
		= & 8 \gamma^{ij} \gamma^{kl} K^{mn} K_{ln} F_{jm} F_{ik} + 2 \gamma^{ij} \gamma^{kl} \gamma^{mn} \left(\partial_u \partial_u \gamma_{ln} \right) F_{jm} F_{ik}\\
		= & 2 \gamma^{ij} \gamma^{kl} \gamma^{mn} \left(\partial_u \partial_u \gamma_{ln} \right) F_{jm} F_{ik}\,.
	\end{split}
\end{equation}

The thirteenth term of Eq. (\ref{rhkk1first}) is
\begin{equation}
	\begin{split}
		& - 4 k^a k^b g^{ce} g^{df} \Gamma^{h}_{\ ab} \Gamma^{g}_{\ hf} F_{eg} F_{cd} = - 4 g^{ce} g^{df} \Gamma^{h}_{\ uu} \Gamma^{g}_{\ hf} F_{eg} F_{cd} = 0\,.
	\end{split}
\end{equation}

The fourteenth term of Eq. (\ref{rhkk1first}) is
\begin{equation}
	\begin{split}
		& - 4 k^a k^b g^{ce} g^{df} \Gamma^{h}_{\ af} \Gamma^{g}_{\ bh} F_{eg} F_{cd} = - 4 g^{ce} g^{df} \Gamma^{h}_{\ uf} \Gamma^{g}_{\ uh} F_{eg} F_{cd}\\
		= & - 4 \Gamma^{h}_{\ uz} \Gamma^{g}_{\ uh} F_{ug} F_{zu} - 4 \gamma^{ij} \Gamma^{h}_{\ uj} \Gamma^{g}_{\ uh} F_{ug} F_{zi} - 4 \gamma^{ij} \gamma^{kl} \Gamma^{h}_{\ ul} \Gamma^{g}_{\ uh} F_{jg} F_{ik}\,.
	\end{split}
\end{equation}
The index $g$ should be further expanded.
\begin{equation}\label{rhkk1firstfourteenth}
	\begin{split}
		& - 4 \Gamma^{h}_{\ uz} \Gamma^{g}_{\ uh} F_{ug} F_{zu} - 4 \gamma^{ij} \Gamma^{h}_{\ uj} \Gamma^{g}_{\ uh} F_{ug} F_{zi} - 4 \gamma^{ij} \gamma^{kl} \Gamma^{h}_{\ ul} \Gamma^{g}_{\ uh} F_{jg} F_{ik}\\
		= & - 4 \Gamma^{h}_{\ uz} \Gamma^{u}_{\ uh} F_{uu} F_{zu} - 4 \Gamma^{h}_{\ uz} \Gamma^{z}_{\ uh} F_{uz} F_{zu} - 4 \Gamma^{h}_{\ uz} \Gamma^{i}_{\ uh} F_{ui} F_{zu}\\
		& - 4 \gamma^{ij} \Gamma^{h}_{\ uj} \Gamma^{u}_{\ uh} F_{uu} F_{zi} - 4 \gamma^{ij} \Gamma^{h}_{\ uj} \Gamma^{z}_{\ uh} F_{uz} F_{zi} - 4 \gamma^{ij} \Gamma^{h}_{\ uj} \Gamma^{k}_{\ uh} F_{uk} F_{zi}\\
		& - 4 \gamma^{ij} \gamma^{kl} \Gamma^{h}_{\ ul} \Gamma^{u}_{\ uh} F_{ju} F_{ik} - 4 \gamma^{ij} \gamma^{kl} \Gamma^{h}_{\ ul} \Gamma^{z}_{\ uh} F_{jz} F_{ik} - 4 \gamma^{ij} \gamma^{kl} \Gamma^{h}_{\ ul} \Gamma^{m}_{\ uh} F_{jm} F_{ik}\\
		= & - 4 \Gamma^{h}_{\ uz} \Gamma^{z}_{\ uh} F_{uz} F_{zu} - 4 \gamma^{ij} \Gamma^{h}_{\ uj} \Gamma^{z}_{\ uh} F_{uz} F_{zi} - 4 \gamma^{ij} \gamma^{kl} \Gamma^{h}_{\ ul} \Gamma^{z}_{\ uh} F_{jz} F_{ik}\\
		& - 4 \gamma^{ij} \gamma^{kl} \Gamma^{h}_{\ ul} \Gamma^{m}_{\ uh} F_{jm} F_{ik}\,.
	\end{split}
\end{equation}
The repeated index $h$ should be further expanded. The first term of Eq. (\ref{rhkk1firstfourteenth}) is 
\begin{equation}
	\begin{split}
		& - 4 \Gamma^{h}_{\ uz} \Gamma^{z}_{\ uh} F_{uz} F_{zu}\\
		= & - 4 \Gamma^{u}_{\ uz} \Gamma^{z}_{\ uu} F_{uz} F_{zu} - 4 \Gamma^{z}_{\ uz} \Gamma^{z}_{\ uz} F_{uz} F_{zu} - 4 \Gamma^{i}_{\ uz} \Gamma^{z}_{\ ui} F_{uz} F_{zu}\\
		= & 0\,.
	\end{split}
\end{equation}
The second term of Eq. (\ref{rhkk1firstfourteenth}) is
\begin{equation}
	\begin{split}
		& - 4 \gamma^{ij} \Gamma^{h}_{\ uj} \Gamma^{z}_{\ uh} F_{uz} F_{zi}\\
		= & - 4 \gamma^{ij} \Gamma^{u}_{\ uj} \Gamma^{z}_{\ uu} F_{uz} F_{zi} - 4 \gamma^{ij} \Gamma^{z}_{\ uj} \Gamma^{z}_{\ uz} F_{uz} F_{zi} - 4 \gamma^{ij} \Gamma^{k}_{\ uj} \Gamma^{z}_{\ uk} F_{uz} F_{zi}\\
		= & 0\,.
	\end{split}
\end{equation}
The third term of Eq. (\ref{rhkk1firstfourteenth}) is
\begin{equation}
	\begin{split}
		& - 4 \gamma^{ij} \gamma^{kl} \Gamma^{h}_{\ ul} \Gamma^{z}_{\ uh} F_{jz} F_{ik}\\
		= & - 4 \gamma^{ij} \gamma^{kl} \Gamma^{u}_{\ ul} \Gamma^{z}_{\ uu} F_{jz} F_{ik} - 4 \gamma^{ij} \gamma^{kl} \Gamma^{z}_{\ ul} \Gamma^{z}_{\ uz} F_{jz} F_{ik} - 4 \gamma^{ij} \gamma^{kl} \Gamma^{m}_{\ ul} \Gamma^{z}_{\ um} F_{jz} F_{ik}\\
		= & 0\,.
	\end{split}
\end{equation}
The fourth term of Eq. (\ref{rhkk1firstfourteenth}) is
\begin{equation}
	\begin{split}
		& - 4 \gamma^{ij} \gamma^{kl} \Gamma^{h}_{\ ul} \Gamma^{m}_{\ uh} F_{jm} F_{ik}\\
		= & - 4 \gamma^{ij} \gamma^{kl} \Gamma^{u}_{\ ul} \Gamma^{m}_{\ uu} F_{jm} F_{ik} - 4 \gamma^{ij} \gamma^{kl} \Gamma^{z}_{\ ul} \Gamma^{m}_{\ uz} F_{jm} F_{ik} - 4 \gamma^{ij} \gamma^{kl} \Gamma^{n}_{\ ul} \Gamma^{m}_{\ un} F_{jm} F_{ik}\\
		= & - \gamma^{ij} \gamma^{kl} \gamma^{no} \gamma^{mp} \left(\partial_u \gamma_{lo} \right) \left(\partial_u \gamma_{np} \right) F_{jm} F_{ik}\,.
	\end{split}
\end{equation}
Therefore, the fourteenth term of Eq. (\ref{rhkk1first}) is obtained as
\begin{equation}
	\begin{split}
		& - 4 k^a k^b g^{ce} g^{df} \Gamma^{h}_{\ af} \Gamma^{g}_{\ bh} F_{eg} F_{cd} = - \gamma^{ij} \gamma^{kl} \gamma^{no} \gamma^{mp} \left(\partial_u \gamma_{lo} \right) \left(\partial_u \gamma_{np} \right) F_{jm} F_{ik}\\
		= & - 4 \gamma^{ij} \gamma^{kl} \gamma^{no} \gamma^{mp} K_{lo} K_{np} F_{jm} F_{ik}\\
		= & 0\,.
	\end{split}
\end{equation}

The fifteenth term of Eq. (\ref{rhkk1first}) is
\begin{equation}
	\begin{split}
		& 4 k^a k^b g^{ce} g^{df} \Gamma^{g}_{\ ah} \Gamma^{h}_{\ bf} F_{eg} F_{cd} = 4 g^{ce} g^{df} \Gamma^{g}_{\ uh} \Gamma^{h}_{\ uf} F_{eg} F_{cd}\\
		= & 4 \Gamma^{g}_{\ uh} \Gamma^{h}_{\ uz} F_{ug} F_{zu} + 4 \gamma^{ij} \Gamma^{g}_{\ uh} \Gamma^{h}_{\ uj} F_{ug} F_{zi} + 4 \gamma^{ij} \gamma^{kl} \Gamma^{g}_{\ uh} \Gamma^{h}_{\ ul} F_{jg} F_{ik}\,.
	\end{split}
\end{equation}
The repeated index $g$ should be further expanded using the metric as
\begin{equation}\label{rhkk1firstfifteenth}
	\begin{split}
		& 4 \Gamma^{g}_{\ uh} \Gamma^{h}_{\ uz} F_{ug} F_{zu} + 4 \gamma^{ij} \Gamma^{g}_{\ uh} \Gamma^{h}_{\ uj} F_{ug} F_{zi} + 4 \gamma^{ij} \gamma^{kl} \Gamma^{g}_{\ uh} \Gamma^{h}_{\ ul} F_{jg} F_{ik}\\
		= & 4 \Gamma^{u}_{\ uh} \Gamma^{h}_{\ uz} F_{uu} F_{zu} + 4 \Gamma^{z}_{\ uh} \Gamma^{h}_{\ uz} F_{uz} F_{zu} + 4 \Gamma^{i}_{\ uh} \Gamma^{h}_{\ uz} F_{ui} F_{zu}\\
		& + 4 \gamma^{ij} \Gamma^{u}_{\ uh} \Gamma^{h}_{\ uj} F_{uu} F_{zi} + 4 \gamma^{ij} \Gamma^{z}_{\ uh} \Gamma^{h}_{\ uj} F_{uz} F_{zi} + 4 \gamma^{ij} \Gamma^{k}_{\ uh} \Gamma^{h}_{\ uj} F_{uk} F_{zi}\\
		& + 4 \gamma^{ij} \gamma^{kl} \Gamma^{u}_{\ uh} \Gamma^{h}_{\ ul} F_{ju} F_{ik} + 4 \gamma^{ij} \gamma^{kl} \Gamma^{z}_{\ uh} \Gamma^{h}_{\ ul} F_{jz} F_{ik} + 4 \gamma^{ij} \gamma^{kl} \Gamma^{m}_{\ uh} \Gamma^{h}_{\ ul} F_{jm} F_{ik}\\
		= & 4 \Gamma^{z}_{\ uh} \Gamma^{h}_{\ uz} F_{uz} F_{zu} + 4 \gamma^{ij} \Gamma^{z}_{\ uh} \Gamma^{h}_{\ uj} F_{uz} F_{zi} + 4 \gamma^{ij} \gamma^{kl} \Gamma^{z}_{\ uh} \Gamma^{h}_{\ ul} F_{jz} F_{ik}\\
		& + 4 \gamma^{ij} \gamma^{kl} \Gamma^{m}_{\ uh} \Gamma^{h}_{\ ul} F_{jm} F_{ik}\,.
	\end{split}
\end{equation}
The repeated index $h$ should be further expanded. The first term of Eq. (\ref{rhkk1firstfifteenth}) is
\begin{equation}
	\begin{split}
		& 4 \Gamma^{z}_{\ uh} \Gamma^{h}_{\ uz} F_{uz} F_{zu}\\
		= & 4 \Gamma^{z}_{\ uu} \Gamma^{u}_{\ uz} F_{uz} F_{zu} + 4 \Gamma^{z}_{\ uz} \Gamma^{z}_{\ uz} F_{uz} F_{zu} + 4 \Gamma^{z}_{\ ui} \Gamma^{i}_{\ uz} F_{uz} F_{zu}\\
		= & 0\,.
	\end{split}
\end{equation}
The second term of Eq. (\ref{rhkk1firstfifteenth}) is
\begin{equation}
	\begin{split}
		& 4 \gamma^{ij} \Gamma^{z}_{\ uh} \Gamma^{h}_{\ uj} F_{uz} F_{zi}\\
		= & 4 \gamma^{ij} \Gamma^{z}_{\ uu} \Gamma^{u}_{\ uj} F_{uz} F_{zi} + 4 \gamma^{ij} \Gamma^{z}_{\ uz} \Gamma^{z}_{\ uj} F_{uz} F_{zi} + 4 \gamma^{ij} \Gamma^{z}_{\ uk} \Gamma^{k}_{\ uj} F_{uz} F_{zi}\\
		= & 0\,.
	\end{split}
\end{equation}
The third term of Eq. (\ref{rhkk1firstfifteenth}) is
\begin{equation}
	\begin{split}
		& 4 \gamma^{ij} \gamma^{kl} \Gamma^{z}_{\ uh} \Gamma^{h}_{\ ul} F_{jz} F_{ik}\\
		= & 4 \gamma^{ij} \gamma^{kl} \Gamma^{z}_{\ uu} \Gamma^{u}_{\ ul} F_{jz} F_{ik} + 4 \gamma^{ij} \gamma^{kl} \Gamma^{z}_{\ uz} \Gamma^{z}_{\ ul} F_{jz} F_{ik} + 4 \gamma^{ij} \gamma^{kl} \Gamma^{z}_{\ um} \Gamma^{m}_{\ ul} F_{jz} F_{ik}\\
		= & 0\,.
	\end{split}
\end{equation}
The fourth term of Eq. (\ref{rhkk1firstfifteenth}) is
\begin{equation}
	\begin{split}
		& 4 \gamma^{ij} \gamma^{kl} \Gamma^{m}_{\ uh} \Gamma^{h}_{\ ul} F_{jm} F_{ik}\\
		= & 4 \gamma^{ij} \gamma^{kl} \Gamma^{m}_{\ uu} \Gamma^{u}_{\ ul} F_{jm} F_{ik} + 4 \gamma^{ij} \gamma^{kl} \Gamma^{m}_{\ uz} \Gamma^{z}_{\ ul} F_{jm} F_{ik} + 4 \gamma^{ij} \gamma^{kl} \Gamma^{m}_{\ un} \Gamma^{n}_{\ ul} F_{jm} F_{ik}\\
		= & \gamma^{ij} \gamma^{kl} \gamma^{mo} \gamma^{np} \left(\partial_u \gamma_{no} \right) \left(\partial_u \gamma_{lp} \right) F_{jm} F_{ik}\,.
	\end{split}
\end{equation}
Therefore, the fifteenth term of Eq. (\ref{rhkk1first}) is obtained as
\begin{equation}
	\begin{split}
		& 4 k^a k^b g^{ce} g^{df} \Gamma^{g}_{\ ah} \Gamma^{h}_{\ bf} F_{eg} F_{cd} = \gamma^{ij} \gamma^{kl} \gamma^{mo} \gamma^{np} \left(\partial_u \gamma_{no} \right) \left(\partial_u \gamma_{lp} \right) F_{jm} F_{ik}\\
		= & 4 \gamma^{ij} \gamma^{kl} \gamma^{mo} \gamma^{np} K_{no} K_{lp} F_{jm} F_{ik}\\
		= & 0\,.
	\end{split}
\end{equation}

The sixteenth term of Eq. (\ref{rhkk1first}) is
\begin{equation}
	\begin{split}
		& 4 k^a k^b g^{ce} g^{df} \Gamma^{g}_{\ bf} \left(\partial_a F_{eg} \right) F_{cd} = 4 g^{ce} g^{df} \Gamma^{g}_{\ uf} \left(\partial_u F_{eg} \right) F_{cd}\\
		= & 4 \Gamma^{g}_{\ uz} \left(\partial_u F_{ug} \right) F_{zu} + 4 \gamma^{ij} \Gamma^{g}_{\ uj} \left(\partial_u F_{ug} \right) F_{zi} + 4 \gamma^{ij} \gamma^{kl} \Gamma^{g}_{\ ul} \left(\partial_u F_{jg} \right) F_{ik}\,.
	\end{split}
\end{equation}
The index $g$ should be further expanded.
\begin{equation}\label{rhkk1firstsixteenth}
	\begin{split}
		&4 \Gamma^{g}_{\ uz} \left(\partial_u F_{ug} \right) F_{zu} + 4 \gamma^{ij} \Gamma^{g}_{\ uj} \left(\partial_u F_{ug} \right) F_{zi} + 4 \gamma^{ij} \gamma^{kl} \Gamma^{g}_{\ ul} \left(\partial_u F_{jg} \right) F_{ik}\\
		= & 4 \Gamma^{u}_{\ uz} \left(\partial_u F_{uu} \right) F_{zu} + 4 \Gamma^{z}_{\ uz} \left(\partial_u F_{uz} \right) F_{zu} + 4 \Gamma^{i}_{\ uz} \left(\partial_u F_{ui} \right) F_{zu}\\
		& + 4 \gamma^{ij} \Gamma^{u}_{\ uj} \left(\partial_u F_{uu} \right) F_{zi} + 4 \gamma^{ij} \Gamma^{z}_{\ uj} \left(\partial_u F_{uz} \right) F_{zi} + 4 \gamma^{ij} \Gamma^{k}_{\ uj} \left(\partial_u F_{uk} \right) F_{zi}\\
		& + 4 \gamma^{ij} \gamma^{kl} \Gamma^{u}_{\ ul} \left(\partial_u F_{ju} \right) F_{ik} + 4 \gamma^{ij} \gamma^{kl} \Gamma^{z}_{\ ul} \left(\partial_u F_{jz} \right) F_{ik} + 4 \gamma^{ij} \gamma^{kl} \Gamma^{m}_{\ ul} \left(\partial_u F_{jm} \right) F_{ik}\\
		= & 4 \Gamma^{z}_{\ uz} \left(\partial_u F_{uz} \right) F_{zu} + 4 \Gamma^{i}_{\ uz} \left(\partial_u F_{ui} \right) F_{zu} + 4 \gamma^{ij} \Gamma^{z}_{\ uj} \left(\partial_u F_{uz} \right) F_{zi}\\
		& + 4 \gamma^{ij} \Gamma^{k}_{\ uj} \left(\partial_u F_{uk} \right) F_{zi} + 4 \gamma^{ij} \gamma^{kl} \Gamma^{u}_{\ ul} \left(\partial_u F_{ju} \right) F_{ik} + 4 \gamma^{ij} \gamma^{kl} \Gamma^{z}_{\ ul} \left(\partial_u F_{jz} \right) F_{ik}\\
		& + 4 \gamma^{ij} \gamma^{kl} \Gamma^{m}_{\ ul} \left(\partial_u F_{jm} \right) F_{ik}\,.
	\end{split}
\end{equation}
Therefore, the sixteenth term of Eq. (\ref{rhkk1first}) is obtained as
\begin{equation}
	\begin{split}
		& 4 k^a k^b g^{ce} g^{df} \Gamma^{g}_{\ bf} \left(\partial_a F_{eg} \right) F_{cd}\\
		= & 2 \gamma^{ij} \beta_j \left(\partial_u F_{ui} \right) F_{zu} + 2 \gamma^{ij} \gamma^{kl} \left(\partial_u \gamma_{jl} \right) \left(\partial_u F_{uk} \right) F_{zi} - 2 \gamma^{ij} \gamma^{kl} \beta_l \left(\partial_u F_{ju} \right) F_{ik}\\
		& + 2 \gamma^{ij} \gamma^{kl} \gamma^{mn} \left(\partial_u \gamma_{ln} \right) \left(\partial_u F_{jm} \right) F_{ik}\\
		= & 2 \gamma^{ij} \beta_j \left(\partial_u F_{ui} \right) F_{zu} + 4 \gamma^{ij} \gamma^{kl} K_{jl} \left(\partial_u F_{uk} \right) F_{zi} - 2 \gamma^{ij} \gamma^{kl} \beta_l \left(\partial_u F_{ju} \right) F_{ik}\\
		& + 4 \gamma^{ij} \gamma^{kl} \gamma^{mn} K_{ln} \left(\partial_u F_{jm} \right) F_{ik}\\
		= & 2 \gamma^{ij} \beta_j \left(\partial_u F_{ui} \right) F_{zu} - 2 \gamma^{ij} \gamma^{kl} \beta_l \left(\partial_u F_{ju} \right) F_{ik}\,.
	\end{split}
\end{equation}

The seventeenth term of Eq. (\ref{rhkk1first}) is
\begin{equation}
	\begin{split}
		& - 4 k^a k^b g^{ce} g^{df} \Gamma^{g}_{\ bf} \Gamma^{h}_{\ ae} F_{hg} F_{cd} = - 4 g^{ce} g^{df} \Gamma^{g}_{\ uf} \Gamma^{h}_{\ ue} F_{hg} F_{cd}\\
		= & - 4 \Gamma^{g}_{\ uu} \Gamma^{h}_{\ uz} F_{hg} F_{uz} - 4 \Gamma^{g}_{\ uz} \Gamma^{h}_{\ uu} F_{hg} F_{zu} - 4 \gamma^{ij} \Gamma^{g}_{\ uj} \Gamma^{h}_{\ uu} F_{hg} F_{zi}\\
		& - 4 \gamma^{ij} \Gamma^{g}_{\ uu} \Gamma^{h}_{\ uj} F_{hg} F_{iz} - 4 \gamma^{ij} \gamma^{kl} \Gamma^{g}_{\ ul} \Gamma^{h}_{\ uj} F_{hg} F_{ik}\\
		= & - 4 \gamma^{ij} \gamma^{kl} \Gamma^{g}_{\ ul} \Gamma^{h}_{\ uj} F_{hg} F_{ik}\,.
	\end{split}
\end{equation}
The repeated index $g$ should be further expanded as
\begin{equation}\label{rhkk1firstseventeenth}
	\begin{split}
		& - 4 \gamma^{ij} \gamma^{kl} \Gamma^{g}_{\ ul} \Gamma^{h}_{\ uj} F_{hg} F_{ik}\\
		= & - 4 \gamma^{ij} \gamma^{kl} \Gamma^{u}_{\ ul} \Gamma^{h}_{\ uj} F_{hu} F_{ik} - 4 \gamma^{ij} \gamma^{kl} \Gamma^{z}_{\ ul} \Gamma^{h}_{\ uj} F_{hz} F_{ik} - 4 \gamma^{ij} \gamma^{kl} \Gamma^{m}_{\ ul} \Gamma^{h}_{\ uj} F_{hm} F_{ik}\,.
	\end{split}
\end{equation}
The repeated index $h$ should be further expanded using the metric. The first term of Eq. (\ref{rhkk1firstseventeenth}) is
\begin{equation}
	\begin{split}
		& - 4 \gamma^{ij} \gamma^{kl} \Gamma^{u}_{\ ul} \Gamma^{h}_{\ uj} F_{hu} F_{ik}\\
		= & - 4 \gamma^{ij} \gamma^{kl} \Gamma^{u}_{\ ul} \Gamma^{u}_{\ uj} F_{uu} F_{ik} - 4 \gamma^{ij} \gamma^{kl} \Gamma^{u}_{\ ul} \Gamma^{z}_{\ uj} F_{zu} F_{ik} - 4 \gamma^{ij} \gamma^{kl} \Gamma^{u}_{\ ul} \Gamma^{m}_{\ uj} F_{mu} F_{ik}\\
		= & 0\,.
	\end{split}
\end{equation}
The second term of Eq. (\ref{rhkk1firstseventeenth}) is
\begin{equation}
	\begin{split}
		- 4 \gamma^{ij} \gamma^{kl} \Gamma^{z}_{\ ul} \Gamma^{h}_{\ uj} F_{hz} F_{ik} = 0\,.
	\end{split}
\end{equation}
The third term of Eq. (\ref{rhkk1firstseventeenth}) is
\begin{equation}
	\begin{split}
		& - 4 \gamma^{ij} \gamma^{kl} \Gamma^{m}_{\ ul} \Gamma^{h}_{\ uj} F_{hm} F_{ik}\\
		= & - 4 \gamma^{ij} \gamma^{kl} \Gamma^{m}_{\ ul} \Gamma^{u}_{\ uj} F_{um} F_{ik} - 4 \gamma^{ij} \gamma^{kl} \Gamma^{m}_{\ ul} \Gamma^{z}_{\ uj} F_{zm} F_{ik} - 4 \gamma^{ij} \gamma^{kl} \Gamma^{m}_{\ ul} \Gamma^{n}_{\ uj} F_{nm} F_{ik}\\
		= & - \gamma^{ij} \gamma^{kl} \gamma^{mo} \gamma^{np} \left(\partial_u \gamma_{lo} \right) \left(\partial_u \gamma_{jp} \right) F_{nm} F_{ik}\,.
	\end{split}
\end{equation}
Therefore, the seventeenth term of Eq. (\ref{rhkk1first}) is obtained as
\begin{equation}
	\begin{split}
		& - 4 k^a k^b g^{ce} g^{df} \Gamma^{g}_{\ bf} \Gamma^{h}_{\ ae} F_{hg} F_{cd} = - \gamma^{ij} \gamma^{kl} \gamma^{mo} \gamma^{np} \left(\partial_u \gamma_{lo} \right) \left(\partial_u \gamma_{jp} \right) F_{nm} F_{ik}\\
		= & - 4 \gamma^{ij} \gamma^{kl} \gamma^{mo} \gamma^{np} K_{lo} K_{jp} F_{nm} F_{ik}\\
		= & 0\,.
	\end{split}
\end{equation}

The eighteenth term of Eq. (\ref{rhkk1first}) is
\begin{equation}
	\begin{split}
		& - 4 k^a k^b g^{ce} g^{df} \Gamma^{g}_{\ bf} \Gamma^{h}_{\ ag} F_{eh} F_{cd} = - 4 g^{ce} g^{df} \Gamma^{g}_{\ uf} \Gamma^{h}_{\ ug} F_{eh} F_{cd}\\
		= & - 4 \Gamma^{g}_{\ uz} \Gamma^{h}_{\ ug} F_{uh} F_{zu} - 4 \gamma^{ij} \Gamma^{g}_{\ uj} \Gamma^{h}_{\ ug} F_{uh} F_{zi} - 4 \gamma^{ij} \gamma^{kl} \Gamma^{g}_{\ ul} \Gamma^{h}_{\ ug} F_{jh} F_{ik}\,.
	\end{split}
\end{equation}
The index $g$ should be further expanded.
\begin{equation}\label{rhkk1firsteighteenth}
	\begin{split}
		& - 4 \Gamma^{g}_{\ uz} \Gamma^{h}_{\ ug} F_{uh} F_{zu} - 4 \gamma^{ij} \Gamma^{g}_{\ uj} \Gamma^{h}_{\ ug} F_{uh} F_{zi} - 4 \gamma^{ij} \gamma^{kl} \Gamma^{g}_{\ ul} \Gamma^{h}_{\ ug} F_{jh} F_{ik}\\
		= & - 4 \Gamma^{u}_{\ uz} \Gamma^{h}_{\ uu} F_{uh} F_{zu} - 4 \Gamma^{z}_{\ uz} \Gamma^{h}_{\ uz} F_{uh} F_{zu} - 4 \Gamma^{i}_{\ uz} \Gamma^{h}_{\ ui} F_{uh} F_{zu}\\
		& - 4 \gamma^{ij} \Gamma^{u}_{\ uj} \Gamma^{h}_{\ uu} F_{uh} F_{zi} - 4 \gamma^{ij} \Gamma^{z}_{\ uj} \Gamma^{h}_{\ uz} F_{uh} F_{zi} - 4 \gamma^{ij} \Gamma^{k}_{\ uj} \Gamma^{h}_{\ uk} F_{uh} F_{zi}\\
		& - 4 \gamma^{ij} \gamma^{kl} \Gamma^{u}_{\ ul} \Gamma^{h}_{\ uu} F_{jh} F_{ik} - 4 \gamma^{ij} \gamma^{kl} \Gamma^{z}_{\ ul} \Gamma^{h}_{\ uz} F_{jh} F_{ik} - 4 \gamma^{ij} \gamma^{kl} \Gamma^{m}_{\ ul} \Gamma^{h}_{\ um} F_{jh} F_{ik}\\
		= & - 4 \Gamma^{z}_{\ uz} \Gamma^{h}_{\ uz} F_{uh} F_{zu} - 4 \Gamma^{i}_{\ uz} \Gamma^{h}_{\ ui} F_{uh} F_{zu} - 4 \gamma^{ij} \Gamma^{z}_{\ uj} \Gamma^{h}_{\ uz} F_{uh} F_{zi}\\
		& - 4 \gamma^{ij} \Gamma^{k}_{\ uj} \Gamma^{h}_{\ uk} F_{uh} F_{zi} - 4 \gamma^{ij} \gamma^{kl} \Gamma^{z}_{\ ul} \Gamma^{h}_{\ uz} F_{jh} F_{ik} - 4 \gamma^{ij} \gamma^{kl} \Gamma^{m}_{\ ul} \Gamma^{h}_{\ um} F_{jh} F_{ik}\,.
	\end{split}
\end{equation}
The repeated index $h$ should be further expanded. The first term of Eq. (\ref{rhkk1firsteighteenth}) is 
\begin{equation}
	\begin{split}
		- 4 \Gamma^{z}_{\ uz} \Gamma^{h}_{\ uz} F_{uh} F_{zu} = 0\,.
	\end{split}
\end{equation}
The second term of Eq. (\ref{rhkk1firsteighteenth}) is 
\begin{equation}
	\begin{split}
		& - 4 \Gamma^{i}_{\ uz} \Gamma^{h}_{\ ui} F_{uh} F_{zu}\\
		= & - 4 \Gamma^{i}_{\ uz} \Gamma^{u}_{\ ui} F_{uu} F_{zu} - 4 \Gamma^{i}_{\ uz} \Gamma^{z}_{\ ui} F_{uz} F_{zu} - 4 \Gamma^{i}_{\ uz} \Gamma^{j}_{\ ui} F_{uj} F_{zu}\\
		= & 0\,.
	\end{split}
\end{equation}
The third term of Eq. (\ref{rhkk1firsteighteenth}) is 
\begin{equation}
	\begin{split}
		- 4 \gamma^{ij} \Gamma^{z}_{\ uj} \Gamma^{h}_{\ uz} F_{uh} F_{zi} = 0\,.
	\end{split}
\end{equation}
The fourth term of Eq. (\ref{rhkk1firsteighteenth}) is 
\begin{equation}
	\begin{split}
		& - 4 \gamma^{ij} \Gamma^{k}_{\ uj} \Gamma^{h}_{\ uk} F_{uh} F_{zi}\\
		= & - 4 \gamma^{ij} \Gamma^{k}_{\ uj} \Gamma^{u}_{\ uk} F_{uu} F_{zi} - 4 \gamma^{ij} \Gamma^{k}_{\ uj} \Gamma^{z}_{\ uk} F_{uz} F_{zi} - 4 \gamma^{ij} \Gamma^{k}_{\ uj} \Gamma^{l}_{\ uk} F_{ul} F_{zi}\\
		= & 0\,.
	\end{split}
\end{equation}
The fifth term of Eq. (\ref{rhkk1firsteighteenth}) is 
\begin{equation}
	\begin{split}
		- 4 \gamma^{ij} \gamma^{kl} \Gamma^{z}_{\ ul} \Gamma^{h}_{\ uz} F_{jh} F_{ik} = 0\,.
	\end{split}
\end{equation}
The sixth term of Eq. (\ref{rhkk1firsteighteenth}) is 
\begin{equation}
	\begin{split}
		& - 4 \gamma^{ij} \gamma^{kl} \Gamma^{m}_{\ ul} \Gamma^{h}_{\ um} F_{jh} F_{ik}\\
		= & - 4 \gamma^{ij} \gamma^{kl} \Gamma^{m}_{\ ul} \Gamma^{u}_{\ um} F_{ju} F_{ik} - 4 \gamma^{ij} \gamma^{kl} \Gamma^{m}_{\ ul} \Gamma^{z}_{\ um} F_{jz} F_{ik} - 4 \gamma^{ij} \gamma^{kl} \Gamma^{m}_{\ ul} \Gamma^{n}_{\ um} F_{jn} F_{ik}\\
		= & - \gamma^{ij} \gamma^{kl} \gamma^{mo} \gamma^{np} \left(\partial_u \gamma_{lo} \right) \left(\partial_u \gamma_{mp} \right) F_{jn} F_{ik}\,.
	\end{split}
\end{equation}
Therefore, the eighteenth term of Eq. (\ref{rhkk1first}) is obtained as
\begin{equation}
	\begin{split}
		& - 4 k^a k^b g^{ce} g^{df} \Gamma^{g}_{\ bf} \Gamma^{h}_{\ ag} F_{eh} F_{cd} = - \gamma^{ij} \gamma^{kl} \gamma^{mo} \gamma^{np} \left(\partial_u \gamma_{lo} \right) \left(\partial_u \gamma_{mp} \right) F_{jn} F_{ik}\\
		= & - 4 \gamma^{ij} \gamma^{kl} \gamma^{mo} \gamma^{np} K_{lo} K_{mp} F_{jn} F_{ik}\\
		= & 0\,.
	\end{split}
\end{equation}

Finally, combining with the nonzero results in the eighteen terms, the first term of Eq.(\ref{rhkk1}) is
\begin{equation}
	\begin{split}
		& - 4 k^a k^b \left(\nabla_a \nabla_b F^{cd} \right) F_{cd}\\
		= & 8 \left(\partial_u \partial_u F_{uz} \right) F_{uz} - 8 \gamma^{ij} \left(\partial_u \partial_u F_{ui} \right) F_{zj} - 4 \gamma^{ij} \gamma^{kl} \left(\partial_u \partial_u F_{il} \right) F_{jk}\\
		& + 2 \gamma^{ij} \beta_j \left(\partial_u F_{iu} \right) F_{uz} - 2 \gamma^{ij} \gamma^{kl} \beta_j \left(\partial_u F_{ul} \right) F_{ik} + 2 \gamma^{ij} \beta_j \left(\partial_u F_{ui} \right) F_{zu}\\
		& - 2 \gamma^{ij} \gamma^{kl} \beta_l \left(\partial_u F_{ju} \right) F_{ik} + 2 \gamma^{ij} \gamma^{kl} \gamma^{mn} \left(\partial_u \partial_u \gamma_{jn} \right) F_{ml} F_{ik} + 2 \gamma^{ij} \beta_j \left(\partial_u F_{iu} \right) F_{uz}\\
		& - 2 \gamma^{ij} \gamma^{kl} \beta_j \left(\partial_u F_{ul} \right) F_{ik} + 2 \gamma^{ij} \gamma^{kl} \gamma^{mn} \left(\partial_u \partial_u \gamma_{ln} \right) F_{jm} F_{ik} + 2 \gamma^{ij} \beta_j \left(\partial_u F_{ui} \right) F_{zu}\\
		& - 2 \gamma^{ij} \gamma^{kl} \beta_l \left(\partial_u F_{ju} \right) F_{ik}\\
		= & 8 \left(\partial_u \partial_u F_{uz} \right) F_{uz} - 8 \gamma^{ij} \left(\partial_u \partial_u F_{ui} \right) F_{zj} - 4 \gamma^{ij} \gamma^{kl} \left(\partial_u \partial_u F_{il} \right) F_{jk}\\
		& - 8 \gamma^{ij} \beta_j \left(\partial_u F_{ui} \right) F_{uz} + 8 \gamma^{ij} \gamma^{kl} \beta_l \left(\partial_u F_{uj} \right) F_{ik} + 4 \gamma^{ij} \gamma^{kl} \gamma^{mn} \left(\partial_u \partial_u \gamma_{jn} \right) F_{ml} F_{ik}\,.
	\end{split}
\end{equation}

After expanding the two covariant derivative operations by the definition, the second term of Eq.(\ref{rhkk1}) is
\begin{equation}\label{rhkk1second}
	\begin{split}
		& - 4 k^a k^b \left(\nabla_b F^{cd} \right) \left(\nabla_a F_{cd} \right) = - 4 k^a k^b g^{ce} g^{df} \left(\nabla_b F_{ef} \right) \left(\nabla_a F_{cd} \right)\\
		= & - 4 k^a k^b g^{ce} g^{df} \left(\partial_b F_{ef} - \Gamma^{g}_{\ be} F_{gf} - \Gamma^{g}_{\ bf} F_{eg} \right) \left(\partial_a F_{cd} - \Gamma^{h}_{\ ac} F_{hd} - \Gamma^{h}_{\ ad} F_{ch} \right)\\
		= & - 4 k^a k^b g^{ce} g^{df} \left(\partial_b F_{ef} \right) \left(\partial_a F_{cd} \right) + 4 k^a k^b g^{ce} g^{df} \left(\partial_b F_{ef} \right) \Gamma^{h}_{\ ac} F_{hd}\\
		& + 4 k^a k^b g^{ce} g^{df} \left(\partial_b F_{ef} \right) \Gamma^{h}_{\ ad} F_{ch} + 4 k^a k^b g^{ce} g^{df} \Gamma^{g}_{\ be} F_{gf} \left(\partial_a F_{cd} \right)\\
		& - 4 k^a k^b g^{ce} g^{df} \Gamma^{g}_{\ be} F_{gf} \Gamma^{h}_{\ ac} F_{hd} - 4 k^a k^b g^{ce} g^{df} \Gamma^{g}_{\ be} F_{gf} \Gamma^{h}_{\ ad} F_{ch}\\
		& + 4 k^a k^b g^{ce} g^{df} \Gamma^{g}_{\ bf} F_{eg} \left(\partial_a F_{cd} \right) - 4 k^a k^b g^{ce} g^{df} \Gamma^{g}_{\ bf} F_{eg} \Gamma^{h}_{\ ac} F_{hd}\\
		& - 4 k^a k^b g^{ce} g^{df} \Gamma^{g}_{\ bf} F_{eg} \Gamma^{h}_{\ ad} F_{ch}\,.
	\end{split}
\end{equation}

The first term of Eq. (\ref{rhkk1second}) is 
\begin{equation}
	\begin{split}
		& - 4 k^a k^b g^{ce} g^{df} \left(\partial_b F_{ef} \right) \left(\partial_a F_{cd} \right) = - 4 g^{ce} g^{df} \left(\partial_u F_{ef} \right) \left(\partial_u F_{cd} \right)\\
		= & - 4 \left(\partial_u F_{zu} \right) \left(\partial_u F_{uz} \right) - 4 \gamma^{ij} \left(\partial_u F_{zj} \right) \left(\partial_u F_{ui} \right) - 4 \left(\partial_u F_{uz} \right) \left(\partial_u F_{zu} \right)\\
		& - 4 \gamma^{ij} \left(\partial_u F_{uj} \right) \left(\partial_u F_{zi} \right) - 4 \gamma^{ij} \left(\partial_u F_{jz} \right) \left(\partial_u F_{iu} \right) - 4 \gamma^{ij} \left(\partial_u F_{ju} \right) \left(\partial_u F_{iz} \right)\\
		& - 4 \gamma^{ij} \gamma^{kl} \left(\partial_u F_{jl} \right) \left(\partial_u F_{ik} \right)\,.
	\end{split}
\end{equation}
Therefore, the first term of Eq. (\ref{rhkk1second}) is obtained as
\begin{equation}
	\begin{split}
		& - 4 k^a k^b g^{ce} g^{df} \left(\partial_b F_{ef} \right) \left(\partial_a F_{cd} \right)\\
		= & - 4 \left(\partial_u F_{zu} \right) \left(\partial_u F_{uz} \right) - 4 \gamma^{ij} \left(\partial_u F_{zj} \right) \left(\partial_u F_{ui} \right) - 4 \left(\partial_u F_{uz} \right) \left(\partial_u F_{zu} \right)\\
		& - 4 \gamma^{ij} \left(\partial_u F_{uj} \right) \left(\partial_u F_{zi} \right) - 4 \gamma^{ij} \left(\partial_u F_{jz} \right) \left(\partial_u F_{iu} \right) - 4 \gamma^{ij} \left(\partial_u F_{ju} \right) \left(\partial_u F_{iz} \right)\\
		& - 4 \gamma^{ij} \gamma^{kl} \left(\partial_u F_{jl} \right) \left(\partial_u F_{ik} \right)\\
		= & 8 \left(\partial_u F_{uz} \right) \left(\partial_u F_{uz} \right) - 16 \gamma^{ij} \left(\partial_u F_{ui} \right) \left(\partial_u F_{zj} \right) - 4 \gamma^{ij} \gamma^{kl} \left(\partial_u F_{jl} \right) \left(\partial_u F_{ik} \right)\,.
	\end{split}
\end{equation}

The second term of Eq. (\ref{rhkk1second}) is
\begin{equation}\label{rhkk1secondsecond}
	\begin{split}
		& 4 k^a k^b g^{ce} g^{df} \left(\partial_b F_{ef} \right) \Gamma^{h}_{\ ac} F_{hd} = 4 g^{ce} g^{df} \left(\partial_u F_{ef} \right) \Gamma^{h}_{\ uc} F_{hd}\\
		= & 4 \left(\partial_u F_{zu} \right) \Gamma^{h}_{\ uu} F_{hz} + 4 \gamma^{ij} \left(\partial_u F_{zj} \right) \Gamma^{h}_{\ uu} F_{hi} + 4 \left(\partial_u F_{uz} \right) \Gamma^{h}_{\ uz} F_{hu}\\
		& + 4 \gamma^{ij} \left(\partial_u F_{uj} \right) \Gamma^{h}_{\ uz} F_{hi} + 4 \gamma^{ij} \left(\partial_u F_{jz} \right) \Gamma^{h}_{\ ui} F_{hu} + 4 \gamma^{ij} \left(\partial_u F_{ju} \right) \Gamma^{h}_{\ ui} F_{hz}\\
		& + 4 \gamma^{ij} \gamma^{kl} \left(\partial_u F_{jl} \right) \Gamma^{h}_{\ ui} F_{hk}\\
		= & 4 \left(\partial_u F_{uz} \right) \Gamma^{h}_{\ uz} F_{hu} + 4 \gamma^{ij} \left(\partial_u F_{uj} \right) \Gamma^{h}_{\ uz} F_{hi} + 4 \gamma^{ij} \left(\partial_u F_{jz} \right) \Gamma^{h}_{\ ui} F_{hu}\\
		& + 4 \gamma^{ij} \left(\partial_u F_{ju} \right) \Gamma^{h}_{\ ui} F_{hz} + 4 \gamma^{ij} \gamma^{kl} \left(\partial_u F_{jl} \right) \Gamma^{h}_{\ ui} F_{hk}\,.
	\end{split}
\end{equation}
The repeated index $h$ should be further expanded. The first term of Eq. (\ref{rhkk1secondsecond}) is
\begin{equation}
	\begin{split}
		& 4 \left(\partial_u F_{uz} \right) \Gamma^{h}_{\ uz} F_{hu}\\
		= & 4 \left(\partial_u F_{uz} \right) \Gamma^{u}_{\ uz} F_{uu} + 4 \left(\partial_u F_{uz} \right) \Gamma^{z}_{\ uz} F_{zu} + 4 \left(\partial_u F_{uz} \right) \Gamma^{i}_{\ uz} F_{iu}\\
		= & 0\,.
	\end{split}
\end{equation}
The second term of Eq. (\ref{rhkk1secondsecond}) is
\begin{equation}
	\begin{split}
		& 4 \gamma^{ij} \left(\partial_u F_{uj} \right) \Gamma^{h}_{\ uz} F_{hi}\\
		= & 4 \gamma^{ij} \left(\partial_u F_{uj} \right) \Gamma^{u}_{\ uz} F_{ui} + 4 \gamma^{ij} \left(\partial_u F_{uj} \right) \Gamma^{z}_{\ uz} F_{zi} + 4 \gamma^{ij} \left(\partial_u F_{uj} \right) \Gamma^{k}_{\ uz} F_{ki}\\
		= & 2 \gamma^{ij} \gamma^{kl} \beta_l \left(\partial_u F_{uj} \right) F_{ki}\,.
	\end{split}
\end{equation}
The third term of Eq. (\ref{rhkk1secondsecond}) is
\begin{equation}
	\begin{split}
		& 4 \gamma^{ij} \left(\partial_u F_{jz} \right) \Gamma^{h}_{\ ui} F_{hu}\\
		= & 4 \gamma^{ij} \left(\partial_u F_{jz} \right) \Gamma^{u}_{\ ui} F_{uu} + 4 \gamma^{ij} \left(\partial_u F_{jz} \right) \Gamma^{z}_{\ ui} F_{zu} + 4 \gamma^{ij} \left(\partial_u F_{jz} \right) \Gamma^{k}_{\ ui} F_{ku}\\
		= & 0\,.
	\end{split}
\end{equation}
The fourth term of Eq. (\ref{rhkk1secondsecond}) is
\begin{equation}
	\begin{split}
		& 4 \gamma^{ij} \left(\partial_u F_{ju} \right) \Gamma^{h}_{\ ui} F_{hz}\\
		= & 4 \gamma^{ij} \left(\partial_u F_{ju} \right) \Gamma^{u}_{\ ui} F_{uz} + 4 \gamma^{ij} \left(\partial_u F_{ju} \right) \Gamma^{z}_{\ ui} F_{zz} + 4 \gamma^{ij} \left(\partial_u F_{ju} \right) \Gamma^{k}_{\ ui} F_{kz}\\
		= & - 2 \gamma^{ij} \beta_i \left(\partial_u F_{ju} \right) F_{uz} + 2 \gamma^{ij} \gamma^{kl} \left(\partial_u F_{ju} \right) \left(\partial_u \gamma_{il} \right) F_{kz}\,.
	\end{split}
\end{equation}
The fifth term of Eq. (\ref{rhkk1secondsecond}) is
\begin{equation}
	\begin{split}
		& 4 \gamma^{ij} \gamma^{kl} \left(\partial_u F_{jl} \right) \Gamma^{h}_{\ ui} F_{hk}\\
		= & 4 \gamma^{ij} \gamma^{kl} \left(\partial_u F_{jl} \right) \Gamma^{u}_{\ ui} F_{uk} + 4 \gamma^{ij} \gamma^{kl} \left(\partial_u F_{jl} \right) \Gamma^{z}_{\ ui} F_{zk} + 4 \gamma^{ij} \gamma^{kl} \left(\partial_u F_{jl} \right) \Gamma^{m}_{\ ui} F_{mk}\\
		= & 2 \gamma^{ij} \gamma^{kl} \gamma^{mn} \left(\partial_u F_{jl} \right) \left(\partial_u \gamma_{in} \right) F_{mk}\,.
	\end{split}
\end{equation}
Therefore, the second term of Eq. (\ref{rhkk1second}) is obtained as
\begin{equation}
	\begin{split}
		& 4 k^a k^b g^{ce} g^{df} \left(\partial_b F_{ef} \right) \Gamma^{h}_{\ ac} F_{hd}\\
		= & 2 \gamma^{ij} \gamma^{kl} \beta_l \left(\partial_u F_{uj} \right) F_{ki} - 2 \gamma^{ij} \beta_i \left(\partial_u F_{ju} \right) F_{uz} + 2 \gamma^{ij} \gamma^{kl} \left(\partial_u F_{ju} \right) \left(\partial_u \gamma_{il} \right) F_{kz}\\
		& + 2 \gamma^{ij} \gamma^{kl} \gamma^{mn} \left(\partial_u F_{jl} \right) \left(\partial_u \gamma_{in} \right) F_{mk}\\
		= & 2 \gamma^{ij} \gamma^{kl} \beta_l \left(\partial_u F_{uj} \right) F_{ki} - 2 \gamma^{ij} \beta_i \left(\partial_u F_{ju} \right) F_{uz} + 4 \gamma^{ij} \gamma^{kl} \left(\partial_u F_{ju} \right) K_{il} F_{kz}\\
		& + 4 \gamma^{ij} \gamma^{kl} \gamma^{mn} \left(\partial_u F_{jl} \right) K_{in} F_{mk}\\
		= & 2 \gamma^{ij} \gamma^{kl} \beta_l \left(\partial_u F_{uj} \right) F_{ki} - 2 \gamma^{ij} \beta_i \left(\partial_u F_{ju} \right) F_{uz}\,.
	\end{split}
\end{equation}

The third term of Eq. (\ref{rhkk1second}) is
\begin{equation}\label{rhkk1secondthird}
	\begin{split}
		& 4 k^a k^b g^{ce} g^{df} \left(\partial_b F_{ef} \right) \Gamma^{h}_{\ ad} F_{ch} = 4 g^{ce} g^{df} \left(\partial_u F_{ef} \right) \Gamma^{h}_{\ ud} F_{ch}\\
		= & 4 \left(\partial_u F_{zu} \right) \Gamma^{h}_{\ uz} F_{uh} + 4 \gamma^{ij} \left(\partial_u F_{zj} \right) \Gamma^{h}_{\ ui} F_{uh} + 4 \left(\partial_u F_{uz} \right) \Gamma^{h}_{\ uu} F_{zh}\\
		& + 4 \gamma^{ij} \left(\partial_u F_{uj} \right) \Gamma^{h}_{\ ui} F_{zh} + 4 \gamma^{ij} \left(\partial_u F_{jz} \right) \Gamma^{h}_{\ uu} F_{ih} + 4 \gamma^{ij} \left(\partial_u F_{ju} \right) \Gamma^{h}_{\ uz} F_{ih}\\
		& + 4 \gamma^{ij} \gamma^{kl} \left(\partial_u F_{jl} \right) \Gamma^{h}_{\ uk} F_{ih}\\
		= & 4 \left(\partial_u F_{zu} \right) \Gamma^{h}_{\ uz} F_{uh} + 4 \gamma^{ij} \left(\partial_u F_{zj} \right) \Gamma^{h}_{\ ui} F_{uh} + 4 \gamma^{ij} \left(\partial_u F_{uj} \right) \Gamma^{h}_{\ ui} F_{zh}\\
		& + 4 \gamma^{ij} \left(\partial_u F_{ju} \right) \Gamma^{h}_{\ uz} F_{ih} + 4 \gamma^{ij} \gamma^{kl} \left(\partial_u F_{jl} \right) \Gamma^{h}_{\ uk} F_{ih}\,.
	\end{split}
\end{equation}
The repeated index $h$ should be further expanded. The first term of Eq. (\ref{rhkk1secondthird}) is 
\begin{equation}
	\begin{split}
		& 4 \left(\partial_u F_{zu} \right) \Gamma^{h}_{\ uz} F_{uh}\\
		= & 4 \left(\partial_u F_{zu} \right) \Gamma^{u}_{\ uz} F_{uu} + 4 \left(\partial_u F_{zu} \right) \Gamma^{z}_{\ uz} F_{uz} + 4 \left(\partial_u F_{zu} \right) \Gamma^{i}_{\ uz} F_{ui}\\
		= & 0\,.
	\end{split}
\end{equation}
The second term of Eq. (\ref{rhkk1secondthird}) is
\begin{equation}
	\begin{split}
		& 4 \gamma^{ij} \left(\partial_u F_{zj} \right) \Gamma^{h}_{\ ui} F_{uh}\\
		= & 4 \gamma^{ij} \left(\partial_u F_{zj} \right) \Gamma^{u}_{\ ui} F_{uu} + 4 \gamma^{ij} \left(\partial_u F_{zj} \right) \Gamma^{z}_{\ ui} F_{uz} + 4 \gamma^{ij} \left(\partial_u F_{zj} \right) \Gamma^{k}_{\ ui} F_{uk}\\
		= & 0\,.
	\end{split}
\end{equation}
The third term of Eq. (\ref{rhkk1secondthird}) is
\begin{equation}
	\begin{split}
		& 4 \gamma^{ij} \left(\partial_u F_{uj} \right) \Gamma^{h}_{\ ui} F_{zh}\\
		= & 4 \gamma^{ij} \left(\partial_u F_{uj} \right) \Gamma^{u}_{\ ui} F_{zu} + 4 \gamma^{ij} \left(\partial_u F_{uj} \right) \Gamma^{z}_{\ ui} F_{zz} + 4 \gamma^{ij} \left(\partial_u F_{uj} \right) \Gamma^{k}_{\ ui} F_{zk}\\
		= & - 2 \gamma^{ij} \beta_i \left(\partial_u F_{uj} \right) F_{zu} + 2 \gamma^{ij} \gamma^{kl} \left(\partial_u F_{uj} \right) \left(\partial_u \gamma_{il} \right) F_{zk}\,.
	\end{split}
\end{equation}
The fourth term of Eq. (\ref{rhkk1secondthird}) is
\begin{equation}
	\begin{split}
		& 4 \gamma^{ij} \left(\partial_u F_{ju} \right) \Gamma^{h}_{\ uz} F_{ih}\\
		= & 4 \gamma^{ij} \left(\partial_u F_{ju} \right) \Gamma^{u}_{\ uz} F_{iu} + 4 \gamma^{ij} \left(\partial_u F_{ju} \right) \Gamma^{z}_{\ uz} F_{iz} + 4 \gamma^{ij} \left(\partial_u F_{ju} \right) \Gamma^{k}_{\ uz} F_{ik}\\
		= & 2 \gamma^{ij} \gamma^{kl} \beta_l \left(\partial_u F_{ju} \right) F_{ik}\,.
	\end{split}
\end{equation}
The fifth term of Eq. (\ref{rhkk1secondthird}) is
\begin{equation}
	\begin{split}
		& 4 \gamma^{ij} \gamma^{kl} \left(\partial_u F_{jl} \right) \Gamma^{h}_{\ uk} F_{ih}\\
		= & 4 \gamma^{ij} \gamma^{kl} \left(\partial_u F_{jl} \right) \Gamma^{u}_{\ uk} F_{iu} + 4 \gamma^{ij} \gamma^{kl} \left(\partial_u F_{jl} \right) \Gamma^{z}_{\ uk} F_{iz} + 4 \gamma^{ij} \gamma^{kl} \left(\partial_u F_{jl} \right) \Gamma^{m}_{\ uk} F_{im}\\
		= & 2 \gamma^{ij} \gamma^{kl} \gamma^{mn} \left(\partial_u F_{jl} \right) \left(\partial_u \gamma_{kn} \right) F_{im}\,.
	\end{split}
\end{equation}
Therefore, the third term of Eq. (\ref{rhkk1second}) is obtained as
\begin{equation}
	\begin{split}
		& 4 k^a k^b g^{ce} g^{df} \left(\partial_b F_{ef} \right) \Gamma^{h}_{\ ad} F_{ch}\\
		= & - 2 \gamma^{ij} \beta_i \left(\partial_u F_{uj} \right) F_{zu} + 2 \gamma^{ij} \gamma^{kl} \left(\partial_u F_{uj} \right) \left(\partial_u \gamma_{il} \right) F_{zk} + 2 \gamma^{ij} \gamma^{kl} \beta_l \left(\partial_u F_{ju} \right) F_{ik}\\
		& + 2 \gamma^{ij} \gamma^{kl} \gamma^{mn} \left(\partial_u F_{jl} \right) \left(\partial_u \gamma_{kn} \right) F_{im}\\
		= & - 2 \gamma^{ij} \beta_i \left(\partial_u F_{uj} \right) F_{zu} + 4 \gamma^{ij} \gamma^{kl} \left(\partial_u F_{uj} \right) K_{il} F_{zk} + 2 \gamma^{ij} \gamma^{kl} \beta_l \left(\partial_u F_{ju} \right) F_{ik}\\
		& + 4 \gamma^{ij} \gamma^{kl} \gamma^{mn} \left(\partial_u F_{jl} \right) K_{kn} F_{im}\\
		= & - 2 \gamma^{ij} \beta_i \left(\partial_u F_{uj} \right) F_{zu} + 2 \gamma^{ij} \gamma^{kl} \beta_l \left(\partial_u F_{ju} \right) F_{ik}\,.
	\end{split}
\end{equation}

The fourth term of Eq. (\ref{rhkk1second}) is
\begin{equation}
	\begin{split}
		& 4 k^a k^b g^{ce} g^{df} \Gamma^{g}_{\ be} F_{gf} \left(\partial_a F_{cd} \right) = 4 g^{ce} g^{df} \Gamma^{g}_{\ ue} F_{gf} \left(\partial_u F_{cd} \right)\\
		= & 4 \Gamma^{g}_{\ uz} F_{gu} \left(\partial_u F_{uz} \right) + 4 \gamma^{ij} \Gamma^{g}_{\ uz} F_{gj} \left(\partial_u F_{ui} \right) + 4 \Gamma^{g}_{\ uu} F_{gz} \left(\partial_u F_{zu} \right)\\
		& + 4 \gamma^{ij} \Gamma^{g}_{\ uu} F_{gj} \left(\partial_u F_{zi} \right) + 4 \gamma^{ij} \Gamma^{g}_{\ uj} F_{gz} \left(\partial_u F_{iu} \right) + 4 \gamma^{ij} \Gamma^{g}_{\ uj} F_{gu} \left(\partial_u F_{iz} \right)\\
		& + 4 \gamma^{ij} \gamma^{kl} \Gamma^{g}_{\ uj} F_{gl} \left(\partial_u F_{ik} \right)\\
		= & 4 \Gamma^{g}_{\ uz} F_{gu} \left(\partial_u F_{uz} \right) + 4 \gamma^{ij} \Gamma^{g}_{\ uz} F_{gj} \left(\partial_u F_{ui} \right) + 4 \gamma^{ij} \Gamma^{g}_{\ uj} F_{gz} \left(\partial_u F_{iu} \right)\\
		& + 4 \gamma^{ij} \Gamma^{g}_{\ uj} F_{gu} \left(\partial_u F_{iz} \right) + 4 \gamma^{ij} \gamma^{kl} \Gamma^{g}_{\ uj} F_{gl} \left(\partial_u F_{ik} \right)\,.
	\end{split}
\end{equation}
The index $g$ should be further expanded.
\begin{equation}\label{rhkk1secondfourth}
	\begin{split}
		& 4 \Gamma^{g}_{\ uz} F_{gu} \left(\partial_u F_{uz} \right) + 4 \gamma^{ij} \Gamma^{g}_{\ uz} F_{gj} \left(\partial_u F_{ui} \right) + 4 \gamma^{ij} \Gamma^{g}_{\ uj} F_{gz} \left(\partial_u F_{iu} \right)\\
		& + 4 \gamma^{ij} \Gamma^{g}_{\ uj} F_{gu} \left(\partial_u F_{iz} \right) + 4 \gamma^{ij} \gamma^{kl} \Gamma^{g}_{\ uj} F_{gl} \left(\partial_u F_{ik} \right)\\
		= & 4 \Gamma^{u}_{\ uz} F_{uu} \left(\partial_u F_{uz} \right) + 4 \Gamma^{z}_{\ uz} F_{zu} \left(\partial_u F_{uz} \right) + 4 \Gamma^{i}_{\ uz} F_{iu} \left(\partial_u F_{uz} \right)\\
		& + 4 \gamma^{ij} \Gamma^{u}_{\ uz} F_{uj} \left(\partial_u F_{ui} \right) + 4 \gamma^{ij} \Gamma^{z}_{\ uz} F_{zj} \left(\partial_u F_{ui} \right) + 4 \gamma^{ij} \Gamma^{k}_{\ uz} F_{kj} \left(\partial_u F_{ui} \right)\\
		& + 4 \gamma^{ij} \Gamma^{u}_{\ uj} F_{uz} \left(\partial_u F_{iu} \right) + 4 \gamma^{ij} \Gamma^{z}_{\ uj} F_{zz} \left(\partial_u F_{iu} \right) + 4 \gamma^{ij} \Gamma^{k}_{\ uj} F_{kz} \left(\partial_u F_{iu} \right)\\
		& + 4 \gamma^{ij} \Gamma^{u}_{\ uj} F_{uu} \left(\partial_u F_{iz} \right) + 4 \gamma^{ij} \Gamma^{z}_{\ uj} F_{zu} \left(\partial_u F_{iz} \right) + 4 \gamma^{ij} \Gamma^{k}_{\ uj} F_{ku} \left(\partial_u F_{iz} \right)\\
		& + 4 \gamma^{ij} \gamma^{kl} \Gamma^{u}_{\ uj} F_{ul} \left(\partial_u F_{ik} \right) + 4 \gamma^{ij} \gamma^{kl} \Gamma^{z}_{\ uj} F_{zl} \left(\partial_u F_{ik} \right) + 4 \gamma^{ij} \gamma^{kl} \Gamma^{m}_{\ uj} F_{ml} \left(\partial_u F_{ik} \right)\\
		= & 4 \Gamma^{z}_{\ uz} F_{zu} \left(\partial_u F_{uz} \right) + 4 \gamma^{ij} \Gamma^{z}_{\ uz} F_{zj} \left(\partial_u F_{ui} \right) + 4 \gamma^{ij} \Gamma^{k}_{\ uz} F_{kj} \left(\partial_u F_{ui} \right)\\
		& + 4 \gamma^{ij} \Gamma^{u}_{\ uj} F_{uz} \left(\partial_u F_{iu} \right) + 4 \gamma^{ij} \Gamma^{k}_{\ uj} F_{kz} \left(\partial_u F_{iu} \right) + 4 \gamma^{ij} \Gamma^{z}_{\ uj} F_{zu} \left(\partial_u F_{iz} \right)\\
		& + 4 \gamma^{ij} \gamma^{kl} \Gamma^{z}_{\ uj} F_{zl} \left(\partial_u F_{ik} \right) + 4 \gamma^{ij} \gamma^{kl} \Gamma^{m}_{\ uj} F_{ml} \left(\partial_u F_{ik} \right)\,.
	\end{split}
\end{equation}
Therefore, the fourth term of Eq. (\ref{rhkk1second}) is obtained as 
\begin{equation}
	\begin{split}
		& 4 k^a k^b g^{ce} g^{df} \Gamma^{g}_{\ be} F_{gf} \left(\partial_a F_{cd} \right)\\
		= & 2 \gamma^{ij} \gamma^{kl} \beta_l F_{kj} \left(\partial_u F_{ui} \right) - 2 \gamma^{ij} \beta_j F_{uz} \left(\partial_u F_{iu} \right) + 2 \gamma^{ij} \gamma^{kl} \left(\partial_u \gamma_{jl} \right) F_{kz} \left(\partial_u F_{iu} \right)\\
		& + 2 \gamma^{ij} \gamma^{kl} \gamma^{mn} \left(\partial_u \gamma_{jn} \right) F_{ml} \left(\partial_u F_{ik} \right)\\
		= & 2 \gamma^{ij} \gamma^{kl} \beta_l F_{kj} \left(\partial_u F_{ui} \right) - 2 \gamma^{ij} \beta_j F_{uz} \left(\partial_u F_{iu} \right) + 4 \gamma^{ij} \gamma^{kl} K_{jl} F_{kz} \left(\partial_u F_{iu} \right)\\
		& + 4 \gamma^{ij} \gamma^{kl} \gamma^{mn} K_{jn} F_{ml} \left(\partial_u F_{ik} \right)\\
		= & 2 \gamma^{ij} \gamma^{kl} \beta_l F_{kj} \left(\partial_u F_{ui} \right) - 2 \gamma^{ij} \beta_j F_{uz} \left(\partial_u F_{iu} \right)\,.
	\end{split}
\end{equation}

The fifth term of Eq. (\ref{rhkk1second}) is
\begin{equation}
	\begin{split}
		& - 4 k^a k^b g^{ce} g^{df} \Gamma^{g}_{\ be} F_{gf} \Gamma^{h}_{\ ac} F_{hd} = - 4 g^{ce} g^{df} \Gamma^{g}_{\ ue} F_{gf} \Gamma^{h}_{\ uc} F_{hd}\\
		= & - 4 \Gamma^{g}_{\ uz} F_{gz} \Gamma^{h}_{\ uu} F_{hu} - 4 \Gamma^{g}_{\ uz} F_{gu} \Gamma^{h}_{\ uu} F_{hz} - 4 \gamma^{ij} \Gamma^{g}_{\ uz} F_{gj} \Gamma^{h}_{\ uu} F_{hi}\\
		& - 4 \Gamma^{g}_{\ uu} F_{gz} \Gamma^{h}_{\ uz} F_{hu} - 4 \Gamma^{g}_{\ uu} F_{gu} \Gamma^{h}_{\ uz} F_{hz} - 4 \gamma^{ij} \Gamma^{g}_{\ uu} F_{gj} \Gamma^{h}_{\ uz} F_{hi}\\
		& - 4 \gamma^{ij} \Gamma^{g}_{\ uj} F_{gz} \Gamma^{h}_{\ ui} F_{hu} - 4 \gamma^{ij} \Gamma^{g}_{\ uj} F_{gu} \Gamma^{h}_{\ ui} F_{hz} - 4 \gamma^{ij} \gamma^{kl} \Gamma^{g}_{\ uj} F_{gl} \Gamma^{h}_{\ ui} F_{hk}\\
		= & - 4 \gamma^{ij} \Gamma^{g}_{\ uj} F_{gz} \Gamma^{h}_{\ ui} F_{hu} - 4 \gamma^{ij} \Gamma^{g}_{\ uj} F_{gu} \Gamma^{h}_{\ ui} F_{hz} - 4 \gamma^{ij} \gamma^{kl} \Gamma^{g}_{\ uj} F_{gl} \Gamma^{h}_{\ ui} F_{hk}\,.
	\end{split}
\end{equation}
The index $g$ should be further expanded.
\begin{equation}\label{rhkk1secondfifth}
	\begin{split}
		& - 4 \gamma^{ij} \Gamma^{g}_{\ uj} F_{gz} \Gamma^{h}_{\ ui} F_{hu} - 4 \gamma^{ij} \Gamma^{g}_{\ uj} F_{gu} \Gamma^{h}_{\ ui} F_{hz} - 4 \gamma^{ij} \gamma^{kl} \Gamma^{g}_{\ uj} F_{gl} \Gamma^{h}_{\ ui} F_{hk}\\
		= & - 4 \gamma^{ij} \Gamma^{u}_{\ uj} F_{uz} \Gamma^{h}_{\ ui} F_{hu} - 4 \gamma^{ij} \Gamma^{z}_{\ uj} F_{zz} \Gamma^{h}_{\ ui} F_{hu} - 4 \gamma^{ij} \Gamma^{k}_{\ uj} F_{kz} \Gamma^{h}_{\ ui} F_{hu}\\
		& - 4 \gamma^{ij} \Gamma^{u}_{\ uj} F_{uu} \Gamma^{h}_{\ ui} F_{hz} - 4 \gamma^{ij} \Gamma^{z}_{\ uj} F_{zu} \Gamma^{h}_{\ ui} F_{hz} - 4 \gamma^{ij} \Gamma^{k}_{\ uj} F_{ku} \Gamma^{h}_{\ ui} F_{hz}\\
		& - 4 \gamma^{ij} \gamma^{kl} \Gamma^{u}_{\ uj} F_{ul} \Gamma^{h}_{\ ui} F_{hk} - 4 \gamma^{ij} \gamma^{kl} \Gamma^{z}_{\ uj} F_{zl} \Gamma^{h}_{\ ui} F_{hk} - 4 \gamma^{ij} \gamma^{kl} \Gamma^{m}_{\ uj} F_{ml} \Gamma^{h}_{\ ui} F_{hk}\\
		= & - 4 \gamma^{ij} \Gamma^{u}_{\ uj} F_{uz} \Gamma^{h}_{\ ui} F_{hu} - 4 \gamma^{ij} \Gamma^{k}_{\ uj} F_{kz} \Gamma^{h}_{\ ui} F_{hu} - 4 \gamma^{ij} \Gamma^{z}_{\ uj} F_{zu} \Gamma^{h}_{\ ui} F_{hz}\\
		& - 4 \gamma^{ij} \gamma^{kl} \Gamma^{z}_{\ uj} F_{zl} \Gamma^{h}_{\ ui} F_{hk} - 4 \gamma^{ij} \gamma^{kl} \Gamma^{m}_{\ uj} F_{ml} \Gamma^{h}_{\ ui} F_{hk}\,.
	\end{split}
\end{equation}
The repeated index $h$ should be further expanded using the metric. The first term of Eq. (\ref{rhkk1secondfifth}) is
\begin{equation}
	\begin{split}
		& - 4 \gamma^{ij} \Gamma^{u}_{\ uj} F_{uz} \Gamma^{h}_{\ ui} F_{hu}\\
		= & - 4 \gamma^{ij} \Gamma^{u}_{\ uj} F_{uz} \Gamma^{u}_{\ ui} F_{uu} - 4 \gamma^{ij} \Gamma^{u}_{\ uj} F_{uz} \Gamma^{z}_{\ ui} F_{zu} - 4 \gamma^{ij} \Gamma^{u}_{\ uj} F_{uz} \Gamma^{k}_{\ ui} F_{ku}\\
		= & 0\,.
	\end{split}
\end{equation}
The second term of Eq. (\ref{rhkk1secondfifth}) is
\begin{equation}
	\begin{split}
		& - 4 \gamma^{ij} \Gamma^{k}_{\ uj} F_{kz} \Gamma^{h}_{\ ui} F_{hu}\\
		= & - 4 \gamma^{ij} \Gamma^{k}_{\ uj} F_{kz} \Gamma^{u}_{\ ui} F_{uu} - 4 \gamma^{ij} \Gamma^{k}_{\ uj} F_{kz} \Gamma^{z}_{\ ui} F_{zu} - 4 \gamma^{ij} \Gamma^{k}_{\ uj} F_{kz} \Gamma^{l}_{\ ui} F_{lu}\\
		= & 0\,.
	\end{split}
\end{equation}
The third term of Eq. (\ref{rhkk1secondfifth}) is
\begin{equation}
	\begin{split}
		- 4 \gamma^{ij} \Gamma^{z}_{\ uj} F_{zu} \Gamma^{h}_{\ ui} F_{hz} = 0\,.
	\end{split}
\end{equation}
The fourth term of Eq. (\ref{rhkk1secondfifth}) is
\begin{equation}
	\begin{split}
		- 4 \gamma^{ij} \gamma^{kl} \Gamma^{z}_{\ uj} F_{zl} \Gamma^{h}_{\ ui} F_{hk} = 0\,.
	\end{split}
\end{equation}
The fifth term of Eq. (\ref{rhkk1secondfifth}) is
\begin{equation}
	\begin{split}
		& - 4 \gamma^{ij} \gamma^{kl} \Gamma^{m}_{\ uj} F_{ml} \Gamma^{h}_{\ ui} F_{hk}\\
		= & - 4 \gamma^{ij} \gamma^{kl} \Gamma^{m}_{\ uj} F_{ml} \Gamma^{u}_{\ ui} F_{uk} - 4 \gamma^{ij} \gamma^{kl} \Gamma^{m}_{\ uj} F_{ml} \Gamma^{z}_{\ ui} F_{zk} - 4 \gamma^{ij} \gamma^{kl} \Gamma^{m}_{\ uj} F_{ml} \Gamma^{n}_{\ ui} F_{nk}\\
		= & - \gamma^{ij} \gamma^{kl} \gamma^{mo} \gamma^{np} \left(\partial_u \gamma_{jo} \right) F_{ml} \left(\partial_u \gamma_{ip} \right) F_{nk}\,.
	\end{split}
\end{equation}
Therefore, the fifth term of Eq. (\ref{rhkk1second}) is obtained as
\begin{equation}
	\begin{split}
		& - 4 k^a k^b g^{ce} g^{df} \Gamma^{g}_{\ be} F_{gf} \Gamma^{h}_{\ ac} F_{hd} = - \gamma^{ij} \gamma^{kl} \gamma^{mo} \gamma^{np} \left(\partial_u \gamma_{jo} \right) F_{ml} \left(\partial_u \gamma_{ip} \right) F_{nk}\\
		= & - 4 \gamma^{ij} \gamma^{kl} \gamma^{mo} \gamma^{np} K_{jo} F_{ml} K_{ip} F_{nk}\\
		= & 0\,.
	\end{split}
\end{equation}

The sixth term of Eq. (\ref{rhkk1second}) is
\begin{equation}
	\begin{split}
		& - 4 k^a k^b g^{ce} g^{df} \Gamma^{g}_{\ be} F_{gf} \Gamma^{h}_{\ ad} F_{ch} = - 4 g^{ce} g^{df} \Gamma^{g}_{\ ue} F_{gf} \Gamma^{h}_{\ ud} F_{ch}\\
		= & - 4 \Gamma^{g}_{\ uz} F_{gz} \Gamma^{h}_{\ uu} F_{uh} - 4 \Gamma^{g}_{\ uz} F_{gu} \Gamma^{h}_{\ uz} F_{uh} - 4 \gamma^{ij} \Gamma^{g}_{\ uz} F_{gj} \Gamma^{h}_{\ ui} F_{uh}\\
		& - 4 \Gamma^{g}_{\ uu} F_{gz} \Gamma^{h}_{\ uu} F_{zh} - 4 \Gamma^{g}_{\ uu} F_{gu} \Gamma^{h}_{\ uz} F_{zh} - 4 \gamma^{ij} \Gamma^{g}_{\ uu} F_{gj} \Gamma^{h}_{\ ui} F_{zh}\\
		& - 4 \gamma^{ij} \Gamma^{g}_{\ uj} F_{gz} \Gamma^{h}_{\ uu} F_{ih} - 4 \gamma^{ij} \Gamma^{g}_{\ uj} F_{gu} \Gamma^{h}_{\ uz} F_{ih} - 4 \gamma^{ij} \gamma^{kl} \Gamma^{g}_{\ uj} F_{gl} \Gamma^{h}_{\ uk} F_{ih}\\
		= & - 4 \Gamma^{g}_{\ uz} F_{gu} \Gamma^{h}_{\ uz} F_{uh} - 4 \gamma^{ij} \Gamma^{g}_{\ uz} F_{gj} \Gamma^{h}_{\ ui} F_{uh} - 4 \gamma^{ij} \Gamma^{g}_{\ uj} F_{gu} \Gamma^{h}_{\ uz} F_{ih}\\
		& - 4 \gamma^{ij} \gamma^{kl} \Gamma^{g}_{\ uj} F_{gl} \Gamma^{h}_{\ uk} F_{ih}\,.
	\end{split}
\end{equation}
The index $g$ should be further expanded.
\begin{equation}\label{rhkk1secondsixth}
	\begin{split}
		& - 4 \Gamma^{g}_{\ uz} F_{gu} \Gamma^{h}_{\ uz} F_{uh} - 4 \gamma^{ij} \Gamma^{g}_{\ uz} F_{gj} \Gamma^{h}_{\ ui} F_{uh} - 4 \gamma^{ij} \Gamma^{g}_{\ uj} F_{gu} \Gamma^{h}_{\ uz} F_{ih}\\
		& - 4 \gamma^{ij} \gamma^{kl} \Gamma^{g}_{\ uj} F_{gl} \Gamma^{h}_{\ uk} F_{ih}\\
		= & - 4 \Gamma^{u}_{\ uz} F_{uu} \Gamma^{h}_{\ uz} F_{uh} - 4 \Gamma^{z}_{\ uz} F_{zu} \Gamma^{h}_{\ uz} F_{uh} - 4 \Gamma^{i}_{\ uz} F_{iu} \Gamma^{h}_{\ uz} F_{uh}\\
		& - 4 \gamma^{ij} \Gamma^{u}_{\ uz} F_{uj} \Gamma^{h}_{\ ui} F_{uh} - 4 \gamma^{ij} \Gamma^{z}_{\ uz} F_{zj} \Gamma^{h}_{\ ui} F_{uh} - 4 \gamma^{ij} \Gamma^{k}_{\ uz} F_{kj} \Gamma^{h}_{\ ui} F_{uh}\\
		& - 4 \gamma^{ij} \Gamma^{u}_{\ uj} F_{uu} \Gamma^{h}_{\ uz} F_{ih} - 4 \gamma^{ij} \Gamma^{z}_{\ uj} F_{zu} \Gamma^{h}_{\ uz} F_{ih} - 4 \gamma^{ij} \Gamma^{k}_{\ uj} F_{ku} \Gamma^{h}_{\ uz} F_{ih}\\
		& - 4 \gamma^{ij} \gamma^{kl} \Gamma^{u}_{\ uj} F_{ul} \Gamma^{h}_{\ uk} F_{ih} - 4 \gamma^{ij} \gamma^{kl} \Gamma^{z}_{\ uj} F_{zl} \Gamma^{h}_{\ uk} F_{ih} - 4 \gamma^{ij} \gamma^{kl} \Gamma^{m}_{\ uj} F_{ml} \Gamma^{h}_{\ uk} F_{ih}\\
		= & - 4 \Gamma^{z}_{\ uz} F_{zu} \Gamma^{h}_{\ uz} F_{uh} - 4 \gamma^{ij} \Gamma^{z}_{\ uz} F_{zj} \Gamma^{h}_{\ ui} F_{uh} - 4 \gamma^{ij} \Gamma^{k}_{\ uz} F_{kj} \Gamma^{h}_{\ ui} F_{uh}\\
		& - 4 \gamma^{ij} \Gamma^{z}_{\ uj} F_{zu} \Gamma^{h}_{\ uz} F_{ih} - 4 \gamma^{ij} \gamma^{kl} \Gamma^{z}_{\ uj} F_{zl} \Gamma^{h}_{\ uk} F_{ih} - 4 \gamma^{ij} \gamma^{kl} \Gamma^{m}_{\ uj} F_{ml} \Gamma^{h}_{\ uk} F_{ih}\,.
	\end{split}
\end{equation}
The repeated index $h$ should be further expanded using the metric. The first term of Eq. (\ref{rhkk1secondsixth}) is 
\begin{equation}
	\begin{split}
		- 4 \Gamma^{z}_{\ uz} F_{zu} \Gamma^{h}_{\ uz} F_{uh} = 0\,.
	\end{split}
\end{equation}
The second term of Eq. (\ref{rhkk1secondsixth}) is 
\begin{equation}
	\begin{split}
		- 4 \gamma^{ij} \Gamma^{z}_{\ uz} F_{zj} \Gamma^{h}_{\ ui} F_{uh} = 0\,.
	\end{split}
\end{equation}
The third term of Eq. (\ref{rhkk1secondsixth}) is 
\begin{equation}
	\begin{split}
		& - 4 \gamma^{ij} \Gamma^{k}_{\ uz} F_{kj} \Gamma^{h}_{\ ui} F_{uh}\\
		= & - 4 \gamma^{ij} \Gamma^{k}_{\ uz} F_{kj} \Gamma^{u}_{\ ui} F_{uu} - 4 \gamma^{ij} \Gamma^{k}_{\ uz} F_{kj} \Gamma^{z}_{\ ui} F_{uz} - 4 \gamma^{ij} \Gamma^{k}_{\ uz} F_{kj} \Gamma^{l}_{\ ui} F_{ul}\\
		= & 0\,.
	\end{split}
\end{equation}
The fourth term of Eq. (\ref{rhkk1secondsixth}) is
\begin{equation}
	\begin{split}
		- 4 \gamma^{ij} \Gamma^{z}_{\ uj} F_{zu} \Gamma^{h}_{\ uz} F_{ih} = 0\,.
	\end{split}
\end{equation} 
The fifth term of Eq. (\ref{rhkk1secondsixth}) is
\begin{equation}
	\begin{split}
		- 4 \gamma^{ij} \gamma^{kl} \Gamma^{z}_{\ uj} F_{zl} \Gamma^{h}_{\ uk} F_{ih} = 0\,.
	\end{split}
\end{equation}
The fifteenth term of Eq. (\ref{rhkk1secondsixth}) is
\begin{equation}
	\begin{split}
		& - 4 \gamma^{ij} \gamma^{kl} \Gamma^{m}_{\ uj} F_{ml} \Gamma^{h}_{\ uk} F_{ih}\\
		= & - 4 \gamma^{ij} \gamma^{kl} \Gamma^{m}_{\ uj} F_{ml} \Gamma^{u}_{\ uk} F_{iu} - 4 \gamma^{ij} \gamma^{kl} \Gamma^{m}_{\ uj} F_{ml} \Gamma^{z}_{\ uk} F_{iz} - 4 \gamma^{ij} \gamma^{kl} \Gamma^{m}_{\ uj} F_{ml} \Gamma^{n}_{\ uk} F_{in}\\
		= & - \gamma^{ij} \gamma^{kl} \gamma^{mo} \gamma^{np} \left(\partial_u \gamma_{jo} \right) F_{ml} \left(\partial_u \gamma_{kp} \right) F_{in}\,. 
	\end{split}
\end{equation}
Therefore, the sixth term of Eq. (\ref{rhkk1second}) is obtained as 
\begin{equation}
	\begin{split}
		& - 4 k^a k^b g^{ce} g^{df} \Gamma^{g}_{\ be} F_{gf} \Gamma^{h}_{\ ad} F_{ch} = - \gamma^{ij} \gamma^{kl} \gamma^{mo} \gamma^{np} \left(\partial_u \gamma_{jo} \right) F_{ml} \left(\partial_u \gamma_{kp} \right) F_{in}\\
		= & - 4 \gamma^{ij} \gamma^{kl} \gamma^{mo} \gamma^{np} K_{jo} F_{ml} K_{kp} F_{in}\\
		= & 0\,.
	\end{split}
\end{equation}

The seventh term of Eq. (\ref{rhkk1second}) is
\begin{equation}
	\begin{split}
		& 4 k^a k^b g^{ce} g^{df} \Gamma^{g}_{\ bf} F_{eg} \left(\partial_a F_{cd} \right) = 4 g^{ce} g^{df} \Gamma^{g}_{\ uf} F_{eg} \left(\partial_u F_{cd} \right)\\
		= & 4 \Gamma^{g}_{\ uu} F_{zg} \left(\partial_u F_{uz} \right) + 4 \gamma^{ij} \Gamma^{g}_{\ uj} F_{zg} \left(\partial_u F_{ui} \right) + 4 \Gamma^{g}_{\ uz} F_{ug} \left(\partial_u F_{zu} \right)\\
		& + 4 \gamma^{ij} \Gamma^{g}_{\ uj} F_{ug} \left(\partial_u F_{zi} \right) + 4 \gamma^{ij} \Gamma^{g}_{\ uz} F_{jg} \left(\partial_u F_{iu} \right) + 4 \gamma^{ij} \Gamma^{g}_{\ uu} F_{jg} \left(\partial_u F_{iz} \right)\\
		& + 4 \gamma^{ij} \gamma^{kl} \Gamma^{g}_{\ ul} F_{jg} \left(\partial_u F_{ik} \right)\\
		= & 4 \gamma^{ij} \Gamma^{g}_{\ uj} F_{zg} \left(\partial_u F_{ui} \right) + 4 \Gamma^{g}_{\ uz} F_{ug} \left(\partial_u F_{zu} \right) + 4 \gamma^{ij} \Gamma^{g}_{\ uj} F_{ug} \left(\partial_u F_{zi} \right)\\
		& + 4 \gamma^{ij} \Gamma^{g}_{\ uz} F_{jg} \left(\partial_u F_{iu} \right) + 4 \gamma^{ij} \gamma^{kl} \Gamma^{g}_{\ ul} F_{jg} \left(\partial_u F_{ik} \right)\,.
	\end{split}
\end{equation}
The index $g$ should be further expanded.
\begin{equation}\label{rhkk1secondseventh}
	\begin{split}
		& 4 \gamma^{ij} \Gamma^{g}_{\ uj} F_{zg} \left(\partial_u F_{ui} \right) + 4 \Gamma^{g}_{\ uz} F_{ug} \left(\partial_u F_{zu} \right) + 4 \gamma^{ij} \Gamma^{g}_{\ uj} F_{ug} \left(\partial_u F_{zi} \right)\\
		& + 4 \gamma^{ij} \Gamma^{g}_{\ uz} F_{jg} \left(\partial_u F_{iu} \right) + 4 \gamma^{ij} \gamma^{kl} \Gamma^{g}_{\ ul} F_{jg} \left(\partial_u F_{ik} \right)\\
		= & 4 \gamma^{ij} \Gamma^{u}_{\ uj} F_{zu} \left(\partial_u F_{ui} \right) + 4 \gamma^{ij} \Gamma^{z}_{\ uj} F_{zz} \left(\partial_u F_{ui} \right) + 4 \gamma^{ij} \Gamma^{k}_{\ uj} F_{zk} \left(\partial_u F_{ui} \right)\\
		& + 4 \Gamma^{u}_{\ uz} F_{uu} \left(\partial_u F_{zu} \right) + 4 \Gamma^{z}_{\ uz} F_{uz} \left(\partial_u F_{zu} \right) + 4 \Gamma^{i}_{\ uz} F_{ui} \left(\partial_u F_{zu} \right)\\
		& + 4 \gamma^{ij} \Gamma^{u}_{\ uj} F_{uu} \left(\partial_u F_{zi} \right) + 4 \gamma^{ij} \Gamma^{z}_{\ uj} F_{uz} \left(\partial_u F_{zi} \right) + 4 \gamma^{ij} \Gamma^{k}_{\ uj} F_{uk} \left(\partial_u F_{zi} \right)\\
		& + 4 \gamma^{ij} \Gamma^{u}_{\ uz} F_{ju} \left(\partial_u F_{iu} \right) + 4 \gamma^{ij} \Gamma^{z}_{\ uz} F_{jz} \left(\partial_u F_{iu} \right) + 4 \gamma^{ij} \Gamma^{k}_{\ uz} F_{jk} \left(\partial_u F_{iu} \right)\\
		& + 4 \gamma^{ij} \gamma^{kl} \Gamma^{u}_{\ ul} F_{ju} \left(\partial_u F_{ik} \right) + 4 \gamma^{ij} \gamma^{kl} \Gamma^{z}_{\ ul} F_{jz} \left(\partial_u F_{ik} \right) + 4 \gamma^{ij} \gamma^{kl} \Gamma^{m}_{\ ul} F_{jm} \left(\partial_u F_{ik} \right)\\
		= & 4 \gamma^{ij} \Gamma^{u}_{\ uj} F_{zu} \left(\partial_u F_{ui} \right) + 4 \gamma^{ij} \Gamma^{k}_{\ uj} F_{zk} \left(\partial_u F_{ui} \right) + 4 \Gamma^{z}_{\ uz} F_{uz} \left(\partial_u F_{zu} \right)\\
		& + 4 \gamma^{ij} \Gamma^{z}_{\ uj} F_{uz} \left(\partial_u F_{zi} \right) + 4 \gamma^{ij} \Gamma^{z}_{\ uz} F_{jz} \left(\partial_u F_{iu} \right) + 4 \gamma^{ij} \Gamma^{k}_{\ uz} F_{jk} \left(\partial_u F_{iu} \right)\\
		& + 4 \gamma^{ij} \gamma^{kl} \Gamma^{z}_{\ ul} F_{jz} \left(\partial_u F_{ik} \right) + 4 \gamma^{ij} \gamma^{kl} \Gamma^{m}_{\ ul} F_{jm} \left(\partial_u F_{ik} \right)\,.
	\end{split}
\end{equation}
Therefore, the seventh term of Eq. (\ref{rhkk1second}) is obtained as 
\begin{equation}
	\begin{split}
		& 4 k^a k^b g^{ce} g^{df} \Gamma^{g}_{\ bf} F_{eg} \left(\partial_a F_{cd} \right)\\
		= & - 2 \gamma^{ij} \beta_j F_{zu} \left(\partial_u F_{ui} \right) + 2 \gamma^{ij} \gamma^{kl} \left(\partial_u \gamma_{jl} \right) F_{zk} \left(\partial_u F_{ui} \right) + 2 \gamma^{ij} \gamma^{kl} \beta_l F_{jk} \left(\partial_u F_{iu} \right)\\
		& + 2 \gamma^{ij} \gamma^{kl} \gamma^{mn} \left(\partial_u \gamma_{ln} \right) F_{jm} \left(\partial_u F_{ik} \right)\\
		= & - 2 \gamma^{ij} \beta_j F_{zu} \left(\partial_u F_{ui} \right) + 4 \gamma^{ij} \gamma^{kl} K_{jl} F_{zk} \left(\partial_u F_{ui} \right) + 2 \gamma^{ij} \gamma^{kl} \beta_l F_{jk} \left(\partial_u F_{iu} \right)\\
		& + 4 \gamma^{ij} \gamma^{kl} \gamma^{mn} K_{ln} F_{jm} \left(\partial_u F_{ik} \right)\\
		= & - 2 \gamma^{ij} \beta_j F_{zu} \left(\partial_u F_{ui} \right) + 2 \gamma^{ij} \gamma^{kl} \beta_l F_{jk} \left(\partial_u F_{iu} \right)\,.
	\end{split}
\end{equation}

The eighth term of Eq. (\ref{rhkk1second}) is
\begin{equation}
	\begin{split}
		& - 4 k^a k^b g^{ce} g^{df} \Gamma^{g}_{\ bf} F_{eg} \Gamma^{h}_{\ ac} F_{hd} = - 4 g^{ce} g^{df} \Gamma^{g}_{\ uf} F_{eg} \Gamma^{h}_{\ uc} F_{hd}\\
		= & - 4 \Gamma^{g}_{\ uz} F_{zg} \Gamma^{h}_{\ uu} F_{hu} - 4 \Gamma^{g}_{\ uu} F_{zg} \Gamma^{h}_{\ uu} F_{hz} - 4 \gamma^{ij} \Gamma^{g}_{\ uj} F_{zg} \Gamma^{h}_{\ uu} F_{hi}\\
		& - 4 \Gamma^{g}_{\ uz} F_{ug} \Gamma^{h}_{\ uz} F_{hu} - 4 \Gamma^{g}_{\ uu} F_{ug} \Gamma^{h}_{\ uz} F_{hz} - 4 \gamma^{ij} \Gamma^{g}_{\ uj} F_{ug} \Gamma^{h}_{\ uz} F_{hi}\\
		& - 4 \gamma^{ij} \Gamma^{g}_{\ uz} F_{jg} \Gamma^{h}_{\ ui} F_{hu} - 4 \gamma^{ij} \Gamma^{g}_{\ uu} F_{jg} \Gamma^{h}_{\ ui} F_{hz} - 4 \gamma^{ij} \gamma^{kl} \Gamma^{g}_{\ ul} F_{jg} \Gamma^{h}_{\ ui} F_{hk}\\
		= & - 4 \Gamma^{g}_{\ uz} F_{ug} \Gamma^{h}_{\ uz} F_{hu} - 4 \gamma^{ij} \Gamma^{g}_{\ uj} F_{ug} \Gamma^{h}_{\ uz} F_{hi} - 4 \gamma^{ij} \Gamma^{g}_{\ uz} F_{jg} \Gamma^{h}_{\ ui} F_{hu}\\
		& - 4 \gamma^{ij} \gamma^{kl} \Gamma^{g}_{\ ul} F_{jg} \Gamma^{h}_{\ ui} F_{hk}\,.
	\end{split}
\end{equation}
The index $g$ should be further expanded.
\begin{equation}\label{rhkk1secondeighth}
	\begin{split}
		& - 4 \Gamma^{g}_{\ uz} F_{ug} \Gamma^{h}_{\ uz} F_{hu} - 4 \gamma^{ij} \Gamma^{g}_{\ uj} F_{ug} \Gamma^{h}_{\ uz} F_{hi} - 4 \gamma^{ij} \Gamma^{g}_{\ uz} F_{jg} \Gamma^{h}_{\ ui} F_{hu}\\
		& - 4 \gamma^{ij} \gamma^{kl} \Gamma^{g}_{\ ul} F_{jg} \Gamma^{h}_{\ ui} F_{hk}\\
		= & - 4 \Gamma^{u}_{\ uz} F_{uu} \Gamma^{h}_{\ uz} F_{hu} - 4 \Gamma^{z}_{\ uz} F_{uz} \Gamma^{h}_{\ uz} F_{hu} - 4 \Gamma^{i}_{\ uz} F_{ui} \Gamma^{h}_{\ uz} F_{hu}\\
		& - 4 \gamma^{ij} \Gamma^{u}_{\ uj} F_{uu} \Gamma^{h}_{\ uz} F_{hi} - 4 \gamma^{ij} \Gamma^{z}_{\ uj} F_{uz} \Gamma^{h}_{\ uz} F_{hi} - 4 \gamma^{ij} \Gamma^{k}_{\ uj} F_{uk} \Gamma^{h}_{\ uz} F_{hi}\\
		& - 4 \gamma^{ij} \Gamma^{u}_{\ uz} F_{ju} \Gamma^{h}_{\ ui} F_{hu} - 4 \gamma^{ij} \Gamma^{z}_{\ uz} F_{jz} \Gamma^{h}_{\ ui} F_{hu} - 4 \gamma^{ij} \Gamma^{k}_{\ uz} F_{jk} \Gamma^{h}_{\ ui} F_{hu}\\
		& - 4 \gamma^{ij} \gamma^{kl} \Gamma^{u}_{\ ul} F_{ju} \Gamma^{h}_{\ ui} F_{hk} - 4 \gamma^{ij} \gamma^{kl} \Gamma^{z}_{\ ul} F_{jz} \Gamma^{h}_{\ ui} F_{hk} - 4 \gamma^{ij} \gamma^{kl} \Gamma^{m}_{\ ul} F_{jm} \Gamma^{h}_{\ ui} F_{hk}\\
		= & - 4 \Gamma^{z}_{\ uz} F_{uz} \Gamma^{h}_{\ uz} F_{hu} - 4 \gamma^{ij} \Gamma^{z}_{\ uj} F_{uz} \Gamma^{h}_{\ uz} F_{hi} - 4 \gamma^{ij} \Gamma^{z}_{\ uz} F_{jz} \Gamma^{h}_{\ ui} F_{hu}\\
		& - 4 \gamma^{ij} \Gamma^{k}_{\ uz} F_{jk} \Gamma^{h}_{\ ui} F_{hu} - 4 \gamma^{ij} \gamma^{kl} \Gamma^{z}_{\ ul} F_{jz} \Gamma^{h}_{\ ui} F_{hk} - 4 \gamma^{ij} \gamma^{kl} \Gamma^{m}_{\ ul} F_{jm} \Gamma^{h}_{\ ui} F_{hk}\,.
	\end{split}
\end{equation}
The repeated index $h$ should be further expanded using the metric. The first term of Eq. (\ref{rhkk1secondeighth}) is
\begin{equation}
	\begin{split}
		- 4 \Gamma^{z}_{\ uz} F_{uz} \Gamma^{h}_{\ uz} F_{hu} = 0\,.
	\end{split}
\end{equation}
The second term of Eq. (\ref{rhkk1secondeighth}) is
\begin{equation}
	\begin{split}
		- 4 \gamma^{ij} \Gamma^{z}_{\ uj} F_{uz} \Gamma^{h}_{\ uz} F_{hi} = 0\,.
	\end{split}
\end{equation}
The third term of Eq. (\ref{rhkk1secondeighth}) is
\begin{equation}
	\begin{split}
		- 4 \gamma^{ij} \Gamma^{z}_{\ uz} F_{jz} \Gamma^{h}_{\ ui} F_{hu} = 0\,.
	\end{split}
\end{equation}
The fourth term of Eq. (\ref{rhkk1secondeighth}) is
\begin{equation}
	\begin{split}
		& - 4 \gamma^{ij} \Gamma^{k}_{\ uz} F_{jk} \Gamma^{h}_{\ ui} F_{hu}\\
		= & - 4 \gamma^{ij} \Gamma^{k}_{\ uz} F_{jk} \Gamma^{u}_{\ ui} F_{uu} - 4 \gamma^{ij} \Gamma^{k}_{\ uz} F_{jk} \Gamma^{z}_{\ ui} F_{zu} - 4 \gamma^{ij} \Gamma^{k}_{\ uz} F_{jk} \Gamma^{l}_{\ ui} F_{lu}\\
		= & 0\,.
	\end{split}
\end{equation}
The fifth term of Eq. (\ref{rhkk1secondeighth}) is
\begin{equation}
	\begin{split}
		- 4 \gamma^{ij} \gamma^{kl} \Gamma^{z}_{\ ul} F_{jz} \Gamma^{h}_{\ ui} F_{hk} = 0\,.
	\end{split}
\end{equation}
The sixth term of Eq. (\ref{rhkk1secondeighth}) is
\begin{equation}
	\begin{split}
		& - 4 \gamma^{ij} \gamma^{kl} \Gamma^{m}_{\ ul} F_{jm} \Gamma^{h}_{\ ui} F_{hk}\\
		= & - 4 \gamma^{ij} \gamma^{kl} \Gamma^{m}_{\ ul} F_{jm} \Gamma^{u}_{\ ui} F_{uk} - 4 \gamma^{ij} \gamma^{kl} \Gamma^{m}_{\ ul} F_{jm} \Gamma^{z}_{\ ui} F_{zk} - 4 \gamma^{ij} \gamma^{kl} \Gamma^{m}_{\ ul} F_{jm} \Gamma^{n}_{\ ui} F_{nk}\\
		= & - \gamma^{ij} \gamma^{kl} \gamma^{mo} \gamma^{np} \left(\partial_u \gamma_{lo} \right) F_{jm} \left(\partial_u \gamma_{ip} \right) F_{nk}\,.
	\end{split}
\end{equation}
Therefore, the eighth term of Eq. (\ref{rhkk1second}) is obtained as 
\begin{equation}
	\begin{split}
		& - 4 k^a k^b g^{ce} g^{df} \Gamma^{g}_{\ bf} F_{eg} \Gamma^{h}_{\ ac} F_{hd} = - \gamma^{ij} \gamma^{kl} \gamma^{mo} \gamma^{np} \left(\partial_u \gamma_{lo} \right) F_{jm} \left(\partial_u \gamma_{ip} \right) F_{nk}\\
		= & - 4 \gamma^{ij} \gamma^{kl} \gamma^{mo} \gamma^{np} K_{lo} F_{jm} K_{ip} F_{nk}\\
		= & 0\,.
	\end{split}
\end{equation}

The ninth term of Eq. (\ref{rhkk1second}) is
\begin{equation}
	\begin{split}
		& - 4 k^a k^b g^{ce} g^{df} \Gamma^{g}_{\ bf} F_{eg} \Gamma^{h}_{\ ad} F_{ch} = - 4 g^{ce} g^{df} \Gamma^{g}_{\ uf} F_{eg} \Gamma^{h}_{\ ud} F_{ch}\\
		= & - 4 \Gamma^{g}_{\ uz} F_{zg} \Gamma^{h}_{\ uu} F_{uh} - 4 \Gamma^{g}_{\ uu} F_{zg} \Gamma^{h}_{\ uz} F_{uh} - 4 \gamma^{ij} \Gamma^{g}_{\ uj} F_{zg} \Gamma^{h}_{\ ui} F_{uh}\\
		& - 4 \Gamma^{g}_{\ uz} F_{ug} \Gamma^{h}_{\ uu} F_{zh} - 4 \Gamma^{g}_{\ uu} F_{ug} \Gamma^{h}_{\ uz} F_{zh} - 4 \gamma^{ij} \Gamma^{g}_{\ uj} F_{ug} \Gamma^{h}_{\ ui} F_{zh}\\
		& - 4 \gamma^{ij} \Gamma^{g}_{\ uz} F_{jg} \Gamma^{h}_{\ uu} F_{ih} - 4 \gamma^{ij} \Gamma^{g}_{\ uu} F_{jg} \Gamma^{h}_{\ uz} F_{ih} - 4 \gamma^{ij} \gamma^{kl} \Gamma^{g}_{\ ul} F_{jg} \Gamma^{h}_{\ uk} F_{ih}\\
		= & - 4 \gamma^{ij} \Gamma^{g}_{\ uj} F_{zg} \Gamma^{h}_{\ ui} F_{uh} - 4 \gamma^{ij} \Gamma^{g}_{\ uj} F_{ug} \Gamma^{h}_{\ ui} F_{zh} - 4 \gamma^{ij} \gamma^{kl} \Gamma^{g}_{\ ul} F_{jg} \Gamma^{h}_{\ uk} F_{ih}\,.
	\end{split}
\end{equation}
The index $g$ should be further expanded.
\begin{equation}\label{rhkk1secondninth}
	\begin{split}
		& - 4 \gamma^{ij} \Gamma^{g}_{\ uj} F_{zg} \Gamma^{h}_{\ ui} F_{uh} - 4 \gamma^{ij} \Gamma^{g}_{\ uj} F_{ug} \Gamma^{h}_{\ ui} F_{zh} - 4 \gamma^{ij} \gamma^{kl} \Gamma^{g}_{\ ul} F_{jg} \Gamma^{h}_{\ uk} F_{ih}\\
		= & - 4 \gamma^{ij} \Gamma^{u}_{\ uj} F_{zu} \Gamma^{h}_{\ ui} F_{uh} - 4 \gamma^{ij} \Gamma^{z}_{\ uj} F_{zz} \Gamma^{h}_{\ ui} F_{uh} - 4 \gamma^{ij} \Gamma^{k}_{\ uj} F_{zk} \Gamma^{h}_{\ ui} F_{uh}\\
		& - 4 \gamma^{ij} \Gamma^{u}_{\ uj} F_{uu} \Gamma^{h}_{\ ui} F_{zh} - 4 \gamma^{ij} \Gamma^{z}_{\ uj} F_{uz} \Gamma^{h}_{\ ui} F_{zh} - 4 \gamma^{ij} \Gamma^{k}_{\ uj} F_{uk} \Gamma^{h}_{\ ui} F_{zh}\\
		& - 4 \gamma^{ij} \gamma^{kl} \Gamma^{u}_{\ ul} F_{ju} \Gamma^{h}_{\ uk} F_{ih} - 4 \gamma^{ij} \gamma^{kl} \Gamma^{z}_{\ ul} F_{jz} \Gamma^{h}_{\ uk} F_{ih} - 4 \gamma^{ij} \gamma^{kl} \Gamma^{m}_{\ ul} F_{jm} \Gamma^{h}_{\ uk} F_{ih}\\
		= & - 4 \gamma^{ij} \Gamma^{u}_{\ uj} F_{zu} \Gamma^{h}_{\ ui} F_{uh} - 4 \gamma^{ij} \Gamma^{k}_{\ uj} F_{zk} \Gamma^{h}_{\ ui} F_{uh} - 4 \gamma^{ij} \Gamma^{z}_{\ uj} F_{uz} \Gamma^{h}_{\ ui} F_{zh}\\
		& - 4 \gamma^{ij} \gamma^{kl} \Gamma^{z}_{\ ul} F_{jz} \Gamma^{h}_{\ uk} F_{ih} - 4 \gamma^{ij} \gamma^{kl} \Gamma^{m}_{\ ul} F_{jm} \Gamma^{h}_{\ uk} F_{ih}\,.
	\end{split}
\end{equation}
The repeated index $h$ should be further expanded. The first term of Eq. (\ref{rhkk1secondninth}) is
\begin{equation}
	\begin{split}
		& - 4 \gamma^{ij} \Gamma^{u}_{\ uj} F_{zu} \Gamma^{h}_{\ ui} F_{uh}\\
		= & - 4 \gamma^{ij} \Gamma^{u}_{\ uj} F_{zu} \Gamma^{u}_{\ ui} F_{uu} - 4 \gamma^{ij} \Gamma^{u}_{\ uj} F_{zu} \Gamma^{z}_{\ ui} F_{uz} - 4 \gamma^{ij} \Gamma^{u}_{\ uj} F_{zu} \Gamma^{k}_{\ ui} F_{uk}\\
		= & 0\,.
	\end{split}
\end{equation}
The second term of Eq. (\ref{rhkk1secondninth}) is
\begin{equation}
	\begin{split}
		& - 4 \gamma^{ij} \Gamma^{k}_{\ uj} F_{zk} \Gamma^{h}_{\ ui} F_{uh}\\
		= & - 4 \gamma^{ij} \Gamma^{k}_{\ uj} F_{zk} \Gamma^{u}_{\ ui} F_{uu} - 4 \gamma^{ij} \Gamma^{k}_{\ uj} F_{zk} \Gamma^{z}_{\ ui} F_{uz} - 4 \gamma^{ij} \Gamma^{k}_{\ uj} F_{zk} \Gamma^{l}_{\ ui} F_{ul}\\
		= & 0\,.
	\end{split}
\end{equation}
The third term of Eq. (\ref{rhkk1secondninth}) is
\begin{equation}
	\begin{split}
		- 4 \gamma^{ij} \Gamma^{z}_{\ uj} F_{uz} \Gamma^{h}_{\ ui} F_{zh} = 0\,.
	\end{split}
\end{equation}
The fourth term of Eq. (\ref{rhkk1secondninth}) is
\begin{equation}
	\begin{split}
		- 4 \gamma^{ij} \gamma^{kl} \Gamma^{z}_{\ ul} F_{jz} \Gamma^{h}_{\ uk} F_{ih} = 0\,.
	\end{split}
\end{equation}
The fifth term of Eq. (\ref{rhkk1secondninth}) is
\begin{equation}
	\begin{split}
		& - 4 \gamma^{ij} \gamma^{kl} \Gamma^{m}_{\ ul} F_{jm} \Gamma^{h}_{\ uk} F_{ih}\\
		= & - 4 \gamma^{ij} \gamma^{kl} \Gamma^{m}_{\ ul} F_{jm} \Gamma^{u}_{\ uk} F_{iu} - 4 \gamma^{ij} \gamma^{kl} \Gamma^{m}_{\ ul} F_{jm} \Gamma^{z}_{\ uk} F_{iz} - 4 \gamma^{ij} \gamma^{kl} \Gamma^{m}_{\ ul} F_{jm} \Gamma^{n}_{\ uk} F_{in}\\
		= & - \gamma^{ij} \gamma^{kl} \gamma^{mo} \gamma^{np} \left(\partial_u \gamma_{lo} \right) F_{jm} \left(\partial_u \gamma_{kp} \right) F_{in}\,.
	\end{split}
\end{equation}
Therefore, the ninth term of Eq. (\ref{rhkk1second}) is obtained as 
\begin{equation}
	\begin{split}
		& - 4 k^a k^b g^{ce} g^{df} \Gamma^{g}_{\ bf} F_{eg} \Gamma^{h}_{\ ad} F_{ch} = - \gamma^{ij} \gamma^{kl} \gamma^{mo} \gamma^{np} \left(\partial_u \gamma_{lo} \right) F_{jm} \left(\partial_u \gamma_{kp} \right) F_{in}\\
		= & - 4 \gamma^{ij} \gamma^{kl} \gamma^{mo} \gamma^{np} K_{lo} F_{jm} K_{kp} F_{in}\\
		= & 0\,.
	\end{split}
\end{equation}

Finally, the second term of Eq. (\ref{rhkk1}) is obtained as 
\begin{equation}
	\begin{split}
		& - 4 k^a k^b \left(\nabla_b F^{cd} \right) \left(\nabla_a F_{cd} \right)\\
		= & 8 \left(\partial_u F_{uz} \right) \left(\partial_u F_{uz} \right) - 16 \gamma^{ij} \left(\partial_u F_{ui} \right) \left(\partial_u F_{zj} \right) - 4 \gamma^{ij} \gamma^{kl} \left(\partial_u F_{jl} \right) \left(\partial_u F_{ik} \right)\\
		& + 2 \gamma^{ij} \gamma^{kl} \beta_l \left(\partial_u F_{uj} \right) F_{ki} - 2 \gamma^{ij} \beta_i \left(\partial_u F_{ju} \right) F_{uz} - 2 \gamma^{ij} \beta_i \left(\partial_u F_{uj} \right) F_{zu}\\
		& + 2 \gamma^{ij} \gamma^{kl} \beta_l \left(\partial_u F_{ju} \right) F_{ik} + 2 \gamma^{ij} \gamma^{kl} \beta_l F_{kj} \left(\partial_u F_{ui} \right) - 2 \gamma^{ij} \beta_j F_{uz} \left(\partial_u F_{iu} \right)\\
		& - 2 \gamma^{ij} \beta_j F_{zu} \left(\partial_u F_{ui} \right) + 2 \gamma^{ij} \gamma^{kl} \beta_l F_{jk} \left(\partial_u F_{iu} \right)\\
		= & 8 \left(\partial_u F_{uz} \right) \left(\partial_u F_{uz} \right) - 16 \gamma^{ij} \left(\partial_u F_{ui} \right) \left(\partial_u F_{zj} \right) - 4 \gamma^{ij} \gamma^{kl} \left(\partial_u F_{jl} \right) \left(\partial_u F_{ik} \right)\\
		& - 8 \gamma^{ij} \gamma^{kl} \beta_l \left(\partial_u F_{uj} \right) F_{ik} + 8 \gamma^{ij} \beta_i \left(\partial_u F_{uj} \right) F_{uz}\,.
	\end{split}
\end{equation}

The third term of Eq.(\ref{rhkk1}) is
\begin{equation}\label{rhkk1third}
	\begin{split}
		& 2 k^a k^b R_{ab} F_{cd} F^{cd} = 2 g^{ce} g^{df} R_{uu} F_{cd} F_{ef}\\
		= & 2 R_{uu} F_{uu} F_{zz} + 2 R_{uu} F_{uz} F_{zu} + 2 \gamma^{ij} R_{uu} F_{ui} F_{zj}\\
		& + 2 R_{uu} F_{zu} F_{uz} + 2 R_{uu} F_{zz} F_{uu} + 2 \gamma^{ij} R_{uu} F_{zi} F_{uj}\\
		& + 2 \gamma^{ij} R_{uu} F_{iu} F_{jz} + 2 \gamma^{ij} R_{uu} F_{iz} F_{ju} + 2 \gamma^{ij} \gamma^{kl} R_{uu} F_{ik} F_{jl}\\
		= & 2 R_{uu} F_{uz} F_{zu} + 2 R_{uu} F_{zu} F_{uz} + 2 \gamma^{ij} \gamma^{kl} R_{uu} F_{ik} F_{jl}\\
		= & 4 R_{uu} F_{uz} F_{zu} + 2 \gamma^{ij} \gamma^{kl} R_{uu} F_{ik} F_{jl}\,.
	\end{split}
\end{equation}
The first term of Eq. (\ref{rhkk1third}) is 
\begin{equation}
	\begin{split}
		& 4 R_{uu} F_{uz} F_{zu}\\
		= & 4 \left[- \frac{1}{2} \left(\partial_u \gamma^{ij} \right) \left(\partial_u \gamma_{ij} \right) - \frac{1}{2} \gamma^{ij} \partial_u^2 \gamma_{ij} - \frac{1}{4} \gamma^{ij} \gamma^{kl} \left(\partial_u \gamma_{ik} \right) \left(\partial_u \gamma_{jl} \right) \right] F_{uz} F_{zu}\\
		= & - 2 \left(\partial_u \gamma^{ij} \right) \left(\partial_u \gamma_{ij} \right) F_{uz} F_{zu} - 2 \gamma^{ij} \left(\partial_u^2 \gamma_{ij} \right) F_{uz} F_{zu} - \gamma^{ij} \gamma^{kl} \left(\partial_u \gamma_{ik} \right) \left(\partial_u \gamma_{jl} \right) F_{uz} F_{zu}\\
		= & - 8 K^{ij} K_{ij} F_{uz} F_{zu} - 2 \gamma^{ij} \left(\partial_u^2 \gamma_{ij} \right) F_{uz} F_{zu} - 4 \gamma^{ij} \gamma^{kl} K_{ik} K_{jl} F_{uz} F_{zu}\\
		= & - 2 \gamma^{ij} \left(\partial_u^2 \gamma_{ij} \right) F_{uz} F_{zu}\,.
	\end{split}
\end{equation}
The second term of Eq. (\ref{rhkk1third}) is 
\begin{equation}
	\begin{split}
		& 2 \gamma^{ij} \gamma^{kl} R_{uu} F_{ik} F_{jl}\\
		= & 2 \gamma^{ij} \gamma^{kl} \left[- \frac{1}{2} \left(\partial_u \gamma^{mn} \right) \left(\partial_u \gamma_{mn} \right) - \frac{1}{2} \gamma^{mn} \partial_u^2 \gamma_{mn} - \frac{1}{4} \gamma^{mn} \gamma^{op} \left(\partial_u \gamma_{mo} \right) \left(\partial_u \gamma_{np} \right) \right] F_{ik} F_{jl}\\
		= & - \gamma^{ij} \gamma^{kl} \left(\partial_u \gamma^{mn} \right) \left(\partial_u \gamma_{mn} \right) F_{ik} F_{jl} - \gamma^{ij} \gamma^{kl} \gamma^{mn} \left(\partial_u^2 \gamma_{mn} \right) F_{ik} F_{jl}\\
		& - \frac{1}{2} \gamma^{ij} \gamma^{kl} \gamma^{mn} \gamma^{op} \left(\partial_u \gamma_{mo} \right) \left(\partial_u \gamma_{np} \right) F_{ik} F_{jl}\\
		= & - 4 \gamma^{ij} \gamma^{kl} K^{mn} K_{mn} F_{ik} F_{jl} - \gamma^{ij} \gamma^{kl} \gamma^{mn} \left(\partial_u^2 \gamma_{mn} \right) F_{ik} F_{jl}\\
		& - 2 \gamma^{ij} \gamma^{kl} \gamma^{mn} \gamma^{op} K_{mo} K_{np} F_{ik} F_{jl}\\
		= & - \gamma^{ij} \gamma^{kl} \gamma^{mn} \left(\partial_u^2 \gamma_{mn} \right) F_{ik} F_{jl}\,.
	\end{split}
\end{equation}

Finally, the third term of Eq.(\ref{rhkk1}) is obtained as
\begin{equation}
	\begin{split}
		2 k^a k^b R_{ab} F_{cd} F^{cd} = - 2 \gamma^{ij} \left(\partial_u^2 \gamma_{ij} \right) F_{uz} F_{zu} - \gamma^{ij} \gamma^{kl} \gamma^{mn} \left(\partial_u^2 \gamma_{mn} \right) F_{ik} F_{jl}\,.
	\end{split}
\end{equation}

The fourth term of Eq.(\ref{rhkk1}) is
\begin{equation}
	\begin{split}
		& - 4 R k^a k^b F_{a}^{\ c} F_{bc} = - 4 R k^a k^b g^{ce} F_{ae} F_{bc} = - 4 R g^{ce} F_{ue} F_{uc}\\
		= & - 4 R g^{ue} F_{ue} F_{uu} - 4 R g^{ze} F_{ue} F_{uz} - 4 R g^{ie} F_{ue} F_{ui}\\
		= & - 4 R g^{uu} F_{uu} F_{uu} - 4 R g^{uz} F_{uz} F_{uu} - 4 R g^{ui} F_{ui} F_{uu}\\
		& - 4 R g^{zu} F_{uu} F_{uz} - 4 R g^{zz} F_{uz} F_{uz} - 4 R g^{zi} F_{ui} F_{uz}\\
		& - 4 R g^{iu} F_{uu} F_{ui} - 4 R g^{iz} F_{uz} F_{ui} - 4 R g^{ij} F_{uj} F_{ui}\\
		= & - 4 R F_{uz} F_{uu} - 4 R F_{uu} F_{uz} - 4 R \gamma^{ij} F_{uj} F_{ui}\\
		= & 0\,.
	\end{split}
\end{equation}

Finally, the fourth term of Eq.(\ref{rhkk1}) is obtained as 
\begin{equation}
	\begin{split}
		- 4 R k^a k^b F_{a}^{\ c} F_{bc} = 0\,.
	\end{split}
\end{equation}

Building on the results derived from the preceding calculations, the expression of $H_{uu}^{(1)}$ under the linear-order constraints of the quantum corrections can be formulated as
\begin{equation}\label{rhkk1result}
	\begin{split}
		H_{uu}^{(1)} = & 8 \left(\partial_u \partial_u F_{uz} \right) F_{uz} - 8 \gamma^{ij} \left(\partial_u \partial_u F_{ui} \right) F_{zj} - 4 \gamma^{ij} \gamma^{kl} \left(\partial_u \partial_u F_{il} \right) F_{jk}\\
		& - 8 \gamma^{ij} \beta_j \left(\partial_u F_{ui} \right) F_{uz} + 8 \gamma^{ij} \gamma^{kl} \beta_l \left(\partial_u F_{uj} \right) F_{ik} + 4 \gamma^{ij} \gamma^{kl} \gamma^{mn} \left(\partial_u \partial_u \gamma_{jn} \right) F_{ml} F_{ik}\\
		& + 8 \left(\partial_u F_{uz} \right) \left(\partial_u F_{uz} \right) - 16 \gamma^{ij} \left(\partial_u F_{ui} \right) \left(\partial_u F_{zj} \right) - 4 \gamma^{ij} \gamma^{kl} \left(\partial_u F_{jl} \right) \left(\partial_u F_{ik} \right)\\
		& - 8 \gamma^{ij} \gamma^{kl} \beta_l \left(\partial_u F_{uj} \right) F_{ik} + 8 \gamma^{ij} \beta_i \left(\partial_u F_{uj} \right) F_{uz} - 2 \gamma^{ij} \left(\partial_u^2 \gamma_{ij} \right) F_{uz} F_{zu}\\
		& - \gamma^{ij} \gamma^{kl} \gamma^{mn} \left(\partial_u^2 \gamma_{mn} \right) F_{ik} F_{jl}\,.
	\end{split}
\end{equation}
The first, seventh, and twelfth terms of Eq. (\ref{rhkk1result}) can be simplified as 
\begin{equation}\label{rhkk1resultfirstseventwelf}
	\begin{split}
		& 8 \left(\partial_u \partial_u F_{uz} \right) F_{uz} + 8 \left(\partial_u F_{uz} \right) \left(\partial_u F_{uz} \right) - 2 \gamma^{ij} \left(\partial_u^2 \gamma_{ij} \right) F_{uz} F_{zu}\\
		= & 4 \partial_u^2 \left(F_{uz} F_{uz} \right) + 2 \gamma^{ij} \left(\partial_u^2 \gamma_{ij} \right) F_{uz} F_{uz}\,.
	\end{split}
\end{equation}
Since the expression for the function $\mathcal{F} \left(u, x \right)$ should be integrated over the cross-section $B(u)$ on the null hypersurface $L$, it is necessary to multiply the expression for $\mathcal{F} \left(u, x \right)$ by the square root of the determinant of the induced metric on the cross-section, denoted as $\sqrt{\gamma (u\,, x)}$. To streamline the formulation of equations in the subsequent derivations, $\sqrt{\gamma (u\,, x)}$ will be denoted in a simplified form as $\sqrt{\gamma}$. After multiplying the induced metric $\sqrt{\gamma}$, the results in Eq. (\ref{rhkk1resultfirstseventwelf}) can be further simplified as 
\begin{equation}\label{rhkk1firstresult}
	\begin{split}
		4 \sqrt{\gamma} \partial_u^2 \left(F_{uz} F_{uz} \right) + 2 \sqrt{\gamma} \gamma^{ij} \left(\partial_u^2 \gamma_{ij} \right) F_{uz} F_{uz} = 4 \partial_u^2 \left(\sqrt{\gamma} F_{uz} F_{uz} \right)\,.
	\end{split}
\end{equation}
The second and eighth terms of Eq. (\ref{rhkk1result}) can be simplified as
\begin{equation}\label{rhkk1resultsecondeighth}
	\begin{split}
		& - 8 \gamma^{ij} \left(\partial_u \partial_u F_{ui} \right) F_{zj} - 16 \gamma^{ij} \left(\partial_u F_{ui} \right) \left(\partial_u F_{zj} \right)\\
		= & - 8 \gamma^{ij} \left(\partial_u^2 F_{ui} \right) F_{zj} - 16 \gamma^{ij} \left(\partial_u F_{ui} \right) \left(\partial_u F_{zj} \right) - 8 \gamma^{ij} F_{ui} \left(\partial_u^2 F_{zj} \right)\\
		= & - 8 \gamma^{ij} \partial_u^2 \left(F_{ui} F_{zj} \right)\,.
	\end{split}
\end{equation}
The second-order derivative of the term $\gamma^{ij} F_{ui} F_{zj}$ with respect to $u$ is calculated as
\begin{equation}\label{partial2ugammafuifzj}
	\begin{split}
		& \partial_u^2 \left(\gamma^{ij} F_{ui} F_{zj} \right) = \partial_u \left[\left(\partial_u \gamma^{ij} \right) F_{ui} F_{zj} + \gamma^{ij} \partial_u \left(F_{ui} F_{zj} \right) \right]\\
		= & \left(\partial_u^2 \gamma^{ij} \right) F_{ui} F_{zj} + \left(\partial_u \gamma^{ij} \right) \partial_u \left(F_{ui} F_{zj} \right) + \left(\partial_u \gamma^{ij} \right) \partial_u \left(F_{ui} F_{zj} \right) + \gamma^{ij} \partial_u^2 \left(F_{ui} F_{zj} \right)\\
		= & \left(\partial_u^2 \gamma^{ij} \right) F_{ui} F_{zj} + 2 K^{ij} \partial_u \left(F_{ui} F_{zj} \right) + 2 K^{ij}\partial_u \left(F_{ui} F_{zj} \right) + \gamma^{ij} \partial_u^2 \left(F_{ui} F_{zj} \right)\\
		= & \left(\partial_u^2 \gamma^{ij} \right) F_{ui} F_{zj} + \gamma^{ij} \partial_u^2 \left(F_{ui} F_{zj} \right)\\
		= & \gamma^{ij} \partial_u^2 \left(F_{ui} F_{zj} \right)\,.
	\end{split}
\end{equation}
Using the result in Eq. (\ref{partial2ugammafuifzj}), the result of the second and eighth terms in Eq. (\ref{rhkk1result}) can be expressed as 
\begin{equation}
	\begin{split}
		- 8 \gamma^{ij} \left(\partial_u \partial_u F_{ui} \right) F_{zj} - 16 \gamma^{ij} \left(\partial_u F_{ui} \right) \left(\partial_u F_{zj} \right) = - 8 \partial_u^2 \left(\gamma^{ij} F_{ui} F_{zj} \right)\,.
	\end{split}
\end{equation}
After incorporating the induced metric $\sqrt{\gamma}$, the result of the second and eighth terms in Eq. (\ref{rhkk1result}) can be further reformulated as 
\begin{equation}\label{rhkk1secondresult}
	\begin{split}
		& - 8 \sqrt{\gamma} \partial_u^2 \left(\gamma^{ij} F_{ui} F_{zj} \right)\\
		= & - 8 \sqrt{\gamma} \partial_u^2 \left(\gamma^{ij} F_{ui} F_{zj} \right) - 4 \sqrt{\gamma} \gamma^{kl} \left(\partial_u^2 \gamma_{kl} \right) \gamma^{ij} F_{ui} F_{zj}\\
		= & - 8 \left[\sqrt{\gamma} \partial_u^2 \left(\gamma^{ij} F_{ui} F_{zj} \right) + \frac{1}{2} \sqrt{\gamma} \gamma^{kl} \left(\partial_u^2 \gamma_{kl} \right) \gamma^{ij} F_{ui} F_{zj} \right]\\
		= & - 8 \partial_u^2 \left(\sqrt{\gamma} \gamma^{ij} F_{ui} F_{zj} \right)\,.
	\end{split}
\end{equation}
The third, sixth, ninth, and thirteenth terms of Eq. (\ref{rhkk1result}) are given as
\begin{equation}\label{thsinithrhkk1result}
	\begin{split}
		& - 4 \gamma^{ij} \gamma^{kl} \left(\partial_u \partial_u F_{il} \right) F_{jk} + 4 \gamma^{ij} \gamma^{kl} \gamma^{mn} \left(\partial_u \partial_u \gamma_{jn} \right) F_{ml} F_{ik} - 4 \gamma^{ij} \gamma^{kl} \left(\partial_u F_{jl} \right) \left(\partial_u F_{ik} \right)\\
		& - \gamma^{ij} \gamma^{kl} \gamma^{mn} \left(\partial_u^2 \gamma_{mn} \right) F_{ik} F_{jl}\,.
	\end{split}
\end{equation}
To simplify these four terms, the second-order derivative of the term $\gamma^{ij} \gamma^{kl} F_{il} F_{jk}$ with respect to $u$ is calculated as
\begin{equation}
	\begin{split}
		& \partial_u^2 \left(\gamma^{ij} \gamma^{kl} F_{ik} F_{jl} \right)\\
		= & \partial_u \left[\left(\partial_u \gamma^{ij} \right) \gamma^{kl} F_{ik} F_{jl} + \gamma^{ij} \left(\partial_u \gamma^{kl} \right) F_{ik} F_{jl} + \gamma^{ij} \gamma^{kl} \left(\partial_u F_{ik} \right) F_{jl} \right.\\
		& \left. + \gamma^{ij} \gamma^{kl} F_{ik} \left(\partial_u F_{jl} \right) \right]\\
		= & \partial_u \left[\left(\partial_u \gamma^{ij} \right) \gamma^{kl} F_{ik} F_{jl} + \gamma^{kl} \left(\partial_u \gamma^{ij} \right) F_{ki} F_{lj} + \gamma^{ij} \gamma^{kl} \left(\partial_u F_{ik} \right) F_{jl} \right.\\
		& \left. + \gamma^{ij} \gamma^{kl} F_{jl} \left(\partial_u F_{ik} \right) \right]\\
		= & \partial_u \left[2 \left(\partial_u \gamma^{ij} \right) \gamma^{kl} F_{ik} F_{jl} + 2 \gamma^{ij} \gamma^{kl} \left(\partial_u F_{ik} \right) F_{jl} \right]\\
		= & 2 \left(\partial_u^2 \gamma^{ij} \right) \gamma^{kl} F_{ik} F_{jl} + 2 \gamma^{ij} \gamma^{kl} \left(\partial_u^2 F_{ik} \right) F_{jl} + 2 \gamma^{ij} \gamma^{kl} \left(\partial_u F_{ik} \right) \left(\partial_u F_{jl} \right)\,.
	\end{split}
\end{equation}
On the other hand, the second-order derivative of the term $\gamma^{ij} \gamma_{jn}$ can be calculated as
\begin{equation}\label{partialu2gijgjn}
	\begin{split}
		& \partial_u^2 \left(\gamma^{ij} \gamma_{jn} \right) = \left(\partial_u^2 \gamma^{ij} \right) \gamma_{jn} + 2 \left(\partial_u \gamma^{ij} \right) \left(\partial_u \gamma_{jn} \right) + \gamma^{ij} \left(\partial_u^2 \gamma_{jn} \right)\\
		= & \left(\partial_u^2 \gamma^{ij} \right) \gamma_{jn} + 8 K^{ij} K_{jn} + \gamma^{ij} \left(\partial_u^2 \gamma_{jn} \right)\\
		= & \left(\partial_u^2 \gamma^{ij} \right) \gamma_{jn} + \gamma^{ij} \left(\partial_u^2 \gamma_{jn} \right)\\
		= & 0\,.
	\end{split}
\end{equation}
From the computational process in Eq. (\ref{partialu2gijgjn}), we can obtain the following relationship
\begin{equation}
	\begin{split}
		\gamma^{ij} \left(\partial_u^2 \gamma_{jn} \right) = - \left(\partial_u^2 \gamma^{ij} \right) \gamma_{jn}\,.
	\end{split}
\end{equation} 
Using this relationship, the second term of Eq. (\ref{thsinithrhkk1result}) can be further expressed as 
\begin{equation}
	\begin{split}
		4 \gamma^{ij} \gamma^{kl} \gamma^{mn} \left(\partial_u \partial_u \gamma_{jn} \right) F_{ml} F_{ik} = - 4 \gamma^{kl} \gamma^{mn} \gamma_{jn} \left(\partial_u^2 \gamma^{ij} \right) F_{ml} F_{ik} = - 4 \left(\partial_u^2 \gamma^{ij} \right) \gamma^{kl} F_{ik} F_{jl}\,.
	\end{split}
\end{equation}
Therefore, Eq. (\ref{thsinithrhkk1result}) can be simplified as 
\begin{equation}
	\begin{split}
		& - 4 \gamma^{ij} \gamma^{kl} \left(\partial_u \partial_u F_{ik} \right) F_{jl} - 4 \left(\partial_u^2 \gamma^{ij} \right) \gamma^{kl} F_{ik} F_{jl} - 4 \gamma^{ij} \gamma^{kl} \left(\partial_u F_{ik} \right) \left(\partial_u F_{jl} \right)\\
		& - \gamma^{ij} \gamma^{kl} \gamma^{mn} \left(\partial_u^2 \gamma_{mn} \right) F_{ik} F_{jl}\\
		= & - 2 \left[2 \gamma^{ij} \gamma^{kl} \left(\partial_u^2 F_{ik} \right) F_{jl} + 2 \left(\partial_u^2 \gamma^{ij} \right) \gamma^{kl} F_{ik} F_{jl} + 2 \gamma^{ij} \gamma^{kl} \left(\partial_u F_{ik} \right) \left(\partial_u F_{jl} \right) \right]\\
		& - \gamma^{ij} \gamma^{kl} \gamma^{mn} \left(\partial_u^2 \gamma_{mn} \right) F_{ik} F_{jl}\\
		= & - 2 \partial_u^2 \left(\gamma^{ij} \gamma^{kl} F_{ik} F_{jl} \right) - \gamma^{ij} \gamma^{kl} \gamma^{mn} \left(\partial_u^2 \gamma_{mn} \right) F_{ik} F_{jl}\,.
	\end{split}
\end{equation}
After multiplying the induced metric $\sqrt{\gamma}$, the above result allows for further calculation and simplification as
\begin{equation}\label{rhkk1thirdresult}
	\begin{split}
		& - 2 \sqrt{\gamma} \partial_u^2 \left(\gamma^{ij} \gamma^{kl} F_{ik} F_{jl} \right) - \sqrt{\gamma} \gamma^{ij} \gamma^{kl} \gamma^{mn} \left(\partial_u^2 \gamma_{mn} \right) F_{ik} F_{jl}\\
		= & - 2 \left[\sqrt{\gamma} \partial_u^2 \left(\gamma^{ij} \gamma^{kl} F_{ik} F_{jl} \right) + \frac{1}{2} \sqrt{\gamma} \gamma^{mn} \left(\partial_u^2 \gamma_{mn} \right) \left(\gamma^{ij} \gamma^{kl} F_{ik} F_{jl} \right) \right]\\
		= & - 2 \partial_u^2 \left(\sqrt{\gamma} \gamma^{ij} \gamma^{kl} F_{ik} F_{jl} \right)\,.
	\end{split}
\end{equation}
The fourth, fifth, tenth and eleventh terms in Eq. (\ref{rhkk1result}) are
\begin{equation}
	\begin{split}
		& - 8 \gamma^{ij} \beta_j \left(\partial_u F_{ui} \right) F_{uz} + 8 \gamma^{ij} \beta_i \left(\partial_u F_{uj} \right) F_{uz} + 8 \gamma^{ij} \gamma^{kl} \beta_l \left(\partial_u F_{uj} \right) F_{ik}\\
		& - 8 \gamma^{ij} \gamma^{kl} \beta_l \left(\partial_u F_{uj} \right) F_{ik}\\
		= & 0\,,
	\end{split}
\end{equation}
where the symmetric property of the two indices $i$ and $j$ in the induced metric $\gamma^{ij}$ has been applied in the final step. Therefore, combining the results from Eqs. (\ref{rhkk1firstresult}), (\ref{rhkk1secondresult}), and (\ref{rhkk1thirdresult}), the specific expression of $H_{uu}^{(1)}$ with the induced metric $\sqrt{\gamma}$ can be explicitly written as
\begin{equation}\label{rhkk1resultfinalwithgamma}
	\begin{split}
		\sqrt{\gamma} H_{uu}^{(1)} = & 4 \partial_u^2 \left(\sqrt{\gamma} F_{uz} F_{uz} \right) - 8 \partial_u^2 \left(\sqrt{\gamma} \gamma^{ij} F_{ui} F_{zj} \right) - 2 \partial_u^2 \left(\sqrt{\gamma} \gamma^{ij} \gamma^{kl} F_{ik} F_{jl} \right)\\
		= & 2 \partial_u^2 \left(2 \sqrt{\gamma} F_{uz} F_{uz} - 4 \sqrt{\gamma} \gamma^{ij} F_{ui} F_{zj} - \sqrt{\gamma} \gamma^{ij} \gamma^{kl} F_{ik} F_{jl} \right)\,.
	\end{split}
\end{equation}

Next, we derive the specific expression of $H_{uu}^{(2)}$. According to the symmetric property of the null vectors $k^a$ and $k^b$, the expression of $H_{uu}^{(2)}$ in Eq. (\ref{ohkk2}) can be simplified as 
\begin{equation}\label{rewrittenhkk2}
	\begin{split}
		H_{uu}^{(2)} = & 2 k^a k^b F_{a}^{\ c} \nabla_d \nabla_b F_{c}^{\ d} + 2 k^a k^b F^{cd} \nabla_d \nabla_a F_{bc} + 2 k^a k^b F_{a}^{\ c} \nabla_d \nabla^d F_{bc}\\
		& + 2 k^a k^b \nabla_a F_{b}^{\ c} \nabla_d F_{c}^{\ d} + 2 k^a k^b \nabla_a F_{cd} \nabla^d F_{b}^{\ c} + 2 k^a k^b \nabla_d F_{bc} \nabla^d F_{a}^{\ c}\\
		& - 2 k^a k^b F_{a}^{\ c} F_b^{\ d} R_{cd}\,.
	\end{split}
\end{equation}

The first term in Eq. (\ref{rewrittenhkk2}) is
\begin{equation}\label{hkk2first}
	\begin{split}
		& 2 k^a k^b F_{a}^{\ c} \nabla_d \nabla_b F_{c}^{\ d} = 2 k^a k^b g^{ce} g^{df} F_{ae} \nabla_d \nabla_b F_{cf}\\
		= & 2 k^a k^b g^{ce} g^{df} F_{ae} \left(\partial_d \partial_b F_{cf}\right) - 2 k^a k^b g^{ce} g^{df} F_{ae} \Gamma^{g}_{\ db} \left(\partial_g F_{cf}\right)\\
		& - 2 k^a k^b g^{ce} g^{df} F_{ae} \Gamma^{g}_{\ dc} \left(\partial_b F_{gf} \right) - 2 k^a k^b g^{ce} g^{df} F_{ae} \Gamma^{g}_{\ df} \left(\partial_b F_{cg} \right)\\
		& - 2 k^a k^b g^{ce} g^{df} F_{ae} \left(\partial_d \Gamma^{g}_{\ bc} \right) F_{gf} + 2 k^a k^b g^{ce} g^{df} F_{ae} \Gamma^{h}_{\ db} \Gamma^{g}_{\ hc} F_{gf}\\
		& + 2 k^a k^b g^{ce} g^{df} F_{ae} \Gamma^{h}_{\ dc} \Gamma^{g}_{\ bh} F_{gf} - 2 k^a k^b g^{ce} g^{df} F_{ae} \Gamma^{g}_{\ dh} \Gamma^{h}_{\ bc} F_{gf}\\
		& - 2 k^a k^b g^{ce} g^{df} F_{ae} \Gamma^{g}_{\ bc} \left(\partial_d F_{gf} \right) + 2 k^a k^b g^{ce} g^{df} F_{ae} \Gamma^{g}_{\ bc} \Gamma^{h}_{\ dg} F_{hf}\\
		& + 2 k^a k^b g^{ce} g^{df} F_{ae} \Gamma^{g}_{\ bc} \Gamma^{h}_{\ df} F_{gh} - 2 k^a k^b g^{ce} g^{df} F_{ae} \left(\partial_d \Gamma^{g}_{\ bf} \right) F_{cg}\\
		& + 2 k^a k^b g^{ce} g^{df} F_{ae} \Gamma^{h}_{\ db} \Gamma^{g}_{\ hf} F_{cg} + 2 k^a k^b g^{ce} g^{df} F_{ae} \Gamma^{h}_{\ df} \Gamma^{g}_{\ bh} F_{cg}\\
		& - 2 k^a k^b g^{ce} g^{df} F_{ae} \Gamma^{g}_{\ dh} \Gamma^{h}_{\ bf} F_{cg} - 2 k^a k^b g^{ce} g^{df} F_{ae} \Gamma^{g}_{\ bf} \left(\partial_d F_{cg} \right)\\
		& + 2 k^a k^b g^{ce} g^{df} F_{ae} \Gamma^{g}_{\ bf} \Gamma^{h}_{\ dc} F_{hg} + 2 k^a k^b g^{ce} g^{df} F_{ae} \Gamma^{g}_{\ bf} \Gamma^{h}_{\ dg} F_{ch}\,.
	\end{split}
\end{equation}

The first term of Eq. (\ref{hkk2first}) is obtained as 
\begin{equation}
	\begin{split}
		& 2 k^a k^b g^{ce} g^{df} F_{ae} \left(\partial_d \partial_b F_{cf} \right) = 2 g^{ce} g^{df} F_{ue} \left(\partial_d \partial_u F_{cf} \right)\\
		= & 2 F_{uz} \left(\partial_u \partial_u F_{uz} \right) + 2 \gamma^{ij} F_{uz} \left(\partial_i \partial_u F_{uj} \right) + 2 \gamma^{ij} F_{uj} \left(\partial_u \partial_u F_{iz} \right)\\
		& + 2 \gamma^{ij} F_{uj} \left(\partial_z \partial_u F_{iu} \right) + 2 \gamma^{ij} \gamma^{kl} F_{uj} \left(\partial_k \partial_u F_{il} \right)\\
		= & 2 F_{uz} \left(\partial_u \partial_u F_{uz} \right) + 2 \gamma^{ij} F_{uz} \left(\partial_i \partial_u F_{uj} \right)\,.
	\end{split}
\end{equation}

The second term of Eq. (\ref{hkk2first}) is 
\begin{equation}
	\begin{split}
		& - 2 k^a k^b g^{ce} g^{df} F_{ae} \Gamma^{g}_{\ db} \left(\partial_g F_{cf} \right) = - 2 g^{ce} g^{df} F_{ue} \Gamma^{g}_{\ du} \left(\partial_g F_{cf} \right)\\
		= & - 2 F_{uz} \Gamma^{g}_{\ uu} \left(\partial_g F_{uz} \right) - 2 \gamma^{ij} F_{uz} \Gamma^{g}_{\ iu} \left(\partial_g F_{uj} \right)\\
		= & - 2 \gamma^{ij} F_{uz} \Gamma^{g}_{\ iu} \left(\partial_g F_{uj} \right)\,.
	\end{split}
\end{equation}
The repeated index $g$ should be expanded as
\begin{equation}\label{hkk2firstsecond}
	\begin{split}
		& - 2 \gamma^{ij} F_{uz} \Gamma^{g}_{\ iu} \left(\partial_g F_{uj} \right)\\
		= & - 2 \gamma^{ij} F_{uz} \Gamma^{u}_{\ iu} \left(\partial_u F_{uj} \right) - 2 \gamma^{ij} F_{uz} \Gamma^{z}_{\ iu} \left(\partial_z F_{uj} \right) - 2 \gamma^{ij} F_{uz} \Gamma^{k}_{\ iu} \left(\partial_k F_{uj} \right)\,.
	\end{split}
\end{equation}
Therefore, the second term of Eq. (\ref{hkk2first}) is obtained as 
\begin{equation}
	\begin{split}
		& - 2 k^a k^b g^{ce} g^{df} F_{ae} \Gamma^{g}_{\ db} \left(\partial_g F_{cf} \right)\\
		= & \gamma^{ij} \beta_i F_{uz} \left(\partial_u F_{uj} \right) - \gamma^{ij} \gamma^{kl} F_{uz} \left(\partial_u \gamma_{il} \right) \left(\partial_k F_{uj} \right)\\
		= & \gamma^{ij} \beta_i F_{uz} \left(\partial_u F_{uj} \right) - 2 \gamma^{ij} \gamma^{kl} F_{uz} K_{il} \left(\partial_k F_{uj} \right)\\
		= & \gamma^{ij} \beta_i F_{uz} \left(\partial_u F_{uj} \right)\,.
	\end{split}
\end{equation}

The third term of Eq. (\ref{hkk2first}) is
\begin{equation}
	\begin{split}
		& - 2 k^a k^b g^{ce} g^{df} F_{ae} \Gamma^{g}_{\ dc} \left(\partial_b F_{gf} \right) = - 2 g^{ce} g^{df} F_{ue} \Gamma^{g}_{\ dc} \left(\partial_u F_{gf} \right)\\
		= & - 2 F_{uz} \Gamma^{g}_{\ uu} \left(\partial_u F_{gz} \right) - 2 F_{uz} \Gamma^{g}_{\ zu} \left(\partial_u F_{gu} \right) - 2 \gamma^{ij} F_{uz} \Gamma^{g}_{\ iu} \left(\partial_u F_{gj} \right)\\
		= & - 2 F_{uz} \Gamma^{g}_{\ zu} \left(\partial_u F_{gu} \right) - 2 \gamma^{ij} F_{uz} \Gamma^{g}_{\ iu} \left(\partial_u F_{gj} \right)\,.
	\end{split}
\end{equation}
The repeated index $g$ is expanded as 
\begin{equation}\label{hkk2firstthird}
	\begin{split}
		& - 2 F_{uz} \Gamma^{g}_{\ zu} \left(\partial_u F_{gu} \right) - 2 \gamma^{ij} F_{uz} \Gamma^{g}_{\ iu} \left(\partial_u F_{gj} \right)\\
		= & - 2 F_{uz} \Gamma^{u}_{\ zu} \left(\partial_u F_{uu} \right) - 2 F_{uz} \Gamma^{z}_{\ zu} \left(\partial_u F_{zu} \right) - 2 F_{uz} \Gamma^{i}_{\ zu} \left(\partial_u F_{iu} \right)\\
		& - 2 \gamma^{ij} F_{uz} \Gamma^{u}_{\ iu} \left(\partial_u F_{uj} \right) - 2 \gamma^{ij} F_{uz} \Gamma^{z}_{\ iu} \left(\partial_u F_{zj} \right) - 2 \gamma^{ij} F_{uz} \Gamma^{k}_{\ iu} \left(\partial_u F_{kj} \right)\\
		= & - 2 F_{uz} \Gamma^{z}_{\ zu} \left(\partial_u F_{zu} \right) - 2 F_{uz} \Gamma^{i}_{\ zu} \left(\partial_u F_{iu} \right) - 2 \gamma^{ij} F_{uz} \Gamma^{u}_{\ iu} \left(\partial_u F_{uj} \right)\\
		& - 2 \gamma^{ij} F_{uz} \Gamma^{z}_{\ iu} \left(\partial_u F_{zj} \right) - 2 \gamma^{ij} F_{uz} \Gamma^{k}_{\ iu} \left(\partial_u F_{kj} \right)\,.
	\end{split}
\end{equation}
Therefore, the third term of Eq. (\ref{hkk2first}) is obtained as 
\begin{equation}
	\begin{split}
		& - 2 k^a k^b g^{ce} g^{df} F_{ae} \Gamma^{g}_{\ dc} \left(\partial_b F_{gf} \right)\\
		= & - \gamma^{ij} \beta_j F_{uz} \left(\partial_u F_{iu} \right) + \gamma^{ij} \beta_i F_{uz} \left(\partial_u F_{uj} \right) - \gamma^{ij} \gamma^{kl} F_{uz} \left(\partial_u \gamma_{il} \right) \left(\partial_u F_{kj} \right)\\
		= & \gamma^{ij} \beta_i F_{uz} \left(\partial_u F_{uj} \right) + \gamma^{ij} \beta_i F_{uz} \left(\partial_u F_{uj} \right) - 2 \gamma^{ij} \gamma^{kl} F_{uz} K_{il} \left(\partial_u F_{kj} \right)\\
		= & 2 \gamma^{ij} \beta_i F_{uz} \left(\partial_u F_{uj} \right)\,.
	\end{split}
\end{equation}

The fourth term of Eq. (\ref{hkk2first}) is
\begin{equation}
	\begin{split}
		& - 2 k^a k^b g^{ce} g^{df} F_{ae} \Gamma^{g}_{\ df} \left(\partial_b F_{cg} \right) = - 2 g^{ce} g^{df} F_{ue} \Gamma^{g}_{\ df} \left(\partial_u F_{cg} \right)\\
		= & - 2 F_{uz} \Gamma^{g}_{\ uz} \left(\partial_u F_{ug} \right) - 2 F_{uz} \Gamma^{g}_{\ zu} \left(\partial_u F_{ug} \right) - 2 \gamma^{ij} F_{uz} \Gamma^{g}_{\ ij} \left(\partial_u F_{ug} \right)\,.
	\end{split}
\end{equation}
The index $g$ should be expanded.
\begin{equation}\label{hkk2firstfourth}
	\begin{split}
		& - 2 F_{uz} \Gamma^{g}_{\ uz} \left(\partial_u F_{ug} \right) - 2 F_{uz} \Gamma^{g}_{\ zu} \left(\partial_u F_{ug} \right) - 2 \gamma^{ij} F_{uz} \Gamma^{g}_{\ ij} \left(\partial_u F_{ug} \right)\\
		= & - 2 F_{uz} \Gamma^{u}_{\ uz} \left(\partial_u F_{uu} \right) - 2 F_{uz} \Gamma^{z}_{\ uz} \left(\partial_u F_{uz} \right) - 2 F_{uz} \Gamma^{i}_{\ uz} \left(\partial_u F_{ui} \right)\\
		& - 2 F_{uz} \Gamma^{u}_{\ zu} \left(\partial_u F_{uu} \right) - 2 F_{uz} \Gamma^{z}_{\ zu} \left(\partial_u F_{uz} \right) - 2 F_{uz} \Gamma^{i}_{\ zu} \left(\partial_u F_{ui} \right)\\
		& - 2 \gamma^{ij} F_{uz} \Gamma^{u}_{\ ij} \left(\partial_u F_{uu} \right) - 2 \gamma^{ij} F_{uz} \Gamma^{z}_{\ ij} \left(\partial_u F_{uz} \right) - 2 \gamma^{ij} F_{uz} \Gamma^{k}_{\ ij} \left(\partial_u F_{uk} \right)\\
		= & - 2 F_{uz} \Gamma^{z}_{\ uz} \left(\partial_u F_{uz} \right) - 2 F_{uz} \Gamma^{i}_{\ uz} \left(\partial_u F_{ui} \right) - 2 F_{uz} \Gamma^{z}_{\ zu} \left(\partial_u F_{uz} \right)\\
		& - 2 F_{uz} \Gamma^{i}_{\ zu} \left(\partial_u F_{ui} \right) - 2 \gamma^{ij} F_{uz} \Gamma^{z}_{\ ij} \left(\partial_u F_{uz} \right) - 2 \gamma^{ij} F_{uz} \Gamma^{k}_{\ ij} \left(\partial_u F_{uk} \right)\,.
	\end{split}
\end{equation}
Therefore, the fourth term of Eq. (\ref{hkk2first}) is obtained as 
\begin{equation}
	\begin{split}
		& - 2 k^a k^b g^{ce} g^{df} F_{ae} \Gamma^{g}_{\ df} \left(\partial_b F_{cg} \right)\\
		= & - \gamma^{ij} \beta_j F_{uz} \left(\partial_u F_{ui} \right) - \gamma^{ij} \beta_j F_{uz} \left(\partial_u F_{ui} \right) + \gamma^{ij} F_{uz} \left(\partial_u \gamma_{ij} \right) \left(\partial_u F_{uz} \right)\\
		& - 2 \gamma^{ij} F_{uz} \hat{\Gamma}^{k}_{\ ij} \left(\partial_u F_{uk} \right)\\
		= & - 2 \gamma^{ij} \beta_j F_{uz} \left(\partial_u F_{ui} \right) + 2 \gamma^{ij} F_{uz} K_{ij} \left(\partial_u F_{uz} \right) - 2 \gamma^{ij} F_{uz} \hat{\Gamma}^{k}_{\ ij} \left(\partial_u F_{uk} \right)\\
		= & - 2 \gamma^{ij} \beta_j F_{uz} \left(\partial_u F_{ui} \right) - 2 \gamma^{ij} F_{uz} \hat{\Gamma}^{k}_{\ ij} \left(\partial_u F_{uk} \right)\,.
	\end{split}
\end{equation}

The fifth term of Eq. (\ref{hkk2first}) is
\begin{equation}
	\begin{split}
		& - 2 k^a k^b g^{ce} g^{df} F_{ae} \left(\partial_d \Gamma^{g}_{\ bc} \right) F_{gf} = - 2 g^{ce} g^{df} F_{ue} \left(\partial_d \Gamma^{g}_{\ uc} \right) F_{gf}\\
		= & - 2 F_{uz} \left(\partial_u \Gamma^{g}_{\ uu} \right) F_{gz} - 2 F_{uz} \left(\partial_z \Gamma^{g}_{\ uu} \right) F_{gu} - 2 \gamma^{ij} F_{uz} \left(\partial_i \Gamma^{g}_{\ uu} \right) F_{gj}\,.
	\end{split}
\end{equation}
The index $g$ should be expanded.
\begin{equation}\label{hkk2firstfifth}
	\begin{split}
		& - 2 F_{uz} \left(\partial_u \Gamma^{g}_{\ uu} \right) F_{gz} - 2 F_{uz} \left(\partial_z \Gamma^{g}_{\ uu} \right) F_{gu} - 2 \gamma^{ij} F_{uz} \left(\partial_i \Gamma^{g}_{\ uu} \right) F_{gj}\\
		= & - 2 F_{uz} \left(\partial_u \Gamma^{u}_{\ uu} \right) F_{uz} - 2 F_{uz} \left(\partial_u \Gamma^{z}_{\ uu} \right) F_{zz} - 2 F_{uz} \left(\partial_u \Gamma^{i}_{\ uu} \right) F_{iz}\\
		& - 2 F_{uz} \left(\partial_z \Gamma^{u}_{\ uu} \right) F_{uu} - 2 F_{uz} \left(\partial_z \Gamma^{z}_{\ uu} \right) F_{zu} - 2 F_{uz} \left(\partial_z \Gamma^{i}_{\ uu} \right) F_{iu}\\
		& - 2 \gamma^{ij} F_{uz} \left(\partial_i \Gamma^{u}_{\ uu} \right) F_{uj} - 2 \gamma^{ij} F_{uz} \left(\partial_i \Gamma^{z}_{\ uu} \right) F_{zj} - 2 \gamma^{ij} F_{uz} \left(\partial_i \Gamma^{k}_{\ uu} \right) F_{kj}\\
		= & - 2 F_{uz} \left(\partial_u \Gamma^{u}_{\ uu} \right) F_{uz} - 2 F_{uz} \left(\partial_u \Gamma^{i}_{\ uu} \right) F_{iz} - 2 F_{uz} \left(\partial_z \Gamma^{z}_{\ uu} \right) F_{zu}\\
		& - 2 \gamma^{ij} F_{uz} \left(\partial_i \Gamma^{z}_{\ uu} \right) F_{zj} - 2 \gamma^{ij} F_{uz} \left(\partial_i \Gamma^{k}_{\ uu} \right) F_{kj}\,.
	\end{split}
\end{equation}
The first term of Eq. (\ref{hkk2firstfifth}) is 
\begin{equation}
	\begin{split}
		& - 2 F_{uz} \left(\partial_u \Gamma^{u}_{\ uu} \right) F_{uz}\\
		= & - 2 F_{uz} \partial_u \left(- z^2 \partial_z \alpha - 2 z \alpha \right) F_{uz}\\
		= & - 2 F_{uz} \left(- z^2 \partial_u \partial_z \alpha - 2 z \partial_u \alpha \right) F_{uz}\\
		= & 0\,.
	\end{split}
\end{equation}
The second term of Eq. (\ref{hkk2firstfifth}) is 
\begin{equation}
	\begin{split}
		& - 2 F_{uz} \left(\partial_u \Gamma^{i}_{\ uu} \right) F_{iz}\\
		= & - 2 F_{uz} \partial_u \left[\left(z \beta^i \right) \left(z^2 \partial_z \alpha + 2 z \alpha \right) + \gamma^{ij} \left(z \partial_u \beta_j - z^2 \partial_j \alpha \right) \right] F_{iz}\\
		= & - 2 F_{uz} \left[\left(z \partial_u \beta^i \right) \left(z^2 \partial_z \alpha + 2 z \alpha \right) + \left(z \beta^i \right) \left(z^2 \partial_u \partial_z \alpha + 2 z \partial_u \alpha \right) \right.\\
		& \left. + \left(\partial_u \gamma^{ij} \right) \left(z \partial_u \beta_j - z^2 \partial_j \alpha \right) + \gamma^{ij} \left(z \partial_u^2 \beta_j - z^2 \partial_u \partial_j \alpha \right) \right] F_{iz}\\
		= & 0\,.
	\end{split}
\end{equation}
The third term of Eq. (\ref{hkk2firstfifth}) is
\begin{equation}
	\begin{split}
		& - 2 F_{uz} \left(\partial_z \Gamma^{z}_{\ uu} \right) F_{zu}\\
		= & - 2 F_{uz} \partial_z \left[z^2 \partial_u \alpha - z^2 \left(\beta^2 - 2 \alpha \right) \left(z^2 \partial_z \alpha + 2 z \alpha \right) - z \beta^i \left(z \partial_u \beta_i - z^2 \partial_i \alpha \right) \right] F_{zu}\\
		= & - 2 F_{uz} \left[2 z \partial_u \alpha + z^2 \partial_z \partial_u \alpha - 2 z \left(\beta^2 - 2 \alpha \right) \left(z^2 \partial_z \alpha + 2 z \alpha \right) \right.\\
		& \left.- z^2 \left(\partial_z \beta^2 - 2 \partial_z \alpha \right) \left(z^2 \partial_z \alpha + 2 z \alpha \right) - z^2 \left(\beta^2 - 2 \alpha \right) \left(2 z \partial_z \alpha + z^2 \partial_z^2 \alpha + 2 \alpha + 2 z \partial_z \alpha \right) \right.\\
		& \left. - \beta^i \left(z \partial_u \beta_i - z^2 \partial_i \alpha \right) - z \left(\partial_z \beta^i \right) \left(z \partial_u \beta_i - z^2 \partial_i \alpha \right) \right.\\
		& \left. - z \beta^i \left(\partial_u \beta_i + z \partial_z \partial_u \beta_i - 2 z \partial_i \alpha - z^2 \partial_z \partial_i \alpha \right) \right] F_{zu}\\
		= & 0\,.
	\end{split}
\end{equation}
The fourth term of Eq. (\ref{hkk2firstfifth}) is
\begin{equation}
	\begin{split}
		& - 2 \gamma^{ij} F_{uz} \left(\partial_i \Gamma^{z}_{\ uu} \right) F_{zj}\\
		= & - 2 \gamma^{ij} F_{uz} \partial_i \left[z^2 \partial_u \alpha - z^2 \left(\beta^2 - 2 \alpha \right) \left(z^2 \partial_z \alpha + 2 z \alpha \right) - z \beta^k \left(z \partial_u \beta_k - z^2 \partial_k \alpha \right) \right] F_{zj}\\
		= & - 2 \gamma^{ij} F_{uz} \left[z^2 \partial_i \partial_u \alpha - z^2 \left(\partial_i \beta^2 - 2 \partial_i \alpha \right) \left(z^2 \partial_z \alpha + 2 z \alpha \right) \right.\\
		& \left. - z^2 \left(\beta^2 - 2 \alpha \right) \left(z^2 \partial_i \partial_z \alpha + 2 z \partial_i \alpha \right) - z \left(\partial_i \beta^k \right) \left(z \partial_u \beta_k - z^2 \partial_k \alpha \right) \right.\\
		& \left. - z \beta^k \left(z \partial_i \partial_u \beta_k - z^2 \partial_i \partial_k \alpha \right) \right] F_{zj}\\
		= & 0\,.
	\end{split}
\end{equation}
The fifth term of Eq. (\ref{hkk2firstfifth}) is
\begin{equation}
	\begin{split}
		& - 2 \gamma^{ij} F_{uz} \left(\partial_i \Gamma^{k}_{\ uu} \right) F_{kj}\\
		= & - 2 \gamma^{ij} F_{uz} \partial_i \left[\left(z \beta^k \right) \left(z^2 \partial_z \alpha + 2 z \alpha \right) + \gamma^{kl} \left(z \partial_u \beta_l - z^2 \partial_l \alpha \right) \right] F_{kj}\\
		= & - 2 \gamma^{ij} F_{uz} \left[\left(z \partial_i \beta^k \right) \left(z^2 \partial_z \alpha + 2 z \alpha \right) + \left(z \beta^k \right) \left(z^2 \partial_i \partial_z \alpha + 2 z \partial_i \alpha \right) \right.\\
		& \left. + \left(\partial_i \gamma^{kl} \right) \left(z \partial_u \beta_l - z^2 \partial_l \alpha \right) + \gamma^{kl} \left(z \partial_i \partial_u \beta_l - z^2 \partial_i \partial_l \alpha \right) \right] F_{kj}\\
		= & 0\,.
	\end{split}
\end{equation}
Therefore, the fifth term of Eq. (\ref{hkk2first}) is obtained as 
\begin{equation}
	\begin{split}
		- 2 k^a k^b g^{ce} g^{df} F_{ae} \left(\partial_d \Gamma^{g}_{\ bc} \right) F_{gf} = 0
	\end{split}
\end{equation}

The sixth term of Eq. (\ref{hkk2first}) is
\begin{equation}
	\begin{split}
		& 2 k^a k^b g^{ce} g^{df} F_{ae} \Gamma^{h}_{\ db} \Gamma^{g}_{\ hc} F_{gf} = 2 g^{ce} g^{df} F_{ue} \Gamma^{h}_{\ du} \Gamma^{g}_{\ hc} F_{gf}\\
		= & 2 F_{uz} \Gamma^{h}_{\ uu} \Gamma^{g}_{\ hu} F_{gz} + 2 F_{uz} \Gamma^{h}_{\ zu} \Gamma^{g}_{\ hu} F_{gu} + 2 \gamma^{ij} F_{uz} \Gamma^{h}_{\ iu} \Gamma^{g}_{\ hu} F_{gj}\\
		= & 2 F_{uz} \Gamma^{h}_{\ zu} \Gamma^{g}_{\ hu} F_{gu} + 2 \gamma^{ij} F_{uz} \Gamma^{h}_{\ iu} \Gamma^{g}_{\ hu} F_{gj}\,.
	\end{split}
\end{equation}
The index $g$ should be expanded.
\begin{equation}\label{hkk2firstsixth}
	\begin{split}
		& 2 F_{uz} \Gamma^{h}_{\ zu} \Gamma^{g}_{\ hu} F_{gu} + 2 \gamma^{ij} F_{uz} \Gamma^{h}_{\ iu} \Gamma^{g}_{\ hu} F_{gj}\\
		= & 2 F_{uz} \Gamma^{h}_{\ zu} \Gamma^{u}_{\ hu} F_{uu} + 2 F_{uz} \Gamma^{h}_{\ zu} \Gamma^{z}_{\ hu} F_{zu} + 2 F_{uz} \Gamma^{h}_{\ zu} \Gamma^{i}_{\ hu} F_{iu}\\
		& + 2 \gamma^{ij} F_{uz} \Gamma^{h}_{\ iu} \Gamma^{u}_{\ hu} F_{uj} + 2 \gamma^{ij} F_{uz} \Gamma^{h}_{\ iu} \Gamma^{z}_{\ hu} F_{zj} + 2 \gamma^{ij} F_{uz} \Gamma^{h}_{\ iu} \Gamma^{k}_{\ hu} F_{kj}\\
		= & 2 F_{uz} \Gamma^{h}_{\ zu} \Gamma^{z}_{\ hu} F_{zu} + 2 \gamma^{ij} F_{uz} \Gamma^{h}_{\ iu} \Gamma^{z}_{\ hu} F_{zj} + 2 \gamma^{ij} F_{uz} \Gamma^{h}_{\ iu} \Gamma^{k}_{\ hu} F_{kj}\,.
	\end{split}
\end{equation}
The repeated index $h$ should be further expanded. The first term of Eq. (\ref{hkk2firstsixth}) is
\begin{equation}
	\begin{split}
		& 2 F_{uz} \Gamma^{h}_{\ zu} \Gamma^{z}_{\ hu} F_{zu}\\
		= & 2 F_{uz} \Gamma^{u}_{\ zu} \Gamma^{z}_{\ uu} F_{zu} + 2 F_{uz} \Gamma^{z}_{\ zu} \Gamma^{z}_{\ zu} F_{zu} + 2 F_{uz} \Gamma^{i}_{\ zu} \Gamma^{z}_{\ iu} F_{zu}\\
		= & 0\,.
	\end{split}
\end{equation}
The second term of Eq. (\ref{hkk2firstsixth}) is
\begin{equation}
	\begin{split}
		& 2 \gamma^{ij} F_{uz} \Gamma^{h}_{\ iu} \Gamma^{z}_{\ hu} F_{zj}\\
		= & 2 \gamma^{ij} F_{uz} \Gamma^{u}_{\ iu} \Gamma^{z}_{\ uu} F_{zj} + 2 \gamma^{ij} F_{uz} \Gamma^{z}_{\ iu} \Gamma^{z}_{\ zu} F_{zj} + 2 \gamma^{ij} F_{uz} \Gamma^{k}_{\ iu} \Gamma^{z}_{\ ku} F_{zj}\\
		= & 0\,.
	\end{split}
\end{equation}
The third term of Eq. (\ref{hkk2firstsixth}) is
\begin{equation}
	\begin{split}
		& 2 \gamma^{ij} F_{uz} \Gamma^{h}_{\ iu} \Gamma^{k}_{\ hu} F_{kj}\\
		= & 2 \gamma^{ij} F_{uz} \Gamma^{u}_{\ iu} \Gamma^{k}_{\ uu} F_{kj} + 2 \gamma^{ij} F_{uz} \Gamma^{z}_{\ iu} \Gamma^{k}_{\ zu} F_{kj} + 2 \gamma^{ij} F_{uz} \Gamma^{l}_{\ iu} \Gamma^{k}_{\ lu} F_{kj}\\
		= & \frac{1}{2} \gamma^{ij} \gamma^{kn} \gamma^{lm} F_{uz} \left(\partial_u \gamma_{im} \right) \left(\partial_u \gamma_{ln} \right) F_{kj}\,.
	\end{split}
\end{equation}
Therefore, the sixth term of Eq. (\ref{hkk2first}) is obtained as
\begin{equation}
	\begin{split}
		& 2 k^a k^b g^{ce} g^{df} F_{ae} \Gamma^{h}_{\ db} \Gamma^{g}_{\ hc} F_{gf} = \frac{1}{2} \gamma^{ij} \gamma^{kn} \gamma^{lm} F_{uz} \left(\partial_u \gamma_{im} \right) \left(\partial_u \gamma_{ln} \right) F_{kj}\\
		= & 2 \gamma^{ij} \gamma^{kn} \gamma^{lm} F_{uz} K_{im} K_{ln} F_{kj}\\
		= & 0\,.
	\end{split}
\end{equation}

The seventh term of Eq. (\ref{hkk2first}) is
\begin{equation}
	\begin{split}
		& 2 k^a k^b g^{ce} g^{df} F_{ae} \Gamma^{h}_{\ dc} \Gamma^{g}_{\ bh} F_{gf} = 2 g^{ce} g^{df} F_{ue} \Gamma^{h}_{\ dc} \Gamma^{g}_{\ uh} F_{gf}\\
		= & 2 F_{uz} \Gamma^{h}_{\ uu} \Gamma^{g}_{\ uh} F_{gz} + 2 F_{uz} \Gamma^{h}_{\ zu} \Gamma^{g}_{\ uh} F_{gu} + 2 \gamma^{ij} F_{uz} \Gamma^{h}_{\ iu} \Gamma^{g}_{\ uh} F_{gj}\\
		= & 2 F_{uz} \Gamma^{h}_{\ zu} \Gamma^{g}_{\ uh} F_{gu} + 2 \gamma^{ij} F_{uz} \Gamma^{h}_{\ iu} \Gamma^{g}_{\ uh} F_{gj}\,.
	\end{split}
\end{equation}
The index $g$ should be expanded.
\begin{equation}\label{hkk2firstseventh}
	\begin{split}
		& 2 F_{uz} \Gamma^{h}_{\ zu} \Gamma^{g}_{\ uh} F_{gu} + 2 \gamma^{ij} F_{uz} \Gamma^{h}_{\ iu} \Gamma^{g}_{\ uh} F_{gj}\\
		= & + 2 F_{uz} \Gamma^{h}_{\ zu} \Gamma^{u}_{\ uh} F_{uu} + 2 F_{uz} \Gamma^{h}_{\ zu} \Gamma^{z}_{\ uh} F_{zu} + 2 F_{uz} \Gamma^{h}_{\ zu} \Gamma^{i}_{\ uh} F_{iu}\\
		& + 2 \gamma^{ij} F_{uz} \Gamma^{h}_{\ iu} \Gamma^{u}_{\ uh} F_{uj} + 2 \gamma^{ij} F_{uz} \Gamma^{h}_{\ iu} \Gamma^{z}_{\ uh} F_{zj} + 2 \gamma^{ij} F_{uz} \Gamma^{h}_{\ iu} \Gamma^{k}_{\ uh} F_{kj}\\
		= & 2 F_{uz} \Gamma^{h}_{\ zu} \Gamma^{z}_{\ uh} F_{zu} + 2 \gamma^{ij} F_{uz} \Gamma^{h}_{\ iu} \Gamma^{z}_{\ uh} F_{zj} + 2 \gamma^{ij} F_{uz} \Gamma^{h}_{\ iu} \Gamma^{k}_{\ uh} F_{kj}\,.
	\end{split}
\end{equation}
The first term of Eq. (\ref{hkk2firstseventh}) is
\begin{equation}
	\begin{split}
		& 2 F_{uz} \Gamma^{h}_{\ zu} \Gamma^{z}_{\ uh} F_{zu}\\
		= & 2 F_{uz} \Gamma^{u}_{\ zu} \Gamma^{z}_{\ uu} F_{zu} + 2 F_{uz} \Gamma^{z}_{\ zu} \Gamma^{z}_{\ uz} F_{zu} + 2 F_{uz} \Gamma^{i}_{\ zu} \Gamma^{z}_{\ ui} F_{zu}\\
		= & 0\,.
	\end{split}
\end{equation}
The second term of Eq. (\ref{hkk2firstseventh}) is
\begin{equation}
	\begin{split}
		& 2 \gamma^{ij} F_{uz} \Gamma^{h}_{\ iu} \Gamma^{z}_{\ uh} F_{zj}\\
		= & 2 \gamma^{ij} F_{uz} \Gamma^{u}_{\ iu} \Gamma^{z}_{\ uu} F_{zj} + 2 \gamma^{ij} F_{uz} \Gamma^{z}_{\ iu} \Gamma^{z}_{\ uz} F_{zj} + 2 \gamma^{ij} F_{uz} \Gamma^{k}_{\ iu} \Gamma^{z}_{\ uk} F_{zj}\\
		= & 0\,.
	\end{split}
\end{equation}
The third term of Eq. (\ref{hkk2firstseventh}) is
\begin{equation}
	\begin{split}
		& 2 \gamma^{ij} F_{uz} \Gamma^{h}_{\ iu} \Gamma^{k}_{\ uh} F_{kj}\\
		= & 2 \gamma^{ij} F_{uz} \Gamma^{u}_{\ iu} \Gamma^{k}_{\ uu} F_{kj} + 2 \gamma^{ij} F_{uz} \Gamma^{z}_{\ iu} \Gamma^{k}_{\ uz} F_{kj} + 2 \gamma^{ij} F_{uz} \Gamma^{l}_{\ iu} \Gamma^{k}_{\ ul} F_{kj}\\
		= & \frac{1}{2} \gamma^{ij} \gamma^{kn} \gamma^{lm} F_{uz} \left(\partial_u \gamma_{im} \right) \left(\partial_u \gamma_{ln} \right) F_{kj}\,.
	\end{split}
\end{equation}
Therefore, the seventh term of Eq. (\ref{hkk2first}) is obtained as
\begin{equation}
	\begin{split}
		& 2 k^a k^b g^{ce} g^{df} F_{ae} \Gamma^{h}_{\ dc} \Gamma^{g}_{\ bh} F_{gf} = \frac{1}{2} \gamma^{ij} \gamma^{kn} \gamma^{lm} F_{uz} \left(\partial_u \gamma_{im} \right) \left(\partial_u \gamma_{ln} \right) F_{kj}\\
		= & 2 \gamma^{ij} \gamma^{kn} \gamma^{lm} F_{uz} K_{im} K_{ln} F_{kj}\\
		= & 0\,.
	\end{split}
\end{equation}

The eighth term of Eq. (\ref{hkk2first}) is
\begin{equation}
	\begin{split}
		& - 2 k^a k^b g^{ce} g^{df} F_{ae} \Gamma^{g}_{\ dh} \Gamma^{h}_{\ bc} F_{gf} = - 2 g^{ce} g^{df} F_{ue} \Gamma^{g}_{\ dh} \Gamma^{h}_{\ uc} F_{gf}\\
		= & - 2 F_{uz} \Gamma^{g}_{\ uh} \Gamma^{h}_{\ uu} F_{gz} - 2 F_{uz} \Gamma^{g}_{\ zh} \Gamma^{h}_{\ uu} F_{gu} - 2 \gamma^{ij} F_{uz} \Gamma^{g}_{\ ih} \Gamma^{h}_{\ uu} F_{gj}\\
		= & 0\,.
	\end{split}
\end{equation}
Therefore, the eighth term of Eq. (\ref{hkk2first}) is obtained as
\begin{equation}
	\begin{split}
		- 2 k^a k^b g^{ce} g^{df} F_{ae} \Gamma^{g}_{\ dh} \Gamma^{h}_{\ bc} F_{gf} = 0\,. 
	\end{split}
\end{equation}

The ninth term of Eq. (\ref{hkk2first}) is
\begin{equation}
	\begin{split}
		& - 2 k^a k^b g^{ce} g^{df} F_{ae} \Gamma^{g}_{\ bc} \left(\partial_d F_{gf} \right) = - 2 g^{ce} g^{df} F_{ue} \Gamma^{g}_{\ uc} \left(\partial_d F_{gf} \right)\\
		= & - 2 F_{uz} \Gamma^{g}_{\ uu} \left(\partial_u F_{gz} \right) - 2 F_{uz} \Gamma^{g}_{\ uu} \left(\partial_z F_{gu} \right) - 2 \gamma^{ij} F_{uz} \Gamma^{g}_{\ uu} \left(\partial_i F_{gj} \right)\\
		= & 0\,.
	\end{split}
\end{equation}
Therefore, the ninth term of Eq. (\ref{hkk2first}) is obtained as
\begin{equation}
	\begin{split}
		- 2 k^a k^b g^{ce} g^{df} F_{ae} \Gamma^{g}_{\ bc} \left(\partial_d F_{gf} \right) = 0\,.
	\end{split}
\end{equation}

The tenth term of Eq. (\ref{hkk2first}) is
\begin{equation}
	\begin{split}
		& 2 k^a k^b g^{ce} g^{df} F_{ae} \Gamma^{g}_{\ bc} \Gamma^{h}_{\ dg} F_{hf} = 2 g^{ce} g^{df} F_{ue} \Gamma^{g}_{\ uc} \Gamma^{h}_{\ dg} F_{hf}\\
		= & 2 F_{uz} \Gamma^{g}_{\ uu} \Gamma^{h}_{\ ug} F_{hz} + 2 F_{uz} \Gamma^{g}_{\ uu} \Gamma^{h}_{\ zg} F_{hu} + 2 \gamma^{ij} F_{uz} \Gamma^{g}_{\ uu} \Gamma^{h}_{\ ig} F_{hj}\\
		= & 0\,.
	\end{split}
\end{equation}
Therefore, the tenth term of Eq. (\ref{hkk2first}) is obtained as 
\begin{equation}
	\begin{split}
		2 k^a k^b g^{ce} g^{df} F_{ae} \Gamma^{g}_{\ bc} \Gamma^{h}_{\ dg} F_{hf} = 0\,.
	\end{split}
\end{equation}

The eleventh term of Eq. (\ref{hkk2first}) is
\begin{equation}
	\begin{split}
		& 2 k^a k^b g^{ce} g^{df} F_{ae} \Gamma^{g}_{\ bc} \Gamma^{h}_{\ df} F_{gh} = 2 g^{ce} g^{df} F_{ue} \Gamma^{g}_{\ uc} \Gamma^{h}_{\ df} F_{gh}\\
		= & 2 F_{uz} \Gamma^{g}_{\ uu} \Gamma^{h}_{\ uz} F_{gh} + 2 F_{uz} \Gamma^{g}_{\ uu} \Gamma^{h}_{\ zu} F_{gh} + 2 \gamma^{ij} F_{uz} \Gamma^{g}_{\ uu} \Gamma^{h}_{\ ij} F_{gh}\\
		= & 0\,.
	\end{split}
\end{equation}
Therefore the eleventh term of Eq. (\ref{hkk2first}) is obtained as 
\begin{equation}
	\begin{split}
		2 k^a k^b g^{ce} g^{df} F_{ae} \Gamma^{g}_{\ bc} \Gamma^{h}_{\ df} F_{gh} = 0\,. 
	\end{split}
\end{equation}

The twelfth term of Eq. (\ref{hkk2first}) is
\begin{equation}
	\begin{split}
		& - 2 k^a k^b g^{ce} g^{df} F_{ae} \left(\partial_d \Gamma^{g}_{\ bf} \right) F_{cg} = - 2 g^{ce} g^{df} F_{ue} \left(\partial_d \Gamma^{g}_{\ uf} \right) F_{cg}\\
		= & - 2 F_{uz} \left(\partial_u \Gamma^{g}_{\ uz} \right) F_{ug} - 2 F_{uz} \left(\partial_z \Gamma^{g}_{\ uu} \right) F_{ug} - 2 \gamma^{ij} F_{uz} \left(\partial_i \Gamma^{g}_{\ uj} \right) F_{ug}\,.
	\end{split}
\end{equation}
The repeated index $g$ should be further expanded as
\begin{equation}\label{hkk2firsttwelfth}
	\begin{split}
		& - 2 F_{uz} \left(\partial_u \Gamma^{g}_{\ uz} \right) F_{ug} - 2 F_{uz} \left(\partial_z \Gamma^{g}_{\ uu} \right) F_{ug} - 2 \gamma^{ij} F_{uz} \left(\partial_i \Gamma^{g}_{\ uj} \right) F_{ug}\\
		= & - 2 F_{uz} \left(\partial_u \Gamma^{u}_{\ uz} \right) F_{uu} - 2 F_{uz} \left(\partial_u \Gamma^{z}_{\ uz} \right) F_{uz} - 2 F_{uz} \left(\partial_u \Gamma^{i}_{\ uz} \right) F_{ui}\\
		& - 2 F_{uz} \left(\partial_z \Gamma^{u}_{\ uu} \right) F_{uu} - 2 F_{uz} \left(\partial_z \Gamma^{z}_{\ uu} \right) F_{uz} - 2 F_{uz} \left(\partial_z \Gamma^{i}_{\ uu} \right) F_{ui}\\
		& - 2 \gamma^{ij} F_{uz} \left(\partial_i \Gamma^{u}_{\ uj} \right) F_{uu} - 2 \gamma^{ij} F_{uz} \left(\partial_i \Gamma^{z}_{\ uj} \right) F_{uz} - 2 \gamma^{ij} F_{uz} \left(\partial_i \Gamma^{k}_{\ uj} \right) F_{uk}\\
		= & - 2 F_{uz} \left(\partial_u \Gamma^{z}_{\ uz} \right) F_{uz} - 2 F_{uz} \left(\partial_z \Gamma^{z}_{\ uu} \right) F_{uz} - 2 \gamma^{ij} F_{uz} \left(\partial_i \Gamma^{z}_{\ uj} \right) F_{uz}\,.
	\end{split}
\end{equation}
The first term of Eq. (\ref{hkk2firsttwelfth}) is 
\begin{equation}
	\begin{split}
		& - 2 F_{uz} \left(\partial_u \Gamma^{z}_{\ uz} \right) F_{uz}\\
		= & - 2 F_{uz} \partial_u \left(z^2 \partial_z \alpha + 2 z \alpha - \frac{1}{2} z \beta^i \beta_i - \frac{1}{2} z^2 \beta^i \partial_z \beta_i \right) F_{uz}\\
		= & - 2 F_{uz} \left[z^2 \partial_u \partial_z \alpha + 2 z \partial_u \alpha - \frac{1}{2} z \partial_u \left(\beta^i \beta_i \right) - \frac{1}{2} z^2 \left(\partial_u \beta^i \right) \left(\partial_z \beta_i \right) - \frac{1}{2} z^2 \beta^i \partial_u \partial_z \beta_i \right]\\
		& \times F_{uz}\\
		= & 0\,.
	\end{split}
\end{equation}
The second term of Eq. (\ref{hkk2firsttwelfth}) is
\begin{equation}
	\begin{split}
		& - 2 F_{uz} \left(\partial_z \Gamma^{z}_{\ uu} \right) F_{uz}\\
		= & - 2 F_{uz} \partial_z \left[z^2 \partial_u \alpha - z^2 \left(\beta^2 - 2 \alpha \right) \left(z^2 \partial_z \alpha + 2 z \alpha \right) - z \beta^i \left(z \partial_u \beta_i - z^2 \partial_i \alpha \right) \right] F_{uz}\\
		= & - 2 F_{uz} \left[2 z \partial_u \alpha + z^2 \partial_z \partial_u \alpha - 2 z \left(\beta^2 - 2 \alpha \right) \left(z^2 \partial_z \alpha + 2 z \alpha \right)\right.\\
		& \left. - z^2 \left(\partial_z \beta^2 - 2 \partial_z \alpha \right) \left(z^2 \partial_z \alpha + 2 z \alpha \right) - z^2 \left(\beta^2 - 2 \alpha \right) \left(2 z \partial_z \alpha + z^2 \partial_z^2 \alpha + 2 \alpha + 2 z \partial_z \alpha \right) \right.\\
		& \left. - \beta^i \left(z \partial_u \beta_i - z^2 \partial_i \alpha \right) - z \left(\partial_z \beta^i \right) \left(z \partial_u \beta_i - z^2 \partial_i \alpha \right) \right.\\
		& \left. - z \beta^i \left(\partial_u \beta_i + z \partial_z \partial_u \beta_i - 2 z \partial_i \alpha - z^2 \partial_z \partial_i \alpha \right) \right] F_{uz}\\
		= & 0\,.
	\end{split}
\end{equation}
The third term of Eq. (\ref{hkk2firsttwelfth}) is
\begin{equation}
	\begin{split}
		& - 2 \gamma^{ij} F_{uz} \left(\partial_i \Gamma^{z}_{\ uj} \right) F_{uz}\\
		= & - 2 \gamma^{ij} F_{uz}\\
		& \times \partial_i \left[z^2 \partial_j \alpha - \frac{1}{2} z^2 \left(\beta^2 - 2 \alpha \right) \left(\beta_j + z \partial_z \beta_j \right) - \frac{1}{2} z \beta^k \left(z \partial_j \beta_k + \partial_u \gamma_{jk} - z \partial_k \beta_j \right) \right] \\
		= & - 2 \gamma^{ij} F_{uz} \left[z^2 \partial_i \partial_j \alpha - \frac{1}{2} z^2 \left(\partial_i \beta^2 - 2 \partial_i \alpha \right) \left(\beta_j + z \partial_z \beta_j \right) \right.\\
		& \left. - \frac{1}{2} z^2 \left(\beta^2 - 2 \alpha \right) \left(\partial_i \beta_j + z \partial_i \partial_z \beta_j \right) - \frac{1}{2} z \left(\partial_i \beta^k \right) \left(z \partial_j \beta_k + \partial_u \gamma_{jk} - z \partial_k \beta_j \right)\right.\\
		& \left. - \frac{1}{2} z \beta^k \left(z \partial_i \partial_j \beta_k + \partial_i \partial_u \gamma_{jk} - z \partial_i \partial_k \beta_j \right) \right] F_{uz}\\
		= & 0\,.
	\end{split}
\end{equation}
Therefore, the twelfth term of Eq. (\ref{hkk2first}) is obtained as 
\begin{equation}
	\begin{split}
		- 2 k^a k^b g^{ce} g^{df} F_{ae} \left(\partial_d \Gamma^{g}_{\ bf} \right) F_{cg} = 0\,.
	\end{split}
\end{equation}

The thirteenth term of Eq. (\ref{hkk2first}) is
\begin{equation}
	\begin{split}
		& 2 k^a k^b g^{ce} g^{df} F_{ae} \Gamma^{h}_{\ db} \Gamma^{g}_{\ hf} F_{cg} = 2 g^{ce} g^{df} F_{ue} \Gamma^{h}_{\ du} \Gamma^{g}_{\ hf} F_{cg}\\
		= & 2 F_{uz} \Gamma^{h}_{\ uu} \Gamma^{g}_{\ hz} F_{ug} + 2 F_{uz} \Gamma^{h}_{\ zu} \Gamma^{g}_{\ hu} F_{ug} + 2 \gamma^{ij} F_{uz} \Gamma^{h}_{\ iu} \Gamma^{g}_{\ hj} F_{ug}\\
		= & 2 F_{uz} \Gamma^{h}_{\ zu} \Gamma^{g}_{\ hu} F_{ug} + 2 \gamma^{ij} F_{uz} \Gamma^{h}_{\ iu} \Gamma^{g}_{\ hj} F_{ug}\,.
	\end{split}
\end{equation}
The index $g$ should be expanded.
\begin{equation}\label{hkk2firstthirteenth}
	\begin{split}
		& 2 F_{uz} \Gamma^{h}_{\ zu} \Gamma^{g}_{\ hu} F_{ug} + 2 \gamma^{ij} F_{uz} \Gamma^{h}_{\ iu} \Gamma^{g}_{\ hj} F_{ug}\\
		= & 2 F_{uz} \Gamma^{h}_{\ zu} \Gamma^{u}_{\ hu} F_{uu} + 2 F_{uz} \Gamma^{h}_{\ zu} \Gamma^{z}_{\ hu} F_{uz} + 2 F_{uz} \Gamma^{h}_{\ zu} \Gamma^{i}_{\ hu} F_{ui}\\
		& + 2 \gamma^{ij} F_{uz} \Gamma^{h}_{\ iu} \Gamma^{u}_{\ hj} F_{uu} + 2 \gamma^{ij} F_{uz} \Gamma^{h}_{\ iu} \Gamma^{z}_{\ hj} F_{uz} + 2 \gamma^{ij} F_{uz} \Gamma^{h}_{\ iu} \Gamma^{k}_{\ hj} F_{uk}\\
		= & 2 F_{uz} \Gamma^{h}_{\ zu} \Gamma^{z}_{\ hu} F_{uz} + 2 \gamma^{ij} F_{uz} \Gamma^{h}_{\ iu} \Gamma^{z}_{\ hj} F_{uz}\,.
	\end{split}
\end{equation}
The repeated index $h$ should be further expanded. The first term of Eq. (\ref{hkk2firstthirteenth}) is
\begin{equation}
	\begin{split}
		& 2 F_{uz} \Gamma^{h}_{\ zu} \Gamma^{z}_{\ hu} F_{uz}\\
		= & 2 F_{uz} \Gamma^{u}_{\ zu} \Gamma^{z}_{\ uu} F_{uz} + 2 F_{uz} \Gamma^{z}_{\ zu} \Gamma^{z}_{\ zu} F_{uz} + 2 F_{uz} \Gamma^{i}_{\ zu} \Gamma^{z}_{\ iu} F_{uz}\\
		= & 0\,.
	\end{split}
\end{equation}
The second term of Eq. (\ref{hkk2firstthirteenth}) is
\begin{equation}
	\begin{split}
		& 2 \gamma^{ij} F_{uz} \Gamma^{h}_{\ iu} \Gamma^{z}_{\ hj} F_{uz}\\
		= & 2 \gamma^{ij} F_{uz} \Gamma^{u}_{\ iu} \Gamma^{z}_{\ uj} F_{uz} + 2 \gamma^{ij} F_{uz} \Gamma^{z}_{\ iu} \Gamma^{z}_{\ zj} F_{uz} + 2 \gamma^{ij} F_{uz} \Gamma^{k}_{\ iu} \Gamma^{z}_{\ kj} F_{uz}\\
		= & - \frac{1}{2} \gamma^{ij} \gamma^{kl} F_{uz} \left(\partial_u \gamma_{il} \right) \left(\partial_u \gamma_{kj} \right) F_{uz}\,.
	\end{split}
\end{equation}
Therefore, the thirteenth term of Eq. (\ref{hkk2first}) is obtained as 
\begin{equation}
	\begin{split}
		& 2 k^a k^b g^{ce} g^{df} F_{ae} \Gamma^{h}_{\ db} \Gamma^{g}_{\ hf} F_{cg} = - \frac{1}{2} \gamma^{ij} \gamma^{kl} F_{uz} \left(\partial_u \gamma_{il} \right) \left(\partial_u \gamma_{kj} \right) F_{uz}\\
		= & - 2 \gamma^{ij} \gamma^{kl} F_{uz} K_{il} K_{kj} F_{uz}\\
		= & 0\,.
	\end{split}
\end{equation}

The fourteenth term of Eq. (\ref{hkk2first}) is
\begin{equation}
	\begin{split}
		& 2 k^a k^b g^{ce} g^{df} F_{ae} \Gamma^{h}_{\ df} \Gamma^{g}_{\ bh} F_{cg} = 2 g^{ce} g^{df} F_{ue} \Gamma^{h}_{\ df} \Gamma^{g}_{\ uh} F_{cg}\\
		= & 4 F_{uz} \Gamma^{h}_{\ uz} \Gamma^{g}_{\ uh} F_{ug} + 2 \gamma^{ij} F_{uz} \Gamma^{h}_{\ ij} \Gamma^{g}_{\ uh} F_{ug}\,.
	\end{split}
\end{equation}
The index $g$ should be further expanded.
\begin{equation}\label{hkk2firstfourteenth}
	\begin{split}
		& 4 F_{uz} \Gamma^{h}_{\ uz} \Gamma^{g}_{\ uh} F_{ug} + 2 \gamma^{ij} F_{uz} \Gamma^{h}_{\ ij} \Gamma^{g}_{\ uh} F_{ug}\\
		= & 4 F_{uz} \Gamma^{h}_{\ uz} \Gamma^{u}_{\ uh} F_{uu} + 4 F_{uz} \Gamma^{h}_{\ uz} \Gamma^{z}_{\ uh} F_{uz} + 4 F_{uz} \Gamma^{h}_{\ uz} \Gamma^{i}_{\ uh} F_{ui}\\
		& + 2 \gamma^{ij} F_{uz} \Gamma^{h}_{\ ij} \Gamma^{u}_{\ uh} F_{uu} + 2 \gamma^{ij} F_{uz} \Gamma^{h}_{\ ij} \Gamma^{z}_{\ uh} F_{uz} + 2 \gamma^{ij} F_{uz} \Gamma^{h}_{\ ij} \Gamma^{k}_{\ uh} F_{uk}\\
		= & 4 F_{uz} \Gamma^{h}_{\ uz} \Gamma^{z}_{\ uh} F_{uz} + 2 \gamma^{ij} F_{uz} \Gamma^{h}_{\ ij} \Gamma^{z}_{\ uh} F_{uz}\,.
	\end{split}
\end{equation}
The first term of Eq. (\ref{hkk2firstfourteenth}) is 
\begin{equation}
	\begin{split}
		& 4 F_{uz} \Gamma^{h}_{\ uz} \Gamma^{z}_{\ uh} F_{uz}\\
		= & 4 F_{uz} \Gamma^{u}_{\ uz} \Gamma^{z}_{\ uu} F_{uz} + 4 F_{uz} \Gamma^{z}_{\ uz} \Gamma^{z}_{\ uz} F_{uz} + 4 F_{uz} \Gamma^{i}_{\ uz} \Gamma^{z}_{\ ui} F_{uz}\\
		= & 0\,.
	\end{split}
\end{equation}
The second term of Eq. (\ref{hkk2firstfourteenth}) is
\begin{equation}
	\begin{split}
		& 2 \gamma^{ij} F_{uz} \Gamma^{h}_{\ ij} \Gamma^{z}_{\ uh} F_{uz}\\
		= & 2 \gamma^{ij} F_{uz} \Gamma^{u}_{\ ij} \Gamma^{z}_{\ uu} F_{uz} + 2 \gamma^{ij} F_{uz} \Gamma^{z}_{\ ij} \Gamma^{z}_{\ uz} F_{uz} + 2 \gamma^{ij} F_{uz} \Gamma^{k}_{\ ij} \Gamma^{z}_{\ uk} F_{uz}\\
		= & 0\,.
	\end{split}
\end{equation}
Therefore, the fourteenth term of Eq. (\ref{hkk2first}) is obtained as 
\begin{equation}
	\begin{split}
		2 k^a k^b g^{ce} g^{df} F_{ae} \Gamma^{h}_{\ df} \Gamma^{g}_{\ bh} F_{cg} = 0\,.
	\end{split}
\end{equation}

The fifteenth term of Eq. (\ref{hkk2first}) is
\begin{equation}
	\begin{split}
		& - 2 k^a k^b g^{ce} g^{df} F_{ae} \Gamma^{g}_{\ dh} \Gamma^{h}_{\ bf} F_{cg} = - 2 g^{ce} g^{df} F_{ue} \Gamma^{g}_{\ dh} \Gamma^{h}_{\ uf} F_{cg}\\
		= & - 2 F_{uz} \Gamma^{g}_{\ uh} \Gamma^{h}_{\ uz} F_{ug} - 2 F_{uz} \Gamma^{g}_{\ zh} \Gamma^{h}_{\ uu} F_{ug} - 2 \gamma^{ij} F_{uz} \Gamma^{g}_{\ ih} \Gamma^{h}_{\ uj} F_{ug}\\
		= & - 2 F_{uz} \Gamma^{g}_{\ uh} \Gamma^{h}_{\ uz} F_{ug} - 2 \gamma^{ij} F_{uz} \Gamma^{g}_{\ ih} \Gamma^{h}_{\ uj} F_{ug}\,.
	\end{split}
\end{equation}
The index $g$ should be further expanded.
\begin{equation}\label{hkk2firstfifteenth}
	\begin{split}
		& - 2 F_{uz} \Gamma^{g}_{\ uh} \Gamma^{h}_{\ uz} F_{ug} - 2 \gamma^{ij} F_{uz} \Gamma^{g}_{\ ih} \Gamma^{h}_{\ uj} F_{ug}\\
		= & - 2 F_{uz} \Gamma^{u}_{\ uh} \Gamma^{h}_{\ uz} F_{uu} - 2 F_{uz} \Gamma^{z}_{\ uh} \Gamma^{h}_{\ uz} F_{uz} - 2 F_{uz} \Gamma^{i}_{\ uh} \Gamma^{h}_{\ uz} F_{ui}\\
		& - 2 \gamma^{ij} F_{uz} \Gamma^{u}_{\ ih} \Gamma^{h}_{\ uj} F_{uu} - 2 \gamma^{ij} F_{uz} \Gamma^{z}_{\ ih} \Gamma^{h}_{\ uj} F_{uz} - 2 \gamma^{ij} F_{uz} \Gamma^{k}_{\ ih} \Gamma^{h}_{\ uj} F_{uk}\\
		= & - 2 F_{uz} \Gamma^{z}_{\ uh} \Gamma^{h}_{\ uz} F_{uz} - 2 \gamma^{ij} F_{uz} \Gamma^{z}_{\ ih} \Gamma^{h}_{\ uj} F_{uz}\,.
	\end{split}
\end{equation}
The first term of Eq. (\ref{hkk2firstfifteenth}) is 
\begin{equation}
	\begin{split}
		& - 2 F_{uz} \Gamma^{z}_{\ uh} \Gamma^{h}_{\ uz} F_{uz}\\
		= & - 2 F_{uz} \Gamma^{z}_{\ uu} \Gamma^{u}_{\ uz} F_{uz} - 2 F_{uz} \Gamma^{z}_{\ uz} \Gamma^{z}_{\ uz} F_{uz} - 2 F_{uz} \Gamma^{z}_{\ ui} \Gamma^{i}_{\ uz} F_{uz}\\
		= & 0\,.
	\end{split}
\end{equation}
The second term of Eq. (\ref{hkk2firstfifteenth}) is
\begin{equation}
	\begin{split}
		& - 2 \gamma^{ij} F_{uz} \Gamma^{z}_{\ ih} \Gamma^{h}_{\ uj} F_{uz}\\
		= & - 2 \gamma^{ij} F_{uz} \Gamma^{z}_{\ iu} \Gamma^{u}_{\ uj} F_{uz} - 2 \gamma^{ij} F_{uz} \Gamma^{z}_{\ iz} \Gamma^{z}_{\ uj} F_{uz} - 2 \gamma^{ij} F_{uz} \Gamma^{z}_{\ ik} \Gamma^{k}_{\ uj} F_{uz}\\
		= & \frac{1}{2} \gamma^{ij} \gamma^{km} F_{uz} \left(\partial_u \gamma_{ik} \right) \left(\partial_u \gamma_{jm} \right) F_{uz}\,.
	\end{split}
\end{equation}
Therefore, the fifteenth term of Eq. (\ref{hkk2first}) is obtained as 
\begin{equation}
	\begin{split}
		& - 2 k^a k^b g^{ce} g^{df} F_{ae} \Gamma^{g}_{\ dh} \Gamma^{h}_{\ bf} F_{cg} = \frac{1}{2} \gamma^{ij} \gamma^{km} F_{uz} \left(\partial_u \gamma_{ik} \right) \left(\partial_u \gamma_{jm} \right) F_{uz}\\
		= & 2 \gamma^{ij} \gamma^{km} F_{uz} K_{ik} K_{jm} F_{uz}\\
		= & 0\,.
	\end{split}
\end{equation}

The sixteenth term of Eq. (\ref{hkk2first}) is
\begin{equation}
	\begin{split}
		& - 2 k^a k^b g^{ce} g^{df} F_{ae} \Gamma^{g}_{\ bf} \left(\partial_d F_{cg} \right) = - 2 g^{ce} g^{df} F_{ue} \Gamma^{g}_{\ uf} \left(\partial_d F_{cg} \right)\\
		= & - 2 F_{uz} \Gamma^{g}_{\ uz} \left(\partial_u F_{ug} \right) - 2 F_{uz} \Gamma^{g}_{\ uu} \left(\partial_z F_{ug} \right) - 2 \gamma^{ij} F_{uz} \Gamma^{g}_{\ uj} \left(\partial_i F_{ug} \right)\\
		= & - 2 F_{uz} \Gamma^{g}_{\ uz} \left(\partial_u F_{ug} \right) - 2 \gamma^{ij} F_{uz} \Gamma^{g}_{\ uj} \left(\partial_i F_{ug} \right)\,.
	\end{split}
\end{equation}
The index $g$ should be further expanded.
\begin{equation}\label{hkk2firstsixteenth}
	\begin{split}
		& - 2 F_{uz} \Gamma^{g}_{\ uz} \left(\partial_u F_{ug} \right) - 2 \gamma^{ij} F_{uz} \Gamma^{g}_{\ uj} \left(\partial_i F_{ug} \right)\\
		= & - 2 F_{uz} \Gamma^{u}_{\ uz} \left(\partial_u F_{uu} \right) - 2 F_{uz} \Gamma^{z}_{\ uz} \left(\partial_u F_{uz} \right) - 2 F_{uz} \Gamma^{i}_{\ uz} \left(\partial_u F_{ui} \right)\\
		& - 2 \gamma^{ij} F_{uz} \Gamma^{u}_{\ uj} \left(\partial_i F_{uu} \right) - 2 \gamma^{ij} F_{uz} \Gamma^{z}_{\ uj} \left(\partial_i F_{uz} \right) - 2 \gamma^{ij} F_{uz} \Gamma^{k}_{\ uj} \left(\partial_i F_{uk} \right)\\
		= & - 2 F_{uz} \Gamma^{z}_{\ uz} \left(\partial_u F_{uz} \right) - 2 F_{uz} \Gamma^{i}_{\ uz} \left(\partial_u F_{ui} \right) - 2 \gamma^{ij} F_{uz} \Gamma^{z}_{\ uj} \left(\partial_i F_{uz} \right)\,.
	\end{split}
\end{equation}
Therefore, the sixteenth term of Eq. (\ref{hkk2first}) is obtained as 
\begin{equation}
	\begin{split}
		- 2 k^a k^b g^{ce} g^{df} F_{ae} \Gamma^{g}_{\ bf} \left(\partial_d F_{cg} \right) = - \gamma^{ij} \beta_j F_{uz} \left(\partial_u F_{ui} \right)\,.
	\end{split}
\end{equation}

The seventeenth term of Eq. (\ref{hkk2first}) is
\begin{equation}
	\begin{split}
		& 2 k^a k^b g^{ce} g^{df} F_{ae} \Gamma^{g}_{\ bf} \Gamma^{h}_{\ dc} F_{hg} = 2 g^{ce} g^{df} F_{ue} \Gamma^{g}_{\ uf} \Gamma^{h}_{\ dc} F_{hg}\\
		= & 2 F_{uz} \Gamma^{g}_{\ uz} \Gamma^{h}_{\ uu} F_{hg} + 2 F_{uz} \Gamma^{g}_{\ uu} \Gamma^{h}_{\ zu} F_{hg} + 2 \gamma^{ij} F_{uz} \Gamma^{g}_{\ uj} \Gamma^{h}_{\ iu} F_{hg}\\
		= & 2 \gamma^{ij} F_{uz} \Gamma^{g}_{\ uj} \Gamma^{h}_{\ iu} F_{hg}\,.
	\end{split}
\end{equation}
The repeated index $g$ should be further expanded as 
\begin{equation}\label{hkk2firstseventeenth}
	\begin{split}
		&2 \gamma^{ij} F_{uz} \Gamma^{g}_{\ uj} \Gamma^{h}_{\ iu} F_{hg}\\
		= & 2 \gamma^{ij} F_{uz} \Gamma^{u}_{\ uj} \Gamma^{h}_{\ iu} F_{hu} + 2 \gamma^{ij} F_{uz} \Gamma^{z}_{\ uj} \Gamma^{h}_{\ iu} F_{hz} + 2 \gamma^{ij} F_{uz} \Gamma^{k}_{\ uj} \Gamma^{h}_{\ iu} F_{hk}\,.
	\end{split}
\end{equation}
The repeated index $h$ should be further expanded. The first term of Eq. (\ref{hkk2firstseventeenth}) is
\begin{equation}
	\begin{split}
		& 2 \gamma^{ij} F_{uz} \Gamma^{u}_{\ uj} \Gamma^{h}_{\ iu} F_{hu}\\
		= & 2 \gamma^{ij} F_{uz} \Gamma^{u}_{\ uj} \Gamma^{u}_{\ iu} F_{uu} + 2 \gamma^{ij} F_{uz} \Gamma^{u}_{\ uj} \Gamma^{z}_{\ iu} F_{zu} + 2 \gamma^{ij} F_{uz} \Gamma^{u}_{\ uj} \Gamma^{k}_{\ iu} F_{ku}\\
		= & 2 \gamma^{ij} F_{uz} \Gamma^{u}_{\ uj} \Gamma^{z}_{\ iu} F_{zu}\\
		= & 0\,.
	\end{split}
\end{equation}
The second term of Eq. (\ref{hkk2firstseventeenth}) is
\begin{equation}
	\begin{split}
		2 \gamma^{ij} F_{uz} \Gamma^{z}_{\ uj} \Gamma^{h}_{\ iu} F_{hz} = 0\,.
	\end{split}
\end{equation}
The third term of Eq. (\ref{hkk2firstseventeenth}) is
\begin{equation}
	\begin{split}
		& 2 \gamma^{ij} F_{uz} \Gamma^{k}_{\ uj} \Gamma^{h}_{\ iu} F_{hk}\\
		= & 2 \gamma^{ij} F_{uz} \Gamma^{k}_{\ uj} \Gamma^{u}_{\ iu} F_{uk} + 2 \gamma^{ij} F_{uz} \Gamma^{k}_{\ uj} \Gamma^{z}_{\ iu} F_{zk} + 2 \gamma^{ij} F_{uz} \Gamma^{k}_{\ uj} \Gamma^{l}_{\ iu} F_{lk}\\
		= & 2 \gamma^{ij} F_{uz} \Gamma^{k}_{\ uj} \Gamma^{z}_{\ iu} F_{zk} + 2 \gamma^{ij} F_{uz} \Gamma^{k}_{\ uj} \Gamma^{l}_{\ iu} F_{lk}\\
		= & \frac{1}{2} \gamma^{ij} \gamma^{km} \gamma^{ln} F_{uz} \left(\partial_u \gamma_{jm} \right) \left(\partial_u \gamma_{in} \right) F_{lk}\,.
	\end{split}
\end{equation}
Therefore, the seventeenth term of Eq. (\ref{hkk2first}) is obtained as 
\begin{equation}
	\begin{split}
		& 2 k^a k^b g^{ce} g^{df} F_{ae} \Gamma^{g}_{\ bf} \Gamma^{h}_{\ dc} F_{hg} = \frac{1}{2} \gamma^{ij} \gamma^{km} \gamma^{ln} F_{uz} \left(\partial_u \gamma_{jm} \right) \left(\partial_u \gamma_{in} \right) F_{lk}\\
		= & 2 \gamma^{ij} \gamma^{km} \gamma^{ln} F_{uz} K_{jm} K_{in} F_{lk}\\
		= & 0\,.
	\end{split}
\end{equation}

The eighteenth term of Eq. (\ref{hkk2first}) is
\begin{equation}
	\begin{split}
		& 2 k^a k^b g^{ce} g^{df} F_{ae} \Gamma^{g}_{\ bf} \Gamma^{h}_{\ dg} F_{ch} = 2 g^{ce} g^{df} F_{ue} \Gamma^{g}_{\ uf} \Gamma^{h}_{\ dg} F_{ch}\\
		= & 2 F_{uz} \Gamma^{g}_{\ uz} \Gamma^{h}_{\ ug} F_{uh} + 2 F_{uz} \Gamma^{g}_{\ uu} \Gamma^{h}_{\ zg} F_{uh} + 2 \gamma^{ij} F_{uz} \Gamma^{g}_{\ uj} \Gamma^{h}_{\ ig} F_{uh}\\
		= & 2 F_{uz} \Gamma^{g}_{\ uz} \Gamma^{h}_{\ ug} F_{uh} + 2 \gamma^{ij} F_{uz} \Gamma^{g}_{\ uj} \Gamma^{h}_{\ ig} F_{uh}\,.
	\end{split}
\end{equation}
The index $g$ should be further expanded.
\begin{equation}\label{hkk2firsteighteenth}
	\begin{split}
		& 2 F_{uz} \Gamma^{g}_{\ uz} \Gamma^{h}_{\ ug} F_{uh} + 2 \gamma^{ij} F_{uz} \Gamma^{g}_{\ uj} \Gamma^{h}_{\ ig} F_{uh}\\
		= & 2 F_{uz} \Gamma^{u}_{\ uz} \Gamma^{h}_{\ uu} F_{uh} + 2 F_{uz} \Gamma^{z}_{\ uz} \Gamma^{h}_{\ uz} F_{uh} + 2 F_{uz} \Gamma^{i}_{\ uz} \Gamma^{h}_{\ ui} F_{uh}\\
		& + 2 \gamma^{ij} F_{uz} \Gamma^{u}_{\ uj} \Gamma^{h}_{\ iu} F_{uh} + 2 \gamma^{ij} F_{uz} \Gamma^{z}_{\ uj} \Gamma^{h}_{\ iz} F_{uh} + 2 \gamma^{ij} F_{uz} \Gamma^{k}_{\ uj} \Gamma^{h}_{\ ik} F_{uh}\\
		= & 2 F_{uz} \Gamma^{z}_{\ uz} \Gamma^{h}_{\ uz} F_{uh} + 2 F_{uz} \Gamma^{i}_{\ uz} \Gamma^{h}_{\ ui} F_{uh} + 2 \gamma^{ij} F_{uz} \Gamma^{u}_{\ uj} \Gamma^{h}_{\ iu} F_{uh}\\
		& + 2 \gamma^{ij} F_{uz} \Gamma^{z}_{\ uj} \Gamma^{h}_{\ iz} F_{uh} + 2 \gamma^{ij} F_{uz} \Gamma^{k}_{\ uj} \Gamma^{h}_{\ ik} F_{uh}\,.
	\end{split}
\end{equation}
The repeated index $h$ should be further expanded. The first term of Eq. (\ref{hkk2firsteighteenth}) is 
\begin{equation}
	\begin{split}
		2 F_{uz} \Gamma^{z}_{\ uz} \Gamma^{h}_{\ uz} F_{uh} = 0\,.
	\end{split}
\end{equation}
The second term of Eq. (\ref{hkk2firsteighteenth}) is
\begin{equation}
	\begin{split}
		& 2 F_{uz} \Gamma^{i}_{\ uz} \Gamma^{h}_{\ ui} F_{uh}\\
		= & 2 F_{uz} \Gamma^{i}_{\ uz} \Gamma^{u}_{\ ui} F_{uu} + 2 F_{uz} \Gamma^{i}_{\ uz} \Gamma^{z}_{\ ui} F_{uz} + 2 F_{uz} \Gamma^{i}_{\ uz} \Gamma^{j}_{\ ui} F_{uj}\\
		= & 2 F_{uz} \Gamma^{i}_{\ uz} \Gamma^{z}_{\ ui} F_{uz}\\
		= & 0\,.
	\end{split}
\end{equation}
The third term of Eq. (\ref{hkk2firsteighteenth}) is
\begin{equation}
	\begin{split}
		& 2 \gamma^{ij} F_{uz} \Gamma^{u}_{\ uj} \Gamma^{h}_{\ iu} F_{uh}\\
		= & 2 \gamma^{ij} F_{uz} \Gamma^{u}_{\ uj} \Gamma^{u}_{\ iu} F_{uu} + 2 \gamma^{ij} F_{uz} \Gamma^{u}_{\ uj} \Gamma^{z}_{\ iu} F_{uz} + 2 \gamma^{ij} F_{uz} \Gamma^{u}_{\ uj} \Gamma^{k}_{\ iu} F_{uk}\\
		= & 2 \gamma^{ij} F_{uz} \Gamma^{u}_{\ uj} \Gamma^{z}_{\ iu} F_{uz}\\
		= & 0\,.
	\end{split}
\end{equation}
The fourth term of Eq. (\ref{hkk2firsteighteenth}) is
\begin{equation}
	\begin{split}
		2 \gamma^{ij} F_{uz} \Gamma^{z}_{\ uj} \Gamma^{h}_{\ iz} F_{uh} = 0\,.
	\end{split}
\end{equation}
The fifth term of Eq. (\ref{hkk2firsteighteenth}) is
\begin{equation}
	\begin{split}
		& 2 \gamma^{ij} F_{uz} \Gamma^{k}_{\ uj} \Gamma^{h}_{\ ik} F_{uh}\\
		= & 2 \gamma^{ij} F_{uz} \Gamma^{k}_{\ uj} \Gamma^{u}_{\ ik} F_{uu} + 2 \gamma^{ij} F_{uz} \Gamma^{k}_{\ uj} \Gamma^{z}_{\ ik} F_{uz} + 2 \gamma^{ij} F_{uz} \Gamma^{k}_{\ uj} \Gamma^{l}_{\ ik} F_{ul}\\
		= & 2 \gamma^{ij} F_{uz} \Gamma^{k}_{\ uj} \Gamma^{z}_{\ ik} F_{uz}\\
		= & - \frac{1}{2} \gamma^{ij} \gamma^{kl} F_{uz} \left(\partial_u \gamma_{jl} \right) \left(\partial_u \gamma_{ik} \right) F_{uz}\,.
	\end{split}
\end{equation}
Therefore, the eighteenth term of Eq. (\ref{hkk2first}) is obtained as 
\begin{equation}
	\begin{split}
		& 2 k^a k^b g^{ce} g^{df} F_{ae} \Gamma^{g}_{\ bf} \Gamma^{h}_{\ dg} F_{ch} = - \frac{1}{2} \gamma^{ij} \gamma^{kl} F_{uz} \left(\partial_u \gamma_{jl} \right) \left(\partial_u \gamma_{ik} \right) F_{uz}\\
		= & - 2 \gamma^{ij} \gamma^{kl} F_{uz} K_{jl} K_{ik} F_{uz}\\
		= & 0\,.
	\end{split}
\end{equation}

Finally, the first term of Eq. (\ref{rewrittenhkk2}) is 
\begin{equation}
	\begin{split}
		& 2 k^a k^b F_{a}^{\ c} \nabla_d \nabla_b F_{c}^{\ d}\\
		= & 2 F_{uz} \left(\partial_u \partial_u F_{uz} \right) + 2 \gamma^{ij} F_{uz} \left(\partial_i \partial_u F_{uj} \right) + \gamma^{ij} \beta_i F_{uz} \left(\partial_u F_{uj} \right)\\
		& + 2 \gamma^{ij} \beta_i F_{uz} \left(\partial_u F_{uj} \right) - 2 \gamma^{ij} \beta_j F_{uz} \left(\partial_u F_{ui} \right) - 2 \gamma^{ij} F_{uz} \hat{\Gamma}^{k}_{\ ij} \left(\partial_u F_{uk} \right)\\
		& - \gamma^{ij} \beta_j F_{uz} \left(\partial_u F_{ui} \right)\\
		= & 2 F_{uz} \left(\partial_u \partial_u F_{uz} \right) + 2 \gamma^{ij} F_{uz} \left[\partial_i \left(\partial_u F_{uj} \right) - \hat{\Gamma}^{k}_{\ ij} \left(\partial_u F_{uk} \right) \right]\\
		= & 2 F_{uz} \left(\partial_u \partial_u F_{uz} \right) + 2 \gamma^{ij} F_{uz} D_i \left(\partial_u F_{uj} \right)\,.
	\end{split}
\end{equation}

The second term of Eq. (\ref{rewrittenhkk2}) is 
\begin{equation}\label{hkk2second}
	\begin{split}
		& 2 k^a k^b F^{cd} \nabla_d \nabla_a F_{bc} = 2 k^a k^b g^{ce} g^{df} F_{ef} \nabla_d \nabla_a F_{bc}\\
		= & 2 k^a k^b g^{ce} g^{df} F_{ef} \left(\partial_d \partial_a F_{bc} \right) - 2 k^a k^b g^{ce} g^{df} F_{ef} \Gamma^{g}_{\ da} \left(\partial_g F_{bc} \right)\\
		& - 2 k^a k^b g^{ce} g^{df} F_{ef} \Gamma^{g}_{\ db} \left(\partial_a F_{gc} \right) - 2 k^a k^b g^{ce} g^{df} F_{ef} \Gamma^{g}_{\ dc} \left(\partial_a F_{bg} \right)\\
		& - 2 k^a k^b g^{ce} g^{df} F_{ef} \left(\partial_d \Gamma^{g}_{\ ab} \right) F_{gc} + 2 k^a k^b g^{ce} g^{df} F_{ef} \Gamma^{h}_{\ da} \Gamma^{g}_{\ hb} F_{gc}\\
		& + 2 k^a k^b g^{ce} g^{df} F_{ef} \Gamma^{h}_{\ db} \Gamma^{g}_{\ ah} F_{gc} - 2 k^a k^b g^{ce} g^{df} F_{ef} \Gamma^{g}_{\ dh} \Gamma^{h}_{\ ab} F_{gc}\\
		& - 2 k^a k^b g^{ce} g^{df} F_{ef} \Gamma^{g}_{\ ab} \left(\partial_d F_{gc} \right) + 2 k^a k^b g^{ce} g^{df} F_{ef} \Gamma^{g}_{\ ab} \Gamma^{h}_{\ dg} F_{hc}\\
		& + 2 k^a k^b g^{ce} g^{df} F_{ef} \Gamma^{g}_{\ ab} \Gamma^{h}_{\ dc} F_{gh} - 2 k^a k^b g^{ce} g^{df} F_{ef} \left(\partial_d \Gamma^{g}_{\ ac} \right) F_{bg}\\
		& + 2 k^a k^b g^{ce} g^{df} F_{ef} \Gamma^{h}_{\ da} \Gamma^{g}_{\ hc} F_{bg} + 2 k^a k^b g^{ce} g^{df} F_{ef} \Gamma^{h}_{\ dc} \Gamma^{g}_{\ ah} F_{bg}\\
		& - 2 k^a k^b g^{ce} g^{df} F_{ef} \Gamma^{g}_{\ dh} \Gamma^{h}_{\ ac} F_{bg} - 2 k^a k^b g^{ce} g^{df} F_{ef} \Gamma^{g}_{\ ac} \left(\partial_d F_{bg} \right)\\
		& + 2 k^a k^b g^{ce} g^{df} F_{ef} \Gamma^{g}_{\ ac} \Gamma^{h}_{\ db} F_{hg} + 2 k^a k^b g^{ce} g^{df} F_{ef} \Gamma^{g}_{\ ac} \Gamma^{h}_{\ dg} F_{bh}\,.
	\end{split}
\end{equation}

The first term of Eq. (\ref{hkk2second}) is 
\begin{equation}
	\begin{split}
		& 2 k^a k^b g^{ce} g^{df} F_{ef} \left(\partial_d \partial_a F_{bc} \right) = 2 g^{ce} g^{df} F_{ef} \left(\partial_d \partial_u F_{uc} \right)\\
		= & 2 F_{uz} \left(\partial_u \partial_u F_{uz} \right) + 2 \gamma^{ij} F_{jz} \left(\partial_u \partial_u F_{ui} \right) + 2 \gamma^{ij} \gamma^{kl} F_{jl} \left(\partial_k \partial_u F_{ui} \right)
	\end{split}
\end{equation}
Therefore, the first term of Eq. (\ref{hkk2second}) is obtained as
\begin{equation}
	\begin{split}
		& 2 k^a k^b g^{ce} g^{df} F_{ef} \left(\partial_d \partial_a F_{bc} \right)\\
		= & 2 F_{uz} \left(\partial_u \partial_u F_{uz} \right) + 2 \gamma^{ij} F_{jz} \left(\partial_u \partial_u F_{ui} \right) + 2 \gamma^{ij} \gamma^{kl} F_{jl} \left(\partial_k \partial_u F_{ui} \right)\,.
	\end{split}
\end{equation}

The second term of Eq. (\ref{hkk2second}) is
\begin{equation}\label{hkk2secondsecond}
	\begin{split}
		& - 2 k^a k^b g^{ce} g^{df} F_{ef} \Gamma^{g}_{\ da} \left(\partial_g F_{bc} \right) = - 2 g^{ce} g^{df} F_{ef} \Gamma^{g}_{\ du} \left(\partial_g F_{uc} \right)\\
		= & - 2 F_{uz} \Gamma^{g}_{\ uu} \left(\partial_g F_{uz} \right) - 2 \gamma^{ij} F_{jz} \Gamma^{g}_{\ uu} \left(\partial_g F_{ui} \right) - 2 \gamma^{ij} \gamma^{kl} F_{jl} \Gamma^{g}_{\ ku} \left(\partial_g F_{ui} \right)\\
		= & - 2 \gamma^{ij} \gamma^{kl} F_{jl} \Gamma^{g}_{\ ku} \left(\partial_g F_{ui} \right)\,.
	\end{split}
\end{equation}
The repeated index $g$ should be further expanded as 
\begin{equation}
	\begin{split}
		& - 2 \gamma^{ij} \gamma^{kl} F_{jl} \Gamma^{g}_{\ ku} \left(\partial_g F_{ui} \right)\\
		= & - 2 \gamma^{ij} \gamma^{kl} F_{jl} \Gamma^{u}_{\ ku} \left(\partial_u F_{ui} \right) - 2 \gamma^{ij} \gamma^{kl} F_{jl} \Gamma^{z}_{\ ku} \left(\partial_z F_{ui} \right) - 2 \gamma^{ij} \gamma^{kl} F_{jl} \Gamma^{m}_{\ ku} \left(\partial_m F_{ui} \right)\\
		= & - 2 \gamma^{ij} \gamma^{kl} F_{jl} \Gamma^{u}_{\ ku} \left(\partial_u F_{ui} \right) - 2 \gamma^{ij} \gamma^{kl} F_{jl} \Gamma^{z}_{\ ku} \left(\partial_z F_{ui} \right)\\
		= & \gamma^{ij} \gamma^{kl} \beta_k F_{jl} \left(\partial_u F_{ui} \right)\,.
	\end{split}
\end{equation}
Therefore, the second term of Eq. (\ref{hkk2second}) is obtained as 
\begin{equation}
	\begin{split}
		- 2 k^a k^b g^{ce} g^{df} F_{ef} \Gamma^{g}_{\ da} \left(\partial_g F_{bc} \right) = \gamma^{ij} \gamma^{kl} \beta_k F_{jl} \left(\partial_u F_{ui} \right)\,.
	\end{split}
\end{equation}

The third term of Eq. (\ref{hkk2second}) is
\begin{equation}
	\begin{split}
		& - 2 k^a k^b g^{ce} g^{df} F_{ef} \Gamma^{g}_{\ db} \left(\partial_a F_{gc} \right) = - 2 g^{ce} g^{df} F_{ef} \Gamma^{g}_{\ du} \left(\partial_u F_{gc} \right)\\
		= & - 2 F_{zu} \Gamma^{g}_{\ zu} \left(\partial_u F_{gu} \right) - 2 \gamma^{ij} F_{zj} \Gamma^{g}_{\ iu} \left(\partial_u F_{gu} \right) - 2 F_{uz} \Gamma^{g}_{\ uu} \left(\partial_u F_{gz} \right)\\
		& - 2 \gamma^{ij} F_{jz} \Gamma^{g}_{\ uu} \left(\partial_u F_{gi} \right) - 2 \gamma^{ij} \gamma^{kl} F_{jl} \Gamma^{g}_{\ ku} \left(\partial_u F_{gi} \right)\\
		= & - 2 F_{zu} \Gamma^{g}_{\ zu} \left(\partial_u F_{gu} \right) - 2 \gamma^{ij} F_{zj} \Gamma^{g}_{\ iu} \left(\partial_u F_{gu} \right) - 2 \gamma^{ij} \gamma^{kl} F_{jl} \Gamma^{g}_{\ ku} \left(\partial_u F_{gi} \right)\,.
	\end{split}
\end{equation}
The index $g$ should be further expanded.
\begin{equation}\label{hkk2secondthird}
	\begin{split}
		& - 2 F_{zu} \Gamma^{g}_{\ zu} \left(\partial_u F_{gu} \right) - 2 \gamma^{ij} F_{zj} \Gamma^{g}_{\ iu} \left(\partial_u F_{gu} \right) - 2 \gamma^{ij} \gamma^{kl} F_{jl} \Gamma^{g}_{\ ku} \left(\partial_u F_{gi} \right)\\
		= & - 2 F_{zu} \Gamma^{u}_{\ zu} \left(\partial_u F_{uu} \right) - 2 F_{zu} \Gamma^{z}_{\ zu} \left(\partial_u F_{zu} \right) - 2 F_{zu} \Gamma^{i}_{\ zu} \left(\partial_u F_{iu} \right)\\
		& - 2 \gamma^{ij} F_{zj} \Gamma^{u}_{\ iu} \left(\partial_u F_{uu} \right) - 2 \gamma^{ij} F_{zj} \Gamma^{z}_{\ iu} \left(\partial_u F_{zu} \right) - 2 \gamma^{ij} F_{zj} \Gamma^{k}_{\ iu} \left(\partial_u F_{ku} \right)\\
		& - 2 \gamma^{ij} \gamma^{kl} F_{jl} \Gamma^{u}_{\ ku} \left(\partial_u F_{ui} \right) - 2 \gamma^{ij} \gamma^{kl} F_{jl} \Gamma^{z}_{\ ku} \left(\partial_u F_{zi} \right) - 2 \gamma^{ij} \gamma^{kl} F_{jl} \Gamma^{m}_{\ ku} \left(\partial_u F_{mi} \right)\\
		= & - 2 F_{zu} \Gamma^{z}_{\ zu} \left(\partial_u F_{zu} \right) - 2 F_{zu} \Gamma^{i}_{\ zu} \left(\partial_u F_{iu} \right) - 2 \gamma^{ij} F_{zj} \Gamma^{z}_{\ iu} \left(\partial_u F_{zu} \right)\\
		& - 2 \gamma^{ij} F_{zj} \Gamma^{k}_{\ iu} \left(\partial_u F_{ku} \right) - 2 \gamma^{ij} \gamma^{kl} F_{jl} \Gamma^{u}_{\ ku} \left(\partial_u F_{ui} \right) - 2 \gamma^{ij} \gamma^{kl} F_{jl} \Gamma^{z}_{\ ku} \left(\partial_u F_{zi} \right)\\
		& - 2 \gamma^{ij} \gamma^{kl} F_{jl} \Gamma^{m}_{\ ku} \left(\partial_u F_{mi} \right)\,.
	\end{split}
\end{equation}
Therefore, the third term of Eq. (\ref{hkk2second}) is obetined as 
\begin{equation}
	\begin{split}
		& - 2 k^a k^b g^{ce} g^{df} F_{ef} \Gamma^{g}_{\ db} \left(\partial_a F_{gc} \right)\\
		= & - \gamma^{ij} \beta_j F_{zu} \left(\partial_u F_{iu} \right) - \gamma^{ij} \gamma^{kl} F_{zj} \left(\partial_u \gamma_{il} \right) \left(\partial_u F_{ku} \right) + \gamma^{ij} \gamma^{kl} \beta_k F_{jl} \left(\partial_u F_{ui} \right)\\
		& - \gamma^{ij} \gamma^{kl} \gamma^{mn} F_{jl} \left(\partial_u \gamma_{kn} \right) \left(\partial_u F_{mi} \right)\\
		= & - \gamma^{ij} \beta_j F_{zu} \left(\partial_u F_{iu} \right) - 2 \gamma^{ij} \gamma^{kl} F_{zj} K_{il} \left(\partial_u F_{ku} \right) + \gamma^{ij} \gamma^{kl} \beta_k F_{jl} \left(\partial_u F_{ui} \right)\\
		& - 2 \gamma^{ij} \gamma^{kl} \gamma^{mn} F_{jl} K_{kn} \left(\partial_u F_{mi} \right)\\
		= & - \gamma^{ij} \beta_j F_{zu} \left(\partial_u F_{iu} \right) + \gamma^{ij} \gamma^{kl} \beta_k F_{jl} \left(\partial_u F_{ui} \right)\,.
	\end{split}
\end{equation}

The fourth term of Eq. (\ref{hkk2second}) is
\begin{equation}
	\begin{split}
		& - 2 k^a k^b g^{ce} g^{df} F_{ef} \Gamma^{g}_{\ dc} \left(\partial_a F_{bg} \right) = - 2 g^{ce} g^{df} F_{ef} \Gamma^{g}_{\ dc} \left(\partial_u F_{ug} \right)\\
		= & - 2 F_{zu} \Gamma^{g}_{\ zu} \left(\partial_u F_{ug} \right) - 2 \gamma^{ij} F_{zj} \Gamma^{g}_{\ iu} \left(\partial_u F_{ug} \right) - 2 F_{uz} \Gamma^{g}_{\ uz} \left(\partial_u F_{ug} \right)\\
		& - 2 \gamma^{ij} F_{jz} \Gamma^{g}_{\ ui} \left(\partial_u F_{ug} \right) - 2 \gamma^{ij} \gamma^{kl} F_{jl} \Gamma^{g}_{\ ki} \left(\partial_u F_{ug} \right)\,.
	\end{split}
\end{equation}
The index $g$ should be further expanded.
\begin{equation}\label{hkk2secondfourth}
	\begin{split}
		& - 2 F_{zu} \Gamma^{g}_{\ zu} \left(\partial_u F_{ug} \right) - 2 \gamma^{ij} F_{zj} \Gamma^{g}_{\ iu} \left(\partial_u F_{ug} \right) - 2 F_{uz} \Gamma^{g}_{\ uz} \left(\partial_u F_{ug} \right)\\
		& - 2 \gamma^{ij} F_{jz} \Gamma^{g}_{\ ui} \left(\partial_u F_{ug} \right) - 2 \gamma^{ij} \gamma^{kl} F_{jl} \Gamma^{g}_{\ ki} \left(\partial_u F_{ug} \right)\\
		= & - 2 F_{zu} \Gamma^{u}_{\ zu} \left(\partial_u F_{uu} \right) - 2 F_{zu} \Gamma^{z}_{\ zu} \left(\partial_u F_{uz} \right) - 2 F_{zu} \Gamma^{i}_{\ zu} \left(\partial_u F_{ui} \right)\\
		& - 2 \gamma^{ij} F_{zj} \Gamma^{u}_{\ iu} \left(\partial_u F_{uu} \right) - 2 \gamma^{ij} F_{zj} \Gamma^{z}_{\ iu} \left(\partial_u F_{uz} \right) - 2 \gamma^{ij} F_{zj} \Gamma^{k}_{\ iu} \left(\partial_u F_{uk} \right)\\
		& - 2 F_{uz} \Gamma^{u}_{\ uz} \left(\partial_u F_{uu} \right) - 2 F_{uz} \Gamma^{z}_{\ uz} \left(\partial_u F_{uz} \right) - 2 F_{uz} \Gamma^{i}_{\ uz} \left(\partial_u F_{ui} \right)\\
		& - 2 \gamma^{ij} F_{jz} \Gamma^{u}_{\ ui} \left(\partial_u F_{uu} \right) - 2 \gamma^{ij} F_{jz} \Gamma^{z}_{\ ui} \left(\partial_u F_{uz} \right) - 2 \gamma^{ij} F_{jz} \Gamma^{k}_{\ ui} \left(\partial_u F_{uk} \right)\\
		& - 2 \gamma^{ij} \gamma^{kl} F_{jl} \Gamma^{u}_{\ ki} \left(\partial_u F_{uu} \right) - 2 \gamma^{ij} \gamma^{kl} F_{jl} \Gamma^{z}_{\ ki} \left(\partial_u F_{uz} \right) - 2 \gamma^{ij} \gamma^{kl} F_{jl} \Gamma^{m}_{\ ki} \left(\partial_u F_{um} \right)\\
		= & - 2 F_{zu} \Gamma^{z}_{\ zu} \left(\partial_u F_{uz} \right) - 2 F_{zu} \Gamma^{i}_{\ zu} \left(\partial_u F_{ui} \right) - 2 \gamma^{ij} F_{zj} \Gamma^{z}_{\ iu} \left(\partial_u F_{uz} \right)\\
		& - 2 \gamma^{ij} F_{zj} \Gamma^{k}_{\ iu} \left(\partial_u F_{uk} \right) - 2 F_{uz} \Gamma^{z}_{\ uz} \left(\partial_u F_{uz} \right) - 2 F_{uz} \Gamma^{i}_{\ uz} \left(\partial_u F_{ui} \right)\\
		& - 2 \gamma^{ij} F_{jz} \Gamma^{z}_{\ ui} \left(\partial_u F_{uz} \right) - 2 \gamma^{ij} F_{jz} \Gamma^{k}_{\ ui} \left(\partial_u F_{uk} \right) - 2 \gamma^{ij} \gamma^{kl} F_{jl} \Gamma^{z}_{\ ki} \left(\partial_u F_{uz} \right)\\
		& - 2 \gamma^{ij} \gamma^{kl} F_{jl} \Gamma^{m}_{\ ki} \left(\partial_u F_{um} \right)\,.
	\end{split}
\end{equation}
Therefore, the fourth term of Eq. (\ref{hkk2second}) is obtained as 
\begin{equation}
	\begin{split}
		& - 2 k^a k^b g^{ce} g^{df} F_{ef} \Gamma^{g}_{\ dc} \left(\partial_a F_{bg} \right)\\
		= & - \gamma^{ij} \beta_j F_{zu} \left(\partial_u F_{ui} \right) - \gamma^{ij} \gamma^{kl} F_{zj} \left(\partial_u \gamma_{il} \right) \left(\partial_u F_{uk} \right) - \gamma^{ij} \beta_j F_{uz} \left(\partial_u F_{ui} \right)\\
		& - \gamma^{ij} \gamma^{kl} F_{jz} \left(\partial_u \gamma_{il} \right) \left(\partial_u F_{uk} \right) + \gamma^{ij} \gamma^{kl} F_{jl} \left(\partial_u \gamma_{ki} \right) \left(\partial_u F_{uz} \right) - 2 \gamma^{ij} \gamma^{kl} F_{jl} \hat{\Gamma}^{m}_{\ ki} \left(\partial_u F_{um} \right)\\
		= & \gamma^{ij} \beta_j F_{uz} \left(\partial_u F_{ui} \right) - \gamma^{ij} \beta_j F_{uz} \left(\partial_u F_{ui} \right) - \gamma^{ij} \gamma^{kl} F_{zj} \left(\partial_u \gamma_{il} \right) \left(\partial_u F_{uk} \right)\\
		& + \gamma^{ij} \gamma^{kl} F_{zj} \left(\partial_u \gamma_{il} \right) \left(\partial_u F_{uk} \right) + 2 \gamma^{ij} \gamma^{kl} F_{jl} K_{ki} \left(\partial_u F_{uz} \right) - 2 \gamma^{ij} \gamma^{kl} F_{jl} \hat{\Gamma}^{m}_{\ ki} \left(\partial_u F_{um} \right)\\
		= & - 2 \gamma^{i(j} \gamma^{|k|l)} F_{[jl]} \hat{\Gamma}^{m}_{\ (ki)} \left(\partial_u F_{um} \right)\\
		= & 0\,.
	\end{split}
\end{equation}

The fifth term of Eq. (\ref{hkk2second}) is
\begin{equation}
	\begin{split}
		& - 2 k^a k^b g^{ce} g^{df} F_{ef} \left(\partial_d \Gamma^{g}_{\ ab} \right) F_{gc} = - 2 g^{ce} g^{df} F_{ef} \left(\partial_d \Gamma^{g}_{\ uu} \right) F_{gc}\\
		= & - 2 F_{zu} \left(\partial_z \Gamma^{g}_{\ uu} \right) F_{gu} - 2 \gamma^{ij} F_{zj} \left(\partial_i \Gamma^{g}_{\ uu} \right) F_{gu} - 2 F_{uz} \left(\partial_u \Gamma^{g}_{\ uu} \right) F_{gz}\\
		& - 2 \gamma^{ij} F_{jz} \left(\partial_u \Gamma^{g}_{\ uu} \right) F_{gi} - 2 \gamma^{ij} \gamma^{kl} F_{jl} \left(\partial_k \Gamma^{g}_{\ uu} \right) F_{gi}\,.
	\end{split}
\end{equation}
The repeated index $g$ should be further expanded as 
\begin{equation}\label{hkk2secondfifth}
	\begin{split}
		& - 2 F_{zu} \left(\partial_z \Gamma^{g}_{\ uu} \right) F_{gu} - 2 \gamma^{ij} F_{zj} \left(\partial_i \Gamma^{g}_{\ uu} \right) F_{gu} - 2 F_{uz} \left(\partial_u \Gamma^{g}_{\ uu} \right) F_{gz}\\
		& - 2 \gamma^{ij} F_{jz} \left(\partial_u \Gamma^{g}_{\ uu} \right) F_{gi} - 2 \gamma^{ij} \gamma^{kl} F_{jl} \left(\partial_k \Gamma^{g}_{\ uu} \right) F_{gi}\\
		= & - 2 F_{zu} \left(\partial_z \Gamma^{u}_{\ uu} \right) F_{uu} - 2 F_{zu} \left(\partial_z \Gamma^{z}_{\ uu} \right) F_{zu} - 2 F_{zu} \left(\partial_z \Gamma^{i}_{\ uu} \right) F_{iu}\\
		& - 2 \gamma^{ij} F_{zj} \left(\partial_i \Gamma^{u}_{\ uu} \right) F_{uu} - 2 \gamma^{ij} F_{zj} \left(\partial_i \Gamma^{z}_{\ uu} \right) F_{zu} - 2 \gamma^{ij} F_{zj} \left(\partial_i \Gamma^{k}_{\ uu} \right) F_{ku}\\
		& - 2 F_{uz} \left(\partial_u \Gamma^{u}_{\ uu} \right) F_{uz} - 2 F_{uz} \left(\partial_u \Gamma^{z}_{\ uu} \right) F_{zz} - 2 F_{uz} \left(\partial_u \Gamma^{i}_{\ uu} \right) F_{iz}\\
		& - 2 \gamma^{ij} F_{jz} \left(\partial_u \Gamma^{u}_{\ uu} \right) F_{ui} - 2 \gamma^{ij} F_{jz} \left(\partial_u \Gamma^{z}_{\ uu} \right) F_{zi} - 2 \gamma^{ij} F_{jz} \left(\partial_u \Gamma^{k}_{\ uu} \right) F_{ki}\\
		& - 2 \gamma^{ij} \gamma^{kl} F_{jl} \left(\partial_k \Gamma^{u}_{\ uu} \right) F_{ui} - 2 \gamma^{ij} \gamma^{kl} F_{jl} \left(\partial_k \Gamma^{z}_{\ uu} \right) F_{zi} - 2 \gamma^{ij} \gamma^{kl} F_{jl} \left(\partial_k \Gamma^{m}_{\ uu} \right) F_{mi}\\
		= & - 2 F_{zu} \left(\partial_z \Gamma^{z}_{\ uu} \right) F_{zu} - 2 \gamma^{ij} F_{zj} \left(\partial_i \Gamma^{z}_{\ uu} \right) F_{zu} - 2 F_{uz} \left(\partial_u \Gamma^{u}_{\ uu} \right) F_{uz}\\
		& - 2 F_{uz} \left(\partial_u \Gamma^{i}_{\ uu} \right) F_{iz} - 2 \gamma^{ij} F_{jz} \left(\partial_u \Gamma^{z}_{\ uu} \right) F_{zi} - 2 \gamma^{ij} F_{jz} \left(\partial_u \Gamma^{k}_{\ uu} \right) F_{ki}\\
		& - 2 \gamma^{ij} \gamma^{kl} F_{jl} \left(\partial_k \Gamma^{z}_{\ uu} \right) F_{zi} - 2 \gamma^{ij} \gamma^{kl} F_{jl} \left(\partial_k \Gamma^{m}_{\ uu} \right) F_{mi}\,.
	\end{split}
\end{equation}
The first term of Eq. (\ref{hkk2secondfifth}) is 
\begin{equation}
	\begin{split}
		& - 2 F_{zu} \left(\partial_z \Gamma^{z}_{\ uu} \right) F_{zu}\\
		= & - 2 F_{zu} \partial_z \left[z^2 \partial_u \alpha - z^2 \left(\beta^2 - 2 \alpha \right) \left(z^2 \partial_z \alpha + 2 z \alpha \right) - z \beta^i \left(z \partial_u \beta_i - z^2 \partial_i \alpha \right) \right] F_{zu}\\
		= & - 2 F_{zu} \left[2 z \partial_u \alpha + z^2 \partial_z \partial_u \alpha - 2 z \left(\beta^2 - 2 \alpha \right) \left(z^2 \partial_z \alpha + 2 z \alpha \right)\right.\\
		& \left. - z^2 \left(\partial_z \beta^2 - 2 \partial_z \alpha \right) \left(z^2 \partial_z \alpha + 2 z \alpha \right) - z^2 \left(\beta^2 - 2 \alpha \right) \left(2 z \partial_z \alpha + z^2 \partial_z^2 \alpha + 2 \alpha + 2 z \partial_z \alpha \right)\right.\\
		& \left. - \beta^i \left(z \partial_u \beta_i - z^2 \partial_i \alpha \right) - z \left(\partial_z \beta^i \right) \left(z \partial_u \beta_i - z^2 \partial_i \alpha \right)\right.\\
		& \left. - z \beta^i \left(\partial_u \beta_i + z \partial_z \partial_u \beta_i - 2 z \partial_i \alpha - z^2 \partial_z \partial_i \alpha \right) \right]\\
		= & 0\,.
	\end{split}
\end{equation}
The second term of Eq. (\ref{hkk2secondfifth}) is
\begin{equation}
	\begin{split}
		& - 2 \gamma^{ij} F_{zj} \left(\partial_i \Gamma^{z}_{\ uu} \right) F_{zu}\\
		= & - 2 \gamma^{ij} F_{zj} \partial_i \left[z^2 \partial_u \alpha - z^2 \left(\beta^2 - 2 \alpha \right) \left(z^2 \partial_z \alpha + 2 z \alpha \right) - z \beta^k \left(z \partial_u \beta_k - z^2 \partial_k \alpha \right) \right] F_{zu}\\
		= & - 2 \gamma^{ij} F_{zj} \left[z^2 \partial_i \partial_u \alpha - z^2 \left(\partial_i \beta^2 - 2 \partial_i \alpha \right) \left(z^2 \partial_z \alpha + 2 z \alpha \right)\right.\\
		& \left. - z^2 \left(\beta^2 - 2 \alpha \right) \left(z^2 \partial_i \partial_z \alpha + 2 z \partial_i \alpha \right) - z \left(\partial_i \beta^k \right) \left(z \partial_u \beta_k - z^2 \partial_k \alpha \right)\right.\\
		& \left. - z \beta^k \left(z \partial_i \partial_u \beta_k - z^2 \partial_i \partial_k \alpha \right) \right] F_{zu}\\
		= & 0\,.
	\end{split}
\end{equation}
The third term of Eq. (\ref{hkk2secondfifth}) is
\begin{equation}
	\begin{split}
		& - 2 F_{uz} \left(\partial_u \Gamma^{u}_{\ uu} \right) F_{uz}\\
		= & - 2 F_{uz} \partial_u \left(- z^2 \partial_z \alpha - 2 z \alpha \right) F_{uz}\\
		= & - 2 F_{uz} \left(- z^2 \partial_u \partial_z \alpha - 2 z \partial_u \alpha \right) F_{uz}\\
		= & 0\,.
	\end{split}
\end{equation}
The fourth term of Eq. (\ref{hkk2secondfifth}) is
\begin{equation}
	\begin{split}
		& - 2 F_{uz} \left(\partial_u \Gamma^{i}_{\ uu} \right) F_{iz}\\
		= & - 2 F_{uz} \partial_u \left[\left(z \beta^i \right) \left(z^2 \partial_z \alpha + 2 z \alpha \right) + \gamma^{ij} \left(z \partial_u \beta_j - z^2 \partial_j \alpha \right) \right] F_{iz}\\
		= & - 2 F_{uz} \left[\left(z \partial_u \beta^i \right) \left(z^2 \partial_z \alpha + 2 z \alpha \right) + \left(z \beta^i \right) \left(z^2 \partial_u \partial_z \alpha + 2 z \partial_u \alpha \right)\right.\\
		& \left. + \left(\partial_u \gamma^{ij} \right) \left(z \partial_u \beta_j - z^2 \partial_j \alpha \right) + \gamma^{ij} \left(z \partial_u \partial_u \beta_j - z^2 \partial_u \partial_j \alpha \right) \right] F_{iz}\\
		= & 0\,.
	\end{split}
\end{equation}
The fifth term of Eq. (\ref{hkk2secondfifth}) is
\begin{equation}
	\begin{split}
		& - 2 \gamma^{ij} F_{jz} \left(\partial_u \Gamma^{z}_{\ uu} \right) F_{zi}\\
		= & - 2 \gamma^{ij} F_{jz} \partial_u \left[z^2 \partial_u \alpha - z^2 \left(\beta^2 - 2 \alpha \right) \left(z^2 \partial_z \alpha + 2 z \alpha \right) - z \beta^k \left(z \partial_u \beta_k - z^2 \partial_k \alpha \right) \right] F_{zi}\\
		= & - 2 \gamma^{ij} F_{jz} \left[z^2 \partial_u^2 \alpha - z^2 \left(\partial_u \beta^2 - 2 \partial_u \alpha \right) \left(z^2 \partial_z \alpha + 2 z \alpha \right)\right.\\
		& \left. - z^2 \left(\beta^2 - 2 \alpha \right) \left(z^2 \partial_u \partial_z \alpha + 2 z \partial_u \alpha \right) - z \left(\partial_u \beta^k \right) \left(z \partial_u \beta_k - z^2 \partial_k \alpha \right)\right.\\
		& \left. - z \beta^k \left(z \partial_u^2 \beta_k - z^2 \partial_u \partial_k \alpha \right) \right] F_{zi}\\
		= & 0\,.
	\end{split}
\end{equation}
The sixth term of Eq. (\ref{hkk2secondfifth}) is
\begin{equation}
	\begin{split}
		& - 2 \gamma^{ij} F_{jz} \left(\partial_u \Gamma^{k}_{\ uu} \right) F_{ki}\\
		= & - 2 \gamma^{ij} F_{jz} \partial_u \left[\left(z \beta^k \right) \left(z^2 \partial_z \alpha + 2 z \alpha \right) + \gamma^{kl} \left(z \partial_u \beta_l - z^2 \partial_l \alpha \right) \right] F_{ki}\\
		= & - 2 \gamma^{ij} F_{jz} \left[\left(z \partial_u \beta^k \right) \left(z^2 \partial_z \alpha + 2 z \alpha \right) + \left(z \beta^k \right) \left(z^2 \partial_u \partial_z \alpha + 2 z \partial_u \alpha \right)\right.\\
		& \left. + \left(\partial_u \gamma^{kl} \right) \left(z \partial_u \beta_l - z^2 \partial_l \alpha \right) + \gamma^{kl} \left(z \partial_u^2 \beta_l - z^2 \partial_u \partial_l \alpha \right) \right] F_{ki}\\
		= & 0\,.
	\end{split}
\end{equation}
The seventh term of Eq. (\ref{hkk2secondfifth}) is
\begin{equation}
	\begin{split}
		& - 2 \gamma^{ij} \gamma^{kl} F_{jl} \left(\partial_k \Gamma^{z}_{\ uu} \right) F_{zi}\\
		= & - 2 \gamma^{ij} \gamma^{kl} F_{jl} \partial_k \left[z^2 \partial_u \alpha - z^2 \left(\beta^2 - 2 \alpha \right) \left(z^2 \partial_z \alpha + 2 z \alpha \right) - z \beta^m \left(z \partial_u \beta_m - z^2 \partial_m \alpha \right) \right] F_{zi}\\
		= & - 2 \gamma^{ij} \gamma^{kl} F_{jl} \left[z^2 \partial_k \partial_u \alpha - z^2 \left(\partial_k \beta^2 - 2 \partial_k \alpha \right) \left(z^2 \partial_z \alpha + 2 z \alpha \right)\right.\\
		& \left. - z^2 \left(\beta^2 - 2 \alpha \right) \left(z^2 \partial_k \partial_z \alpha + 2 z \partial_k \alpha \right) - z \left(\partial_k \beta^m \right) \left(z \partial_u \beta_m - z^2 \partial_m \alpha \right)\right.\\
		& \left. - z \beta^m \left(z \partial_k \partial_u \beta_m - z^2 \partial_k \partial_m \alpha \right) \right] F_{zi}\\
		= & 0\,.
	\end{split}
\end{equation}
The eighth term of Eq. (\ref{hkk2secondfifth}) is
\begin{equation}
	\begin{split}
		& - 2 \gamma^{ij} \gamma^{kl} F_{jl} \left(\partial_k \Gamma^{m}_{\ uu} \right) F_{mi}\\
		= & - 2 \gamma^{ij} \gamma^{kl} F_{jl} \partial_k \left[\left(z \beta^m \right) \left(z^2 \partial_z \alpha + 2 z \alpha \right) + \gamma^{mn} \left(z \partial_u \beta_n - z^2 \partial_n \alpha \right) \right] F_{mi}\\
		= & - 2 \gamma^{ij} \gamma^{kl} F_{jl} \left[\left(z \partial_k \beta^m \right) \left(z^2 \partial_z \alpha + 2 z \alpha \right) + \left(z \beta^m \right) \left(z^2 \partial_k \partial_z \alpha + 2 z \partial_k \alpha \right)\right.\\
		& \left. + \left(\partial_k \gamma^{mn} \right) \left(z \partial_u \beta_n - z^2 \partial_n \alpha \right) + \gamma^{mn} \left(z \partial_k \partial_u \beta_n - z^2 \partial_k \partial_n \alpha \right) \right] F_{mi}\\
		= & 0\,.
	\end{split}
\end{equation}
Therefore, the fifth term of Eq. (\ref{hkk2second}) is obtained as
\begin{equation}
	\begin{split}
		- 2 k^a k^b g^{ce} g^{df} F_{ef} \left(\partial_d \Gamma^{g}_{\ ab} \right) F_{gc} = 0\,.
	\end{split}
\end{equation}

The sixth term of Eq. (\ref{hkk2second}) is
\begin{equation}
	\begin{split}
		& 2 k^a k^b g^{ce} g^{df} F_{ef} \Gamma^{h}_{\ da} \Gamma^{g}_{\ hb} F_{gc} = 2 g^{ce} g^{df} F_{ef} \Gamma^{h}_{\ du} \Gamma^{g}_{\ hu} F_{gc}\\
		= & 2 F_{zu} \Gamma^{h}_{\ zu} \Gamma^{g}_{\ hu} F_{gu} + 2 \gamma^{ij} F_{zj} \Gamma^{h}_{\ iu} \Gamma^{g}_{\ hu} F_{gu} + 2 F_{uz} \Gamma^{h}_{\ uu} \Gamma^{g}_{\ hu} F_{gz}\\
		& + 2 \gamma^{ij} F_{jz} \Gamma^{h}_{\ uu} \Gamma^{g}_{\ hu} F_{gi} + 2 \gamma^{ij} \gamma^{kl} F_{jl} \Gamma^{h}_{\ ku} \Gamma^{g}_{\ hu} F_{gi}\\
		= & 2 F_{zu} \Gamma^{h}_{\ zu} \Gamma^{g}_{\ hu} F_{gu} + 2 \gamma^{ij} F_{zj} \Gamma^{h}_{\ iu} \Gamma^{g}_{\ hu} F_{gu} + 2 \gamma^{ij} \gamma^{kl} F_{jl} \Gamma^{h}_{\ ku} \Gamma^{g}_{\ hu} F_{gi}\,.
	\end{split}
\end{equation}
The index $g$ should be further expanded.
\begin{equation}\label{hkk2secondsixth}
	\begin{split}
		& 2 F_{zu} \Gamma^{h}_{\ zu} \Gamma^{g}_{\ hu} F_{gu} + 2 \gamma^{ij} F_{zj} \Gamma^{h}_{\ iu} \Gamma^{g}_{\ hu} F_{gu} + 2 \gamma^{ij} \gamma^{kl} F_{jl} \Gamma^{h}_{\ ku} \Gamma^{g}_{\ hu} F_{gi}\\
		= & 2 F_{zu} \Gamma^{h}_{\ zu} \Gamma^{u}_{\ hu} F_{uu} + 2 F_{zu} \Gamma^{h}_{\ zu} \Gamma^{z}_{\ hu} F_{zu} + 2 F_{zu} \Gamma^{h}_{\ zu} \Gamma^{i}_{\ hu} F_{iu}\\
		& + 2 \gamma^{ij} F_{zj} \Gamma^{h}_{\ iu} \Gamma^{u}_{\ hu} F_{uu} + 2 \gamma^{ij} F_{zj} \Gamma^{h}_{\ iu} \Gamma^{z}_{\ hu} F_{zu} + 2 \gamma^{ij} F_{zj} \Gamma^{h}_{\ iu} \Gamma^{k}_{\ hu} F_{ku}\\
		& + 2 \gamma^{ij} \gamma^{kl} F_{jl} \Gamma^{h}_{\ ku} \Gamma^{u}_{\ hu} F_{ui} + 2 \gamma^{ij} \gamma^{kl} F_{jl} \Gamma^{h}_{\ ku} \Gamma^{z}_{\ hu} F_{zi} + 2 \gamma^{ij} \gamma^{kl} F_{jl} \Gamma^{h}_{\ ku} \Gamma^{m}_{\ hu} F_{mi}\\
		= & 2 F_{zu} \Gamma^{h}_{\ zu} \Gamma^{z}_{\ hu} F_{zu} + 2 \gamma^{ij} F_{zj} \Gamma^{h}_{\ iu} \Gamma^{z}_{\ hu} F_{zu} + 2 \gamma^{ij} \gamma^{kl} F_{jl} \Gamma^{h}_{\ ku} \Gamma^{z}_{\ hu} F_{zi}\\
		& + 2 \gamma^{ij} \gamma^{kl} F_{jl} \Gamma^{h}_{\ ku} \Gamma^{m}_{\ hu} F_{mi}\,.
	\end{split}
\end{equation}
The repeated index $h$ should be further expanded. The first term of Eq. (\ref{hkk2secondsixth}) is 
\begin{equation}
	\begin{split}
		& 2 F_{zu} \Gamma^{h}_{\ zu} \Gamma^{z}_{\ hu} F_{zu}\\
		= & 2 F_{zu} \Gamma^{u}_{\ zu} \Gamma^{z}_{\ uu} F_{zu} + 2 F_{zu} \Gamma^{z}_{\ zu} \Gamma^{z}_{\ zu} F_{zu} + 2 F_{zu} \Gamma^{i}_{\ zu} \Gamma^{z}_{\ iu} F_{zu}\\
		= & 0\,.
	\end{split}
\end{equation}
The second term of Eq. (\ref{hkk2secondsixth}) is 
\begin{equation}
	\begin{split}
		& 2 \gamma^{ij} F_{zj} \Gamma^{h}_{\ iu} \Gamma^{z}_{\ hu} F_{zu}\\
		= & 2 \gamma^{ij} F_{zj} \Gamma^{u}_{\ iu} \Gamma^{z}_{\ uu} F_{zu} + 2 \gamma^{ij} F_{zj} \Gamma^{z}_{\ iu} \Gamma^{z}_{\ zu} F_{zu} + 2 \gamma^{ij} F_{zj} \Gamma^{k}_{\ iu} \Gamma^{z}_{\ ku} F_{zu}\\
		= & 0\,.
	\end{split}
\end{equation}
The third term of Eq. (\ref{hkk2secondsixth}) is
\begin{equation}
	\begin{split}
		& 2 \gamma^{ij} \gamma^{kl} F_{jl} \Gamma^{h}_{\ ku} \Gamma^{z}_{\ hu} F_{zi}\\
		= & 2 \gamma^{ij} \gamma^{kl} F_{jl} \Gamma^{u}_{\ ku} \Gamma^{z}_{\ uu} F_{zi} + 2 \gamma^{ij} \gamma^{kl} F_{jl} \Gamma^{z}_{\ ku} \Gamma^{z}_{\ zu} F_{zi} + 2 \gamma^{ij} \gamma^{kl} F_{jl} \Gamma^{m}_{\ ku} \Gamma^{z}_{\ mu} F_{zi}\\
		= & 0\,.
	\end{split}
\end{equation}
The fourth term of Eq. (\ref{hkk2secondsixth}) is
\begin{equation}
	\begin{split}
		& 2 \gamma^{ij} \gamma^{kl} F_{jl} \Gamma^{h}_{\ ku} \Gamma^{m}_{\ hu} F_{mi}\\
		= & 2 \gamma^{ij} \gamma^{kl} F_{jl} \Gamma^{u}_{\ ku} \Gamma^{m}_{\ uu} F_{mi} + 2 \gamma^{ij} \gamma^{kl} F_{jl} \Gamma^{z}_{\ ku} \Gamma^{m}_{\ zu} F_{mi} + 2 \gamma^{ij} \gamma^{kl} F_{jl} \Gamma^{n}_{\ ku} \Gamma^{m}_{\ nu} F_{mi}\\
		= & \frac{1}{2} \gamma^{ij} \gamma^{kl} \gamma^{no} \gamma^{mp} F_{jl} \left(\partial_u \gamma_{ko} \right) \left(\partial_u \gamma_{np} \right) F_{mi}\,.
	\end{split}
\end{equation}
Therefore, the sixth term of Eq. (\ref{hkk2second}) is obtained as 
\begin{equation}
	\begin{split}
		& 2 k^a k^b g^{ce} g^{df} F_{ef} \Gamma^{h}_{\ da} \Gamma^{g}_{\ hb} F_{gc} = \frac{1}{2} \gamma^{ij} \gamma^{kl} \gamma^{no} \gamma^{mp} F_{jl} \left(\partial_u \gamma_{ko} \right) \left(\partial_u \gamma_{np} \right) F_{mi}\\
		= & 2 \gamma^{ij} \gamma^{kl} \gamma^{no} \gamma^{mp} F_{jl} K_{ko} K_{np} F_{mi}\\
		= & 0\,.
	\end{split}
\end{equation}

The seventh term of Eq. (\ref{hkk2second}) is 
\begin{equation}
	\begin{split}
		& 2 k^a k^b g^{ce} g^{df} F_{ef} \Gamma^{h}_{\ db} \Gamma^{g}_{\ ah} F_{gc} = 2 g^{ce} g^{df} F_{ef} \Gamma^{h}_{\ du} \Gamma^{g}_{\ uh} F_{gc}\\
		= & 2 F_{zu} \Gamma^{h}_{\ zu} \Gamma^{g}_{\ uh} F_{gu} + 2 \gamma^{ij} F_{zj} \Gamma^{h}_{\ iu} \Gamma^{g}_{\ uh} F_{gu} + 2 F_{uz} \Gamma^{h}_{\ uu} \Gamma^{g}_{\ uh} F_{gz}\\
		& + 2 \gamma^{ij} F_{jz} \Gamma^{h}_{\ uu} \Gamma^{g}_{\ uh} F_{gi} + 2 \gamma^{ij} \gamma^{kl} F_{jl} \Gamma^{h}_{\ ku} \Gamma^{g}_{\ uh} F_{gi}\\
		= & 2 F_{zu} \Gamma^{h}_{\ zu} \Gamma^{g}_{\ uh} F_{gu} + 2 \gamma^{ij} F_{zj} \Gamma^{h}_{\ iu} \Gamma^{g}_{\ uh} F_{gu} + 2 \gamma^{ij} \gamma^{kl} F_{jl} \Gamma^{h}_{\ ku} \Gamma^{g}_{\ uh} F_{gi}\,.
	\end{split}
\end{equation}
The index $g$ should be further expanded.
\begin{equation}\label{hkk2secondseventh}
	\begin{split}
		& 2 F_{zu} \Gamma^{h}_{\ zu} \Gamma^{g}_{\ uh} F_{gu} + 2 \gamma^{ij} F_{zj} \Gamma^{h}_{\ iu} \Gamma^{g}_{\ uh} F_{gu} + 2 \gamma^{ij} \gamma^{kl} F_{jl} \Gamma^{h}_{\ ku} \Gamma^{g}_{\ uh} F_{gi}\\
		= & 2 F_{zu} \Gamma^{h}_{\ zu} \Gamma^{u}_{\ uh} F_{uu} + 2 F_{zu} \Gamma^{h}_{\ zu} \Gamma^{z}_{\ uh} F_{zu} + 2 F_{zu} \Gamma^{h}_{\ zu} \Gamma^{i}_{\ uh} F_{iu}\\
		& + 2 \gamma^{ij} F_{zj} \Gamma^{h}_{\ iu} \Gamma^{u}_{\ uh} F_{uu} + 2 \gamma^{ij} F_{zj} \Gamma^{h}_{\ iu} \Gamma^{z}_{\ uh} F_{zu} + 2 \gamma^{ij} F_{zj} \Gamma^{h}_{\ iu} \Gamma^{k}_{\ uh} F_{ku}\\
		& + 2 \gamma^{ij} \gamma^{kl} F_{jl} \Gamma^{h}_{\ ku} \Gamma^{u}_{\ uh} F_{ui} + 2 \gamma^{ij} \gamma^{kl} F_{jl} \Gamma^{h}_{\ ku} \Gamma^{z}_{\ uh} F_{zi} + 2 \gamma^{ij} \gamma^{kl} F_{jl} \Gamma^{h}_{\ ku} \Gamma^{m}_{\ uh} F_{mi}\\
		= & 2 F_{zu} \Gamma^{h}_{\ zu} \Gamma^{z}_{\ uh} F_{zu} + 2 \gamma^{ij} F_{zj} \Gamma^{h}_{\ iu} \Gamma^{z}_{\ uh} F_{zu} + 2 \gamma^{ij} \gamma^{kl} F_{jl} \Gamma^{h}_{\ ku} \Gamma^{z}_{\ uh} F_{zi}\\
		& + 2 \gamma^{ij} \gamma^{kl} F_{jl} \Gamma^{h}_{\ ku} \Gamma^{m}_{\ uh} F_{mi}\,.
	\end{split}
\end{equation}
The repeated index $h$ should be further expanded. The first term of Eq. (\ref{hkk2secondseventh}) is 
\begin{equation}
	\begin{split}
		& 2 F_{zu} \Gamma^{h}_{\ zu} \Gamma^{z}_{\ uh} F_{zu}\\
		= & 2 F_{zu} \Gamma^{u}_{\ zu} \Gamma^{z}_{\ uu} F_{zu} + 2 F_{zu} \Gamma^{z}_{\ zu} \Gamma^{z}_{\ uz} F_{zu} + 2 F_{zu} \Gamma^{i}_{\ zu} \Gamma^{z}_{\ ui} F_{zu}\\
		= & 0\,.
	\end{split}
\end{equation}
The second term of Eq. (\ref{hkk2secondseventh}) is 
\begin{equation}
	\begin{split}
		& 2 \gamma^{ij} F_{zj} \Gamma^{h}_{\ iu} \Gamma^{z}_{\ uh} F_{zu}\\
		= & 2 \gamma^{ij} F_{zj} \Gamma^{u}_{\ iu} \Gamma^{z}_{\ uu} F_{zu} + 2 \gamma^{ij} F_{zj} \Gamma^{z}_{\ iu} \Gamma^{z}_{\ uz} F_{zu} + 2 \gamma^{ij} F_{zj} \Gamma^{k}_{\ iu} \Gamma^{z}_{\ uk} F_{zu}\\
		= & 0\,.
	\end{split}
\end{equation}
The third term of Eq. (\ref{hkk2secondseventh}) is
\begin{equation}
	\begin{split}
		& 2 \gamma^{ij} \gamma^{kl} F_{jl} \Gamma^{h}_{\ ku} \Gamma^{z}_{\ uh} F_{zi}\\
		= & 2 \gamma^{ij} \gamma^{kl} F_{jl} \Gamma^{u}_{\ ku} \Gamma^{z}_{\ uu} F_{zi} + 2 \gamma^{ij} \gamma^{kl} F_{jl} \Gamma^{z}_{\ ku} \Gamma^{z}_{\ uz} F_{zi} + 2 \gamma^{ij} \gamma^{kl} F_{jl} \Gamma^{m}_{\ ku} \Gamma^{z}_{\ um} F_{zi}\\
		= & 0\,.
	\end{split}
\end{equation}
The fourth term of Eq. (\ref{hkk2secondseventh}) is
\begin{equation}
	\begin{split}
		& 2 \gamma^{ij} \gamma^{kl} F_{jl} \Gamma^{h}_{\ ku} \Gamma^{m}_{\ uh} F_{mi}\\
		= & 2 \gamma^{ij} \gamma^{kl} F_{jl} \Gamma^{u}_{\ ku} \Gamma^{m}_{\ uu} F_{mi} + 2 \gamma^{ij} \gamma^{kl} F_{jl} \Gamma^{z}_{\ ku} \Gamma^{m}_{\ uz} F_{mi} + 2 \gamma^{ij} \gamma^{kl} F_{jl} \Gamma^{n}_{\ ku} \Gamma^{m}_{\ un} F_{mi}\\
		= & \frac{1}{2} \gamma^{ij} \gamma^{kl} \gamma^{mp} \gamma^{no} F_{jl} \left(\partial_u \gamma_{ko} \right) \left(\partial_u \gamma_{np} \right) F_{mi}\,.
	\end{split}
\end{equation}
Therefore, the seventh term of Eq. (\ref{hkk2second}) is obtained as 
\begin{equation}
	\begin{split}
		& 2 k^a k^b g^{ce} g^{df} F_{ef} \Gamma^{h}_{\ db} \Gamma^{g}_{\ ah} F_{gc} = \frac{1}{2} \gamma^{ij} \gamma^{kl} \gamma^{mp} \gamma^{no} F_{jl} \left(\partial_u \gamma_{ko} \right) \left(\partial_u \gamma_{np} \right) F_{mi}\\
		= & 2 \gamma^{ij} \gamma^{kl} \gamma^{mp} \gamma^{no} F_{jl} K_{ko} K_{np} F_{mi}\\
		= & 0\,.
	\end{split}
\end{equation}

The eighth term of Eq. (\ref{hkk2second}) is 
\begin{equation}
	\begin{split}
		- 2 k^a k^b g^{ce} g^{df} F_{ef} \Gamma^{g}_{\ dh} \Gamma^{h}_{\ ab} F_{gc} = - 2 g^{ce} g^{df} F_{ef} \Gamma^{g}_{\ dh} \Gamma^{h}_{\ uu} F_{gc} = 0\,.
	\end{split}
\end{equation}
Therefore, the eighth term of Eq. (\ref{hkk2second}) is obtained as 
\begin{equation}
	\begin{split}
		- 2 k^a k^b g^{ce} g^{df} F_{ef} \Gamma^{g}_{\ dh} \Gamma^{h}_{\ ab} F_{gc} = 0\,.
	\end{split}
\end{equation}

The ninth term of Eq. (\ref{hkk2second}) is 
\begin{equation}
	\begin{split}
		- 2 k^a k^b g^{ce} g^{df} F_{ef} \Gamma^{g}_{\ ab} \left(\partial_d F_{gc} \right) = - 2 g^{ce} g^{df} F_{ef} \Gamma^{g}_{\ uu} \left(\partial_d F_{gc} \right) = 0\,.
	\end{split}
\end{equation}
Therefore, the ninth term of Eq. (\ref{hkk2second}) is obtained as 
\begin{equation}
	\begin{split}
		- 2 k^a k^b g^{ce} g^{df} F_{ef} \Gamma^{g}_{\ ab} \left(\partial_d F_{gc} \right) = 0\,.
	\end{split}
\end{equation}

The tenth term of Eq. (\ref{hkk2second}) is
\begin{equation}
	\begin{split}
		2 k^a k^b g^{ce} g^{df} F_{ef} \Gamma^{g}_{\ ab} \Gamma^{h}_{\ dg} F_{hc} = 2 g^{ce} g^{df} F_{ef} \Gamma^{g}_{\ uu} \Gamma^{h}_{\ dg} F_{hc} = 0\,.
	\end{split}
\end{equation}
Therefore, the tenth term of Eq. (\ref{hkk2second}) is obtained as 
\begin{equation}
	\begin{split}
		2 k^a k^b g^{ce} g^{df} F_{ef} \Gamma^{g}_{\ ab} \Gamma^{h}_{\ dg} F_{hc} = 0\,.
	\end{split}
\end{equation}

The eleventh term of Eq. (\ref{hkk2second}) is
\begin{equation}
	\begin{split}
		& 2 k^a k^b g^{ce} g^{df} F_{ef} \Gamma^{g}_{\ ab} \Gamma^{h}_{\ dc} F_{gh} = 2 g^{ce} g^{df} F_{ef} \Gamma^{g}_{\ uu} \Gamma^{h}_{\ dc} F_{gh} = 0\,.
	\end{split}
\end{equation}
Therefore, the eleventh term of Eq. (\ref{hkk2second}) is obtained as
\begin{equation}
	\begin{split}
		2 k^a k^b g^{ce} g^{df} F_{ef} \Gamma^{g}_{\ ab} \Gamma^{h}_{\ dc} F_{gh} = 0\,.
	\end{split}
\end{equation}

The twelfth term of Eq. (\ref{hkk2second}) is
\begin{equation}
	\begin{split}
		& - 2 k^a k^b g^{ce} g^{df} F_{ef} \left(\partial_d \Gamma^{g}_{\ ac} \right) F_{bg} = - 2 g^{ce} g^{df} F_{ef} \left(\partial_d \Gamma^{g}_{\ uc} \right) F_{ug}\\
		= & - 2 F_{zu} \left(\partial_z \Gamma^{g}_{\ uu} \right) F_{ug} - 2 \gamma^{ij} F_{zj} \left(\partial_i \Gamma^{g}_{\ uu} \right) F_{ug} - 2 F_{uz} \left(\partial_u \Gamma^{g}_{\ uz} \right) F_{ug}\\
		& - 2 \gamma^{ij} F_{jz} \left(\partial_u \Gamma^{g}_{\ ui} \right) F_{ug} - 2 \gamma^{ij} \gamma^{kl} F_{jl} \left(\partial_k \Gamma^{g}_{\ ui} \right) F_{ug}\,.
	\end{split}
\end{equation}
The index $g$ should be further expanded.
\begin{equation}\label{hkk2secondtwelfth}
	\begin{split}
		& - 2 F_{zu} \left(\partial_z \Gamma^{g}_{\ uu} \right) F_{ug} - 2 \gamma^{ij} F_{zj} \left(\partial_i \Gamma^{g}_{\ uu} \right) F_{ug} - 2 F_{uz} \left(\partial_u \Gamma^{g}_{\ uz} \right) F_{ug}\\
		& - 2 \gamma^{ij} F_{jz} \left(\partial_u \Gamma^{g}_{\ ui} \right) F_{ug} - 2 \gamma^{ij} \gamma^{kl} F_{jl} \left(\partial_k \Gamma^{g}_{\ ui} \right) F_{ug}\\
		= & - 2 F_{zu} \left(\partial_z \Gamma^{u}_{\ uu} \right) F_{uu} - 2 F_{zu} \left(\partial_z \Gamma^{z}_{\ uu} \right) F_{uz} - 2 F_{zu} \left(\partial_z \Gamma^{i}_{\ uu} \right) F_{ui}\\
		& - 2 \gamma^{ij} F_{zj} \left(\partial_i \Gamma^{u}_{\ uu} \right) F_{uu} - 2 \gamma^{ij} F_{zj} \left(\partial_i \Gamma^{z}_{\ uu} \right) F_{uz} - 2 \gamma^{ij} F_{zj} \left(\partial_i \Gamma^{k}_{\ uu} \right) F_{uk}\\
		& - 2 F_{uz} \left(\partial_u \Gamma^{u}_{\ uz} \right) F_{uu} - 2 F_{uz} \left(\partial_u \Gamma^{z}_{\ uz} \right) F_{uz} - 2 F_{uz} \left(\partial_u \Gamma^{i}_{\ uz} \right) F_{ui}\\
		& - 2 \gamma^{ij} F_{jz} \left(\partial_u \Gamma^{u}_{\ ui} \right) F_{uu} - 2 \gamma^{ij} F_{jz} \left(\partial_u \Gamma^{z}_{\ ui} \right) F_{uz} - 2 \gamma^{ij} F_{jz} \left(\partial_u \Gamma^{k}_{\ ui} \right) F_{uk}\\
		& - 2 \gamma^{ij} \gamma^{kl} F_{jl} \left(\partial_k \Gamma^{u}_{\ ui} \right) F_{uu} - 2 \gamma^{ij} \gamma^{kl} F_{jl} \left(\partial_k \Gamma^{z}_{\ ui} \right) F_{uz} - 2 \gamma^{ij} \gamma^{kl} F_{jl} \left(\partial_k \Gamma^{m}_{\ ui} \right) F_{um}\\
		= & - 2 F_{zu} \left(\partial_z \Gamma^{z}_{\ uu} \right) F_{uz} - 2 \gamma^{ij} F_{zj} \left(\partial_i \Gamma^{z}_{\ uu} \right) F_{uz} - 2 F_{uz} \left(\partial_u \Gamma^{z}_{\ uz} \right) F_{uz}\\
		& - 2 \gamma^{ij} F_{jz} \left(\partial_u \Gamma^{z}_{\ ui} \right) F_{uz} - 2 \gamma^{ij} \gamma^{kl} F_{jl} \left(\partial_k \Gamma^{z}_{\ ui} \right) F_{uz}\,.
	\end{split}
\end{equation}
The first term of Eq. (\ref{hkk2secondtwelfth}) is 
\begin{equation}
	\begin{split}
		& - 2 F_{zu} \left(\partial_z \Gamma^{z}_{\ uu} \right) F_{uz}\\
		= & - 2 F_{zu} \partial_z \left[z^2 \partial_u \alpha - z^2 \left(\beta^2 - 2 \alpha \right) \left(z^2 \partial_z \alpha + 2 z \alpha \right) - z \beta^i \left(z \partial_u \beta_i - z^2 \partial_i \alpha \right) \right] F_{uz}\\
		= & - 2 F_{zu} \left[2 z \partial_u \alpha + z^2 \partial_z \partial_u \alpha - 2 z \left(\beta^2 - 2 \alpha \right) \left(z^2 \partial_z \alpha + 2 z \alpha \right)\right.\\
		& \left. - z^2 \left(\partial_z \beta^2 - 2 \partial_z \alpha \right) \left(z^2 \partial_z \alpha + 2 z \alpha \right)\right.\\
		& \left. - z^2 \left(\beta^2 - 2 \alpha \right) \left(2 z \partial_z \alpha + z^2 \partial_z^2 \alpha + 2 \alpha + 2 z \partial_z \alpha \right)\right.\\
		& \left. - \beta^i \left(z \partial_u \beta_i - z^2 \partial_i \alpha \right) - z \left(\partial_z \beta^i \right) \left(z \partial_u \beta_i - z^2 \partial_i \alpha \right)\right.\\
		& \left. - z \beta^i \left(\partial_u \beta_i + z \partial_z \partial_u \beta_i - 2 z \partial_i \alpha - z^2 \partial_z \partial_i \alpha \right) \right] F_{uz}\\
		= & 0\,.
	\end{split}
\end{equation}
The second term of Eq. (\ref{hkk2secondtwelfth}) is 
\begin{equation}
	\begin{split}
		& - 2 \gamma^{ij} F_{zj} \left(\partial_i \Gamma^{z}_{\ uu} \right) F_{uz}\\
		= & - 2 \gamma^{ij} F_{zj} \partial_i \left[z^2 \partial_u \alpha - z^2 \left(\beta^2 - 2 \alpha \right) \left(z^2 \partial_z \alpha + 2 z \alpha \right) - z \beta^k \left(z \partial_u \beta_k - z^2 \partial_k \alpha \right) \right] F_{uz}\\
		= & - 2 \gamma^{ij} F_{zj} \left[z^2 \partial_i \partial_u \alpha - z^2 \left(\partial_i \beta^2 - 2 \partial_i \alpha \right) \left(z^2 \partial_z \alpha + 2 z \alpha \right)\right.\\
		& \left. - z^2 \left(\beta^2 - 2 \alpha \right) \left(z^2 \partial_i \partial_z \alpha + 2 z \partial_i \alpha \right) - z \left(\partial_i \beta^k \right) \left(z \partial_u \beta_k - z^2 \partial_k \alpha \right)\right.\\
		& \left. - z \beta^k \left(z \partial_i \partial_u \beta_k - z^2 \partial_i \partial_k \alpha \right)\right] F_{uz}\\
		= & 0\,.
	\end{split}
\end{equation}
The third term of Eq. (\ref{hkk2secondtwelfth}) is 
\begin{equation}
	\begin{split}
		& - 2 F_{uz} \left(\partial_u \Gamma^{z}_{\ uz} \right) F_{uz}\\
		= & - 2 F_{uz} \partial_u \left(z^2 \partial_z \alpha + 2 z \alpha - \frac{1}{2} z \beta^i \beta_i - \frac{1}{2} z^2 \beta^i \partial_z \beta_i \right) F_{uz}\\
		= & - 2 F_{uz} \left[z^2 \partial_u \partial_z \alpha + 2 z \partial_u \alpha - \frac{1}{2} z \partial_u \left(\beta^i \beta_i\right) - \frac{1}{2} z^2 \left(\partial_u \beta^i \right) \partial_z \beta_i - \frac{1}{2} z^2 \beta^i \partial_u \partial_z \beta_i \right] F_{uz}\\
		= & 0\,.
	\end{split}
\end{equation}
The fourth term of Eq. (\ref{hkk2secondtwelfth}) is
\begin{equation}
	\begin{split}
		& - 2 \gamma^{ij} F_{jz} \left(\partial_u \Gamma^{z}_{\ ui} \right) F_{uz}\\
		= & - 2 \gamma^{ij} F_{jz}\\
		& \times \partial_u \left[z^2 \partial_i \alpha - \frac{1}{2} z^2 \left(\beta^2 - 2 \alpha \right) \left(\beta_i + z \partial_z \beta_i \right) - \frac{1}{2} z \beta^k \left(z \partial_i \beta_k + \partial_u \gamma_{ik} - z \partial_k \beta_i \right) \right] F_{uz}\\
		= & - 2 \gamma^{ij} F_{jz} \left[z^2 \partial_u \partial_i \alpha - \frac{1}{2} z^2 \left(\partial_u \beta^2 - 2 \partial_u \alpha \right) \left(\beta_i + z \partial_z \beta_i \right)\right.\\
		& \left. - \frac{1}{2} z^2 \left(\beta^2 - 2 \alpha \right) \left(\partial_u \beta_i + z \partial_u \partial_z \beta_i \right) - \frac{1}{2} z \left(\partial_u \beta^k \right) \left(z \partial_i \beta_k + \partial_u \gamma_{ik} - z \partial_k \beta_i \right)\right.\\
		& \left. - \frac{1}{2} z \beta^k \left(z \partial_u \partial_i \beta_k + \partial_u^2 \gamma_{ik} - z \partial_u \partial_k \beta_i \right) \right] F_{uz}\\
		= & 0\,.
	\end{split}
\end{equation}
The fifth term of Eq. (\ref{hkk2secondtwelfth}) is
\begin{equation}
	\begin{split}
		& - 2 \gamma^{ij} \gamma^{kl} F_{jl} \left(\partial_k \Gamma^{z}_{\ ui} \right) F_{uz}\\
		= & - 2 \gamma^{ij} \gamma^{kl} F_{jl}\\
		& \times \partial_k \left[z^2 \partial_i \alpha - \frac{1}{2} z^2 \left(\beta^2 - 2 \alpha \right) \left(\beta_i + z \partial_z \beta_i \right) - \frac{1}{2} z \beta^m \left(z \partial_i \beta_m + \partial_u \gamma_{im} - z \partial_m \beta_i \right) \right] F_{uz}\\
		= & - 2 \gamma^{ij} \gamma^{kl} F_{jl} \left[z^2 \partial_k \partial_i \alpha - \frac{1}{2} z^2 \left(\partial_k \beta^2 - 2 \partial_k \alpha \right) \left(\beta_i + z \partial_z \beta_i \right)\right.\\
		& \left. - \frac{1}{2} z^2 \left(\beta^2 - 2 \alpha \right) \left(\partial_k \beta_i + z \partial_k \partial_z \beta_i \right) - \frac{1}{2} z \left(\partial_k \beta^m \right) \left(z \partial_i \beta_m + \partial_u \gamma_{im} - z \partial_m \beta_i \right)\right.\\
		& \left. - \frac{1}{2} z \beta^m \left(z \partial_k \partial_i \beta_m + \partial_k \partial_u \gamma_{im} - z \partial_k \partial_m \beta_i \right) \right] F_{uz}\\
		= & 0\,.
	\end{split}
\end{equation}
Therefore, the twelfth term of Eq. (\ref{hkk2second}) is obtained as 
\begin{equation}
	\begin{split}
		- 2 k^a k^b g^{ce} g^{df} F_{ef} \left(\partial_d \Gamma^{g}_{\ ac} \right) F_{bg} = 0\,.
	\end{split}
\end{equation}

The thirteenth term of Eq. (\ref{hkk2second}) is
\begin{equation}
	\begin{split}
		& 2 k^a k^b g^{ce} g^{df} F_{ef} \Gamma^{h}_{\ da} \Gamma^{g}_{\ hc} F_{bg} = 2 g^{ce} g^{df} F_{ef} \Gamma^{h}_{\ du} \Gamma^{g}_{\ hc} F_{ug}\\
		= & 2 F_{zu} \Gamma^{h}_{\ zu} \Gamma^{g}_{\ hu} F_{ug} + 2 \gamma^{ij} F_{zj} \Gamma^{h}_{\ iu} \Gamma^{g}_{\ hu} F_{ug} + 2 F_{uz} \Gamma^{h}_{\ uu} \Gamma^{g}_{\ hz} F_{ug}\\
		& + 2 \gamma^{ij} F_{jz} \Gamma^{h}_{\ uu} \Gamma^{g}_{\ hi} F_{ug} + 2 \gamma^{ij} \gamma^{kl} F_{jl} \Gamma^{h}_{\ ku} \Gamma^{g}_{\ hi} F_{ug}\\
		= & 2 F_{zu} \Gamma^{h}_{\ zu} \Gamma^{g}_{\ hu} F_{ug} + 2 \gamma^{ij} F_{zj} \Gamma^{h}_{\ iu} \Gamma^{g}_{\ hu} F_{ug} + 2 \gamma^{ij} \gamma^{kl} F_{jl} \Gamma^{h}_{\ ku} \Gamma^{g}_{\ hi} F_{ug}\,.
	\end{split}
\end{equation}
The index $g$ should be further expanded.
\begin{equation}\label{hkk2secondthirteenth}
	\begin{split}
		& 2 F_{zu} \Gamma^{h}_{\ zu} \Gamma^{g}_{\ hu} F_{ug} + 2 \gamma^{ij} F_{zj} \Gamma^{h}_{\ iu} \Gamma^{g}_{\ hu} F_{ug} + 2 \gamma^{ij} \gamma^{kl} F_{jl} \Gamma^{h}_{\ ku} \Gamma^{g}_{\ hi} F_{ug}\\
		= & 2 F_{zu} \Gamma^{h}_{\ zu} \Gamma^{u}_{\ hu} F_{uu} + 2 F_{zu} \Gamma^{h}_{\ zu} \Gamma^{z}_{\ hu} F_{uz} + 2 F_{zu} \Gamma^{h}_{\ zu} \Gamma^{i}_{\ hu} F_{ui}\\
		& + 2 \gamma^{ij} F_{zj} \Gamma^{h}_{\ iu} \Gamma^{u}_{\ hu} F_{uu} + 2 \gamma^{ij} F_{zj} \Gamma^{h}_{\ iu} \Gamma^{z}_{\ hu} F_{uz} + 2 \gamma^{ij} F_{zj} \Gamma^{h}_{\ iu} \Gamma^{k}_{\ hu} F_{uk}\\
		& + 2 \gamma^{ij} \gamma^{kl} F_{jl} \Gamma^{h}_{\ ku} \Gamma^{u}_{\ hi} F_{uu} + 2 \gamma^{ij} \gamma^{kl} F_{jl} \Gamma^{h}_{\ ku} \Gamma^{z}_{\ hi} F_{uz} + 2 \gamma^{ij} \gamma^{kl} F_{jl} \Gamma^{h}_{\ ku} \Gamma^{m}_{\ hi} F_{um}\\
		= & 2 F_{zu} \Gamma^{h}_{\ zu} \Gamma^{z}_{\ hu} F_{uz} + 2 \gamma^{ij} F_{zj} \Gamma^{h}_{\ iu} \Gamma^{z}_{\ hu} F_{uz} + 2 \gamma^{ij} \gamma^{kl} F_{jl} \Gamma^{h}_{\ ku} \Gamma^{z}_{\ hi} F_{uz}\,.
	\end{split}
\end{equation}
The first term of Eq. (\ref{hkk2secondthirteenth}) is 
\begin{equation}
	\begin{split}
		& 2 F_{zu} \Gamma^{h}_{\ zu} \Gamma^{z}_{\ hu} F_{uz}\\
		= & 2 F_{zu} \Gamma^{u}_{\ zu} \Gamma^{z}_{\ uu} F_{uz} + 2 F_{zu} \Gamma^{z}_{\ zu} \Gamma^{z}_{\ zu} F_{uz} + 2 F_{zu} \Gamma^{i}_{\ zu} \Gamma^{z}_{\ iu} F_{uz}\\
		= & 0\,.
	\end{split}
\end{equation}
The second term of Eq. (\ref{hkk2secondthirteenth}) is
\begin{equation}
	\begin{split}
		& 2 \gamma^{ij} F_{zj} \Gamma^{h}_{\ iu} \Gamma^{z}_{\ hu} F_{uz}\\
		= & 2 \gamma^{ij} F_{zj} \Gamma^{u}_{\ iu} \Gamma^{z}_{\ uu} F_{uz} + 2 \gamma^{ij} F_{zj} \Gamma^{z}_{\ iu} \Gamma^{z}_{\ zu} F_{uz} + 2 \gamma^{ij} F_{zj} \Gamma^{k}_{\ iu} \Gamma^{z}_{\ ku} F_{uz}\\
		= & 0\,.
	\end{split}
\end{equation}
The third term of Eq. (\ref{hkk2secondthirteenth}) is
\begin{equation}
	\begin{split}
		& 2 \gamma^{ij} \gamma^{kl} F_{jl} \Gamma^{h}_{\ ku} \Gamma^{z}_{\ hi} F_{uz}\\
		= & 2 \gamma^{ij} \gamma^{kl} F_{jl} \Gamma^{u}_{\ ku} \Gamma^{z}_{\ ui} F_{uz} + 2 \gamma^{ij} \gamma^{kl} F_{jl} \Gamma^{z}_{\ ku} \Gamma^{z}_{\ zi} F_{uz} + 2 \gamma^{ij} \gamma^{kl} F_{jl} \Gamma^{m}_{\ ku} \Gamma^{z}_{\ mi} F_{uz}\\
		= & - \frac{1}{2} \gamma^{ij} \gamma^{kl} \gamma^{mn} F_{jl} \left(\partial_u \gamma_{kn} \right) \left(\partial_u \gamma_{mi} \right) F_{uz}\,.
	\end{split}
\end{equation}
Therefore, the thirteenth term of Eq. (\ref{hkk2second}) is obtained as 
\begin{equation}
	\begin{split}
		& 2 k^a k^b g^{ce} g^{df} F_{ef} \Gamma^{h}_{\ da} \Gamma^{g}_{\ hc} F_{bg} = - \frac{1}{2} \gamma^{ij} \gamma^{kl} \gamma^{mn} F_{jl} \left(\partial_u \gamma_{kn} \right) \left(\partial_u \gamma_{mi} \right) F_{uz}\\
		= & - 2 \gamma^{ij} \gamma^{kl} \gamma^{mn} F_{jl} K_{kn} K_{mi} F_{uz}\\
		= & 0\,.
	\end{split}
\end{equation}

The fourteenth term of Eq. (\ref{hkk2second}) is
\begin{equation}
	\begin{split}
		& 2 k^a k^b g^{ce} g^{df} F_{ef} \Gamma^{h}_{\ dc} \Gamma^{g}_{\ ah} F_{bg} = 2 g^{ce} g^{df} F_{ef} \Gamma^{h}_{\ dc} \Gamma^{g}_{\ uh} F_{ug}\\
		= & 2 F_{zu} \Gamma^{h}_{\ zu} \Gamma^{g}_{\ uh} F_{ug} + 2 \gamma^{ij} F_{zj} \Gamma^{h}_{\ iu} \Gamma^{g}_{\ uh} F_{ug} + 2 F_{uz} \Gamma^{h}_{\ uz} \Gamma^{g}_{\ uh} F_{ug}\\
		& + 2 \gamma^{ij} F_{jz} \Gamma^{h}_{\ ui} \Gamma^{g}_{\ uh} F_{ug} + 2 \gamma^{ij} \gamma^{kl} F_{jl} \Gamma^{h}_{\ ki} \Gamma^{g}_{\ uh} F_{ug}\,.
	\end{split}
\end{equation}
The index $g$ should be further expanded.
\begin{equation}\label{hkk2secondfourteenth}
	\begin{split}
		& 2 F_{zu} \Gamma^{h}_{\ zu} \Gamma^{g}_{\ uh} F_{ug} + 2 \gamma^{ij} F_{zj} \Gamma^{h}_{\ iu} \Gamma^{g}_{\ uh} F_{ug} + 2 F_{uz} \Gamma^{h}_{\ uz} \Gamma^{g}_{\ uh} F_{ug}\\
		& + 2 \gamma^{ij} F_{jz} \Gamma^{h}_{\ ui} \Gamma^{g}_{\ uh} F_{ug} + 2 \gamma^{ij} \gamma^{kl} F_{jl} \Gamma^{h}_{\ ki} \Gamma^{g}_{\ uh} F_{ug}\\
		= & 2 F_{zu} \Gamma^{h}_{\ zu} \Gamma^{u}_{\ uh} F_{uu} + 2 F_{zu} \Gamma^{h}_{\ zu} \Gamma^{z}_{\ uh} F_{uz} + 2 F_{zu} \Gamma^{h}_{\ zu} \Gamma^{i}_{\ uh} F_{ui}\\
		& + 2 \gamma^{ij} F_{zj} \Gamma^{h}_{\ iu} \Gamma^{u}_{\ uh} F_{uu} + 2 \gamma^{ij} F_{zj} \Gamma^{h}_{\ iu} \Gamma^{z}_{\ uh} F_{uz} + 2 \gamma^{ij} F_{zj} \Gamma^{h}_{\ iu} \Gamma^{k}_{\ uh} F_{uk}\\
		& + 2 F_{uz} \Gamma^{h}_{\ uz} \Gamma^{u}_{\ uh} F_{uu} + 2 F_{uz} \Gamma^{h}_{\ uz} \Gamma^{z}_{\ uh} F_{uz} + 2 F_{uz} \Gamma^{h}_{\ uz} \Gamma^{i}_{\ uh} F_{ui}\\
		& + 2 \gamma^{ij} F_{jz} \Gamma^{h}_{\ ui} \Gamma^{u}_{\ uh} F_{uu} + 2 \gamma^{ij} F_{jz} \Gamma^{h}_{\ ui} \Gamma^{z}_{\ uh} F_{uz} + 2 \gamma^{ij} F_{jz} \Gamma^{h}_{\ ui} \Gamma^{k}_{\ uh} F_{uk}\\
		& + 2 \gamma^{ij} \gamma^{kl} F_{jl} \Gamma^{h}_{\ ki} \Gamma^{u}_{\ uh} F_{uu} + 2 \gamma^{ij} \gamma^{kl} F_{jl} \Gamma^{h}_{\ ki} \Gamma^{z}_{\ uh} F_{uz} + 2 \gamma^{ij} \gamma^{kl} F_{jl} \Gamma^{h}_{\ ki} \Gamma^{m}_{\ uh} F_{um}\\
		= & 2 F_{zu} \Gamma^{h}_{\ zu} \Gamma^{z}_{\ uh} F_{uz} + 2 \gamma^{ij} F_{zj} \Gamma^{h}_{\ iu} \Gamma^{z}_{\ uh} F_{uz} + 2 F_{uz} \Gamma^{h}_{\ uz} \Gamma^{z}_{\ uh} F_{uz}\\
		& + 2 \gamma^{ij} F_{jz} \Gamma^{h}_{\ ui} \Gamma^{z}_{\ uh} F_{uz} + 2 \gamma^{ij} \gamma^{kl} F_{jl} \Gamma^{h}_{\ ki} \Gamma^{z}_{\ uh} F_{uz}\,.
	\end{split}
\end{equation}
The first term of Eq. (\ref{hkk2secondfourteenth}) is 
\begin{equation}
	\begin{split}
		& 2 F_{zu} \Gamma^{h}_{\ zu} \Gamma^{z}_{\ uh} F_{uz}\\
		= & 2 F_{zu} \Gamma^{u}_{\ zu} \Gamma^{z}_{\ uu} F_{uz} + 2 F_{zu} \Gamma^{z}_{\ zu} \Gamma^{z}_{\ uz} F_{uz} + 2 F_{zu} \Gamma^{i}_{\ zu} \Gamma^{z}_{\ ui} F_{uz}\\
		= & 0\,.
	\end{split}
\end{equation}
The second term of Eq. (\ref{hkk2secondfourteenth}) is 
\begin{equation}
	\begin{split}
		& 2 \gamma^{ij} F_{zj} \Gamma^{h}_{\ iu} \Gamma^{z}_{\ uh} F_{uz}\\
		= & 2 \gamma^{ij} F_{zj} \Gamma^{u}_{\ iu} \Gamma^{z}_{\ uu} F_{uz} + 2 \gamma^{ij} F_{zj} \Gamma^{z}_{\ iu} \Gamma^{z}_{\ uz} F_{uz} + 2 \gamma^{ij} F_{zj} \Gamma^{k}_{\ iu} \Gamma^{z}_{\ uk} F_{uz}\\
		= & 0\,.
	\end{split}
\end{equation}
The third term of Eq. (\ref{hkk2secondfourteenth}) is
\begin{equation}
	\begin{split}
		& 2 F_{uz} \Gamma^{h}_{\ uz} \Gamma^{z}_{\ uh} F_{uz}\\
		= & 2 F_{uz} \Gamma^{u}_{\ uz} \Gamma^{z}_{\ uu} F_{uz} + 2 F_{uz} \Gamma^{z}_{\ uz} \Gamma^{z}_{\ uz} F_{uz} + 2 F_{uz} \Gamma^{i}_{\ uz} \Gamma^{z}_{\ ui} F_{uz}\\
		= & 0\,.
	\end{split}
\end{equation}
The fourth term of Eq. (\ref{hkk2secondfourteenth}) is
\begin{equation}
	\begin{split}
		& 2 \gamma^{ij} F_{jz} \Gamma^{h}_{\ ui} \Gamma^{z}_{\ uh} F_{uz}\\
		= & 2 \gamma^{ij} F_{jz} \Gamma^{u}_{\ ui} \Gamma^{z}_{\ uu} F_{uz} + 2 \gamma^{ij} F_{jz} \Gamma^{z}_{\ ui} \Gamma^{z}_{\ uz} F_{uz} + 2 \gamma^{ij} F_{jz} \Gamma^{k}_{\ ui} \Gamma^{z}_{\ uk} F_{uz}\\
		= & 0\,.
	\end{split}
\end{equation}
The fifth term of Eq. (\ref{hkk2secondfourteenth}) is
\begin{equation}
	\begin{split}
		& 2 \gamma^{ij} \gamma^{kl} F_{jl} \Gamma^{h}_{\ ki} \Gamma^{z}_{\ uh} F_{uz}\\
		= & 2 \gamma^{ij} \gamma^{kl} F_{jl} \Gamma^{u}_{\ ki} \Gamma^{z}_{\ uu} F_{uz} + 2 \gamma^{ij} \gamma^{kl} F_{jl} \Gamma^{z}_{\ ki} \Gamma^{z}_{\ uz} F_{uz} + 2 \gamma^{ij} \gamma^{kl} F_{jl} \Gamma^{m}_{\ ki} \Gamma^{z}_{\ um} F_{uz}\\
		= & 0\,.
	\end{split}
\end{equation}
Therefore, the fourteenth term of Eq. (\ref{hkk2second}) is obtained as
\begin{equation}
	\begin{split}
		2 k^a k^b g^{ce} g^{df} F_{ef} \Gamma^{h}_{\ dc} \Gamma^{g}_{\ ah} F_{bg} = 0\,.
	\end{split}
\end{equation}

The fifteenth term of Eq. (\ref{hkk2second}) is
\begin{equation}
	\begin{split}
		& - 2 k^a k^b g^{ce} g^{df} F_{ef} \Gamma^{g}_{\ dh} \Gamma^{h}_{\ ac} F_{bg} = - 2 g^{ce} g^{df} F_{ef} \Gamma^{g}_{\ dh} \Gamma^{h}_{\ uc} F_{ug}\\
		= & - 2 F_{uz} \Gamma^{g}_{\ uh} \Gamma^{h}_{\ uz} F_{ug} - 2 \gamma^{ij} F_{jz} \Gamma^{g}_{\ uh} \Gamma^{h}_{\ ui} F_{ug} - 2 \gamma^{ij} \gamma^{kl} F_{jl} \Gamma^{g}_{\ kh} \Gamma^{h}_{\ ui} F_{ug}\,.
	\end{split}
\end{equation}
The index $g$ should be further expanded.
\begin{equation}\label{hkk2secondfifteenth}
	\begin{split}
		& - 2 F_{uz} \Gamma^{g}_{\ uh} \Gamma^{h}_{\ uz} F_{ug} - 2 \gamma^{ij} F_{jz} \Gamma^{g}_{\ uh} \Gamma^{h}_{\ ui} F_{ug} - 2 \gamma^{ij} \gamma^{kl} F_{jl} \Gamma^{g}_{\ kh} \Gamma^{h}_{\ ui} F_{ug}\\
		= & - 2 F_{uz} \Gamma^{u}_{\ uh} \Gamma^{h}_{\ uz} F_{uu} - 2 F_{uz} \Gamma^{z}_{\ uh} \Gamma^{h}_{\ uz} F_{uz} - 2 F_{uz} \Gamma^{i}_{\ uh} \Gamma^{h}_{\ uz} F_{ui}\\
		& - 2 \gamma^{ij} F_{jz} \Gamma^{u}_{\ uh} \Gamma^{h}_{\ ui} F_{uu} - 2 \gamma^{ij} F_{jz} \Gamma^{z}_{\ uh} \Gamma^{h}_{\ ui} F_{uz} - 2 \gamma^{ij} F_{jz} \Gamma^{k}_{\ uh} \Gamma^{h}_{\ ui} F_{uk}\\
		& - 2 \gamma^{ij} \gamma^{kl} F_{jl} \Gamma^{u}_{\ kh} \Gamma^{h}_{\ ui} F_{uu} - 2 \gamma^{ij} \gamma^{kl} F_{jl} \Gamma^{z}_{\ kh} \Gamma^{h}_{\ ui} F_{uz} - 2 \gamma^{ij} \gamma^{kl} F_{jl} \Gamma^{m}_{\ kh} \Gamma^{h}_{\ ui} F_{um}\\
		= & - 2 F_{uz} \Gamma^{z}_{\ uh} \Gamma^{h}_{\ uz} F_{uz} - 2 \gamma^{ij} F_{jz} \Gamma^{z}_{\ uh} \Gamma^{h}_{\ ui} F_{uz} - 2 \gamma^{ij} \gamma^{kl} F_{jl} \Gamma^{z}_{\ kh} \Gamma^{h}_{\ ui} F_{uz}\,.
	\end{split}
\end{equation}
The repeated index $h$ should be further expanded. The first term of Eq. (\ref{hkk2secondfifteenth}) is
\begin{equation}
	\begin{split}
		& - 2 F_{uz} \Gamma^{z}_{\ uh} \Gamma^{h}_{\ uz} F_{uz}\\
		= & - 2 F_{uz} \Gamma^{z}_{\ uu} \Gamma^{u}_{\ uz} F_{uz} - 2 F_{uz} \Gamma^{z}_{\ uz} \Gamma^{z}_{\ uz} F_{uz} - 2 F_{uz} \Gamma^{z}_{\ ui} \Gamma^{i}_{\ uz} F_{uz}\\
		= & 0\,.
	\end{split}
\end{equation}
The second term of Eq. (\ref{hkk2secondfifteenth}) is
\begin{equation}
	\begin{split}
		& - 2 \gamma^{ij} F_{jz} \Gamma^{z}_{\ uh} \Gamma^{h}_{\ ui} F_{uz}\\
		= & - 2 \gamma^{ij} F_{jz} \Gamma^{z}_{\ uu} \Gamma^{u}_{\ ui} F_{uz} - 2 \gamma^{ij} F_{jz} \Gamma^{z}_{\ uz} \Gamma^{z}_{\ ui} F_{uz} - 2 \gamma^{ij} F_{jz} \Gamma^{z}_{\ uk} \Gamma^{k}_{\ ui} F_{uz}\\
		= & 0\,.
	\end{split}
\end{equation}
The third term of Eq. (\ref{hkk2secondfifteenth}) is
\begin{equation}
	\begin{split}
		& - 2 \gamma^{ij} \gamma^{kl} F_{jl} \Gamma^{z}_{\ kh} \Gamma^{h}_{\ ui} F_{uz}\\
		= & - 2 \gamma^{ij} \gamma^{kl} F_{jl} \Gamma^{z}_{\ ku} \Gamma^{u}_{\ ui} F_{uz} - 2 \gamma^{ij} \gamma^{kl} F_{jl} \Gamma^{z}_{\ kz} \Gamma^{z}_{\ ui} F_{uz} - 2 \gamma^{ij} \gamma^{kl} F_{jl} \Gamma^{z}_{\ km} \Gamma^{m}_{\ ui} F_{uz}\\
		= & \frac{1}{2} \gamma^{ij} \gamma^{kl} \gamma^{mo} F_{jl} \left(\partial_u \gamma_{km} \right) \left(\partial_u \gamma_{io} \right) F_{uz}\,.
	\end{split}
\end{equation}
Therefore, the fifteenth term of Eq. (\ref{hkk2second}) is obtained as
\begin{equation}
	\begin{split}
		& - 2 k^a k^b g^{ce} g^{df} F_{ef} \Gamma^{g}_{\ dh} \Gamma^{h}_{\ ac} F_{bg} = \frac{1}{2} \gamma^{ij} \gamma^{kl} \gamma^{mo} F_{jl} \left(\partial_u \gamma_{km} \right) \left(\partial_u \gamma_{io} \right) F_{uz}\\
		= & 2 \gamma^{ij} \gamma^{kl} \gamma^{mo} F_{jl} K_{km} K_{io} F_{uz}\\
		= & 0\,.
	\end{split}
\end{equation}

The sixteenth term of Eq. (\ref{hkk2second}) is
\begin{equation}
	\begin{split}
		& - 2 k^a k^b g^{ce} g^{df} F_{ef} \Gamma^{g}_{\ ac} \left(\partial_d F_{bg} \right) = - 2 g^{ce} g^{df} F_{ef} \Gamma^{g}_{\ uc} \left(\partial_d F_{ug} \right)\\
		= & - 2 F_{zu} \Gamma^{g}_{\ uu} \left(\partial_z F_{ug} \right) - 2 \gamma^{ij} F_{zj} \Gamma^{g}_{\ uu} \left(\partial_i F_{ug} \right) - 2 F_{uz} \Gamma^{g}_{\ uz} \left(\partial_u F_{ug} \right)\\
		& - 2 \gamma^{ij} F_{jz} \Gamma^{g}_{\ ui} \left(\partial_u F_{ug} \right) - 2 \gamma^{ij} \gamma^{kl} F_{jl} \Gamma^{g}_{\ ui} \left(\partial_k F_{ug} \right)\\
		= & - 2 F_{uz} \Gamma^{g}_{\ uz} \left(\partial_u F_{ug} \right) - 2 \gamma^{ij} F_{jz} \Gamma^{g}_{\ ui} \left(\partial_u F_{ug} \right) - 2 \gamma^{ij} \gamma^{kl} F_{jl} \Gamma^{g}_{\ ui} \left(\partial_k F_{ug} \right)\,.
	\end{split}
\end{equation}
The index $g$ should be further expanded.
\begin{equation}\label{hkk2secondsixteenth}
	\begin{split}
		& - 2 F_{uz} \Gamma^{g}_{\ uz} \left(\partial_u F_{ug} \right) - 2 \gamma^{ij} F_{jz} \Gamma^{g}_{\ ui} \left(\partial_u F_{ug} \right) - 2 \gamma^{ij} \gamma^{kl} F_{jl} \Gamma^{g}_{\ ui} \left(\partial_k F_{ug} \right)\\
		= & - 2 F_{uz} \Gamma^{u}_{\ uz} \left(\partial_u F_{uu} \right) - 2 F_{uz} \Gamma^{z}_{\ uz} \left(\partial_u F_{uz} \right) - 2 F_{uz} \Gamma^{i}_{\ uz} \left(\partial_u F_{ui} \right)\\
		& - 2 \gamma^{ij} F_{jz} \Gamma^{u}_{\ ui} \left(\partial_u F_{uu} \right) - 2 \gamma^{ij} F_{jz} \Gamma^{z}_{\ ui} \left(\partial_u F_{uz} \right) - 2 \gamma^{ij} F_{jz} \Gamma^{k}_{\ ui} \left(\partial_u F_{uk} \right)\\
		& - 2 \gamma^{ij} \gamma^{kl} F_{jl} \Gamma^{u}_{\ ui} \left(\partial_k F_{uu} \right) - 2 \gamma^{ij} \gamma^{kl} F_{jl} \Gamma^{z}_{\ ui} \left(\partial_k F_{uz} \right) - 2 \gamma^{ij} \gamma^{kl} F_{jl} \Gamma^{m}_{\ ui} \left(\partial_k F_{um} \right)\\
		= & - 2 F_{uz} \Gamma^{z}_{\ uz} \left(\partial_u F_{uz} \right) - 2 F_{uz} \Gamma^{i}_{\ uz} \left(\partial_u F_{ui} \right) - 2 \gamma^{ij} F_{jz} \Gamma^{z}_{\ ui} \left(\partial_u F_{uz} \right)\\
		& - 2 \gamma^{ij} F_{jz} \Gamma^{k}_{\ ui} \left(\partial_u F_{uk} \right) - 2 \gamma^{ij} \gamma^{kl} F_{jl} \Gamma^{z}_{\ ui} \left(\partial_k F_{uz} \right)\,.
	\end{split}
\end{equation}
Therefore, the sixteenth term of Eq. (\ref{hkk2second}) is obtained as 
\begin{equation}
	\begin{split}
		& - 2 k^a k^b g^{ce} g^{df} F_{ef} \Gamma^{g}_{\ ac} \left(\partial_d F_{bg} \right)\\
		= & - \gamma^{ij} \beta_j F_{uz} \left(\partial_u F_{ui} \right) - \gamma^{ij} \gamma^{kl} F_{jz} \left(\partial_u \gamma_{il} \right) \left(\partial_u F_{uk} \right)\\
		= & - \gamma^{ij} \beta_j F_{uz} \left(\partial_u F_{ui} \right) - 2 \gamma^{ij} \gamma^{kl} F_{jz} K_{il} \left(\partial_u F_{uk} \right)\\
		= & - \gamma^{ij} \beta_j F_{uz} \left(\partial_u F_{ui} \right)\,.
	\end{split}
\end{equation}

The seventeenth term of Eq. (\ref{hkk2second}) is 
\begin{equation}
	\begin{split}
		& 2 k^a k^b g^{ce} g^{df} F_{ef} \Gamma^{g}_{\ ac} \Gamma^{h}_{\ db} F_{hg} = 2 g^{ce} g^{df} F_{ef} \Gamma^{g}_{\ uc} \Gamma^{h}_{\ du} F_{hg}\\
		= & 2 F_{zu} \Gamma^{g}_{\ uu} \Gamma^{h}_{\ zu} F_{hg} + 2 \gamma^{ij} F_{zj} \Gamma^{g}_{\ uu} \Gamma^{h}_{\ iu} F_{hg} + 2 F_{uz} \Gamma^{g}_{\ uz} \Gamma^{h}_{\ uu} F_{hg}\\
		& + 2 \gamma^{ij} F_{jz} \Gamma^{g}_{\ ui} \Gamma^{h}_{\ uu} F_{hg} + 2 \gamma^{ij} \gamma^{kl} F_{jl} \Gamma^{g}_{\ ui} \Gamma^{h}_{\ ku} F_{hg}\\
		= & 2 \gamma^{ij} \gamma^{kl} F_{jl} \Gamma^{g}_{\ ui} \Gamma^{h}_{\ ku} F_{hg}\,.
	\end{split}
\end{equation}
The index $g$ should be further expanded.
\begin{equation}\label{hkk2secondseventeenth}
	\begin{split}
		& 2 \gamma^{ij} \gamma^{kl} F_{jl} \Gamma^{g}_{\ ui} \Gamma^{h}_{\ ku} F_{hg}\\
		= & 2 \gamma^{ij} \gamma^{kl} F_{jl} \Gamma^{u}_{\ ui} \Gamma^{h}_{\ ku} F_{hu} + 2 \gamma^{ij} \gamma^{kl} F_{jl} \Gamma^{z}_{\ ui} \Gamma^{h}_{\ ku} F_{hz} + 2 \gamma^{ij} \gamma^{kl} F_{jl} \Gamma^{m}_{\ ui} \Gamma^{h}_{\ ku} F_{hm}\,.
	\end{split}
\end{equation}
The repeated index $h$ should be further expanded. The first term of Eq. (\ref{hkk2secondseventeenth}) is
\begin{equation}
	\begin{split}
		& 2 \gamma^{ij} \gamma^{kl} F_{jl} \Gamma^{u}_{\ ui} \Gamma^{h}_{\ ku} F_{hu}\\
		= & 2 \gamma^{ij} \gamma^{kl} F_{jl} \Gamma^{u}_{\ ui} \Gamma^{u}_{\ ku} F_{uu} + 2 \gamma^{ij} \gamma^{kl} F_{jl} \Gamma^{u}_{\ ui} \Gamma^{z}_{\ ku} F_{zu} + 2 \gamma^{ij} \gamma^{kl} F_{jl} \Gamma^{u}_{\ ui} \Gamma^{m}_{\ ku} F_{mu}\\
		= & 2 \gamma^{ij} \gamma^{kl} F_{jl} \Gamma^{u}_{\ ui} \Gamma^{z}_{\ ku} F_{zu}\\
		= & 0\,.
	\end{split}
\end{equation}
The second term of Eq. (\ref{hkk2secondseventeenth}) is
\begin{equation}
	\begin{split}
		2 \gamma^{ij} \gamma^{kl} F_{jl} \Gamma^{z}_{\ ui} \Gamma^{h}_{\ ku} F_{hz} = 0\,.
	\end{split}
\end{equation}
The third term of Eq. (\ref{hkk2secondseventeenth}) is
\begin{equation}
	\begin{split}
		& 2 \gamma^{ij} \gamma^{kl} F_{jl} \Gamma^{m}_{\ ui} \Gamma^{h}_{\ ku} F_{hm}\\
		= & 2 \gamma^{ij} \gamma^{kl} F_{jl} \Gamma^{m}_{\ ui} \Gamma^{u}_{\ ku} F_{um} + 2 \gamma^{ij} \gamma^{kl} F_{jl} \Gamma^{m}_{\ ui} \Gamma^{z}_{\ ku} F_{zm} + 2 \gamma^{ij} \gamma^{kl} F_{jl} \Gamma^{m}_{\ ui} \Gamma^{n}_{\ ku} F_{nm}\\
		= & 2 \gamma^{ij} \gamma^{kl} F_{jl} \Gamma^{m}_{\ ui} \Gamma^{z}_{\ ku} F_{zm} + 2 \gamma^{ij} \gamma^{kl} F_{jl} \Gamma^{m}_{\ ui} \Gamma^{n}_{\ ku} F_{nm}\\
		= & \frac{1}{2} \gamma^{ij} \gamma^{kl} \gamma^{mo} \gamma^{np} F_{jl} \left(\partial_u \gamma_{io} \right) \left(\partial_u \gamma_{kp} \right) F_{nm}\,.
	\end{split}
\end{equation}
Therefore, the seventeenth term of Eq. (\ref{hkk2second}) is obtained as
\begin{equation}
	\begin{split}
		& 2 k^a k^b g^{ce} g^{df} F_{ef} \Gamma^{g}_{\ ac} \Gamma^{h}_{\ db} F_{hg} = \frac{1}{2} \gamma^{ij} \gamma^{kl} \gamma^{mo} \gamma^{np} F_{jl} \left(\partial_u \gamma_{io} \right) \left(\partial_u \gamma_{kp} \right) F_{nm}\\
		= & 2 \gamma^{ij} \gamma^{kl} \gamma^{mo} \gamma^{np} F_{jl} K_{io} K_{kp} F_{nm}\\
		= & 0
	\end{split}
\end{equation}

The eighteenth term of Eq. (\ref{hkk2second}) is
\begin{equation}
	\begin{split}
		& 2 k^a k^b g^{ce} g^{df} F_{ef} \Gamma^{g}_{\ ac} \Gamma^{h}_{\ dg} F_{bh} = 2 g^{ce} g^{df} F_{ef} \Gamma^{g}_{\ uc} \Gamma^{h}_{\ dg} F_{uh}\\
		= & 2 F_{zu} \Gamma^{g}_{\ uu} \Gamma^{h}_{\ zg} F_{uh} + 2 \gamma^{ij} F_{zj} \Gamma^{g}_{\ uu} \Gamma^{h}_{\ ig} F_{uh} + 2 F_{uz} \Gamma^{g}_{\ uz} \Gamma^{h}_{\ ug} F_{uh}\\
		& + 2 \gamma^{ij} F_{jz} \Gamma^{g}_{\ ui} \Gamma^{h}_{\ ug} F_{uh} + 2 \gamma^{ij} \gamma^{kl} F_{jl} \Gamma^{g}_{\ ui} \Gamma^{h}_{\ kg} F_{uh}\\
		= & 2 F_{uz} \Gamma^{g}_{\ uz} \Gamma^{h}_{\ ug} F_{uh} + 2 \gamma^{ij} F_{jz} \Gamma^{g}_{\ ui} \Gamma^{h}_{\ ug} F_{uh} + 2 \gamma^{ij} \gamma^{kl} F_{jl} \Gamma^{g}_{\ ui} \Gamma^{h}_{\ kg} F_{uh}\,.
	\end{split}
\end{equation}
The index $g$ should be further expanded.
\begin{equation}\label{hkk2secondeighteenth}
	\begin{split}
		& 2 F_{uz} \Gamma^{g}_{\ uz} \Gamma^{h}_{\ ug} F_{uh} + 2 \gamma^{ij} F_{jz} \Gamma^{g}_{\ ui} \Gamma^{h}_{\ ug} F_{uh} + 2 \gamma^{ij} \gamma^{kl} F_{jl} \Gamma^{g}_{\ ui} \Gamma^{h}_{\ kg} F_{uh}\\
		= & 2 F_{uz} \Gamma^{u}_{\ uz} \Gamma^{h}_{\ uu} F_{uh} + 2 F_{uz} \Gamma^{z}_{\ uz} \Gamma^{h}_{\ uz} F_{uh} + 2 F_{uz} \Gamma^{i}_{\ uz} \Gamma^{h}_{\ ui} F_{uh}\\
		& + 2 \gamma^{ij} F_{jz} \Gamma^{u}_{\ ui} \Gamma^{h}_{\ uu} F_{uh} + 2 \gamma^{ij} F_{jz} \Gamma^{z}_{\ ui} \Gamma^{h}_{\ uz} F_{uh} + 2 \gamma^{ij} F_{jz} \Gamma^{k}_{\ ui} \Gamma^{h}_{\ uk} F_{uh}\\
		& + 2 \gamma^{ij} \gamma^{kl} F_{jl} \Gamma^{u}_{\ ui} \Gamma^{h}_{\ ku} F_{uh} + 2 \gamma^{ij} \gamma^{kl} F_{jl} \Gamma^{z}_{\ ui} \Gamma^{h}_{\ kz} F_{uh} + 2 \gamma^{ij} \gamma^{kl} F_{jl} \Gamma^{m}_{\ ui} \Gamma^{h}_{\ km} F_{uh}\\
		= & 2 F_{uz} \Gamma^{z}_{\ uz} \Gamma^{h}_{\ uz} F_{uh} + 2 F_{uz} \Gamma^{i}_{\ uz} \Gamma^{h}_{\ ui} F_{uh} + 2 \gamma^{ij} F_{jz} \Gamma^{z}_{\ ui} \Gamma^{h}_{\ uz} F_{uh}\\
		& + 2 \gamma^{ij} F_{jz} \Gamma^{k}_{\ ui} \Gamma^{h}_{\ uk} F_{uh} + 2 \gamma^{ij} \gamma^{kl} F_{jl} \Gamma^{u}_{\ ui} \Gamma^{h}_{\ ku} F_{uh} + 2 \gamma^{ij} \gamma^{kl} F_{jl} \Gamma^{z}_{\ ui} \Gamma^{h}_{\ kz} F_{uh}\\
		& + 2 \gamma^{ij} \gamma^{kl} F_{jl} \Gamma^{m}_{\ ui} \Gamma^{h}_{\ km} F_{uh}\,.
	\end{split}
\end{equation}
The repeated index $h$ should be further expanded. The first term of Eq. (\ref{hkk2secondeighteenth}) is
\begin{equation}
	\begin{split}
		2 F_{uz} \Gamma^{z}_{\ uz} \Gamma^{h}_{\ uz} F_{uh} = 0\,.
	\end{split}
\end{equation}
The second term of Eq. (\ref{hkk2secondeighteenth}) is
\begin{equation}
	\begin{split}
		& 2 F_{uz} \Gamma^{i}_{\ uz} \Gamma^{h}_{\ ui} F_{uh}\\
		= & 2 F_{uz} \Gamma^{i}_{\ uz} \Gamma^{u}_{\ ui} F_{uu} + 2 F_{uz} \Gamma^{i}_{\ uz} \Gamma^{z}_{\ ui} F_{uz} + 2 F_{uz} \Gamma^{i}_{\ uz} \Gamma^{j}_{\ ui} F_{uj}\\
		= & 2 F_{uz} \Gamma^{i}_{\ uz} \Gamma^{z}_{\ ui} F_{uz}\\
		= & 0\,.
	\end{split}
\end{equation}
The third term of Eq. (\ref{hkk2secondeighteenth}) is
\begin{equation}
	\begin{split}
		2 \gamma^{ij} F_{jz} \Gamma^{z}_{\ ui} \Gamma^{h}_{\ uz} F_{uh} = 0\,.
	\end{split}
\end{equation}
The fourth term of Eq. (\ref{hkk2secondeighteenth}) is
\begin{equation}
	\begin{split}
		& 2 \gamma^{ij} F_{jz} \Gamma^{k}_{\ ui} \Gamma^{h}_{\ uk} F_{uh}\\
		= & 2 \gamma^{ij} F_{jz} \Gamma^{k}_{\ ui} \Gamma^{u}_{\ uk} F_{uu} + 2 \gamma^{ij} F_{jz} \Gamma^{k}_{\ ui} \Gamma^{z}_{\ uk} F_{uz} + 2 \gamma^{ij} F_{jz} \Gamma^{k}_{\ ui} \Gamma^{l}_{\ uk} F_{ul}\\
		= & 2 \gamma^{ij} F_{jz} \Gamma^{k}_{\ ui} \Gamma^{z}_{\ uk} F_{uz}\\
		= & 0\,.
	\end{split}
\end{equation}
The fifth term of Eq. (\ref{hkk2secondeighteenth}) is
\begin{equation}
	\begin{split}
		& 2 \gamma^{ij} \gamma^{kl} F_{jl} \Gamma^{u}_{\ ui} \Gamma^{h}_{\ ku} F_{uh}\\
		= & 2 \gamma^{ij} \gamma^{kl} F_{jl} \Gamma^{u}_{\ ui} \Gamma^{u}_{\ ku} F_{uu} + 2 \gamma^{ij} \gamma^{kl} F_{jl} \Gamma^{u}_{\ ui} \Gamma^{z}_{\ ku} F_{uz} + 2 \gamma^{ij} \gamma^{kl} F_{jl} \Gamma^{u}_{\ ui} \Gamma^{m}_{\ ku} F_{um}\\
		= & 2 \gamma^{ij} \gamma^{kl} F_{jl} \Gamma^{u}_{\ ui} \Gamma^{z}_{\ ku} F_{uz}\\
		= & 0\,.
	\end{split}
\end{equation}
The sixth term of Eq. (\ref{hkk2secondeighteenth}) is
\begin{equation}
	\begin{split}
		2 \gamma^{ij} \gamma^{kl} F_{jl} \Gamma^{z}_{\ ui} \Gamma^{h}_{\ kz} F_{uh} = 0\,.
	\end{split}
\end{equation}
The seventh term of Eq. (\ref{hkk2secondeighteenth}) is
\begin{equation}
	\begin{split}
		& 2 \gamma^{ij} \gamma^{kl} F_{jl} \Gamma^{m}_{\ ui} \Gamma^{h}_{\ km} F_{uh}\\
		= & 2 \gamma^{ij} \gamma^{kl} F_{jl} \Gamma^{m}_{\ ui} \Gamma^{u}_{\ km} F_{uu} + 2 \gamma^{ij} \gamma^{kl} F_{jl} \Gamma^{m}_{\ ui} \Gamma^{z}_{\ km} F_{uz} + 2 \gamma^{ij} \gamma^{kl} F_{jl} \Gamma^{m}_{\ ui} \Gamma^{n}_{\ km} F_{un}\\
		= & 2 \gamma^{ij} \gamma^{kl} F_{jl} \Gamma^{m}_{\ ui} \Gamma^{z}_{\ km} F_{uz}\\
		= & - \frac{1}{2} \gamma^{ij} \gamma^{kl} \gamma^{mn} F_{jl} \left(\partial_u \gamma_{in} \right) \left(\partial_u \gamma_{km} \right) F_{uz}\,.
	\end{split}
\end{equation}
Therefore, the eighteenth term of Eq. (\ref{hkk2second}) is obtained as
\begin{equation}
	\begin{split}
		& 2 k^a k^b g^{ce} g^{df} F_{ef} \Gamma^{g}_{\ ac} \Gamma^{h}_{\ dg} F_{bh} = - \frac{1}{2} \gamma^{ij} \gamma^{kl} \gamma^{mn} F_{jl} \left(\partial_u \gamma_{in} \right) \left(\partial_u \gamma_{km} \right) F_{uz}\\
		= & - 2 \gamma^{ij} \gamma^{kl} \gamma^{mn} F_{jl} K_{in} K_{km} F_{uz}\\
		= & 0\,.
	\end{split}
\end{equation}

Finally, the second term of Eq. (\ref{rewrittenhkk2}) is 
\begin{equation}
	\begin{split}
		& 2 k^a k^b F^{cd} \nabla_d \nabla_a F_{bc}\\
		= & 2 F_{uz} \left(\partial_u \partial_u F_{uz} \right) + 2 \gamma^{ij} F_{jz} \left(\partial_u \partial_u F_{ui} \right) + 2 \gamma^{ij} \gamma^{kl} F_{jl} \left(\partial_k \partial_u F_{ui} \right)\\
		& + \gamma^{ij} \gamma^{kl} \beta_k F_{jl} \left(\partial_u F_{ui} \right) - \gamma^{ij} \beta_j F_{zu} \left(\partial_u F_{iu} \right) + \gamma^{ij} \gamma^{kl} \beta_k F_{jl} \left(\partial_u F_{ui} \right)\\
		& - \gamma^{ij} \beta_j F_{uz} \left(\partial_u F_{ui} \right)\\
		= & 2 F_{uz} \left(\partial_u \partial_u F_{uz} \right) + 2 \gamma^{ij} F_{jz} \left(\partial_u \partial_u F_{ui} \right) + 2 \gamma^{ij} \gamma^{kl} F_{jl} \left(\partial_k \partial_u F_{ui} \right)\\
		& + 2 \gamma^{ij} \gamma^{kl} \beta_k F_{jl} \left(\partial_u F_{ui} \right) - 2 \gamma^{ij} \beta_j F_{uz} \left(\partial_u F_{ui} \right)\,.
	\end{split}
\end{equation}

The third term of Eq. (\ref{rewrittenhkk2}) is
\begin{equation}\label{hkk2third}
	\begin{split}
		& 2 k^a k^b F_{a}^{\ c} \nabla_d \nabla^d F_{bc} = 2 k^a k^b g^{ce} g^{df} F_{ae} \nabla_d \nabla_f F_{bc}\\
		= & 2 k^a k^b g^{ce} g^{df} F_{ae} \left(\partial_d \partial_f F_{bc} \right) - 2 k^a k^b g^{ce} g^{df} F_{ae} \Gamma^{g}_{\ df} \left(\partial_g F_{bc} \right)\\
		& - 2 k^a k^b g^{ce} g^{df} F_{ae} \Gamma^{g}_{\ db} \left(\partial_f F_{gc} \right) - 2 k^a k^b g^{ce} g^{df} F_{ae} \Gamma^{g}_{\ dc} \left(\partial_f F_{bg} \right)\\
		& - 2 k^a k^b g^{ce} g^{df} F_{ae} \left(\partial_d \Gamma^{g}_{\ fb} \right) F_{gc} + 2 k^a k^b g^{ce} g^{df} F_{ae} \Gamma^{h}_{\ df} \Gamma^{g}_{\ hb} F_{gc}\\
		& + 2 k^a k^b g^{ce} g^{df} F_{ae} \Gamma^{h}_{\ db} \Gamma^{g}_{\ fh} F_{gc} - 2 k^a k^b g^{ce} g^{df} F_{ae} \Gamma^{g}_{\ dh} \Gamma^{h}_{\ fb} F_{gc}\\
		& - 2 k^a k^b g^{ce} g^{df} F_{ae} \Gamma^{g}_{\ fb} \left(\partial_d F_{gc} \right) + 2 k^a k^b g^{ce} g^{df} F_{ae} \Gamma^{g}_{\ fb} \Gamma^{h}_{\ dg} F_{hc}\\
		& + 2 k^a k^b g^{ce} g^{df} F_{ae} \Gamma^{g}_{\ fb} \Gamma^{h}_{\ dc} F_{gh} - 2 k^a k^b g^{ce} g^{df} F_{ae} \left(\partial_d \Gamma^{g}_{\ fc} \right) F_{bg}\\
		& + 2 k^a k^b g^{ce} g^{df} F_{ae} \Gamma^{h}_{\ df} \Gamma^{g}_{\ hc} F_{bg} + 2 k^a k^b g^{ce} g^{df} F_{ae} \Gamma^{h}_{\ dc} \Gamma^{g}_{\ fh} F_{bg}\\
		& - 2 k^a k^b g^{ce} g^{df} F_{ae} \Gamma^{g}_{\ dh} \Gamma^{h}_{\ fc} F_{bg} - 2 k^a k^b g^{ce} g^{df} F_{ae} \Gamma^{g}_{\ fc} \left(\partial_d F_{bg} \right)\\
		& + 2 k^a k^b g^{ce} g^{df} F_{ae} \Gamma^{g}_{\ fc} \Gamma^{h}_{\ db} F_{hg} + 2 k^a k^b g^{ce} g^{df} F_{ae} \Gamma^{g}_{\ fc} \Gamma^{h}_{\ dg} F_{bh}\,.
	\end{split}
\end{equation}

The first term of Eq. (\ref{hkk2third}) is 
\begin{equation}
	\begin{split}
		& 2 k^a k^b g^{ce} g^{df} F_{ae} \left(\partial_d \partial_f F_{bc} \right) = 2 g^{ce} g^{df} F_{ue} \left(\partial_d \partial_f F_{uc} \right)\\
		= & 2 \gamma^{ij} g^{df} F_{uj} \left(\partial_d \partial_f F_{ui} \right)\\
		= & 0\,.
	\end{split}
\end{equation}
Therefore, the first term of Eq. (\ref{hkk2third}) is obtained as
\begin{equation}
	\begin{split}
		2 k^a k^b g^{ce} g^{df} F_{ae} \left(\partial_d \partial_f F_{bc} \right) = 0\,.
	\end{split}
\end{equation}

The second term of Eq. (\ref{hkk2third}) is
\begin{equation}
	\begin{split}
		& - 2 k^a k^b g^{ce} g^{df} F_{ae} \Gamma^{g}_{\ df} \left(\partial_g F_{bc} \right) = - 2 g^{ce} g^{df} F_{ue} \Gamma^{g}_{\ df} \left(\partial_g F_{uc} \right)\\
		= & - 2 \gamma^{ij} g^{df} F_{uj} \Gamma^{g}_{\ df} \left(\partial_g F_{ui} \right)\\
		= & 0\,.
	\end{split}
\end{equation}
Therefore, the second term of Eq. (\ref{hkk2third}) is obtained as
\begin{equation}
	\begin{split}
		- 2 k^a k^b g^{ce} g^{df} F_{ae} \Gamma^{g}_{\ df} \left(\partial_g F_{bc} \right) = 0\,.
	\end{split}
\end{equation}

The third term of Eq. (\ref{hkk2third}) is
\begin{equation}
	\begin{split}
		& - 2 k^a k^b g^{ce} g^{df} F_{ae} \Gamma^{g}_{\ db} \left(\partial_f F_{gc} \right) = - 2 g^{ce} g^{df} F_{ue} \Gamma^{g}_{\ du} \left(\partial_f F_{gc} \right)\\
		= & - 2 F_{uz} \Gamma^{g}_{\ uu} \left(\partial_z F_{gu} \right) - 2 F_{uz} \Gamma^{g}_{\ zu} \left(\partial_u F_{gu} \right) - 2 \gamma^{ij} F_{uz} \Gamma^{g}_{\ iu} \left(\partial_j F_{gu} \right)\\
		= & - 2 F_{uz} \Gamma^{g}_{\ zu} \left(\partial_u F_{gu} \right) - 2 \gamma^{ij} F_{uz} \Gamma^{g}_{\ iu} \left(\partial_j F_{gu} \right)\,.
	\end{split}
\end{equation}
The index $g$ should be further expanded.
\begin{equation}\label{hkk2thirdthird}
	\begin{split}
		& - 2 F_{uz} \Gamma^{g}_{\ zu} \left(\partial_u F_{gu} \right) - 2 \gamma^{ij} F_{uz} \Gamma^{g}_{\ iu} \left(\partial_j F_{gu} \right)\\
		= & - 2 F_{uz} \Gamma^{u}_{\ zu} \left(\partial_u F_{uu} \right) - 2 F_{uz} \Gamma^{z}_{\ zu} \left(\partial_u F_{zu} \right) - 2 F_{uz} \Gamma^{i}_{\ zu} \left(\partial_u F_{iu} \right)\\
		& - 2 \gamma^{ij} F_{uz} \Gamma^{u}_{\ iu} \left(\partial_j F_{uu} \right) - 2 \gamma^{ij} F_{uz} \Gamma^{z}_{\ iu} \left(\partial_j F_{zu} \right) - 2 \gamma^{ij} F_{uz} \Gamma^{k}_{\ iu} \left(\partial_j F_{ku} \right)\\
		= & - 2 F_{uz} \Gamma^{z}_{\ zu} \left(\partial_u F_{zu} \right) - 2 F_{uz} \Gamma^{i}_{\ zu} \left(\partial_u F_{iu} \right) - 2 \gamma^{ij} F_{uz} \Gamma^{z}_{\ iu} \left(\partial_j F_{zu} \right)\\
		& - 2 \gamma^{ij} F_{uz} \Gamma^{k}_{\ iu} \left(\partial_j F_{ku} \right)\,.
	\end{split}
\end{equation}
Therefore, the third term of Eq. (\ref{hkk2third}) is obtained as
\begin{equation}
	\begin{split}
		- 2 k^a k^b g^{ce} g^{df} F_{ae} \Gamma^{g}_{\ db} \left(\partial_f F_{gc} \right) = - \gamma^{ij} \beta_j F_{uz} \left(\partial_u F_{iu} \right)\,.
	\end{split}
\end{equation}

The fourth term of Eq. (\ref{hkk2third}) is
\begin{equation}
	\begin{split}
		& - 2 k^a k^b g^{ce} g^{df} F_{ae} \Gamma^{g}_{\ dc} \left(\partial_f F_{bg} \right) = - 2 g^{ce} g^{df} F_{ue} \Gamma^{g}_{\ dc} \left(\partial_f F_{ug} \right)\\
		= & - 2 F_{uz} \Gamma^{g}_{\ uu} \left(\partial_z F_{ug} \right) - 2 F_{uz} \Gamma^{g}_{\ zu} \left(\partial_u F_{ug} \right) - 2 \gamma^{ij} F_{uz} \Gamma^{g}_{\ iu} \left(\partial_j F_{ug} \right)\\
		= & - 2 F_{uz} \Gamma^{g}_{\ zu} \left(\partial_u F_{ug} \right) - 2 \gamma^{ij} F_{uz} \Gamma^{g}_{\ iu} \left(\partial_j F_{ug} \right)\,.
	\end{split}
\end{equation}
The index $g$ should be further expanded.
\begin{equation}\label{hkk2thirdfourth}
	\begin{split}
		& - 2 F_{uz} \Gamma^{g}_{\ zu} \left(\partial_u F_{ug} \right) - 2 \gamma^{ij} F_{uz} \Gamma^{g}_{\ iu} \left(\partial_j F_{ug} \right)\\
		= & - 2 F_{uz} \Gamma^{u}_{\ zu} \left(\partial_u F_{uu} \right) - 2 F_{uz} \Gamma^{z}_{\ zu} \left(\partial_u F_{uz} \right) - 2 F_{uz} \Gamma^{i}_{\ zu} \left(\partial_u F_{ui} \right)\\
		& - 2 \gamma^{ij} F_{uz} \Gamma^{u}_{\ iu} \left(\partial_j F_{uu} \right) - 2 \gamma^{ij} F_{uz} \Gamma^{z}_{\ iu} \left(\partial_j F_{uz} \right) - 2 \gamma^{ij} F_{uz} \Gamma^{k}_{\ iu} \left(\partial_j F_{uk} \right)\\
		= & - 2 F_{uz} \Gamma^{z}_{\ zu} \left(\partial_u F_{uz} \right) - 2 F_{uz} \Gamma^{i}_{\ zu} \left(\partial_u F_{ui} \right) - 2 \gamma^{ij} F_{uz} \Gamma^{z}_{\ iu} \left(\partial_j F_{uz} \right)\,.
	\end{split}
\end{equation}
Therefore, the fourth term of Eq. (\ref{hkk2third}) is obtained as
\begin{equation}
	\begin{split}
		- 2 k^a k^b g^{ce} g^{df} F_{ae} \Gamma^{g}_{\ dc} \left(\partial_f F_{bg} \right) = - \gamma^{ij} \beta_j F_{uz} \left(\partial_u F_{ui} \right)\,.
	\end{split}
\end{equation}

The fifth term of Eq. (\ref{hkk2third}) is
\begin{equation}
	\begin{split}
		& - 2 k^a k^b g^{ce} g^{df} F_{ae} \left(\partial_d \Gamma^{g}_{\ fb} \right) F_{gc} = - 2 g^{ce} g^{df} F_{ue} \left(\partial_d \Gamma^{g}_{\ fu} \right) F_{gc}\\
		= & - 2 F_{uz} \left(\partial_u \Gamma^{g}_{\ zu} \right) F_{gu} - 2 F_{uz} \left(\partial_z \Gamma^{g}_{\ uu} \right) F_{gu} - 2 \gamma^{ij} F_{uz} \left(\partial_i \Gamma^{g}_{\ ju} \right) F_{gu}\,.
	\end{split}
\end{equation}
The index $g$ should be further expanded.
\begin{equation}\label{hkk2thirdfifth}
	\begin{split}
		& - 2 F_{uz} \left(\partial_u \Gamma^{g}_{\ zu} \right) F_{gu} - 2 F_{uz} \left(\partial_z \Gamma^{g}_{\ uu} \right) F_{gu} - 2 \gamma^{ij} F_{uz} \left(\partial_i \Gamma^{g}_{\ ju} \right) F_{gu}\\
		= & - 2 F_{uz} \left(\partial_u \Gamma^{u}_{\ zu} \right) F_{uu} - 2 F_{uz} \left(\partial_u \Gamma^{z}_{\ zu} \right) F_{zu} - 2 F_{uz} \left(\partial_u \Gamma^{i}_{\ zu} \right) F_{iu}\\
		& - 2 F_{uz} \left(\partial_z \Gamma^{u}_{\ uu} \right) F_{uu} - 2 F_{uz} \left(\partial_z \Gamma^{z}_{\ uu} \right) F_{zu} - 2 F_{uz} \left(\partial_z \Gamma^{i}_{\ uu} \right) F_{iu}\\
		& - 2 \gamma^{ij} F_{uz} \left(\partial_i \Gamma^{u}_{\ ju} \right) F_{uu} - 2 \gamma^{ij} F_{uz} \left(\partial_i \Gamma^{z}_{\ ju} \right) F_{zu} - 2 \gamma^{ij} F_{uz} \left(\partial_i \Gamma^{k}_{\ ju} \right) F_{ku}\\
		= & - 2 F_{uz} \left(\partial_u \Gamma^{z}_{\ zu} \right) F_{zu} - 2 F_{uz} \left(\partial_z \Gamma^{z}_{\ uu} \right) F_{zu} - 2 \gamma^{ij} F_{uz} \left(\partial_i \Gamma^{z}_{\ ju} \right) F_{zu}\,. 
	\end{split}
\end{equation}
The first term of Eq. (\ref{hkk2thirdfifth}) is 
\begin{equation}
	\begin{split}
		& - 2 F_{uz} \left(\partial_u \Gamma^{z}_{\ zu} \right) F_{zu}\\
		= & - 2 F_{uz} \partial_u \left(z^2 \partial_z \alpha + 2 z \alpha - \frac{1}{2} z \beta^i \beta_i - \frac{1}{2} z^2 \beta^i \partial_z \beta_i \right) F_{zu}\\
		= & - 2 F_{uz} \left[z^2 \partial_u \partial_z \alpha + 2 z \partial_u \alpha - \frac{1}{2} z \partial_u \left(\beta^i \beta_i \right) - \frac{1}{2} z^2 \left(\partial_u \beta^i \right) \partial_z \beta_i - \frac{1}{2} z^2 \beta^i \partial_u \partial_z \beta_i \right] F_{zu}\\
		= & 0\,.
	\end{split}
\end{equation}
The second term of Eq. (\ref{hkk2thirdfifth}) is 
\begin{equation}
	\begin{split}
		& - 2 F_{uz} \left(\partial_z \Gamma^{z}_{\ uu} \right) F_{zu}\\
		= & - 2 F_{uz} \partial_z \left[z^2 \partial_u \alpha - z^2 \left(\beta^2 - 2 \alpha \right) \left(z^2 \partial_z \alpha + 2 z \alpha \right) - z \beta^i \left(z \partial_u \beta_i - z^2 \partial_i \alpha \right) \right] F_{zu}\\
		= & - 2 F_{uz} \left[2 z \partial_u \alpha + z^2 \partial_z \partial_u \alpha - 2 z \left(\beta^2 - 2 \alpha \right) \left(z^2 \partial_z \alpha + 2 z \alpha \right)\right.\\
		& \left. - z^2 \left(\partial_z \beta^2 - 2 \partial_z \alpha \right) \left(z^2 \partial_z \alpha + 2 z \alpha \right) - z^2 \left(\beta^2 - 2 \alpha \right) \left(2 z \partial_z \alpha + z^2 \partial_z^2 \alpha + 2 \alpha + 2 z \partial_z \alpha \right)\right.\\
		& \left. - \beta^i \left(z \partial_u \beta_i - z^2 \partial_i \alpha \right) - z \left(\partial_z \beta^i \right) \left(z \partial_u \beta_i - z^2 \partial_i \alpha \right)\right.\\
		& \left. - z \beta^i \left(\partial_u \beta_i + z \partial_z \partial_u \beta_i - 2 z \partial_i \alpha - z^2 \partial_z \partial_i \alpha \right) \right] F_{zu}\\
		= & 0\,.
	\end{split}
\end{equation}
The third term of Eq. (\ref{hkk2thirdfifth}) is 
\begin{equation}
	\begin{split}
		& - 2 \gamma^{ij} F_{uz} \left(\partial_i \Gamma^{z}_{\ ju} \right) F_{zu}\\
		= & - 2 \gamma^{ij} F_{uz} \partial_i \left[z^2 \partial_j \alpha - \frac{1}{2} z^2 \left(\beta^2 - 2 \alpha \right) \left(\beta_j + z \partial_z \beta_j \right) - \frac{1}{2} z \beta^k \left(z \partial_j \beta_k + \partial_u \gamma_{jk} - z \partial_k \beta_j \right) \right]\\
		& \times F_{zu}\\
		= & - 2 \gamma^{ij} F_{uz} \left[z^2 \partial_i \partial_j \alpha - \frac{1}{2} z^2 \left(\partial_i \beta^2 - 2 \partial_i \alpha \right) \left(\beta_j + z \partial_z \beta_j \right)\right.\\
		& \left. - \frac{1}{2} z^2 \left(\beta^2 - 2 \alpha \right) \left(\partial_i \beta_j + z \partial_i \partial_z \beta_j \right) - \frac{1}{2} z \left(\partial_i \beta^k \right) \left(z \partial_j \beta_k + \partial_u \gamma_{jk} - z \partial_k \beta_j \right)\right.\\
		& \left. - \frac{1}{2} z \beta^k \left(z \partial_i \partial_j \beta_k + \partial_i \partial_u \gamma_{jk} - z \partial_i \partial_k \beta_j \right) \right] F_{zu}\\
		= & 0\,.
	\end{split}
\end{equation}
Therefore, the fifth term of Eq. (\ref{hkk2third}) is obtained as 
\begin{equation}
	\begin{split}
		- 2 k^a k^b g^{ce} g^{df} F_{ae} \left(\partial_d \Gamma^{g}_{\ fb} \right) F_{gc} = 0\,.
	\end{split}
\end{equation}

The sixth term of Eq. (\ref{hkk2third}) is
\begin{equation}
	\begin{split}
		& 2 k^a k^b g^{ce} g^{df} F_{ae} \Gamma^{h}_{\ df} \Gamma^{g}_{\ hb} F_{gc} = 2 g^{ce} g^{df} F_{ue} \Gamma^{h}_{\ df} \Gamma^{g}_{\ hu} F_{gc}\\
		= & 2 F_{uz} \Gamma^{h}_{\ uz} \Gamma^{g}_{\ hu} F_{gu} + 2 F_{uz} \Gamma^{h}_{\ zu} \Gamma^{g}_{\ hu} F_{gu} + 2 \gamma^{ij} F_{uz} \Gamma^{h}_{\ ij} \Gamma^{g}_{\ hu} F_{gu}\\
		= & 4 F_{uz} \Gamma^{h}_{\ uz} \Gamma^{g}_{\ hu} F_{gu} + 2 \gamma^{ij} F_{uz} \Gamma^{h}_{\ ij} \Gamma^{g}_{\ hu} F_{gu}\,.
	\end{split}
\end{equation}
The index $g$ should be further expanded.
\begin{equation}\label{hkk2thirdsixth}
	\begin{split}
		& 4 F_{uz} \Gamma^{h}_{\ uz} \Gamma^{g}_{\ hu} F_{gu} + 2 \gamma^{ij} F_{uz} \Gamma^{h}_{\ ij} \Gamma^{g}_{\ hu} F_{gu}\\
		= & 4 F_{uz} \Gamma^{h}_{\ uz} \Gamma^{u}_{\ hu} F_{uu} + 4 F_{uz} \Gamma^{h}_{\ uz} \Gamma^{z}_{\ hu} F_{zu} + 4 F_{uz} \Gamma^{h}_{\ uz} \Gamma^{i}_{\ hu} F_{iu}\\
		& + 2 \gamma^{ij} F_{uz} \Gamma^{h}_{\ ij} \Gamma^{u}_{\ hu} F_{uu} + 2 \gamma^{ij} F_{uz} \Gamma^{h}_{\ ij} \Gamma^{z}_{\ hu} F_{zu} + 2 \gamma^{ij} F_{uz} \Gamma^{h}_{\ ij} \Gamma^{k}_{\ hu} F_{ku}\\
		= & 4 F_{uz} \Gamma^{h}_{\ uz} \Gamma^{z}_{\ hu} F_{zu} + 2 \gamma^{ij} F_{uz} \Gamma^{h}_{\ ij} \Gamma^{z}_{\ hu} F_{zu}\,.
	\end{split}
\end{equation}
The repeated index $h$ should be further expanded. The first term of Eq. (\ref{hkk2thirdsixth}) is
\begin{equation}
	\begin{split}
		& 4 F_{uz} \Gamma^{h}_{\ uz} \Gamma^{z}_{\ hu} F_{zu}\\
		= & 4 F_{uz} \Gamma^{u}_{\ uz} \Gamma^{z}_{\ uu} F_{zu} + 4 F_{uz} \Gamma^{z}_{\ uz} \Gamma^{z}_{\ zu} F_{zu} + 4 F_{uz} \Gamma^{i}_{\ uz} \Gamma^{z}_{\ iu} F_{zu}\\
		= & 0\,.
	\end{split}
\end{equation}
The second term of Eq. (\ref{hkk2thirdsixth}) is
\begin{equation}
	\begin{split}
		& 2 \gamma^{ij} F_{uz} \Gamma^{h}_{\ ij} \Gamma^{z}_{\ hu} F_{zu}\\
		= & 2 \gamma^{ij} F_{uz} \Gamma^{u}_{\ ij} \Gamma^{z}_{\ uu} F_{zu} + 2 \gamma^{ij} F_{uz} \Gamma^{z}_{\ ij} \Gamma^{z}_{\ zu} F_{zu} + 2 \gamma^{ij} F_{uz} \Gamma^{k}_{\ ij} \Gamma^{z}_{\ ku} F_{zu}\\
		= & 0\,.
	\end{split}
\end{equation}
Therefore, the sixth term of Eq. (\ref{hkk2third}) is obtained as 
\begin{equation}
	\begin{split}
		2 k^a k^b g^{ce} g^{df} F_{ae} \Gamma^{h}_{\ df} \Gamma^{g}_{\ hb} F_{gc} = 0\,.
	\end{split}
\end{equation}

The seventh term of Eq. (\ref{hkk2third}) is
\begin{equation}
	\begin{split}
		& 2 k^a k^b g^{ce} g^{df} F_{ae} \Gamma^{h}_{\ db} \Gamma^{g}_{\ fh} F_{gc} = 2 g^{ce} g^{df} F_{ue} \Gamma^{h}_{\ du} \Gamma^{g}_{\ fh} F_{gc}\\
		= & 2 F_{uz} \Gamma^{h}_{\ uu} \Gamma^{g}_{\ zh} F_{gu} + 2 F_{uz} \Gamma^{h}_{\ zu} \Gamma^{g}_{\ uh} F_{gu} + 2 \gamma^{ij} F_{uz} \Gamma^{h}_{\ iu} \Gamma^{g}_{\ jh} F_{gu}\\
		= & 2 F_{uz} \Gamma^{h}_{\ zu} \Gamma^{g}_{\ uh} F_{gu} + 2 \gamma^{ij} F_{uz} \Gamma^{h}_{\ iu} \Gamma^{g}_{\ jh} F_{gu}\,.
	\end{split}
\end{equation}
The index $g$ should be further expanded.
\begin{equation}\label{hkk2thirdseventh}
	\begin{split}
		& 2 F_{uz} \Gamma^{h}_{\ zu} \Gamma^{g}_{\ uh} F_{gu} + 2 \gamma^{ij} F_{uz} \Gamma^{h}_{\ iu} \Gamma^{g}_{\ jh} F_{gu}\\
		= & 2 F_{uz} \Gamma^{h}_{\ zu} \Gamma^{u}_{\ uh} F_{uu} + 2 F_{uz} \Gamma^{h}_{\ zu} \Gamma^{z}_{\ uh} F_{zu} + 2 F_{uz} \Gamma^{h}_{\ zu} \Gamma^{i}_{\ uh} F_{iu}\\
		& + 2 \gamma^{ij} F_{uz} \Gamma^{h}_{\ iu} \Gamma^{u}_{\ jh} F_{uu} + 2 \gamma^{ij} F_{uz} \Gamma^{h}_{\ iu} \Gamma^{z}_{\ jh} F_{zu} + 2 \gamma^{ij} F_{uz} \Gamma^{h}_{\ iu} \Gamma^{k}_{\ jh} F_{ku}\\
		= & 2 F_{uz} \Gamma^{h}_{\ zu} \Gamma^{z}_{\ uh} F_{zu} + 2 \gamma^{ij} F_{uz} \Gamma^{h}_{\ iu} \Gamma^{z}_{\ jh} F_{zu}\,.
	\end{split}
\end{equation}
The repeated index $h$ should be further expanded. The first term of Eq. (\ref{hkk2thirdseventh}) is 
\begin{equation}
	\begin{split}
		& 2 F_{uz} \Gamma^{h}_{\ zu} \Gamma^{z}_{\ uh} F_{zu}\\
		= & 2 F_{uz} \Gamma^{u}_{\ zu} \Gamma^{z}_{\ uu} F_{zu} + 2 F_{uz} \Gamma^{z}_{\ zu} \Gamma^{z}_{\ uz} F_{zu} + 2 F_{uz} \Gamma^{i}_{\ zu} \Gamma^{z}_{\ ui} F_{zu}\\
		= & 0\,.
	\end{split}
\end{equation}
The second term of Eq. (\ref{hkk2thirdseventh}) is 
\begin{equation}
	\begin{split}
		& 2 \gamma^{ij} F_{uz} \Gamma^{h}_{\ iu} \Gamma^{z}_{\ jh} F_{zu}\\
		= & 2 \gamma^{ij} F_{uz} \Gamma^{u}_{\ iu} \Gamma^{z}_{\ ju} F_{zu} + 2 \gamma^{ij} F_{uz} \Gamma^{z}_{\ iu} \Gamma^{z}_{\ jz} F_{zu} + 2 \gamma^{ij} F_{uz} \Gamma^{k}_{\ iu} \Gamma^{z}_{\ jk} F_{zu}\\
		= & - \frac{1}{2} \gamma^{ij} \gamma^{kl} F_{uz} \left(\partial_u \gamma_{il} \right) \left(\partial_u \gamma_{jk} \right) F_{zu}\,.
	\end{split}
\end{equation}
Therefore, the seventh term of Eq. (\ref{hkk2third}) is obtained as 
\begin{equation}
	\begin{split}
		& 2 k^a k^b g^{ce} g^{df} F_{ae} \Gamma^{h}_{\ db} \Gamma^{g}_{\ fh} F_{gc} = - \frac{1}{2} \gamma^{ij} \gamma^{kl} F_{uz} \left(\partial_u \gamma_{il} \right) \left(\partial_u \gamma_{jk} \right) F_{zu}\\
		= & - 2 \gamma^{ij} \gamma^{kl} F_{uz} K_{il} K_{jk} F_{zu}\\
		= & 0\,.
	\end{split}
\end{equation}

The eighth term of Eq. (\ref{hkk2third}) is
\begin{equation}
	\begin{split}
		& - 2 k^a k^b g^{ce} g^{df} F_{ae} \Gamma^{g}_{\ dh} \Gamma^{h}_{\ fb} F_{gc} = - 2 g^{ce} g^{df} F_{ue} \Gamma^{g}_{\ dh} \Gamma^{h}_{\ fu} F_{gc}\\
		= & - 2 F_{uz} \Gamma^{g}_{\ uh} \Gamma^{h}_{\ zu} F_{gu} - 2 F_{uz} \Gamma^{g}_{\ zh} \Gamma^{h}_{\ uu} F_{gu} - 2 \gamma^{ij} F_{uz} \Gamma^{g}_{\ ih} \Gamma^{h}_{\ ju} F_{gu}\\
		= & - 2 F_{uz} \Gamma^{g}_{\ uh} \Gamma^{h}_{\ zu} F_{gu} - 2 \gamma^{ij} F_{uz} \Gamma^{g}_{\ ih} \Gamma^{h}_{\ ju} F_{gu}\,.
	\end{split}
\end{equation}
The index $g$ should be further expanded.
\begin{equation}\label{hkk2thirdeighth}
	\begin{split}
		& - 2 F_{uz} \Gamma^{g}_{\ uh} \Gamma^{h}_{\ zu} F_{gu} - 2 \gamma^{ij} F_{uz} \Gamma^{g}_{\ ih} \Gamma^{h}_{\ ju} F_{gu}\\
		= & - 2 F_{uz} \Gamma^{u}_{\ uh} \Gamma^{h}_{\ zu} F_{uu} - 2 F_{uz} \Gamma^{z}_{\ uh} \Gamma^{h}_{\ zu} F_{zu} - 2 F_{uz} \Gamma^{i}_{\ uh} \Gamma^{h}_{\ zu} F_{iu}\\
		& - 2 \gamma^{ij} F_{uz} \Gamma^{u}_{\ ih} \Gamma^{h}_{\ ju} F_{uu} - 2 \gamma^{ij} F_{uz} \Gamma^{z}_{\ ih} \Gamma^{h}_{\ ju} F_{zu} - 2 \gamma^{ij} F_{uz} \Gamma^{k}_{\ ih} \Gamma^{h}_{\ ju} F_{ku}\\
		= & - 2 F_{uz} \Gamma^{z}_{\ uh} \Gamma^{h}_{\ zu} F_{zu} - 2 \gamma^{ij} F_{uz} \Gamma^{z}_{\ ih} \Gamma^{h}_{\ ju} F_{zu}\,.
	\end{split}
\end{equation}
The first term of Eq. (\ref{hkk2thirdeighth}) is 
\begin{equation}
	\begin{split}
		& - 2 F_{uz} \Gamma^{z}_{\ uh} \Gamma^{h}_{\ zu} F_{zu}\\
		= & - 2 F_{uz} \Gamma^{z}_{\ uu} \Gamma^{u}_{\ zu} F_{zu} - 2 F_{uz} \Gamma^{z}_{\ uz} \Gamma^{z}_{\ zu} F_{zu} - 2 F_{uz} \Gamma^{z}_{\ ui} \Gamma^{i}_{\ zu} F_{zu}\\
		= & 0\,.
	\end{split}
\end{equation}
The second term of Eq. (\ref{hkk2thirdeighth}) is 
\begin{equation}
	\begin{split}
		& - 2 \gamma^{ij} F_{uz} \Gamma^{z}_{\ ih} \Gamma^{h}_{\ ju} F_{zu}\\
		= & - 2 \gamma^{ij} F_{uz} \Gamma^{z}_{\ iu} \Gamma^{u}_{\ ju} F_{zu} - 2 \gamma^{ij} F_{uz} \Gamma^{z}_{\ iz} \Gamma^{z}_{\ ju} F_{zu} - 2 \gamma^{ij} F_{uz} \Gamma^{z}_{\ ik} \Gamma^{k}_{\ ju} F_{zu}\\
		= & \frac{1}{2} \gamma^{ij} \gamma^{km} F_{uz} \left(\partial_u \gamma_{ik} \right) \left(\partial_u \gamma_{jm} \right) F_{zu}\,.
	\end{split}
\end{equation}
Therefore, the eighth term of Eq. (\ref{hkk2third}) is obtained as 
\begin{equation}
	\begin{split}
		& - 2 k^a k^b g^{ce} g^{df} F_{ae} \Gamma^{g}_{\ dh} \Gamma^{h}_{\ fb} F_{gc} = \frac{1}{2} \gamma^{ij} \gamma^{km} F_{uz} \left(\partial_u \gamma_{ik} \right) \left(\partial_u \gamma_{jm} \right) F_{zu}\\
		= & 2 \gamma^{ij} \gamma^{km} F_{uz} K_{ik} K_{jm} F_{zu}\\
		= & 0\,.
	\end{split}
\end{equation}

The ninth term of Eq. (\ref{hkk2third}) is
\begin{equation}
	\begin{split}
		& - 2 k^a k^b g^{ce} g^{df} F_{ae} \Gamma^{g}_{\ fb} \left(\partial_d F_{gc} \right) = - 2 g^{ce} g^{df} F_{ue} \Gamma^{g}_{\ fu} \left(\partial_d F_{gc} \right)\\
		= & - 2 F_{uz} \Gamma^{g}_{\ zu} \left(\partial_u F_{gu} \right) - 2 F_{uz} \Gamma^{g}_{\ uu} \left(\partial_z F_{gu} \right) - 2 \gamma^{ij} F_{uz} \Gamma^{g}_{\ ju} \left(\partial_i F_{gu} \right)\\
		= & - 2 F_{uz} \Gamma^{g}_{\ zu} \left(\partial_u F_{gu} \right) - 2 \gamma^{ij} F_{uz} \Gamma^{g}_{\ ju} \left(\partial_i F_{gu} \right)\,.
	\end{split}
\end{equation}
The index $g$ should be further expanded.
\begin{equation}\label{hkk2thirdninth}
	\begin{split}
		& - 2 F_{uz} \Gamma^{g}_{\ zu} \left(\partial_u F_{gu} \right) - 2 \gamma^{ij} F_{uz} \Gamma^{g}_{\ ju} \left(\partial_i F_{gu} \right)\\\
		= & - 2 F_{uz} \Gamma^{u}_{\ zu} \left(\partial_u F_{uu} \right) - 2 F_{uz} \Gamma^{z}_{\ zu} \left(\partial_u F_{zu} \right) - 2 F_{uz} \Gamma^{i}_{\ zu} \left(\partial_u F_{iu} \right)\\
		& - 2 \gamma^{ij} F_{uz} \Gamma^{u}_{\ ju} \left(\partial_i F_{uu} \right) - 2 \gamma^{ij} F_{uz} \Gamma^{z}_{\ ju} \left(\partial_i F_{zu} \right) - 2 \gamma^{ij} F_{uz} \Gamma^{k}_{\ ju} \left(\partial_i F_{ku} \right)\\
		= & - 2 F_{uz} \Gamma^{z}_{\ zu} \left(\partial_u F_{zu} \right) - 2 F_{uz} \Gamma^{i}_{\ zu} \left(\partial_u F_{iu} \right) - 2 \gamma^{ij} F_{uz} \Gamma^{z}_{\ ju} \left(\partial_i F_{zu} \right)\,.
	\end{split}
\end{equation}
Therefore, the ninth term of Eq. (\ref{hkk2third}) is obtained as 
\begin{equation}
	\begin{split}
		- 2 k^a k^b g^{ce} g^{df} F_{ae} \Gamma^{g}_{\ fb} \left(\partial_d F_{gc} \right) = - \gamma^{ij} \beta_j F_{uz} \left(\partial_u F_{iu} \right)\,.
	\end{split}
\end{equation}

The tenth term of Eq. (\ref{hkk2third}) is
\begin{equation}
	\begin{split}
		& 2 k^a k^b g^{ce} g^{df} F_{ae} \Gamma^{g}_{\ fb} \Gamma^{h}_{\ dg} F_{hc} = 2 g^{ce} g^{df} F_{ue} \Gamma^{g}_{\ fu} \Gamma^{h}_{\ dg} F_{hc}\\
		= & 2 F_{uz} \Gamma^{g}_{\ zu} \Gamma^{h}_{\ ug} F_{hu} + 2 F_{uz} \Gamma^{g}_{\ uu} \Gamma^{h}_{\ zg} F_{hu} + 2 \gamma^{ij} F_{uz} \Gamma^{g}_{\ ju} \Gamma^{h}_{\ ig} F_{hu}\\
		= & 2 F_{uz} \Gamma^{g}_{\ zu} \Gamma^{h}_{\ ug} F_{hu} + 2 \gamma^{ij} F_{uz} \Gamma^{g}_{\ ju} \Gamma^{h}_{\ ig} F_{hu}\,.
	\end{split}
\end{equation}
The index $g$ should be further expanded.
\begin{equation}\label{hkk2thirdtenth}
	\begin{split}
		& 2 F_{uz} \Gamma^{g}_{\ zu} \Gamma^{h}_{\ ug} F_{hu} + 2 \gamma^{ij} F_{uz} \Gamma^{g}_{\ ju} \Gamma^{h}_{\ ig} F_{hu}\\
		= & 2 F_{uz} \Gamma^{u}_{\ zu} \Gamma^{h}_{\ uu} F_{hu} + 2 F_{uz} \Gamma^{z}_{\ zu} \Gamma^{h}_{\ uz} F_{hu} + 2 F_{uz} \Gamma^{i}_{\ zu} \Gamma^{h}_{\ ui} F_{hu}\\
		& + 2 \gamma^{ij} F_{uz} \Gamma^{u}_{\ ju} \Gamma^{h}_{\ iu} F_{hu} + 2 \gamma^{ij} F_{uz} \Gamma^{z}_{\ ju} \Gamma^{h}_{\ iz} F_{hu} + 2 \gamma^{ij} F_{uz} \Gamma^{k}_{\ ju} \Gamma^{h}_{\ ik} F_{hu}\\
		= & 2 F_{uz} \Gamma^{z}_{\ zu} \Gamma^{h}_{\ uz} F_{hu} + 2 F_{uz} \Gamma^{i}_{\ zu} \Gamma^{h}_{\ ui} F_{hu} + 2 \gamma^{ij} F_{uz} \Gamma^{u}_{\ ju} \Gamma^{h}_{\ iu} F_{hu}\\
		& + 2 \gamma^{ij} F_{uz} \Gamma^{z}_{\ ju} \Gamma^{h}_{\ iz} F_{hu} + 2 \gamma^{ij} F_{uz} \Gamma^{k}_{\ ju} \Gamma^{h}_{\ ik} F_{hu}\,.
	\end{split}
\end{equation}
The repeated index $h$ should be further expanded. The first term of Eq. (\ref{hkk2thirdtenth}) is
\begin{equation}
	\begin{split}
		2 F_{uz} \Gamma^{z}_{\ zu} \Gamma^{h}_{\ uz} F_{hu} = 0\,.
	\end{split}
\end{equation}
The second term of Eq. (\ref{hkk2thirdtenth}) is
\begin{equation}
	\begin{split}
		& 2 F_{uz} \Gamma^{i}_{\ zu} \Gamma^{h}_{\ ui} F_{hu}\\
		= & 2 F_{uz} \Gamma^{i}_{\ zu} \Gamma^{u}_{\ ui} F_{uu} + 2 F_{uz} \Gamma^{i}_{\ zu} \Gamma^{z}_{\ ui} F_{zu} + 2 F_{uz} \Gamma^{i}_{\ zu} \Gamma^{j}_{\ ui} F_{ju}\\
		= & 2 F_{uz} \Gamma^{i}_{\ zu} \Gamma^{z}_{\ ui} F_{zu}\\
		= & 0\,.
	\end{split}
\end{equation}
The third term of Eq. (\ref{hkk2thirdtenth}) is
\begin{equation}
	\begin{split}
		& 2 \gamma^{ij} F_{uz} \Gamma^{u}_{\ ju} \Gamma^{h}_{\ iu} F_{hu}\\
		= & 2 \gamma^{ij} F_{uz} \Gamma^{u}_{\ ju} \Gamma^{u}_{\ iu} F_{uu} + 2 \gamma^{ij} F_{uz} \Gamma^{u}_{\ ju} \Gamma^{z}_{\ iu} F_{zu} + 2 \gamma^{ij} F_{uz} \Gamma^{u}_{\ ju} \Gamma^{k}_{\ iu} F_{ku}\\
		= & 2 \gamma^{ij} F_{uz} \Gamma^{u}_{\ ju} \Gamma^{z}_{\ iu} F_{zu}\\
		= & 0\,.
	\end{split}
\end{equation}
The fourth term of Eq. (\ref{hkk2thirdtenth}) is
\begin{equation}
	\begin{split}
		2 \gamma^{ij} F_{uz} \Gamma^{z}_{\ ju} \Gamma^{h}_{\ iz} F_{hu} = 0\,.
	\end{split}
\end{equation}
The fifth term of Eq. (\ref{hkk2thirdtenth}) is
\begin{equation}
	\begin{split}
		& 2 \gamma^{ij} F_{uz} \Gamma^{k}_{\ ju} \Gamma^{h}_{\ ik} F_{hu}\\
		= & 2 \gamma^{ij} F_{uz} \Gamma^{k}_{\ ju} \Gamma^{u}_{\ ik} F_{uu} + 2 \gamma^{ij} F_{uz} \Gamma^{k}_{\ ju} \Gamma^{z}_{\ ik} F_{zu} + 2 \gamma^{ij} F_{uz} \Gamma^{k}_{\ ju} \Gamma^{l}_{\ ik} F_{lu}\\
		= & 2 \gamma^{ij} F_{uz} \Gamma^{k}_{\ ju} \Gamma^{z}_{\ ik} F_{zu}\\
		= & - \frac{1}{2} \gamma^{ij} \gamma^{kl} F_{uz} \left(\partial_u \gamma_{jl} \right) \left(\partial_u \gamma_{ik} \right) F_{zu}\,.
	\end{split}
\end{equation}
Therefore, the tenth term of Eq. (\ref{hkk2third}) is obtained as 
\begin{equation}
	\begin{split}
		& 2 k^a k^b g^{ce} g^{df} F_{ae} \Gamma^{g}_{\ fb} \Gamma^{h}_{\ dg} F_{hc} = - \frac{1}{2} \gamma^{ij} \gamma^{kl} F_{uz} \left(\partial_u \gamma_{jl} \right) \left(\partial_u \gamma_{ik} \right) F_{zu}\\
		= & - 2 \gamma^{ij} \gamma^{kl} F_{uz} K_{jl} K_{ik} F_{zu}\\
		= & 0\,.
	\end{split}
\end{equation}

The eleventh term of Eq. (\ref{hkk2third}) is
\begin{equation}
	\begin{split}
		& 2 k^a k^b g^{ce} g^{df} F_{ae} \Gamma^{g}_{\ fb} \Gamma^{h}_{\ dc} F_{gh} = 2 g^{ce} g^{df} F_{ue} \Gamma^{g}_{\ fu} \Gamma^{h}_{\ dc} F_{gh}\\
		= & 2 F_{uz} \Gamma^{g}_{\ zu} \Gamma^{h}_{\ uu} F_{gh} + 2 F_{uz} \Gamma^{g}_{\ uu} \Gamma^{h}_{\ zu} F_{gh} + 2 \gamma^{ij} F_{uz} \Gamma^{g}_{\ ju} \Gamma^{h}_{\ iu} F_{gh}\\
		= & 2 \gamma^{ij} F_{uz} \Gamma^{g}_{\ ju} \Gamma^{h}_{\ iu} F_{gh}\,.
	\end{split}
\end{equation}
The index $g$ should be further expanded.
\begin{equation}\label{hkk2thirdeleventh}
	\begin{split}
		& 2 \gamma^{ij} F_{uz} \Gamma^{g}_{\ ju} \Gamma^{h}_{\ iu} F_{gh}\\
		= & 2 \gamma^{ij} F_{uz} \Gamma^{u}_{\ ju} \Gamma^{h}_{\ iu} F_{uh} + 2 \gamma^{ij} F_{uz} \Gamma^{z}_{\ ju} \Gamma^{h}_{\ iu} F_{zh} + 2 \gamma^{ij} F_{uz} \Gamma^{k}_{\ ju} \Gamma^{h}_{\ iu} F_{kh}\,.
	\end{split}
\end{equation}
The first term of Eq. (\ref{hkk2thirdeleventh}) is
\begin{equation}
	\begin{split}
		& 2 \gamma^{ij} F_{uz} \Gamma^{u}_{\ ju} \Gamma^{h}_{\ iu} F_{uh}\\
		= & 2 \gamma^{ij} F_{uz} \Gamma^{u}_{\ ju} \Gamma^{u}_{\ iu} F_{uu} + 2 \gamma^{ij} F_{uz} \Gamma^{u}_{\ ju} \Gamma^{z}_{\ iu} F_{uz} + 2 \gamma^{ij} F_{uz} \Gamma^{u}_{\ ju} \Gamma^{k}_{\ iu} F_{uk}\\
		= & 2 \gamma^{ij} F_{uz} \Gamma^{u}_{\ ju} \Gamma^{z}_{\ iu} F_{uz}\\
		= & 0\,.
	\end{split}
\end{equation}
The second term of Eq. (\ref{hkk2thirdeleventh}) is
\begin{equation}
	\begin{split}
		2 \gamma^{ij} F_{uz} \Gamma^{z}_{\ ju} \Gamma^{h}_{\ iu} F_{zh} = 0\,.
	\end{split}
\end{equation}
The third term of Eq. (\ref{hkk2thirdeleventh}) is
\begin{equation}
	\begin{split}
		& 2 \gamma^{ij} F_{uz} \Gamma^{k}_{\ ju} \Gamma^{h}_{\ iu} F_{kh}\\
		= & 2 \gamma^{ij} F_{uz} \Gamma^{k}_{\ ju} \Gamma^{u}_{\ iu} F_{ku} + 2 \gamma^{ij} F_{uz} \Gamma^{k}_{\ ju} \Gamma^{z}_{\ iu} F_{kz} + 2 \gamma^{ij} F_{uz} \Gamma^{k}_{\ ju} \Gamma^{l}_{\ iu} F_{kl}\\
		= & 2 \gamma^{ij} F_{uz} \Gamma^{k}_{\ ju} \Gamma^{z}_{\ iu} F_{kz} + 2 \gamma^{ij} F_{uz} \Gamma^{k}_{\ ju} \Gamma^{l}_{\ iu} F_{kl}\\
		= & \frac{1}{2} \gamma^{ij} \gamma^{km} \gamma^{ln} F_{uz} \left(\partial_u \gamma_{jm} \right) \left(\partial_u \gamma_{in} \right) F_{kl}\,.
	\end{split}
\end{equation}
Therefore, the eleventh term of Eq. (\ref{hkk2third}) is obtained as 
\begin{equation}
	\begin{split}
		& 2 k^a k^b g^{ce} g^{df} F_{ae} \Gamma^{g}_{\ fb} \Gamma^{h}_{\ dc} F_{gh} = \frac{1}{2} \gamma^{ij} \gamma^{km} \gamma^{ln} F_{uz} \left(\partial_u \gamma_{jm} \right) \left(\partial_u \gamma_{in} \right) F_{kl}\\
		= & 2 \gamma^{ij} \gamma^{km} \gamma^{ln} F_{uz} K_{jm} K_{in} F_{kl}\\
		= & 0\,.
	\end{split}
\end{equation}

The twelfth term of Eq. (\ref{hkk2third}) is
\begin{equation}
	\begin{split}
		& - 2 k^a k^b g^{ce} g^{df} F_{ae} \left(\partial_d \Gamma^{g}_{\ fc} \right) F_{bg} = - 2 g^{ce} g^{df} F_{ue} \left(\partial_d \Gamma^{g}_{\ fc} \right) F_{ug}\\
		= & - 2 F_{uz} \left(\partial_u \Gamma^{g}_{\ zu} \right) F_{ug} - 2 F_{uz} \left(\partial_z \Gamma^{g}_{\ uu} \right) F_{ug} - 2 \gamma^{ij} F_{uz} \left(\partial_i \Gamma^{g}_{\ ju} \right) F_{ug}\,.
	\end{split}
\end{equation}
The index $g$ should be further expanded.
\begin{equation}\label{hkk2thirdtwelfth}
	\begin{split}
		& - 2 F_{uz} \left(\partial_u \Gamma^{g}_{\ zu} \right) F_{ug} - 2 F_{uz} \left(\partial_z \Gamma^{g}_{\ uu} \right) F_{ug} - 2 \gamma^{ij} F_{uz} \left(\partial_i \Gamma^{g}_{\ ju} \right) F_{ug}\\
		= & - 2 F_{uz} \left(\partial_u \Gamma^{u}_{\ zu} \right) F_{uu} - 2 F_{uz} \left(\partial_u \Gamma^{z}_{\ zu} \right) F_{uz} - 2 F_{uz} \left(\partial_u \Gamma^{i}_{\ zu} \right) F_{ui}\\
		& - 2 F_{uz} \left(\partial_z \Gamma^{u}_{\ uu} \right) F_{uu} - 2 F_{uz} \left(\partial_z \Gamma^{z}_{\ uu} \right) F_{uz} - 2 F_{uz} \left(\partial_z \Gamma^{i}_{\ uu} \right) F_{ui}\\
		& - 2 \gamma^{ij} F_{uz} \left(\partial_i \Gamma^{u}_{\ ju} \right) F_{uu} - 2 \gamma^{ij} F_{uz} \left(\partial_i \Gamma^{z}_{\ ju} \right) F_{uz} - 2 \gamma^{ij} F_{uz} \left(\partial_i \Gamma^{k}_{\ ju} \right) F_{uk}\\
		= & - 2 F_{uz} \left(\partial_u \Gamma^{z}_{\ zu} \right) F_{uz} - 2 F_{uz} \left(\partial_z \Gamma^{z}_{\ uu} \right) F_{uz} - 2 \gamma^{ij} F_{uz} \left(\partial_i \Gamma^{z}_{\ ju} \right) F_{uz}\,.
	\end{split}
\end{equation}
The first term of Eq. (\ref{hkk2thirdtwelfth}) is 
\begin{equation}
	\begin{split}
		& - 2 F_{uz} \left(\partial_u \Gamma^{z}_{\ zu} \right) F_{uz}\\
		= & - 2 F_{uz} \partial_u \left(z^2 \partial_z \alpha + 2 z \alpha - \frac{1}{2} z \beta^i \beta_i - \frac{1}{2} z^2 \beta^i \partial_z \beta_i \right) F_{uz}\\
		= & - 2 F_{uz} \left[z^2 \partial_u \partial_z \alpha + 2 z \partial_u \alpha - \frac{1}{2} z \partial_u \left(\beta^i \beta_i \right) - \frac{1}{2} z^2 \left(\partial_u \beta^i \right) \partial_z \beta_i - \frac{1}{2} z^2 \beta^i \partial_u \partial_z \beta_i \right] F_{uz}\\
		= & 0\,.
	\end{split}
\end{equation}
The second term of Eq. (\ref{hkk2thirdtwelfth}) is 
\begin{equation}
	\begin{split}
		& - 2 F_{uz} \left(\partial_z \Gamma^{z}_{\ uu} \right) F_{uz}\\
		= & - 2 F_{uz} \partial_z \left[z^2 \partial_u \alpha - z^2 \left(\beta^2 - 2 \alpha \right) \left(z^2 \partial_z \alpha + 2 z \alpha \right) - z \beta^i \left(z \partial_u \beta_i - z^2 \partial_i \alpha \right) \right] F_{uz}\\
		= & - 2 F_{uz} \left[2 z \partial_u \alpha + z^2 \partial_z \partial_u \alpha - 2 z \left(\beta^2 - 2 \alpha \right) \left(z^2 \partial_z \alpha + 2 z \alpha \right)\right.\\
		& \left. - z^2 \left(\partial_z \beta^2 - 2 \partial_z \alpha \right) \left(z^2 \partial_z \alpha + 2 z \alpha \right) - z^2 \left(\beta^2 - 2 \alpha \right) \left(2 z \partial_z \alpha + z^2 \partial_z^2 \alpha + 2 \alpha + 2 z \partial_z \alpha \right)\right.\\
		& \left. - \beta^i \left(z \partial_u \beta_i - z^2 \partial_i \alpha \right) - z \left(\partial_z \beta^i \right) \left(z \partial_u \beta_i - z^2 \partial_i \alpha \right)\right.\\
		& \left. - z \beta^i \left(\partial_u \beta_i + z \partial_z \partial_u \beta_i - 2 z \partial_i \alpha - z^2 \partial_z \partial_i \alpha \right) \right] F_{uz}\\
		= & 0\,.
	\end{split}
\end{equation}
The third term of Eq. (\ref{hkk2thirdtwelfth}) is 
\begin{equation}
	\begin{split}
		& - 2 \gamma^{ij} F_{uz} \left(\partial_i \Gamma^{z}_{\ ju} \right) F_{uz}\\
		= & - 2 \gamma^{ij} F_{uz} \partial_i \left[z^2 \partial_j \alpha - \frac{1}{2} z^2 \left(\beta^2 - 2 \alpha \right) \left(\beta_j + z \partial_z \beta_j \right) - \frac{1}{2} z \beta^k \left(z \partial_j \beta_k + \partial_u \gamma_{jk} - z \partial_k \beta_j \right) \right]\\
		& \times F_{uz}\\
		= & - 2 \gamma^{ij} F_{uz} \left[z^2 \partial_i \partial_j \alpha - \frac{1}{2} z^2 \left(\partial_i \beta^2 - 2 \partial_i \alpha \right) \left(\beta_j + z \partial_z \beta_j \right)\right.\\
		& \left. - \frac{1}{2} z^2 \left(\beta^2 - 2 \alpha \right) \left(\partial_i \beta_j + z \partial_i \partial_z \beta_j \right) - \frac{1}{2} z \left(\partial_i \beta^k \right) \left(z \partial_j \beta_k + \partial_u \gamma_{jk} - z \partial_k \beta_j \right)\right.\\
		& \left. - \frac{1}{2} z \beta^k \left(z \partial_i \partial_j \beta_k + \partial_i \partial_u \gamma_{jk} - z \partial_i \partial_k \beta_j \right) \right] F_{uz}\\
		= & 0\,.
	\end{split}
\end{equation}
Therefore, the twelfth term of Eq. (\ref{hkk2third}) is obtained as 
\begin{equation}
	\begin{split}
		- 2 k^a k^b g^{ce} g^{df} F_{ae} \left(\partial_d \Gamma^{g}_{\ fc} \right) F_{bg} = 0\,.
	\end{split}
\end{equation}

The thirteenth term of Eq. (\ref{hkk2third}) is
\begin{equation}
	\begin{split}
		& 2 k^a k^b g^{ce} g^{df} F_{ae} \Gamma^{h}_{\ df} \Gamma^{g}_{\ hc} F_{bg} = 2 g^{ce} g^{df} F_{ue} \Gamma^{h}_{\ df} \Gamma^{g}_{\ hc} F_{ug}\\
		= & 2 F_{uz} \Gamma^{h}_{\ uz} \Gamma^{g}_{\ hu} F_{ug} + 2 F_{uz} \Gamma^{h}_{\ zu} \Gamma^{g}_{\ hu} F_{ug} + 2 \gamma^{ij} F_{uz} \Gamma^{h}_{\ ij} \Gamma^{g}_{\ hu} F_{ug}\\
		= & 4 F_{uz} \Gamma^{h}_{\ uz} \Gamma^{g}_{\ hu} F_{ug} + 2 \gamma^{ij} F_{uz} \Gamma^{h}_{\ ij} \Gamma^{g}_{\ hu} F_{ug}\,.
	\end{split}
\end{equation}
The index $g$ should be further expanded.
\begin{equation}\label{hkk2thirdthirteenth}
	\begin{split}
		& 4 F_{uz} \Gamma^{h}_{\ uz} \Gamma^{g}_{\ hu} F_{ug} + 2 \gamma^{ij} F_{uz} \Gamma^{h}_{\ ij} \Gamma^{g}_{\ hu} F_{ug}\\
		= & 4 F_{uz} \Gamma^{h}_{\ uz} \Gamma^{u}_{\ hu} F_{uu} + 4 F_{uz} \Gamma^{h}_{\ uz} \Gamma^{z}_{\ hu} F_{uz} + 4 F_{uz} \Gamma^{h}_{\ uz} \Gamma^{i}_{\ hu} F_{ui}\\
		& + 2 \gamma^{ij} F_{uz} \Gamma^{h}_{\ ij} \Gamma^{u}_{\ hu} F_{uu} + 2 \gamma^{ij} F_{uz} \Gamma^{h}_{\ ij} \Gamma^{z}_{\ hu} F_{uz} + 2 \gamma^{ij} F_{uz} \Gamma^{h}_{\ ij} \Gamma^{k}_{\ hu} F_{uk}\\
		= & 4 F_{uz} \Gamma^{h}_{\ uz} \Gamma^{z}_{\ hu} F_{uz} + 2 \gamma^{ij} F_{uz} \Gamma^{h}_{\ ij} \Gamma^{z}_{\ hu} F_{uz}\,.
	\end{split}
\end{equation}
The repeated index $h$ should be further expanded. The first term of Eq. (\ref{hkk2thirdthirteenth}) is 
\begin{equation}
	\begin{split}
		& 4 F_{uz} \Gamma^{h}_{\ uz} \Gamma^{z}_{\ hu} F_{uz}\\
		= & 4 F_{uz} \Gamma^{u}_{\ uz} \Gamma^{z}_{\ uu} F_{uz} + 4 F_{uz} \Gamma^{z}_{\ uz} \Gamma^{z}_{\ zu} F_{uz} + 4 F_{uz} \Gamma^{i}_{\ uz} \Gamma^{z}_{\ iu} F_{uz}\\
		= & 0\,.
	\end{split}
\end{equation}
The second term of Eq. (\ref{hkk2thirdthirteenth}) is
\begin{equation}
	\begin{split}
		& 2 \gamma^{ij} F_{uz} \Gamma^{h}_{\ ij} \Gamma^{z}_{\ hu} F_{uz}\\
		= & 2 \gamma^{ij} F_{uz} \Gamma^{u}_{\ ij} \Gamma^{z}_{\ uu} F_{uz} + 2 \gamma^{ij} F_{uz} \Gamma^{z}_{\ ij} \Gamma^{z}_{\ zu} F_{uz} + 2 \gamma^{ij} F_{uz} \Gamma^{k}_{\ ij} \Gamma^{z}_{\ ku} F_{uz}\\
		= & 0\,.
	\end{split}
\end{equation}
Therefore, the thirteenth term of Eq. (\ref{hkk2third}) is obtained as 
\begin{equation}
	\begin{split}
		2 k^a k^b g^{ce} g^{df} F_{ae} \Gamma^{h}_{\ df} \Gamma^{g}_{\ hc} F_{bg} = 0\,.
	\end{split}
\end{equation}

The fourteenth term of Eq. (\ref{hkk2third}) is
\begin{equation}
	\begin{split}
		& 2 k^a k^b g^{ce} g^{df} F_{ae} \Gamma^{h}_{\ dc} \Gamma^{g}_{\ fh} F_{bg} = 2 g^{ce} g^{df} F_{ue} \Gamma^{h}_{\ dc} \Gamma^{g}_{\ fh} F_{ug}\\
		= & 2 F_{uz} \Gamma^{h}_{\ uu} \Gamma^{g}_{\ zh} F_{ug} + 2 F_{uz} \Gamma^{h}_{\ zu} \Gamma^{g}_{\ uh} F_{ug} + 2 \gamma^{ij} F_{uz} \Gamma^{h}_{\ iu} \Gamma^{g}_{\ jh} F_{ug}\\
		= & 2 F_{uz} \Gamma^{h}_{\ zu} \Gamma^{g}_{\ uh} F_{ug} + 2 \gamma^{ij} F_{uz} \Gamma^{h}_{\ iu} \Gamma^{g}_{\ jh} F_{ug}\,.
	\end{split}
\end{equation}
The index $g$ should be further expanded.
\begin{equation}\label{hkk2thirdfourteenth}
	\begin{split}
		& 2 F_{uz} \Gamma^{h}_{\ zu} \Gamma^{g}_{\ uh} F_{ug} + 2 \gamma^{ij} F_{uz} \Gamma^{h}_{\ iu} \Gamma^{g}_{\ jh} F_{ug}\\
		= & 2 F_{uz} \Gamma^{h}_{\ zu} \Gamma^{u}_{\ uh} F_{uu} + 2 F_{uz} \Gamma^{h}_{\ zu} \Gamma^{z}_{\ uh} F_{uz} + 2 F_{uz} \Gamma^{h}_{\ zu} \Gamma^{i}_{\ uh} F_{ui}\\
		& + 2 \gamma^{ij} F_{uz} \Gamma^{h}_{\ iu} \Gamma^{u}_{\ jh} F_{uu} + 2 \gamma^{ij} F_{uz} \Gamma^{h}_{\ iu} \Gamma^{z}_{\ jh} F_{uz} + 2 \gamma^{ij} F_{uz} \Gamma^{h}_{\ iu} \Gamma^{k}_{\ jh} F_{uk}\\
		= & 2 F_{uz} \Gamma^{h}_{\ zu} \Gamma^{z}_{\ uh} F_{uz} + 2 \gamma^{ij} F_{uz} \Gamma^{h}_{\ iu} \Gamma^{z}_{\ jh} F_{uz}\,.
	\end{split}
\end{equation}
The first term of Eq. (\ref{hkk2thirdfourteenth}) is
\begin{equation}
	\begin{split}
		& 2 F_{uz} \Gamma^{h}_{\ zu} \Gamma^{z}_{\ uh} F_{uz}\\
		= & 2 F_{uz} \Gamma^{u}_{\ zu} \Gamma^{z}_{\ uu} F_{uz} + 2 F_{uz} \Gamma^{z}_{\ zu} \Gamma^{z}_{\ uz} F_{uz} + 2 F_{uz} \Gamma^{i}_{\ zu} \Gamma^{z}_{\ ui} F_{uz}\\
		= & 0\,.
	\end{split}
\end{equation}
The second term of Eq. (\ref{hkk2thirdfourteenth}) is
\begin{equation}
	\begin{split}
		& 2 \gamma^{ij} F_{uz} \Gamma^{h}_{\ iu} \Gamma^{z}_{\ jh} F_{uz}\\
		= & 2 \gamma^{ij} F_{uz} \Gamma^{u}_{\ iu} \Gamma^{z}_{\ ju} F_{uz} + 2 \gamma^{ij} F_{uz} \Gamma^{z}_{\ iu} \Gamma^{z}_{\ jz} F_{uz} + 2 \gamma^{ij} F_{uz} \Gamma^{k}_{\ iu} \Gamma^{z}_{\ jk} F_{uz}\\
		= & - \frac{1}{2} \gamma^{ij} \gamma^{kl} F_{uz} \left(\partial_u \gamma_{il} \right) \left(\partial_u \gamma_{jk} \right) F_{uz}\,.
	\end{split}
\end{equation}
Therefore, the fourteenth term of Eq. (\ref{hkk2third}) is obtained as 
\begin{equation}
	\begin{split}
		& 2 k^a k^b g^{ce} g^{df} F_{ae} \Gamma^{h}_{\ dc} \Gamma^{g}_{\ fh} F_{bg} = - \frac{1}{2} \gamma^{ij} \gamma^{kl} F_{uz} \left(\partial_u \gamma_{il} \right) \left(\partial_u \gamma_{jk} \right) F_{uz}\\
		= & - 2 \gamma^{ij} \gamma^{kl} F_{uz} K_{il} K_{jk} F_{uz}\\
		= & 0\,.
	\end{split}
\end{equation}

The fifteenth term of Eq. (\ref{hkk2third}) is
\begin{equation}
	\begin{split}
		& - 2 k^a k^b g^{ce} g^{df} F_{ae} \Gamma^{g}_{\ dh} \Gamma^{h}_{\ fc} F_{bg} = - 2 g^{ce} g^{df} F_{ue} \Gamma^{g}_{\ dh} \Gamma^{h}_{\ fc} F_{ug}\\
		= & - 2 F_{uz} \Gamma^{g}_{\ uh} \Gamma^{h}_{\ zu} F_{ug} - 2 F_{uz} \Gamma^{g}_{\ zh} \Gamma^{h}_{\ uu} F_{ug} - 2 \gamma^{ij} F_{uz} \Gamma^{g}_{\ ih} \Gamma^{h}_{\ ju} F_{ug}\\
		= & - 2 F_{uz} \Gamma^{g}_{\ uh} \Gamma^{h}_{\ zu} F_{ug} - 2 \gamma^{ij} F_{uz} \Gamma^{g}_{\ ih} \Gamma^{h}_{\ ju} F_{ug}\,.
	\end{split}
\end{equation}
The index $g$ should be further expanded.
\begin{equation}\label{hkk2thirdfifteenth}
	\begin{split}
		& - 2 F_{uz} \Gamma^{g}_{\ uh} \Gamma^{h}_{\ zu} F_{ug} - 2 \gamma^{ij} F_{uz} \Gamma^{g}_{\ ih} \Gamma^{h}_{\ ju} F_{ug}\\
		= & - 2 F_{uz} \Gamma^{u}_{\ uh} \Gamma^{h}_{\ zu} F_{uu} - 2 F_{uz} \Gamma^{z}_{\ uh} \Gamma^{h}_{\ zu} F_{uz} - 2 F_{uz} \Gamma^{i}_{\ uh} \Gamma^{h}_{\ zu} F_{ui}\\
		& - 2 \gamma^{ij} F_{uz} \Gamma^{u}_{\ ih} \Gamma^{h}_{\ ju} F_{uu} - 2 \gamma^{ij} F_{uz} \Gamma^{z}_{\ ih} \Gamma^{h}_{\ ju} F_{uz} - 2 \gamma^{ij} F_{uz} \Gamma^{k}_{\ ih} \Gamma^{h}_{\ ju} F_{uk}\\
		= & - 2 F_{uz} \Gamma^{z}_{\ uh} \Gamma^{h}_{\ zu} F_{uz} - 2 \gamma^{ij} F_{uz} \Gamma^{z}_{\ ih} \Gamma^{h}_{\ ju} F_{uz}\,. 
	\end{split}
\end{equation}
The first term of Eq. (\ref{hkk2thirdfifteenth}) is 
\begin{equation}
	\begin{split}
		& - 2 F_{uz} \Gamma^{z}_{\ uh} \Gamma^{h}_{\ zu} F_{uz}\\
		= & - 2 F_{uz} \Gamma^{z}_{\ uu} \Gamma^{u}_{\ zu} F_{uz} - 2 F_{uz} \Gamma^{z}_{\ uz} \Gamma^{z}_{\ zu} F_{uz} - 2 F_{uz} \Gamma^{z}_{\ ui} \Gamma^{i}_{\ zu} F_{uz}\\
		= & 0\,.
	\end{split}
\end{equation}
The second term of Eq. (\ref{hkk2thirdfifteenth}) is
\begin{equation}
	\begin{split}
		& - 2 \gamma^{ij} F_{uz} \Gamma^{z}_{\ ih} \Gamma^{h}_{\ ju} F_{uz}\\
		= & - 2 \gamma^{ij} F_{uz} \Gamma^{z}_{\ iu} \Gamma^{u}_{\ ju} F_{uz} - 2 \gamma^{ij} F_{uz} \Gamma^{z}_{\ iz} \Gamma^{z}_{\ ju} F_{uz} - 2 \gamma^{ij} F_{uz} \Gamma^{z}_{\ ik} \Gamma^{k}_{\ ju} F_{uz}\\
		= & \frac{1}{2} \gamma^{ij} \gamma^{km} F_{uz} \left(\partial_u \gamma_{ik} \right) \left(\partial_u \gamma_{jm} \right) F_{uz}\,.
	\end{split}
\end{equation}
Therefore, the fifteenth term of Eq. (\ref{hkk2third}) is obtained as
\begin{equation}
	\begin{split}
		& - 2 k^a k^b g^{ce} g^{df} F_{ae} \Gamma^{g}_{\ dh} \Gamma^{h}_{\ fc} F_{bg} = \frac{1}{2} \gamma^{ij} \gamma^{km} F_{uz} \left(\partial_u \gamma_{ik} \right) \left(\partial_u \gamma_{jm} \right) F_{uz}\\
		= & 2 \gamma^{ij} \gamma^{km} F_{uz} K_{ik} K_{jm} F_{uz}\\
		= & 0\,.
	\end{split}
\end{equation}

The sixteenth term of Eq. (\ref{hkk2third}) is
\begin{equation}
	\begin{split}
		& - 2 k^a k^b g^{ce} g^{df} F_{ae} \Gamma^{g}_{\ fc} \left(\partial_d F_{bg} \right) = - 2 g^{ce} g^{df} F_{ue} \Gamma^{g}_{\ fc} \left(\partial_d F_{ug} \right)\\
		= & - 2 F_{uz} \Gamma^{g}_{\ zu} \left(\partial_u F_{ug} \right) - 2 F_{uz} \Gamma^{g}_{\ uu} \left(\partial_z F_{ug} \right) - 2 \gamma^{ij} F_{uz} \Gamma^{g}_{\ ju} \left(\partial_i F_{ug} \right)\\
		= & - 2 F_{uz} \Gamma^{g}_{\ zu} \left(\partial_u F_{ug} \right) - 2 \gamma^{ij} F_{uz} \Gamma^{g}_{\ ju} \left(\partial_i F_{ug} \right)\,.
	\end{split}
\end{equation}
The index $g$ should be further expanded.
\begin{equation}\label{hkk2thirdsixteenth}
	\begin{split}
		& - 2 F_{uz} \Gamma^{g}_{\ zu} \left(\partial_u F_{ug} \right) - 2 \gamma^{ij} F_{uz} \Gamma^{g}_{\ ju} \left(\partial_i F_{ug} \right)\\
		= & - 2 F_{uz} \Gamma^{u}_{\ zu} \left(\partial_u F_{uu} \right) - 2 F_{uz} \Gamma^{z}_{\ zu} \left(\partial_u F_{uz} \right) - 2 F_{uz} \Gamma^{i}_{\ zu} \left(\partial_u F_{ui} \right)\\
		& - 2 \gamma^{ij} F_{uz} \Gamma^{u}_{\ ju} \left(\partial_i F_{uu} \right) - 2 \gamma^{ij} F_{uz} \Gamma^{z}_{\ ju} \left(\partial_i F_{uz} \right) - 2 \gamma^{ij} F_{uz} \Gamma^{k}_{\ ju} \left(\partial_i F_{uk} \right)\\
		= & - 2 F_{uz} \Gamma^{z}_{\ zu} \left(\partial_u F_{uz} \right) - 2 F_{uz} \Gamma^{i}_{\ zu} \left(\partial_u F_{ui} \right) - 2 \gamma^{ij} F_{uz} \Gamma^{z}_{\ ju} \left(\partial_i F_{uz} \right)\,.
	\end{split}
\end{equation}
Therefore, the sixteenth term of Eq. (\ref{hkk2third}) is obtained as
\begin{equation}
	\begin{split}
		- 2 k^a k^b g^{ce} g^{df} F_{ae} \Gamma^{g}_{\ fc} \left(\partial_d F_{bg} \right) = - \gamma^{ij} \beta_j F_{uz} \left(\partial_u F_{ui} \right)\,.
	\end{split}
\end{equation}

The seventeenth term of Eq. (\ref{hkk2third}) is
\begin{equation}
	\begin{split}
		& 2 k^a k^b g^{ce} g^{df} F_{ae} \Gamma^{g}_{\ fc} \Gamma^{h}_{\ db} F_{hg} = 2 g^{ce} g^{df} F_{ue} \Gamma^{g}_{\ fc} \Gamma^{h}_{\ du} F_{hg}\\
		= & 2 F_{uz} \Gamma^{g}_{\ zu} \Gamma^{h}_{\ uu} F_{hg} + 2 F_{uz} \Gamma^{g}_{\ uu} \Gamma^{h}_{\ zu} F_{hg} + 2 \gamma^{ij} F_{uz} \Gamma^{g}_{\ ju} \Gamma^{h}_{\ iu} F_{hg}\\
		= & 2 \gamma^{ij} F_{uz} \Gamma^{g}_{\ ju} \Gamma^{h}_{\ iu} F_{hg}\,.
	\end{split}
\end{equation}
The index $g$ should be further expanded.
\begin{equation}\label{hkk2thirdseventeenth}
	\begin{split}
		& 2 \gamma^{ij} F_{uz} \Gamma^{g}_{\ ju} \Gamma^{h}_{\ iu} F_{hg}\\
		= & 2 \gamma^{ij} F_{uz} \Gamma^{u}_{\ ju} \Gamma^{h}_{\ iu} F_{hu} + 2 \gamma^{ij} F_{uz} \Gamma^{z}_{\ ju} \Gamma^{h}_{\ iu} F_{hz} + 2 \gamma^{ij} F_{uz} \Gamma^{k}_{\ ju} \Gamma^{h}_{\ iu} F_{hk}\,.
	\end{split}
\end{equation}
The repeated index $h$ should be further expanded. The first term of Eq. (\ref{hkk2thirdseventeenth}) is
\begin{equation}
	\begin{split}
		& 2 \gamma^{ij} F_{uz} \Gamma^{u}_{\ ju} \Gamma^{h}_{\ iu} F_{hu}\\
		= & 2 \gamma^{ij} F_{uz} \Gamma^{u}_{\ ju} \Gamma^{u}_{\ iu} F_{uu} + 2 \gamma^{ij} F_{uz} \Gamma^{u}_{\ ju} \Gamma^{z}_{\ iu} F_{zu} + 2 \gamma^{ij} F_{uz} \Gamma^{u}_{\ ju} \Gamma^{k}_{\ iu} F_{ku}\\
		= & 2 \gamma^{ij} F_{uz} \Gamma^{u}_{\ ju} \Gamma^{z}_{\ iu} F_{zu}\\
		= & 0\,.
	\end{split}
\end{equation}
The second term of Eq. (\ref{hkk2thirdseventeenth}) is
\begin{equation}
	\begin{split}
		2 \gamma^{ij} F_{uz} \Gamma^{z}_{\ ju} \Gamma^{h}_{\ iu} F_{hz} = 0\,.
	\end{split}
\end{equation}
The third term of Eq. (\ref{hkk2thirdseventeenth}) is
\begin{equation}
	\begin{split}
		& 2 \gamma^{ij} F_{uz} \Gamma^{k}_{\ ju} \Gamma^{h}_{\ iu} F_{hk}\\
		= & 2 \gamma^{ij} F_{uz} \Gamma^{k}_{\ ju} \Gamma^{u}_{\ iu} F_{uk} + 2 \gamma^{ij} F_{uz} \Gamma^{k}_{\ ju} \Gamma^{z}_{\ iu} F_{zk} + 2 \gamma^{ij} F_{uz} \Gamma^{k}_{\ ju} \Gamma^{l}_{\ iu} F_{lk}\\
		= & 2 \gamma^{ij} F_{uz} \Gamma^{k}_{\ ju} \Gamma^{z}_{\ iu} F_{zk} + 2 \gamma^{ij} F_{uz} \Gamma^{k}_{\ ju} \Gamma^{l}_{\ iu} F_{lk}\\
		= & \frac{1}{2} \gamma^{ij} \gamma^{km} \gamma^{ln} F_{uz} \left(\partial_u \gamma_{jm} \right) \left(\partial_u \gamma_{in} \right) F_{lk}\,.
	\end{split}
\end{equation}
Therefore, the seventeenth term of Eq. (\ref{hkk2third}) is obtained as 
\begin{equation}
	\begin{split}
		& 2 k^a k^b g^{ce} g^{df} F_{ae} \Gamma^{g}_{\ fc} \Gamma^{h}_{\ db} F_{hg} = \frac{1}{2} \gamma^{ij} \gamma^{km} \gamma^{ln} F_{uz} \left(\partial_u \gamma_{jm} \right) \left(\partial_u \gamma_{in} \right) F_{lk}\\
		= & 2 \gamma^{ij} \gamma^{km} \gamma^{ln} F_{uz} K_{jm} K_{in} F_{lk}\\
		= & 0\,.
	\end{split}
\end{equation}

The eighteenth term of Eq. (\ref{hkk2third}) is
\begin{equation}
	\begin{split}
		& 2 k^a k^b g^{ce} g^{df} F_{ae} \Gamma^{g}_{\ fc} \Gamma^{h}_{\ dg} F_{bh} = 2 g^{ce} g^{df} F_{ue} \Gamma^{g}_{\ fc} \Gamma^{h}_{\ dg} F_{uh}\\
		= & 2 F_{uz} \Gamma^{g}_{\ zu} \Gamma^{h}_{\ ug} F_{uh} + 2 F_{uz} \Gamma^{g}_{\ uu} \Gamma^{h}_{\ zg} F_{uh} + 2 \gamma^{ij} F_{uz} \Gamma^{g}_{\ ju} \Gamma^{h}_{\ ig} F_{uh}\\
		= & 2 F_{uz} \Gamma^{g}_{\ zu} \Gamma^{h}_{\ ug} F_{uh} + 2 \gamma^{ij} F_{uz} \Gamma^{g}_{\ ju} \Gamma^{h}_{\ ig} F_{uh}\,.
	\end{split}
\end{equation}
The index $g$ should be further expanded.
\begin{equation}\label{hkk2thirdeighteenth}
	\begin{split}
		& 2 F_{uz} \Gamma^{g}_{\ zu} \Gamma^{h}_{\ ug} F_{uh} + 2 \gamma^{ij} F_{uz} \Gamma^{g}_{\ ju} \Gamma^{h}_{\ ig} F_{uh}\\
		= & 2 F_{uz} \Gamma^{u}_{\ zu} \Gamma^{h}_{\ uu} F_{uh} + 2 F_{uz} \Gamma^{z}_{\ zu} \Gamma^{h}_{\ uz} F_{uh} + 2 F_{uz} \Gamma^{i}_{\ zu} \Gamma^{h}_{\ ui} F_{uh}\\
		& + 2 \gamma^{ij} F_{uz} \Gamma^{u}_{\ ju} \Gamma^{h}_{\ iu} F_{uh} + 2 \gamma^{ij} F_{uz} \Gamma^{z}_{\ ju} \Gamma^{h}_{\ iz} F_{uh} + 2 \gamma^{ij} F_{uz} \Gamma^{k}_{\ ju} \Gamma^{h}_{\ ik} F_{uh}\\
		= & 2 F_{uz} \Gamma^{z}_{\ zu} \Gamma^{h}_{\ uz} F_{uh} + 2 F_{uz} \Gamma^{i}_{\ zu} \Gamma^{h}_{\ ui} F_{uh} + 2 \gamma^{ij} F_{uz} \Gamma^{u}_{\ ju} \Gamma^{h}_{\ iu} F_{uh}\\
		& + 2 \gamma^{ij} F_{uz} \Gamma^{z}_{\ ju} \Gamma^{h}_{\ iz} F_{uh} + 2 \gamma^{ij} F_{uz} \Gamma^{k}_{\ ju} \Gamma^{h}_{\ ik} F_{uh}\,.
	\end{split}
\end{equation}
The repeated index $h$ should be further expanded. The first term of Eq. (\ref{hkk2thirdeighteenth}) is
\begin{equation}
	\begin{split}
		2 F_{uz} \Gamma^{z}_{\ zu} \Gamma^{h}_{\ uz} F_{uh} = 0\,.
	\end{split}
\end{equation}
The second term of Eq. (\ref{hkk2thirdeighteenth}) is
\begin{equation}
	\begin{split}
		& 2 F_{uz} \Gamma^{i}_{\ zu} \Gamma^{h}_{\ ui} F_{uh}\\
		= & 2 F_{uz} \Gamma^{i}_{\ zu} \Gamma^{u}_{\ ui} F_{uu} + 2 F_{uz} \Gamma^{i}_{\ zu} \Gamma^{z}_{\ ui} F_{uz} + 2 F_{uz} \Gamma^{i}_{\ zu} \Gamma^{j}_{\ ui} F_{uj}\\
		= & 2 F_{uz} \Gamma^{i}_{\ zu} \Gamma^{z}_{\ ui} F_{uz}\\
		= & 0\,.
	\end{split}
\end{equation}
The third term of Eq. (\ref{hkk2thirdeighteenth}) is
\begin{equation}
	\begin{split}
		& 2 \gamma^{ij} F_{uz} \Gamma^{u}_{\ ju} \Gamma^{h}_{\ iu} F_{uh}\\
		= & 2 \gamma^{ij} F_{uz} \Gamma^{u}_{\ ju} \Gamma^{u}_{\ iu} F_{uu} + 2 \gamma^{ij} F_{uz} \Gamma^{u}_{\ ju} \Gamma^{z}_{\ iu} F_{uz} + 2 \gamma^{ij} F_{uz} \Gamma^{u}_{\ ju} \Gamma^{k}_{\ iu} F_{uk}\\
		= & 2 \gamma^{ij} F_{uz} \Gamma^{u}_{\ ju} \Gamma^{z}_{\ iu} F_{uz}\\
		= & 0\,.
	\end{split}
\end{equation}
The fourth term of Eq. (\ref{hkk2thirdeighteenth}) is
\begin{equation}
	\begin{split}
		2 \gamma^{ij} F_{uz} \Gamma^{z}_{\ ju} \Gamma^{h}_{\ iz} F_{uh} = 0\,.
	\end{split}
\end{equation}
The fifth term of Eq. (\ref{hkk2thirdeighteenth}) is
\begin{equation}
	\begin{split}
		& 2 \gamma^{ij} F_{uz} \Gamma^{k}_{\ ju} \Gamma^{h}_{\ ik} F_{uh}\\
		= & 2 \gamma^{ij} F_{uz} \Gamma^{k}_{\ ju} \Gamma^{u}_{\ ik} F_{uu} + 2 \gamma^{ij} F_{uz} \Gamma^{k}_{\ ju} \Gamma^{z}_{\ ik} F_{uz} + 2 \gamma^{ij} F_{uz} \Gamma^{k}_{\ ju} \Gamma^{l}_{\ ik} F_{ul}\\
		= & 2 \gamma^{ij} F_{uz} \Gamma^{k}_{\ ju} \Gamma^{z}_{\ ik} F_{uz}\\
		= & - \frac{1}{2} \gamma^{ij} \gamma^{kl} F_{uz} \left(\partial_u \gamma_{jl} \right) \left(\partial_u \gamma_{ik} \right) F_{uz}\,.
	\end{split}
\end{equation}
Therefore, the eighteenth term of Eq. (\ref{hkk2third}) is obtained as 
\begin{equation}
	\begin{split}
		& 2 k^a k^b g^{ce} g^{df} F_{ae} \Gamma^{g}_{\ fc} \Gamma^{h}_{\ dg} F_{bh} = - \frac{1}{2} \gamma^{ij} \gamma^{kl} F_{uz} \left(\partial_u \gamma_{jl} \right) \left(\partial_u \gamma_{ik} \right) F_{uz}\\
		= & - 2 \gamma^{ij} \gamma^{kl} F_{uz} K_{jl} K_{ik} F_{uz}\\
		= & 0\,.
	\end{split}
\end{equation}

Finally, the third term of Eq. (\ref{rewrittenhkk2}) is
\begin{equation}
	\begin{split}
		& 2 k^a k^b F_{a}^{\ c} \nabla_d \nabla^d F_{bc}\\
		= & - \gamma^{ij} \beta_j F_{uz} \left(\partial_u F_{iu} \right) - \gamma^{ij} \beta_j F_{uz} \left(\partial_u F_{ui} \right) - \gamma^{ij} \beta_j F_{uz} \left(\partial_u F_{iu} \right) - \gamma^{ij} \beta_j F_{uz} \left(\partial_u F_{ui} \right)\\
		= & \gamma^{ij} \beta_j F_{uz} \left(\partial_u F_{ui} \right) - \gamma^{ij} \beta_j F_{uz} \left(\partial_u F_{ui} \right) + \gamma^{ij} \beta_j F_{uz} \left(\partial_u F_{ui} \right) - \gamma^{ij} \beta_j F_{uz} \left(\partial_u F_{ui} \right)\\
		= & 0\,.
	\end{split}
\end{equation}

The fourth term of Eq. (\ref{rewrittenhkk2}) is
\begin{equation}\label{hkk2fourth}
	\begin{split}
		& 2 k^a k^b \nabla_a F_{b}^{\ c} \nabla_d F_{c}^{\ d} = 2 k^a k^b g^{ce} g^{df} \nabla_a F_{be} \nabla_d F_{cf}\\
		= & 2 k^a k^b g^{ce} g^{df} \left(\partial_a F_{be} \right) \left(\partial_d F_{cf} \right) - 2 k^a k^b g^{ce} g^{df} \left(\partial_a F_{be} \right) \Gamma^{h}_{\ dc} F_{hf}\\
		& - 2 k^a k^b g^{ce} g^{df} \left(\partial_a F_{be} \right) \Gamma^{h}_{\ df} F_{ch} - 2 k^a k^b g^{ce} g^{df} \Gamma^{g}_{\ ab} F_{ge} \left(\partial_d F_{cf} \right)\\
		& + 2 k^a k^b g^{ce} g^{df} \Gamma^{g}_{\ ab} F_{ge} \Gamma^{h}_{\ dc} F_{hf} + 2 k^a k^b g^{ce} g^{df} \Gamma^{g}_{\ ab} F_{ge} \Gamma^{h}_{\ df} F_{ch}\\
		& - 2 k^a k^b g^{ce} g^{df} \Gamma^{g}_{\ ae} F_{bg} \left(\partial_d F_{cf} \right) + 2 k^a k^b g^{ce} g^{df} \Gamma^{g}_{\ ae} F_{bg} \Gamma^{h}_{\ dc} F_{hf}\\
		& + 2 k^a k^b g^{ce} g^{df} \Gamma^{g}_{\ ae} F_{bg} \Gamma^{h}_{\ df} F_{ch}\,.
	\end{split}
\end{equation}

The first term of Eq. (\ref{hkk2fourth}) is 
\begin{equation}
	\begin{split}
		& 2 k^a k^b g^{ce} g^{df} \left(\partial_a F_{be} \right) \left(\partial_d F_{cf} \right) = 2 g^{ce} g^{df} \left(\partial_u F_{ue} \right) \left(\partial_d F_{cf} \right)\\
		= & 2 \left(\partial_u F_{uz} \right) \left(\partial_u F_{uz} \right) + 2 \gamma^{ij} \left(\partial_u F_{uz} \right) \left(\partial_i F_{uj} \right) + 2 \gamma^{ij} \left(\partial_u F_{uj} \right) \left(\partial_u F_{iz} \right)\\
		& + 2 \gamma^{ij} \left(\partial_u F_{uj} \right) \left(\partial_z F_{iu} \right) + 2 \gamma^{ij} \gamma^{kl} \left(\partial_u F_{uj} \right) \left(\partial_k F_{il} \right)\\
		= & 2 \left(\partial_u F_{uz} \right) \left(\partial_u F_{uz} \right) + 2 \gamma^{ij} \left(\partial_u F_{uj} \right) \left(\partial_u F_{iz} \right) + 2 \gamma^{ij} \left(\partial_u F_{uj} \right) \left(\partial_z F_{iu} \right)\\
		& + 2 \gamma^{ij} \gamma^{kl} \left(\partial_u F_{uj} \right) \left(\partial_k F_{il} \right)\,.
	\end{split}
\end{equation}
Therefore, the first term of Eq. (\ref{hkk2fourth}) is obtained as 
\begin{equation}
	\begin{split}
		& 2 k^a k^b g^{ce} g^{df} \left(\partial_a F_{be} \right) \left(\partial_d F_{cf} \right)\\
		= & 2 \left(\partial_u F_{uz} \right) \left(\partial_u F_{uz} \right) + 2 \gamma^{ij} \left(\partial_u F_{uj} \right) \left(\partial_u F_{iz} \right) + 2 \gamma^{ij} \left(\partial_u F_{uj} \right) \left(\partial_z F_{iu} \right)\\
		& + 2 \gamma^{ij} \gamma^{kl} \left(\partial_u F_{uj} \right) \left(\partial_k F_{il} \right)\,.
	\end{split}
\end{equation}

The second term of Eq. (\ref{hkk2fourth}) is
\begin{equation}\label{hkk2fourthsecond}
	\begin{split}
		& - 2 k^a k^b g^{ce} g^{df} \left(\partial_a F_{be} \right) \Gamma^{h}_{\ dc} F_{hf} = - 2 g^{ce} g^{df} \left(\partial_u F_{ue} \right) \Gamma^{h}_{\ dc} F_{hf}\\
		= & - 2 \left(\partial_u F_{uz} \right) \Gamma^{h}_{\ uu} F_{hz} - 2 \left(\partial_u F_{uz} \right) \Gamma^{h}_{\ zu} F_{hu} - 2 \gamma^{ij} \left(\partial_u F_{uz} \right) \Gamma^{h}_{\ iu} F_{hj}\\
		& - 2 \gamma^{ij} \left(\partial_u F_{uj} \right) \Gamma^{h}_{\ ui} F_{hz} - 2 \gamma^{ij} \left(\partial_u F_{uj} \right) \Gamma^{h}_{\ zi} F_{hu} - 2 \gamma^{ij} \gamma^{kl} \left(\partial_u F_{uj} \right) \Gamma^{h}_{\ ki} F_{hl}\\
		= & - 2 \left(\partial_u F_{uz} \right) \Gamma^{h}_{\ zu} F_{hu} - 2 \gamma^{ij} \left(\partial_u F_{uz} \right) \Gamma^{h}_{\ iu} F_{hj} - 2 \gamma^{ij} \left(\partial_u F_{uj} \right) \Gamma^{h}_{\ ui} F_{hz}\\
		& - 2 \gamma^{ij} \left(\partial_u F_{uj} \right) \Gamma^{h}_{\ zi} F_{hu} - 2 \gamma^{ij} \gamma^{kl} \left(\partial_u F_{uj} \right) \Gamma^{h}_{\ ki} F_{hl}\,.
	\end{split}
\end{equation}
The repeated index $h$ should be further expanded. The first term of Eq. (\ref{hkk2fourthsecond}) is
\begin{equation}
	\begin{split}
		& - 2 \left(\partial_u F_{uz} \right) \Gamma^{h}_{\ zu} F_{hu}\\
		= & - 2 \left(\partial_u F_{uz} \right) \Gamma^{u}_{\ zu} F_{uu} - 2 \left(\partial_u F_{uz} \right) \Gamma^{z}_{\ zu} F_{zu} - 2 \left(\partial_u F_{uz} \right) \Gamma^{i}_{\ zu} F_{iu}\\
		= & - 2 \left(\partial_u F_{uz} \right) \Gamma^{z}_{\ zu} F_{zu}\\
		= & 0\,.
	\end{split}
\end{equation}
The second term of Eq. (\ref{hkk2fourthsecond}) is
\begin{equation}
	\begin{split}
		& - 2 \gamma^{ij} \left(\partial_u F_{uz} \right) \Gamma^{h}_{\ iu} F_{hj}\\
		= & - 2 \gamma^{ij} \left(\partial_u F_{uz} \right) \Gamma^{u}_{\ iu} F_{uj} - 2 \gamma^{ij} \left(\partial_u F_{uz} \right) \Gamma^{z}_{\ iu} F_{zj} - 2 \gamma^{ij} \left(\partial_u F_{uz} \right) \Gamma^{k}_{\ iu} F_{kj}\\
		= & - 2 \gamma^{ij} \left(\partial_u F_{uz} \right) \Gamma^{z}_{\ iu} F_{zj} - 2 \gamma^{ij} \left(\partial_u F_{uz} \right) \Gamma^{k}_{\ iu} F_{kj}\\
		= & - \gamma^{ij} \gamma^{kl} \left(\partial_u F_{uz} \right) \left(\partial_u \gamma_{il} \right) F_{kj}\,.
	\end{split}
\end{equation}
The third term of Eq. (\ref{hkk2fourthsecond}) is
\begin{equation}
	\begin{split}
		& - 2 \gamma^{ij} \left(\partial_u F_{uj} \right) \Gamma^{h}_{\ ui} F_{hz}\\
		= & - 2 \gamma^{ij} \left(\partial_u F_{uj} \right) \Gamma^{u}_{\ ui} F_{uz} - 2 \gamma^{ij} \left(\partial_u F_{uj} \right) \Gamma^{z}_{\ ui} F_{zz} - 2 \gamma^{ij} \left(\partial_u F_{uj} \right) \Gamma^{k}_{\ ui} F_{kz}\\
		= & - 2 \gamma^{ij} \left(\partial_u F_{uj} \right) \Gamma^{u}_{\ ui} F_{uz} - 2 \gamma^{ij} \left(\partial_u F_{uj} \right) \Gamma^{k}_{\ ui} F_{kz}\\
		= & \gamma^{ij} \beta_i \left(\partial_u F_{uj} \right) F_{uz} - \gamma^{ij} \gamma^{kl} \left(\partial_u F_{uj} \right) \left(\partial_u \gamma_{il} \right) F_{kz}\,.
	\end{split}
\end{equation}
The fourth term of Eq. (\ref{hkk2fourthsecond}) is
\begin{equation}
	\begin{split}
		& - 2 \gamma^{ij} \left(\partial_u F_{uj} \right) \Gamma^{h}_{\ zi} F_{hu}\\
		= & - 2 \gamma^{ij} \left(\partial_u F_{uj} \right) \Gamma^{u}_{\ zi} F_{uu} - 2 \gamma^{ij} \left(\partial_u F_{uj} \right) \Gamma^{z}_{\ zi} F_{zu} - 2 \gamma^{ij} \left(\partial_u F_{uj} \right) \Gamma^{k}_{\ zi} F_{ku}\\
		= & - 2 \gamma^{ij} \left(\partial_u F_{uj} \right) \Gamma^{z}_{\ zi} F_{zu}\\
		= & - \gamma^{ij} \beta_i \left(\partial_u F_{uj} \right) F_{zu}\,.
	\end{split}
\end{equation}
The fifth term of Eq. (\ref{hkk2fourthsecond}) is
\begin{equation}
	\begin{split}
		& - 2 \gamma^{ij} \gamma^{kl} \left(\partial_u F_{uj} \right) \Gamma^{h}_{\ ki} F_{hl}\\
		= & - 2 \gamma^{ij} \gamma^{kl} \left(\partial_u F_{uj} \right) \Gamma^{u}_{\ ki} F_{ul} - 2 \gamma^{ij} \gamma^{kl} \left(\partial_u F_{uj} \right) \Gamma^{z}_{\ ki} F_{zl} - 2 \gamma^{ij} \gamma^{kl} \left(\partial_u F_{uj} \right) \Gamma^{m}_{\ ki} F_{ml}\\
		= & - 2 \gamma^{ij} \gamma^{kl} \left(\partial_u F_{uj} \right) \Gamma^{z}_{\ ki} F_{zl} - 2 \gamma^{ij} \gamma^{kl} \left(\partial_u F_{uj} \right) \Gamma^{m}_{\ ki} F_{ml}\\
		= & \gamma^{ij} \gamma^{kl} \left(\partial_u F_{uj} \right) \left(\partial_u \gamma_{ki} \right) F_{zl} - 2 \gamma^{ij} \gamma^{kl} \left(\partial_u F_{uj} \right) \hat{\Gamma}^{m}_{\ ki} F_{ml}\,.
	\end{split}
\end{equation}
Therefore, the second term of Eq. (\ref{hkk2fourth}) is obtained as 
\begin{equation}
	\begin{split}
		& - 2 k^a k^b g^{ce} g^{df} \left(\partial_a F_{be} \right) \Gamma^{h}_{\ dc} F_{hf}\\
		= & - \gamma^{ij} \gamma^{kl} \left(\partial_u F_{uz} \right) \left(\partial_u \gamma_{il} \right) F_{kj} + \gamma^{ij} \beta_i \left(\partial_u F_{uj} \right) F_{uz} - \gamma^{ij} \gamma^{kl} \left(\partial_u F_{uj} \right) \left(\partial_u \gamma_{il} \right) F_{kz}\\
		& - \gamma^{ij} \beta_i \left(\partial_u F_{uj} \right) F_{zu} + \gamma^{ij} \gamma^{kl} \left(\partial_u F_{uj} \right) \left(\partial_u \gamma_{ki} \right) F_{zl} - 2 \gamma^{ij} \gamma^{kl} \left(\partial_u F_{uj} \right) \hat{\Gamma}^{m}_{\ ki} F_{ml}\\
		= & - 2 \gamma^{ij} \gamma^{kl} \left(\partial_u F_{uz} \right) K_{il} F_{kj} + \gamma^{ij} \beta_i \left(\partial_u F_{uj} \right) F_{uz} - 2 \gamma^{ij} \gamma^{kl} \left(\partial_u F_{uj} \right) K_{il} F_{kz}\\
		& - \gamma^{ij} \beta_i \left(\partial_u F_{uj} \right) F_{zu} + 2 \gamma^{ij} \gamma^{kl} \left(\partial_u F_{uj} \right) K_{ki} F_{zl} - 2 \gamma^{ij} \gamma^{kl} \left(\partial_u F_{uj} \right) \hat{\Gamma}^{m}_{\ ki} F_{ml}\\
		= & \gamma^{ij} \beta_i \left(\partial_u F_{uj} \right) F_{uz} - \gamma^{ij} \beta_i \left(\partial_u F_{uj} \right) F_{zu} - 2 \gamma^{ij} \gamma^{kl} \left(\partial_u F_{uj} \right) \hat{\Gamma}^{m}_{\ ki} F_{ml}\,.
	\end{split}
\end{equation}

The third term of Eq. (\ref{hkk2fourth}) is
\begin{equation}\label{hkk2fourththird}
	\begin{split}
		& - 2 k^a k^b g^{ce} g^{df} \left(\partial_a F_{be} \right) \Gamma^{h}_{\ df} F_{ch} = - 2 g^{ce} g^{df} \left(\partial_u F_{ue} \right) \Gamma^{h}_{\ df} F_{ch}\\
		= & - 4 \left(\partial_u F_{uz} \right) \Gamma^{h}_{\ uz} F_{uh} - 2 \gamma^{ij} \left(\partial_u F_{uz} \right) \Gamma^{h}_{\ ij} F_{uh} - 4 \gamma^{ij} \left(\partial_u F_{uj} \right) \Gamma^{h}_{\ uz} F_{ih}\\
		& - 2 \gamma^{ij} \gamma^{kl} \left(\partial_u F_{uj} \right) \Gamma^{h}_{\ kl} F_{ih}\,.
	\end{split}
\end{equation}
The repeated index $h$ should be further expanded. The first term of Eq. (\ref{hkk2fourththird}) is 
\begin{equation}
	\begin{split}
		& - 4 \left(\partial_u F_{uz} \right) \Gamma^{h}_{\ uz} F_{uh}\\
		= & - 4 \left(\partial_u F_{uz} \right) \Gamma^{u}_{\ uz} F_{uu} - 4 \left(\partial_u F_{uz} \right) \Gamma^{z}_{\ uz} F_{uz} - 4 \left(\partial_u F_{uz} \right) \Gamma^{i}_{\ uz} F_{ui}\\
		= & - 4 \left(\partial_u F_{uz} \right) \Gamma^{z}_{\ uz} F_{uz}\\
		= & 0\,.
	\end{split}
\end{equation}
The second term of Eq. (\ref{hkk2fourththird}) is
\begin{equation}
	\begin{split}
		& - 2 \gamma^{ij} \left(\partial_u F_{uz} \right) \Gamma^{h}_{\ ij} F_{uh}\\
		= & - 2 \gamma^{ij} \left(\partial_u F_{uz} \right) \Gamma^{u}_{\ ij} F_{uu} - 2 \gamma^{ij} \left(\partial_u F_{uz} \right) \Gamma^{z}_{\ ij} F_{uz} - 2 \gamma^{ij} \left(\partial_u F_{uz} \right) \Gamma^{k}_{\ ij} F_{uk}\\
		= & - 2 \gamma^{ij} \left(\partial_u F_{uz} \right) \Gamma^{z}_{\ ij} F_{uz}\\
		= & \gamma^{ij} \left(\partial_u F_{uz} \right) \left(\partial_u \gamma_{ij} \right) F_{uz}\,.
	\end{split}
\end{equation}
The third term of Eq. (\ref{hkk2fourththird}) is
\begin{equation}
	\begin{split}
		& - 4 \gamma^{ij} \left(\partial_u F_{uj} \right) \Gamma^{h}_{\ uz} F_{ih}\\
		= & - 4 \gamma^{ij} \left(\partial_u F_{uj} \right) \Gamma^{u}_{\ uz} F_{iu} - 4 \gamma^{ij} \left(\partial_u F_{uj} \right) \Gamma^{z}_{\ uz} F_{iz} - 4 \gamma^{ij} \left(\partial_u F_{uj} \right) \Gamma^{k}_{\ uz} F_{ik}\\
		= & - 4 \gamma^{ij} \left(\partial_u F_{uj} \right) \Gamma^{z}_{\ uz} F_{iz} - 4 \gamma^{ij} \left(\partial_u F_{uj} \right) \Gamma^{k}_{\ uz} F_{ik}\\
		= & - 2 \gamma^{ij} \gamma^{kl} \beta_l \left(\partial_u F_{uj} \right) F_{ik}\,.
	\end{split}
\end{equation}
The fourth term of Eq. (\ref{hkk2fourththird}) is
\begin{equation}
	\begin{split}
		& - 2 \gamma^{ij} \gamma^{kl} \left(\partial_u F_{uj} \right) \Gamma^{h}_{\ kl} F_{ih}\\
		= & - 2 \gamma^{ij} \gamma^{kl} \left(\partial_u F_{uj} \right) \Gamma^{u}_{\ kl} F_{iu} - 2 \gamma^{ij} \gamma^{kl} \left(\partial_u F_{uj} \right) \Gamma^{z}_{\ kl} F_{iz} - 2 \gamma^{ij} \gamma^{kl} \left(\partial_u F_{uj} \right) \Gamma^{m}_{\ kl} F_{im}\\
		= & - 2 \gamma^{ij} \gamma^{kl} \left(\partial_u F_{uj} \right) \Gamma^{z}_{\ kl} F_{iz} - 2 \gamma^{ij} \gamma^{kl} \left(\partial_u F_{uj} \right) \Gamma^{m}_{\ kl} F_{im}\\
		= & \gamma^{ij} \gamma^{kl} \left(\partial_u F_{uj} \right) \left(\partial_u \gamma_{kl} \right) F_{iz} - 2 \gamma^{ij} \gamma^{kl} \left(\partial_u F_{uj} \right) \hat{\Gamma}^{m}_{\ kl} F_{im}\,.
	\end{split}
\end{equation}
Therefore, the third term of Eq. (\ref{hkk2fourth}) is obtained as 
\begin{equation}
	\begin{split}
		& - 2 k^a k^b g^{ce} g^{df} \left(\partial_a F_{be} \right) \Gamma^{h}_{\ df} F_{ch}\\
		= & \gamma^{ij} \left(\partial_u F_{uz} \right) \left(\partial_u \gamma_{ij} \right) F_{uz} - 2 \gamma^{ij} \gamma^{kl} \beta_l \left(\partial_u F_{uj} \right) F_{ik} + \gamma^{ij} \gamma^{kl} \left(\partial_u F_{uj} \right) \left(\partial_u \gamma_{kl} \right) F_{iz}\\
		& - 2 \gamma^{ij} \gamma^{kl} \left(\partial_u F_{uj} \right) \hat{\Gamma}^{m}_{\ kl} F_{im}\\
		= & 2 \gamma^{ij} \left(\partial_u F_{uz} \right) K_{ij} F_{uz} - 2 \gamma^{ij} \gamma^{kl} \beta_l \left(\partial_u F_{uj} \right) F_{ik} + 2 \gamma^{ij} \gamma^{kl} \left(\partial_u F_{uj} \right) K_{kl} F_{iz}\\
		& - 2 \gamma^{ij} \gamma^{kl} \left(\partial_u F_{uj} \right) \hat{\Gamma}^{m}_{\ kl} F_{im}\\
		= & - 2 \gamma^{ij} \gamma^{kl} \beta_l \left(\partial_u F_{uj} \right) F_{ik} - 2 \gamma^{ij} \gamma^{kl} \left(\partial_u F_{uj} \right) \hat{\Gamma}^{m}_{\ kl} F_{im}\,.
	\end{split}
\end{equation}

The fourth term of Eq. (\ref{hkk2fourth}) is
\begin{equation}
	\begin{split}
		- 2 k^a k^b g^{ce} g^{df} \Gamma^{g}_{\ ab} F_{ge} \left(\partial_d F_{cf} \right) = - 2 g^{ce} g^{df} \Gamma^{g}_{\ uu} F_{ge} \left(\partial_d F_{cf} \right) = 0\,.
	\end{split}
\end{equation}
Therefore, the fourth term of Eq. (\ref{hkk2fourth}) is obtained as
\begin{equation}
	\begin{split}
		- 2 k^a k^b g^{ce} g^{df} \Gamma^{g}_{\ ab} F_{ge} \left(\partial_d F_{cf} \right) = 0\,.
	\end{split}
\end{equation}

The fifth term of Eq. (\ref{hkk2fourth}) is
\begin{equation}
	\begin{split}
		2 k^a k^b g^{ce} g^{df} \Gamma^{g}_{\ ab} F_{ge} \Gamma^{h}_{\ dc} F_{hf} = 2 g^{ce} g^{df} \Gamma^{g}_{\ uu} F_{ge} \Gamma^{h}_{\ dc} F_{hf} = 0\,.
	\end{split}
\end{equation}
Therefore, the fifth term of Eq. (\ref{hkk2fourth}) is obtained as
\begin{equation}
	\begin{split}
		2 k^a k^b g^{ce} g^{df} \Gamma^{g}_{\ ab} F_{ge} \Gamma^{h}_{\ dc} F_{hf} = 0\,.
	\end{split}
\end{equation}

The sixth term of Eq. (\ref{hkk2fourth}) is
\begin{equation}
	\begin{split}
		2 k^a k^b g^{ce} g^{df} \Gamma^{g}_{\ ab} F_{ge} \Gamma^{h}_{\ df} F_{ch} = 2 g^{ce} g^{df} \Gamma^{g}_{\ uu} F_{ge} \Gamma^{h}_{\ df} F_{ch} = 0\,.
	\end{split}
\end{equation}
Therefore, the sixth term of Eq. (\ref{hkk2fourth}) is obtained as 
\begin{equation}
	\begin{split}
		2 k^a k^b g^{ce} g^{df} \Gamma^{g}_{\ ab} F_{ge} \Gamma^{h}_{\ df} F_{ch} = 0\,.
	\end{split}
\end{equation}

The seventh term of Eq. (\ref{hkk2fourth}) is
\begin{equation}
	\begin{split}
		& - 2 k^a k^b g^{ce} g^{df} \Gamma^{g}_{\ ae} F_{bg} \left(\partial_d F_{cf} \right) = - 2 g^{ce} g^{df} \Gamma^{g}_{\ ue} F_{ug} \left(\partial_d F_{cf} \right)\\
		= & - 2 \Gamma^{g}_{\ uz} F_{ug} \left(\partial_u F_{uz} \right) - 2 \gamma^{ij} \Gamma^{g}_{\ uz} F_{ug} \left(\partial_i F_{uj} \right) - 2 \Gamma^{g}_{\ uu} F_{ug} \left(\partial_z F_{zu} \right)\\
		& - 2 \gamma^{ij} \Gamma^{g}_{\ uu} F_{ug} \left(\partial_i F_{zj} \right) - 2 \gamma^{ij} \Gamma^{g}_{\ uj} F_{ug} \left(\partial_u F_{iz} \right) - 2 \gamma^{ij} \Gamma^{g}_{\ uj} F_{ug} \left(\partial_z F_{iu} \right)\\
		& - 2 \gamma^{ij} \gamma^{kl} \Gamma^{g}_{\ uj} F_{ug} \left(\partial_k F_{il} \right)\\
		= & - 2 \Gamma^{g}_{\ uz} F_{ug} \left(\partial_u F_{uz} \right) - 2 \gamma^{ij} \Gamma^{g}_{\ uj} F_{ug} \left(\partial_u F_{iz} \right) - 2 \gamma^{ij} \Gamma^{g}_{\ uj} F_{ug} \left(\partial_z F_{iu} \right)\\
		& - 2 \gamma^{ij} \gamma^{kl} \Gamma^{g}_{\ uj} F_{ug} \left(\partial_k F_{il} \right)\,.
	\end{split}
\end{equation}
The index $g$ should be further expanded.
\begin{equation}\label{hkk2fourthseventh}
	\begin{split}
		& - 2 \Gamma^{g}_{\ uz} F_{ug} \left(\partial_u F_{uz} \right) - 2 \gamma^{ij} \Gamma^{g}_{\ uj} F_{ug} \left(\partial_u F_{iz} \right) - 2 \gamma^{ij} \Gamma^{g}_{\ uj} F_{ug} \left(\partial_z F_{iu} \right)\\
		& - 2 \gamma^{ij} \gamma^{kl} \Gamma^{g}_{\ uj} F_{ug} \left(\partial_k F_{il} \right)\\
		= & - 2 \Gamma^{u}_{\ uz} F_{uu} \left(\partial_u F_{uz} \right) - 2 \Gamma^{z}_{\ uz} F_{uz} \left(\partial_u F_{uz} \right) - 2 \Gamma^{i}_{\ uz} F_{ui} \left(\partial_u F_{uz} \right)\\
		& - 2 \gamma^{ij} \Gamma^{u}_{\ uj} F_{uu} \left(\partial_u F_{iz} \right) - 2 \gamma^{ij} \Gamma^{z}_{\ uj} F_{uz} \left(\partial_u F_{iz} \right) - 2 \gamma^{ij} \Gamma^{k}_{\ uj} F_{uk} \left(\partial_u F_{iz} \right)\\
		& - 2 \gamma^{ij} \Gamma^{u}_{\ uj} F_{uu} \left(\partial_z F_{iu} \right) - 2 \gamma^{ij} \Gamma^{z}_{\ uj} F_{uz} \left(\partial_z F_{iu} \right) - 2 \gamma^{ij} \Gamma^{k}_{\ uj} F_{uk} \left(\partial_z F_{iu} \right)\\
		& - 2 \gamma^{ij} \gamma^{kl} \Gamma^{u}_{\ uj} F_{uu} \left(\partial_k F_{il} \right) - 2 \gamma^{ij} \gamma^{kl} \Gamma^{z}_{\ uj} F_{uz} \left(\partial_k F_{il} \right) - 2 \gamma^{ij} \gamma^{kl} \Gamma^{m}_{\ uj} F_{um} \left(\partial_k F_{il} \right)\\
		= & - 2 \Gamma^{z}_{\ uz} F_{uz} \left(\partial_u F_{uz} \right) - 2 \gamma^{ij} \Gamma^{z}_{\ uj} F_{uz} \left(\partial_u F_{iz} \right) - 2 \gamma^{ij} \Gamma^{z}_{\ uj} F_{uz} \left(\partial_z F_{iu} \right)\\
		& - 2 \gamma^{ij} \gamma^{kl} \Gamma^{z}_{\ uj} F_{uz} \left(\partial_k F_{il} \right)\,.
	\end{split}
\end{equation}
Therefore, the seventh term of Eq. (\ref{hkk2fourth}) is obtained as 
\begin{equation}
	\begin{split}
		- 2 k^a k^b g^{ce} g^{df} \Gamma^{g}_{\ ae} F_{bg} \left(\partial_d F_{cf} \right) = 0\,.
	\end{split}
\end{equation}

The eighth term of Eq. (\ref{hkk2fourth}) is
\begin{equation}
	\begin{split}
		& 2 k^a k^b g^{ce} g^{df} \Gamma^{g}_{\ ae} F_{bg} \Gamma^{h}_{\ dc} F_{hf} = 2 g^{ce} g^{df} \Gamma^{g}_{\ ue} F_{ug} \Gamma^{h}_{\ dc} F_{hf}\\
		= & 2 \Gamma^{g}_{\ uz} F_{ug} \Gamma^{h}_{\ uu} F_{hz} + 2 \Gamma^{g}_{\ uz} F_{ug} \Gamma^{h}_{\ zu} F_{hu} + 2 \gamma^{ij} \Gamma^{g}_{\ uz} F_{ug} \Gamma^{h}_{\ iu} F_{hj}\\
		& + 2 \Gamma^{g}_{\ uu} F_{ug} \Gamma^{h}_{\ uz} F_{hz} + 2 \Gamma^{g}_{\ uu} F_{ug} \Gamma^{h}_{\ zz} F_{hu} + 2 \gamma^{ij} \Gamma^{g}_{\ uu} F_{ug} \Gamma^{h}_{\ iz} F_{hj}\\
		& + 2 \gamma^{ij} \Gamma^{g}_{\ uj} F_{ug} \Gamma^{h}_{\ ui} F_{hz} + 2 \gamma^{ij} \Gamma^{g}_{\ uj} F_{ug} \Gamma^{h}_{\ zi} F_{hu} + 2 \gamma^{ij} \gamma^{kl} \Gamma^{g}_{\ uj} F_{ug} \Gamma^{h}_{\ ki} F_{hl}\\
		= & 2 \Gamma^{g}_{\ uz} F_{ug} \Gamma^{h}_{\ zu} F_{hu} + 2 \gamma^{ij} \Gamma^{g}_{\ uz} F_{ug} \Gamma^{h}_{\ iu} F_{hj} + 2 \gamma^{ij} \Gamma^{g}_{\ uj} F_{ug} \Gamma^{h}_{\ ui} F_{hz}\\
		& + 2 \gamma^{ij} \Gamma^{g}_{\ uj} F_{ug} \Gamma^{h}_{\ zi} F_{hu} + 2 \gamma^{ij} \gamma^{kl} \Gamma^{g}_{\ uj} F_{ug} \Gamma^{h}_{\ ki} F_{hl}\,.
	\end{split}
\end{equation}
The index $g$ should be further expanded.
\begin{equation}\label{hkk2fourtheighth}
	\begin{split}
		& 2 \Gamma^{g}_{\ uz} F_{ug} \Gamma^{h}_{\ zu} F_{hu} + 2 \gamma^{ij} \Gamma^{g}_{\ uz} F_{ug} \Gamma^{h}_{\ iu} F_{hj} + 2 \gamma^{ij} \Gamma^{g}_{\ uj} F_{ug} \Gamma^{h}_{\ ui} F_{hz}\\
		& + 2 \gamma^{ij} \Gamma^{g}_{\ uj} F_{ug} \Gamma^{h}_{\ zi} F_{hu} + 2 \gamma^{ij} \gamma^{kl} \Gamma^{g}_{\ uj} F_{ug} \Gamma^{h}_{\ ki} F_{hl}\\
		= & 2 \Gamma^{u}_{\ uz} F_{uu} \Gamma^{h}_{\ zu} F_{hu} + 2 \Gamma^{z}_{\ uz} F_{uz} \Gamma^{h}_{\ zu} F_{hu} + 2 \Gamma^{i}_{\ uz} F_{ui} \Gamma^{h}_{\ zu} F_{hu}\\
		& + 2 \gamma^{ij} \Gamma^{u}_{\ uz} F_{uu} \Gamma^{h}_{\ iu} F_{hj} + 2 \gamma^{ij} \Gamma^{z}_{\ uz} F_{uz} \Gamma^{h}_{\ iu} F_{hj} + 2 \gamma^{ij} \Gamma^{k}_{\ uz} F_{uk} \Gamma^{h}_{\ iu} F_{hj}\\
		& + 2 \gamma^{ij} \Gamma^{u}_{\ uj} F_{uu} \Gamma^{h}_{\ ui} F_{hz} + 2 \gamma^{ij} \Gamma^{z}_{\ uj} F_{uz} \Gamma^{h}_{\ ui} F_{hz} + 2 \gamma^{ij} \Gamma^{k}_{\ uj} F_{uk} \Gamma^{h}_{\ ui} F_{hz}\\
		& + 2 \gamma^{ij} \Gamma^{u}_{\ uj} F_{uu} \Gamma^{h}_{\ zi} F_{hu} + 2 \gamma^{ij} \Gamma^{z}_{\ uj} F_{uz} \Gamma^{h}_{\ zi} F_{hu} + 2 \gamma^{ij} \Gamma^{k}_{\ uj} F_{uk} \Gamma^{h}_{\ zi} F_{hu}\\
		& + 2 \gamma^{ij} \gamma^{kl} \Gamma^{u}_{\ uj} F_{uu} \Gamma^{h}_{\ ki} F_{hl} + 2 \gamma^{ij} \gamma^{kl} \Gamma^{z}_{\ uj} F_{uz} \Gamma^{h}_{\ ki} F_{hl} + 2 \gamma^{ij} \gamma^{kl} \Gamma^{m}_{\ uj} F_{um} \Gamma^{h}_{\ ki} F_{hl}\\
		= & 2 \Gamma^{z}_{\ uz} F_{uz} \Gamma^{h}_{\ zu} F_{hu} + 2 \gamma^{ij} \Gamma^{z}_{\ uz} F_{uz} \Gamma^{h}_{\ iu} F_{hj} + 2 \gamma^{ij} \Gamma^{z}_{\ uj} F_{uz} \Gamma^{h}_{\ ui} F_{hz}\\
		& + 2 \gamma^{ij} \Gamma^{z}_{\ uj} F_{uz} \Gamma^{h}_{\ zi} F_{hu} + 2 \gamma^{ij} \gamma^{kl} \Gamma^{z}_{\ uj} F_{uz} \Gamma^{h}_{\ ki} F_{hl}\,.
	\end{split}
\end{equation}
The repeated index $h$ should be further expanded. The first term of Eq. (\ref{hkk2fourtheighth}) is
\begin{equation}
	\begin{split}
		2 \Gamma^{z}_{\ uz} F_{uz} \Gamma^{h}_{\ zu} F_{hu} = 0\,.
	\end{split}
\end{equation}
The second term of Eq. (\ref{hkk2fourtheighth}) is
\begin{equation}
	\begin{split}
		2 \gamma^{ij} \Gamma^{z}_{\ uz} F_{uz} \Gamma^{h}_{\ iu} F_{hj} = 0\,.
	\end{split}
\end{equation}
The third term of Eq. (\ref{hkk2fourtheighth}) is
\begin{equation}
	\begin{split}
		2 \gamma^{ij} \Gamma^{z}_{\ uj} F_{uz} \Gamma^{h}_{\ ui} F_{hz} = 0\,.
	\end{split}
\end{equation}
The fourth term of Eq. (\ref{hkk2fourtheighth}) is
\begin{equation}
	\begin{split}
		2 \gamma^{ij} \Gamma^{z}_{\ uj} F_{uz} \Gamma^{h}_{\ zi} F_{hu} = 0\,.
	\end{split}
\end{equation}
The fifth term of Eq. (\ref{hkk2fourtheighth}) is
\begin{equation}
	\begin{split}
		2 \gamma^{ij} \gamma^{kl} \Gamma^{z}_{\ uj} F_{uz} \Gamma^{h}_{\ ki} F_{hl} = 0\,.
	\end{split}
\end{equation}
Therefore, the eighth term of Eq. (\ref{hkk2fourth}) is obtained as 
\begin{equation}
	\begin{split}
		2 k^a k^b g^{ce} g^{df} \Gamma^{g}_{\ ae} F_{bg} \Gamma^{h}_{\ dc} F_{hf} = 0\,.
	\end{split}
\end{equation}

The ninth term of Eq. (\ref{hkk2fourth}) is
\begin{equation}
	\begin{split}
		& 2 k^a k^b g^{ce} g^{df} \Gamma^{g}_{\ ae} F_{bg} \Gamma^{h}_{\ df} F_{ch} = 2 g^{ce} g^{df} \Gamma^{g}_{\ ue} F_{ug} \Gamma^{h}_{\ df} F_{ch}\\
		= & 2 \Gamma^{g}_{\ uz} F_{ug} \Gamma^{h}_{\ uz} F_{uh} + 2 \Gamma^{g}_{\ uz} F_{ug} \Gamma^{h}_{\ zu} F_{uh} + 2 \gamma^{ij} \Gamma^{g}_{\ uz} F_{ug} \Gamma^{h}_{\ ij} F_{uh}\\
		& + 2 \Gamma^{g}_{\ uu} F_{ug} \Gamma^{h}_{\ uz} F_{zh} + 2 \Gamma^{g}_{\ uu} F_{ug} \Gamma^{h}_{\ zu} F_{zh} + 2 \gamma^{ij} \Gamma^{g}_{\ uu} F_{ug} \Gamma^{h}_{\ ij} F_{zh}\\
		& + 2 \gamma^{ij} \Gamma^{g}_{\ uj} F_{ug} \Gamma^{h}_{\ uz} F_{ih} + 2 \gamma^{ij} \Gamma^{g}_{\ uj} F_{ug} \Gamma^{h}_{\ zu} F_{ih} + 2 \gamma^{ij} \gamma^{kl} \Gamma^{g}_{\ uj} F_{ug} \Gamma^{h}_{\ kl} F_{ih}\\
		= & 2 \Gamma^{g}_{\ uz} F_{ug} \Gamma^{h}_{\ uz} F_{uh} + 2 \Gamma^{g}_{\ uz} F_{ug} \Gamma^{h}_{\ zu} F_{uh} + 2 \gamma^{ij} \Gamma^{g}_{\ uz} F_{ug} \Gamma^{h}_{\ ij} F_{uh}\\
		& + 2 \gamma^{ij} \Gamma^{g}_{\ uj} F_{ug} \Gamma^{h}_{\ uz} F_{ih} + 2 \gamma^{ij} \Gamma^{g}_{\ uj} F_{ug} \Gamma^{h}_{\ zu} F_{ih} + 2 \gamma^{ij} \gamma^{kl} \Gamma^{g}_{\ uj} F_{ug} \Gamma^{h}_{\ kl} F_{ih}\,.
	\end{split}
\end{equation}
The index $g$ should be further expanded.
\begin{equation}\label{hkk2fourthninth}
	\begin{split}
		& 2 \Gamma^{g}_{\ uz} F_{ug} \Gamma^{h}_{\ uz} F_{uh} + 2 \Gamma^{g}_{\ uz} F_{ug} \Gamma^{h}_{\ zu} F_{uh} + 2 \gamma^{ij} \Gamma^{g}_{\ uz} F_{ug} \Gamma^{h}_{\ ij} F_{uh}\\
		& + 2 \gamma^{ij} \Gamma^{g}_{\ uj} F_{ug} \Gamma^{h}_{\ uz} F_{ih} + 2 \gamma^{ij} \Gamma^{g}_{\ uj} F_{ug} \Gamma^{h}_{\ zu} F_{ih} + 2 \gamma^{ij} \gamma^{kl} \Gamma^{g}_{\ uj} F_{ug} \Gamma^{h}_{\ kl} F_{ih}\\
		= & 2 \Gamma^{u}_{\ uz} F_{uu} \Gamma^{h}_{\ uz} F_{uh} + 2 \Gamma^{z}_{\ uz} F_{uz} \Gamma^{h}_{\ uz} F_{uh} + 2 \Gamma^{i}_{\ uz} F_{ui} \Gamma^{h}_{\ uz} F_{uh}\\
		& + 2 \Gamma^{u}_{\ uz} F_{uu} \Gamma^{h}_{\ zu} F_{uh} + 2 \Gamma^{z}_{\ uz} F_{uz} \Gamma^{h}_{\ zu} F_{uh} + 2 \Gamma^{i}_{\ uz} F_{ui} \Gamma^{h}_{\ zu} F_{uh}\\
		& + 2 \gamma^{ij} \Gamma^{u}_{\ uz} F_{uu} \Gamma^{h}_{\ ij} F_{uh} + 2 \gamma^{ij} \Gamma^{z}_{\ uz} F_{uz} \Gamma^{h}_{\ ij} F_{uh} + 2 \gamma^{ij} \Gamma^{k}_{\ uz} F_{uk} \Gamma^{h}_{\ ij} F_{uh}\\
		& + 2 \gamma^{ij} \Gamma^{u}_{\ uj} F_{uu} \Gamma^{h}_{\ uz} F_{ih} + 2 \gamma^{ij} \Gamma^{z}_{\ uj} F_{uz} \Gamma^{h}_{\ uz} F_{ih} + 2 \gamma^{ij} \Gamma^{k}_{\ uj} F_{uk} \Gamma^{h}_{\ uz} F_{ih}\\
		& + 2 \gamma^{ij} \Gamma^{u}_{\ uj} F_{uu} \Gamma^{h}_{\ zu} F_{ih} + 2 \gamma^{ij} \Gamma^{z}_{\ uj} F_{uz} \Gamma^{h}_{\ zu} F_{ih} + 2 \gamma^{ij} \Gamma^{k}_{\ uj} F_{uk} \Gamma^{h}_{\ zu} F_{ih}\\
		& + 2 \gamma^{ij} \gamma^{kl} \Gamma^{u}_{\ uj} F_{uu} \Gamma^{h}_{\ kl} F_{ih} + 2 \gamma^{ij} \gamma^{kl} \Gamma^{z}_{\ uj} F_{uz} \Gamma^{h}_{\ kl} F_{ih} + 2 \gamma^{ij} \gamma^{kl} \Gamma^{m}_{\ uj} F_{um} \Gamma^{h}_{\ kl} F_{ih}\\
		= & 2 \Gamma^{z}_{\ uz} F_{uz} \Gamma^{h}_{\ uz} F_{uh} + 2 \Gamma^{z}_{\ uz} F_{uz} \Gamma^{h}_{\ zu} F_{uh} + 2 \gamma^{ij} \Gamma^{z}_{\ uz} F_{uz} \Gamma^{h}_{\ ij} F_{uh}\\
		& + 2 \gamma^{ij} \Gamma^{z}_{\ uj} F_{uz} \Gamma^{h}_{\ uz} F_{ih} + 2 \gamma^{ij} \Gamma^{z}_{\ uj} F_{uz} \Gamma^{h}_{\ zu} F_{ih} + 2 \gamma^{ij} \gamma^{kl} \Gamma^{z}_{\ uj} F_{uz} \Gamma^{h}_{\ kl} F_{ih}\,.
	\end{split}
\end{equation}
The repeated index $h$ should be further expanded. The first term of Eq. (\ref{hkk2fourthninth}) is 
\begin{equation}
	\begin{split}
		2 \Gamma^{z}_{\ uz} F_{uz} \Gamma^{h}_{\ uz} F_{uh} = 0\,.
	\end{split}
\end{equation}
The second term of Eq. (\ref{hkk2fourthninth}) is 
\begin{equation}
	\begin{split}
		2 \Gamma^{z}_{\ uz} F_{uz} \Gamma^{h}_{\ zu} F_{uh} = 0\,.
	\end{split}
\end{equation}
The third term of Eq. (\ref{hkk2fourthninth}) is
\begin{equation}
	\begin{split}
		2 \gamma^{ij} \Gamma^{z}_{\ uz} F_{uz} \Gamma^{h}_{\ ij} F_{uh} = 0\,.
	\end{split}
\end{equation}
The fourth term of Eq. (\ref{hkk2fourthninth}) is
\begin{equation}
	\begin{split}
		2 \gamma^{ij} \Gamma^{z}_{\ uj} F_{uz} \Gamma^{h}_{\ uz} F_{ih} = 0\,.
	\end{split}
\end{equation}
The fifth term of Eq. (\ref{hkk2fourthninth}) is
\begin{equation}
	\begin{split}
		2 \gamma^{ij} \Gamma^{z}_{\ uj} F_{uz} \Gamma^{h}_{\ zu} F_{ih} = 0\,.
	\end{split}
\end{equation}
The sixth term of Eq. (\ref{hkk2fourthninth}) is
\begin{equation}
	\begin{split}
		2 \gamma^{ij} \gamma^{kl} \Gamma^{z}_{\ uj} F_{uz} \Gamma^{h}_{\ kl} F_{ih} = 0\,.
	\end{split}
\end{equation}
Therefore, the ninth term of Eq. (\ref{hkk2fourth}) is
\begin{equation}
	\begin{split}
		2 k^a k^b g^{ce} g^{df} \Gamma^{g}_{\ ae} F_{bg} \Gamma^{h}_{\ df} F_{ch} = 0\,.
	\end{split}
\end{equation}

Finally, the fourth term of Eq. (\ref{rewrittenhkk2}) is
\begin{equation}
	\begin{split}
		& 2 k^a k^b \nabla_a F_{b}^{\ c} \nabla_d F_{c}^{\ d}\\
		= & 2 \left(\partial_u F_{uz} \right) \left(\partial_u F_{uz} \right) + 2 \gamma^{ij} \left(\partial_u F_{uj} \right) \left(\partial_u F_{iz} \right) + 2 \gamma^{ij} \left(\partial_u F_{uj} \right) \left(\partial_z F_{iu} \right)\\
		& + 2 \gamma^{ij} \gamma^{kl} \left(\partial_u F_{uj} \right) \left(\partial_k F_{il} \right) + \gamma^{ij} \beta_i \left(\partial_u F_{uj} \right) F_{uz} - \gamma^{ij} \beta_i \left(\partial_u F_{uj} \right) F_{zu}\\
		& - 2 \gamma^{ij} \gamma^{kl} \left(\partial_u F_{uj} \right) \hat{\Gamma}^{m}_{\ ki} F_{ml} - 2 \gamma^{ij} \gamma^{kl} \beta_l \left(\partial_u F_{uj} \right) F_{ik} - 2 \gamma^{ij} \gamma^{kl} \left(\partial_u F_{uj} \right) \hat{\Gamma}^{m}_{\ kl} F_{im}\\
		= & 2 \left(\partial_u F_{uz} \right) \left(\partial_u F_{uz} \right) + 2 \gamma^{ij} \left(\partial_u F_{uj} \right) \left(\partial_u F_{iz} \right) + 2 \gamma^{ij} \left(\partial_u F_{uj} \right) \left(\partial_z F_{iu} \right)\\
		& + 2 \gamma^{ij} \gamma^{kl} \left(\partial_u F_{uj} \right) \left(\partial_k F_{il} \right) + 2 \gamma^{ij} \beta_i \left(\partial_u F_{uj} \right) F_{uz} - 2 \gamma^{ij} \gamma^{kl} \left(\partial_u F_{uj} \right) \hat{\Gamma}^{m}_{\ ki} F_{ml}\\
		& - 2 \gamma^{ij} \gamma^{kl} \beta_l \left(\partial_u F_{uj} \right) F_{ik} - 2 \gamma^{ij} \gamma^{kl} \left(\partial_u F_{uj} \right) \hat{\Gamma}^{m}_{\ kl} F_{im}\,.
	\end{split}
\end{equation}

The fifth term of Eq. (\ref{rewrittenhkk2}) is
\begin{equation}\label{hkk2fifth}
	\begin{split}
		& 2 k^a k^b \nabla_a F_{cd} \nabla^d F_{b}^{\ c} = 2 k^a k^b g^{ce} g^{df} \nabla_a F_{cd} \nabla_f F_{be}\\
		= & 2 k^a k^b g^{ce} g^{df} \left(\partial_a F_{cd} \right) \left(\partial_f F_{be} \right) - 2 k^a k^b g^{ce} g^{df} \left(\partial_a F_{cd} \right) \Gamma^{h}_{\ fb} F_{he}\\
		& - 2 k^a k^b g^{ce} g^{df} \left(\partial_a F_{cd} \right) \Gamma^{h}_{\ fe} F_{bh} - 2 k^a k^b g^{ce} g^{df} \Gamma^{g}_{\ ac} F_{gd} \left(\partial_f F_{be} \right)\\
		& + 2 k^a k^b g^{ce} g^{df} \Gamma^{g}_{\ ac} F_{gd} \Gamma^{h}_{\ fb} F_{he} + 2 k^a k^b g^{ce} g^{df} \Gamma^{g}_{\ ac} F_{gd} \Gamma^{h}_{\ fe} F_{bh}\\
		& - 2 k^a k^b g^{ce} g^{df} \Gamma^{g}_{\ ad} F_{cg} \left(\partial_f F_{be} \right) + 2 k^a k^b g^{ce} g^{df} \Gamma^{g}_{\ ad} F_{cg} \Gamma^{h}_{\ fb} F_{he}\\
		& + 2 k^a k^b g^{ce} g^{df} \Gamma^{g}_{\ ad} F_{cg} \Gamma^{h}_{\ fe} F_{bh}\,.
	\end{split}
\end{equation}

The first term of Eq. (\ref{hkk2fifth}) is 
\begin{equation}
	\begin{split}
		& 2 k^a k^b g^{ce} g^{df} \left(\partial_a F_{cd} \right) \left(\partial_f F_{be} \right) = 2 g^{ce} g^{df} \left(\partial_u F_{cd} \right) \left(\partial_f F_{ue} \right)\\
		= & 2 \left(\partial_u F_{uz} \right) \left(\partial_u F_{uz} \right) + 2 \gamma^{ij} \left(\partial_u F_{ui} \right) \left(\partial_j F_{uz} \right) + 2 \gamma^{ij} \left(\partial_u F_{iu} \right) \left(\partial_z F_{uj} \right)\\
		& + 2 \gamma^{ij} \left(\partial_u F_{iz} \right) \left(\partial_u F_{uj} \right) + 2 \gamma^{ij} \gamma^{kl} \left(\partial_u F_{ik} \right) \left(\partial_l F_{uj} \right)\\
		= & 2 \left(\partial_u F_{uz} \right) \left(\partial_u F_{uz} \right) + 2 \gamma^{ij} \left(\partial_u F_{ui} \right) \left(\partial_j F_{uz} \right) + 2 \gamma^{ij} \left(\partial_u F_{iu} \right) \left(\partial_z F_{uj} \right)\\
		& + 2 \gamma^{ij} \left(\partial_u F_{iz} \right) \left(\partial_u F_{uj} \right)\,.
	\end{split}
\end{equation}
Therefore, the first term of Eq. (\ref{hkk2fifth}) is obtained as 
\begin{equation}
	\begin{split}
		& 2 k^a k^b g^{ce} g^{df} \left(\partial_a F_{cd} \right) \left(\partial_f F_{be} \right)\\
		= & 2 \left(\partial_u F_{uz} \right) \left(\partial_u F_{uz} \right) + 2 \gamma^{ij} \left(\partial_u F_{ui} \right) \left(\partial_j F_{uz} \right) + 2 \gamma^{ij} \left(\partial_u F_{iu} \right) \left(\partial_z F_{uj} \right)\\
		& + 2 \gamma^{ij} \left(\partial_u F_{iz} \right) \left(\partial_u F_{uj} \right)\,.
	\end{split}
\end{equation}

The second term of Eq. (\ref{hkk2fifth}) is
\begin{equation}\label{hkk2fifthsecond}
	\begin{split}
		& - 2 k^a k^b g^{ce} g^{df} \left(\partial_a F_{cd} \right) \Gamma^{h}_{\ fb} F_{he} = - 2 g^{ce} g^{df} \left(\partial_u F_{cd} \right) \Gamma^{h}_{\ fu} F_{he}\\
		= & - 2 \left(\partial_u F_{uz} \right) \Gamma^{h}_{\ uu} F_{hz} - 2 \gamma^{ij} \left(\partial_u F_{ui} \right) \Gamma^{h}_{\ ju} F_{hz} - 2 \left(\partial_u F_{zu} \right) \Gamma^{h}_{\ zu} F_{hu}\\
		& - 2 \gamma^{ij} \left(\partial_u F_{zi} \right) \Gamma^{h}_{\ ju} F_{hu} - 2 \gamma^{ij} \left(\partial_u F_{iu} \right) \Gamma^{h}_{\ zu} F_{hj} - 2 \gamma^{ij} \left(\partial_u F_{iz} \right) \Gamma^{h}_{\ uu} F_{hj}\\
		& - 2 \gamma^{ij} \gamma^{kl} \left(\partial_u F_{ik} \right) \Gamma^{h}_{\ lu} F_{hj}\\
		= & - 2 \gamma^{ij} \left(\partial_u F_{ui} \right) \Gamma^{h}_{\ ju} F_{hz} - 2 \left(\partial_u F_{zu} \right) \Gamma^{h}_{\ zu} F_{hu} - 2 \gamma^{ij} \left(\partial_u F_{zi} \right) \Gamma^{h}_{\ ju} F_{hu}\\
		& - 2 \gamma^{ij} \left(\partial_u F_{iu} \right) \Gamma^{h}_{\ zu} F_{hj} - 2 \gamma^{ij} \gamma^{kl} \left(\partial_u F_{ik} \right) \Gamma^{h}_{\ lu} F_{hj}\,.
	\end{split}
\end{equation}
The repeated index $h$ should be further expanded. The first term of Eq. (\ref{hkk2fifthsecond}) is
\begin{equation}
	\begin{split}
		& - 2 \gamma^{ij} \left(\partial_u F_{ui} \right) \Gamma^{h}_{\ ju} F_{hz}\\
		= & - 2 \gamma^{ij} \left(\partial_u F_{ui} \right) \Gamma^{u}_{\ ju} F_{uz} - 2 \gamma^{ij} \left(\partial_u F_{ui} \right) \Gamma^{z}_{\ ju} F_{zz} - 2 \gamma^{ij} \left(\partial_u F_{ui} \right) \Gamma^{k}_{\ ju} F_{kz}\\
		= & - 2 \gamma^{ij} \left(\partial_u F_{ui} \right) \Gamma^{u}_{\ ju} F_{uz} - 2 \gamma^{ij} \left(\partial_u F_{ui} \right) \Gamma^{k}_{\ ju} F_{kz}\\
		= & \gamma^{ij} \beta_j \left(\partial_u F_{ui} \right) F_{uz} - \gamma^{ij} \gamma^{kl} \left(\partial_u F_{ui} \right) \left(\partial_u \gamma_{jl} \right) F_{kz}\,.
	\end{split}
\end{equation}
The second term of Eq. (\ref{hkk2fifthsecond}) is
\begin{equation}
	\begin{split}
		& - 2 \left(\partial_u F_{zu} \right) \Gamma^{h}_{\ zu} F_{hu}\\
		= & - 2 \left(\partial_u F_{zu} \right) \Gamma^{u}_{\ zu} F_{uu} - 2 \left(\partial_u F_{zu} \right) \Gamma^{z}_{\ zu} F_{zu} - 2 \left(\partial_u F_{zu} \right) \Gamma^{i}_{\ zu} F_{iu}\\
		= & - 2 \left(\partial_u F_{zu} \right) \Gamma^{z}_{\ zu} F_{zu}\\
		= & 0\,.
	\end{split}
\end{equation}
The third term of Eq. (\ref{hkk2fifthsecond}) is
\begin{equation}
	\begin{split}
		& - 2 \gamma^{ij} \left(\partial_u F_{zi} \right) \Gamma^{h}_{\ ju} F_{hu}\\
		= & - 2 \gamma^{ij} \left(\partial_u F_{zi} \right) \Gamma^{u}_{\ ju} F_{uu} - 2 \gamma^{ij} \left(\partial_u F_{zi} \right) \Gamma^{z}_{\ ju} F_{zu} - 2 \gamma^{ij} \left(\partial_u F_{zi} \right) \Gamma^{k}_{\ ju} F_{ku}\\
		= & - 2 \gamma^{ij} \left(\partial_u F_{zi} \right) \Gamma^{z}_{\ ju} F_{zu}\\
		= & 0\,.
	\end{split}
\end{equation}
The fourth term of Eq. (\ref{hkk2fifthsecond}) is
\begin{equation}
	\begin{split}
		& - 2 \gamma^{ij} \left(\partial_u F_{iu} \right) \Gamma^{h}_{\ zu} F_{hj}\\
		= & - 2 \gamma^{ij} \left(\partial_u F_{iu} \right) \Gamma^{u}_{\ zu} F_{uj} - 2 \gamma^{ij} \left(\partial_u F_{iu} \right) \Gamma^{z}_{\ zu} F_{zj} - 2 \gamma^{ij} \left(\partial_u F_{iu} \right) \Gamma^{k}_{\ zu} F_{kj}\\
		= & - 2 \gamma^{ij} \left(\partial_u F_{iu} \right) \Gamma^{z}_{\ zu} F_{zj} - 2 \gamma^{ij} \left(\partial_u F_{iu} \right) \Gamma^{k}_{\ zu} F_{kj}\\
		= & - \gamma^{ij} \gamma^{kl} \beta_l \left(\partial_u F_{iu} \right) F_{kj}\,.
	\end{split}
\end{equation}
The fifth term of Eq. (\ref{hkk2fifthsecond}) is
\begin{equation}
	\begin{split}
		& - 2 \gamma^{ij} \gamma^{kl} \left(\partial_u F_{ik} \right) \Gamma^{h}_{\ lu} F_{hj}\\
		= & - 2 \gamma^{ij} \gamma^{kl} \left(\partial_u F_{ik} \right) \Gamma^{u}_{\ lu} F_{uj} - 2 \gamma^{ij} \gamma^{kl} \left(\partial_u F_{ik} \right) \Gamma^{z}_{\ lu} F_{zj} - 2 \gamma^{ij} \gamma^{kl} \left(\partial_u F_{ik} \right) \Gamma^{m}_{\ lu} F_{mj}\\
		= & - 2 \gamma^{ij} \gamma^{kl} \left(\partial_u F_{ik} \right) \Gamma^{z}_{\ lu} F_{zj} - 2 \gamma^{ij} \gamma^{kl} \left(\partial_u F_{ik} \right) \Gamma^{m}_{\ lu} F_{mj}\\
		= & - \gamma^{ij} \gamma^{kl} \gamma^{mn} \left(\partial_u F_{ik} \right) \left(\partial_u \gamma_{ln} \right) F_{mj}\,.
	\end{split}
\end{equation}
Therefore, the second term of Eq. (\ref{hkk2fifth}) is obtained as 
\begin{equation}
	\begin{split}
		& - 2 k^a k^b g^{ce} g^{df} \left(\partial_a F_{cd} \right) \Gamma^{h}_{\ fb} F_{he}\\
		= & \gamma^{ij} \beta_j \left(\partial_u F_{ui} \right) F_{uz} - \gamma^{ij} \gamma^{kl} \left(\partial_u F_{ui} \right) \left(\partial_u \gamma_{jl} \right) F_{kz} - \gamma^{ij} \gamma^{kl} \beta_l \left(\partial_u F_{iu} \right) F_{kj}\\
		& - \gamma^{ij} \gamma^{kl} \gamma^{mn} \left(\partial_u F_{ik} \right) \left(\partial_u \gamma_{ln} \right) F_{mj}\\
		= & \gamma^{ij} \beta_j \left(\partial_u F_{ui} \right) F_{uz} - 2 \gamma^{ij} \gamma^{kl} \left(\partial_u F_{ui} \right) K_{jl} F_{kz} - \gamma^{ij} \gamma^{kl} \beta_l \left(\partial_u F_{iu} \right) F_{kj}\\
		& - 2 \gamma^{ij} \gamma^{kl} \gamma^{mn} \left(\partial_u F_{ik} \right) K_{ln} F_{mj}\\
		= & \gamma^{ij} \beta_j \left(\partial_u F_{ui} \right) F_{uz} - \gamma^{ij} \gamma^{kl} \beta_l \left(\partial_u F_{iu} \right) F_{kj}\,.
	\end{split}
\end{equation}

The third term of Eq. (\ref{hkk2fifth}) is
\begin{equation}\label{hkk2fifththird}
	\begin{split}
		& - 2 k^a k^b g^{ce} g^{df} \left(\partial_a F_{cd} \right) \Gamma^{h}_{\ fe} F_{bh} = - 2 g^{ce} g^{df} \left(\partial_u F_{cd} \right) \Gamma^{h}_{\ fe} F_{uh}\\
		= & - 2 \left(\partial_u F_{uz} \right) \Gamma^{h}_{\ uz} F_{uh} - 2 \gamma^{ij} \left(\partial_u F_{ui} \right) \Gamma^{h}_{\ jz} F_{uh} - 2 \left(\partial_u F_{zu} \right) \Gamma^{h}_{\ zu} F_{uh}\\
		& - 2 \gamma^{ij} \left(\partial_u F_{zi} \right) \Gamma^{h}_{\ ju} F_{uh} - 2 \gamma^{ij} \left(\partial_u F_{iu} \right) \Gamma^{h}_{\ zj} F_{uh} - 2 \gamma^{ij} \left(\partial_u F_{iz} \right) \Gamma^{h}_{\ uj} F_{uh}\\
		& - 2 \gamma^{ij} \gamma^{kl} \left(\partial_u F_{ik} \right) \Gamma^{h}_{\ lj} F_{uh}\,.
	\end{split}
\end{equation}
The first term of Eq. (\ref{hkk2fifththird}) is 
\begin{equation}
	\begin{split}
		& - 2 \left(\partial_u F_{uz} \right) \Gamma^{h}_{\ uz} F_{uh}\\
		= & - 2 \left(\partial_u F_{uz} \right) \Gamma^{u}_{\ uz} F_{uu} - 2 \left(\partial_u F_{uz} \right) \Gamma^{z}_{\ uz} F_{uz} - 2 \left(\partial_u F_{uz} \right) \Gamma^{i}_{\ uz} F_{ui}\\
		= & - 2 \left(\partial_u F_{uz} \right) \Gamma^{z}_{\ uz} F_{uz}\\
		= & 0\,.
	\end{split}
\end{equation}
The second term of Eq. (\ref{hkk2fifththird}) is 
\begin{equation}
	\begin{split}
		& - 2 \gamma^{ij} \left(\partial_u F_{ui} \right) \Gamma^{h}_{\ jz} F_{uh}\\
		= & - 2 \gamma^{ij} \left(\partial_u F_{ui} \right) \Gamma^{u}_{\ jz} F_{uu} - 2 \gamma^{ij} \left(\partial_u F_{ui} \right) \Gamma^{z}_{\ jz} F_{uz} - 2 \gamma^{ij} \left(\partial_u F_{ui} \right) \Gamma^{k}_{\ jz} F_{uk}\\
		= & - 2 \gamma^{ij} \left(\partial_u F_{ui} \right) \Gamma^{z}_{\ jz} F_{uz}\\
		= & - \gamma^{ij} \beta_j \left(\partial_u F_{ui} \right) F_{uz}\,.
	\end{split}
\end{equation}
The third term of Eq. (\ref{hkk2fifththird}) is 
\begin{equation}
	\begin{split}
		& - 2 \left(\partial_u F_{zu} \right) \Gamma^{h}_{\ zu} F_{uh}\\
		= & - 2 \left(\partial_u F_{zu} \right) \Gamma^{u}_{\ zu} F_{uu} - 2 \left(\partial_u F_{zu} \right) \Gamma^{z}_{\ zu} F_{uz} - 2 \left(\partial_u F_{zu} \right) \Gamma^{i}_{\ zu} F_{ui}\\
		= & - 2 \left(\partial_u F_{zu} \right) \Gamma^{z}_{\ zu} F_{uz}\\
		= & 0\,.
	\end{split}
\end{equation}
The fourth term of Eq. (\ref{hkk2fifththird}) is
\begin{equation}
	\begin{split}
		& - 2 \gamma^{ij} \left(\partial_u F_{zi} \right) \Gamma^{h}_{\ ju} F_{uh}\\
		= & - 2 \gamma^{ij} \left(\partial_u F_{zi} \right) \Gamma^{u}_{\ ju} F_{uu} - 2 \gamma^{ij} \left(\partial_u F_{zi} \right) \Gamma^{z}_{\ ju} F_{uz} - 2 \gamma^{ij} \left(\partial_u F_{zi} \right) \Gamma^{k}_{\ ju} F_{uk}\\
		= & - 2 \gamma^{ij} \left(\partial_u F_{zi} \right) \Gamma^{z}_{\ ju} F_{uz}\\
		= & 0\,.
	\end{split}
\end{equation}
The fifth term of Eq. (\ref{hkk2fifththird}) is
\begin{equation}
	\begin{split}
		& - 2 \gamma^{ij} \left(\partial_u F_{iu} \right) \Gamma^{h}_{\ zj} F_{uh}\\
		= & - 2 \gamma^{ij} \left(\partial_u F_{iu} \right) \Gamma^{u}_{\ zj} F_{uu} - 2 \gamma^{ij} \left(\partial_u F_{iu} \right) \Gamma^{z}_{\ zj} F_{uz} - 2 \gamma^{ij} \left(\partial_u F_{iu} \right) \Gamma^{k}_{\ zj} F_{uk}\\
		= & - 2 \gamma^{ij} \left(\partial_u F_{iu} \right) \Gamma^{z}_{\ zj} F_{uz}\\
		= & - \gamma^{ij} \beta_j \left(\partial_u F_{iu} \right) F_{uz}\,.
	\end{split}
\end{equation}
The sixth term of Eq. (\ref{hkk2fifththird}) is
\begin{equation}
	\begin{split}
		& - 2 \gamma^{ij} \left(\partial_u F_{iz} \right) \Gamma^{h}_{\ uj} F_{uh}\\
		= & - 2 \gamma^{ij} \left(\partial_u F_{iz} \right) \Gamma^{u}_{\ uj} F_{uu} - 2 \gamma^{ij} \left(\partial_u F_{iz} \right) \Gamma^{z}_{\ uj} F_{uz} - 2 \gamma^{ij} \left(\partial_u F_{iz} \right) \Gamma^{k}_{\ uj} F_{uk}\\
		= & - 2 \gamma^{ij} \left(\partial_u F_{iz} \right) \Gamma^{z}_{\ uj} F_{uz}\\
		= & 0\,.
	\end{split}
\end{equation}
The seventh term of Eq. (\ref{hkk2fifththird}) is
\begin{equation}
	\begin{split}
		& - 2 \gamma^{ij} \gamma^{kl} \left(\partial_u F_{ik} \right) \Gamma^{h}_{\ lj} F_{uh}\\
		= & - 2 \gamma^{ij} \gamma^{kl} \left(\partial_u F_{ik} \right) \Gamma^{u}_{\ lj} F_{uu} - 2 \gamma^{ij} \gamma^{kl} \left(\partial_u F_{ik} \right) \Gamma^{z}_{\ lj} F_{uz} - 2 \gamma^{ij} \gamma^{kl} \left(\partial_u F_{ik} \right) \Gamma^{m}_{\ lj} F_{um}\\
		= & - 2 \gamma^{ij} \gamma^{kl} \left(\partial_u F_{ik} \right) \Gamma^{z}_{\ lj} F_{uz}\\
		= & \gamma^{ij} \gamma^{kl} \left(\partial_u F_{ik} \right) \left(\partial_u \gamma_{lj} \right) F_{uz}\,.
	\end{split}
\end{equation}
Therefore, the third term of Eq. (\ref{hkk2fifth}) is obtained as 
\begin{equation}
	\begin{split}
		& - 2 k^a k^b g^{ce} g^{df} \left(\partial_a F_{cd} \right) \Gamma^{h}_{\ fe} F_{bh}\\
		= & - \gamma^{ij} \beta_j \left(\partial_u F_{ui} \right) F_{uz} - \gamma^{ij} \beta_j \left(\partial_u F_{iu} \right) F_{uz} + \gamma^{ij} \gamma^{kl} \left(\partial_u F_{ik} \right) \left(\partial_u \gamma_{lj} \right) F_{uz}\\
		= & - \gamma^{ij} \beta_j \left(\partial_u F_{ui} \right) F_{uz} + \gamma^{ij} \beta_j \left(\partial_u F_{ui} \right) F_{uz} + 2 \gamma^{ij} \gamma^{kl} \left(\partial_u F_{ik} \right) K_{lj} F_{uz}\\
		= & 0\,.
	\end{split}
\end{equation}

The fourth term of Eq. (\ref{hkk2fifth}) is
\begin{equation}
	\begin{split}
		& - 2 k^a k^b g^{ce} g^{df} \Gamma^{g}_{\ ac} F_{gd} \left(\partial_f F_{be} \right) = - 2 g^{ce} g^{df} \Gamma^{g}_{\ uc} F_{gd} \left(\partial_f F_{ue} \right)\\
		= & - 2 \Gamma^{g}_{\ uu} F_{gu} \left(\partial_z F_{uz} \right) - 2 \Gamma^{g}_{\ uu} F_{gz} \left(\partial_u F_{uz} \right) - 2 \gamma^{ij} \Gamma^{g}_{\ uu} F_{gi} \left(\partial_j F_{uz} \right)\\
		& - 2 \gamma^{ij} \Gamma^{g}_{\ ui} F_{gu} \left(\partial_z F_{uj} \right) - 2 \gamma^{ij} \Gamma^{g}_{\ ui} F_{gz} \left(\partial_u F_{uj} \right) - 2 \gamma^{ij} \gamma^{kl} \Gamma^{g}_{\ ui} F_{gk} \left(\partial_l F_{uj} \right)\\
		= & - 2 \gamma^{ij} \Gamma^{g}_{\ ui} F_{gu} \left(\partial_z F_{uj} \right) - 2 \gamma^{ij} \Gamma^{g}_{\ ui} F_{gz} \left(\partial_u F_{uj} \right) - 2 \gamma^{ij} \gamma^{kl} \Gamma^{g}_{\ ui} F_{gk} \left(\partial_l F_{uj} \right)\,.
	\end{split}
\end{equation}
The index $g$ should be further expanded.
\begin{equation}\label{hkk2fifthfourth}
	\begin{split}
		& - 2 \gamma^{ij} \Gamma^{g}_{\ ui} F_{gu} \left(\partial_z F_{uj} \right) - 2 \gamma^{ij} \Gamma^{g}_{\ ui} F_{gz} \left(\partial_u F_{uj} \right) - 2 \gamma^{ij} \gamma^{kl} \Gamma^{g}_{\ ui} F_{gk} \left(\partial_l F_{uj} \right)\\
		= & - 2 \gamma^{ij} \Gamma^{u}_{\ ui} F_{uu} \left(\partial_z F_{uj} \right) - 2 \gamma^{ij} \Gamma^{z}_{\ ui} F_{zu} \left(\partial_z F_{uj} \right) - 2 \gamma^{ij} \Gamma^{k}_{\ ui} F_{ku} \left(\partial_z F_{uj} \right)\\
		& - 2 \gamma^{ij} \Gamma^{u}_{\ ui} F_{uz} \left(\partial_u F_{uj} \right) - 2 \gamma^{ij} \Gamma^{z}_{\ ui} F_{zz} \left(\partial_u F_{uj} \right) - 2 \gamma^{ij} \Gamma^{k}_{\ ui} F_{kz} \left(\partial_u F_{uj} \right)\\
		& - 2 \gamma^{ij} \gamma^{kl} \Gamma^{u}_{\ ui} F_{uk} \left(\partial_l F_{uj} \right) - 2 \gamma^{ij} \gamma^{kl} \Gamma^{z}_{\ ui} F_{zk} \left(\partial_l F_{uj} \right) - 2 \gamma^{ij} \gamma^{kl} \Gamma^{m}_{\ ui} F_{mk} \left(\partial_l F_{uj} \right)\\
		= & - 2 \gamma^{ij} \Gamma^{z}_{\ ui} F_{zu} \left(\partial_z F_{uj} \right) - 2 \gamma^{ij} \Gamma^{u}_{\ ui} F_{uz} \left(\partial_u F_{uj} \right) - 2 \gamma^{ij} \Gamma^{k}_{\ ui} F_{kz} \left(\partial_u F_{uj} \right)\\
		& - 2 \gamma^{ij} \gamma^{kl} \Gamma^{z}_{\ ui} F_{zk} \left(\partial_l F_{uj} \right) - 2 \gamma^{ij} \gamma^{kl} \Gamma^{m}_{\ ui} F_{mk} \left(\partial_l F_{uj} \right)\,.
	\end{split}
\end{equation}
Therefore, the fourth term of Eq. (\ref{hkk2fifth}) is obtained as 
\begin{equation}
	\begin{split}
		& - 2 k^a k^b g^{ce} g^{df} \Gamma^{g}_{\ ac} F_{gd} \left(\partial_f F_{be} \right)\\
		= & \gamma^{ij} \beta_i F_{uz} \left(\partial_u F_{uj} \right) - \gamma^{ij} \gamma^{kl} \left(\partial_u \gamma_{il} \right) F_{kz} \left(\partial_u F_{uj} \right) - \gamma^{ij} \gamma^{kl} \gamma^{mn} \left(\partial_u \gamma_{in} \right) F_{mk} \left(\partial_l F_{uj} \right)\\
		= & \gamma^{ij} \beta_i F_{uz} \left(\partial_u F_{uj} \right) - 2 \gamma^{ij} \gamma^{kl} K_{il} F_{kz} \left(\partial_u F_{uj} \right) - 2 \gamma^{ij} \gamma^{kl} \gamma^{mn} K_{in} F_{mk} \left(\partial_l F_{uj} \right)\\
		= & \gamma^{ij} \beta_i F_{uz} \left(\partial_u F_{uj} \right)\,.
	\end{split}
\end{equation}

The fifth term of Eq. (\ref{hkk2fifth}) is
\begin{equation}
	\begin{split}
		& 2 k^a k^b g^{ce} g^{df} \Gamma^{g}_{\ ac} F_{gd} \Gamma^{h}_{\ fb} F_{he} = 2 g^{ce} g^{df} \Gamma^{g}_{\ uc} F_{gd} \Gamma^{h}_{\ fu} F_{he}\\
		= & 2 \Gamma^{g}_{\ uu} F_{gu} \Gamma^{h}_{\ zu} F_{hz} + 2 \Gamma^{g}_{\ uu} F_{gz} \Gamma^{h}_{\ uu} F_{hz} + 2 \gamma^{ij} \Gamma^{g}_{\ uu} F_{gi} \Gamma^{h}_{\ ju} F_{hz}\\
		& + 2 \Gamma^{g}_{\ uz} F_{gu} \Gamma^{h}_{\ zu} F_{hu} + 2 \Gamma^{g}_{\ uz} F_{gz} \Gamma^{h}_{\ uu} F_{hu} + 2 \gamma^{ij} \Gamma^{g}_{\ uz} F_{gi} \Gamma^{h}_{\ ju} F_{hu}\\
		& + 2 \gamma^{ij} \Gamma^{g}_{\ ui} F_{gu} \Gamma^{h}_{\ zu} F_{hj} + 2 \gamma^{ij} \Gamma^{g}_{\ ui} F_{gz} \Gamma^{h}_{\ uu} F_{hj} + 2 \gamma^{ij} \gamma^{kl} \Gamma^{g}_{\ ui} F_{gk} \Gamma^{h}_{\ lu} F_{hj}\\
		= & 2 \Gamma^{g}_{\ uz} F_{gu} \Gamma^{h}_{\ zu} F_{hu} + 2 \gamma^{ij} \Gamma^{g}_{\ uz} F_{gi} \Gamma^{h}_{\ ju} F_{hu} + 2 \gamma^{ij} \Gamma^{g}_{\ ui} F_{gu} \Gamma^{h}_{\ zu} F_{hj}\\
		& + 2 \gamma^{ij} \gamma^{kl} \Gamma^{g}_{\ ui} F_{gk} \Gamma^{h}_{\ lu} F_{hj}\,.
	\end{split}
\end{equation}
The index $g$ should be further expanded.
\begin{equation}\label{hkk2fifthfifth}
	\begin{split}
		& 2 \Gamma^{g}_{\ uz} F_{gu} \Gamma^{h}_{\ zu} F_{hu} + 2 \gamma^{ij} \Gamma^{g}_{\ uz} F_{gi} \Gamma^{h}_{\ ju} F_{hu} + 2 \gamma^{ij} \Gamma^{g}_{\ ui} F_{gu} \Gamma^{h}_{\ zu} F_{hj}\\
		& + 2 \gamma^{ij} \gamma^{kl} \Gamma^{g}_{\ ui} F_{gk} \Gamma^{h}_{\ lu} F_{hj}\\
		= & 2 \Gamma^{u}_{\ uz} F_{uu} \Gamma^{h}_{\ zu} F_{hu} + 2 \Gamma^{z}_{\ uz} F_{zu} \Gamma^{h}_{\ zu} F_{hu} + 2 \Gamma^{i}_{\ uz} F_{iu} \Gamma^{h}_{\ zu} F_{hu}\\
		& + 2 \gamma^{ij} \Gamma^{u}_{\ uz} F_{ui} \Gamma^{h}_{\ ju} F_{hu} + 2 \gamma^{ij} \Gamma^{z}_{\ uz} F_{zi} \Gamma^{h}_{\ ju} F_{hu} + 2 \gamma^{ij} \Gamma^{k}_{\ uz} F_{ki} \Gamma^{h}_{\ ju} F_{hu}\\
		& + 2 \gamma^{ij} \Gamma^{u}_{\ ui} F_{uu} \Gamma^{h}_{\ zu} F_{hj} + 2 \gamma^{ij} \Gamma^{z}_{\ ui} F_{zu} \Gamma^{h}_{\ zu} F_{hj} + 2 \gamma^{ij} \Gamma^{k}_{\ ui} F_{ku} \Gamma^{h}_{\ zu} F_{hj}\\
		& + 2 \gamma^{ij} \gamma^{kl} \Gamma^{u}_{\ ui} F_{uk} \Gamma^{h}_{\ lu} F_{hj} + 2 \gamma^{ij} \gamma^{kl} \Gamma^{z}_{\ ui} F_{zk} \Gamma^{h}_{\ lu} F_{hj} + 2 \gamma^{ij} \gamma^{kl} \Gamma^{m}_{\ ui} F_{mk} \Gamma^{h}_{\ lu} F_{hj}\\
		= & 2 \Gamma^{z}_{\ uz} F_{zu} \Gamma^{h}_{\ zu} F_{hu} + 2 \gamma^{ij} \Gamma^{z}_{\ uz} F_{zi} \Gamma^{h}_{\ ju} F_{hu} + 2 \gamma^{ij} \Gamma^{k}_{\ uz} F_{ki} \Gamma^{h}_{\ ju} F_{hu}\\
		& + 2 \gamma^{ij} \Gamma^{z}_{\ ui} F_{zu} \Gamma^{h}_{\ zu} F_{hj} + 2 \gamma^{ij} \gamma^{kl} \Gamma^{z}_{\ ui} F_{zk} \Gamma^{h}_{\ lu} F_{hj} + 2 \gamma^{ij} \gamma^{kl} \Gamma^{m}_{\ ui} F_{mk} \Gamma^{h}_{\ lu} F_{hj}\,.
	\end{split}
\end{equation}
The repeated index $h$ should be further expanded. The first term of Eq. (\ref{hkk2fifthfifth}) is
\begin{equation}
	\begin{split}
		2 \Gamma^{z}_{\ uz} F_{zu} \Gamma^{h}_{\ zu} F_{hu} = 0\,.
	\end{split}
\end{equation}
The second term of Eq. (\ref{hkk2fifthfifth}) is
\begin{equation}
	\begin{split}
		2 \gamma^{ij} \Gamma^{z}_{\ uz} F_{zi} \Gamma^{h}_{\ ju} F_{hu} = 0\,.
	\end{split}
\end{equation}
The third term of Eq. (\ref{hkk2fifthfifth}) is
\begin{equation}
	\begin{split}
		& 2 \gamma^{ij} \Gamma^{k}_{\ uz} F_{ki} \Gamma^{h}_{\ ju} F_{hu}\\
		= & 2 \gamma^{ij} \Gamma^{k}_{\ uz} F_{ki} \Gamma^{u}_{\ ju} F_{uu} + 2 \gamma^{ij} \Gamma^{k}_{\ uz} F_{ki} \Gamma^{z}_{\ ju} F_{zu} + 2 \gamma^{ij} \Gamma^{k}_{\ uz} F_{ki} \Gamma^{l}_{\ ju} F_{lu}\\
		= & 2 \gamma^{ij} \Gamma^{k}_{\ uz} F_{ki} \Gamma^{z}_{\ ju} F_{zu}\\
		= & 0\,.
	\end{split}
\end{equation}
The fourth term of Eq. (\ref{hkk2fifthfifth}) is
\begin{equation}
	\begin{split}
		2 \gamma^{ij} \Gamma^{z}_{\ ui} F_{zu} \Gamma^{h}_{\ zu} F_{hj} = 0\,.
	\end{split}
\end{equation}
The fifth term of Eq. (\ref{hkk2fifthfifth}) is
\begin{equation}
	\begin{split}
		2 \gamma^{ij} \gamma^{kl} \Gamma^{z}_{\ ui} F_{zk} \Gamma^{h}_{\ lu} F_{hj} = 0\,.
	\end{split}
\end{equation}
The sixth term of Eq. (\ref{hkk2fifthfifth}) is
\begin{equation}
	\begin{split}
		& 2 \gamma^{ij} \gamma^{kl} \Gamma^{m}_{\ ui} F_{mk} \Gamma^{h}_{\ lu} F_{hj}\\
		= & 2 \gamma^{ij} \gamma^{kl} \Gamma^{m}_{\ ui} F_{mk} \Gamma^{u}_{\ lu} F_{uj} + 2 \gamma^{ij} \gamma^{kl} \Gamma^{m}_{\ ui} F_{mk} \Gamma^{z}_{\ lu} F_{zj} + 2 \gamma^{ij} \gamma^{kl} \Gamma^{m}_{\ ui} F_{mk} \Gamma^{n}_{\ lu} F_{nj}\\
		= & 2 \gamma^{ij} \gamma^{kl} \Gamma^{m}_{\ ui} F_{mk} \Gamma^{z}_{\ lu} F_{zj} + 2 \gamma^{ij} \gamma^{kl} \Gamma^{m}_{\ ui} F_{mk} \Gamma^{n}_{\ lu} F_{nj}\\
		= & \frac{1}{2} \gamma^{ij} \gamma^{mo} \gamma^{np} \gamma^{kl} \left(\partial_u \gamma_{io} \right) F_{mk} \left(\partial_u \gamma_{lp} \right) F_{nj}\,.
	\end{split}
\end{equation}
Therefore, the fifth term of Eq. (\ref{hkk2fifth}) is obtained as 
\begin{equation}
	\begin{split}
		& 2 k^a k^b g^{ce} g^{df} \Gamma^{g}_{\ ac} F_{gd} \Gamma^{h}_{\ fb} F_{he} = \frac{1}{2} \gamma^{ij} \gamma^{mo} \gamma^{np} \gamma^{kl} \left(\partial_u \gamma_{io} \right) F_{mk} \left(\partial_u \gamma_{lp} \right) F_{nj}\\
		= & 2 \gamma^{ij} \gamma^{mo} \gamma^{np} \gamma^{kl} K_{io} F_{mk} K_{lp} F_{nj}\\
		= & 0\,.
	\end{split}
\end{equation}

The sixth term of Eq. (\ref{hkk2fifth}) is
\begin{equation}
	\begin{split}
		& 2 k^a k^b g^{ce} g^{df} \Gamma^{g}_{\ ac} F_{gd} \Gamma^{h}_{\ fe} F_{bh} = 2 g^{ce} g^{df} \Gamma^{g}_{\ uc} F_{gd} \Gamma^{h}_{\ fe} F_{uh}\\
		= & 2 \Gamma^{g}_{\ uu} F_{gu} \Gamma^{h}_{\ zz} F_{uh} + 2 \Gamma^{g}_{\ uu} F_{gz} \Gamma^{h}_{\ uz} F_{uh} + 2 \gamma^{ij} \Gamma^{g}_{\ uu} F_{gi} \Gamma^{h}_{\ jz} F_{uh}\\
		& + 2 \Gamma^{g}_{\ uz} F_{gu} \Gamma^{h}_{\ zu} F_{uh} + 2 \Gamma^{g}_{\ uz} F_{gz} \Gamma^{h}_{\ uu} F_{uh} + 2 \gamma^{ij} \Gamma^{g}_{\ uz} F_{gi} \Gamma^{h}_{\ ju} F_{uh}\\
		& + 2 \gamma^{ij} \Gamma^{g}_{\ ui} F_{gu} \Gamma^{h}_{\ zj} F_{uh} + 2 \gamma^{ij} \Gamma^{g}_{\ ui} F_{gz} \Gamma^{h}_{\ uj} F_{uh} + 2 \gamma^{ij} \gamma^{kl} \Gamma^{g}_{\ ui} F_{gk} \Gamma^{h}_{\ lj} F_{uh}\\
		= & 2 \Gamma^{g}_{\ uz} F_{gu} \Gamma^{h}_{\ zu} F_{uh} + 2 \gamma^{ij} \Gamma^{g}_{\ uz} F_{gi} \Gamma^{h}_{\ ju} F_{uh} + 2 \gamma^{ij} \Gamma^{g}_{\ ui} F_{gu} \Gamma^{h}_{\ zj} F_{uh}\\
		& + 2 \gamma^{ij} \Gamma^{g}_{\ ui} F_{gz} \Gamma^{h}_{\ uj} F_{uh} + 2 \gamma^{ij} \gamma^{kl} \Gamma^{g}_{\ ui} F_{gk} \Gamma^{h}_{\ lj} F_{uh}\,.
	\end{split}
\end{equation}
The index $g$ should be further expanded.
\begin{equation}\label{hkk2fifthsixth}
	\begin{split}
		& 2 \Gamma^{g}_{\ uz} F_{gu} \Gamma^{h}_{\ zu} F_{uh} + 2 \gamma^{ij} \Gamma^{g}_{\ uz} F_{gi} \Gamma^{h}_{\ ju} F_{uh} + 2 \gamma^{ij} \Gamma^{g}_{\ ui} F_{gu} \Gamma^{h}_{\ zj} F_{uh}\\
		& + 2 \gamma^{ij} \Gamma^{g}_{\ ui} F_{gz} \Gamma^{h}_{\ uj} F_{uh} + 2 \gamma^{ij} \gamma^{kl} \Gamma^{g}_{\ ui} F_{gk} \Gamma^{h}_{\ lj} F_{uh}\\
		= & 2 \Gamma^{u}_{\ uz} F_{uu} \Gamma^{h}_{\ zu} F_{uh} + 2 \Gamma^{z}_{\ uz} F_{zu} \Gamma^{h}_{\ zu} F_{uh} + 2 \Gamma^{i}_{\ uz} F_{iu} \Gamma^{h}_{\ zu} F_{uh}\\
		& + 2 \gamma^{ij} \Gamma^{u}_{\ uz} F_{ui} \Gamma^{h}_{\ ju} F_{uh} + 2 \gamma^{ij} \Gamma^{z}_{\ uz} F_{zi} \Gamma^{h}_{\ ju} F_{uh} + 2 \gamma^{ij} \Gamma^{k}_{\ uz} F_{ki} \Gamma^{h}_{\ ju} F_{uh}\\
		& + 2 \gamma^{ij} \Gamma^{u}_{\ ui} F_{uu} \Gamma^{h}_{\ zj} F_{uh} + 2 \gamma^{ij} \Gamma^{z}_{\ ui} F_{zu} \Gamma^{h}_{\ zj} F_{uh} + 2 \gamma^{ij} \Gamma^{k}_{\ ui} F_{ku} \Gamma^{h}_{\ zj} F_{uh}\\
		& + 2 \gamma^{ij} \Gamma^{u}_{\ ui} F_{uz} \Gamma^{h}_{\ uj} F_{uh} + 2 \gamma^{ij} \Gamma^{z}_{\ ui} F_{zz} \Gamma^{h}_{\ uj} F_{uh} + 2 \gamma^{ij} \Gamma^{k}_{\ ui} F_{kz} \Gamma^{h}_{\ uj} F_{uh}\\
		& + 2 \gamma^{ij} \gamma^{kl} \Gamma^{u}_{\ ui} F_{uk} \Gamma^{h}_{\ lj} F_{uh} + 2 \gamma^{ij} \gamma^{kl} \Gamma^{z}_{\ ui} F_{zk} \Gamma^{h}_{\ lj} F_{uh} + 2 \gamma^{ij} \gamma^{kl} \Gamma^{m}_{\ ui} F_{mk} \Gamma^{h}_{\ lj} F_{uh}\\
		= & 2 \Gamma^{z}_{\ uz} F_{zu} \Gamma^{h}_{\ zu} F_{uh} + 2 \gamma^{ij} \Gamma^{z}_{\ uz} F_{zi} \Gamma^{h}_{\ ju} F_{uh} + 2 \gamma^{ij} \Gamma^{k}_{\ uz} F_{ki} \Gamma^{h}_{\ ju} F_{uh}\\
		& + 2 \gamma^{ij} \Gamma^{z}_{\ ui} F_{zu} \Gamma^{h}_{\ zj} F_{uh} + 2 \gamma^{ij} \Gamma^{u}_{\ ui} F_{uz} \Gamma^{h}_{\ uj} F_{uh} + 2 \gamma^{ij} \Gamma^{k}_{\ ui} F_{kz} \Gamma^{h}_{\ uj} F_{uh}\\
		& + 2 \gamma^{ij} \gamma^{kl} \Gamma^{z}_{\ ui} F_{zk} \Gamma^{h}_{\ lj} F_{uh} + 2 \gamma^{ij} \gamma^{kl} \Gamma^{m}_{\ ui} F_{mk} \Gamma^{h}_{\ lj} F_{uh}\,.
	\end{split}
\end{equation}
The repeated index $h$ should be further expanded. The first term of Eq. (\ref{hkk2fifthsixth}) is
\begin{equation}
	\begin{split}
		2 \Gamma^{z}_{\ uz} F_{zu} \Gamma^{h}_{\ zu} F_{uh} = 0\,.
	\end{split}
\end{equation}
The second term of Eq. (\ref{hkk2fifthsixth}) is
\begin{equation}
	\begin{split}
		2 \gamma^{ij} \Gamma^{z}_{\ uz} F_{zi} \Gamma^{h}_{\ ju} F_{uh} = 0\,.
	\end{split}
\end{equation}
The third term of Eq. (\ref{hkk2fifthsixth}) is
\begin{equation}
	\begin{split}
		& 2 \gamma^{ij} \Gamma^{k}_{\ uz} F_{ki} \Gamma^{h}_{\ ju} F_{uh}\\
		= & 2 \gamma^{ij} \Gamma^{k}_{\ uz} F_{ki} \Gamma^{u}_{\ ju} F_{uu} + 2 \gamma^{ij} \Gamma^{k}_{\ uz} F_{ki} \Gamma^{z}_{\ ju} F_{uz} + 2 \gamma^{ij} \Gamma^{k}_{\ uz} F_{ki} \Gamma^{l}_{\ ju} F_{ul}\\
		= & 2 \gamma^{ij} \Gamma^{k}_{\ uz} F_{ki} \Gamma^{z}_{\ ju} F_{uz}\\
		= & 0\,.
	\end{split}
\end{equation}
The fourth term of Eq. (\ref{hkk2fifthsixth}) is
\begin{equation}
	\begin{split}
		2 \gamma^{ij} \Gamma^{z}_{\ ui} F_{zu} \Gamma^{h}_{\ zj} F_{uh} = 0\,.
	\end{split}
\end{equation}
The fifth term of Eq. (\ref{hkk2fifthsixth}) is
\begin{equation}
	\begin{split}
		& 2 \gamma^{ij} \Gamma^{u}_{\ ui} F_{uz} \Gamma^{h}_{\ uj} F_{uh}\\
		= & 2 \gamma^{ij} \Gamma^{u}_{\ ui} F_{uz} \Gamma^{u}_{\ uj} F_{uu} + 2 \gamma^{ij} \Gamma^{u}_{\ ui} F_{uz} \Gamma^{z}_{\ uj} F_{uz} + 2 \gamma^{ij} \Gamma^{u}_{\ ui} F_{uz} \Gamma^{k}_{\ uj} F_{uk}\\
		= & 2 \gamma^{ij} \Gamma^{u}_{\ ui} F_{uz} \Gamma^{z}_{\ uj} F_{uz}\\
		= & 0\,.
	\end{split}
\end{equation}
The sixth term of Eq. (\ref{hkk2fifthsixth}) is
\begin{equation}
	\begin{split}
		& 2 \gamma^{ij} \Gamma^{k}_{\ ui} F_{kz} \Gamma^{h}_{\ uj} F_{uh}\\
		= & 2 \gamma^{ij} \Gamma^{k}_{\ ui} F_{kz} \Gamma^{u}_{\ uj} F_{uu} + 2 \gamma^{ij} \Gamma^{k}_{\ ui} F_{kz} \Gamma^{z}_{\ uj} F_{uz} + 2 \gamma^{ij} \Gamma^{k}_{\ ui} F_{kz} \Gamma^{l}_{\ uj} F_{ul}\\
		= & 2 \gamma^{ij} \Gamma^{k}_{\ ui} F_{kz} \Gamma^{z}_{\ uj} F_{uz}\\
		= & 0\,.
	\end{split}
\end{equation}
The seventh term of Eq. (\ref{hkk2fifthsixth}) is
\begin{equation}
	\begin{split}
		2 \gamma^{ij} \gamma^{kl} \Gamma^{z}_{\ ui} F_{zk} \Gamma^{h}_{\ lj} F_{uh} = 0\,.
	\end{split}
\end{equation}
The eighth term of Eq. (\ref{hkk2fifthsixth}) is
\begin{equation}
	\begin{split}
		& 2 \gamma^{ij} \gamma^{kl} \Gamma^{m}_{\ ui} F_{mk} \Gamma^{h}_{\ lj} F_{uh}\\
		= & 2 \gamma^{ij} \gamma^{kl} \Gamma^{m}_{\ ui} F_{mk} \Gamma^{u}_{\ lj} F_{uu} + 2 \gamma^{ij} \gamma^{kl} \Gamma^{m}_{\ ui} F_{mk} \Gamma^{z}_{\ lj} F_{uz} + 2 \gamma^{ij} \gamma^{kl} \Gamma^{m}_{\ ui} F_{mk} \Gamma^{n}_{\ lj} F_{un}\\
		= & 2 \gamma^{ij} \gamma^{kl} \Gamma^{m}_{\ ui} F_{mk} \Gamma^{z}_{\ lj} F_{uz}\\
		= & - \frac{1}{2} \gamma^{ij} \gamma^{kl} \gamma^{mn} \left(\partial_u \gamma_{in} \right) F_{mk} \left(\partial_u \gamma_{lj} \right) F_{uz}\,.
	\end{split}
\end{equation}
Therefore, the sixth term of Eq. (\ref{hkk2fifth}) is obtained as 
\begin{equation}
	\begin{split}
		& 2 k^a k^b g^{ce} g^{df} \Gamma^{g}_{\ ac} F_{gd} \Gamma^{h}_{\ fe} F_{bh} = - \frac{1}{2} \gamma^{ij} \gamma^{kl} \gamma^{mn} \left(\partial_u \gamma_{in} \right) F_{mk} \left(\partial_u \gamma_{lj} \right) F_{uz}\\
		= & - 2 \gamma^{ij} \gamma^{kl} \gamma^{mn} K_{in} F_{mk} K_{lj} F_{uz}\\
		= & 0\,.
	\end{split}
\end{equation}

The seventh term of Eq. (\ref{hkk2fifth}) is
\begin{equation}
	\begin{split}
		& - 2 k^a k^b g^{ce} g^{df} \Gamma^{g}_{\ ad} F_{cg} \left(\partial_f F_{be} \right) = - 2 g^{ce} g^{df} \Gamma^{g}_{\ ud} F_{cg} \left(\partial_f F_{ue} \right)\\
		= & - 2 \Gamma^{g}_{\ uu} F_{ug} \left(\partial_z F_{uz} \right) - 2 \Gamma^{g}_{\ uz} F_{ug} \left(\partial_u F_{uz} \right) - 2 \gamma^{ij} \Gamma^{g}_{\ ui} F_{ug} \left(\partial_j F_{uz} \right)\\
		& - 2 \gamma^{ij} \Gamma^{g}_{\ uu} F_{ig} \left(\partial_z F_{uj} \right) - 2 \gamma^{ij} \Gamma^{g}_{\ uz} F_{ig} \left(\partial_u F_{uj} \right) - 2 \gamma^{ij} \gamma^{kl} \Gamma^{g}_{\ uk} F_{ig} \left(\partial_l F_{uj} \right)\\
		= & - 2 \Gamma^{g}_{\ uz} F_{ug} \left(\partial_u F_{uz} \right) - 2 \gamma^{ij} \Gamma^{g}_{\ ui} F_{ug} \left(\partial_j F_{uz} \right) - 2 \gamma^{ij} \Gamma^{g}_{\ uz} F_{ig} \left(\partial_u F_{uj} \right)\\
		& - 2 \gamma^{ij} \gamma^{kl} \Gamma^{g}_{\ uk} F_{ig} \left(\partial_l F_{uj} \right)\,.
	\end{split}
\end{equation}
The index $g$ should be further expanded.
\begin{equation}\label{hkk2fifthseventh}
	\begin{split}
		& - 2 \Gamma^{g}_{\ uz} F_{ug} \left(\partial_u F_{uz} \right) - 2 \gamma^{ij} \Gamma^{g}_{\ ui} F_{ug} \left(\partial_j F_{uz} \right) - 2 \gamma^{ij} \Gamma^{g}_{\ uz} F_{ig} \left(\partial_u F_{uj} \right)\\
		& - 2 \gamma^{ij} \gamma^{kl} \Gamma^{g}_{\ uk} F_{ig} \left(\partial_l F_{uj} \right)\\
		= & - 2 \Gamma^{u}_{\ uz} F_{uu} \left(\partial_u F_{uz} \right) - 2 \Gamma^{z}_{\ uz} F_{uz} \left(\partial_u F_{uz} \right) - 2 \Gamma^{i}_{\ uz} F_{ui} \left(\partial_u F_{uz} \right)\\
		& - 2 \gamma^{ij} \Gamma^{u}_{\ ui} F_{uu} \left(\partial_j F_{uz} \right) - 2 \gamma^{ij} \Gamma^{z}_{\ ui} F_{uz} \left(\partial_j F_{uz} \right) - 2 \gamma^{ij} \Gamma^{k}_{\ ui} F_{uk} \left(\partial_j F_{uz} \right)\\
		& - 2 \gamma^{ij} \Gamma^{u}_{\ uz} F_{iu} \left(\partial_u F_{uj} \right) - 2 \gamma^{ij} \Gamma^{z}_{\ uz} F_{iz} \left(\partial_u F_{uj} \right) - 2 \gamma^{ij} \Gamma^{k}_{\ uz} F_{ik} \left(\partial_u F_{uj} \right)\\
		& - 2 \gamma^{ij} \gamma^{kl} \Gamma^{u}_{\ uk} F_{iu} \left(\partial_l F_{uj} \right) - 2 \gamma^{ij} \gamma^{kl} \Gamma^{z}_{\ uk} F_{iz} \left(\partial_l F_{uj} \right) - 2 \gamma^{ij} \gamma^{kl} \Gamma^{m}_{\ uk} F_{im} \left(\partial_l F_{uj} \right)\\
		= & - 2 \Gamma^{z}_{\ uz} F_{uz} \left(\partial_u F_{uz} \right) - 2 \gamma^{ij} \Gamma^{z}_{\ ui} F_{uz} \left(\partial_j F_{uz} \right) - 2 \gamma^{ij} \Gamma^{z}_{\ uz} F_{iz} \left(\partial_u F_{uj} \right)\\
		& - 2 \gamma^{ij} \Gamma^{k}_{\ uz} F_{ik} \left(\partial_u F_{uj} \right) - 2 \gamma^{ij} \gamma^{kl} \Gamma^{z}_{\ uk} F_{iz} \left(\partial_l F_{uj} \right) - 2 \gamma^{ij} \gamma^{kl} \Gamma^{m}_{\ uk} F_{im} \left(\partial_l F_{uj} \right)\,.
	\end{split}
\end{equation}
Therefore, the seventh term of Eq. (\ref{hkk2fifth}) is obtained as 
\begin{equation}
	\begin{split}
		& - 2 k^a k^b g^{ce} g^{df} \Gamma^{g}_{\ ad} F_{cg} \left(\partial_f F_{be} \right)\\
		= & - \gamma^{ij} \gamma^{kl} \beta_l F_{ik} \left(\partial_u F_{uj} \right) - \gamma^{ij} \gamma^{kl} \gamma^{mn} \left(\partial_u \gamma_{kn} \right) F_{im} \left(\partial_l F_{uj} \right)\\
		= & - \gamma^{ij} \gamma^{kl} \beta_l F_{ik} \left(\partial_u F_{uj} \right) - 2 \gamma^{ij} \gamma^{kl} \gamma^{mn} K_{kn} F_{im} \left(\partial_l F_{uj} \right)\\
		= & - \gamma^{ij} \gamma^{kl} \beta_l F_{ik} \left(\partial_u F_{uj} \right)\,.
	\end{split}
\end{equation}

The eighth term of Eq. (\ref{hkk2fifth}) is
\begin{equation}
	\begin{split}
		& 2 k^a k^b g^{ce} g^{df} \Gamma^{g}_{\ ad} F_{cg} \Gamma^{h}_{\ fb} F_{he} = 2 g^{ce} g^{df} \Gamma^{g}_{\ ud} F_{cg} \Gamma^{h}_{\ fu} F_{he}\\
		= & 2 \Gamma^{g}_{\ uu} F_{ug} \Gamma^{h}_{\ zu} F_{hz} + 2 \Gamma^{g}_{\ uz} F_{ug} \Gamma^{h}_{\ uu} F_{hz} + 2 \gamma^{ij} \Gamma^{g}_{\ ui} F_{ug} \Gamma^{h}_{\ ju} F_{hz}\\
		& + 2 \Gamma^{g}_{\ uu} F_{zg} \Gamma^{h}_{\ zu} F_{hu} + 2 \Gamma^{g}_{\ uz} F_{zg} \Gamma^{h}_{\ uu} F_{hu} + 2 \gamma^{ij} \Gamma^{g}_{\ ui} F_{zg} \Gamma^{h}_{\ ju} F_{hu}\\
		& + 2 \gamma^{ij} \Gamma^{g}_{\ uu} F_{ig} \Gamma^{h}_{\ zu} F_{hj} + 2 \gamma^{ij} \Gamma^{g}_{\ uz} F_{ig} \Gamma^{h}_{\ uu} F_{hj} + 2 \gamma^{ij} \gamma^{kl} \Gamma^{g}_{\ uk} F_{ig} \Gamma^{h}_{\ lu} F_{hj}\\
		= & 2 \gamma^{ij} \Gamma^{g}_{\ ui} F_{ug} \Gamma^{h}_{\ ju} F_{hz} + 2 \gamma^{ij} \Gamma^{g}_{\ ui} F_{zg} \Gamma^{h}_{\ ju} F_{hu} + 2 \gamma^{ij} \gamma^{kl} \Gamma^{g}_{\ uk} F_{ig} \Gamma^{h}_{\ lu} F_{hj}\,.
	\end{split}
\end{equation}
The index $g$ should be further expanded.
\begin{equation}\label{hkk2fiftheighth}
	\begin{split}
		& 2 \gamma^{ij} \Gamma^{g}_{\ ui} F_{ug} \Gamma^{h}_{\ ju} F_{hz} + 2 \gamma^{ij} \Gamma^{g}_{\ ui} F_{zg} \Gamma^{h}_{\ ju} F_{hu} + 2 \gamma^{ij} \gamma^{kl} \Gamma^{g}_{\ uk} F_{ig} \Gamma^{h}_{\ lu} F_{hj}\\
		= & 2 \gamma^{ij} \Gamma^{u}_{\ ui} F_{uu} \Gamma^{h}_{\ ju} F_{hz} + 2 \gamma^{ij} \Gamma^{z}_{\ ui} F_{uz} \Gamma^{h}_{\ ju} F_{hz} + 2 \gamma^{ij} \Gamma^{k}_{\ ui} F_{uk} \Gamma^{h}_{\ ju} F_{hz}\\
		& + 2 \gamma^{ij} \Gamma^{u}_{\ ui} F_{zu} \Gamma^{h}_{\ ju} F_{hu} + 2 \gamma^{ij} \Gamma^{z}_{\ ui} F_{zz} \Gamma^{h}_{\ ju} F_{hu} + 2 \gamma^{ij} \Gamma^{k}_{\ ui} F_{zk} \Gamma^{h}_{\ ju} F_{hu}\\
		& + 2 \gamma^{ij} \gamma^{kl} \Gamma^{u}_{\ uk} F_{iu} \Gamma^{h}_{\ lu} F_{hj} + 2 \gamma^{ij} \gamma^{kl} \Gamma^{z}_{\ uk} F_{iz} \Gamma^{h}_{\ lu} F_{hj} + 2 \gamma^{ij} \gamma^{kl} \Gamma^{m}_{\ uk} F_{im} \Gamma^{h}_{\ lu} F_{hj}\\
		= & 2 \gamma^{ij} \Gamma^{z}_{\ ui} F_{uz} \Gamma^{h}_{\ ju} F_{hz} + 2 \gamma^{ij} \Gamma^{u}_{\ ui} F_{zu} \Gamma^{h}_{\ ju} F_{hu} + 2 \gamma^{ij} \Gamma^{k}_{\ ui} F_{zk} \Gamma^{h}_{\ ju} F_{hu}\\
		& + 2 \gamma^{ij} \gamma^{kl} \Gamma^{z}_{\ uk} F_{iz} \Gamma^{h}_{\ lu} F_{hj} + 2 \gamma^{ij} \gamma^{kl} \Gamma^{m}_{\ uk} F_{im} \Gamma^{h}_{\ lu} F_{hj}\,.
	\end{split}
\end{equation}
The repeated index $h$ should be further expanded. The first term of Eq. (\ref{hkk2fiftheighth}) is
\begin{equation}
	\begin{split}
		2 \gamma^{ij} \Gamma^{z}_{\ ui} F_{uz} \Gamma^{h}_{\ ju} F_{hz} = 0\,.
	\end{split}
\end{equation}
The second term of Eq. (\ref{hkk2fiftheighth}) is
\begin{equation}
	\begin{split}
		& 2 \gamma^{ij} \Gamma^{u}_{\ ui} F_{zu} \Gamma^{h}_{\ ju} F_{hu}\\
		= & 2 \gamma^{ij} \Gamma^{u}_{\ ui} F_{zu} \Gamma^{u}_{\ ju} F_{uu} + 2 \gamma^{ij} \Gamma^{u}_{\ ui} F_{zu} \Gamma^{z}_{\ ju} F_{zu} + 2 \gamma^{ij} \Gamma^{u}_{\ ui} F_{zu} \Gamma^{k}_{\ ju} F_{ku}\\
		= & 2 \gamma^{ij} \Gamma^{u}_{\ ui} F_{zu} \Gamma^{z}_{\ ju} F_{zu}\\
		= & 0\,.
	\end{split}
\end{equation}
The third term of Eq. (\ref{hkk2fiftheighth}) is
\begin{equation}
	\begin{split}
		& 2 \gamma^{ij} \Gamma^{k}_{\ ui} F_{zk} \Gamma^{h}_{\ ju} F_{hu}\\
		= & 2 \gamma^{ij} \Gamma^{k}_{\ ui} F_{zk} \Gamma^{u}_{\ ju} F_{uu} + 2 \gamma^{ij} \Gamma^{k}_{\ ui} F_{zk} \Gamma^{z}_{\ ju} F_{zu} + 2 \gamma^{ij} \Gamma^{k}_{\ ui} F_{zk} \Gamma^{l}_{\ ju} F_{lu}\\
		= & 2 \gamma^{ij} \Gamma^{k}_{\ ui} F_{zk} \Gamma^{z}_{\ ju} F_{zu}\\
		= & 0\,.
	\end{split}
\end{equation}
The fourth term of Eq. (\ref{hkk2fiftheighth}) is
\begin{equation}
	\begin{split}
		2 \gamma^{ij} \gamma^{kl} \Gamma^{z}_{\ uk} F_{iz} \Gamma^{h}_{\ lu} F_{hj} = 0\,.
	\end{split}
\end{equation}
The fifteenth term of Eq. (\ref{hkk2fiftheighth}) is
\begin{equation}
	\begin{split}
		& 2 \gamma^{ij} \gamma^{kl} \Gamma^{m}_{\ uk} F_{im} \Gamma^{h}_{\ lu} F_{hj}\\
		= & 2 \gamma^{ij} \gamma^{kl} \Gamma^{m}_{\ uk} F_{im} \Gamma^{u}_{\ lu} F_{uj} + 2 \gamma^{ij} \gamma^{kl} \Gamma^{m}_{\ uk} F_{im} \Gamma^{z}_{\ lu} F_{zj} + 2 \gamma^{ij} \gamma^{kl} \Gamma^{m}_{\ uk} F_{im} \Gamma^{n}_{\ lu} F_{nj}\\
		= & 2 \gamma^{ij} \gamma^{kl} \Gamma^{m}_{\ uk} F_{im} \Gamma^{z}_{\ lu} F_{zj} + 2 \gamma^{ij} \gamma^{kl} \Gamma^{m}_{\ uk} F_{im} \Gamma^{n}_{\ lu} F_{nj}\\
		= & \frac{1}{2} \gamma^{ij} \gamma^{kl} \gamma^{mo} \gamma^{np} \left(\partial_u \gamma_{ko} \right) F_{im} \left(\partial_u \gamma_{lp} \right) F_{nj}\,.
	\end{split}
\end{equation}
Therefore, the eighth term of Eq. (\ref{hkk2fifth}) is obtained as 
\begin{equation}
	\begin{split}
		& 2 k^a k^b g^{ce} g^{df} \Gamma^{g}_{\ ad} F_{cg} \Gamma^{h}_{\ fb} F_{he} = \frac{1}{2} \gamma^{ij} \gamma^{kl} \gamma^{mo} \gamma^{np} \left(\partial_u \gamma_{ko} \right) F_{im} \left(\partial_u \gamma_{lp} \right) F_{nj}\\
		= & 2 \gamma^{ij} \gamma^{kl} \gamma^{mo} \gamma^{np} K_{ko} F_{im} K_{lp} F_{nj}\\
		= & 0\,.
	\end{split}
\end{equation}

The ninth term of Eq. (\ref{hkk2fifth}) is
\begin{equation}
	\begin{split}
		& 2 k^a k^b g^{ce} g^{df} \Gamma^{g}_{\ ad} F_{cg} \Gamma^{h}_{\ fe} F_{bh} = 2 g^{ce} g^{df} \Gamma^{g}_{\ ud} F_{cg} \Gamma^{h}_{\ fe} F_{uh}\\
		= & 2 \Gamma^{g}_{\ uu} F_{ug} \Gamma^{h}_{\ zz} F_{uh} + 2 \Gamma^{g}_{\ uz} F_{ug} \Gamma^{h}_{\ uz} F_{uh} + 2 \gamma^{ij} \Gamma^{g}_{\ ui} F_{ug} \Gamma^{h}_{\ jz} F_{uh}\\
		& + 2 \Gamma^{g}_{\ uu} F_{zg} \Gamma^{h}_{\ zu} F_{uh} + 2 \Gamma^{g}_{\ uz} F_{zg} \Gamma^{h}_{\ uu} F_{uh} + 2 \gamma^{ij} \Gamma^{g}_{\ ui} F_{zg} \Gamma^{h}_{\ ju} F_{uh}\\
		& + 2 \gamma^{ij} \Gamma^{g}_{\ uu} F_{ig} \Gamma^{h}_{\ zj} F_{uh} + 2 \gamma^{ij} \Gamma^{g}_{\ uz} F_{ig} \Gamma^{h}_{\ uj} F_{uh} + 2 \gamma^{ij} \gamma^{kl} \Gamma^{g}_{\ uk} F_{ig} \Gamma^{h}_{\ lj} F_{uh}\\
		= & 2 \Gamma^{g}_{\ uz} F_{ug} \Gamma^{h}_{\ uz} F_{uh} + 2 \gamma^{ij} \Gamma^{g}_{\ ui} F_{ug} \Gamma^{h}_{\ jz} F_{uh} + 2 \gamma^{ij} \Gamma^{g}_{\ ui} F_{zg} \Gamma^{h}_{\ ju} F_{uh}\\
		& + 2 \gamma^{ij} \Gamma^{g}_{\ uz} F_{ig} \Gamma^{h}_{\ uj} F_{uh} + 2 \gamma^{ij} \gamma^{kl} \Gamma^{g}_{\ uk} F_{ig} \Gamma^{h}_{\ lj} F_{uh}\,.
	\end{split}
\end{equation}
The index $g$ should be further expanded.
\begin{equation}\label{hkk2fifthninth}
	\begin{split}
		& 2 \Gamma^{g}_{\ uz} F_{ug} \Gamma^{h}_{\ uz} F_{uh} + 2 \gamma^{ij} \Gamma^{g}_{\ ui} F_{ug} \Gamma^{h}_{\ jz} F_{uh} + 2 \gamma^{ij} \Gamma^{g}_{\ ui} F_{zg} \Gamma^{h}_{\ ju} F_{uh}\\
		& + 2 \gamma^{ij} \Gamma^{g}_{\ uz} F_{ig} \Gamma^{h}_{\ uj} F_{uh} + 2 \gamma^{ij} \gamma^{kl} \Gamma^{g}_{\ uk} F_{ig} \Gamma^{h}_{\ lj} F_{uh}\\
		= & 2 \Gamma^{u}_{\ uz} F_{uu} \Gamma^{h}_{\ uz} F_{uh} + 2 \Gamma^{z}_{\ uz} F_{uz} \Gamma^{h}_{\ uz} F_{uh} + 2 \Gamma^{i}_{\ uz} F_{ui} \Gamma^{h}_{\ uz} F_{uh}\\
		& + 2 \gamma^{ij} \Gamma^{u}_{\ ui} F_{uu} \Gamma^{h}_{\ jz} F_{uh} + 2 \gamma^{ij} \Gamma^{z}_{\ ui} F_{uz} \Gamma^{h}_{\ jz} F_{uh} + 2 \gamma^{ij} \Gamma^{k}_{\ ui} F_{uk} \Gamma^{h}_{\ jz} F_{uh}\\
		& + 2 \gamma^{ij} \Gamma^{u}_{\ ui} F_{zu} \Gamma^{h}_{\ ju} F_{uh} + 2 \gamma^{ij} \Gamma^{z}_{\ ui} F_{zz} \Gamma^{h}_{\ ju} F_{uh} + 2 \gamma^{ij} \Gamma^{k}_{\ ui} F_{zk} \Gamma^{h}_{\ ju} F_{uh}\\
		& + 2 \gamma^{ij} \Gamma^{u}_{\ uz} F_{iu} \Gamma^{h}_{\ uj} F_{uh} + 2 \gamma^{ij} \Gamma^{z}_{\ uz} F_{iz} \Gamma^{h}_{\ uj} F_{uh} + 2 \gamma^{ij} \Gamma^{k}_{\ uz} F_{ik} \Gamma^{h}_{\ uj} F_{uh}\\
		& + 2 \gamma^{ij} \gamma^{kl} \Gamma^{u}_{\ uk} F_{iu} \Gamma^{h}_{\ lj} F_{uh} + 2 \gamma^{ij} \gamma^{kl} \Gamma^{z}_{\ uk} F_{iz} \Gamma^{h}_{\ lj} F_{uh} + 2 \gamma^{ij} \gamma^{kl} \Gamma^{m}_{\ uk} F_{im} \Gamma^{h}_{\ lj} F_{uh}\\
		= & 2 \Gamma^{z}_{\ uz} F_{uz} \Gamma^{h}_{\ uz} F_{uh} + 2 \gamma^{ij} \Gamma^{z}_{\ ui} F_{uz} \Gamma^{h}_{\ jz} F_{uh} + 2 \gamma^{ij} \Gamma^{u}_{\ ui} F_{zu} \Gamma^{h}_{\ ju} F_{uh}\\
		& + 2 \gamma^{ij} \Gamma^{k}_{\ ui} F_{zk} \Gamma^{h}_{\ ju} F_{uh} + 2 \gamma^{ij} \Gamma^{z}_{\ uz} F_{iz} \Gamma^{h}_{\ uj} F_{uh} + 2 \gamma^{ij} \Gamma^{k}_{\ uz} F_{ik} \Gamma^{h}_{\ uj} F_{uh}\\
		& + 2 \gamma^{ij} \gamma^{kl} \Gamma^{z}_{\ uk} F_{iz} \Gamma^{h}_{\ lj} F_{uh} + 2 \gamma^{ij} \gamma^{kl} \Gamma^{m}_{\ uk} F_{im} \Gamma^{h}_{\ lj} F_{uh}\,.
	\end{split}
\end{equation}
The repeated index $h$ should be further expanded. The first term of Eq. (\ref{hkk2fifthninth}) is 
\begin{equation}
	\begin{split}
		2 \Gamma^{z}_{\ uz} F_{uz} \Gamma^{h}_{\ uz} F_{uh} = 0\,.
	\end{split}
\end{equation}
The second term of Eq. (\ref{hkk2fifthninth}) is 
\begin{equation}
	\begin{split}
		2 \gamma^{ij} \Gamma^{z}_{\ ui} F_{uz} \Gamma^{h}_{\ jz} F_{uh} = 0\,.
	\end{split}
\end{equation}
The third term of Eq. (\ref{hkk2fifthninth}) is
\begin{equation}
	\begin{split}
		& 2 \gamma^{ij} \Gamma^{u}_{\ ui} F_{zu} \Gamma^{h}_{\ ju} F_{uh}\\
		= & 2 \gamma^{ij} \Gamma^{u}_{\ ui} F_{zu} \Gamma^{u}_{\ ju} F_{uu} + 2 \gamma^{ij} \Gamma^{u}_{\ ui} F_{zu} \Gamma^{z}_{\ ju} F_{uz} + 2 \gamma^{ij} \Gamma^{u}_{\ ui} F_{zu} \Gamma^{k}_{\ ju} F_{uk}\\
		= & 2 \gamma^{ij} \Gamma^{u}_{\ ui} F_{zu} \Gamma^{z}_{\ ju} F_{uz}\\
		= & 0\,.
	\end{split}
\end{equation}
The fourth term of Eq. (\ref{hkk2fifthninth}) is
\begin{equation}
	\begin{split}
		& 2 \gamma^{ij} \Gamma^{k}_{\ ui} F_{zk} \Gamma^{h}_{\ ju} F_{uh}\\
		= & 2 \gamma^{ij} \Gamma^{k}_{\ ui} F_{zk} \Gamma^{u}_{\ ju} F_{uu} + 2 \gamma^{ij} \Gamma^{k}_{\ ui} F_{zk} \Gamma^{z}_{\ ju} F_{uz} + 2 \gamma^{ij} \Gamma^{k}_{\ ui} F_{zk} \Gamma^{l}_{\ ju} F_{ul}\\
		= & 2 \gamma^{ij} \Gamma^{k}_{\ ui} F_{zk} \Gamma^{z}_{\ ju} F_{uz}\\
		= & 0\,.
	\end{split}
\end{equation}
The fifth term of Eq. (\ref{hkk2fifthninth}) is
\begin{equation}
	\begin{split}
		2 \gamma^{ij} \Gamma^{z}_{\ uz} F_{iz} \Gamma^{h}_{\ uj} F_{uh} = 0\,.
	\end{split}
\end{equation}
The sixth term of Eq. (\ref{hkk2fifthninth}) is
\begin{equation}
	\begin{split}
		& 2 \gamma^{ij} \Gamma^{k}_{\ uz} F_{ik} \Gamma^{h}_{\ uj} F_{uh}\\
		= & 2 \gamma^{ij} \Gamma^{k}_{\ uz} F_{ik} \Gamma^{u}_{\ uj} F_{uu} + 2 \gamma^{ij} \Gamma^{k}_{\ uz} F_{ik} \Gamma^{z}_{\ uj} F_{uz} + 2 \gamma^{ij} \Gamma^{k}_{\ uz} F_{ik} \Gamma^{l}_{\ uj} F_{ul}\\
		= & 2 \gamma^{ij} \Gamma^{k}_{\ uz} F_{ik} \Gamma^{z}_{\ uj} F_{uz}\\
		= & 0\,.
	\end{split}
\end{equation}
The seventh term of Eq. (\ref{hkk2fifthninth}) is
\begin{equation}
	\begin{split}
		2 \gamma^{ij} \gamma^{kl} \Gamma^{z}_{\ uk} F_{iz} \Gamma^{h}_{\ lj} F_{uh} = 0\,.
	\end{split}
\end{equation}
The fifteenth term of Eq. (\ref{hkk2fifthninth}) is
\begin{equation}
	\begin{split}
		& 2 \gamma^{ij} \gamma^{kl} \Gamma^{m}_{\ uk} F_{im} \Gamma^{h}_{\ lj} F_{uh}\\
		= & 2 \gamma^{ij} \gamma^{kl} \Gamma^{m}_{\ uk} F_{im} \Gamma^{u}_{\ lj} F_{uu} + 2 \gamma^{ij} \gamma^{kl} \Gamma^{m}_{\ uk} F_{im} \Gamma^{z}_{\ lj} F_{uz} + 2 \gamma^{ij} \gamma^{kl} \Gamma^{m}_{\ uk} F_{im} \Gamma^{n}_{\ lj} F_{un}\\
		= & 2 \gamma^{ij} \gamma^{kl} \Gamma^{m}_{\ uk} F_{im} \Gamma^{z}_{\ lj} F_{uz}\\
		= & - \frac{1}{2} \gamma^{ij} \gamma^{kl} \gamma^{mn} \left(\partial_u \gamma_{kn} \right) F_{im} \left(\partial_u \gamma_{lj} \right) F_{uz}\,.
	\end{split}
\end{equation}
Therefore, the ninth term of Eq. (\ref{hkk2fifth}) is obtained as 
\begin{equation}
	\begin{split}
		& 2 k^a k^b g^{ce} g^{df} \Gamma^{g}_{\ ad} F_{cg} \Gamma^{h}_{\ fe} F_{bh} = - \frac{1}{2} \gamma^{ij} \gamma^{kl} \gamma^{mn} \left(\partial_u \gamma_{kn} \right) F_{im} \left(\partial_u \gamma_{lj} \right) F_{uz}\\
		= & - 2 \gamma^{ij} \gamma^{kl} \gamma^{mn} K_{kn} F_{im} K_{lj} F_{uz}\\
		= & 0\,.
	\end{split}
\end{equation}

Finally, the fifth term of Eq. (\ref{rewrittenhkk2}) is
\begin{equation}
	\begin{split}
		& 2 k^a k^b \nabla_a F_{cd} \nabla^d F_{b}^{\ c}\\
		= & 2 \left(\partial_u F_{uz} \right) \left(\partial_u F_{uz} \right) + 2 \gamma^{ij} \left(\partial_u F_{ui} \right) \left(\partial_j F_{uz} \right) + 2 \gamma^{ij} \left(\partial_u F_{iu} \right) \left(\partial_z F_{uj} \right)\\
		& + 2 \gamma^{ij} \left(\partial_u F_{iz} \right) \left(\partial_u F_{uj} \right) + \gamma^{ij} \beta_j \left(\partial_u F_{ui} \right) F_{uz} - \gamma^{ij} \gamma^{kl} \beta_l \left(\partial_u F_{iu} \right) F_{kj}\\
		& + \gamma^{ij} \beta_i F_{uz} \left(\partial_u F_{uj} \right) - \gamma^{ij} \gamma^{kl} \beta_l F_{ik} \left(\partial_u F_{uj} \right)\\
		= & 2 \left(\partial_u F_{uz} \right) \left(\partial_u F_{uz} \right) + 2 \gamma^{ij} \left(\partial_u F_{ui} \right) \left(\partial_j F_{uz} \right) + 2 \gamma^{ij} \left(\partial_u F_{iu} \right) \left(\partial_z F_{uj} \right)\\
		& + 2 \gamma^{ij} \left(\partial_u F_{iz} \right) \left(\partial_u F_{uj} \right) + 2 \gamma^{ij} \beta_i F_{uz} \left(\partial_u F_{uj} \right) + 2 \gamma^{ij} \gamma^{kl} \beta_l \left(\partial_u F_{ui} \right) F_{kj}\,.
	\end{split}
\end{equation}

The sixth term of Eq. (\ref{rewrittenhkk2}) is
\begin{equation}\label{hkk2sixth}
	\begin{split}
		& 2 k^a k^b \nabla_d F_{bc} \nabla^d F_{a}^{\ c} = 2 k^a k^b g^{ce} g^{df} \nabla_d F_{bc} \nabla_f F_{ae}\\
		= & 2 k^a k^b g^{ce} g^{df} \left(\partial_d F_{bc} \right) \left(\partial_f F_{ae} \right) - 2 k^a k^b g^{ce} g^{df} \left(\partial_d F_{bc} \right) \Gamma^{h}_{\ fa} F_{he}\\
		& - 2 k^a k^b g^{ce} g^{df} \left(\partial_d F_{bc} \right) \Gamma^{h}_{\ fe} F_{ah} - 2 k^a k^b g^{ce} g^{df} \Gamma^{g}_{\ db} F_{gc} \left(\partial_f F_{ae} \right)\\
		& + 2 k^a k^b g^{ce} g^{df} \Gamma^{g}_{\ db} F_{gc} \Gamma^{h}_{\ fa} F_{he} + 2 k^a k^b g^{ce} g^{df} \Gamma^{g}_{\ db} F_{gc} \Gamma^{h}_{\ fe} F_{ah}\\
		& - 2 k^a k^b g^{ce} g^{df} \Gamma^{g}_{\ dc} F_{bg} \left(\partial_f F_{ae} \right) + 2 k^a k^b g^{ce} g^{df} \Gamma^{g}_{\ dc} F_{bg} \Gamma^{h}_{\ fa} F_{he}\\
		& + 2 k^a k^b g^{ce} g^{df} \Gamma^{g}_{\ dc} F_{bg} \Gamma^{h}_{\ fe} F_{ah}\,.
	\end{split}
\end{equation}

The first term of Eq. (\ref{hkk2sixth}) is 
\begin{equation}
	\begin{split}
		& 2 k^a k^b g^{ce} g^{df} \left(\partial_d F_{bc} \right) \left(\partial_f F_{ae} \right) = 2 g^{ce} g^{df} \left(\partial_d F_{uc} \right) \left(\partial_f F_{ue} \right)\\
		= & 2 \gamma^{ij} \left(\partial_u F_{ui} \right) \left(\partial_z F_{uj} \right) + 2 \gamma^{ij} \left(\partial_z F_{ui} \right) \left(\partial_u F_{uj} \right) + 2 \gamma^{ij} \gamma^{kl} \left(\partial_k F_{ui} \right) \left(\partial_l F_{uj} \right)\\
		= & 2 \gamma^{ij} \left(\partial_u F_{ui} \right) \left(\partial_z F_{uj} \right) + 2 \gamma^{ij} \left(\partial_z F_{ui} \right) \left(\partial_u F_{uj} \right)\,.
	\end{split}
\end{equation}
Therefore, the first term of Eq. (\ref{hkk2sixth}) is obtained as 
\begin{equation}
	\begin{split}
		& 2 k^a k^b g^{ce} g^{df} \left(\partial_d F_{bc} \right) \left(\partial_f F_{ae} \right)\\
		= & 2 \gamma^{ij} \left(\partial_u F_{ui} \right) \left(\partial_z F_{uj} \right) + 2 \gamma^{ij} \left(\partial_z F_{ui} \right) \left(\partial_u F_{uj} \right)\,.
	\end{split}
\end{equation}

The second term of Eq. (\ref{hkk2sixth}) is
\begin{equation}\label{hkk2sixthsecond}
	\begin{split}
		& - 2 k^a k^b g^{ce} g^{df} \left(\partial_d F_{bc} \right) \Gamma^{h}_{\ fa} F_{he} = - 2 g^{ce} g^{df} \left(\partial_d F_{uc} \right) \Gamma^{h}_{\ fu} F_{he}\\
		= & - 2 \left(\partial_u F_{uz} \right) \Gamma^{h}_{\ zu} F_{hu} - 2 \left(\partial_z F_{uz} \right) \Gamma^{h}_{\ uu} F_{hu} - 2 \gamma^{ij} \left(\partial_i F_{uz} \right) \Gamma^{h}_{\ ju} F_{hu}\\
		& - 2 \gamma^{ij} \left(\partial_u F_{ui} \right) \Gamma^{h}_{\ zu} F_{hj} - 2 \gamma^{ij} \left(\partial_z F_{ui} \right) \Gamma^{h}_{\ uu} F_{hj} - 2 \gamma^{ij} \gamma^{kl} \left(\partial_k F_{ui} \right) \Gamma^{h}_{\ lu} F_{hj}\\
		= & - 2 \left(\partial_u F_{uz} \right) \Gamma^{h}_{\ zu} F_{hu} - 2 \gamma^{ij} \left(\partial_i F_{uz} \right) \Gamma^{h}_{\ ju} F_{hu} - 2 \gamma^{ij} \left(\partial_u F_{ui} \right) \Gamma^{h}_{\ zu} F_{hj}\,.
	\end{split}
\end{equation}
The repeated index $h$ should be further expanded. The first term of Eq. (\ref{hkk2sixthsecond}) is 
\begin{equation}
	\begin{split}
		& - 2 \left(\partial_u F_{uz} \right) \Gamma^{h}_{\ zu} F_{hu}\\
		= & - 2 \left(\partial_u F_{uz} \right) \Gamma^{u}_{\ zu} F_{uu} - 2 \left(\partial_u F_{uz} \right) \Gamma^{z}_{\ zu} F_{zu} - 2 \left(\partial_u F_{uz} \right) \Gamma^{i}_{\ zu} F_{iu}\\
		= & - 2 \left(\partial_u F_{uz} \right) \Gamma^{z}_{\ zu} F_{zu}\\
		= & 0\,.
	\end{split}
\end{equation}
The second term of Eq. (\ref{hkk2sixthsecond}) is 
\begin{equation}
	\begin{split}
		& - 2 \gamma^{ij} \left(\partial_i F_{uz} \right) \Gamma^{h}_{\ ju} F_{hu}\\
		= & - 2 \gamma^{ij} \left(\partial_i F_{uz} \right) \Gamma^{u}_{\ ju} F_{uu} - 2 \gamma^{ij} \left(\partial_i F_{uz} \right) \Gamma^{z}_{\ ju} F_{zu} - 2 \gamma^{ij} \left(\partial_i F_{uz} \right) \Gamma^{k}_{\ ju} F_{ku}\\
		= & - 2 \gamma^{ij} \left(\partial_i F_{uz} \right) \Gamma^{z}_{\ ju} F_{zu}\\
		= & 0\,.
	\end{split}
\end{equation}
The third term of Eq. (\ref{hkk2sixthsecond}) is
\begin{equation}
	\begin{split}
		& - 2 \gamma^{ij} \left(\partial_u F_{ui} \right) \Gamma^{h}_{\ zu} F_{hj}\\
		= & - 2 \gamma^{ij} \left(\partial_u F_{ui} \right) \Gamma^{u}_{\ zu} F_{uj} - 2 \gamma^{ij} \left(\partial_u F_{ui} \right) \Gamma^{z}_{\ zu} F_{zj} - 2 \gamma^{ij} \left(\partial_u F_{ui} \right) \Gamma^{k}_{\ zu} F_{kj}\\
		= & - 2 \gamma^{ij} \left(\partial_u F_{ui} \right) \Gamma^{z}_{\ zu} F_{zj} - 2 \gamma^{ij} \left(\partial_u F_{ui} \right) \Gamma^{k}_{\ zu} F_{kj}\\
		= & - \gamma^{ij} \gamma^{kl} \beta_l \left(\partial_u F_{ui} \right) F_{kj}\,.
	\end{split}
\end{equation}
Therefore, the second term of Eq. (\ref{hkk2sixth}) is obtained as
\begin{equation}
	\begin{split}
		- 2 k^a k^b g^{ce} g^{df} \left(\partial_d F_{bc} \right) \Gamma^{h}_{\ fa} F_{he} = - \gamma^{ij} \gamma^{kl} \beta_l \left(\partial_u F_{ui} \right) F_{kj}\,.
	\end{split}
\end{equation}

The third term of Eq. (\ref{hkk2sixth}) is
\begin{equation}\label{hkk2sixththird}
	\begin{split}
		& - 2 k^a k^b g^{ce} g^{df} \left(\partial_d F_{bc} \right) \Gamma^{h}_{\ fe} F_{ah} = - 2 g^{ce} g^{df} \left(\partial_d F_{uc} \right) \Gamma^{h}_{\ fe} F_{uh}\\
		= & - 2 \left(\partial_u F_{uz} \right) \Gamma^{h}_{\ zu} F_{uh} - 2 \left(\partial_z F_{uz} \right) \Gamma^{h}_{\ uu} F_{uh} - 2 \gamma^{ij} \left(\partial_i F_{uz} \right) \Gamma^{h}_{\ ju} F_{uh}\\
		& - 2 \gamma^{ij} \left(\partial_u F_{ui} \right) \Gamma^{h}_{\ zj} F_{uh} - 2 \gamma^{ij} \left(\partial_z F_{ui} \right) \Gamma^{h}_{\ uj} F_{uh} - 2 \gamma^{ij} \gamma^{kl} \left(\partial_k F_{ui} \right) \Gamma^{h}_{\ lj} F_{uh}\\
		= & - 2 \left(\partial_u F_{uz} \right) \Gamma^{h}_{\ zu} F_{uh} - 2 \gamma^{ij} \left(\partial_i F_{uz} \right) \Gamma^{h}_{\ ju} F_{uh} - 2 \gamma^{ij} \left(\partial_u F_{ui} \right) \Gamma^{h}_{\ zj} F_{uh}\\
		& - 2 \gamma^{ij} \left(\partial_z F_{ui} \right) \Gamma^{h}_{\ uj} F_{uh}\,.
	\end{split}
\end{equation}
The repeated index $h$ should be further expanded. The first term of Eq. (\ref{hkk2sixththird}) is 
\begin{equation}
	\begin{split}
		& - 2 \left(\partial_u F_{uz} \right) \Gamma^{h}_{\ zu} F_{uh}\\
		= & - 2 \left(\partial_u F_{uz} \right) \Gamma^{u}_{\ zu} F_{uu} - 2 \left(\partial_u F_{uz} \right) \Gamma^{z}_{\ zu} F_{uz} - 2 \left(\partial_u F_{uz} \right) \Gamma^{i}_{\ zu} F_{ui}\\
		= & - 2 \left(\partial_u F_{uz} \right) \Gamma^{z}_{\ zu} F_{uz}\\
		= & 0\,.
	\end{split}
\end{equation}
The second term of Eq. (\ref{hkk2sixththird}) is 
\begin{equation}
	\begin{split}
		& - 2 \gamma^{ij} \left(\partial_i F_{uz} \right) \Gamma^{h}_{\ ju} F_{uh}\\
		= & - 2 \gamma^{ij} \left(\partial_i F_{uz} \right) \Gamma^{u}_{\ ju} F_{uu} - 2 \gamma^{ij} \left(\partial_i F_{uz} \right) \Gamma^{z}_{\ ju} F_{uz} - 2 \gamma^{ij} \left(\partial_i F_{uz} \right) \Gamma^{k}_{\ ju} F_{uk}\\
		= & - 2 \gamma^{ij} \left(\partial_i F_{uz} \right) \Gamma^{z}_{\ ju} F_{uz}\\
		= & 0\,.
	\end{split}
\end{equation}
The third term of Eq. (\ref{hkk2sixththird}) is 
\begin{equation}
	\begin{split}
		& - 2 \gamma^{ij} \left(\partial_u F_{ui} \right) \Gamma^{h}_{\ zj} F_{uh}\\
		= & - 2 \gamma^{ij} \left(\partial_u F_{ui} \right) \Gamma^{u}_{\ zj} F_{uu} - 2 \gamma^{ij} \left(\partial_u F_{ui} \right) \Gamma^{z}_{\ zj} F_{uz} - 2 \gamma^{ij} \left(\partial_u F_{ui} \right) \Gamma^{k}_{\ zj} F_{uk}\\
		= & - 2 \gamma^{ij} \left(\partial_u F_{ui} \right) \Gamma^{z}_{\ zj} F_{uz}\\
		= & - \gamma^{ij} \beta_j \left(\partial_u F_{ui} \right) F_{uz}\,.
	\end{split}
\end{equation}
The fourth term of Eq. (\ref{hkk2sixththird}) is 
\begin{equation}
	\begin{split}
		& - 2 \gamma^{ij} \left(\partial_z F_{ui} \right) \Gamma^{h}_{\ uj} F_{uh}\\
		= & - 2 \gamma^{ij} \left(\partial_z F_{ui} \right) \Gamma^{u}_{\ uj} F_{uu} - 2 \gamma^{ij} \left(\partial_z F_{ui} \right) \Gamma^{z}_{\ uj} F_{uz} - 2 \gamma^{ij} \left(\partial_z F_{ui} \right) \Gamma^{k}_{\ uj} F_{uk}\\
		= & - 2 \gamma^{ij} \left(\partial_z F_{ui} \right) \Gamma^{z}_{\ uj} F_{uz}\\
		= & 0\,.
	\end{split}
\end{equation}
Therefore, the third term of Eq. (\ref{hkk2sixth}) is obtained as 
\begin{equation}
	\begin{split}
		- 2 k^a k^b g^{ce} g^{df} \left(\partial_d F_{bc} \right) \Gamma^{h}_{\ fe} F_{ah} = - \gamma^{ij} \beta_j \left(\partial_u F_{ui} \right) F_{uz}\,.
	\end{split}
\end{equation}

The fourth term of Eq. (\ref{hkk2sixth}) is
\begin{equation}
	\begin{split}
		& - 2 k^a k^b g^{ce} g^{df} \Gamma^{g}_{\ db} F_{gc} \left(\partial_f F_{ae} \right) = - 2 g^{ce} g^{df} \Gamma^{g}_{\ du} F_{gc} \left(\partial_f F_{ue} \right)\\
		= & - 2 \Gamma^{g}_{\ uu} F_{gu} \left(\partial_z F_{uz} \right) - 2 \Gamma^{g}_{\ zu} F_{gu} \left(\partial_u F_{uz} \right) - 2 \gamma^{ij} \Gamma^{g}_{\ iu} F_{gu} \left(\partial_j F_{uz} \right)\\
		& - 2 \gamma^{ij} \Gamma^{g}_{\ uu} F_{gi} \left(\partial_z F_{uj} \right) - 2 \gamma^{ij} \Gamma^{g}_{\ zu} F_{gi} \left(\partial_u F_{uj} \right) - 2 \gamma^{ij} \gamma^{kl} \Gamma^{g}_{\ ku} F_{gi} \left(\partial_l F_{uj} \right)\\
		= & - 2 \Gamma^{g}_{\ zu} F_{gu} \left(\partial_u F_{uz} \right) - 2 \gamma^{ij} \Gamma^{g}_{\ iu} F_{gu} \left(\partial_j F_{uz} \right) - 2 \gamma^{ij} \Gamma^{g}_{\ zu} F_{gi} \left(\partial_u F_{uj} \right)\,.
	\end{split}
\end{equation}
The index $g$ should be further expanded.
\begin{equation}\label{hkk2sixthfourth}
	\begin{split}
		& - 2 \Gamma^{g}_{\ zu} F_{gu} \left(\partial_u F_{uz} \right) - 2 \gamma^{ij} \Gamma^{g}_{\ iu} F_{gu} \left(\partial_j F_{uz} \right) - 2 \gamma^{ij} \Gamma^{g}_{\ zu} F_{gi} \left(\partial_u F_{uj} \right)\\
		= & - 2 \Gamma^{u}_{\ zu} F_{uu} \left(\partial_u F_{uz} \right) - 2 \Gamma^{z}_{\ zu} F_{zu} \left(\partial_u F_{uz} \right) - 2 \Gamma^{i}_{\ zu} F_{iu} \left(\partial_u F_{uz} \right)\\
		& - 2 \gamma^{ij} \Gamma^{u}_{\ iu} F_{uu} \left(\partial_j F_{uz} \right) - 2 \gamma^{ij} \Gamma^{z}_{\ iu} F_{zu} \left(\partial_j F_{uz} \right) - 2 \gamma^{ij} \Gamma^{k}_{\ iu} F_{ku} \left(\partial_j F_{uz} \right)\\
		& - 2 \gamma^{ij} \Gamma^{u}_{\ zu} F_{ui} \left(\partial_u F_{uj} \right) - 2 \gamma^{ij} \Gamma^{z}_{\ zu} F_{zi} \left(\partial_u F_{uj} \right) - 2 \gamma^{ij} \Gamma^{k}_{\ zu} F_{ki} \left(\partial_u F_{uj} \right)\\
		= & - 2 \Gamma^{z}_{\ zu} F_{zu} \left(\partial_u F_{uz} \right) - 2 \gamma^{ij} \Gamma^{z}_{\ iu} F_{zu} \left(\partial_j F_{uz} \right) - 2 \gamma^{ij} \Gamma^{z}_{\ zu} F_{zi} \left(\partial_u F_{uj} \right)\\
		& - 2 \gamma^{ij} \Gamma^{k}_{\ zu} F_{ki} \left(\partial_u F_{uj} \right)\,.
	\end{split}
\end{equation}
Therefore, the fourth term of Eq. (\ref{hkk2sixth}) is obtained as 
\begin{equation}
	\begin{split}
		- 2 k^a k^b g^{ce} g^{df} \Gamma^{g}_{\ db} F_{gc} \left(\partial_f F_{ae} \right) = - \gamma^{ij} \gamma^{kl} \beta_l F_{ki} \left(\partial_u F_{uj} \right)\,.
	\end{split}
\end{equation}

The fifth term of Eq. (\ref{hkk2sixth}) is
\begin{equation}
	\begin{split}
		& 2 k^a k^b g^{ce} g^{df} \Gamma^{g}_{\ db} F_{gc} \Gamma^{h}_{\ fa} F_{he} = 2 g^{ce} g^{df} \Gamma^{g}_{\ du} F_{gc} \Gamma^{h}_{\ fu} F_{he}\\
		= & 2 \Gamma^{g}_{\ uu} F_{gu} \Gamma^{h}_{\ zu} F_{hz} + 2 \Gamma^{g}_{\ zu} F_{gu} \Gamma^{h}_{\ uu} F_{hz} + 2 \gamma^{ij} \Gamma^{g}_{\ iu} F_{gu} \Gamma^{h}_{\ ju} F_{hz}\\
		& + 2 \Gamma^{g}_{\ uu} F_{gz} \Gamma^{h}_{\ zu} F_{hu} + 2 \Gamma^{g}_{\ zu} F_{gz} \Gamma^{h}_{\ uu} F_{hu} + 2 \gamma^{ij} \Gamma^{g}_{\ iu} F_{gz} \Gamma^{h}_{\ ju} F_{hu}\\
		& + 2 \gamma^{ij} \Gamma^{g}_{\ uu} F_{gi} \Gamma^{h}_{\ zu} F_{hj} + 2 \gamma^{ij} \Gamma^{g}_{\ zu} F_{gi} \Gamma^{h}_{\ uu} F_{hj} + 2 \gamma^{ij} \gamma^{kl} \Gamma^{g}_{\ ku} F_{gi} \Gamma^{h}_{\ lu} F_{hj}\\
		= & 2 \gamma^{ij} \Gamma^{g}_{\ iu} F_{gu} \Gamma^{h}_{\ ju} F_{hz} + 2 \gamma^{ij} \Gamma^{g}_{\ iu} F_{gz} \Gamma^{h}_{\ ju} F_{hu} + 2 \gamma^{ij} \gamma^{kl} \Gamma^{g}_{\ ku} F_{gi} \Gamma^{h}_{\ lu} F_{hj}\,.
	\end{split}
\end{equation}
The index $g$ should be further expanded.
\begin{equation}\label{hkk2sixthfifth}
	\begin{split}
		& 2 \gamma^{ij} \Gamma^{g}_{\ iu} F_{gu} \Gamma^{h}_{\ ju} F_{hz} + 2 \gamma^{ij} \Gamma^{g}_{\ iu} F_{gz} \Gamma^{h}_{\ ju} F_{hu} + 2 \gamma^{ij} \gamma^{kl} \Gamma^{g}_{\ ku} F_{gi} \Gamma^{h}_{\ lu} F_{hj}\\
		= & 2 \gamma^{ij} \Gamma^{u}_{\ iu} F_{uu} \Gamma^{h}_{\ ju} F_{hz} + 2 \gamma^{ij} \Gamma^{z}_{\ iu} F_{zu} \Gamma^{h}_{\ ju} F_{hz} + 2 \gamma^{ij} \Gamma^{k}_{\ iu} F_{ku} \Gamma^{h}_{\ ju} F_{hz}\\
		& + 2 \gamma^{ij} \Gamma^{u}_{\ iu} F_{uz} \Gamma^{h}_{\ ju} F_{hu} + 2 \gamma^{ij} \Gamma^{z}_{\ iu} F_{zz} \Gamma^{h}_{\ ju} F_{hu} + 2 \gamma^{ij} \Gamma^{k}_{\ iu} F_{kz} \Gamma^{h}_{\ ju} F_{hu}\\
		& + 2 \gamma^{ij} \gamma^{kl} \Gamma^{u}_{\ ku} F_{ui} \Gamma^{h}_{\ lu} F_{hj} + 2 \gamma^{ij} \gamma^{kl} \Gamma^{z}_{\ ku} F_{zi} \Gamma^{h}_{\ lu} F_{hj} + 2 \gamma^{ij} \gamma^{kl} \Gamma^{m}_{\ ku} F_{mi} \Gamma^{h}_{\ lu} F_{hj}\\
		= & 2 \gamma^{ij} \Gamma^{z}_{\ iu} F_{zu} \Gamma^{h}_{\ ju} F_{hz} + 2 \gamma^{ij} \Gamma^{u}_{\ iu} F_{uz} \Gamma^{h}_{\ ju} F_{hu} + 2 \gamma^{ij} \Gamma^{k}_{\ iu} F_{kz} \Gamma^{h}_{\ ju} F_{hu}\\
		& + 2 \gamma^{ij} \gamma^{kl} \Gamma^{z}_{\ ku} F_{zi} \Gamma^{h}_{\ lu} F_{hj} + 2 \gamma^{ij} \gamma^{kl} \Gamma^{m}_{\ ku} F_{mi} \Gamma^{h}_{\ lu} F_{hj}\,.
	\end{split}
\end{equation}
The repeated index $h$ should be further expanded. The first term of Eq. (\ref{hkk2sixthfifth}) is
\begin{equation}
	\begin{split}
		2 \gamma^{ij} \Gamma^{z}_{\ iu} F_{zu} \Gamma^{h}_{\ ju} F_{hz} = 0\,.
	\end{split}
\end{equation}
The second term of Eq. (\ref{hkk2sixthfifth}) is
\begin{equation}
	\begin{split}
		& 2 \gamma^{ij} \Gamma^{u}_{\ iu} F_{uz} \Gamma^{h}_{\ ju} F_{hu}\\
		= & 2 \gamma^{ij} \Gamma^{u}_{\ iu} F_{uz} \Gamma^{u}_{\ ju} F_{uu} + 2 \gamma^{ij} \Gamma^{u}_{\ iu} F_{uz} \Gamma^{z}_{\ ju} F_{zu} + 2 \gamma^{ij} \Gamma^{u}_{\ iu} F_{uz} \Gamma^{k}_{\ ju} F_{ku}\\
		= & 2 \gamma^{ij} \Gamma^{u}_{\ iu} F_{uz} \Gamma^{z}_{\ ju} F_{zu}\\
		= & 0\,.
	\end{split}
\end{equation}
The third term of Eq. (\ref{hkk2sixthfifth}) is
\begin{equation}
	\begin{split}
		& 2 \gamma^{ij} \Gamma^{k}_{\ iu} F_{kz} \Gamma^{h}_{\ ju} F_{hu}\\
		= & 2 \gamma^{ij} \Gamma^{k}_{\ iu} F_{kz} \Gamma^{u}_{\ ju} F_{uu} + 2 \gamma^{ij} \Gamma^{k}_{\ iu} F_{kz} \Gamma^{z}_{\ ju} F_{zu} + 2 \gamma^{ij} \Gamma^{k}_{\ iu} F_{kz} \Gamma^{l}_{\ ju} F_{lu}\\
		= & 2 \gamma^{ij} \Gamma^{k}_{\ iu} F_{kz} \Gamma^{z}_{\ ju} F_{zu}\\
		= & 0\,.
	\end{split}
\end{equation}
The fourth term of Eq. (\ref{hkk2sixthfifth}) is
\begin{equation}
	\begin{split}
		2 \gamma^{ij} \gamma^{kl} \Gamma^{z}_{\ ku} F_{zi} \Gamma^{h}_{\ lu} F_{hj} = 0\,.
	\end{split}
\end{equation}
The fifth term of Eq. (\ref{hkk2sixthfifth}) is
\begin{equation}
	\begin{split}
		& 2 \gamma^{ij} \gamma^{kl} \Gamma^{m}_{\ ku} F_{mi} \Gamma^{h}_{\ lu} F_{hj}\\
		= & 2 \gamma^{ij} \gamma^{kl} \Gamma^{m}_{\ ku} F_{mi} \Gamma^{u}_{\ lu} F_{uj} + 2 \gamma^{ij} \gamma^{kl} \Gamma^{m}_{\ ku} F_{mi} \Gamma^{z}_{\ lu} F_{zj} + 2 \gamma^{ij} \gamma^{kl} \Gamma^{m}_{\ ku} F_{mi} \Gamma^{n}_{\ lu} F_{nj}\\
		= & 2 \gamma^{ij} \gamma^{kl} \Gamma^{m}_{\ ku} F_{mi} \Gamma^{z}_{\ lu} F_{zj} + 2 \gamma^{ij} \gamma^{kl} \Gamma^{m}_{\ ku} F_{mi} \Gamma^{n}_{\ lu} F_{nj}\\
		= & \frac{1}{2} \gamma^{ij} \gamma^{kl} \gamma^{mo} \gamma^{np} \left(\partial_u \gamma_{ko} \right) \left(\partial_u \gamma_{lp} \right) F_{nj}\,.
	\end{split}
\end{equation}
Therefore, the fifth term of Eq. (\ref{hkk2sixth}) is obtained as 
\begin{equation}
	\begin{split}
		& 2 k^a k^b g^{ce} g^{df} \Gamma^{g}_{\ db} F_{gc} \Gamma^{h}_{\ fa} F_{he} = \frac{1}{2} \gamma^{ij} \gamma^{kl} \gamma^{mo} \gamma^{np} \left(\partial_u \gamma_{ko} \right) \left(\partial_u \gamma_{lp} \right) F_{nj}\\
		= & 2 \gamma^{ij} \gamma^{kl} \gamma^{mo} \gamma^{np} K_{ko} K_{lp} F_{nj}\\
		= & 0\,.
	\end{split}
\end{equation}

The sixth term of Eq. (\ref{hkk2sixth}) is
\begin{equation}
	\begin{split}
		& 2 k^a k^b g^{ce} g^{df} \Gamma^{g}_{\ db} F_{gc} \Gamma^{h}_{\ fe} F_{ah} = 2 g^{ce} g^{df} \Gamma^{g}_{\ du} F_{gc} \Gamma^{h}_{\ fe} F_{uh}\\
		= & 2 \Gamma^{g}_{\ uu} F_{gu} \Gamma^{h}_{\ zz} F_{uh} + 2 \Gamma^{g}_{\ zu} F_{gu} \Gamma^{h}_{\ uz} F_{uh} + 2 \gamma^{ij} \Gamma^{g}_{\ iu} F_{gu} \Gamma^{h}_{\ jz} F_{uh}\\
		& + 2 \Gamma^{g}_{\ uu} F_{gz} \Gamma^{h}_{\ zu} F_{uh} + 2 \Gamma^{g}_{\ zu} F_{gz} \Gamma^{h}_{\ uu} F_{uh} + 2 \gamma^{ij} \Gamma^{g}_{\ iu} F_{gz} \Gamma^{h}_{\ ju} F_{uh}\\
		& + 2 \gamma^{ij} \Gamma^{g}_{\ uu} F_{gi} \Gamma^{h}_{\ zj} F_{uh} + 2 \gamma^{ij} \Gamma^{g}_{\ zu} F_{gi} \Gamma^{h}_{\ uj} F_{uh} + 2 \gamma^{ij} \gamma^{kl} \Gamma^{g}_{\ ku} F_{gi} \Gamma^{h}_{\ lj} F_{uh}\\
		= & 2 \Gamma^{g}_{\ zu} F_{gu} \Gamma^{h}_{\ uz} F_{uh} + 2 \gamma^{ij} \Gamma^{g}_{\ iu} F_{gu} \Gamma^{h}_{\ jz} F_{uh} + 2 \gamma^{ij} \Gamma^{g}_{\ iu} F_{gz} \Gamma^{h}_{\ ju} F_{uh}\\
		& + 2 \gamma^{ij} \Gamma^{g}_{\ zu} F_{gi} \Gamma^{h}_{\ uj} F_{uh} + 2 \gamma^{ij} \gamma^{kl} \Gamma^{g}_{\ ku} F_{gi} \Gamma^{h}_{\ lj} F_{uh}\,.
	\end{split}
\end{equation}
The index $g$ should be further expanded.
\begin{equation}\label{hkk2sixthsixth}
	\begin{split}
		& 2 \Gamma^{g}_{\ zu} F_{gu} \Gamma^{h}_{\ uz} F_{uh} + 2 \gamma^{ij} \Gamma^{g}_{\ iu} F_{gu} \Gamma^{h}_{\ jz} F_{uh} + 2 \gamma^{ij} \Gamma^{g}_{\ iu} F_{gz} \Gamma^{h}_{\ ju} F_{uh}\\
		& + 2 \gamma^{ij} \Gamma^{g}_{\ zu} F_{gi} \Gamma^{h}_{\ uj} F_{uh} + 2 \gamma^{ij} \gamma^{kl} \Gamma^{g}_{\ ku} F_{gi} \Gamma^{h}_{\ lj} F_{uh}\\
		= & 2 \Gamma^{u}_{\ zu} F_{uu} \Gamma^{h}_{\ uz} F_{uh} + 2 \Gamma^{z}_{\ zu} F_{zu} \Gamma^{h}_{\ uz} F_{uh} + 2 \Gamma^{i}_{\ zu} F_{iu} \Gamma^{h}_{\ uz} F_{uh}\\
		& + 2 \gamma^{ij} \Gamma^{u}_{\ iu} F_{uu} \Gamma^{h}_{\ jz} F_{uh} + 2 \gamma^{ij} \Gamma^{z}_{\ iu} F_{zu} \Gamma^{h}_{\ jz} F_{uh} + 2 \gamma^{ij} \Gamma^{k}_{\ iu} F_{ku} \Gamma^{h}_{\ jz} F_{uh}\\
		& + 2 \gamma^{ij} \Gamma^{u}_{\ iu} F_{uz} \Gamma^{h}_{\ ju} F_{uh} + 2 \gamma^{ij} \Gamma^{z}_{\ iu} F_{zz} \Gamma^{h}_{\ ju} F_{uh} + 2 \gamma^{ij} \Gamma^{k}_{\ iu} F_{kz} \Gamma^{h}_{\ ju} F_{uh}\\
		& + 2 \gamma^{ij} \Gamma^{u}_{\ zu} F_{ui} \Gamma^{h}_{\ uj} F_{uh} + 2 \gamma^{ij} \Gamma^{z}_{\ zu} F_{zi} \Gamma^{h}_{\ uj} F_{uh} + 2 \gamma^{ij} \Gamma^{k}_{\ zu} F_{ki} \Gamma^{h}_{\ uj} F_{uh}\\
		& + 2 \gamma^{ij} \gamma^{kl} \Gamma^{u}_{\ ku} F_{ui} \Gamma^{h}_{\ lj} F_{uh} + 2 \gamma^{ij} \gamma^{kl} \Gamma^{z}_{\ ku} F_{zi} \Gamma^{h}_{\ lj} F_{uh} + 2 \gamma^{ij} \gamma^{kl} \Gamma^{m}_{\ ku} F_{mi} \Gamma^{h}_{\ lj} F_{uh}\\
		= & 2 \Gamma^{z}_{\ zu} F_{zu} \Gamma^{h}_{\ uz} F_{uh} + 2 \gamma^{ij} \Gamma^{z}_{\ iu} F_{zu} \Gamma^{h}_{\ jz} F_{uh} + 2 \gamma^{ij} \Gamma^{u}_{\ iu} F_{uz} \Gamma^{h}_{\ ju} F_{uh}\\
		& + 2 \gamma^{ij} \Gamma^{k}_{\ iu} F_{kz} \Gamma^{h}_{\ ju} F_{uh} + 2 \gamma^{ij} \Gamma^{z}_{\ zu} F_{zi} \Gamma^{h}_{\ uj} F_{uh} + 2 \gamma^{ij} \Gamma^{k}_{\ zu} F_{ki} \Gamma^{h}_{\ uj} F_{uh}\\
		& + 2 \gamma^{ij} \gamma^{kl} \Gamma^{z}_{\ ku} F_{zi} \Gamma^{h}_{\ lj} F_{uh} + 2 \gamma^{ij} \gamma^{kl} \Gamma^{m}_{\ ku} F_{mi} \Gamma^{h}_{\ lj} F_{uh}\,.
	\end{split}
\end{equation}
The repeated index $h$ should be further expanded. The first term of Eq. (\ref{hkk2sixthsixth}) is
\begin{equation}
	\begin{split}
		2 \Gamma^{z}_{\ zu} F_{zu} \Gamma^{h}_{\ uz} F_{uh} = 0\,.
	\end{split}
\end{equation}
The second term of Eq. (\ref{hkk2sixthsixth}) is
\begin{equation}
	\begin{split}
		2 \gamma^{ij} \Gamma^{z}_{\ iu} F_{zu} \Gamma^{h}_{\ jz} F_{uh} = 0\,.
	\end{split}
\end{equation}
The third term of Eq. (\ref{hkk2sixthsixth}) is
\begin{equation}
	\begin{split}
		& 2 \gamma^{ij} \Gamma^{u}_{\ iu} F_{uz} \Gamma^{h}_{\ ju} F_{uh}\\
		= & 2 \gamma^{ij} \Gamma^{u}_{\ iu} F_{uz} \Gamma^{u}_{\ ju} F_{uu} + 2 \gamma^{ij} \Gamma^{u}_{\ iu} F_{uz} \Gamma^{z}_{\ ju} F_{uz} + 2 \gamma^{ij} \Gamma^{u}_{\ iu} F_{uz} \Gamma^{k}_{\ ju} F_{uk}\\
		= & 2 \gamma^{ij} \Gamma^{u}_{\ iu} F_{uz} \Gamma^{z}_{\ ju} F_{uz}\\
		= & 0\,.
	\end{split}
\end{equation}
The fourth term of Eq. (\ref{hkk2sixthsixth}) is
\begin{equation}
	\begin{split}
		& 2 \gamma^{ij} \Gamma^{k}_{\ iu} F_{kz} \Gamma^{h}_{\ ju} F_{uh}\\
		= & 2 \gamma^{ij} \Gamma^{k}_{\ iu} F_{kz} \Gamma^{u}_{\ ju} F_{uu} + 2 \gamma^{ij} \Gamma^{k}_{\ iu} F_{kz} \Gamma^{z}_{\ ju} F_{uz} + 2 \gamma^{ij} \Gamma^{k}_{\ iu} F_{kz} \Gamma^{l}_{\ ju} F_{ul}\\
		= & 2 \gamma^{ij} \Gamma^{k}_{\ iu} F_{kz} \Gamma^{z}_{\ ju} F_{uz}\\
		= & 0\,.
	\end{split}
\end{equation}
The fifth term of Eq. (\ref{hkk2sixthsixth}) is
\begin{equation}
	\begin{split}
		2 \gamma^{ij} \Gamma^{z}_{\ zu} F_{zi} \Gamma^{h}_{\ uj} F_{uh} = 0\,.
	\end{split}
\end{equation}
The sixth term of Eq. (\ref{hkk2sixthsixth}) is
\begin{equation}
	\begin{split}
		& 2 \gamma^{ij} \Gamma^{k}_{\ zu} F_{ki} \Gamma^{h}_{\ uj} F_{uh}\\
		= & 2 \gamma^{ij} \Gamma^{k}_{\ zu} F_{ki} \Gamma^{u}_{\ uj} F_{uu} + 2 \gamma^{ij} \Gamma^{k}_{\ zu} F_{ki} \Gamma^{z}_{\ uj} F_{uz} + 2 \gamma^{ij} \Gamma^{k}_{\ zu} F_{ki} \Gamma^{l}_{\ uj} F_{ul}\\
		= & 2 \gamma^{ij} \Gamma^{k}_{\ zu} F_{ki} \Gamma^{z}_{\ uj} F_{uz}\\
		= & 0\,.
	\end{split}
\end{equation}
The seventh term of Eq. (\ref{hkk2sixthsixth}) is
\begin{equation}
	\begin{split}
		& 2 \gamma^{ij} \gamma^{kl} \Gamma^{z}_{\ ku} F_{zi} \Gamma^{h}_{\ lj} F_{uh}\\
		= & 2 \gamma^{ij} \gamma^{kl} \Gamma^{z}_{\ ku} F_{zi} \Gamma^{u}_{\ lj} F_{uu} + 2 \gamma^{ij} \gamma^{kl} \Gamma^{z}_{\ ku} F_{zi} \Gamma^{z}_{\ lj} F_{uz} + 2 \gamma^{ij} \gamma^{kl} \Gamma^{z}_{\ ku} F_{zi} \Gamma^{m}_{\ lj} F_{um}\\
		= & 2 \gamma^{ij} \gamma^{kl} \Gamma^{z}_{\ ku} F_{zi} \Gamma^{z}_{\ lj} F_{uz}\\
		= & 0\,.
	\end{split}
\end{equation}
The fifteenth term of Eq. (\ref{hkk2sixthsixth}) is
\begin{equation}
	\begin{split}
		& 2 \gamma^{ij} \gamma^{kl} \Gamma^{m}_{\ ku} F_{mi} \Gamma^{h}_{\ lj} F_{uh}\\
		= & 2 \gamma^{ij} \gamma^{kl} \Gamma^{m}_{\ ku} F_{mi} \Gamma^{u}_{\ lj} F_{uu} + 2 \gamma^{ij} \gamma^{kl} \Gamma^{m}_{\ ku} F_{mi} \Gamma^{z}_{\ lj} F_{uz} + 2 \gamma^{ij} \gamma^{kl} \Gamma^{m}_{\ ku} F_{mi} \Gamma^{n}_{\ lj} F_{un}\\
		= & 2 \gamma^{ij} \gamma^{kl} \Gamma^{m}_{\ ku} F_{mi} \Gamma^{z}_{\ lj} F_{uz}\\
		= & - \frac{1}{2} \gamma^{ij} \gamma^{kl} \gamma^{mn} \left(\partial_u \gamma_{kn} \right) F_{mi} \left(\partial_u \gamma_{lj} \right) F_{uz}\,.
	\end{split}
\end{equation}
Therefore, the sixth term of Eq. (\ref{hkk2sixth}) is obtained as 
\begin{equation}
	\begin{split}
		& 2 k^a k^b g^{ce} g^{df} \Gamma^{g}_{\ db} F_{gc} \Gamma^{h}_{\ fe} F_{ah} = - \frac{1}{2} \gamma^{ij} \gamma^{kl} \gamma^{mn} \left(\partial_u \gamma_{kn} \right) F_{mi} \left(\partial_u \gamma_{lj} \right) F_{uz}\\
		= & - 2 \gamma^{ij} \gamma^{kl} \gamma^{mn} K_{kn} F_{mi} K_{lj} F_{uz}\\
		= & 0\,.
	\end{split}
\end{equation}

The seventh term of Eq. (\ref{hkk2sixth}) is
\begin{equation}
	\begin{split}
		& - 2 k^a k^b g^{ce} g^{df} \Gamma^{g}_{\ dc} F_{bg} \left(\partial_f F_{ae} \right) = - 2 g^{ce} g^{df} \Gamma^{g}_{\ dc} F_{ug} \left(\partial_f F_{ue} \right)\\
		= & - 2 \Gamma^{g}_{\ uu} F_{ug} \left(\partial_z F_{uz} \right) - 2 \Gamma^{g}_{\ zu} F_{ug} \left(\partial_u F_{uz} \right) - 2 \gamma^{ij} \Gamma^{g}_{\ iu} F_{ug} \left(\partial_j F_{uz} \right)\\
		& - 2 \gamma^{ij} \Gamma^{g}_{\ ui} F_{ug} \left(\partial_z F_{uj} \right) - 2 \gamma^{ij} \Gamma^{g}_{\ zi} F_{ug} \left(\partial_u F_{uj} \right) - 2 \gamma^{ij} \gamma^{kl} \Gamma^{g}_{\ ki} F_{ug} \left(\partial_l F_{uj} \right)\\
		= & - 2 \Gamma^{g}_{\ zu} F_{ug} \left(\partial_u F_{uz} \right) - 2 \gamma^{ij} \Gamma^{g}_{\ iu} F_{ug} \left(\partial_j F_{uz} \right) - 2 \gamma^{ij} \Gamma^{g}_{\ ui} F_{ug} \left(\partial_z F_{uj} \right)\\
		& - 2 \gamma^{ij} \Gamma^{g}_{\ zi} F_{ug} \left(\partial_u F_{uj} \right)\,.
	\end{split}
\end{equation}
The index $g$ should be further expanded.
\begin{equation}\label{hkk2sixthseventh}
	\begin{split}
		& - 2 \Gamma^{g}_{\ zu} F_{ug} \left(\partial_u F_{uz} \right) - 2 \gamma^{ij} \Gamma^{g}_{\ iu} F_{ug} \left(\partial_j F_{uz} \right) - 2 \gamma^{ij} \Gamma^{g}_{\ ui} F_{ug} \left(\partial_z F_{uj} \right)\\
		& - 2 \gamma^{ij} \Gamma^{g}_{\ zi} F_{ug} \left(\partial_u F_{uj} \right)\\
		= & - 2 \Gamma^{u}_{\ zu} F_{uu} \left(\partial_u F_{uz} \right) - 2 \Gamma^{z}_{\ zu} F_{uz} \left(\partial_u F_{uz} \right) - 2 \Gamma^{i}_{\ zu} F_{ui} \left(\partial_u F_{uz} \right)\\
		& - 2 \gamma^{ij} \Gamma^{u}_{\ iu} F_{uu} \left(\partial_j F_{uz} \right) - 2 \gamma^{ij} \Gamma^{z}_{\ iu} F_{uz} \left(\partial_j F_{uz} \right) - 2 \gamma^{ij} \Gamma^{k}_{\ iu} F_{uk} \left(\partial_j F_{uz} \right)\\
		& - 2 \gamma^{ij} \Gamma^{u}_{\ ui} F_{uu} \left(\partial_z F_{uj} \right) - 2 \gamma^{ij} \Gamma^{z}_{\ ui} F_{uz} \left(\partial_z F_{uj} \right) - 2 \gamma^{ij} \Gamma^{k}_{\ ui} F_{uk} \left(\partial_z F_{uj} \right)\\
		& - 2 \gamma^{ij} \Gamma^{u}_{\ zi} F_{uu} \left(\partial_u F_{uj} \right) - 2 \gamma^{ij} \Gamma^{z}_{\ zi} F_{uz} \left(\partial_u F_{uj} \right) - 2 \gamma^{ij} \Gamma^{k}_{\ zi} F_{uk} \left(\partial_u F_{uj} \right)\\
		= & - 2 \Gamma^{z}_{\ zu} F_{uz} \left(\partial_u F_{uz} \right) - 2 \gamma^{ij} \Gamma^{z}_{\ iu} F_{uz} \left(\partial_j F_{uz} \right) - 2 \gamma^{ij} \Gamma^{z}_{\ ui} F_{uz} \left(\partial_z F_{uj} \right)\\
		& - 2 \gamma^{ij} \Gamma^{z}_{\ zi} F_{uz} \left(\partial_u F_{uj} \right)\,.
	\end{split}
\end{equation}
Therefore, the seventh term of Eq. (\ref{hkk2sixth}) is obtained as
\begin{equation}
	\begin{split}
		- 2 k^a k^b g^{ce} g^{df} \Gamma^{g}_{\ dc} F_{bg} \left(\partial_f F_{ae} \right) = - \gamma^{ij} \beta_i F_{uz} \left(\partial_u F_{uj} \right)\,.
	\end{split}
\end{equation}

The eighth term of Eq. (\ref{hkk2sixth}) is
\begin{equation}
	\begin{split}
		& 2 k^a k^b g^{ce} g^{df} \Gamma^{g}_{\ dc} F_{bg} \Gamma^{h}_{\ fa} F_{he} = 2 g^{ce} g^{df} \Gamma^{g}_{\ dc} F_{ug} \Gamma^{h}_{\ fu} F_{he}\\
		= & 2 \Gamma^{g}_{\ uu} F_{ug} \Gamma^{h}_{\ zu} F_{hz} + 2 \Gamma^{g}_{\ zu} F_{ug} \Gamma^{h}_{\ uu} F_{hz} + 2 \gamma^{ij} \Gamma^{g}_{\ iu} F_{ug} \Gamma^{h}_{\ ju} F_{hz}\\
		& + 2 \Gamma^{g}_{\ uz} F_{ug} \Gamma^{h}_{\ zu} F_{hu} + 2 \Gamma^{g}_{\ zz} F_{ug} \Gamma^{h}_{\ uu} F_{hu} + 2 \gamma^{ij} \Gamma^{g}_{\ iz} F_{ug} \Gamma^{h}_{\ ju} F_{hu}\\
		& + 2 \gamma^{ij} \Gamma^{g}_{\ ui} F_{ug} \Gamma^{h}_{\ zu} F_{hj} + 2 \gamma^{ij} \Gamma^{g}_{\ zi} F_{ug} \Gamma^{h}_{\ uu} F_{hj} + 2 \gamma^{ij} \gamma^{kl} \Gamma^{g}_{\ ki} F_{ug} \Gamma^{h}_{\ lu} F_{hj}\\
		= & 2 \gamma^{ij} \Gamma^{g}_{\ iu} F_{ug} \Gamma^{h}_{\ ju} F_{hz} + 2 \Gamma^{g}_{\ uz} F_{ug} \Gamma^{h}_{\ zu} F_{hu} + 2 \gamma^{ij} \Gamma^{g}_{\ iz} F_{ug} \Gamma^{h}_{\ ju} F_{hu}\\
		& + 2 \gamma^{ij} \Gamma^{g}_{\ ui} F_{ug} \Gamma^{h}_{\ zu} F_{hj} + 2 \gamma^{ij} \gamma^{kl} \Gamma^{g}_{\ ki} F_{ug} \Gamma^{h}_{\ lu} F_{hj}\,.
	\end{split}
\end{equation}
The index $g$ should be further expanded.
\begin{equation}\label{hkk2sixtheighth}
	\begin{split}
		& 2 \gamma^{ij} \Gamma^{g}_{\ iu} F_{ug} \Gamma^{h}_{\ ju} F_{hz} + 2 \Gamma^{g}_{\ uz} F_{ug} \Gamma^{h}_{\ zu} F_{hu} + 2 \gamma^{ij} \Gamma^{g}_{\ iz} F_{ug} \Gamma^{h}_{\ ju} F_{hu}\\
		& + 2 \gamma^{ij} \Gamma^{g}_{\ ui} F_{ug} \Gamma^{h}_{\ zu} F_{hj} + 2 \gamma^{ij} \gamma^{kl} \Gamma^{g}_{\ ki} F_{ug} \Gamma^{h}_{\ lu} F_{hj}\\
		= & 2 \gamma^{ij} \Gamma^{u}_{\ iu} F_{uu} \Gamma^{h}_{\ ju} F_{hz} + 2 \gamma^{ij} \Gamma^{z}_{\ iu} F_{uz} \Gamma^{h}_{\ ju} F_{hz} + 2 \gamma^{ij} \Gamma^{k}_{\ iu} F_{uk} \Gamma^{h}_{\ ju} F_{hz}\\
		& + 2 \Gamma^{u}_{\ uz} F_{uu} \Gamma^{h}_{\ zu} F_{hu} + 2 \Gamma^{z}_{\ uz} F_{uz} \Gamma^{h}_{\ zu} F_{hu} + 2 \Gamma^{i}_{\ uz} F_{ui} \Gamma^{h}_{\ zu} F_{hu}\\
		& + 2 \gamma^{ij} \Gamma^{u}_{\ iz} F_{uu} \Gamma^{h}_{\ ju} F_{hu} + 2 \gamma^{ij} \Gamma^{z}_{\ iz} F_{uz} \Gamma^{h}_{\ ju} F_{hu} + 2 \gamma^{ij} \Gamma^{k}_{\ iz} F_{uk} \Gamma^{h}_{\ ju} F_{hu}\\
		& + 2 \gamma^{ij} \Gamma^{u}_{\ ui} F_{uu} \Gamma^{h}_{\ zu} F_{hj} + 2 \gamma^{ij} \Gamma^{z}_{\ ui} F_{uz} \Gamma^{h}_{\ zu} F_{hj} + 2 \gamma^{ij} \Gamma^{k}_{\ ui} F_{uk} \Gamma^{h}_{\ zu} F_{hj}\\
		& + 2 \gamma^{ij} \gamma^{kl} \Gamma^{u}_{\ ki} F_{uu} \Gamma^{h}_{\ lu} F_{hj} + 2 \gamma^{ij} \gamma^{kl} \Gamma^{z}_{\ ki} F_{uz} \Gamma^{h}_{\ lu} F_{hj} + 2 \gamma^{ij} \gamma^{kl} \Gamma^{m}_{\ ki} F_{um} \Gamma^{h}_{\ lu} F_{hj}\\
		= & 2 \gamma^{ij} \Gamma^{z}_{\ iu} F_{uz} \Gamma^{h}_{\ ju} F_{hz} + 2 \Gamma^{z}_{\ uz} F_{uz} \Gamma^{h}_{\ zu} F_{hu} + 2 \gamma^{ij} \Gamma^{z}_{\ iz} F_{uz} \Gamma^{h}_{\ ju} F_{hu}\\
		& + 2 \gamma^{ij} \Gamma^{z}_{\ ui} F_{uz} \Gamma^{h}_{\ zu} F_{hj} + 2 \gamma^{ij} \gamma^{kl} \Gamma^{z}_{\ ki} F_{uz} \Gamma^{h}_{\ lu} F_{hj}\,.
	\end{split}
\end{equation}
The repeated index $h$ should be further expanded. The first term of Eq. (\ref{hkk2sixtheighth}) is 
\begin{equation}
	\begin{split}
		2 \gamma^{ij} \Gamma^{z}_{\ iu} F_{uz} \Gamma^{h}_{\ ju} F_{hz} = 0\,.
	\end{split}
\end{equation}
The second term of Eq. (\ref{hkk2sixtheighth}) is 
\begin{equation}
	\begin{split}
		2 \Gamma^{z}_{\ uz} F_{uz} \Gamma^{h}_{\ zu} F_{hu} = 0\,.
	\end{split}
\end{equation}
The third term of Eq. (\ref{hkk2sixtheighth}) is 
\begin{equation}
	\begin{split}
		& 2 \gamma^{ij} \Gamma^{z}_{\ iz} F_{uz} \Gamma^{h}_{\ ju} F_{hu}\\
		= & 2 \gamma^{ij} \Gamma^{z}_{\ iz} F_{uz} \Gamma^{u}_{\ ju} F_{uu} + 2 \gamma^{ij} \Gamma^{z}_{\ iz} F_{uz} \Gamma^{z}_{\ ju} F_{zu} + 2 \gamma^{ij} \Gamma^{z}_{\ iz} F_{uz} \Gamma^{k}_{\ ju} F_{ku}\\
		= & 2 \gamma^{ij} \Gamma^{z}_{\ iz} F_{uz} \Gamma^{z}_{\ ju} F_{zu}\\
		= & 0\,.
	\end{split}
\end{equation}
The fourth term of Eq. (\ref{hkk2sixtheighth}) is 
\begin{equation}
	\begin{split}
		2 \gamma^{ij} \Gamma^{z}_{\ ui} F_{uz} \Gamma^{h}_{\ zu} F_{hj} = 0\,.
	\end{split}
\end{equation}
The fifth term of Eq. (\ref{hkk2sixtheighth}) is
\begin{equation}
	\begin{split}
		& 2 \gamma^{ij} \gamma^{kl} \Gamma^{z}_{\ ki} F_{uz} \Gamma^{h}_{\ lu} F_{hj}\\
		= & 2 \gamma^{ij} \gamma^{kl} \Gamma^{z}_{\ ki} F_{uz} \Gamma^{u}_{\ lu} F_{uj} + 2 \gamma^{ij} \gamma^{kl} \Gamma^{z}_{\ ki} F_{uz} \Gamma^{z}_{\ lu} F_{zj} + 2 \gamma^{ij} \gamma^{kl} \Gamma^{z}_{\ ki} F_{uz} \Gamma^{m}_{\ lu} F_{mj}\\
		= & 2 \gamma^{ij} \gamma^{kl} \Gamma^{z}_{\ ki} F_{uz} \Gamma^{z}_{\ lu} F_{zj} + 2 \gamma^{ij} \gamma^{kl} \Gamma^{z}_{\ ki} F_{uz} \Gamma^{m}_{\ lu} F_{mj}\\
		= & - \frac{1}{2} \gamma^{ij} \gamma^{kl} \gamma^{mo} \left(\partial_u \gamma_{ki} \right) \left(\partial_u \gamma_{lo} \right) F_{mj}
	\end{split}
\end{equation}
Therefore, the eighth term of Eq. (\ref{hkk2sixth}) is obtained as 
\begin{equation}
	\begin{split}
		& 2 k^a k^b g^{ce} g^{df} \Gamma^{g}_{\ dc} F_{bg} \Gamma^{h}_{\ fa} F_{he} = - \frac{1}{2} \gamma^{ij} \gamma^{kl} \gamma^{mo} \left(\partial_u \gamma_{ki} \right) \left(\partial_u \gamma_{lo} \right) F_{mj}\\
		= & - 2 \gamma^{ij} \gamma^{kl} \gamma^{mo} K_{ki} K_{lo} F_{mj}\\
		= & 0\,.
	\end{split}
\end{equation}

The ninth term of Eq. (\ref{hkk2sixth}) is
\begin{equation}
	\begin{split}
		& 2 k^a k^b g^{ce} g^{df} \Gamma^{g}_{\ dc} F_{bg} \Gamma^{h}_{\ fe} F_{ah} = 2 g^{ce} g^{df} \Gamma^{g}_{\ dc} F_{ug} \Gamma^{h}_{\ fe} F_{uh}\\
		= & 2 \Gamma^{g}_{\ uu} F_{ug} \Gamma^{h}_{\ zz} F_{uh} + 2 \Gamma^{g}_{\ zu} F_{ug} \Gamma^{h}_{\ uz} F_{uh} + 2 \gamma^{ij} \Gamma^{g}_{\ iu} F_{ug} \Gamma^{h}_{\ jz} F_{uh}\\
		& + 2 \Gamma^{g}_{\ uz} F_{ug} \Gamma^{h}_{\ zu} F_{uh} + 2 \Gamma^{g}_{\ zz} F_{ug} \Gamma^{h}_{\ uu} F_{uh} + 2 \gamma^{ij} \Gamma^{g}_{\ iz} F_{ug} \Gamma^{h}_{\ ju} F_{uh}\\
		& + 2 \gamma^{ij} \Gamma^{g}_{\ ui} F_{ug} \Gamma^{h}_{\ zj} F_{uh} + 2 \gamma^{ij} \Gamma^{g}_{\ zi} F_{ug} \Gamma^{h}_{\ uj} F_{uh} + 2 \gamma^{ij} \gamma^{kl} \Gamma^{g}_{\ ki} F_{ug} \Gamma^{h}_{\ lj} F_{uh}\\
		= & 2 \Gamma^{g}_{\ zu} F_{ug} \Gamma^{h}_{\ uz} F_{uh} + 2 \gamma^{ij} \Gamma^{g}_{\ iu} F_{ug} \Gamma^{h}_{\ jz} F_{uh} + 2 \Gamma^{g}_{\ uz} F_{ug} \Gamma^{h}_{\ zu} F_{uh}\\
		& + 2 \gamma^{ij} \Gamma^{g}_{\ iz} F_{ug} \Gamma^{h}_{\ ju} F_{uh} + 2 \gamma^{ij} \Gamma^{g}_{\ ui} F_{ug} \Gamma^{h}_{\ zj} F_{uh} + 2 \gamma^{ij} \Gamma^{g}_{\ zi} F_{ug} \Gamma^{h}_{\ uj} F_{uh}\\
		& + 2 \gamma^{ij} \gamma^{kl} \Gamma^{g}_{\ ki} F_{ug} \Gamma^{h}_{\ lj} F_{uh}\,.
	\end{split}
\end{equation}
The index $g$ should be expanded.
\begin{equation}\label{hkk2sixthninth}
	\begin{split}
		& 2 \Gamma^{g}_{\ zu} F_{ug} \Gamma^{h}_{\ uz} F_{uh} + 2 \gamma^{ij} \Gamma^{g}_{\ iu} F_{ug} \Gamma^{h}_{\ jz} F_{uh} + 2 \Gamma^{g}_{\ uz} F_{ug} \Gamma^{h}_{\ zu} F_{uh}\\
		& + 2 \gamma^{ij} \Gamma^{g}_{\ iz} F_{ug} \Gamma^{h}_{\ ju} F_{uh} + 2 \gamma^{ij} \Gamma^{g}_{\ ui} F_{ug} \Gamma^{h}_{\ zj} F_{uh} + 2 \gamma^{ij} \Gamma^{g}_{\ zi} F_{ug} \Gamma^{h}_{\ uj} F_{uh}\\
		& + 2 \gamma^{ij} \gamma^{kl} \Gamma^{g}_{\ ki} F_{ug} \Gamma^{h}_{\ lj} F_{uh}\\
		= & 2 \Gamma^{u}_{\ zu} F_{uu} \Gamma^{h}_{\ uz} F_{uh} + 2 \Gamma^{z}_{\ zu} F_{uz} \Gamma^{h}_{\ uz} F_{uh} + 2 \Gamma^{i}_{\ zu} F_{ui} \Gamma^{h}_{\ uz} F_{uh}\\
		& + 2 \gamma^{ij} \Gamma^{u}_{\ iu} F_{uu} \Gamma^{h}_{\ jz} F_{uh} + 2 \gamma^{ij} \Gamma^{z}_{\ iu} F_{uz} \Gamma^{h}_{\ jz} F_{uh} + 2 \gamma^{ij} \Gamma^{k}_{\ iu} F_{uk} \Gamma^{h}_{\ jz} F_{uh}\\
		& + 2 \Gamma^{u}_{\ uz} F_{uu} \Gamma^{h}_{\ zu} F_{uh} + 2 \Gamma^{z}_{\ uz} F_{uz} \Gamma^{h}_{\ zu} F_{uh} + 2 \Gamma^{i}_{\ uz} F_{ui} \Gamma^{h}_{\ zu} F_{uh}\\
		& + 2 \gamma^{ij} \Gamma^{u}_{\ iz} F_{uu} \Gamma^{h}_{\ ju} F_{uh} + 2 \gamma^{ij} \Gamma^{z}_{\ iz} F_{uz} \Gamma^{h}_{\ ju} F_{uh} + 2 \gamma^{ij} \Gamma^{k}_{\ iz} F_{uk} \Gamma^{h}_{\ ju} F_{uh}\\
		& + 2 \gamma^{ij} \Gamma^{u}_{\ ui} F_{uu} \Gamma^{h}_{\ zj} F_{uh} + 2 \gamma^{ij} \Gamma^{z}_{\ ui} F_{uz} \Gamma^{h}_{\ zj} F_{uh} + 2 \gamma^{ij} \Gamma^{k}_{\ ui} F_{uk} \Gamma^{h}_{\ zj} F_{uh}\\
		& + 2 \gamma^{ij} \Gamma^{u}_{\ zi} F_{uu} \Gamma^{h}_{\ uj} F_{uh} + 2 \gamma^{ij} \Gamma^{z}_{\ zi} F_{uz} \Gamma^{h}_{\ uj} F_{uh} + 2 \gamma^{ij} \Gamma^{k}_{\ zi} F_{uk} \Gamma^{h}_{\ uj} F_{uh}\\
		& + 2 \gamma^{ij} \gamma^{kl} \Gamma^{u}_{\ ki} F_{uu} \Gamma^{h}_{\ lj} F_{uh} + 2 \gamma^{ij} \gamma^{kl} \Gamma^{z}_{\ ki} F_{uz} \Gamma^{h}_{\ lj} F_{uh} + 2 \gamma^{ij} \gamma^{kl} \Gamma^{m}_{\ ki} F_{um} \Gamma^{h}_{\ lj} F_{uh}\\
		= & 2 \Gamma^{z}_{\ zu} F_{uz} \Gamma^{h}_{\ uz} F_{uh} + 2 \gamma^{ij} \Gamma^{z}_{\ iu} F_{uz} \Gamma^{h}_{\ jz} F_{uh} + 2 \Gamma^{z}_{\ uz} F_{uz} \Gamma^{h}_{\ zu} F_{uh}\\
		& + 2 \gamma^{ij} \Gamma^{z}_{\ iz} F_{uz} \Gamma^{h}_{\ ju} F_{uh} + 2 \gamma^{ij} \Gamma^{z}_{\ ui} F_{uz} \Gamma^{h}_{\ zj} F_{uh} + 2 \gamma^{ij} \Gamma^{z}_{\ zi} F_{uz} \Gamma^{h}_{\ uj} F_{uh}\\
		& + 2 \gamma^{ij} \gamma^{kl} \Gamma^{z}_{\ ki} F_{uz} \Gamma^{h}_{\ lj} F_{uh}\,.
	\end{split}
\end{equation}
The repeated index $h$ should be further expanded. The first term of Eq. (\ref{hkk2sixthninth}) is 
\begin{equation}
	\begin{split}
		2 \Gamma^{z}_{\ zu} F_{uz} \Gamma^{h}_{\ uz} F_{uh} = 0\,.
	\end{split}
\end{equation}
The second term of Eq. (\ref{hkk2sixthninth}) is 
\begin{equation}
	\begin{split}
		2 \gamma^{ij} \Gamma^{z}_{\ iu} F_{uz} \Gamma^{h}_{\ jz} F_{uh} = 0\,.
	\end{split}
\end{equation}
The third term of Eq. (\ref{hkk2sixthninth}) is
\begin{equation}
	\begin{split}
		2 \Gamma^{z}_{\ uz} F_{uz} \Gamma^{h}_{\ zu} F_{uh} = 0\,.
	\end{split}
\end{equation}
The fourth term of Eq. (\ref{hkk2sixthninth}) is
\begin{equation}
	\begin{split}
		& 2 \gamma^{ij} \Gamma^{z}_{\ iz} F_{uz} \Gamma^{h}_{\ ju} F_{uh}\\
		= & 2 \gamma^{ij} \Gamma^{z}_{\ iz} F_{uz} \Gamma^{u}_{\ ju} F_{uu} + 2 \gamma^{ij} \Gamma^{z}_{\ iz} F_{uz} \Gamma^{z}_{\ ju} F_{uz} + 2 \gamma^{ij} \Gamma^{z}_{\ iz} F_{uz} \Gamma^{k}_{\ ju} F_{uk}\\
		= & 2 \gamma^{ij} \Gamma^{z}_{\ iz} F_{uz} \Gamma^{z}_{\ ju} F_{uz}\\
		= & 0\,.
	\end{split}
\end{equation}
The fifth term of Eq. (\ref{hkk2sixthninth}) is
\begin{equation}
	\begin{split}
		2 \gamma^{ij} \Gamma^{z}_{\ ui} F_{uz} \Gamma^{h}_{\ zj} F_{uh} = 0\,.
	\end{split}
\end{equation}
The sixth term of Eq. (\ref{hkk2sixthninth}) is
\begin{equation}
	\begin{split}
		& 2 \gamma^{ij} \Gamma^{z}_{\ zi} F_{uz} \Gamma^{h}_{\ uj} F_{uh}\\
		= & 2 \gamma^{ij} \Gamma^{z}_{\ zi} F_{uz} \Gamma^{u}_{\ uj} F_{uu} + 2 \gamma^{ij} \Gamma^{z}_{\ zi} F_{uz} \Gamma^{z}_{\ uj} F_{uz} + 2 \gamma^{ij} \Gamma^{z}_{\ zi} F_{uz} \Gamma^{k}_{\ uj} F_{uk}\\
		= & 2 \gamma^{ij} \Gamma^{z}_{\ zi} F_{uz} \Gamma^{z}_{\ uj} F_{uz}\\
		= & 0\,.
	\end{split}
\end{equation}
The seventh term of Eq. (\ref{hkk2sixthninth}) is
\begin{equation}
	\begin{split}
		& 2 \gamma^{ij} \gamma^{kl} \Gamma^{z}_{\ ki} F_{uz} \Gamma^{h}_{\ lj} F_{uh}\\
		= & 2 \gamma^{ij} \gamma^{kl} \Gamma^{z}_{\ ki} F_{uz} \Gamma^{u}_{\ lj} F_{uu} + 2 \gamma^{ij} \gamma^{kl} \Gamma^{z}_{\ ki} F_{uz} \Gamma^{z}_{\ lj} F_{uz} + 2 \gamma^{ij} \gamma^{kl} \Gamma^{z}_{\ ki} F_{uz} \Gamma^{m}_{\ lj} F_{um}\\
		= & 2 \gamma^{ij} \gamma^{kl} \Gamma^{z}_{\ ki} F_{uz} \Gamma^{z}_{\ lj} F_{uz}\\
		= & \frac{1}{2} \gamma^{ij} \gamma^{kl} \left(\partial_u \gamma_{ki} \right) F_{uz} \left(\partial_u \gamma_{lj} \right) F_{uz}\,.
	\end{split}
\end{equation}
Therefore, the ninth term of Eq. (\ref{hkk2sixth}) is obtained as 
\begin{equation}
	\begin{split}
		& 2 k^a k^b g^{ce} g^{df} \Gamma^{g}_{\ dc} F_{bg} \Gamma^{h}_{\ fe} F_{ah} = \frac{1}{2} \gamma^{ij} \gamma^{kl} \left(\partial_u \gamma_{ki} \right) F_{uz} \left(\partial_u \gamma_{lj} \right) F_{uz}\\
		= & 2 \gamma^{ij} \gamma^{kl} K_{ki} F_{uz} K_{lj} F_{uz}\\
		= & 0\,.
	\end{split}
\end{equation}

Finally, the sixth term of Eq. (\ref{rewrittenhkk2}) is
\begin{equation}
	\begin{split}
		& 2 k^a k^b \nabla_d F_{bc} \nabla^d F_{a}^{\ c}\\
		= & 2 \gamma^{ij} \left(\partial_u F_{ui} \right) \left(\partial_z F_{uj} \right) + 2 \gamma^{ij} \left(\partial_z F_{ui} \right) \left(\partial_u F_{uj} \right) - \gamma^{ij} \gamma^{kl} \beta_l \left(\partial_u F_{ui} \right) F_{kj}\\
		& - \gamma^{ij} \beta_j \left(\partial_u F_{ui} \right) F_{uz} - \gamma^{ij} \gamma^{kl} \beta_l F_{ki} \left(\partial_u F_{uj} \right) - \gamma^{ij} \beta_i F_{uz} \left(\partial_u F_{uj} \right)\\
		= & 2 \gamma^{ij} \left(\partial_u F_{ui} \right) \left(\partial_z F_{uj} \right) + 2 \gamma^{ij} \left(\partial_u F_{ui} \right) \left(\partial_z F_{uj} \right) - \gamma^{ij} \gamma^{kl} \beta_l \left(\partial_u F_{ui} \right) F_{kj}\\
		& - \gamma^{ij} \beta_i F_{uz} \left(\partial_u F_{uj} \right) - \gamma^{ij} \gamma^{kl} \beta_l F_{ki} \left(\partial_u F_{uj} \right) - \gamma^{ij} \beta_i F_{uz} \left(\partial_u F_{uj} \right)\\
		= & 4 \gamma^{ij} \left(\partial_u F_{ui} \right) \left(\partial_z F_{uj} \right) - 2 \gamma^{ij} \gamma^{kl} \beta_l \left(\partial_u F_{ui} \right) F_{kj} - 2 \gamma^{ij} \beta_i F_{uz} \left(\partial_u F_{uj} \right)\,.
	\end{split}
\end{equation}

The seventh term of Eq. (\ref{rewrittenhkk2}) is
\begin{equation}
	\begin{split}
		& - 2 k^a k^b F_{a}^{\ c} F_b^{\ d} R_{cd} = - 2 k^a k^b g^{ce} g^{df} F_{ae} F_{bf} R_{cd} = - 2 g^{ce} g^{df} F_{ue} F_{uf} R_{cd}\\
		= & - 2 F_{uz} F_{uz} R_{uu}\\
		= & - 2 F_{uz} F_{uz} \left[- \frac{1}{2} \left(\partial_u \gamma^{ij} \right) \left(\partial_u \gamma_{ij} \right) - \frac{1}{2} \gamma^{ij} \partial_u^2 \gamma_{ij} - \frac{1}{4} \gamma^{ij} \gamma^{kl} \left(\partial_u \gamma_{ik} \right) \left(\partial_u \gamma_{jl} \right) \right]\\
		= & F_{uz} F_{uz} \left(\partial_u \gamma^{ij} \right) \left(\partial_u \gamma_{ij} \right) + F_{uz} F_{uz} \gamma^{ij} \partial_u^2 \gamma_{ij} + \frac{1}{2} \gamma^{ij} \gamma^{kl} F_{uz} F_{uz} \left(\partial_u \gamma_{ik} \right) \left(\partial_u \gamma_{jl} \right)\\
		= & 4 F_{uz} F_{uz} K^{ij} K_{ij} + F_{uz} F_{uz} \gamma^{ij} \partial_u^2 \gamma_{ij} + 2 \gamma^{ij} \gamma^{kl} F_{uz} F_{uz} K_{ik} K_{jl}\\
		= & F_{uz} F_{uz} \gamma^{ij} \partial_u^2 \gamma_{ij}\,.
	\end{split}
\end{equation}
Finally, The seventh term of Eq. (\ref{rewrittenhkk2}) is
\begin{equation}
	\begin{split}
		& F_{uz} F_{uz} \left(\partial_u \gamma^{ij} \right) \left(\partial_u \gamma_{ij} \right) + F_{uz} F_{uz} \gamma^{ij} \partial_u^2 \gamma_{ij} + \frac{1}{2} \gamma^{ij} \gamma^{kl} F_{uz} F_{uz} \left(\partial_u \gamma_{ik} \right) \left(\partial_u \gamma_{jl} \right)\\
		= & 4 F_{uz} F_{uz} K^{ij} K_{ij} + F_{uz} F_{uz} \gamma^{ij} \partial_u^2 \gamma_{ij} + 2 \gamma^{ij} \gamma^{kl} F_{uz} F_{uz} K_{ik} K_{jl}\\
		= & F_{uz} F_{uz} \gamma^{ij} \partial_u^2 \gamma_{ij}\,.
	\end{split}
\end{equation}

Based on the results derived from the preceding calculations, the expression for $H_{uu}^{(2)}$ under the linear-order constraints of the quantum corrections can be represented as
\begin{equation}\label{hkk2final}
	\begin{split}
		H_{uu}^{(2)} = & 2 F_{uz} \left(\partial_u \partial_u F_{uz} \right) + 2 \gamma^{ij} F_{uz} D_i \left(\partial_u F_{uj} \right) + 2 F_{uz} \left(\partial_u \partial_u F_{uz} \right)\\
		& + 2 \gamma^{ij} F_{jz} \left(\partial_u \partial_u F_{ui} \right) + 2 \gamma^{ij} \gamma^{kl} F_{jl} \left(\partial_k \partial_u F_{ui} \right) + 2 \gamma^{ij} \gamma^{kl} \beta_k F_{jl} \left(\partial_u F_{ui} \right)\\
		& - 2 \gamma^{ij} \beta_j F_{uz} \left(\partial_u F_{ui} \right) + 2 \left(\partial_u F_{uz} \right) \left(\partial_u F_{uz} \right) + 2 \gamma^{ij} \left(\partial_u F_{uj} \right) \left(\partial_u F_{iz} \right)\\
		& + 2 \gamma^{ij} \left(\partial_u F_{uj} \right) \left(\partial_z F_{iu} \right) + 2 \gamma^{ij} \gamma^{kl} \left(\partial_u F_{uj} \right) \left(\partial_k F_{il} \right) + 2 \gamma^{ij} \beta_i \left(\partial_u F_{uj} \right) F_{uz}\\
		& - 2 \gamma^{ij} \gamma^{kl} \left(\partial_u F_{uj} \right) \hat{\Gamma}^{m}_{\ ki} F_{ml} - 2 \gamma^{ij} \gamma^{kl} \beta_l \left(\partial_u F_{uj} \right) F_{ik} - 2 \gamma^{ij} \gamma^{kl} \left(\partial_u F_{uj} \right) \hat{\Gamma}^{m}_{\ kl} F_{im}\\
		& + 2 \left(\partial_u F_{uz} \right) \left(\partial_u F_{uz} \right) + 2 \gamma^{ij} \left(\partial_u F_{ui} \right) \left(\partial_j F_{uz} \right) + 2 \gamma^{ij} \left(\partial_u F_{iu} \right) \left(\partial_z F_{uj} \right)\\
		& + 2 \gamma^{ij} \left(\partial_u F_{iz} \right) \left(\partial_u F_{uj} \right) + 2 \gamma^{ij} \beta_i F_{uz} \left(\partial_u F_{uj} \right) + 2 \gamma^{ij} \gamma^{kl} \beta_l \left(\partial_u F_{ui} \right) F_{kj}\\
		& + 4 \gamma^{ij} \left(\partial_u F_{ui} \right) \left(\partial_z F_{uj} \right) - 2 \gamma^{ij} \gamma^{kl} \beta_l \left(\partial_u F_{ui} \right) F_{kj} - 2 \gamma^{ij} \beta_i F_{uz} \left(\partial_u F_{uj} \right)\\
		& + F_{uz} F_{uz} \gamma^{ij} \partial_u^2 \gamma_{ij}\,.
	\end{split}
\end{equation}
The first, third, eighth, and sixteenth terms in Eq. (\ref{hkk2final}) are calculated as
\begin{equation}
	\begin{split}
		& 2 F_{uz} \left(\partial_u \partial_u F_{uz} \right) + 2 F_{uz} \left(\partial_u \partial_u F_{uz} \right) + 2 \left(\partial_u F_{uz} \right) \left(\partial_u F_{uz} \right) + 2 \left(\partial_u F_{uz} \right) \left(\partial_u F_{uz} \right)\\
		= & 4 F_{uz} \left(\partial_u \partial_u F_{uz} \right) + 4 \left(\partial_u F_{uz} \right) \left(\partial_u F_{uz} \right)\\
		= & 2 \partial_u^2 \left(F_{uz} F_{uz} \right)\,.
	\end{split}
\end{equation}
By combining the above result with the last term in Eq. (\ref{hkk2final}) and multiplying it by the square root of the induced metric, we obtain
\begin{equation}\label{hkk2finalresultfirstthirdeight}
	\begin{split}
		& \sqrt{\gamma} \left[2 \partial_u^2 \left(F_{uz} F_{uz} \right) + F_{uz} F_{uz} \gamma^{ij} \partial_u^2 \gamma_{ij} \right]\\
		= & 2 \left[\sqrt{\gamma} \partial_u^2 \left(F_{uz} F_{uz} \right) + \frac{1}{2} \sqrt{\gamma} F_{uz} F_{uz} \gamma^{ij} \partial_u^2 \gamma_{ij} \right]\\
		= & 2 \partial_u^2 \left(\sqrt{\gamma} F_{uz} F_{uz} \right)\,.
	\end{split}
\end{equation}
The second and seventeenth terms in Eq. (\ref{hkk2final}) are calculated as 
\begin{equation}
	\begin{split}
		& 2 \gamma^{ij} F_{uz} D_i \left(\partial_u F_{uj} \right) + 2 \gamma^{ij} \left(\partial_u F_{ui} \right) \left(\partial_j F_{uz} \right)\\
		= & 2 \gamma^{ij} F_{uz} D_i \left(\partial_u F_{uj} \right) + 2 \gamma^{ij} \left(\partial_u F_{ui} \right) \left(D_j F_{uz} \right)\\
		= & 2 \gamma^{ij} D_i \left[F_{uz} \left(\partial_u F_{uj} \right) \right]\,.
	\end{split}
\end{equation}
The result derived from these two terms represents a total spatial derivative, which can be disregarded directly because the cross-section of the null hypersurface $L$ is assumed to be compact. Furthermore, the fourth, ninth, and nineteenth terms in Eq. (\ref{hkk2final}) are evaluated as
\begin{equation}\label{hkk2fourthninthnineteenth}
	\begin{split}
		& 2 \gamma^{ij} F_{jz} \left(\partial_u \partial_u F_{ui} \right) + 2 \gamma^{ij} \left(\partial_u F_{uj} \right) \left(\partial_u F_{iz} \right) + 2 \gamma^{ij} \left(\partial_u F_{iz} \right) \left(\partial_u F_{uj} \right)\\
		= & - 2 \gamma^{ij} F_{zj} \left(\partial_u \partial_u F_{ui} \right) - 2 \gamma^{ij} \left(\partial_u F_{uj} \right) \left(\partial_u F_{zi} \right) - 2 \gamma^{ij} \left(\partial_u F_{zi} \right) \left(\partial_u F_{uj} \right)\\
		= & - 2 \gamma^{ij} F_{zj} \left(\partial_u \partial_u F_{ui} \right) - 4 \gamma^{ij} \left(\partial_u F_{ui} \right) \left(\partial_u F_{zj} \right)\\
		= & - 2 \gamma^{ij} F_{zj} \left(\partial_u \partial_u F_{ui} \right) - 4 \gamma^{ij} \left(\partial_u F_{ui} \right) \left(\partial_u F_{zj} \right) - 2 \gamma^{ij} F_{ui} \left(\partial_u^2 F_{zj} \right)\\
		= & - 2 \gamma^{ij} \partial_u^2 \left(F_{ui} F_{zj} \right)\,.
	\end{split}
\end{equation}
The fifth, eleventh, thirteenth, and fifteenth terms in Eq. (\ref{hkk2final}) can be given as 
\begin{equation}\label{hkk2fifthelevenththirteenthfifteenth}
	\begin{split}
		& 2 \gamma^{ij} \gamma^{kl} F_{jl} \left(\partial_k \partial_u F_{ui} \right) + 2 \gamma^{ij} \gamma^{kl} \left(\partial_u F_{uj} \right) \left(\partial_k F_{il} \right) - 2 \gamma^{ij} \gamma^{kl} \left(\partial_u F_{uj} \right) \hat{\Gamma}^{m}_{\ ki} F_{ml}\\
		& - 2 \gamma^{ij} \gamma^{kl} \left(\partial_u F_{uj} \right) \hat{\Gamma}^{m}_{\ kl} F_{im}\\
		= & 2 \gamma^{ij} \gamma^{kl} F_{jl} \left(\partial_k \partial_u F_{ui} \right) + 2 \gamma^{ij} \gamma^{kl} \left(\partial_u F_{uj} \right) \left(\partial_k F_{il} - \hat{\Gamma}^{m}_{\ ki} F_{ml} - \hat{\Gamma}^{m}_{\ kl} F_{im} \right)\\
		= & 2 \gamma^{ij} \gamma^{kl} F_{jl} \left(\partial_k \partial_u F_{ui} \right) + 2 \gamma^{ij} \gamma^{kl} \left(\partial_u F_{uj} \right) \left(D_k F_{il} \right)\\
		= & 2 \gamma^{ij} \gamma^{kl} F_{jl} \left(\partial_k \partial_u F_{ui} \right) - 2 \gamma^{i(j} \gamma^{|k|l)} F_{[jl]} \hat{\Gamma}^{m}_{\ (ki)} \left(\partial_u F_{um} \right) + 2 \gamma^{ij} \gamma^{kl} \left(\partial_u F_{uj} \right) \left(D_k F_{il} \right)\\
		= & 2 \gamma^{ij} \gamma^{kl} F_{il} \left(D_k \partial_u F_{uj} \right) + 2 \gamma^{ij} \gamma^{kl} \left(\partial_u F_{uj} \right) \left(D_k F_{il} \right)\\
		= & 2 \gamma^{kl} D_k \left[\gamma^{ij} F_{il} \left(\partial_u F_{uj} \right) \right]\,,
	\end{split}
\end{equation}
where we have used the fact
\begin{equation}
	\begin{split}
		- 2 \gamma^{i(j} \gamma^{|k|l)} F_{[jl]} \hat{\Gamma}^{m}_{\ (ki)} \left(\partial_u F_{um} \right) = 0
	\end{split}
\end{equation}
in the third step. The result in Eq. (\ref{hkk2fifthelevenththirteenthfifteenth}) can also be disregarded due to the compactness of the cross-section. Utilizing the symmetry properties of the indices in the induced metric, the sixth, fourteenth, twenty-first, and twenty-third terms in Eq. (\ref{hkk2final}) can be simplified as 
\begin{equation}
	\begin{split}
		& 2 \gamma^{ij} \gamma^{kl} \beta_k F_{jl} \left(\partial_u F_{ui} \right) - 2 \gamma^{ij} \gamma^{kl} \beta_l \left(\partial_u F_{uj} \right) F_{ik} + 2 \gamma^{ij} \gamma^{kl} \beta_l \left(\partial_u F_{ui} \right) F_{kj}\\
		& - 2 \gamma^{ij} \gamma^{kl} \beta_l \left(\partial_u F_{ui} \right) F_{kj}\\
		= & 2 \gamma^{ij} \gamma^{kl} \beta_l F_{ik} \left(\partial_u F_{uj} \right) - 2 \gamma^{ij} \gamma^{kl} \beta_l \left(\partial_u F_{uj} \right) F_{ik} + 2 \gamma^{ij} \gamma^{kl} \beta_l \left(\partial_u F_{ui} \right) F_{kj}\\
		& - 2 \gamma^{ij} \gamma^{kl} \beta_l \left(\partial_u F_{ui} \right) F_{kj}\\
		= & 0\,.
	\end{split}
\end{equation}
The seventh, twelfth, twentieth, and twenty-fourth terms in Eq. (\ref{hkk2final}) can be simplified as 
\begin{equation}
	\begin{split}
		& - 2 \gamma^{ij} \beta_j F_{uz} \left(\partial_u F_{ui} \right) + 2 \gamma^{ij} \beta_i \left(\partial_u F_{uj} \right) F_{uz} + 2 \gamma^{ij} \beta_i F_{uz} \left(\partial_u F_{uj} \right)\\
		& - 2 \gamma^{ij} \beta_i F_{uz} \left(\partial_u F_{uj} \right)\\
		= & 0\,.
	\end{split}
\end{equation}
The tenth, eighteenth, and twenty-second terms in Eq. (\ref{hkk2final}) are calculated as 
\begin{equation}
	\begin{split}
		& 2 \gamma^{ij} \left(\partial_u F_{uj} \right) \left(\partial_z F_{iu} \right) + 2 \gamma^{ij} \left(\partial_u F_{iu} \right) \left(\partial_z F_{uj} \right) + 4 \gamma^{ij} \left(\partial_u F_{ui} \right) \left(\partial_z F_{uj} \right)\\
		= & - 2 \gamma^{ij} \left(\partial_u F_{ui} \right) \left(\partial_z F_{uj} \right) - 2 \gamma^{ij} \left(\partial_u F_{ui} \right) \left(\partial_z F_{uj} \right) + 4 \gamma^{ij} \left(\partial_u F_{ui} \right) \left(\partial_z F_{uj} \right)\\
		= & 0\,.
	\end{split}
\end{equation}
Therefore, based on the results in Eqs. (\ref{hkk2finalresultfirstthirdeight}) and (\ref{hkk2fourthninthnineteenth}), the expression for $H_{uu}^{(2)}$ with the square root of the induced metric $\sqrt{\gamma}$ can be further written as 
\begin{equation}\label{sqrtgammahkk2}
	\begin{split}
		\sqrt{\gamma} H_{uu}^{(2)} = 2 \partial_u^2 \left(\sqrt{\gamma} F_{uz} F_{uz} \right) - 2 \sqrt{\gamma} \gamma^{ij} \partial_u^2 \left(F_{ui} F_{zj} \right)\,.
	\end{split}
\end{equation}
The second term of Eq. (\ref{sqrtgammahkk2}) is further written as 
\begin{equation}
	\begin{split}
		& - 2 \sqrt{\gamma} \gamma^{ij} \partial_u^2 \left(F_{ui} F_{zj} \right) = - 2 \sqrt{\gamma} \partial_u^2 \left(\gamma^{ij} F_{ui} F_{zj} \right) + 2 \left(\partial_u^2 \gamma^{ij} \right) F_{ui} F_{zj}\\
		= & - 2 \sqrt{\gamma} \partial_u^2 \left(\gamma^{ij} F_{ui} F_{zj} \right)\,.
	\end{split}
\end{equation}
Therefore, the expression of $H_{uu}^{(2)}$ in Eq. (\ref{sqrtgammahkk2}) can be further given as  
\begin{equation}\label{hkk2finalresultwithgamma}
	\begin{split}
		\sqrt{\gamma} H_{uu}^{(2)} = & 2 \partial_u^2 \left(\sqrt{\gamma} F_{uz} F_{uz} \right) - 2 \sqrt{\gamma} \gamma^{ij} \partial_u^2 \left(F_{ui} F_{zj} \right)\\
		= & 2 \partial_u^2 \left(\sqrt{\gamma} F_{uz} F_{uz} \right) - 2 \sqrt{\gamma} \partial_u^2 \left(\gamma^{ij} F_{ui} F_{zj} \right) - \sqrt{\gamma} \gamma^{kl} \left(\partial_u^2 \gamma_{kl} \right) \gamma^{ij} F_{ui} F_{zj}\\
		= & 2 \partial_u^2 \left(\sqrt{\gamma} F_{uz} F_{uz} \right) - 2 \partial_u^2 \left(\sqrt{\gamma} \gamma^{ij} F_{ui} F_{zj} \right)\\
		= & 2 \partial_u^2 \left(\sqrt{\gamma} F_{uz} F_{uz} - \sqrt{\gamma} \gamma^{ij} F_{ui} F_{zj} \right)\,.
	\end{split}
\end{equation}

Subsequently, the specific expression of $H_{uu}^{(3)}$ in Eq. (\ref{ohkk3}) under the linear-order constraints of the quantum corrections should be further calculated. According to the relationship between the covariant derivative and the Riemann curvature tensor
\begin{equation}
	\begin{split}
		\left(\nabla_c \nabla_d - \nabla_d \nabla_c \right) F_{a}^{\ d} = R_{cda}^{\ \ \ e} F_{e}^{\ d} - R_{cde}^{\ \ \ d} F_{a}^{\ e}\,,
	\end{split}
\end{equation}
the first and sixth terms in Eq. (\ref{ohkk3}) can be given as 
\begin{equation}
	\begin{split}
		& 2 k^a k^b F_{b}^{\ c} \nabla_c \nabla_d F_{a}^{\ d} + 2 k^a k^b F_{b}^{\ c} \nabla_d \nabla_c F_{a}^{\ d}\\
		= & 4 k^a k^b F_{b}^{\ c} \nabla_c \nabla_d F_{a}^{\ d} + 2 k^a k^b F_{b}^{\ c} R_{ce} F_{a}^{\ e} - 2 k^a k^b F_{b}^{\ c} R_{cda}^{\ \ \ e} F_{e}^{\ d}\,.
	\end{split}
\end{equation}
Similarly, the fourth and seventh terms in Eq. (\ref{ohkk3}) can also be calculated as 
\begin{equation}
	\begin{split}
		& 2 k^a k^b F_{a}^{\ c} \nabla_c \nabla_d F_{b}^{\ d} + 2 k^a k^b F_{a} ^{\ c} \nabla_d \nabla_c F_{b}^{\ d}\\
		= & 4 k^a k^b F_{a}^{\ c} \nabla_c \nabla_d F_{b}^{\ d} + 2 k^a k^b F_{a}^{\ c} R_{ce} F_{b}^{\ e} - 2 k^a k^b F_{a}^{\ c} R_{cdb}^{\ \ \ e} F_{e}^{\ d}\,.
	\end{split}
\end{equation}
Therefore, the expression of $H_{uu}^{(3)}$ in Eq. (\ref{ohkk3}) can be rewritten as
\begin{equation}\label{rehkk3}
	\begin{split}
		H_{uu}^{(3)} = & 4 k^a k^b F_{b}^{\ c} \nabla_c \nabla_d F_{a}^{\ d} + 4 k^a k^b F_{a}^{\ c} \nabla_c \nabla_d F_{b}^{\ d} + 4 k^a k^b \nabla_c F_{bd} \nabla^d F_{a}^{\ c}  \\
		& + 4 k^a k^b \nabla_c F_{a}^{\ c} \nabla_d F_{b}^{\ d} + 2 k^a k^b F_{a}^{\ c} R_{ce} F_{b}^{\ e} + 2 k^a k^b F_{b}^{\ c} R_{ce} F_{a}^{\ e} \\
		& - 2 k^a k^b F_{a}^{\ c} R_{cdb}^{\ \ \ e} F_{e}^{\ d} - 2 k^a k^b F_{b}^{\ c} R_{cda}^{\ \ \ e} F_{e}^{\ d} - k^a k^b R_{acde} F_{b}^{\ c} F^{de} \\
		& - k^a k^b R_{bcde} F_{a}^{\ c} F^{de}\\
		= & 8 k^a k^b F_{b}^{\ c} \nabla_c \nabla_d F_{a}^{\ d} + 4 k^a k^b \nabla_c F_{bd} \nabla^d F_{a}^{\ c} + 4 k^a k^b \nabla_c F_{a}^{\ c} \nabla_d F_{b}^{\ d}\\
		& + 4 k^a k^b F_{a}^{\ c} R_{ce} F_{b}^{\ e} - 4 k^a k^b F_{b}^{\ c} R_{cda}^{\ \ \ e} F_{e}^{\ d} - 2 k^a k^b R_{acde} F_{b}^{\ c} F^{de}\,.
	\end{split}
\end{equation}

The first term in Eq. (\ref{rehkk3}) is
\begin{equation}\label{hkk3first}
	\begin{split}
		& 8 k^a k^b F_{b}^{\ c} \nabla_c \nabla_d F_{a}^{\ d} = 8 k^a k^b g^{ce} g^{df} F_{be} \nabla_c \nabla_d F_{af}\\
		= & 8 k^a k^b g^{ce} g^{df} F_{be} \partial_c \partial_d F_{af} - 8 k^a k^b g^{ce} g^{df} F_{be} \Gamma^{g}_{\ cd} \partial_g F_{af}\\
		& - 8 k^a k^b g^{ce} g^{df} F_{be} \Gamma^{g}_{\ ca} \partial_d F_{gf} - 8 k^a k^b g^{ce} g^{df} F_{be} \Gamma^{g}_{\ cf} \partial_d F_{ag}\\
		& - 8 k^a k^b g^{ce} g^{df} F_{be} \left(\partial_c \Gamma^{g}_{\ da} \right) F_{gf} + 8 k^a k^b g^{ce} g^{df} F_{be} \Gamma^{h}_{\ cd} \Gamma^{g}_{\ ha} F_{gf}\\
		& + 8 k^a k^b g^{ce} g^{df} F_{be} \Gamma^{h}_{\ ca} \Gamma^{g}_{\ dh} F_{gf} - 8 k^a k^b g^{ce} g^{df} F_{be} \Gamma^{g}_{\ ch} \Gamma^{h}_{\ da} F_{gf}\\
		& - 8 k^a k^b g^{ce} g^{df} F_{be} \Gamma^{g}_{\ da} \partial_c F_{gf} + 8 k^a k^b g^{ce} g^{df} F_{be} \Gamma^{g}_{\ da} \Gamma^{h}_{\ cg} F_{hf}\\
		&  + 8 k^a k^b g^{ce} g^{df} F_{be} \Gamma^{g}_{\ da} \Gamma^{h}_{\ cf} F_{gh} - 8 k^a k^b g^{ce} g^{df} F_{be} \left(\partial_c \Gamma^{g}_{\ df} \right) F_{ag}\\
		& + 8 k^a k^b g^{ce} g^{df} F_{be} \Gamma^{h}_{\ cd} \Gamma^{g}_{\ hf} F_{ag} + 8 k^a k^b g^{ce} g^{df} F_{be} \Gamma^{h}_{\ cf} \Gamma^{g}_{\ dh} F_{ag}\\
		& - 8 k^a k^b g^{ce} g^{df} F_{be} \Gamma^{g}_{\ ch} \Gamma^{h}_{\ df} F_{ag} - 8 k^a k^b g^{ce} g^{df} F_{be} \Gamma^{g}_{df} \partial_c F_{ag}\\
		& + 8 k^a k^b g^{ce} g^{df} F_{be} \Gamma^{g}_{df} \Gamma^{h}_{\ ca} F_{hg} + 8 k^a k^b g^{ce} g^{df} F_{be} \Gamma^{g}_{df} \Gamma^{h}_{\ cg} F_{ah}\,.
	\end{split}
\end{equation}

The first term of Eq. (\ref{hkk3first}) is obtained as 
\begin{equation}
	\begin{split}
		& 8 k^a k^b g^{ce} g^{df} F_{be} \partial_c \partial_d F_{af} = 8 g^{ce} g^{df} F_{ue} \partial_c \partial_d F_{uf}\\
		= & 8 F_{uz} \partial_u^2 F_{uz} + 8 \gamma^{ij} F_{uz} \partial_u \partial_i F_{uj} + 8 \gamma^{ij} F_{uj} \partial_i \partial_u F_{uz}\\
		& + 8 \gamma^{ij} \gamma^{kl} F_{uj} \partial_i \partial_k F_{ul}\\
		= & 8 F_{uz} \partial_u^2 F_{uz} + 8 \gamma^{ij} F_{uz} \partial_u \partial_i F_{uj}\,.
	\end{split}
\end{equation}

The second term of Eq. (\ref{hkk3first}) is 
\begin{equation}\label{secondtermhkk3}
	\begin{split}
		& - 8 k^a k^b g^{ce} g^{df} F_{be} \Gamma^{g}_{\ cd} \partial_g F_{af} = - 8 g^{ce} g^{df} F_{ue} \Gamma^{g}_{\ cd} \partial_g F_{uf}\\
		= & - 8 F_{uz} \Gamma^{g}_{\ uu} \partial_g F_{uz} - 8 \gamma^{ij} F_{uz} \Gamma^{g}_{\ ui} \partial_g F_{uj} - 8 \gamma^{ij} F_{uj} \Gamma^{g}_{\ iu} \partial_g F_{uz}\\
		& - 8 \gamma^{ij} \gamma^{kl} F_{uj} \Gamma^{g}_{\ ik} \partial_g F_{ul}\\
		= & - 8 \gamma^{ij} F_{uz} \Gamma^{g}_{\ ui} \partial_g F_{uj}\,.
	\end{split}
\end{equation}
The repeated index $g$ should be expanded by the metric as
\begin{equation}
	\begin{split}
		& - 8 \gamma^{ij} F_{uz} \Gamma^{g}_{\ ui} \partial_g F_{uj}\\
		= & - 8 \gamma^{ij} F_{uz} \Gamma^{u}_{\ ui} \partial_u F_{uj} - 8 \gamma^{ij} F_{uz} \Gamma^{z}_{\ ui} \partial_z F_{uj} - 8 \gamma^{ij} F_{uz} \Gamma^{k}_{\ ui} \partial_k F_{uj}\,.
	\end{split}
\end{equation}
Therefore, the second term of Eq. (\ref{hkk3first}) is obtained as 
\begin{equation}
	\begin{split}
		& - 8 k^a k^b g^{ce} g^{df} F_{be} \Gamma^{g}_{\ cd} \partial_g F_{af}\\
		= & 4 \gamma^{ij} \beta_i F_{uz} \partial_u F_{uj} - 4 \gamma^{i j} \gamma^{k l} \left(\partial_u \gamma_{i l} \right) F_{uz} \partial_k F_{uj}\\
		= & 4 \gamma^{ij} \beta_i F_{uz} \partial_u F_{uj} - 8 \gamma^{i j} \gamma^{k l} K_{i l} F_{uz} \partial_k F_{uj}\\
		= & 4 \gamma^{ij} \beta_i F_{uz} \partial_u F_{uj}\,.
	\end{split}
\end{equation}

The third term of Eq. (\ref{hkk3first}) is obtained as 
\begin{equation}
	\begin{split}
		& - 8 k^a k^b g^{ce} g^{df} F_{be} \Gamma^{g}_{\ ca} \partial_d F_{gf} = - 8 g^{ce} g^{df} F_{ue} \Gamma^{g}_{\ cu} \partial_d F_{gf}\\
		= & - 8 F_{uz} \Gamma^{g}_{\ uu} \partial_u F_{gz} - 8 F_{uz} \Gamma^{g}_{\ uu} \partial_z F_{gu} - 8 \gamma^{ij} F_{uz} \Gamma^{g}_{\ uu} \partial_i F_{gj}\\
		& - 8 \gamma^{ij} F_{uj} \Gamma^{g}_{\ iu} \partial_u F_{gz} - 8 \gamma^{ij} F_{uj} \Gamma^{g}_{\ iu} \partial_z F_{gu} - 8 \gamma^{ij} \gamma^{kl} F_{uj} \Gamma^{g}_{\ iu} \partial_k F_{gl}\\
		= & 0\,.
	\end{split}
\end{equation}

The fourth term of Eq. (\ref{hkk3first}) is 
\begin{equation}
	\begin{split}
		& - 8 k^a k^b g^{ce} g^{df} F_{be} \Gamma^{g}_{\ cf} \partial_d F_{ag} = - 8 g^{ce} g^{df} F_{ue} \Gamma^{g}_{\ cf} \partial_d F_{ug}\\
		= & - 8 F_{uz} \Gamma^{g}_{\ uz} \partial_u F_{ug} - 8 F_{uz} \Gamma^{g}_{\ uu} \partial_z F_{ug} - 8 \gamma^{ij} F_{uz} \Gamma^{g}_{\ uj} \partial_i F_{ug}\\
		& - 8 \gamma^{ij} F_{uj} \Gamma^{g}_{\ iz} \partial_u F_{ug} - 8 \gamma^{ij} F_{uj} \Gamma^{g}_{\ iu} \partial_z F_{ug} - 8 \gamma^{ij} \gamma^{kl} F_{uj} \Gamma^{g}_{\ il} \partial_k F_{ug}\\
		= & - 8 F_{uz} \Gamma^{g}_{\ uz} \partial_u F_{ug} - 8 \gamma^{ij} F_{uz} \Gamma^{g}_{\ uj} \partial_i F_{ug}\,.
	\end{split}
\end{equation}
The repeated index $g$ should be further expanded as 
\begin{equation}
	\begin{split}
		& - 8 F_{uz} \Gamma^{g}_{\ uz} \partial_u F_{ug} - 8 \gamma^{ij} F_{uz} \Gamma^{g}_{\ uj} \partial_i F_{ug}\\
		= & - 8 F_{uz} \Gamma^{u}_{\ uz} \partial_u F_{uu} - 8 F_{uz} \Gamma^{z}_{\ uz} \partial_u F_{uz} - 8 F_{uz} \Gamma^{i}_{\ uz} \partial_u F_{ui}\\
		& - 8 \gamma^{ij} F_{uz} \Gamma^{u}_{\ uj} \partial_i F_{uu} - 8 \gamma^{ij} F_{uz} \Gamma^{z}_{\ uj} \partial_i F_{uz} - 8 \gamma^{ij} F_{uz} \Gamma^{k}_{\ uj} \partial_i F_{uk}\\
		= & - 8 F_{uz} \Gamma^{z}_{\ uz} \partial_u F_{uz} - 8 F_{uz} \Gamma^{i}_{\ uz} \partial_u F_{ui} - 8 \gamma^{ij} F_{uz} \Gamma^{z}_{\ uj} \partial_i F_{uz}\\
		& - 8 \gamma^{ij} F_{uz} \Gamma^{k}_{\ uj} \partial_i F_{uk}\,.
	\end{split}
\end{equation}
Therefore, the fourth term of Eq. (\ref{hkk3first}) is obtained as 
\begin{equation}
	\begin{split}
		& - 8 k^a k^b g^{ce} g^{df} F_{be} \Gamma^{g}_{\ cf} \partial_d F_{ag}\\
		= & - 4 \gamma^{ij} \beta_j F_{uz} \partial_u F_{ui} - 4 \gamma^{i j} \gamma^{k l} F_{uz} \left(\partial_u \gamma_{jl} \right) \partial_i F_{uk}\\
		= & - 4 \gamma^{ij} \beta_j F_{uz} \partial_u F_{ui} - 8 \gamma^{i j} \gamma^{k l} F_{uz} K_{jl} \partial_i F_{uk}\\
		= & - 4 \gamma^{ij} \beta_j F_{uz} \partial_u F_{ui}\,.
	\end{split}
\end{equation}

The fifth term of Eq. (\ref{hkk3first}) is 
\begin{equation}
	\begin{split}
		& - 8 k^a k^b g^{ce} g^{df} F_{be} \left(\partial_c \Gamma^{g}_{\ da} \right) F_{gf} = - 8 g^{ce} g^{df} F_{ue} \left(\partial_c \Gamma^{g}_{\ du} \right) F_{gf}\\
		= & - 8 F_{uz} \left(\partial_u \Gamma^{g}_{\ uu} \right) F_{gz} - 8 F_{uz} \left(\partial_u \Gamma^{g}_{\ zu} \right) F_{gu} - 8 \gamma^{ij} F_{uz} \left(\partial_u \Gamma^{g}_{\ iu} \right) F_{gj}\\
		& - 8 \gamma^{ij} F_{uj} \left(\partial_i \Gamma^{g}_{\ uu} \right) F_{gz} - 8 \gamma^{ij} F_{uj} \left(\partial_i \Gamma^{g}_{\ zu} \right) F_{gu} - 8 \gamma^{ij} \gamma^{kl} F_{uj} \left(\partial_i \Gamma^{g}_{\ ku} \right) F_{gl}\\
		= & - 8 F_{uz} \left(\partial_u \Gamma^{g}_{\ uu} \right) F_{gz} - 8 F_{uz} \left(\partial_u \Gamma^{g}_{\ zu} \right) F_{gu} - 8 \gamma^{ij} F_{uz} \left(\partial_u \Gamma^{g}_{\ iu} \right) F_{gj}\,.
	\end{split}
\end{equation}
The index $g$ should be further expanded as 
\begin{equation}\label{hkk3firstfifth}
	\begin{split}
		& - 8 F_{uz} \left(\partial_u \Gamma^{g}_{\ uu} \right) F_{gz} - 8 F_{uz} \left(\partial_u \Gamma^{g}_{\ zu} \right) F_{gu} - 8 \gamma^{ij} F_{uz} \left(\partial_u \Gamma^{g}_{\ iu} \right) F_{gj}\\
		= & - 8 F_{uz} \left(\partial_u \Gamma^{u}_{\ uu} \right) F_{uz} - 8 F_{uz} \left(\partial_u \Gamma^{z}_{\ uu} \right) F_{zz} - 8 F_{uz} \left(\partial_u \Gamma^{i}_{\ uu} \right) F_{iz}\\
		& - 8 F_{uz} \left(\partial_u \Gamma^{u}_{\ zu} \right) F_{uu} - 8 F_{uz} \left(\partial_u \Gamma^{z}_{\ zu} \right) F_{zu} - 8 F_{uz} \left(\partial_u \Gamma^{i}_{\ zu} \right) F_{iu}\\
		& - 8 \gamma^{ij} F_{uz} \left(\partial_u \Gamma^{u}_{\ iu} \right) F_{uj} - 8 \gamma^{ij} F_{uz} \left(\partial_u \Gamma^{z}_{\ iu} \right) F_{zj} - 8 \gamma^{ij} F_{uz} \left(\partial_u \Gamma^{k}_{\ iu} \right) F_{kj}\\
		= & - 8 F_{uz} \left(\partial_u \Gamma^{u}_{\ uu} \right) F_{uz} - 8 F_{uz} \left(\partial_u \Gamma^{i}_{\ uu} \right) F_{iz} - 8 F_{uz} \left(\partial_u \Gamma^{z}_{\ zu} \right) F_{zu}\\
		& - 8 \gamma^{ij} F_{uz} \left(\partial_u \Gamma^{z}_{\ iu} \right) F_{zj} - 8 \gamma^{ij} F_{uz} \left(\partial_u \Gamma^{k}_{\ iu} \right) F_{kj}\,.
	\end{split}
\end{equation}
The first term of Eq. (\ref{hkk3firstfifth}) is 
\begin{equation}
	\begin{split}
		& - 8 F_{uz} \left(\partial_u \Gamma^{u}_{\ uu} \right) F_{uz}\\
		= & - 8 F_{uz} \partial_u \left(- z^2 \partial_z \alpha - 2 z \alpha \right) F_{uz}\\
		= & - 8 F_{uz} \left(- z^2 \partial_u \partial_z \alpha - 2 z \partial_u \alpha \right) F_{uz}\\
		= & 0\,.
	\end{split}
\end{equation}
The second term of Eq. (\ref{hkk3firstfifth}) is
\begin{equation}
	\begin{split}
		& - 8 F_{uz} \left(\partial_u \Gamma^{i}_{\ uu} \right) F_{iz}\\
		= & - 8 F_{uz} \partial_u \left[\left(z \beta^i \right) \left(z^2 \partial_z \alpha + 2 z \alpha \right) + \gamma^{i j} \left(z \partial_u \beta_j - z^2 \partial_j \alpha \right) \right] F_{iz}\\
		= & - 8 F_{uz} \left[\left(z \partial_u \beta^i \right) \left(z^2 \partial_z \alpha + 2 z \alpha \right) + \left(z \beta^i \right) \left(z^2 \partial_u \partial_z \alpha + 2 z \partial_u \alpha \right) \right.\\
		& \left. + \left(\partial_u \gamma^{i j} \right) \left(z \partial_u \beta_j - z^2 \partial_j \alpha \right) + \gamma^{i j} \left(z \partial_u^2 \beta_j - z^2 \partial_u \partial_j \alpha \right) \right] F_{iz}\\
		= & 0\,.
	\end{split}
\end{equation}
The third term of Eq. (\ref{hkk3firstfifth}) is
\begin{equation}
	\begin{split}
		& - 8 F_{uz} \left(\partial_u \Gamma^{z}_{\ zu} \right) F_{zu}\\
		= & - 8 F_{uz} \partial_u \left(z^2 \partial_z \alpha + 2 z \alpha - \frac{1}{2} z \beta^i \beta_i - \frac{1}{2} z^2 \beta^i \partial_z \beta_i \right) F_{zu}\\
		= & - 8 F_{uz} \left[z^2 \partial_u \partial_z \alpha + 2 z \partial_u \alpha - z \beta_i \partial_u \beta^i - \frac{1}{2} z^2 \beta^i \partial_u \partial_z \beta_i - \frac{1}{2} z^2 \left(\partial_u \beta^i \right) \partial_z \beta_i \right] F_{zu}\\
		= & 0\,.
	\end{split}
\end{equation}
The fourth term of Eq. (\ref{hkk3firstfifth}) is
\begin{equation}
	\begin{split}
		& - 8 \gamma^{ij} F_{uz} \left(\partial_u \Gamma^{z}_{\ iu} \right) F_{zj}\\
		= & - 8 \gamma^{ij} F_{uz} \partial_u \left[z^2 \partial_i \alpha - \frac{1}{2} z^2 \left(\beta^2 - 2 \alpha \right) \left(\beta_i + z \partial_z \beta_i \right) \right.\\
		& \left. - \frac{1}{2} z \beta^j \left(z \partial_i \beta_j + \partial_u \gamma_{ij} - z \partial_j \beta_i \right) \right] F_{zj}\\
		= & - 8 \gamma^{ij} F_{uz} \left[z^2 \partial_u \partial_i \alpha - \frac{1}{2} z^2 \left(\beta_i + z \partial_z \beta_i \right) \partial_u \left(\beta^2 - 2 \alpha \right) \right.\\
		& \left. - \frac{1}{2} z^2 \left(\beta^2 - 2 \alpha \right) \partial_u \left(\beta_i + z \partial_z \beta_i \right) - \frac{1}{2} z \left(\partial_u \beta^j \right) \left(z \partial_i \beta_j + \partial_u \gamma_{ij} - z \partial_j \beta_i \right) \right.\\
		& \left. - \frac{1}{2} z \beta^j \partial_u \left(z \partial_i \beta_j + \partial_u \gamma_{ij} - z \partial_j \beta_i \right) \right] F_{zj}\\
		= & 0\,.
	\end{split}
\end{equation}
The fifth term of Eq. (\ref{hkk3firstfifth}) is
\begin{equation}
	\begin{split}
		& - 8 \gamma^{ij} F_{uz} \left(\partial_u \Gamma^{k}_{\ iu} \right) F_{kj}\\
		= & - 8 \gamma^{ij} F_{uz} \partial_u \left[\frac{1}{2} \left(z \beta^k \right) \left(\beta_i + z \partial_z \beta_i \right) + \frac{1}{2} \gamma^{kl} \left(z \partial_i \beta_l + \partial_u \gamma_{il} - z \partial_l \beta_i \right) \right] F_{kj}\\
		= & - 8 \gamma^{ij} F_{uz} \left[\frac{1}{2} \left(z \partial_u \beta^k \right) \left(\beta_i + z \partial_z \beta_i \right) + \frac{1}{2} \left(z \beta^k \right) \left(\partial_u \beta_i + z \partial_u \partial_z \beta_i \right) \right.\\
		& \left. + \frac{1}{2} \left(\partial_u \gamma^{kl} \right) \left(z \partial_i \beta_l + \partial_u \gamma_{il} - z \partial_l \beta_i \right) + \frac{1}{2} \gamma^{kl} \left(z \partial_u \partial_i \beta_l + \partial_u^2 \gamma_{il} - z \partial_u \partial_l \beta_i \right) \right] F_{kj}\\
		= & - 4 \gamma^{ij} F_{uz} \left[\left(\partial_u \gamma^{kl} \right) \left(\partial_u \gamma_{il} \right) + \gamma^{kl} \left(\partial_u^2 \gamma_{il} \right) \right] F_{kj}\\
		= & - 4 \gamma^{ij} F_{uz} \left(\partial_u \gamma^{kl} \right) \left(\partial_u \gamma_{il} \right) F_{kj} - 4 \gamma^{ij} \gamma^{kl} F_{uz} \left(\partial_u^2 \gamma_{il} \right) F_{kj}\,.
	\end{split}
\end{equation}
Therefore, the fifth term of Eq. (\ref{hkk3first}) is obtained as 
\begin{equation}
	\begin{split}
		& - 8 k^a k^b g^{ce} g^{df} F_{be} \left(\partial_c \Gamma^{g}_{\ da} \right) F_{gf}\\
		= & - 4 \gamma^{ij} F_{uz} \left(\partial_u \gamma^{kl} \right) \left(\partial_u \gamma_{il} \right) F_{kj} - 4 \gamma^{ij} \gamma^{kl} F_{uz} \left(\partial_u^2 \gamma_{il} \right) F_{kj}\\
		= & - 16 \gamma^{ij} F_{uz} K^{kl} K_{il} F_{kj} - 8 \gamma^{ij} \gamma^{kl} F_{uz} \left(\partial_u K_{il} \right) F_{kj}\\
		= & - 8 \gamma^{ij} \gamma^{kl} F_{uz} \left(\partial_u K_{il} \right) F_{kj}\,.
	\end{split}
\end{equation}

The sixth term of Eq. (\ref{hkk3first}) is
\begin{equation}
	\begin{split}
		& 8 k^a k^b g^{ce} g^{df} F_{be} \Gamma^{h}_{\ cd} \Gamma^{g}_{\ ha} F_{gf} = 8 g^{ce} g^{df} F_{ue} \Gamma^{h}_{\ cd} \Gamma^{g}_{\ hu} F_{gf}\\
		= & 8 F_{uz} \Gamma^{h}_{\ uu} \Gamma^{g}_{\ hu} F_{gz} + 8 F_{uz} \Gamma^{h}_{\ uz} \Gamma^{g}_{\ hu} F_{gu} + 8 \gamma^{ij} F_{uz} \Gamma^{h}_{\ ui} \Gamma^{g}_{\ hu} F_{gj}\\
		& + 8 \gamma^{ij} F_{uj} \Gamma^{h}_{\ iu} \Gamma^{g}_{\ hu} F_{gz} + 8 \gamma^{ij} F_{uj} \Gamma^{h}_{\ iz} \Gamma^{g}_{\ hu} F_{gu} + 8 \gamma^{ij} \gamma^{kl} F_{uj} \Gamma^{h}_{\ ik} \Gamma^{g}_{\ hu} F_{gl}\\
		= & 8 F_{uz} \Gamma^{h}_{\ uz} \Gamma^{g}_{\ hu} F_{gu} + 8 \gamma^{ij} F_{uz} \Gamma^{h}_{\ ui} \Gamma^{g}_{\ hu} F_{gj}\,.
	\end{split}
\end{equation}
The the index $g$ should be further expanded as 
\begin{equation}\label{hkk3firstsixth}
	\begin{split}
		& 8 F_{uz} \Gamma^{h}_{\ uz} \Gamma^{g}_{\ hu} F_{gu} + 8 \gamma^{ij} F_{uz} \Gamma^{h}_{\ ui} \Gamma^{g}_{\ hu} F_{gj}\\
		= & 8 F_{uz} \Gamma^{h}_{\ uz} \Gamma^{u}_{\ hu} F_{uu} + 8 F_{uz} \Gamma^{h}_{\ uz} \Gamma^{z}_{\ hu} F_{zu} + 8 F_{uz} \Gamma^{h}_{\ uz} \Gamma^{i}_{\ hu} F_{iu}\\
		& + 8 \gamma^{ij} F_{uz} \Gamma^{h}_{\ ui} \Gamma^{u}_{\ hu} F_{uj} + 8 \gamma^{ij} F_{uz} \Gamma^{h}_{\ ui} \Gamma^{z}_{\ hu} F_{zj} + 8 \gamma^{ij} F_{uz} \Gamma^{h}_{\ ui} \Gamma^{k}_{\ hu} F_{kj}\\
		= & 8 F_{uz} \Gamma^{h}_{\ uz} \Gamma^{z}_{\ hu} F_{zu} + 8 \gamma^{ij} F_{uz} \Gamma^{h}_{\ ui} \Gamma^{z}_{\ hu} F_{zj} + 8 \gamma^{ij} F_{uz} \Gamma^{h}_{\ ui} \Gamma^{k}_{\ hu} F_{kj}\,.
	\end{split}
\end{equation}
The repeated index $h$ should be further expanded. The first term of Eq. (\ref{hkk3firstsixth}) is
\begin{equation}
	\begin{split}
		& 8 F_{uz} \Gamma^{h}_{\ uz} \Gamma^{z}_{\ hu} F_{zu}\\
		= & 8 F_{uz} \Gamma^{u}_{\ uz} \Gamma^{z}_{\ uu} F_{zu} + 8 F_{uz} \Gamma^{z}_{\ uz} \Gamma^{z}_{\ zu} F_{zu} + 8 F_{uz} \Gamma^{i}_{\ uz} \Gamma^{z}_{\ iu} F_{zu}\\
		= & 0\,.
	\end{split}
\end{equation}
The second term of Eq. (\ref{hkk3firstsixth}) is
\begin{equation}
	\begin{split}
		& 8 \gamma^{ij} F_{uz} \Gamma^{h}_{\ ui} \Gamma^{z}_{\ hu} F_{zj}\\
		= & 8 \gamma^{ij} F_{uz} \Gamma^{u}_{\ ui} \Gamma^{z}_{\ uu} F_{zj} + 8 \gamma^{ij} F_{uz} \Gamma^{z}_{\ ui} \Gamma^{z}_{\ zu} F_{zj} + 8 \gamma^{ij} F_{uz} \Gamma^{k}_{\ ui} \Gamma^{z}_{\ ku} F_{zj}\\
		= & 0\,.
	\end{split}
\end{equation}
The third term of Eq. (\ref{hkk3firstsixth}) is
\begin{equation}
	\begin{split}
		& 8 \gamma^{ij} F_{uz} \Gamma^{h}_{\ ui} \Gamma^{k}_{\ hu} F_{kj}\\
		= & 8 \gamma^{ij} F_{uz} \Gamma^{u}_{\ ui} \Gamma^{k}_{\ uu} F_{kj} + 8 \gamma^{ij} F_{uz} \Gamma^{z}_{\ ui} \Gamma^{k}_{\ zu} F_{kj} + 8 \gamma^{ij} F_{uz} \Gamma^{l}_{\ ui} \Gamma^{k}_{\ lu} F_{kj}\\
		= & 2 \gamma^{ij} \gamma^{lm} \gamma^{kn} F_{uz} \left(\partial_u \gamma_{im} \right) \left(\partial_u \gamma_{ln} \right) F_{kj}\,.
	\end{split}
\end{equation}
Therefore, the sixth term of Eq. (\ref{hkk3first}) is obtained as 
\begin{equation}
	\begin{split}
		& 8 k^a k^b g^{ce} g^{df} F_{be} \Gamma^{h}_{\ cd} \Gamma^{g}_{\ ha} F_{gf}\\
		= & 2 \gamma^{ij} \gamma^{lm} \gamma^{kn} F_{uz} \left(\partial_u \gamma_{im} \right) \left(\partial_u \gamma_{ln} \right) F_{kj}\\
		= & 8 \gamma^{ij} \gamma^{lm} \gamma^{kn} F_{uz} K_{im} K_{ln} F_{kj}\\
		= & 0\,.
	\end{split}
\end{equation}

The seventh term of Eq. (\ref{hkk3first}) is obtained as 
\begin{equation}
	\begin{split}
		& 8 k^a k^b g^{ce} g^{df} F_{be} \Gamma^{h}_{\ ca} \Gamma^{g}_{\ dh} F_{gf} = 8 g^{ce} g^{df} F_{ue} \Gamma^{h}_{\ cu} \Gamma^{g}_{\ dh} F_{gf}\\
		= & 8 F_{uz} \Gamma^{h}_{\ uu} \Gamma^{g}_{\ uh} F_{gz} + 8 F_{uz} \Gamma^{h}_{\ uu} \Gamma^{g}_{\ zh} F_{gu} + 8 \gamma^{ij} F_{uz} \Gamma^{h}_{\ uu} \Gamma^{g}_{\ ih} F_{gj}\\
		& + 8 \gamma^{ij} F_{uj} \Gamma^{h}_{\ iu} \Gamma^{g}_{\ uh} F_{gz} + 8 \gamma^{ij} F_{uj} \Gamma^{h}_{\ iu} \Gamma^{g}_{\ zh} F_{gu} + 8 \gamma^{ij} \gamma^{kl} F_{uj} \Gamma^{h}_{\ iu} \Gamma^{g}_{\ kh} F_{gl}\\
		= & 0\,.
	\end{split}
\end{equation}

The eighth term of Eq. (\ref{hkk3first}) is
\begin{equation}
	\begin{split}
		& - 8 k^a k^b g^{ce} g^{df} F_{be} \Gamma^{g}_{\ ch} \Gamma^{h}_{\ da} F_{gf} = - 8 g^{ce} g^{df} F_{ue} \Gamma^{g}_{\ ch} \Gamma^{h}_{\ du} F_{gf}\\
		= & - 8 F_{uz} \Gamma^{g}_{\ uh} \Gamma^{h}_{\ uu} F_{gz} - 8 F_{uz} \Gamma^{g}_{\ uh} \Gamma^{h}_{\ zu} F_{gu} - 8 \gamma^{ij} F_{uz} \Gamma^{g}_{\ uh} \Gamma^{h}_{\ iu} F_{gj}\\
		& - 8 \gamma^{ij} F_{uj} \Gamma^{g}_{\ ih} \Gamma^{h}_{\ uu} F_{gz} - 8 \gamma^{ij} F_{uj} \Gamma^{g}_{\ ih} \Gamma^{h}_{\ zu} F_{gu} - 8 \gamma^{ij} \gamma^{kl} F_{uj} \Gamma^{g}_{\ ih} \Gamma^{h}_{\ ku} F_{gl}\\
		= & - 8 F_{uz} \Gamma^{g}_{\ uh} \Gamma^{h}_{\ zu} F_{gu} - 8 \gamma^{ij} F_{uz} \Gamma^{g}_{\ uh} \Gamma^{h}_{\ iu} F_{gj}\,.
	\end{split}
\end{equation}
The index $g$ should be further expanded.
\begin{equation}\label{hkk3firsteighth}
	\begin{split}
		& - 8 F_{uz} \Gamma^{g}_{\ uh} \Gamma^{h}_{\ zu} F_{gu} - 8 \gamma^{ij} F_{uz} \Gamma^{g}_{\ uh} \Gamma^{h}_{\ iu} F_{gj}\\
		= & - 8 F_{uz} \Gamma^{u}_{\ uh} \Gamma^{h}_{\ zu} F_{uu} - 8 F_{uz} \Gamma^{z}_{\ uh} \Gamma^{h}_{\ zu} F_{zu} - 8 F_{uz} \Gamma^{i}_{\ uh} \Gamma^{h}_{\ zu} F_{iu}\\
		& - 8 \gamma^{ij} F_{uz} \Gamma^{u}_{\ uh} \Gamma^{h}_{\ iu} F_{uj} - 8 \gamma^{ij} F_{uz} \Gamma^{z}_{\ uh} \Gamma^{h}_{\ iu} F_{zj} - 8 \gamma^{ij} F_{uz} \Gamma^{k}_{\ uh} \Gamma^{h}_{\ iu} F_{kj}\\
		= & - 8 F_{uz} \Gamma^{z}_{\ uh} \Gamma^{h}_{\ zu} F_{zu} - 8 \gamma^{ij} F_{uz} \Gamma^{z}_{\ uh} \Gamma^{h}_{\ iu} F_{zj} - 8 \gamma^{ij} F_{uz} \Gamma^{k}_{\ uh} \Gamma^{h}_{\ iu} F_{kj}\,.
	\end{split}
\end{equation}
The repeated index $h$ should be further expanded. The first term of Eq. (\ref{hkk3firsteighth}) is
\begin{equation}
	\begin{split}
		& - 8 F_{uz} \Gamma^{z}_{\ uh} \Gamma^{h}_{\ zu} F_{zu}\\
		= & - 8 F_{uz} \Gamma^{z}_{\ uu} \Gamma^{u}_{\ zu} F_{zu} - 8 F_{uz} \Gamma^{z}_{\ uz} \Gamma^{z}_{\ zu} F_{zu} - 8 F_{uz} \Gamma^{z}_{\ ui} \Gamma^{i}_{\ zu} F_{zu}\\
		= & 0\,.
	\end{split}
\end{equation}
The second term of Eq. (\ref{hkk3firsteighth}) is
\begin{equation}
	\begin{split}
		& - 8 \gamma^{ij} F_{uz} \Gamma^{z}_{\ uh} \Gamma^{h}_{\ iu} F_{zj}\\
		= & - 8 \gamma^{ij} F_{uz} \Gamma^{z}_{\ uu} \Gamma^{u}_{\ iu} F_{zj} - 8 \gamma^{ij} F_{uz} \Gamma^{z}_{\ uz} \Gamma^{z}_{\ iu} F_{zj} - 8 \gamma^{ij} F_{uz} \Gamma^{z}_{\ uk} \Gamma^{k}_{\ iu} F_{zj}\\
		= & 0\,.
	\end{split}
\end{equation}
The third term of Eq. (\ref{hkk3firsteighth}) is
\begin{equation}
	\begin{split}
		& - 8 \gamma^{ij} F_{uz} \Gamma^{k}_{\ uh} \Gamma^{h}_{\ iu} F_{kj}\\
		= & - 8 \gamma^{ij} F_{uz} \Gamma^{k}_{\ uu} \Gamma^{u}_{\ iu} F_{kj} - 8 \gamma^{ij} F_{uz} \Gamma^{k}_{\ uz} \Gamma^{z}_{\ iu} F_{kj} - 8 \gamma^{ij} F_{uz} \Gamma^{k}_{\ ul} \Gamma^{l}_{\ iu} F_{kj}\\
		= & - 2 \gamma^{ij} \gamma^{km} \gamma^{ln} F_{uz} \left(\partial_u \gamma_{lm} \right) \left(\partial_u \gamma_{in} \right) F_{kj}\,.
	\end{split}
\end{equation}
Therefore, the eighth term of Eq. (\ref{hkk3first}) is obtained as 
\begin{equation}
	\begin{split}
		& - 8 k^a k^b g^{ce} g^{df} F_{be} \Gamma^{g}_{\ ch} \Gamma^{h}_{\ da} F_{gf}\\
		= & - 2 \gamma^{ij} \gamma^{km} \gamma^{ln} F_{uz} \left(\partial_u \gamma_{lm} \right) \left(\partial_u \gamma_{in} \right) F_{kj}\\
		= & - 8 \gamma^{ij} \gamma^{km} \gamma^{ln} F_{uz} K_{lm} K_{in} F_{kj}\\
		= & 0\,.
	\end{split}
\end{equation}

The ninth term of of Eq. (\ref{hkk3first}) is
\begin{equation}
	\begin{split}
		& - 8 k^a k^b g^{ce} g^{df} F_{be} \Gamma^{g}_{\ da} \partial_c F_{gf} = - 8 g^{ce} g^{df} F_{ue} \Gamma^{g}_{\ du} \partial_c F_{gf}\\
		= & - 8 F_{uz} \Gamma^{g}_{\ uu} \partial_u F_{gz} - 8 F_{uz} \Gamma^{g}_{\ zu} \partial_u F_{gu} - 8 \gamma^{ij} F_{uz} \Gamma^{g}_{\ iu} \partial_u F_{gj}\\
		& - 8 \gamma^{ij} F_{uj} \Gamma^{g}_{\ uu} \partial_i F_{gz} - 8 \gamma^{ij} F_{uj} \Gamma^{g}_{\ zu} \partial_i F_{gu} - 8 \gamma^{ij} \gamma^{kl} F_{uj} \Gamma^{g}_{\ ku} \partial_i F_{gl}\\
		= & - 8 F_{uz} \Gamma^{g}_{\ zu} \partial_u F_{gu} - 8 \gamma^{ij} F_{uz} \Gamma^{g}_{\ iu} \partial_u F_{gj}\,.
	\end{split}
\end{equation}
The index $g$ should be further expanded.
\begin{equation}\label{hkk3firstninth}
	\begin{split}
		& - 8 F_{uz} \Gamma^{g}_{\ zu} \partial_u F_{gu} - 8 \gamma^{ij} F_{uz} \Gamma^{g}_{\ iu} \partial_u F_{gj}\\
		= 
		& - 8 F_{uz} \Gamma^{u}_{\ zu} \partial_u F_{uu} - 8 F_{uz} \Gamma^{z}_{\ zu} \partial_u F_{zu} - 8 F_{uz} \Gamma^{i}_{\ zu} \partial_u F_{iu}\\
		& - 8 \gamma^{ij} F_{uz} \Gamma^{u}_{\ iu} \partial_u F_{uj} - 8 \gamma^{ij} F_{uz} \Gamma^{z}_{\ iu} \partial_u F_{zj} - 8 \gamma^{ij} F_{uz} \Gamma^{k}_{\ iu} \partial_u F_{kj}\\
		= & - 8 F_{uz} \Gamma^{z}_{\ zu} \partial_u F_{zu} - 8 F_{uz} \Gamma^{i}_{\ zu} \partial_u F_{iu} - 8 \gamma^{ij} F_{uz} \Gamma^{u}_{\ iu} \partial_u F_{uj}\\
		& - 8 \gamma^{ij} F_{uz} \Gamma^{z}_{\ iu} \partial_u F_{zj} - 8 \gamma^{ij} F_{uz} \Gamma^{k}_{\ iu} \partial_u F_{kj}\,.
	\end{split}
\end{equation}
Therefore, the ninth term of of Eq. (\ref{hkk3first}) is obtained as
\begin{equation}
	\begin{split}
		& - 8 k^a k^b g^{ce} g^{df} F_{be} \Gamma^{g}_{\ da} \partial_c F_{gf}\\
		= & - 4 \gamma^{ij} \beta_j F_{uz} \partial_u F_{iu} + 4 \gamma^{ij} \beta_i F_{uz} \partial_u F_{uj}\\
		& - 4 \gamma^{ij} \gamma^{kl} F_{uz} \left(\partial_u \gamma_{il} \right) \partial_u F_{kj}\\
		= & - 4 \gamma^{ij} \beta_j F_{uz} \partial_u F_{iu} + 4 \gamma^{ij} \beta_i F_{uz} \partial_u F_{uj}\\
		& - 8 \gamma^{ij} \gamma^{kl} F_{uz} K_{il} \partial_u F_{kj}\\
		= & 8 \gamma^{ij} \beta_i F_{uz} \partial_u F_{uj}\,.
	\end{split}
\end{equation}

The tenth term of of Eq. (\ref{hkk3first}) is
\begin{equation}
	\begin{split}
		& 8 k^a k^b g^{ce} g^{df} F_{be} \Gamma^{g}_{\ da} \Gamma^{h}_{\ cg} F_{hf} = 8 g^{ce} g^{df} F_{ue} \Gamma^{g}_{\ du} \Gamma^{h}_{\ cg} F_{hf}\\ 
		= & 8 F_{uz} \Gamma^{g}_{\ uu} \Gamma^{h}_{\ ug} F_{hz} + 8 F_{uz} \Gamma^{g}_{\ zu} \Gamma^{h}_{\ ug} F_{hu} + 8 \gamma^{ij} F_{uz} \Gamma^{g}_{\ iu} \Gamma^{h}_{\ ug} F_{hj}\\
		& + 8 \gamma^{ij} F_{uj} \Gamma^{g}_{\ uu} \Gamma^{h}_{\ ig} F_{hz} + 8 \gamma^{ij} F_{uj} \Gamma^{g}_{\ zu} \Gamma^{h}_{\ ig} F_{hu} + 8 \gamma^{ij} \gamma^{kl} F_{uj} \Gamma^{g}_{\ ku} \Gamma^{h}_{\ ig} F_{hl}\\
		= & 8 F_{uz} \Gamma^{g}_{\ zu} \Gamma^{h}_{\ ug} F_{hu} + 8 \gamma^{ij} F_{uz} \Gamma^{g}_{\ iu} \Gamma^{h}_{\ ug} F_{hj}\,.
	\end{split}
\end{equation}
The index $g$ should be further expanded.
\begin{equation}\label{hkk3firsttenth}
	\begin{split}
		& 8 F_{uz} \Gamma^{g}_{\ zu} \Gamma^{h}_{\ ug} F_{hu} + 8 \gamma^{ij} F_{uz} \Gamma^{g}_{\ iu} \Gamma^{h}_{\ ug} F_{hj}\\
		= & 8 F_{uz} \Gamma^{u}_{\ zu} \Gamma^{h}_{\ uu} F_{hu} + 8 F_{uz} \Gamma^{z}_{\ zu} \Gamma^{h}_{\ uz} F_{hu} + 8 F_{uz} \Gamma^{i}_{\ zu} \Gamma^{h}_{\ ui} F_{hu}\\
		& + 8 \gamma^{ij} F_{uz} \Gamma^{u}_{\ iu} \Gamma^{h}_{\ uu} F_{hj} + 8 \gamma^{ij} F_{uz} \Gamma^{z}_{\ iu} \Gamma^{h}_{\ uz} F_{hj} + 8 \gamma^{ij} F_{uz} \Gamma^{k}_{\ iu} \Gamma^{h}_{\ uk} F_{hj}\\
		= & 8 F_{uz} \Gamma^{z}_{\ zu} \Gamma^{h}_{\ uz} F_{hu} + 8 F_{uz} \Gamma^{i}_{\ zu} \Gamma^{h}_{\ ui} F_{hu} + 8 \gamma^{ij} F_{uz} \Gamma^{z}_{\ iu} \Gamma^{h}_{\ uz} F_{hj}\\
		& + 8 \gamma^{ij} F_{uz} \Gamma^{k}_{\ iu} \Gamma^{h}_{\ uk} F_{hj}\,.
	\end{split}
\end{equation}
The repeated index $h$ should be further expanded. The first term of Eq. (\ref{hkk3firsttenth}) is
\begin{equation}
	\begin{split}
		8 F_{uz} \Gamma^{z}_{\ zu} \Gamma^{h}_{\ uz} F_{hu} = 0\,.
	\end{split}
\end{equation}
The second term of Eq. (\ref{hkk3firsttenth}) is
\begin{equation}
	\begin{split}
		& 8 F_{uz} \Gamma^{i}_{\ zu} \Gamma^{h}_{\ ui} F_{hu}\\
		= & 8 F_{uz} \Gamma^{i}_{\ zu} \Gamma^{u}_{\ ui} F_{uu} + 8 F_{uz} \Gamma^{i}_{\ zu} \Gamma^{z}_{\ ui} F_{zu} + 8 F_{uz} \Gamma^{i}_{\ zu} \Gamma^{j}_{\ ui} F_{ju}\\
		= & 8 F_{uz} \Gamma^{i}_{\ zu} \Gamma^{z}_{\ ui} F_{zu} + 8 F_{uz} \Gamma^{i}_{\ zu} \Gamma^{j}_{\ ui} F_{ju}\\
		= & 2 \gamma^{ik} \gamma^{jl} \beta_k F_{uz} \left(\partial_u \gamma_{il} \right) F_{ju}\,.
	\end{split}
\end{equation}
The third term of Eq. (\ref{hkk3firsttenth}) is
\begin{equation}
	\begin{split}
		8 \gamma^{ij} F_{uz} \Gamma^{z}_{\ iu} \Gamma^{h}_{\ uz} F_{hj} = 0\,.
	\end{split}
\end{equation}
The fourth term of Eq. (\ref{hkk3firsttenth}) is
\begin{equation}
	\begin{split}
		& 8 \gamma^{ij} F_{uz} \Gamma^{k}_{\ iu} \Gamma^{h}_{\ uk} F_{hj}\\
		= & 8 \gamma^{ij} F_{uz} \Gamma^{k}_{\ iu} \Gamma^{u}_{\ uk} F_{uj} + 8 \gamma^{ij} F_{uz} \Gamma^{k}_{\ iu} \Gamma^{z}_{\ uk} F_{zj} + 8 \gamma^{ij} F_{uz} \Gamma^{k}_{\ iu} \Gamma^{l}_{\ uk} F_{lj}\\
		= & - 2 \gamma^{ij} \gamma^{kl} \beta_k F_{uz} \left(\partial_u \gamma_{il} \right) F_{uj} + 2 \gamma^{ij} \gamma^{km} \gamma^{ln} F_{uz} \left(\partial_u \gamma_{im} \right) \left(\partial_u \gamma_{kn} \right) F_{lj}\,.
	\end{split}
\end{equation}
Therefore, the tenth term of Eq. (\ref{hkk3first}) is obtained as 
\begin{equation}
	\begin{split}
		& 8 k^a k^b g^{ce} g^{df} F_{be} \Gamma^{g}_{\ da} \Gamma^{h}_{\ cg} F_{hf}\\ 
		= & - 2 \gamma^{ij} \gamma^{kl} \beta_k F_{uz} \left(\partial_u \gamma_{il} \right) F_{uj} + 2 \gamma^{ij} \gamma^{km} \gamma^{ln} F_{uz} \left(\partial_u \gamma_{im} \right) \left(\partial_u \gamma_{kn} \right) F_{lj}\\
		= & - 4 \gamma^{ij} \gamma^{kl} \beta_k F_{uz} K_{il} F_{uj} + 8 \gamma^{ij} \gamma^{km} \gamma^{ln} F_{uz} K_{im} K_{kn} F_{lj}\\
		= & 0\,.
	\end{split}
\end{equation}

The eleventh term of Eq. (\ref{hkk3first}) is
\begin{equation}
	\begin{split}
		& 8 k^a k^b g^{ce} g^{df} F_{be} \Gamma^{g}_{\ da} \Gamma^{h}_{\ cf} F_{gh} = 8 g^{ce} g^{df} F_{ue} \Gamma^{g}_{\ du} \Gamma^{h}_{\ cf} F_{gh}\\
		= & 8 F_{uz} \Gamma^{g}_{\ uu} \Gamma^{h}_{\ uz} F_{gh} + 8 F_{uz} \Gamma^{g}_{\ zu} \Gamma^{h}_{\ uu} F_{gh} + 8 \gamma^{ij} F_{uz} \Gamma^{g}_{\ iu} \Gamma^{h}_{\ uj} F_{gh}\\
		& + 8 \gamma^{ij} F_{uj} \Gamma^{g}_{\ uu} \Gamma^{h}_{\ iz} F_{gh} + 8 \gamma^{ij} F_{uj} \Gamma^{g}_{\ zu} \Gamma^{h}_{\ iu} F_{gh} + 8 \gamma^{ij} \gamma^{kl} F_{uj} \Gamma^{g}_{\ ku} \Gamma^{h}_{\ il} F_{gh}\\
		= & 8 \gamma^{ij} F_{uz} \Gamma^{g}_{\ iu} \Gamma^{h}_{\ uj} F_{gh}\,.
	\end{split}
\end{equation}
The index $g$ should be further expanded.
\begin{equation}\label{hkk3firstelevrnth}
	\begin{split}
		& 8 \gamma^{ij} F_{uz} \Gamma^{g}_{\ iu} \Gamma^{h}_{\ uj} F_{gh}\\
		= & 8 \gamma^{ij} F_{uz} \Gamma^{u}_{\ iu} \Gamma^{h}_{\ uj} F_{uh} + 8 \gamma^{ij} F_{uz} \Gamma^{z}_{\ iu} \Gamma^{h}_{\ uj} F_{zh} + 8 \gamma^{ij} F_{uz} \Gamma^{k}_{\ iu} \Gamma^{h}_{\ uj} F_{kh}\,.
	\end{split}
\end{equation}
The repeated index $h$ should be further expanded. The first term of Eq. (\ref{hkk3firstelevrnth}) is 
\begin{equation}
	\begin{split}
		& 8 \gamma^{ij} F_{uz} \Gamma^{u}_{\ iu} \Gamma^{h}_{\ uj} F_{uh}\\
		= & 8 \gamma^{ij} F_{uz} \Gamma^{u}_{\ iu} \Gamma^{u}_{\ uj} F_{uu} + 8 \gamma^{ij} F_{uz} \Gamma^{u}_{\ iu} \Gamma^{z}_{\ uj} F_{uz} + 8 \gamma^{ij} F_{uz} \Gamma^{u}_{\ iu} \Gamma^{k}_{\ uj} F_{uk}\\
		= & 8 \gamma^{ij} F_{uz} \Gamma^{u}_{\ iu} \Gamma^{z}_{\ uj} F_{uz} + 8 \gamma^{ij} F_{uz} \Gamma^{u}_{\ iu} \Gamma^{k}_{\ uj} F_{uk}\\
		= & - 2 \gamma^{ij} \gamma^{kl} \beta_i F_{uz} \left(\partial_u \gamma_{jl} \right) F_{uk}\,.
	\end{split}
\end{equation}
The second term of Eq. (\ref{hkk3firstelevrnth}) is
\begin{equation}
	\begin{split}
		8 \gamma^{ij} F_{uz} \Gamma^{z}_{\ iu} \Gamma^{h}_{\ uj} F_{zh} = 0\,.
	\end{split}
\end{equation}
The third term of Eq. (\ref{hkk3firstelevrnth}) is
\begin{equation}
	\begin{split}
		& 8 \gamma^{ij} F_{uz} \Gamma^{k}_{\ iu} \Gamma^{h}_{\ uj} F_{kh}\\
		= & 8 \gamma^{ij} F_{uz} \Gamma^{k}_{\ iu} \Gamma^{u}_{\ uj} F_{ku} + 8 \gamma^{ij} F_{uz} \Gamma^{k}_{\ iu} \Gamma^{z}_{\ uj} F_{kz} + 8 \gamma^{ij} F_{uz} \Gamma^{k}_{\ iu} \Gamma^{l}_{\ uj} F_{kl}\\
		= & - 2 \gamma^{ij} \gamma^{kl} \beta_j F_{uz} \left(\partial_u \gamma_{il} \right) F_{ku} + 2 \gamma^{ij} \gamma^{km} \gamma^{ln} F_{uz} \left(\partial_u \gamma_{im} \right) \left(\partial_u \gamma_{jn} \right) F_{kl}\,.
	\end{split}
\end{equation}
Therefore, the eleventh term of Eq. (\ref{hkk3first}) is obtained as 
\begin{equation}
	\begin{split}
		& 8 k^a k^b g^{ce} g^{df} F_{be} \Gamma^{g}_{\ da} \Gamma^{h}_{\ cf} F_{gh}\\
		= & - 2 \gamma^{ij} \gamma^{kl} \beta_i F_{uz} \left(\partial_u \gamma_{jl} \right) F_{uk} - 2 \gamma^{ij} \gamma^{kl} \beta_j F_{uz} \left(\partial_u \gamma_{il} \right) F_{ku}\\
		& + 2 \gamma^{ij} \gamma^{km} \gamma^{ln} F_{uz} \left(\partial_u \gamma_{im} \right) \left(\partial_u \gamma_{jn} \right) F_{kl}\\
		= & - 4 \gamma^{ij} \gamma^{kl} \beta_i F_{uz} K_{jl} F_{uk} - 4 \gamma^{ij} \gamma^{kl} \beta_j F_{uz} K_{il} F_{ku}\\
		& + 8 \gamma^{ij} \gamma^{km} \gamma^{ln} F_{uz} K_{im} K_{jn} F_{kl}\\
		= & 0\,.
	\end{split}
\end{equation}

The twelfth term of Eq. (\ref{hkk3first}) is
\begin{equation}
	\begin{split}
		& - 8 k^a k^b g^{ce} g^{df} F_{be} \left(\partial_c \Gamma^{g}_{\ df} \right) F_{ag} = - 8 g^{ce} g^{df} F_{ue} \left(\partial_c \Gamma^{g}_{\ df} \right) F_{ug}\\
		= & - 16 F_{uz} \left(\partial_u \Gamma^{g}_{\ uz} \right) F_{ug} - 8 \gamma^{ij} F_{uz} \left(\partial_u \Gamma^{g}_{\ ij} \right) F_{ug} - 16 \gamma^{ij} F_{uj} \left(\partial_i \Gamma^{g}_{\ uz} \right) F_{ug}\\
		& - 8 \gamma^{ij} \gamma^{kl} F_{uj} \left(\partial_i \Gamma^{g}_{\ kl} \right) F_{ug}\\
		= & - 16 F_{uz} \left(\partial_u \Gamma^{g}_{\ uz} \right) F_{ug} - 8 \gamma^{ij} F_{uz} \left(\partial_u \Gamma^{g}_{\ ij} \right) F_{ug}\,.
	\end{split}
\end{equation}
The index $g$ should be further expanded.
\begin{equation}\label{hkk3firsttwelfth}
	\begin{split}
		& - 16 F_{uz} \left(\partial_u \Gamma^{g}_{\ uz} \right) F_{ug} - 8 \gamma^{ij} F_{uz} \left(\partial_u \Gamma^{g}_{\ ij} \right) F_{ug}\\
		= & - 16 F_{uz} \left(\partial_u \Gamma^{u}_{\ uz} \right) F_{uu} - 16 F_{uz} \left(\partial_u \Gamma^{z}_{\ uz} \right) F_{uz} - 16 F_{uz} \left(\partial_u \Gamma^{i}_{\ uz} \right) F_{ui}\\
		& - 8 \gamma^{ij} F_{uz} \left(\partial_u \Gamma^{u}_{\ ij} \right) F_{uu} - 8 \gamma^{ij} F_{uz} \left(\partial_u \Gamma^{z}_{\ ij} \right) F_{uz} - 8 \gamma^{ij} F_{uz} \left(\partial_u \Gamma^{k}_{\ ij} \right) F_{uk}\\
		= & - 16 F_{uz} \left(\partial_u \Gamma^{z}_{\ uz} \right) F_{uz} - 8 \gamma^{ij} F_{uz} \left(\partial_u \Gamma^{z}_{\ ij} \right) F_{uz}\,.
	\end{split}
\end{equation}
The first term of Eq. (\ref{hkk3firsttwelfth}) is 
\begin{equation}
	\begin{split}
		& - 16 F_{uz} \left(\partial_u \Gamma^{z}_{\ uz} \right) F_{uz}\\
		= & - 16 F_{uz} \partial_u \left(z^2 \partial_z \alpha + 2 z \alpha - \frac{1}{2} z \beta^i \beta_i - \frac{1}{2} z^2 \beta^i \partial_z \beta_i \right) F_{uz}\\
		= & - 16 F_{uz} \left[z^2 \partial_u \partial_z \alpha + 2 z \partial_u \alpha - \frac{1}{2} z \partial_u \left(\beta^i \beta_i \right) - \frac{1}{2} z^2 \partial_u \left(\beta^i \partial_z \beta_i \right) \right] F_{uz}\\
		= & 0\,.
	\end{split}
\end{equation}
The second term of Eq. (\ref{hkk3firsttwelfth}) is
\begin{equation}
	\begin{split}
		& - 8 \gamma^{ij} F_{uz} \left(\partial_u \Gamma^{z}_{\ ij} \right) F_{uz}\\
		= & - 8 \gamma^{ij} F_{uz} \partial_u \left[\frac{1}{2} \left(z \partial_j \beta_i + z \partial_i \beta_j - \partial_u \gamma_{ij} \right) - \frac{1}{2} z^2 \left(\beta^2 - 2 \alpha \right) \left(\partial_z \gamma_{i j} \right) \right.\\
		& \left. - \frac{1}{2} \left(z \beta^k \right) \left(\partial_j \gamma_{k i} + \partial_i \gamma_{j k} - \partial_k \gamma_{i j} \right) \right] F_{uz}\\
		= & - 4 \gamma^{ij} F_{uz} \left[\left(z \partial_u \partial_j \beta_i + z \partial_u \partial_i \beta_j - \partial_u \partial_u \gamma_{ij} \right) - z^2 \left(2 \beta \partial_u \beta - 2 \partial_u \alpha \right) \left(\partial_z \gamma_{i j} \right) \right.\\
		& \left. - z^2 \left(\beta^2 - 2 \alpha \right) \left(\partial_u \partial_z \gamma_{i j} \right) - \left(z \partial_u \beta^k \right) \left(\partial_j \gamma_{k i} + \partial_i \gamma_{j k} - \partial_k \gamma_{i j} \right) \right.\\
		& \left. - \left(z \beta^k \right) \left(\partial_u \partial_j \gamma_{k i} + \partial_u \partial_i \gamma_{j k} - \partial_u \partial_k \gamma_{i j} \right) \right] F_{uz}\\
		= & 4 \gamma^{ij} F_{uz} \left(\partial_u^2 \gamma_{ij} \right) F_{uz}\,.
	\end{split}
\end{equation}
Therefore, the twelfth term of Eq. (\ref{hkk3first}) is obtained as 
\begin{equation}
	\begin{split}
		- 8 k^a k^b g^{ce} g^{df} F_{be} \left(\partial_c \Gamma^{g}_{\ df} \right) F_{ag} = 4 \gamma^{ij} F_{uz} \left(\partial_u^2 \gamma_{ij} \right) F_{uz}\,.
	\end{split}
\end{equation}

The thirteenth term of Eq. (\ref{hkk3first}) is
\begin{equation}
	\begin{split}
		& 8 k^a k^b g^{ce} g^{df} F_{be} \Gamma^{h}_{\ cd} \Gamma^{g}_{\ hf} F_{ag} = 8 g^{ce} g^{df} F_{ue} \Gamma^{h}_{\ cd} \Gamma^{g}_{\ hf} F_{ug}\\
		= & 8 F_{uz} \Gamma^{h}_{\ uu} \Gamma^{g}_{\ hz} F_{ug} + 8 F_{uz} \Gamma^{h}_{\ uz} \Gamma^{g}_{\ hu} F_{ug} + 8 \gamma^{ij} F_{uz} \Gamma^{h}_{\ ui} \Gamma^{g}_{\ hj} F_{ug}\\
		& + 8 \gamma^{ij} F_{uj} \Gamma^{h}_{\ iu} \Gamma^{g}_{\ hz} F_{ug} + 8 \gamma^{ij} F_{uj} \Gamma^{h}_{\ iz} \Gamma^{g}_{\ hu} F_{ug} + 8 \gamma^{ij} \gamma^{kl} F_{uj} \Gamma^{h}_{\ ik} \Gamma^{g}_{\ hl} F_{ug}\\
		= & 8 F_{uz} \Gamma^{h}_{\ uz} \Gamma^{g}_{\ hu} F_{ug} + 8 \gamma^{ij} F_{uz} \Gamma^{h}_{\ ui} \Gamma^{g}_{\ hj} F_{ug}\,.
	\end{split}
\end{equation}
The index $g$ should be further expanded.
\begin{equation}\label{hkk3firstthirteenth}
	\begin{split}
		& 8 F_{uz} \Gamma^{h}_{\ uz} \Gamma^{g}_{\ hu} F_{ug} + 8 \gamma^{ij} F_{uz} \Gamma^{h}_{\ ui} \Gamma^{g}_{\ hj} F_{ug}\\
		= & 8 F_{uz} \Gamma^{h}_{\ uz} \Gamma^{u}_{\ hu} F_{uu} + 8 F_{uz} \Gamma^{h}_{\ uz} \Gamma^{z}_{\ hu} F_{uz} + 8 F_{uz} \Gamma^{h}_{\ uz} \Gamma^{i}_{\ hu} F_{ui}\\
		& + 8 \gamma^{ij} F_{uz} \Gamma^{h}_{\ ui} \Gamma^{u}_{\ hj} F_{uu} + 8 \gamma^{ij} F_{uz} \Gamma^{h}_{\ ui} \Gamma^{z}_{\ hj} F_{uz} + 8 \gamma^{ij} F_{uz} \Gamma^{h}_{\ ui} \Gamma^{k}_{\ hj} F_{uk}\\
		= & 8 F_{uz} \Gamma^{h}_{\ uz} \Gamma^{z}_{\ hu} F_{uz} + 8 \gamma^{ij} F_{uz} \Gamma^{h}_{\ ui} \Gamma^{z}_{\ hj} F_{uz}\,.
	\end{split}
\end{equation}
The repeated index $h$ should be further expanded. The first term of Eq. (\ref{hkk3firstthirteenth}) is
\begin{equation}
	\begin{split}
		& 8 F_{uz} \Gamma^{h}_{\ uz} \Gamma^{z}_{\ hu} F_{uz}\\
		= & 8 F_{uz} \Gamma^{u}_{\ uz} \Gamma^{z}_{\ uu} F_{uz} + 8 F_{uz} \Gamma^{z}_{\ uz} \Gamma^{z}_{\ zu} F_{uz} + 8 F_{uz} \Gamma^{i}_{\ uz} \Gamma^{z}_{\ iu} F_{uz}\\
		= & 0\,.
	\end{split}
\end{equation}
The second term of Eq. (\ref{hkk3firstthirteenth}) is
\begin{equation}
	\begin{split}
		& 8 \gamma^{ij} F_{uz} \Gamma^{h}_{\ ui} \Gamma^{z}_{\ hj} F_{uz}\\
		= & 8 \gamma^{ij} F_{uz} \Gamma^{u}_{\ ui} \Gamma^{z}_{\ uj} F_{uz} + 8 \gamma^{ij} F_{uz} \Gamma^{z}_{\ ui} \Gamma^{z}_{\ zj} F_{uz} + 8 \gamma^{ij} F_{uz} \Gamma^{k}_{\ ui} \Gamma^{z}_{\ kj} F_{uz}\\
		= & - 2 \gamma^{ij} \gamma^{kl} F_{uz} \left(\partial_u \gamma_{il} \right) \left(\partial_u \gamma_{kj} \right) F_{uz}\,.
	\end{split}
\end{equation}
Therefore, the thirteenth term of Eq. (\ref{hkk3first}) is obtained as 
\begin{equation}
	\begin{split}
		& 8 k^a k^b g^{ce} g^{df} F_{be} \Gamma^{h}_{\ cd} \Gamma^{g}_{\ hf} F_{ag}\\
		= & - 2 \gamma^{ij} \gamma^{kl} F_{uz} \left(\partial_u \gamma_{il} \right) \left(\partial_u \gamma_{kj} \right) F_{uz}\\
		= & - 8 \gamma^{ij} \gamma^{kl} F_{uz} K_{il} K_{kj} F_{uz}\\
		= & 0\,.
	\end{split}
\end{equation}

The fourteenth term of of Eq. (\ref{hkk3first}) is 
\begin{equation}
	\begin{split}
		& 8 k^a k^b g^{ce} g^{df} F_{be} \Gamma^{h}_{\ cf} \Gamma^{g}_{\ dh} F_{ag} = 8 g^{ce} g^{df} F_{ue} \Gamma^{h}_{\ cf} \Gamma^{g}_{\ dh} F_{ug}\\
		= & 8 F_{uz} \Gamma^{h}_{\ uz} \Gamma^{g}_{\ uh} F_{ug} + 8 F_{uz} \Gamma^{h}_{\ uu} \Gamma^{g}_{\ zh} F_{ug} + 8 \gamma^{ij} F_{uz} \Gamma^{h}_{\ uj} \Gamma^{g}_{\ ih} F_{ug}\\
		& + 8 \gamma^{ij} F_{uj} \Gamma^{h}_{\ iz} \Gamma^{g}_{\ uh} F_{ug} + 8 \gamma^{ij} F_{uj} \Gamma^{h}_{\ iu} \Gamma^{g}_{\ zh} F_{ug} + 8 \gamma^{ij} \gamma^{kl} F_{uj} \Gamma^{h}_{\ il} \Gamma^{g}_{\ kh} F_{ug}\\
		= & 8 F_{uz} \Gamma^{h}_{\ uz} \Gamma^{g}_{\ uh} F_{ug} + 8 \gamma^{ij} F_{uz} \Gamma^{h}_{\ uj} \Gamma^{g}_{\ ih} F_{ug}\,.
	\end{split}
\end{equation}
The index $g$ should be further expanded.
\begin{equation}\label{hkk3firstfourteenth}
	\begin{split}
		& 8 F_{uz} \Gamma^{h}_{\ uz} \Gamma^{g}_{\ uh} F_{ug} + 8 \gamma^{ij} F_{uz} \Gamma^{h}_{\ uj} \Gamma^{g}_{\ ih} F_{ug}\\
		= & 8 F_{uz} \Gamma^{h}_{\ uz} \Gamma^{u}_{\ uh} F_{uu} + 8 F_{uz} \Gamma^{h}_{\ uz} \Gamma^{z}_{\ uh} F_{uz} + 8 F_{uz} \Gamma^{h}_{\ uz} \Gamma^{i}_{\ uh} F_{ui}\\
		& + 8 \gamma^{ij} F_{uz} \Gamma^{h}_{\ uj} \Gamma^{u}_{\ ih} F_{uu} + 8 \gamma^{ij} F_{uz} \Gamma^{h}_{\ uj} \Gamma^{z}_{\ ih} F_{uz} + 8 \gamma^{ij} F_{uz} \Gamma^{h}_{\ uj} \Gamma^{k}_{\ ih} F_{uk}\\
		= & 8 F_{uz} \Gamma^{h}_{\ uz} \Gamma^{z}_{\ uh} F_{uz} + 8 \gamma^{ij} F_{uz} \Gamma^{h}_{\ uj} \Gamma^{z}_{\ ih} F_{uz}\,.
	\end{split}
\end{equation}
The repeated index $h$ should be further expanded. The first term of Eq. (\ref{hkk3firstfourteenth}) is 
\begin{equation}
	\begin{split}
		& 8 F_{uz} \Gamma^{h}_{\ uz} \Gamma^{z}_{\ uh} F_{uz}\\
		= & 8 F_{uz} \Gamma^{u}_{\ uz} \Gamma^{z}_{\ uu} F_{uz} + 8 F_{uz} \Gamma^{z}_{\ uz} \Gamma^{z}_{\ uz} F_{uz} + 8 F_{uz} \Gamma^{i}_{\ uz} \Gamma^{z}_{\ ui} F_{uz}\\
		= & 0\,.
	\end{split}
\end{equation}
The second term of Eq. (\ref{hkk3firstfourteenth}) is
\begin{equation}
	\begin{split}
		& 8 \gamma^{ij} F_{uz} \Gamma^{h}_{\ uj} \Gamma^{z}_{\ ih} F_{uz}\\
		= & 8 \gamma^{ij} F_{uz} \Gamma^{u}_{\ uj} \Gamma^{z}_{\ iu} F_{uz} + 8 \gamma^{ij} F_{uz} \Gamma^{z}_{\ uj} \Gamma^{z}_{\ iz} F_{uz} + 8 \gamma^{ij} F_{uz} \Gamma^{k}_{\ uj} \Gamma^{z}_{\ ik} F_{uz}\\
		= & - 2 \gamma^{ij} \gamma^{kl} F_{uz} \left(\partial_u \gamma_{jl} \right) \left(\partial_u \gamma_{ik} \right) F_{uz}\,.
	\end{split}
\end{equation}
Therefore, the fourteenth term of Eq. (\ref{hkk3first}) is obtained as 
\begin{equation}
	\begin{split}
		& 8 k^a k^b g^{ce} g^{df} F_{be} \Gamma^{h}_{\ cf} \Gamma^{g}_{\ dh} F_{ag}\\
		= & - 2 \gamma^{ij} \gamma^{kl} F_{uz} \left(\partial_u \gamma_{jl} \right) \left(\partial_u \gamma_{ik} \right) F_{uz}\\
		= & - 8 \gamma^{ij} \gamma^{kl} F_{uz} K_{jl} K_{ik} F_{uz}\\
		= & 0\,.
	\end{split}
\end{equation}

The fifteenth term of Eq. (\ref{hkk3first}) is 
\begin{equation}
	\begin{split}
		& - 8 k^a k^b g^{ce} g^{df} F_{be} \Gamma^{g}_{\ ch} \Gamma^{h}_{\ df} F_{ag} = - 8 g^{ce} g^{df} F_{ue} \Gamma^{g}_{\ ch} \Gamma^{h}_{\ df} F_{ug}\\
		= & - 8 F_{uz} \Gamma^{g}_{\ uh} \Gamma^{h}_{\ uz} F_{ug} - 8 F_{uz} \Gamma^{g}_{\ uh} \Gamma^{h}_{\ zu} F_{ug} - 8 \gamma^{ij} F_{uz} \Gamma^{g}_{\ uh} \Gamma^{h}_{\ ij} F_{ug}\\
		& - 8 \gamma^{ij} F_{uj} \Gamma^{g}_{\ ih} \Gamma^{h}_{\ uz} F_{ug} - 8 \gamma^{ij} F_{uj} \Gamma^{g}_{\ ih} \Gamma^{h}_{\ zu} F_{ug} - 8 \gamma^{ij} \gamma^{kl} F_{uj} \Gamma^{g}_{\ ih} \Gamma^{h}_{\ kl} F_{ug}\\
		= & - 8 F_{uz} \Gamma^{g}_{\ uh} \Gamma^{h}_{\ uz} F_{ug} - 8 F_{uz} \Gamma^{g}_{\ uh} \Gamma^{h}_{\ zu} F_{ug} - 8 \gamma^{ij} F_{uz} \Gamma^{g}_{\ uh} \Gamma^{h}_{\ ij} F_{ug}\,.
	\end{split}
\end{equation}
The index $g$ should be further expanded.
\begin{equation}\label{hkk3firstfifteenth}
	\begin{split}
		&  - 8 F_{uz} \Gamma^{g}_{\ uh} \Gamma^{h}_{\ uz} F_{ug} - 8 F_{uz} \Gamma^{g}_{\ uh} \Gamma^{h}_{\ zu} F_{ug} - 8 \gamma^{ij} F_{uz} \Gamma^{g}_{\ uh} \Gamma^{h}_{\ ij} F_{ug}\\
		= & - 8 F_{uz} \Gamma^{u}_{\ uh} \Gamma^{h}_{\ uz} F_{uu} - 8 F_{uz} \Gamma^{z}_{\ uh} \Gamma^{h}_{\ uz} F_{uz} - 8 F_{uz} \Gamma^{i}_{\ uh} \Gamma^{h}_{\ uz} F_{ui}\\
		& - 8 F_{uz} \Gamma^{u}_{\ uh} \Gamma^{h}_{\ zu} F_{uu} - 8 F_{uz} \Gamma^{z}_{\ uh} \Gamma^{h}_{\ zu} F_{uz} - 8 F_{uz} \Gamma^{i}_{\ uh} \Gamma^{h}_{\ zu} F_{ui}\\
		& - 8 \gamma^{ij} F_{uz} \Gamma^{u}_{\ uh} \Gamma^{h}_{\ ij} F_{uu} - 8 \gamma^{ij} F_{uz} \Gamma^{z}_{\ uh} \Gamma^{h}_{\ ij} F_{uz} - 8 \gamma^{ij} F_{uz} \Gamma^{k}_{\ uh} \Gamma^{h}_{\ ij} F_{uk}\\
		= & - 8 F_{uz} \Gamma^{z}_{\ uh} \Gamma^{h}_{\ uz} F_{uz} - 8 F_{uz} \Gamma^{z}_{\ uh} \Gamma^{h}_{\ zu} F_{uz} - 8 \gamma^{ij} F_{uz} \Gamma^{z}_{\ uh} \Gamma^{h}_{\ ij} F_{uz}\,.
	\end{split}
\end{equation}
The repeated index $h$ should be further expanded. The first term of Eq. (\ref{hkk3firstfifteenth}) is 
\begin{equation}
	\begin{split}
		& - 8 F_{uz} \Gamma^{z}_{\ uh} \Gamma^{h}_{\ uz} F_{uz}\\
		= & - 8 F_{uz} \Gamma^{z}_{\ uu} \Gamma^{u}_{\ uz} F_{uz} - 8 F_{uz} \Gamma^{z}_{\ uz} \Gamma^{z}_{\ uz} F_{uz} - 8 F_{uz} \Gamma^{z}_{\ ui} \Gamma^{i}_{\ uz} F_{uz}\\
		= & 0\,.
	\end{split}
\end{equation}
The second term of Eq. (\ref{hkk3firstfifteenth}) is 
\begin{equation}
	\begin{split}
		& - 8 F_{uz} \Gamma^{z}_{\ uh} \Gamma^{h}_{\ zu} F_{uz}\\
		= & - 8 F_{uz} \Gamma^{z}_{\ uu} \Gamma^{u}_{\ zu} F_{uz} - 8 F_{uz} \Gamma^{z}_{\ uz} \Gamma^{z}_{\ zu} F_{uz} - 8 F_{uz} \Gamma^{z}_{\ ui} \Gamma^{i}_{\ zu} F_{uz}\\
		= & 0\,.
	\end{split}
\end{equation}
The third term of Eq. (\ref{hkk3firstfifteenth}) is 
\begin{equation}
	\begin{split}
		& - 8 \gamma^{ij} F_{uz} \Gamma^{z}_{\ uh} \Gamma^{h}_{\ ij} F_{uz}\\
		= & - 8 \gamma^{ij} F_{uz} \Gamma^{z}_{\ uu} \Gamma^{u}_{\ ij} F_{uz} - 8 \gamma^{ij} F_{uz} \Gamma^{z}_{\ uz} \Gamma^{z}_{\ ij} F_{uz} - 8 \gamma^{ij} F_{uz} \Gamma^{z}_{\ uk} \Gamma^{k}_{\ ij} F_{uz}\\
		= & 0\,.
	\end{split}
\end{equation}
Therefore, the fifteenth term of Eq. (\ref{hkk3first}) is obtained as 
\begin{equation}
	\begin{split}
		- 8 k^a k^b g^{ce} g^{df} F_{be} \Gamma^{g}_{\ ch} \Gamma^{h}_{\ df} F_{ag} = 0\,.
	\end{split}
\end{equation}

The sixteenth term of Eq. (\ref{hkk3first}) is
\begin{equation}
	\begin{split}
		& - 8 k^a k^b g^{ce} g^{df} F_{be} \Gamma^{g}_{df} \partial_c F_{ag} = - 8 g^{ce} g^{df} F_{ue} \Gamma^{g}_{df} \partial_c F_{ug}\\
		= & - 16 F_{uz} \Gamma^{g}_{uz} \partial_u F_{ug} - 8 \gamma^{ij} F_{uz} \Gamma^{g}_{ij} \partial_u F_{ug} - 16 \gamma^{ij} F_{uj} \Gamma^{g}_{uz} \partial_i F_{ug}\\
		& - 8 \gamma^{ij} \gamma^{kl} F_{uj} \Gamma^{g}_{kl} \partial_i F_{ug}\\
		= & - 16 F_{uz} \Gamma^{g}_{uz} \partial_u F_{ug} - 8 \gamma^{ij} F_{uz} \Gamma^{g}_{ij} \partial_u F_{ug}
	\end{split}
\end{equation}
The index $g$ should be further expanded.
\begin{equation}\label{hkk3firstsixteenth}
	\begin{split}
		& - 16 F_{uz} \Gamma^{g}_{uz} \partial_u F_{ug} - 8 \gamma^{ij} F_{uz} \Gamma^{g}_{ij} \partial_u F_{ug}\\
		= & - 16 F_{uz} \Gamma^{u}_{uz} \partial_u F_{uu} - 16 F_{uz} \Gamma^{z}_{uz} \partial_u F_{uz} - 16 F_{uz} \Gamma^{i}_{uz} \partial_u F_{ui}\\
		& - 8 \gamma^{ij} F_{uz} \Gamma^{u}_{ij} \partial_u F_{uu} - 8 \gamma^{ij} F_{uz} \Gamma^{z}_{ij} \partial_u F_{uz} - 8 \gamma^{ij} F_{uz} \Gamma^{k}_{ij} \partial_u F_{uk}\\
		= & - 16 F_{uz} \Gamma^{z}_{uz} \partial_u F_{uz} - 16 F_{uz} \Gamma^{i}_{uz} \partial_u F_{ui} - 8 \gamma^{ij} F_{uz} \Gamma^{z}_{ij} \partial_u F_{uz}\\
		& - 8 \gamma^{ij} F_{uz} \Gamma^{k}_{ij} \partial_u F_{uk}
	\end{split}
\end{equation}
The first term of Eq. (\ref{hkk3firstsixteenth}) is 
\begin{equation}
	\begin{split}
		- 16 F_{uz} \Gamma^{z}_{uz} \partial_u F_{uz} = 0\,.
	\end{split}
\end{equation}
The second term of Eq. (\ref{hkk3firstsixteenth}) is 
\begin{equation}
	\begin{split}
		- 16 F_{uz} \Gamma^{i}_{uz} \partial_u F_{ui} = - 8 \gamma^{ij} \beta_j F_{uz} \partial_u F_{ui}\,.
	\end{split}
\end{equation}
The third term of Eq. (\ref{hkk3firstsixteenth}) is 
\begin{equation}
	\begin{split}
		- 8 \gamma^{ij} F_{uz} \Gamma^{z}_{ij} \partial_u F_{uz} = 4 \gamma^{ij} F_{uz} \left(\partial_u \gamma_{ij} \right) \partial_u F_{uz}\,.
	\end{split}
\end{equation}
The fourth term of Eq. (\ref{hkk3firstsixteenth}) is 
\begin{equation}
	\begin{split}
		- 8 \gamma^{ij} F_{uz} \Gamma^{k}_{ij} \partial_u F_{uk} = - 8 \gamma^{ij} F_{uz} \hat{\Gamma}^{k}_{\ ij} \partial_u F_{uk}\,.
	\end{split}
\end{equation}
Therefore, the sixteenth term of Eq. (\ref{hkk3first}) is obtained as 
\begin{equation}
	\begin{split}
		& - 8 k^a k^b g^{ce} g^{df} F_{be} \Gamma^{g}_{df} \partial_c F_{ag}\\
		= & - 8 \gamma^{ij} \beta_j F_{uz} \partial_u F_{ui} + 4 \gamma^{ij} F_{uz} \left(\partial_u \gamma_{ij} \right) \partial_u F_{uz}\\
		& - 8 \gamma^{ij} F_{uz} \hat{\Gamma}^{k}_{\ ij} \partial_u F_{uk}\\
		= & - 8 \gamma^{ij} \beta_j F_{uz} \partial_u F_{ui} + 8 \gamma^{ij} F_{uz} K_{ij} \partial_u F_{uz}\\
		& - 8 \gamma^{ij} F_{uz} \hat{\Gamma}^{k}_{\ ij} \partial_u F_{uk}\\
		= & - 8 \gamma^{ij} \beta_j F_{uz} \partial_u F_{ui} - 8 \gamma^{ij} F_{uz} \hat{\Gamma}^{k}_{\ ij} \partial_u F_{uk}\,.
	\end{split}
\end{equation}

The seventeenth term of Eq. (\ref{hkk3first}) is obtained as 
\begin{equation}
	\begin{split}
		& 8 k^a k^b g^{ce} g^{df} F_{be} \Gamma^{g}_{df} \Gamma^{h}_{\ ca} F_{hg} = 8 g^{ce} g^{df} F_{ue} \Gamma^{g}_{df} \Gamma^{h}_{\ cu} F_{hg}\\
		= & 16 F_{uz} \Gamma^{g}_{uz} \Gamma^{h}_{\ uu} F_{hg} + 8 \gamma^{ij} F_{uz} \Gamma^{g}_{ij} \Gamma^{h}_{\ uu} F_{hg} + 16 \gamma^{ij} F_{uj} \Gamma^{g}_{uz} \Gamma^{h}_{\ iu} F_{hg}\\
		& + 8 \gamma^{ij} \gamma^{kl} F_{uj} \Gamma^{g}_{kl} \Gamma^{h}_{\ iu} F_{hg}\\
		= & 0\,.
	\end{split}
\end{equation}

The eighteenth term of Eq. (\ref{hkk3first}) is
\begin{equation}
	\begin{split}
		& 8 k^a k^b g^{ce} g^{df} F_{be} \Gamma^{g}_{df} \Gamma^{h}_{\ cg} F_{ah} = 8 g^{ce} g^{df} F_{ue} \Gamma^{g}_{df} \Gamma^{h}_{\ cg} F_{uh}\\
		= & 16 F_{uz} \Gamma^{g}_{uz} \Gamma^{h}_{\ ug} F_{uh} + 8 \gamma^{ij} F_{uz} \Gamma^{g}_{ij} \Gamma^{h}_{\ ug} F_{uh} + 16 \gamma^{ij} F_{uj} \Gamma^{g}_{uz} \Gamma^{h}_{\ ig} F_{uh}\\
		& + 8 \gamma^{ij} \gamma^{kl} F_{uj} \Gamma^{g}_{kl} \Gamma^{h}_{\ ig} F_{uh}\\
		= & 16 F_{uz} \Gamma^{g}_{uz} \Gamma^{h}_{\ ug} F_{uh} + 8 \gamma^{ij} F_{uz} \Gamma^{g}_{ij} \Gamma^{h}_{\ ug} F_{uh}\,.
	\end{split}
\end{equation}
The index $g$ should be further expanded.
\begin{equation}\label{hkk3firsteighteenth}
	\begin{split}
		& 16 F_{uz} \Gamma^{g}_{uz} \Gamma^{h}_{\ ug} F_{uh} + 8 \gamma^{ij} F_{uz} \Gamma^{g}_{ij} \Gamma^{h}_{\ ug} F_{uh}\\
		= & 16 F_{uz} \Gamma^{u}_{uz} \Gamma^{h}_{\ uu} F_{uh} + 16 F_{uz} \Gamma^{z}_{uz} \Gamma^{h}_{\ uz} F_{uh} + 16 F_{uz} \Gamma^{i}_{uz} \Gamma^{h}_{\ ui} F_{uh}\\
		& + 8 \gamma^{ij} F_{uz} \Gamma^{u}_{ij} \Gamma^{h}_{\ uu} F_{uh} + 8 \gamma^{ij} F_{uz} \Gamma^{z}_{ij} \Gamma^{h}_{\ uz} F_{uh} + 8 \gamma^{ij} F_{uz} \Gamma^{k}_{ij} \Gamma^{h}_{\ uk} F_{uh}\\
		= & 16 F_{uz} \Gamma^{z}_{uz} \Gamma^{h}_{\ uz} F_{uh} + 16 F_{uz} \Gamma^{i}_{uz} \Gamma^{h}_{\ ui} F_{uh} + 8 \gamma^{ij} F_{uz} \Gamma^{z}_{ij} \Gamma^{h}_{\ uz} F_{uh}\\
		& + 8 \gamma^{ij} F_{uz} \Gamma^{k}_{ij} \Gamma^{h}_{\ uk} F_{uh}\,.
	\end{split}
\end{equation}
The repeated index $h$ should be further expanded. The first term of Eq. (\ref{hkk3firsteighteenth}) is 
\begin{equation}
	\begin{split}
		16 F_{uz} \Gamma^{z}_{uz} \Gamma^{h}_{\ uz} F_{uh} = 0\,.
	\end{split}
\end{equation}
The second term of Eq. (\ref{hkk3firsteighteenth}) is
\begin{equation}
	\begin{split}
		& 16 F_{uz} \Gamma^{i}_{uz} \Gamma^{h}_{\ ui} F_{uh}\\
		= & 16 F_{uz} \Gamma^{i}_{uz} \Gamma^{u}_{\ ui} F_{uu} + 16 F_{uz} \Gamma^{i}_{uz} \Gamma^{z}_{\ ui} F_{uz} + 16 F_{uz} \Gamma^{i}_{uz} \Gamma^{j}_{\ ui} F_{uj}\\
		= & 16 F_{uz} \Gamma^{i}_{uz} \Gamma^{z}_{\ ui} F_{uz} + 16 F_{uz} \Gamma^{i}_{uz} \Gamma^{j}_{\ ui} F_{uj}\\
		= & 4 \gamma^{ik} \gamma^{jl} \beta_k F_{uz} \left(\partial_u \gamma_{il} \right) F_{uj}\,.
	\end{split}
\end{equation}
The third term of Eq. (\ref{hkk3firsteighteenth}) is
\begin{equation}
	\begin{split}
		& 8 \gamma^{ij} F_{uz} \Gamma^{z}_{ij} \Gamma^{h}_{\ uz} F_{uh}\\
		= & 8 \gamma^{ij} F_{uz} \Gamma^{z}_{ij} \Gamma^{u}_{\ uz} F_{uu} + 8 \gamma^{ij} F_{uz} \Gamma^{z}_{ij} \Gamma^{z}_{\ uz} F_{uz} + 8 \gamma^{ij} F_{uz} \Gamma^{z}_{ij} \Gamma^{k}_{\ uz} F_{uk}\\
		= & 8 \gamma^{ij} F_{uz} \Gamma^{z}_{ij} \Gamma^{z}_{\ uz} F_{uz} + 8 \gamma^{ij} F_{uz} \Gamma^{z}_{ij} \Gamma^{k}_{\ uz} F_{uk}\\
		= & - 2 \gamma^{ij} \gamma^{km} \beta_m F_{uz} \left(\partial_u \gamma_{ij} \right) F_{uk}\,.
	\end{split}
\end{equation}
The fourth term of Eq. (\ref{hkk3firsteighteenth}) is
\begin{equation}
	\begin{split}
		& 8 \gamma^{ij} F_{uz} \Gamma^{k}_{ij} \Gamma^{h}_{\ uk} F_{uh}\\
		= & 8 \gamma^{ij} F_{uz} \Gamma^{k}_{ij} \Gamma^{u}_{\ uk} F_{uu} + 8 \gamma^{ij} F_{uz} \Gamma^{k}_{ij} \Gamma^{z}_{\ uk} F_{uz} + 8 \gamma^{ij} F_{uz} \Gamma^{k}_{ij} \Gamma^{l}_{\ uk} F_{ul}\\
		= & 8 \gamma^{ij} F_{uz} \Gamma^{k}_{ij} \Gamma^{z}_{\ uk} F_{uz} + 8 \gamma^{ij} F_{uz} \Gamma^{k}_{ij} \Gamma^{l}_{\ uk} F_{ul}\\
		= & 4 \gamma^{ij} \gamma^{lm} F_{uz} \hat{\Gamma}^{k}_{\ ij} \left(\partial_u \gamma_{km} \right) F_{ul}\,.
	\end{split}
\end{equation}
Therefore, the eighteenth term of Eq. (\ref{hkk3first}) is obtained as 
\begin{equation}
	\begin{split}
		& 8 k^a k^b g^{ce} g^{df} F_{be} \Gamma^{g}_{df} \Gamma^{h}_{\ cg} F_{ah}\\
		= & 4 \gamma^{ik} \gamma^{jl} \beta_k F_{uz} \left(\partial_u \gamma_{il} \right) F_{uj} - 2 \gamma^{ij} \gamma^{km} \beta_m F_{uz} \left(\partial_u \gamma_{ij} \right) F_{uk}\\
		& + 4 \gamma^{ij} \gamma^{lm} F_{uz} \hat{\Gamma}^{k}_{\ ij} \left(\partial_u \gamma_{km} \right) F_{ul}\\
		= & 0\,.
	\end{split}
\end{equation}

Finally, the first term of Eq. (\ref{rehkk3}) is
\begin{equation}
	\begin{split}
		& 8 k^a k^b F_{b}^{\ c} \nabla_c \nabla_d F_{a}^{\ d}\\
		= & 8 F_{uz} \partial_u^2 F_{uz} + 8 \gamma^{ij} F_{uz} \partial_u \partial_i F_{uj} + 4 \gamma^{ij} \beta_i F_{uz} \partial_u F_{uj} \\
		& - 4 \gamma^{ij} \beta_j F_{uz} \partial_u F_{ui} - 8 \gamma^{ij} \gamma^{kl} F_{uz} \left(\partial_u K_{il} \right) F_{kj} + 8 \gamma^{ij} \beta_i F_{uz} \partial_u F_{uj}\\
		& + 4 \gamma^{ij} F_{uz} \left(\partial_u^2 \gamma_{ij} \right) F_{uz} - 8 \gamma^{ij} \beta_j F_{uz} \partial_u F_{ui} - 8 \gamma^{ij} F_{uz} \hat{\Gamma}^{k}_{\ ij} \partial_u F_{uk}\\
		= & 8 F_{uz} \partial_u^2 F_{uz} + 8 \gamma^{ij} F_{uz} \partial_u \partial_i F_{uj} - 8 \gamma^{ij} \gamma^{kl} F_{uz} \left(\partial_u K_{il} \right) F_{kj} + 4 \gamma^{ij} F_{uz} \left(\partial_u^2 \gamma_{ij} \right) F_{uz}\\
		& - 8 \gamma^{ij} F_{uz} \hat{\Gamma}^{k}_{\ ij} \partial_u F_{uk}\,.
	\end{split}
\end{equation}

The second term in Eq. (\ref{rehkk3}) is
\begin{equation}\label{hkk3second}
	\begin{split}
		& 4 k^a k^b \nabla_c F_{bd} \nabla^d F_{a}^{\ c} = 4 k^a k^b g^{ce} g^{df} \nabla_c F_{bf} \nabla_d F_{ae}\\
		= & 4 k^a k^b g^{ce} g^{df} \left(\partial_c F_{bf} \right) \partial_d F_{ae} - 4 k^a k^b g^{ce} g^{df} \left(\partial_c F_{bf} \right) \Gamma^{h}_{\ da} F_{he}\\
		& - 4 k^a k^b g^{ce} g^{df} \left(\partial_c F_{bf} \right) \Gamma^{h}_{\ de} F_{ah} - 4 k^a k^b g^{ce} g^{df} \Gamma^{g}_{\ cb} F_{gf} \partial_d F_{ae} \\
		& + 4 k^a k^b g^{ce} g^{df} \Gamma^{g}_{\ cb} F_{gf} \Gamma^{h}_{\ da} F_{he} + 4 k^a k^b g^{ce} g^{df} \Gamma^{g}_{\ cb} F_{gf} \Gamma^{h}_{\ de} F_{ah}\\
		& - 4 k^a k^b g^{ce} g^{df} \Gamma^{g}_{\ cf} F_{bg} \partial_d F_{ae} + 4 k^a k^b g^{ce} g^{df} \Gamma^{g}_{\ cf} F_{bg} \Gamma^{h}_{\ da} F_{he}\\
		& + 4 k^a k^b g^{ce} g^{df} \Gamma^{g}_{\ cf} F_{bg} \Gamma^{h}_{\ de} F_{ah}\,.
	\end{split}
\end{equation}

The first term in Eq. (\ref{hkk3second}) is obtained as 
\begin{equation}
	\begin{split}
		& 4 k^a k^b g^{ce} g^{df} \left(\partial_c F_{bf} \right) \partial_d F_{ae} = 4 g^{ce} g^{df} \left(\partial_c F_{uf} \right) \partial_d F_{ue}\\
		= & 4 \left(\partial_u F_{uz} \right) \partial_u F_{uz} + 4 \gamma^{ij} \left(\partial_u F_{uj} \right) \partial_i F_{uz} + 4 \gamma^{ij} \left(\partial_i F_{uz} \right) \partial_u F_{uj}\\
		& + 4 \gamma^{ij} \gamma^{kl} \left(\partial_i F_{ul} \right) \partial_k F_{uj}\\
		= & 4 \left(\partial_u F_{uz} \right) \partial_u F_{uz} + 4 \gamma^{ij} \left(\partial_u F_{uj} \right) \partial_i F_{uz} + 4 \gamma^{ij} \left(\partial_i F_{uz} \right) \partial_u F_{uj}\,.
	\end{split}
\end{equation}

The second term in Eq. (\ref{hkk3second}) is 
\begin{equation}\label{hkk3secondsecond}
	\begin{split}
		& - 4 k^a k^b g^{ce} g^{df} \left(\partial_c F_{bf} \right) \Gamma^{h}_{\ da} F_{he} = - 4 g^{ce} g^{df} \left(\partial_c F_{uf} \right) \Gamma^{h}_{\ du} F_{he}\\
		= & - 4 \left(\partial_u F_{uz} \right) \Gamma^{h}_{\ uu} F_{hz} - 4 \gamma^{ij} \left(\partial_u F_{uj} \right) \Gamma^{h}_{\ iu} F_{hz} - 4 \left(\partial_z F_{uz} \right) \Gamma^{h}_{\ uu} F_{hu}\\
		& - 4 \gamma^{ij} \left(\partial_z F_{uj} \right) \Gamma^{h}_{\ iu} F_{hu} - 4 \gamma^{ij} \left(\partial_i F_{uz} \right) \Gamma^{h}_{\ uu} F_{hj} - 4 \gamma^{ij} \gamma^{kl} \left(\partial_i F_{ul} \right) \Gamma^{h}_{\ ku} F_{hj}\\
		= & - 4 \gamma^{ij} \left(\partial_u F_{uj} \right) \Gamma^{h}_{\ iu} F_{hz} - 4 \gamma^{ij} \left(\partial_z F_{uj} \right) \Gamma^{h}_{\ iu} F_{hu} - 4 \gamma^{ij} \gamma^{kl} \left(\partial_i F_{ul} \right) \Gamma^{h}_{\ ku} F_{hj}\,.
	\end{split}
\end{equation}
The repeated index $h$ should be further expanded. The first term of Eq. (\ref{hkk3secondsecond}) is
\begin{equation}
	\begin{split}
		& - 4 \gamma^{ij} \left(\partial_u F_{uj} \right) \Gamma^{h}_{\ iu} F_{hz}\\
		= & - 4 \gamma^{ij} \left(\partial_u F_{uj} \right) \Gamma^{u}_{\ iu} F_{uz} - 4 \gamma^{ij} \left(\partial_u F_{uj} \right) \Gamma^{z}_{\ iu} F_{zz} - 4 \gamma^{ij} \left(\partial_u F_{uj} \right) \Gamma^{k}_{\ iu} F_{kz}\\
		= & - 4 \gamma^{ij} \left(\partial_u F_{uj} \right) \Gamma^{u}_{\ iu} F_{uz} - 4 \gamma^{ij} \left(\partial_u F_{uj} \right) \Gamma^{k}_{\ iu} F_{kz}\\
		= & 2 \gamma^{ij} \beta_i \left(\partial_u F_{uj} \right) F_{uz} - 2 \gamma^{ij} \gamma^{kl} \left(\partial_u F_{uj} \right) \left(\partial_u \gamma_{il} \right) F_{kz}\,.
	\end{split}
\end{equation}
The second term of Eq. (\ref{hkk3secondsecond}) is
\begin{equation}
	\begin{split}
		& - 4 \gamma^{ij} \left(\partial_z F_{uj} \right) \Gamma^{h}_{\ iu} F_{hu}\\
		= & - 4 \gamma^{ij} \left(\partial_z F_{uj} \right) \Gamma^{u}_{\ iu} F_{uu} - 4 \gamma^{ij} \left(\partial_z F_{uj} \right) \Gamma^{z}_{\ iu} F_{zu} - 4 \gamma^{ij} \left(\partial_z F_{uj} \right) \Gamma^{k}_{\ iu} F_{ku}\\
		= & - 4 \gamma^{ij} \left(\partial_z F_{uj} \right) \Gamma^{z}_{\ iu} F_{zu} - 4 \gamma^{ij} \left(\partial_z F_{uj} \right) \Gamma^{k}_{\ iu} F_{ku}\\
		= &  - 2 \gamma^{ij} \gamma^{kl} \left(\partial_z F_{uj} \right) \left(\partial_u \gamma_{il} \right) F_{ku}\,.
	\end{split}
\end{equation}
The third term of Eq. (\ref{hkk3secondsecond}) is
\begin{equation}
	\begin{split}
		& - 4 \gamma^{ij} \gamma^{kl} \left(\partial_i F_{ul} \right) \Gamma^{h}_{\ ku} F_{hj}\\
		= & - 4 \gamma^{ij} \gamma^{kl} \left(\partial_i F_{ul} \right) \Gamma^{u}_{\ ku} F_{uj} - 4 \gamma^{ij} \gamma^{kl} \left(\partial_i F_{ul} \right) \Gamma^{z}_{\ ku} F_{zj} - 4 \gamma^{ij} \gamma^{kl} \left(\partial_i F_{ul} \right) \Gamma^{m}_{\ ku} F_{mj}\\
		= & 2 \gamma^{ij} \gamma^{kl} \beta_k \left(\partial_i F_{ul} \right) F_{uj} - 2 \gamma^{ij} \gamma^{kl} \gamma^{mn} \left(\partial_i F_{ul} \right) \left(\partial_u \gamma_{kn} \right) F_{mj}\,.
	\end{split}
\end{equation}
Therefore, the second term in Eq. (\ref{hkk3second}) is obtained as 
\begin{equation}
	\begin{split}
		& - 4 k^a k^b g^{ce} g^{df} \left(\partial_c F_{bf} \right) \Gamma^{h}_{\ da} F_{he}\\
		= & 2 \gamma^{ij} \beta_i \left(\partial_u F_{uj} \right) F_{uz} - 2 \gamma^{ij} \gamma^{kl} \left(\partial_u F_{uj} \right) \left(\partial_u \gamma_{il} \right) F_{kz}\\
		& - 2 \gamma^{ij} \gamma^{kl} \left(\partial_z F_{uj} \right) \left(\partial_u \gamma_{il} \right) F_{ku} + 2 \gamma^{ij} \gamma^{kl} \beta_k \left(\partial_i F_{ul} \right) F_{uj}\\
		& - 2 \gamma^{ij} \gamma^{kl} \gamma^{mn} \left(\partial_i F_{ul} \right) \left(\partial_u \gamma_{kn} \right) F_{mj}\\
		= & 2 \gamma^{ij} \beta_i \left(\partial_u F_{uj} \right) F_{uz} - 4 \gamma^{ij} \gamma^{kl} \left(\partial_u F_{uj} \right) K_{il} F_{kz}\\
		& - 4 \gamma^{ij} \gamma^{kl} \left(\partial_z F_{uj} \right) K_{il} F_{ku} + 2 \gamma^{ij} \gamma^{kl} \beta_k \left(\partial_i F_{ul} \right) F_{uj}\\
		& - 4 \gamma^{ij} \gamma^{kl} \gamma^{mn} \left(\partial_i F_{ul} \right) K_{kn} F_{mj}\\
		= & 2 \gamma^{ij} \beta_i \left(\partial_u F_{uj} \right) F_{uz}\,.
	\end{split}
\end{equation}

The third term in Eq. (\ref{hkk3second}) is 
\begin{equation}\label{hkk3secondthird}
	\begin{split}
		& - 4 k^a k^b g^{ce} g^{df} \left(\partial_c F_{bf} \right) \Gamma^{h}_{\ de} F_{ah} = - 4 g^{ce} g^{df} \left(\partial_c F_{uf} \right) \Gamma^{h}_{\ de} F_{uh}\\
		= & - 4 \left(\partial_u F_{uz} \right) \Gamma^{h}_{\ uz} F_{uh} - 4 \gamma^{ij} \left(\partial_u F_{uj} \right) \Gamma^{h}_{\ iz} F_{uh} - 4 \left(\partial_z F_{uz} \right) \Gamma^{h}_{\ uu} F_{uh}\\
		& - 4 \gamma^{ij} \left(\partial_z F_{uj} \right) \Gamma^{h}_{\ iu} F_{uh} - 4 \gamma^{ij} \left(\partial_i F_{uz} \right) \Gamma^{h}_{\ uj} F_{uh} - 4 \gamma^{ij} \gamma^{kl} \left(\partial_i F_{ul} \right) \Gamma^{h}_{\ kj} F_{uh}\\
		= & - 4 \left(\partial_u F_{uz} \right) \Gamma^{h}_{\ uz} F_{uh} - 4 \gamma^{ij} \left(\partial_u F_{uj} \right) \Gamma^{h}_{\ iz} F_{uh} - 4 \gamma^{ij} \left(\partial_z F_{uj} \right) \Gamma^{h}_{\ iu} F_{uh}\\
		& - 4 \gamma^{ij} \left(\partial_i F_{uz} \right) \Gamma^{h}_{\ uj} F_{uh} - 4 \gamma^{ij} \gamma^{kl} \left(\partial_i F_{ul} \right) \Gamma^{h}_{\ kj} F_{uh}\,.
	\end{split}
\end{equation}
The repeated index $h$ should be further expanded. The first term of Eq. (\ref{hkk3secondthird}) is 
\begin{equation}
	\begin{split}
		& - 4 \left(\partial_u F_{uz} \right) \Gamma^{h}_{\ uz} F_{uh}\\
		= & - 4 \left(\partial_u F_{uz} \right) \Gamma^{u}_{\ uz} F_{uu} - 4 \left(\partial_u F_{uz} \right) \Gamma^{z}_{\ uz} F_{uz} - 4 \left(\partial_u F_{uz} \right) \Gamma^{i}_{\ uz} F_{ui}\\
		= & - 4 \left(\partial_u F_{uz} \right) \Gamma^{z}_{\ uz} F_{uz} - 4 \left(\partial_u F_{uz} \right) \Gamma^{i}_{\ uz} F_{ui}\\
		= & - 2 \gamma^{ij} \beta_j \left(\partial_u F_{uz} \right) F_{ui}\,.
	\end{split}
\end{equation}
The second term of Eq. (\ref{hkk3secondthird}) is
\begin{equation}
	\begin{split}
		& - 4 \gamma^{ij} \left(\partial_u F_{uj} \right) \Gamma^{h}_{\ iz} F_{uh}\\
		= & - 4 \gamma^{ij} \left(\partial_u F_{uj} \right) \Gamma^{u}_{\ iz} F_{uu} - 4 \gamma^{ij} \left(\partial_u F_{uj} \right) \Gamma^{z}_{\ iz} F_{uz} - 4 \gamma^{ij} \left(\partial_u F_{uj} \right) \Gamma^{k}_{\ iz} F_{uk}\\
		= & - 4 \gamma^{ij} \left(\partial_u F_{uj} \right) \Gamma^{z}_{\ iz} F_{uz} - 4 \gamma^{ij} \left(\partial_u F_{uj} \right) \Gamma^{k}_{\ iz} F_{uk}\\
		= & - 2 \gamma^{ij} \beta_i \left(\partial_u F_{uj} \right) F_{uz} - 2 \gamma^{ij} \gamma^{kl} \left(\partial_u F_{uj} \right) \left(\partial_z \gamma_{il} \right) F_{uk}\,.
	\end{split}
\end{equation}
The third term of Eq. (\ref{hkk3secondthird}) is
\begin{equation}
	\begin{split}
		& - 4 \gamma^{ij} \left(\partial_z F_{uj} \right) \Gamma^{h}_{\ iu} F_{uh}\\
		= & - 4 \gamma^{ij} \left(\partial_z F_{uj} \right) \Gamma^{u}_{\ iu} F_{uu} - 4 \gamma^{ij} \left(\partial_z F_{uj} \right) \Gamma^{z}_{\ iu} F_{uz} - 4 \gamma^{ij} \left(\partial_z F_{uj} \right) \Gamma^{k}_{\ iu} F_{uk}\\
		= & - 4 \gamma^{ij} \left(\partial_z F_{uj} \right) \Gamma^{z}_{\ iu} F_{uz} - 4 \gamma^{ij} \left(\partial_z F_{uj} \right) \Gamma^{k}_{\ iu} F_{uk}\\
		= & - 2 \gamma^{ij} \gamma^{kl} \left(\partial_z F_{uj} \right) \left(\partial_u \gamma_{il} \right) F_{uk}\,.
	\end{split}
\end{equation}
The fourth term of Eq. (\ref{hkk3secondthird}) is
\begin{equation}
	\begin{split}
		& - 4 \gamma^{ij} \left(\partial_i F_{uz} \right) \Gamma^{h}_{\ uj} F_{uh}\\
		= & - 4 \gamma^{ij} \left(\partial_i F_{uz} \right) \Gamma^{u}_{\ uj} F_{uu} - 4 \gamma^{ij} \left(\partial_i F_{uz} \right) \Gamma^{z}_{\ uj} F_{uz} - 4 \gamma^{ij} \left(\partial_i F_{uz} \right) \Gamma^{k}_{\ uj} F_{uk}\\
		= & - 4 \gamma^{ij} \left(\partial_i F_{uz} \right) \Gamma^{z}_{\ uj} F_{uz} - 4 \gamma^{ij} \left(\partial_i F_{uz} \right) \Gamma^{k}_{\ uj} F_{uk}\\
		= & - 2 \gamma^{ij} \gamma^{kl} \left(\partial_i F_{uz} \right) \left(\partial_u \gamma_{jl} \right) F_{uk}\,.
	\end{split}
\end{equation}
The fifth term of Eq. (\ref{hkk3secondthird}) is
\begin{equation}
	\begin{split}
		& - 4 \gamma^{ij} \gamma^{kl} \left(\partial_i F_{ul} \right) \Gamma^{h}_{\ kj} F_{uh}\\
		= & - 4 \gamma^{ij} \gamma^{kl} \left(\partial_i F_{ul} \right) \Gamma^{u}_{\ kj} F_{uu} - 4 \gamma^{ij} \gamma^{kl} \left(\partial_i F_{ul} \right) \Gamma^{z}_{\ kj} F_{uz} - 4 \gamma^{ij} \gamma^{kl} \left(\partial_i F_{ul} \right) \Gamma^{m}_{\ kj} F_{um}\\
		= & - 4 \gamma^{ij} \gamma^{kl} \left(\partial_i F_{ul} \right) \Gamma^{z}_{\ kj} F_{uz} - 4 \gamma^{ij} \gamma^{kl} \left(\partial_i F_{ul} \right) \Gamma^{m}_{\ kj} F_{um}\\
		= & 2 \gamma^{ij} \gamma^{kl} \left(\partial_i F_{ul} \right) \left(\partial_u \gamma_{kj} \right) F_{uz} - 4 \gamma^{ij} \gamma^{kl} \left(\partial_i F_{ul} \right) \hat{\Gamma}^{m}_{\ kj} F_{um}\,.
	\end{split}
\end{equation}
Therefore, the third term in Eq. (\ref{hkk3second}) is obtained as 
\begin{equation}
	\begin{split}
		& - 4 k^a k^b g^{ce} g^{df} \left(\partial_c F_{bf} \right) \Gamma^{h}_{\ de} F_{ah}\\
		= & - 2 \gamma^{ij} \beta_j \left(\partial_u F_{uz} \right) F_{ui} - 2 \gamma^{ij} \beta_i \left(\partial_u F_{uj} \right) F_{uz}\\
		& - 2 \gamma^{ij} \gamma^{kl} \left(\partial_u F_{uj} \right) \left(\partial_z \gamma_{il} \right) F_{uk} - 2 \gamma^{ij} \gamma^{kl} \left(\partial_z F_{uj} \right) \left(\partial_u \gamma_{il} \right) F_{uk}\\
		& - 2 \gamma^{ij} \gamma^{kl} \left(\partial_i F_{uz} \right) \left(\partial_u \gamma_{jl} \right) F_{uk} + 2 \gamma^{ij} \gamma^{kl} \left(\partial_i F_{ul} \right) \left(\partial_u \gamma_{kj} \right) F_{uz}\\
		& - 4 \gamma^{ij} \gamma^{kl} \left(\partial_i F_{ul} \right) \hat{\Gamma}^{m}_{\ kj} F_{um}\\
		= & - 2 \gamma^{ij} \beta_j \left(\partial_u F_{uz} \right) F_{ui} - 2 \gamma^{ij} \beta_i \left(\partial_u F_{uj} \right) F_{uz}\\
		& - 2 \gamma^{ij} \gamma^{kl} \left(\partial_u F_{uj} \right) \left(\partial_z \gamma_{il} \right) F_{uk} - 4 \gamma^{ij} \gamma^{kl} \left(\partial_z F_{uj} \right) K_{il} F_{uk}\\
		& - 4 \gamma^{ij} \gamma^{kl} \left(\partial_i F_{uz} \right) K_{jl} F_{uk} + 4 \gamma^{ij} \gamma^{kl} \left(\partial_i F_{ul} \right) K_{kj} F_{uz}\\
		& - 4 \gamma^{ij} \gamma^{kl} \left(\partial_i F_{ul} \right) \hat{\Gamma}^{m}_{\ kj} F_{um}\\
		= & - 2 \gamma^{ij} \beta_i \left(\partial_u F_{uj} \right) F_{uz}\,.
	\end{split}
\end{equation}

The fourth term in Eq. (\ref{hkk3second}) is
\begin{equation}
	\begin{split}
		& - 4 k^a k^b g^{ce} g^{df} \Gamma^{g}_{\ cb} F_{gf} \partial_d F_{ae} = - 4 g^{ce} g^{df} \Gamma^{g}_{\ cu} F_{gf} \partial_d F_{ue}\\
		= & - 4 \Gamma^{g}_{\ uu} F_{gz} \partial_u F_{uz} - 4 \Gamma^{g}_{\ uu} F_{gu} \partial_z F_{uz} - 4 \gamma^{ij} \Gamma^{g}_{\ uu} F_{g j} \partial_i F_{uz}\\
		& - 4 \gamma^{i j} \Gamma^{g}_{\ iu} F_{gz} \partial_u F_{u j} - 4 \gamma^{i j} \Gamma^{g}_{\ iu} F_{gu} \partial_z F_{u j} - 4 \gamma^{i j} \gamma^{k l} \Gamma^{g}_{\ iu} F_{g l} \partial_k F_{u j}\\
		= & - 4 \gamma^{i j} \Gamma^{g}_{\ iu} F_{gz} \partial_u F_{u j} - 4 \gamma^{i j} \Gamma^{g}_{\ iu} F_{gu} \partial_z F_{u j}\,.
	\end{split}
\end{equation}
The index $g$ should be further expanded.
\begin{equation}\label{hkk3secondfourth}
	\begin{split}
		& - 4 \gamma^{i j} \Gamma^{g}_{\ iu} F_{gz} \partial_u F_{u j} - 4 \gamma^{i j} \Gamma^{g}_{\ iu} F_{gu} \partial_z F_{u j}\\
		= & - 4 \gamma^{i j} \Gamma^{u}_{\ iu} F_{uz} \partial_u F_{u j} - 4 \gamma^{i j} \Gamma^{z}_{\ iu} F_{zz} \partial_u F_{u j} - 4 \gamma^{i j} \Gamma^{k}_{\ iu} F_{kz} \partial_u F_{u j}\\
		& - 4 \gamma^{i j} \Gamma^{u}_{\ iu} F_{uu} \partial_z F_{u j} - 4 \gamma^{i j} \Gamma^{z}_{\ iu} F_{zu} \partial_z F_{u j} - 4 \gamma^{i j} \Gamma^{k}_{\ iu} F_{ku} \partial_z F_{u j}\\
		= & - 4 \gamma^{i j} \Gamma^{u}_{\ iu} F_{uz} \partial_u F_{u j} - 4 \gamma^{i j} \Gamma^{k}_{\ iu} F_{kz} \partial_u F_{u j} - 4 \gamma^{i j} \Gamma^{z}_{\ iu} F_{zu} \partial_z F_{u j}\\
		& - 4 \gamma^{i j} \Gamma^{k}_{\ iu} F_{ku} \partial_z F_{u j}\,.
	\end{split}
\end{equation}
Therefore, the fourth term in Eq. (\ref{hkk3second}) is obtained as 
\begin{equation}
	\begin{split}
		& - 4 k^a k^b g^{ce} g^{df} \Gamma^{g}_{\ cb} F_{gf} \partial_d F_{ae}\\
		= & 2 \gamma^{ij} \beta_i F_{uz} \partial_u F_{u j} - 2 \gamma^{i j} \gamma^{kl} \left(\partial_u \gamma_{il} \right) F_{kz} \partial_u F_{uj}\\
		& - 2 \gamma^{ij} \gamma^{kl} \left(\partial_u \gamma_{il} \right) F_{ku} \partial_z F_{uj}\\
		= & 2 \gamma^{ij} \beta_i F_{uz} \partial_u F_{u j} - 4 \gamma^{i j} \gamma^{kl} K_{il} F_{kz} \partial_u F_{uj}\\
		& - 4 \gamma^{ij} \gamma^{kl} K_{il} F_{ku} \partial_z F_{uj}\\
		= & 2 \gamma^{ij} \beta_i F_{uz} \partial_u F_{u j}\,.
	\end{split}
\end{equation}

The fifth term in Eq. (\ref{hkk3second}) is
\begin{equation}
	\begin{split}
		& 4 k^a k^b g^{ce} g^{df} \Gamma^{g}_{\ cb} F_{gf} \Gamma^{h}_{\ da} F_{he} = 4 g^{ce} g^{df} \Gamma^{g}_{\ cu} F_{gf} \Gamma^{h}_{\ du} F_{he}\\
		= & 4 \Gamma^{g}_{\ uu} F_{gz} \Gamma^{h}_{\ uu} F_{hz} + 4 \Gamma^{g}_{\ uu} F_{gu} \Gamma^{h}_{\ zu} F_{hz} + 4 \gamma^{ij} \Gamma^{g}_{\ uu} F_{gj} \Gamma^{h}_{\ iu} F_{hz}\\
		& + 4 \Gamma^{g}_{\ zu} F_{gz} \Gamma^{h}_{\ uu} F_{hu} + 4 \Gamma^{g}_{\ zu} F_{gu} \Gamma^{h}_{\ zu} F_{hu} + 4 \gamma^{ij} \Gamma^{g}_{\ zu} F_{gj} \Gamma^{h}_{\ iu} F_{hu}\\
		& + 4 \gamma^{ij} \Gamma^{g}_{\ iu} F_{gz} \Gamma^{h}_{\ uu} F_{hj} + 4 \gamma^{ij} \Gamma^{g}_{\ iu} F_{gu} \Gamma^{h}_{\ zu} F_{hj} + 4 \gamma^{ij} \gamma^{kl} \Gamma^{g}_{\ iu} F_{gl} \Gamma^{h}_{\ ku} F_{hj}\\
		= & 4 \Gamma^{g}_{\ zu} F_{gu} \Gamma^{h}_{\ zu} F_{hu} + 4 \gamma^{ij} \Gamma^{g}_{\ zu} F_{gj} \Gamma^{h}_{\ iu} F_{hu} + 4 \gamma^{ij} \Gamma^{g}_{\ iu} F_{gu} \Gamma^{h}_{\ zu} F_{hj}\\
		& + 4 \gamma^{ij} \gamma^{kl} \Gamma^{g}_{\ iu} F_{gl} \Gamma^{h}_{\ ku} F_{hj}\,.
	\end{split}
\end{equation}
The index $g$ should be further expanded.
\begin{equation}\label{hkk3secondfifth}
	\begin{split}
		& 4 \Gamma^{g}_{\ zu} F_{gu} \Gamma^{h}_{\ zu} F_{hu} + 4 \gamma^{ij} \Gamma^{g}_{\ zu} F_{gj} \Gamma^{h}_{\ iu} F_{hu} + 4 \gamma^{ij} \Gamma^{g}_{\ iu} F_{gu} \Gamma^{h}_{\ zu} F_{hj}\\
		& + 4 \gamma^{ij} \gamma^{kl} \Gamma^{g}_{\ iu} F_{gl} \Gamma^{h}_{\ ku} F_{hj}\\
		= & 4 \Gamma^{u}_{\ zu} F_{uu} \Gamma^{h}_{\ zu} F_{hu} + 4 \Gamma^{z}_{\ zu} F_{zu} \Gamma^{h}_{\ zu} F_{hu} + 4 \Gamma^{i}_{\ zu} F_{iu} \Gamma^{h}_{\ zu} F_{hu}\\
		& + 4 \gamma^{ij} \Gamma^{u}_{\ zu} F_{uj} \Gamma^{h}_{\ iu} F_{hu} + 4 \gamma^{ij} \Gamma^{z}_{\ zu} F_{zj} \Gamma^{h}_{\ iu} F_{hu} + 4 \gamma^{ij} \Gamma^{k}_{\ zu} F_{kj} \Gamma^{h}_{\ iu} F_{hu}\\
		& + 4 \gamma^{ij} \Gamma^{u}_{\ iu} F_{uu} \Gamma^{h}_{\ zu} F_{hj} + 4 \gamma^{ij} \Gamma^{z}_{\ iu} F_{zu} \Gamma^{h}_{\ zu} F_{hj} + 4 \gamma^{ij} \Gamma^{k}_{\ iu} F_{ku} \Gamma^{h}_{\ zu} F_{hj}\\
		& + 4 \gamma^{ij} \gamma^{kl} \Gamma^{u}_{\ iu} F_{ul} \Gamma^{h}_{\ ku} F_{hj} + 4 \gamma^{ij} \gamma^{kl} \Gamma^{z}_{\ iu} F_{zl} \Gamma^{h}_{\ ku} F_{hj} + 4 \gamma^{ij} \gamma^{kl} \Gamma^{m}_{\ iu} F_{ml} \Gamma^{h}_{\ ku} F_{hj}\\
		= & 4 \Gamma^{z}_{\ zu} F_{zu} \Gamma^{h}_{\ zu} F_{hu} + 4 \gamma^{ij} \Gamma^{z}_{\ zu} F_{zj} \Gamma^{h}_{\ iu} F_{hu} + 4 \gamma^{ij} \Gamma^{k}_{\ zu} F_{kj} \Gamma^{h}_{\ iu} F_{hu}\\
		& + 4 \gamma^{ij} \Gamma^{z}_{\ iu} F_{zu} \Gamma^{h}_{\ zu} F_{hj} + 4 \gamma^{ij} \gamma^{kl} \Gamma^{z}_{\ iu} F_{zl} \Gamma^{h}_{\ ku} F_{hj} + 4 \gamma^{ij} \gamma^{kl} \Gamma^{m}_{\ iu} F_{ml} \Gamma^{h}_{\ ku} F_{hj}\,.
	\end{split}
\end{equation}
The repeated index $h$ should be further expanded. The first term of Eq. (\ref{hkk3secondfifth}) is
\begin{equation}
	\begin{split}
		4 \Gamma^{z}_{\ zu} F_{zu} \Gamma^{h}_{\ zu} F_{hu} = 0\,.
	\end{split}
\end{equation}
The second term of Eq. (\ref{hkk3secondfifth}) is
\begin{equation}
	\begin{split}
		4 \gamma^{ij} \Gamma^{z}_{\ zu} F_{zj} \Gamma^{h}_{\ iu} F_{hu} = 0\,.
	\end{split}
\end{equation}
The third term of Eq. (\ref{hkk3secondfifth}) is
\begin{equation}
	\begin{split}
		& 4 \gamma^{ij} \Gamma^{k}_{\ zu} F_{kj} \Gamma^{h}_{\ iu} F_{hu}\\
		= & 4 \gamma^{ij} \Gamma^{k}_{\ zu} F_{kj} \Gamma^{u}_{\ iu} F_{uu} + 4 \gamma^{ij} \Gamma^{k}_{\ zu} F_{kj} \Gamma^{z}_{\ iu} F_{zu} + 4 \gamma^{ij} \Gamma^{k}_{\ zu} F_{kj} \Gamma^{l}_{\ iu} F_{lu}\\
		= & 4 \gamma^{ij} \Gamma^{k}_{\ zu} F_{kj} \Gamma^{z}_{\ iu} F_{zu} + 4 \gamma^{ij} \Gamma^{k}_{\ zu} F_{kj} \Gamma^{l}_{\ iu} F_{lu}\\
		= & \gamma^{ij} \gamma^{km} \gamma^{ln} \beta_m F_{kj} \left(\partial_u \gamma_{in} \right) F_{lu}\,.
	\end{split}
\end{equation}
The fourth term of Eq. (\ref{hkk3secondfifth}) is
\begin{equation}
	\begin{split}
		4 \gamma^{ij} \Gamma^{z}_{\ iu} F_{zu} \Gamma^{h}_{\ zu} F_{hj} = 0\,.
	\end{split}
\end{equation}
The fifth term of Eq. (\ref{hkk3secondfifth}) is
\begin{equation}
	\begin{split}
		4 \gamma^{ij} \gamma^{kl} \Gamma^{z}_{\ iu} F_{zl} \Gamma^{h}_{\ ku} F_{hj} = 0\,.
	\end{split}
\end{equation}
The sixth term of Eq. (\ref{hkk3secondfifth}) is
\begin{equation}
	\begin{split}
		& 4 \gamma^{ij} \gamma^{kl} \Gamma^{m}_{\ iu} F_{ml} \Gamma^{h}_{\ ku} F_{hj}\\
		= & 4 \gamma^{ij} \gamma^{kl} \Gamma^{m}_{\ iu} F_{ml} \Gamma^{u}_{\ ku} F_{uj} + 4 \gamma^{ij} \gamma^{kl} \Gamma^{m}_{\ iu} F_{ml} \Gamma^{z}_{\ ku} F_{zj} + 4 \gamma^{ij} \gamma^{kl} \Gamma^{m}_{\ iu} F_{ml} \Gamma^{n}_{\ ku} F_{nj}\\
		= & - \gamma^{ij} \gamma^{kl} \gamma^{mn} \beta_k \left(\partial_u \gamma_{in} \right) F_{ml} F_{uj} + \gamma^{ij} \gamma^{kl} \gamma^{mo} \gamma^{np} \left(\partial_u \gamma_{io} \right) F_{ml} \left(\partial_u \gamma_{kp} \right) F_{nj}\,.
	\end{split}
\end{equation}
Therefore, the fifth term in Eq. (\ref{hkk3second}) is obtained as 
\begin{equation}
	\begin{split}
		& 4 k^a k^b g^{ce} g^{df} \Gamma^{g}_{\ cb} F_{gf} \Gamma^{h}_{\ da} F_{he}\\
		= & \gamma^{ij} \gamma^{km} \gamma^{ln} \beta_m F_{kj} \left(\partial_u \gamma_{in} \right) F_{lu} - \gamma^{ij} \gamma^{kl} \gamma^{mn} \beta_k \left(\partial_u \gamma_{in} \right) F_{ml} F_{uj}\\
		& + \gamma^{ij} \gamma^{kl} \gamma^{mo} \gamma^{np} \left(\partial_u \gamma_{io} \right) F_{ml} \left(\partial_u \gamma_{kp} \right) F_{nj}\\
		= & 2 \gamma^{ij} \gamma^{km} \gamma^{ln} \beta_m F_{kj} K_{in} F_{lu} - 2 \gamma^{ij} \gamma^{kl} \gamma^{mn} \beta_k K_{in} F_{ml} F_{uj}\\
		& + 4 \gamma^{ij} \gamma^{kl} \gamma^{mo} \gamma^{np} K_{io} F_{ml} K_{kp} F_{nj}\\
		= & 0\,.
	\end{split}
\end{equation}

The sixth term in Eq. (\ref{hkk3second}) is
\begin{equation}
	\begin{split}
		& 4 k^a k^b g^{ce} g^{df} \Gamma^{g}_{\ cb} F_{gf} \Gamma^{h}_{\ de} F_{ah} = 4 g^{ce} g^{df} \Gamma^{g}_{\ cu} F_{gf} \Gamma^{h}_{\ de} F_{uh}\\
		= & 4 \Gamma^{g}_{\ uu} F_{gz} \Gamma^{h}_{\ uz} F_{uh} + 4 \Gamma^{g}_{\ uu} F_{gu} \Gamma^{h}_{\ zz} F_{uh} + 4 \gamma^{ij} \Gamma^{g}_{\ uu} F_{gj} \Gamma^{h}_{\ iz} F_{uh}\\
		& + 4 \Gamma^{g}_{\ zu} F_{gz} \Gamma^{h}_{\ uu} F_{uh} + 4 \Gamma^{g}_{\ zu} F_{gu} \Gamma^{h}_{\ zu} F_{uh} + 4 \gamma^{ij} \Gamma^{g}_{\ zu} F_{gj} \Gamma^{h}_{\ iu} F_{uh}\\
		& + 4 \gamma^{ij} \Gamma^{g}_{\ iu} F_{gz} \Gamma^{h}_{\ uj} F_{uh} + 4 \gamma^{ij} \Gamma^{g}_{\ iu} F_{gu} \Gamma^{h}_{\ zj} F_{uh} + 4 \gamma^{ij} \gamma^{kl} \Gamma^{g}_{\ iu} F_{gl} \Gamma^{h}_{\ kj} F_{uh}\\
		= & 4 \Gamma^{g}_{\ zu} F_{gu} \Gamma^{h}_{\ zu} F_{uh} + 4 \gamma^{ij} \Gamma^{g}_{\ zu} F_{gj} \Gamma^{h}_{\ iu} F_{uh} + 4 \gamma^{ij} \Gamma^{g}_{\ iu} F_{gz} \Gamma^{h}_{\ uj} F_{uh}\\
		& + 4 \gamma^{ij} \Gamma^{g}_{\ iu} F_{gu} \Gamma^{h}_{\ zj} F_{uh} + 4 \gamma^{ij} \gamma^{kl} \Gamma^{g}_{\ iu} F_{gl} \Gamma^{h}_{\ kj} F_{uh}\,.
	\end{split}
\end{equation}
The index $g$ should be further expanded.
\begin{equation}\label{hkk3secondsixth}
	\begin{split}
		& 4 \Gamma^{g}_{\ zu} F_{gu} \Gamma^{h}_{\ zu} F_{uh} + 4 \gamma^{ij} \Gamma^{g}_{\ zu} F_{gj} \Gamma^{h}_{\ iu} F_{uh} + 4 \gamma^{ij} \Gamma^{g}_{\ iu} F_{gz} \Gamma^{h}_{\ uj} F_{uh}\\
		& + 4 \gamma^{ij} \Gamma^{g}_{\ iu} F_{gu} \Gamma^{h}_{\ zj} F_{uh} + 4 \gamma^{ij} \gamma^{kl} \Gamma^{g}_{\ iu} F_{gl} \Gamma^{h}_{\ kj} F_{uh}\\
		= & 4 \Gamma^{u}_{\ zu} F_{uu} \Gamma^{h}_{\ zu} F_{uh} + 4 \Gamma^{z}_{\ zu} F_{zu} \Gamma^{h}_{\ zu} F_{uh} + 4 \Gamma^{i}_{\ zu} F_{iu} \Gamma^{h}_{\ zu} F_{uh}\\
		& + 4 \gamma^{ij} \Gamma^{u}_{\ zu} F_{uj} \Gamma^{h}_{\ iu} F_{uh} + 4 \gamma^{ij} \Gamma^{z}_{\ zu} F_{zj} \Gamma^{h}_{\ iu} F_{uh} + 4 \gamma^{ij} \Gamma^{k}_{\ zu} F_{kj} \Gamma^{h}_{\ iu} F_{uh}\\
		& + 4 \gamma^{ij} \Gamma^{u}_{\ iu} F_{uz} \Gamma^{h}_{\ uj} F_{uh} + 4 \gamma^{ij} \Gamma^{z}_{\ iu} F_{zz} \Gamma^{h}_{\ uj} F_{uh} + 4 \gamma^{ij} \Gamma^{k}_{\ iu} F_{kz} \Gamma^{h}_{\ uj} F_{uh}\\
		& + 4 \gamma^{ij} \Gamma^{u}_{\ iu} F_{uu} \Gamma^{h}_{\ zj} F_{uh} + 4 \gamma^{ij} \Gamma^{z}_{\ iu} F_{zu} \Gamma^{h}_{\ zj} F_{uh} + 4 \gamma^{ij} \Gamma^{k}_{\ iu} F_{ku} \Gamma^{h}_{\ zj} F_{uh}\\
		& + 4 \gamma^{ij} \gamma^{kl} \Gamma^{u}_{\ iu} F_{ul} \Gamma^{h}_{\ kj} F_{uh} + 4 \gamma^{ij} \gamma^{kl} \Gamma^{z}_{\ iu} F_{zl} \Gamma^{h}_{\ kj} F_{uh} + 4 \gamma^{ij} \gamma^{kl} \Gamma^{m}_{\ iu} F_{ml} \Gamma^{h}_{\ kj} F_{uh}\\
		= & 4 \Gamma^{z}_{\ zu} F_{zu} \Gamma^{h}_{\ zu} F_{uh} + 4 \gamma^{ij} \Gamma^{z}_{\ zu} F_{zj} \Gamma^{h}_{\ iu} F_{uh} + 4 \gamma^{ij} \Gamma^{k}_{\ zu} F_{kj} \Gamma^{h}_{\ iu} F_{uh}\\
		& + 4 \gamma^{ij} \Gamma^{u}_{\ iu} F_{uz} \Gamma^{h}_{\ uj} F_{uh} + 4 \gamma^{ij} \Gamma^{k}_{\ iu} F_{kz} \Gamma^{h}_{\ uj} F_{uh} + 4 \gamma^{ij} \Gamma^{z}_{\ iu} F_{zu} \Gamma^{h}_{\ zj} F_{uh}\\
		& + 4 \gamma^{ij} \gamma^{kl} \Gamma^{z}_{\ iu} F_{zl} \Gamma^{h}_{\ kj} F_{uh} + 4 \gamma^{ij} \gamma^{kl} \Gamma^{m}_{\ iu} F_{ml} \Gamma^{h}_{\ kj} F_{uh}\,.
	\end{split}
\end{equation}
The repeated index $h$ should be further expanded. The first term of Eq. (\ref{hkk3secondsixth}) is
\begin{equation}
	\begin{split}
		4 \Gamma^{z}_{\ zu} F_{zu} \Gamma^{h}_{\ zu} F_{uh} = 0\,.
	\end{split}
\end{equation}
The second term of Eq. (\ref{hkk3secondsixth}) is
\begin{equation}
	\begin{split}
		4 \gamma^{ij} \Gamma^{z}_{\ zu} F_{zj} \Gamma^{h}_{\ iu} F_{uh} = 0\,.
	\end{split}
\end{equation}
The third term of Eq. (\ref{hkk3secondsixth}) is
\begin{equation}
	\begin{split}
		& 4 \gamma^{ij} \Gamma^{k}_{\ zu} F_{kj} \Gamma^{h}_{\ iu} F_{uh}\\
		= & 4 \gamma^{ij} \Gamma^{k}_{\ zu} F_{kj} \Gamma^{u}_{\ iu} F_{uu} + 4 \gamma^{ij} \Gamma^{k}_{\ zu} F_{kj} \Gamma^{z}_{\ iu} F_{uz} + 4 \gamma^{ij} \Gamma^{k}_{\ zu} F_{kj} \Gamma^{l}_{\ iu} F_{ul}\\
		= & 4 \gamma^{ij} \Gamma^{k}_{\ zu} F_{kj} \Gamma^{z}_{\ iu} F_{uz} + 4 \gamma^{ij} \Gamma^{k}_{\ zu} F_{kj} \Gamma^{l}_{\ iu} F_{ul}\\
		= & \gamma^{ij} \gamma^{km} \gamma^{ln} \beta_m F_{kj} \left(\partial_u \gamma_{in} \right) F_{ul}\,.
	\end{split}
\end{equation}
The fourth term of Eq. (\ref{hkk3secondsixth}) is
\begin{equation}
	\begin{split}
		& 4 \gamma^{ij} \Gamma^{u}_{\ iu} F_{uz} \Gamma^{h}_{\ uj} F_{uh}\\
		= & 4 \gamma^{ij} \Gamma^{u}_{\ iu} F_{uz} \Gamma^{u}_{\ uj} F_{uu} + 4 \gamma^{ij} \Gamma^{u}_{\ iu} F_{uz} \Gamma^{z}_{\ uj} F_{uz} + 4 \gamma^{ij} \Gamma^{u}_{\ iu} F_{uz} \Gamma^{k}_{\ uj} F_{uk}\\
		= & 4 \gamma^{ij} \Gamma^{u}_{\ iu} F_{uz} \Gamma^{z}_{\ uj} F_{uz} + 4 \gamma^{ij} \Gamma^{u}_{\ iu} F_{uz} \Gamma^{k}_{\ uj} F_{uk}\\
		= & - \gamma^{ij} \gamma^{kl} \beta_i F_{uz} \left(\partial_u \gamma_{jl} \right) F_{uk}\,.
	\end{split}
\end{equation}
The fifth term of Eq. (\ref{hkk3secondsixth}) is
\begin{equation}
	\begin{split}
		& 4 \gamma^{ij} \Gamma^{k}_{\ iu} F_{kz} \Gamma^{h}_{\ uj} F_{uh}\\
		= & 4 \gamma^{ij} \Gamma^{k}_{\ iu} F_{kz} \Gamma^{u}_{\ uj} F_{uu} + 4 \gamma^{ij} \Gamma^{k}_{\ iu} F_{kz} \Gamma^{z}_{\ uj} F_{uz} + 4 \gamma^{ij} \Gamma^{k}_{\ iu} F_{kz} \Gamma^{l}_{\ uj} F_{ul}\\
		= & 4 \gamma^{ij} \Gamma^{k}_{\ iu} F_{kz} \Gamma^{z}_{\ uj} F_{uz} + 4 \gamma^{ij} \Gamma^{k}_{\ iu} F_{kz} \Gamma^{l}_{\ uj} F_{ul}\\
		= & \gamma^{ij} \gamma^{km} \gamma^{ln} \left(\partial_u \gamma_{im} \right) F_{kz} \left(\partial_u \gamma_{jn} \right) F_{ul}\,.
	\end{split}
\end{equation}
The sixth term of Eq. (\ref{hkk3secondsixth}) is
\begin{equation}
	\begin{split}
		4 \gamma^{ij} \Gamma^{z}_{\ iu} F_{zu} \Gamma^{h}_{\ zj} F_{uh} = 0\,.
	\end{split}
\end{equation}
The seventh term of Eq. (\ref{hkk3secondsixth}) is
\begin{equation}
	\begin{split}
		4 \gamma^{ij} \gamma^{kl} \Gamma^{z}_{\ iu} F_{zl} \Gamma^{h}_{\ kj} F_{uh} = 0\,.
	\end{split}
\end{equation}
The eighth term of Eq. (\ref{hkk3secondsixth}) is
\begin{equation}
	\begin{split}
		& 4 \gamma^{ij} \gamma^{kl} \Gamma^{m}_{\ iu} F_{ml} \Gamma^{h}_{\ kj} F_{uh}\\
		= & 4 \gamma^{ij} \gamma^{kl} \Gamma^{m}_{\ iu} F_{ml} \Gamma^{u}_{\ kj} F_{uu} + 4 \gamma^{ij} \gamma^{kl} \Gamma^{m}_{\ iu} F_{ml} \Gamma^{z}_{\ kj} F_{uz} + 4 \gamma^{ij} \gamma^{kl} \Gamma^{m}_{\ iu} F_{ml} \Gamma^{n}_{\ kj} F_{un}\\
		= & 4 \gamma^{ij} \gamma^{kl} \Gamma^{m}_{\ iu} F_{ml} \Gamma^{z}_{\ kj} F_{uz} + 4 \gamma^{ij} \gamma^{kl} \Gamma^{m}_{\ iu} F_{ml} \Gamma^{n}_{\ kj} F_{un}\\
		= & - \gamma^{ij} \gamma^{kl} \gamma^{mn} \left(\partial_u \gamma_{in} \right) F_{ml} \left(\partial_u \gamma_{kj} \right) F_{uz} + 2 \gamma^{ij} \gamma^{kl} \gamma^{mo} \left(\partial_u \gamma_{io} \right) F_{ml} \hat{\Gamma}^{n}_{\ kj} F_{un}\,.
	\end{split}
\end{equation}
Therefore, the sixth term in Eq. (\ref{hkk3second}) is obtained as 
\begin{equation}
	\begin{split}
		& 4 k^a k^b g^{ce} g^{df} \Gamma^{g}_{\ cb} F_{gf} \Gamma^{h}_{\ de} F_{ah}\\
		= & \gamma^{ij} \gamma^{km} \gamma^{ln} \beta_m F_{kj} \left(\partial_u \gamma_{in} \right) F_{ul} - \gamma^{ij} \gamma^{kl} \beta_i F_{uz} \left(\partial_u \gamma_{jl} \right) F_{uk}\\
		& + \gamma^{ij} \gamma^{km} \gamma^{ln} \left(\partial_u \gamma_{im} \right) F_{kz} \left(\partial_u \gamma_{jn} \right) F_{ul} - \gamma^{ij} \gamma^{kl} \gamma^{mn} \left(\partial_u \gamma_{in} \right) F_{ml} \left(\partial_u \gamma_{kj} \right) F_{uz}\\
		& + 2 \gamma^{ij} \gamma^{kl} \gamma^{mo} \left(\partial_u \gamma_{io} \right) F_{ml} \hat{\Gamma}^{n}_{\ kj} F_{un}\\
		= & 2 \gamma^{ij} \gamma^{km} \gamma^{ln} \beta_m F_{kj} K_{in} F_{ul} - 2 \gamma^{ij} \gamma^{kl} \beta_i F_{uz} K_{jl} F_{uk}\\
		& + 4 \gamma^{ij} \gamma^{km} \gamma^{ln} K_{im} F_{kz} K_{jn} F_{ul} - 4 \gamma^{ij} \gamma^{kl} \gamma^{mn} K_{in} F_{ml} K_{kj} F_{uz}\\
		& + 4 \gamma^{ij} \gamma^{kl} \gamma^{mo} K_{io} F_{ml} \hat{\Gamma}^{n}_{\ kj} F_{un}\\
		= & 0\,.
	\end{split}
\end{equation}

The seventh term in Eq. (\ref{hkk3second}) is
\begin{equation}
	\begin{split}
		& - 4 k^a k^b g^{ce} g^{df} \Gamma^{g}_{\ cf} F_{bg} \partial_d F_{ae} = - 4 g^{ce} g^{df} \Gamma^{g}_{\ cf} F_{ug} \partial_d F_{ue}\\
		= & - 4 \Gamma^{g}_{\ uz} F_{ug} \partial_u F_{uz} - 4 \Gamma^{g}_{\ uu} F_{ug} \partial_z F_{uz} - 4 \gamma^{ij} \Gamma^{g}_{\ uj} F_{ug} \partial_i F_{uz}\\
		& - 4 \gamma^{ij} \Gamma^{g}_{\ iz} F_{ug} \partial_u F_{uj} - 4 \gamma^{ij} \Gamma^{g}_{\ iu} F_{ug} \partial_z F_{uj} - 4 \gamma^{ij} \gamma^{kl} \Gamma^{g}_{\ il} F_{ug} \partial_k F_{uj}\\
		= & - 4 \Gamma^{g}_{\ uz} F_{ug} \partial_u F_{uz} - 4 \gamma^{ij} \Gamma^{g}_{\ uj} F_{ug} \partial_i F_{uz} - 4 \gamma^{ij} \Gamma^{g}_{\ iz} F_{ug} \partial_u F_{uj}\\
		& - 4 \gamma^{ij} \Gamma^{g}_{\ iu} F_{ug} \partial_z F_{uj} - 4 \gamma^{ij} \gamma^{kl} \Gamma^{g}_{\ il} F_{ug} \partial_k F_{uj}\,.
	\end{split}
\end{equation}
The index $g$ should be further expanded.
\begin{equation}\label{hkk3secondseventh}
	\begin{split}
		& - 4 \Gamma^{g}_{\ uz} F_{ug} \partial_u F_{uz} - 4 \gamma^{ij} \Gamma^{g}_{\ uj} F_{ug} \partial_i F_{uz} - 4 \gamma^{ij} \Gamma^{g}_{\ iz} F_{ug} \partial_u F_{uj}\\
		& - 4 \gamma^{ij} \Gamma^{g}_{\ iu} F_{ug} \partial_z F_{uj} - 4 \gamma^{ij} \gamma^{kl} \Gamma^{g}_{\ il} F_{ug} \partial_k F_{uj}\\
		= & - 4 \Gamma^{u}_{\ uz} F_{uu} \partial_u F_{uz} - 4 \Gamma^{z}_{\ uz} F_{uz} \partial_u F_{uz} - 4 \Gamma^{i}_{\ uz} F_{ui} \partial_u F_{uz}\\
		& - 4 \gamma^{ij} \Gamma^{u}_{\ uj} F_{uu} \partial_i F_{uz} - 4 \gamma^{ij} \Gamma^{z}_{\ uj} F_{uz} \partial_i F_{uz} - 4 \gamma^{ij} \Gamma^{k}_{\ uj} F_{uk} \partial_i F_{uz}\\
		& - 4 \gamma^{ij} \Gamma^{u}_{\ iz} F_{uu} \partial_u F_{uj} - 4 \gamma^{ij} \Gamma^{z}_{\ iz} F_{uz} \partial_u F_{uj} - 4 \gamma^{ij} \Gamma^{k}_{\ iz} F_{uk} \partial_u F_{uj}\\
		& - 4 \gamma^{ij} \Gamma^{u}_{\ iu} F_{uu} \partial_z F_{uj} - 4 \gamma^{ij} \Gamma^{z}_{\ iu} F_{uz} \partial_z F_{uj} - 4 \gamma^{ij} \Gamma^{k}_{\ iu} F_{uk} \partial_z F_{uj}\\
		& - 4 \gamma^{ij} \gamma^{kl} \Gamma^{u}_{\ il} F_{uu} \partial_k F_{uj} - 4 \gamma^{ij} \gamma^{kl} \Gamma^{z}_{\ il} F_{uz} \partial_k F_{uj} - 4 \gamma^{ij} \gamma^{kl} \Gamma^{m}_{\ il} F_{um} \partial_k F_{uj}\\
		= & - 4 \Gamma^{z}_{\ uz} F_{uz} \partial_u F_{uz} - 4 \gamma^{ij} \Gamma^{z}_{\ uj} F_{uz} \partial_i F_{uz} - 4 \gamma^{ij} \Gamma^{z}_{\ iz} F_{uz} \partial_u F_{uj}\\
		& - 4 \gamma^{ij} \Gamma^{z}_{\ iu} F_{uz} \partial_z F_{uj} - 4 \gamma^{ij} \gamma^{kl} \Gamma^{z}_{\ il} F_{uz} \partial_k F_{uj}\,.
	\end{split}
\end{equation}
Therefore, the seventh term in Eq. (\ref{hkk3second}) is obtained as 
\begin{equation}
	\begin{split}
		& - 4 k^a k^b g^{ce} g^{df} \Gamma^{g}_{\ cf} F_{bg} \partial_d F_{ae}\\
		= & - 2 \gamma^{ij} \beta_i F_{uz} \partial_u F_{uj} + 2 \gamma^{ij} \gamma^{kl} \left(\partial_u \gamma_{il} \right) F_{uz} \partial_k F_{uj}\\
		= & - 2 \gamma^{ij} \beta_i F_{uz} \partial_u F_{uj} + 4 \gamma^{ij} \gamma^{kl} K_{il} F_{uz} \partial_k F_{uj}\\
		= & - 2 \gamma^{ij} \beta_i F_{uz} \partial_u F_{uj}\,.
	\end{split}
\end{equation}

The eighth term of Eq. (\ref{hkk3second}) is
\begin{equation}
	\begin{split}
		& 4 k^a k^b g^{ce} g^{df} \Gamma^{g}_{\ cf} F_{bg} \Gamma^{h}_{\ da} F_{he} = 4 g^{ce} g^{df} \Gamma^{g}_{\ cf} F_{ug} \Gamma^{h}_{\ du} F_{he}\\
		= & 4 \Gamma^{g}_{\ uz} F_{ug} \Gamma^{h}_{\ uu} F_{hz} + 4 \Gamma^{g}_{\ uu} F_{ug} \Gamma^{h}_{\ zu} F_{hz} + 4 \gamma^{ij} \Gamma^{g}_{\ uj} F_{ug} \Gamma^{h}_{\ iu} F_{hz}\\
		& + 4 \Gamma^{g}_{\ zz} F_{ug} \Gamma^{h}_{\ uu} F_{hu} + 4 \Gamma^{g}_{\ zu} F_{ug} \Gamma^{h}_{\ zu} F_{hu} + 4 \gamma^{ij} \Gamma^{g}_{\ zj} F_{ug} \Gamma^{h}_{\ iu} F_{hu}\\
		& + 4 \gamma^{ij} \Gamma^{g}_{\ iz} F_{ug} \Gamma^{h}_{\ uu} F_{hj} + 4 \gamma^{ij} \Gamma^{g}_{\ iu} F_{ug} \Gamma^{h}_{\ zu} F_{hj} + 4 \gamma^{ij} \gamma^{kl} \Gamma^{g}_{\ il} F_{ug} \Gamma^{h}_{\ ku} F_{hj}\\
		= & 4 \gamma^{ij} \Gamma^{g}_{\ uj} F_{ug} \Gamma^{h}_{\ iu} F_{hz} + 4 \Gamma^{g}_{\ zu} F_{ug} \Gamma^{h}_{\ zu} F_{hu} + 4 \gamma^{ij} \Gamma^{g}_{\ zj} F_{ug} \Gamma^{h}_{\ iu} F_{hu}\\
		& + 4 \gamma^{ij} \Gamma^{g}_{\ iu} F_{ug} \Gamma^{h}_{\ zu} F_{hj} + 4 \gamma^{ij} \gamma^{kl} \Gamma^{g}_{\ il} F_{ug} \Gamma^{h}_{\ ku} F_{hj}\,.
	\end{split}
\end{equation}
The index $g$ should be further expanded.
\begin{equation}\label{hkk3secondeighth}
	\begin{split}
		& 4 \gamma^{ij} \Gamma^{g}_{\ uj} F_{ug} \Gamma^{h}_{\ iu} F_{hz} + 4 \Gamma^{g}_{\ zu} F_{ug} \Gamma^{h}_{\ zu} F_{hu} + 4 \gamma^{ij} \Gamma^{g}_{\ zj} F_{ug} \Gamma^{h}_{\ iu} F_{hu}\\
		& + 4 \gamma^{ij} \Gamma^{g}_{\ iu} F_{ug} \Gamma^{h}_{\ zu} F_{hj} + 4 \gamma^{ij} \gamma^{kl} \Gamma^{g}_{\ il} F_{ug} \Gamma^{h}_{\ ku} F_{hj}\\
		= & 4 \gamma^{ij} \Gamma^{u}_{\ uj} F_{uu} \Gamma^{h}_{\ iu} F_{hz} + 4 \gamma^{ij} \Gamma^{z}_{\ uj} F_{uz} \Gamma^{h}_{\ iu} F_{hz} + 4 \gamma^{ij} \Gamma^{k}_{\ uj} F_{uk} \Gamma^{h}_{\ iu} F_{hz}\\
		& + 4 \Gamma^{u}_{\ zu} F_{uu} \Gamma^{h}_{\ zu} F_{hu} + 4 \Gamma^{z}_{\ zu} F_{uz} \Gamma^{h}_{\ zu} F_{hu} + 4 \Gamma^{i}_{\ zu} F_{ui} \Gamma^{h}_{\ zu} F_{hu}\\
		& + 4 \gamma^{ij} \Gamma^{u}_{\ zj} F_{uu} \Gamma^{h}_{\ iu} F_{hu} + 4 \gamma^{ij} \Gamma^{z}_{\ zj} F_{uz} \Gamma^{h}_{\ iu} F_{hu} + 4 \gamma^{ij} \Gamma^{k}_{\ zj} F_{uk} \Gamma^{h}_{\ iu} F_{hu}\\
		& + 4 \gamma^{ij} \Gamma^{u}_{\ iu} F_{uu} \Gamma^{h}_{\ zu} F_{hj} + 4 \gamma^{ij} \Gamma^{z}_{\ iu} F_{uz} \Gamma^{h}_{\ zu} F_{hj} + 4 \gamma^{ij} \Gamma^{k}_{\ iu} F_{uk} \Gamma^{h}_{\ zu} F_{hj}\\
		& + 4 \gamma^{ij} \gamma^{kl} \Gamma^{u}_{\ il} F_{uu} \Gamma^{h}_{\ ku} F_{hj} + 4 \gamma^{ij} \gamma^{kl} \Gamma^{z}_{\ il} F_{uz} \Gamma^{h}_{\ ku} F_{hj} + 4 \gamma^{ij} \gamma^{kl} \Gamma^{m}_{\ il} F_{um} \Gamma^{h}_{\ ku} F_{hj}\\
		= & 4 \gamma^{ij} \Gamma^{z}_{\ uj} F_{uz} \Gamma^{h}_{\ iu} F_{hz} + 4 \Gamma^{z}_{\ zu} F_{uz} \Gamma^{h}_{\ zu} F_{hu} + 4 \gamma^{ij} \Gamma^{z}_{\ zj} F_{uz} \Gamma^{h}_{\ iu} F_{hu}\\
		& + 4 \gamma^{ij} \Gamma^{z}_{\ iu} F_{uz} \Gamma^{h}_{\ zu} F_{hj} + 4 \gamma^{ij} \gamma^{kl} \Gamma^{z}_{\ il} F_{uz} \Gamma^{h}_{\ ku} F_{hj}\,.
	\end{split}
\end{equation}
The repeated index $h$ should be further expanded. The first term of Eq. (\ref{hkk3secondeighth}) is
\begin{equation}
	\begin{split}
		4 \gamma^{ij} \Gamma^{z}_{\ uj} F_{uz} \Gamma^{h}_{\ iu} F_{hz} = 0\,.
	\end{split}
\end{equation}
The second term of Eq. (\ref{hkk3secondeighth}) is
\begin{equation}
	\begin{split}
		4 \Gamma^{z}_{\ zu} F_{uz} \Gamma^{h}_{\ zu} F_{hu} = 0\,.
	\end{split}
\end{equation}
The third term of Eq. (\ref{hkk3secondeighth}) is
\begin{equation}
	\begin{split}
		& 4 \gamma^{ij} \Gamma^{z}_{\ zj} F_{uz} \Gamma^{h}_{\ iu} F_{hu}\\
		= & 4 \gamma^{ij} \Gamma^{z}_{\ zj} F_{uz} \Gamma^{u}_{\ iu} F_{uu} + 4 \gamma^{ij} \Gamma^{z}_{\ zj} F_{uz} \Gamma^{z}_{\ iu} F_{zu} + 4 \gamma^{ij} \Gamma^{z}_{\ zj} F_{uz} \Gamma^{k}_{\ iu} F_{ku}\\
		= & 4 \gamma^{ij} \Gamma^{z}_{\ zj} F_{uz} \Gamma^{z}_{\ iu} F_{zu} + 4 \gamma^{ij} \Gamma^{z}_{\ zj} F_{uz} \Gamma^{k}_{\ iu} F_{ku}\\
		= & \gamma^{ij} \gamma^{km} \beta_j F_{uz} \left(\partial_u \gamma_{im} \right) F_{ku}\,.
	\end{split}
\end{equation}
The fourth term of Eq. (\ref{hkk3secondeighth}) is
\begin{equation}
	\begin{split}
		4 \gamma^{ij} \Gamma^{z}_{\ iu} F_{uz} \Gamma^{h}_{\ zu} F_{hj} = 0\,.
	\end{split}
\end{equation}
The fifth term of Eq. (\ref{hkk3secondeighth}) is
\begin{equation}
	\begin{split}
		& 4 \gamma^{ij} \gamma^{kl} \Gamma^{z}_{\ il} F_{uz} \Gamma^{h}_{\ ku} F_{hj}\\
		= & 4 \gamma^{ij} \gamma^{kl} \Gamma^{z}_{\ il} F_{uz} \Gamma^{u}_{\ ku} F_{uj} + 4 \gamma^{ij} \gamma^{kl} \Gamma^{z}_{\ il} F_{uz} \Gamma^{z}_{\ ku} F_{zj} + 4 \gamma^{ij} \gamma^{kl} \Gamma^{z}_{\ il} F_{uz} \Gamma^{m}_{\ ku} F_{mj}\\
		= & \gamma^{ij} \gamma^{kl} \beta_k \left(\partial_u \gamma_{il} \right) F_{uz} F_{uj} - \gamma^{ij} \gamma^{kl} \gamma^{mo} \left(\partial_u \gamma_{il} \right) F_{uz} \left(\partial_u \gamma_{ko} \right) F_{mj}\,.
	\end{split}
\end{equation}
Therefore, the eight term of Eq. (\ref{hkk3second}) is obtained as 
\begin{equation}
	\begin{split}
		& 4 k^a k^b g^{ce} g^{df} \Gamma^{g}_{\ cf} F_{bg} \Gamma^{h}_{\ da} F_{he}\\
		= & \gamma^{ij} \gamma^{km} \beta_j F_{uz} \left(\partial_u \gamma_{im} \right) F_{ku} + \gamma^{ij} \gamma^{kl} \beta_k \left(\partial_u \gamma_{il} \right) F_{uz} F_{uj}\\
		& - \gamma^{ij} \gamma^{kl} \gamma^{mo} \left(\partial_u \gamma_{il} \right) F_{uz} \left(\partial_u \gamma_{ko} \right) F_{mj}\\
		= & 2 \gamma^{ij} \gamma^{km} \beta_j F_{uz} K_{im} F_{ku} + 2 \gamma^{ij} \gamma^{kl} \beta_k K_{il} F_{uz} F_{uj}\\
		& - 4 \gamma^{ij} \gamma^{kl} \gamma^{mo} K_{il} F_{uz} K_{ko} F_{mj}\\
		= & 0\,.
	\end{split}
\end{equation}

The ninth term of Eq. (\ref{hkk3second}) is
\begin{equation}
	\begin{split}
		& 4 k^a k^b g^{ce} g^{df} \Gamma^{g}_{\ cf} F_{bg} \Gamma^{h}_{\ de} F_{ah} = 4 g^{ce} g^{df} \Gamma^{g}_{\ cf} F_{ug} \Gamma^{h}_{\ de} F_{uh}\\
		= & 4 \Gamma^{g}_{\ uz} F_{ug} \Gamma^{h}_{\ uz} F_{uh} + 4 \Gamma^{g}_{\ uu} F_{ug} \Gamma^{h}_{\ zz} F_{uh} + 4 \gamma^{ij} \Gamma^{g}_{\ uj} F_{ug} \Gamma^{h}_{\ iz} F_{uh}\\
		& + 4 \Gamma^{g}_{\ zz} F_{ug} \Gamma^{h}_{\ uu} F_{uh} + 4 \Gamma^{g}_{\ zu} F_{ug} \Gamma^{h}_{\ zu} F_{uh} + 4 \gamma^{ij} \Gamma^{g}_{\ zj} F_{ug} \Gamma^{h}_{\ iu} F_{uh}\\
		& + 4 \gamma^{ij} \Gamma^{g}_{\ iz} F_{ug} \Gamma^{h}_{\ uj} F_{uh} + 4 \gamma^{ij} \Gamma^{g}_{\ iu} F_{ug} \Gamma^{h}_{\ zj} F_{uh} + 4 \gamma^{ij} \gamma^{kl} \Gamma^{g}_{\ il} F_{ug} \Gamma^{h}_{\ kj} F_{uh}\\
		= & 4 \Gamma^{g}_{\ uz} F_{ug} \Gamma^{h}_{\ uz} F_{uh} + 4 \gamma^{ij} \Gamma^{g}_{\ uj} F_{ug} \Gamma^{h}_{\ iz} F_{uh} + 4 \Gamma^{g}_{\ zu} F_{ug} \Gamma^{h}_{\ zu} F_{uh}\\
		& + 4 \gamma^{ij} \Gamma^{g}_{\ zj} F_{ug} \Gamma^{h}_{\ iu} F_{uh} + 4 \gamma^{ij} \Gamma^{g}_{\ iz} F_{ug} \Gamma^{h}_{\ uj} F_{uh} + 4 \gamma^{ij} \Gamma^{g}_{\ iu} F_{ug} \Gamma^{h}_{\ zj} F_{uh}\\
		& + 4 \gamma^{ij} \gamma^{kl} \Gamma^{g}_{\ il} F_{ug} \Gamma^{h}_{\ kj} F_{uh}\,.
	\end{split}
\end{equation}
The index $g$ should be further expanded.
\begin{equation}\label{hkk3secondninth}
	\begin{split}
		& 4 \Gamma^{g}_{\ uz} F_{ug} \Gamma^{h}_{\ uz} F_{uh} + 4 \gamma^{ij} \Gamma^{g}_{\ uj} F_{ug} \Gamma^{h}_{\ iz} F_{uh} + 4 \Gamma^{g}_{\ zu} F_{ug} \Gamma^{h}_{\ zu} F_{uh}\\
		& + 4 \gamma^{ij} \Gamma^{g}_{\ zj} F_{ug} \Gamma^{h}_{\ iu} F_{uh} + 4 \gamma^{ij} \Gamma^{g}_{\ iz} F_{ug} \Gamma^{h}_{\ uj} F_{uh} + 4 \gamma^{ij} \Gamma^{g}_{\ iu} F_{ug} \Gamma^{h}_{\ zj} F_{uh}\\
		& + 4 \gamma^{ij} \gamma^{kl} \Gamma^{g}_{\ il} F_{ug} \Gamma^{h}_{\ kj} F_{uh}\\
		= & 4 \Gamma^{u}_{\ uz} F_{uu} \Gamma^{h}_{\ uz} F_{uh} + 4 \Gamma^{z}_{\ uz} F_{uz} \Gamma^{h}_{\ uz} F_{uh} + 4 \Gamma^{i}_{\ uz} F_{ui} \Gamma^{h}_{\ uz} F_{uh}\\
		& + 4 \gamma^{ij} \Gamma^{u}_{\ uj} F_{uu} \Gamma^{h}_{\ iz} F_{uh} + 4 \gamma^{ij} \Gamma^{z}_{\ uj} F_{uz} \Gamma^{h}_{\ iz} F_{uh} + 4 \gamma^{ij} \Gamma^{k}_{\ uj} F_{uk} \Gamma^{h}_{\ iz} F_{uh}\\
		& + 4 \Gamma^{u}_{\ zu} F_{uu} \Gamma^{h}_{\ zu} F_{uh} + 4 \Gamma^{z}_{\ zu} F_{uz} \Gamma^{h}_{\ zu} F_{uh} + 4 \Gamma^{i}_{\ zu} F_{ui} \Gamma^{h}_{\ zu} F_{uh}\\
		& + 4 \gamma^{ij} \Gamma^{u}_{\ zj} F_{uu} \Gamma^{h}_{\ iu} F_{uh} + 4 \gamma^{ij} \Gamma^{z}_{\ zj} F_{uz} \Gamma^{h}_{\ iu} F_{uh} + 4 \gamma^{ij} \Gamma^{k}_{\ zj} F_{uk} \Gamma^{h}_{\ iu} F_{uh}\\
		& + 4 \gamma^{ij} \Gamma^{u}_{\ iz} F_{uu} \Gamma^{h}_{\ uj} F_{uh} + 4 \gamma^{ij} \Gamma^{z}_{\ iz} F_{uz} \Gamma^{h}_{\ uj} F_{uh} + 4 \gamma^{ij} \Gamma^{k}_{\ iz} F_{uk} \Gamma^{h}_{\ uj} F_{uh}\\
		& + 4 \gamma^{ij} \Gamma^{u}_{\ iu} F_{uu} \Gamma^{h}_{\ zj} F_{uh} + 4 \gamma^{ij} \Gamma^{z}_{\ iu} F_{uz} \Gamma^{h}_{\ zj} F_{uh} + 4 \gamma^{ij} \Gamma^{k}_{\ iu} F_{uk} \Gamma^{h}_{\ zj} F_{uh}\\
		& + 4 \gamma^{ij} \gamma^{kl} \Gamma^{u}_{\ il} F_{uu} \Gamma^{h}_{\ kj} F_{uh} + 4 \gamma^{ij} \gamma^{kl} \Gamma^{z}_{\ il} F_{uz} \Gamma^{h}_{\ kj} F_{uh} + 4 \gamma^{ij} \gamma^{kl} \Gamma^{m}_{\ il} F_{um} \Gamma^{h}_{\ kj} F_{uh}\\
		= & 4 \Gamma^{z}_{\ uz} F_{uz} \Gamma^{h}_{\ uz} F_{uh} + 4 \gamma^{ij} \Gamma^{z}_{\ uj} F_{uz} \Gamma^{h}_{\ iz} F_{uh} + 4 \Gamma^{z}_{\ zu} F_{uz} \Gamma^{h}_{\ zu} F_{uh}\\
		& + 4 \gamma^{ij} \Gamma^{z}_{\ zj} F_{uz} \Gamma^{h}_{\ iu} F_{uh} + 4 \gamma^{ij} \Gamma^{z}_{\ iz} F_{uz} \Gamma^{h}_{\ uj} F_{uh} + 4 \gamma^{ij} \Gamma^{z}_{\ iu} F_{uz} \Gamma^{h}_{\ zj} F_{uh}\\
		& + 4 \gamma^{ij} \gamma^{kl} \Gamma^{z}_{\ il} F_{uz} \Gamma^{h}_{\ kj} F_{uh}\,.
	\end{split}
\end{equation}
The repeated index $h$ should be further expanded. The frist term of Eq. (\ref{hkk3secondninth}) is 
\begin{equation}
	\begin{split}
		4 \Gamma^{z}_{\ uz} F_{uz} \Gamma^{h}_{\ uz} F_{uh} = 0\,.
	\end{split}
\end{equation}
The second term of Eq. (\ref{hkk3secondninth}) is
\begin{equation}
	\begin{split}
		4 \gamma^{ij} \Gamma^{z}_{\ uj} F_{uz} \Gamma^{h}_{\ iz} F_{uh}= 0\,.
	\end{split}
\end{equation}
The third term of Eq. (\ref{hkk3secondninth}) is
\begin{equation}
	\begin{split}
		4 \Gamma^{z}_{\ zu} F_{uz} \Gamma^{h}_{\ zu} F_{uh} = 0\,.
	\end{split}
\end{equation}
The fourth term of Eq. (\ref{hkk3secondninth}) is
\begin{equation}
	\begin{split}
		& 4 \gamma^{ij} \Gamma^{z}_{\ zj} F_{uz} \Gamma^{h}_{\ iu} F_{uh}\\
		= & 4 \gamma^{ij} \Gamma^{z}_{\ zj} F_{uz} \Gamma^{u}_{\ iu} F_{uu} + 4 \gamma^{ij} \Gamma^{z}_{\ zj} F_{uz} \Gamma^{z}_{\ iu} F_{uz} + 4 \gamma^{ij} \Gamma^{z}_{\ zj} F_{uz} \Gamma^{k}_{\ iu} F_{uk}\\
		= & 4 \gamma^{ij} \Gamma^{z}_{\ zj} F_{uz} \Gamma^{z}_{\ iu} F_{uz} + 4 \gamma^{ij} \Gamma^{z}_{\ zj} F_{uz} \Gamma^{k}_{\ iu} F_{uk}\\
		= & \gamma^{ij} \gamma^{km} \beta_j F_{uz} \left(\partial_u \gamma_{im} \right) F_{uk}\,.
	\end{split}
\end{equation}
The fifth term of Eq. (\ref{hkk3secondninth}) is
\begin{equation}
	\begin{split}
		& 4 \gamma^{ij} \Gamma^{z}_{\ iz} F_{uz} \Gamma^{h}_{\ uj} F_{uh}\\
		= & 4 \gamma^{ij} \Gamma^{z}_{\ iz} F_{uz} \Gamma^{u}_{\ uj} F_{uu} + 4 \gamma^{ij} \Gamma^{z}_{\ iz} F_{uz} \Gamma^{z}_{\ uj} F_{uz} + 4 \gamma^{ij} \Gamma^{z}_{\ iz} F_{uz} \Gamma^{k}_{\ uj} F_{uk}\\
		= & 4 \gamma^{ij} \Gamma^{z}_{\ iz} F_{uz} \Gamma^{z}_{\ uj} F_{uz} + 4 \gamma^{ij} \Gamma^{z}_{\ iz} F_{uz} \Gamma^{k}_{\ uj} F_{uk}\\
		= & \gamma^{ij} \gamma^{km} \beta_i F_{uz} \left(\partial_u \gamma_{jm} \right) F_{uk}\,.
	\end{split}
\end{equation}
The sixth term of Eq. (\ref{hkk3secondninth}) is
\begin{equation}
	\begin{split}
		4 \gamma^{ij} \Gamma^{z}_{\ iu} F_{uz} \Gamma^{h}_{\ zj} F_{uh} = 0\,.
	\end{split}
\end{equation}
The seventh term of Eq. (\ref{hkk3secondninth}) is
\begin{equation}
	\begin{split}
		& 4 \gamma^{ij} \gamma^{kl} \Gamma^{z}_{\ il} F_{uz} \Gamma^{h}_{\ kj} F_{uh}\\
		= & 4 \gamma^{ij} \gamma^{kl} \Gamma^{z}_{\ il} F_{uz} \Gamma^{u}_{\ kj} F_{uu} + 4 \gamma^{ij} \gamma^{kl} \Gamma^{z}_{\ il} F_{uz} \Gamma^{z}_{\ kj} F_{uz} + 4 \gamma^{ij} \gamma^{kl} \Gamma^{z}_{\ il} F_{uz} \Gamma^{m}_{\ kj} F_{um}\\
		= & 4 \gamma^{ij} \gamma^{kl} \Gamma^{z}_{\ il} F_{uz} \Gamma^{z}_{\ kj} F_{uz} + 4 \gamma^{ij} \gamma^{kl} \Gamma^{z}_{\ il} F_{uz} \Gamma^{m}_{\ kj} F_{um}\\
		= & \gamma^{ij} \gamma^{kl} \left(\partial_u \gamma_{il} \right) F_{uz} \left(\partial_u \gamma_{kj} \right) F_{uz} - 2 \gamma^{ij} \gamma^{kl} \left(\partial_u \gamma_{il} \right) F_{uz} \hat{\Gamma}^{m}_{\ kj} F_{um}\,.
	\end{split}
\end{equation}
Therefore, the ninth term of Eq. (\ref{hkk3second}) is obtained as 
\begin{equation}
	\begin{split}
		& 4 k^a k^b g^{ce} g^{df} \Gamma^{g}_{\ cf} F_{bg} \Gamma^{h}_{\ de} F_{ah}\\
		= & \gamma^{ij} \gamma^{km} \beta_j F_{uz} \left(\partial_u \gamma_{im} \right) F_{uk} + \gamma^{ij} \gamma^{km} \beta_i F_{uz} \left(\partial_u \gamma_{jm} \right) F_{uk}\\
		& + \gamma^{ij} \gamma^{kl} \left(\partial_u \gamma_{il} \right) F_{uz} \left(\partial_u \gamma_{kj} \right) F_{uz} - 2 \gamma^{ij} \gamma^{kl} \left(\partial_u \gamma_{il} \right) F_{uz} \hat{\Gamma}^{m}_{\ kj} F_{um}\\
		= & 2 \gamma^{ij} \gamma^{km} \beta_j F_{uz} K_{im} F_{uk} + 2 \gamma^{ij} \gamma^{km} \beta_i F_{uz} K_{jm} F_{uk}\\
		& + 4 \gamma^{ij} \gamma^{kl} K_{il} F_{uz} K_{kj} F_{uz} - 4 \gamma^{ij} \gamma^{kl} K_{il} F_{uz} \hat{\Gamma}^{m}_{\ kj} F_{um}\\
		= & 0\,.
	\end{split}
\end{equation}

Finally, the second term of Eq. (\ref{rehkk3}) is 
\begin{equation}
	\begin{split}
		& 4 k^a k^b \nabla_c F_{bd} \nabla^d F_{a}^{\ c}\\
		= & 4 \left(\partial_u F_{uz} \right) \partial_u F_{uz} + 4 \gamma^{ij} \left(\partial_u F_{uj} \right) \partial_i F_{uz} + 4 \gamma^{ij} \left(\partial_i F_{uz} \right) \partial_u F_{uj}\\
		& + 2 \gamma^{ij} \beta_i \left(\partial_u F_{uj} \right) F_{uz} - 2 \gamma^{ij} \beta_i \left(\partial_u F_{uj} \right) F_{uz} + 2 \gamma^{ij} \beta_i F_{uz} \partial_u F_{uj}\\
		& - 2 \gamma^{ij} \beta_i F_{uz} \partial_u F_{uj}\\
		= & 4 \left(\partial_u F_{uz} \right) \partial_u F_{uz} + 8 \gamma^{ij} \left(\partial_u F_{uj} \right) \partial_i F_{uz}\,.
	\end{split}
\end{equation}

The third term in Eq. (\ref{rehkk3}) is 
\begin{equation}\label{hkk3third}
	\begin{split}
		& 4 k^a k^b \nabla_c F_{a}^{\ c} \nabla_d F_{b}^{\ d}\\
		= & 4 k^a k^b g^{ce} g^{df} \left(\partial_c F_{ae} \right) \partial_d F_{bf} - 4 k^a k^b g^{ce} g^{df} \left(\partial_c F_{ae} \right) \Gamma^{h}_{\ db} F_{hf}\\
		& - 4 k^a k^b g^{ce} g^{df} \left(\partial_c F_{ae} \right) \Gamma^{h}_{\ df} F_{bh} - 4 k^a k^b g^{ce} g^{df} \Gamma^{g}_{\ ca} F_{ge} \partial_d F_{bf}\\
		& + 4 k^a k^b g^{ce} g^{df} \Gamma^{g}_{\ ca} F_{ge} \Gamma^{h}_{\ db} F_{hf} + 4 k^a k^b g^{ce} g^{df} \Gamma^{g}_{\ ca} F_{ge} \Gamma^{h}_{\ df} F_{bh}\\
		& - 4 k^a k^b g^{ce} g^{df} \Gamma^{g}_{\ ce} F_{ag} \partial_d F_{bf} + 4 k^a k^b g^{ce} g^{df} \Gamma^{g}_{\ ce} F_{ag} \Gamma^{h}_{\ db} F_{hf}\\
		& + 4 k^a k^b g^{ce} g^{df} \Gamma^{g}_{\ ce} F_{ag} \Gamma^{h}_{\ df} F_{bh}\,.
	\end{split}
\end{equation}

The first term in Eq. (\ref{hkk3third}) is obtained as 
\begin{equation}
	\begin{split}
		& 4 k^a k^b g^{ce} g^{df} \left(\partial_c F_{ae} \right) \partial_d F_{bf} = 4 g^{ce} g^{df} \left(\partial_c F_{ue} \right) \partial_d F_{uf}\\
		= & 4 \left(\partial_u F_{uz} \right) \partial_u F_{uz} + 4 \gamma^{ij} \left(\partial_u F_{uz} \right) \partial_i F_{uj} + 4 \gamma^{ij} \left(\partial_i F_{uj} \right) \partial_u F_{uz}\\
		& + 4 \gamma^{ij} \gamma^{kl} \left(\partial_i F_{uj} \right) \partial_k F_{ul}\\
		= & 4 \left(\partial_u F_{uz} \right) \partial_u F_{uz}\,.
	\end{split}
\end{equation}

The second term in Eq. (\ref{hkk3third}) is 
\begin{equation}\label{hkk3thirdsecond}
	\begin{split}
		& - 4 k^a k^b g^{ce} g^{df} \left(\partial_c F_{ae} \right) \Gamma^{h}_{\ db} F_{hf} = - 4 g^{ce} g^{df} \left(\partial_c F_{ue} \right) \Gamma^{h}_{\ du} F_{hf}\\
		= & - 4 \left(\partial_u F_{uz} \right) \Gamma^{h}_{\ uu} F_{hz} - 4 \left(\partial_u F_{uz} \right) \Gamma^{h}_{\ zu} F_{hu} - 4 \gamma^{ij} \left(\partial_u F_{uz} \right) \Gamma^{h}_{\ iu} F_{hj}\\
		& - 4 \gamma^{ij} \left(\partial_i F_{uj} \right) \Gamma^{h}_{\ uu} F_{hz} - 4 \gamma^{ij} \left(\partial_i F_{uj} \right) \Gamma^{h}_{\ zu} F_{hu} - 4 \gamma^{ij} \gamma^{kl} \left(\partial_i F_{uj} \right) \Gamma^{h}_{\ ku} F_{hl}\\
		= & - 4 \left(\partial_u F_{uz} \right) \Gamma^{h}_{\ zu} F_{hu} - 4 \gamma^{ij} \left(\partial_u F_{uz} \right) \Gamma^{h}_{\ iu} F_{hj}\,.
	\end{split}
\end{equation}
The repeated index $h$ should be further expanded. The second term of Eq. (\ref{hkk3thirdsecond}) is
\begin{equation}
	\begin{split}
		& - 4 \left(\partial_u F_{uz} \right) \Gamma^{h}_{\ zu} F_{hu}\\
		= & - 4 \left(\partial_u F_{uz} \right) \Gamma^{u}_{\ zu} F_{uu} - 4 \left(\partial_u F_{uz} \right) \Gamma^{z}_{\ zu} F_{zu} - 4 \left(\partial_u F_{uz} \right) \Gamma^{i}_{\ zu} F_{iu}\\
		= & - 4 \left(\partial_u F_{uz} \right) \Gamma^{z}_{\ zu} F_{zu} - 4 \left(\partial_u F_{uz} \right) \Gamma^{i}_{\ zu} F_{iu}\\
		= & - 2 \gamma^{ij} \beta_j \left(\partial_u F_{uz} \right) F_{iu}\,.
	\end{split}
\end{equation}
The second term of Eq. (\ref{hkk3thirdsecond}) is
\begin{equation}
	\begin{split}
		& - 4 \gamma^{ij} \left(\partial_u F_{uz} \right) \Gamma^{h}_{\ iu} F_{hj}\\
		= & - 4 \gamma^{ij} \left(\partial_u F_{uz} \right) \Gamma^{u}_{\ iu} F_{uj} - 4 \gamma^{ij} \left(\partial_u F_{uz} \right) \Gamma^{z}_{\ iu} F_{zj} - 4 \gamma^{ij} \left(\partial_u F_{uz} \right) \Gamma^{k}_{\ iu} F_{kj}\\
		= & 2 \gamma^{ij} \beta_i \left(\partial_u F_{uz} \right) F_{uj} - 2 \gamma^{ij} \gamma^{kl} \left(\partial_u F_{uz} \right) \left(\partial_u \gamma_{il} \right) F_{kj}\,.
	\end{split}
\end{equation}
Therefore, the second term in Eq. (\ref{hkk3third}) is obtained as
\begin{equation}
	\begin{split}
		& - 4 k^a k^b g^{ce} g^{df} \left(\partial_c F_{ae} \right) \Gamma^{h}_{\ db} F_{hf}\\
		= & - 2 \gamma^{ij} \beta_j \left(\partial_u F_{uz} \right) F_{iu} + 2 \gamma^{ij} \beta_i \left(\partial_u F_{uz} \right) F_{uj}\\
		& - 2 \gamma^{ij} \gamma^{kl} \left(\partial_u F_{uz} \right) \left(\partial_u \gamma_{il} \right) F_{kj}\\
		= & - 2 \gamma^{ij} \beta_j \left(\partial_u F_{uz} \right) F_{iu} + 2 \gamma^{ij} \beta_i \left(\partial_u F_{uz} \right) F_{uj}\\
		& - 4 \gamma^{ij} \gamma^{kl} \left(\partial_u F_{uz} \right) K_{il} F_{kj}\\
		= & 0\,.
	\end{split}
\end{equation}

The third term of Eq. (\ref{hkk3third}) is
\begin{equation}\label{hkk3thirdthird}
	\begin{split}
		& - 4 k^a k^b g^{ce} g^{df} \left(\partial_c F_{ae} \right) \Gamma^{h}_{\ df} F_{bh} = - 4 g^{ce} g^{df} \left(\partial_c F_{ue} \right) \Gamma^{h}_{\ df} F_{uh}\\
		= & - 4 \left(\partial_u F_{uz} \right) \Gamma^{h}_{\ uz} F_{uh} - 4 \left(\partial_u F_{uz} \right) \Gamma^{h}_{\ zu} F_{uh} - 4 \gamma^{ij} \left(\partial_u F_{uz} \right) \Gamma^{h}_{\ ij} F_{uh}\\
		& - 4 \gamma^{ij} \left(\partial_i F_{uj} \right) \Gamma^{h}_{\ uz} F_{uh} - 4 \gamma^{ij} \left(\partial_i F_{uj} \right) \Gamma^{h}_{\ zu} F_{uh} - 4 \gamma^{ij} \gamma^{kl} \left(\partial_i F_{uj} \right) \Gamma^{h}_{\ kl} F_{uh}\\
		= & - 8 \left(\partial_u F_{uz} \right) \Gamma^{h}_{\ uz} F_{uh} - 4 \gamma^{ij} \left(\partial_u F_{uz} \right) \Gamma^{h}_{\ ij} F_{uh}\,.
	\end{split}
\end{equation}
The repeated index $h$ should be further expanded. The first term in Eq. (\ref{hkk3thirdthird}) is 
\begin{equation}
	\begin{split}
		& - 8 \left(\partial_u F_{uz} \right) \Gamma^{h}_{\ uz} F_{uh}\\
		= & - 8 \left(\partial_u F_{uz} \right) \Gamma^{u}_{\ uz} F_{uu} - 8 \left(\partial_u F_{uz} \right) \Gamma^{z}_{\ uz} F_{uz} - 8 \left(\partial_u F_{uz} \right) \Gamma^{i}_{\ uz} F_{ui}\\
		= & - 8 \left(\partial_u F_{uz} \right) \Gamma^{z}_{\ uz} F_{uz} - 8 \left(\partial_u F_{uz} \right) \Gamma^{i}_{\ uz} F_{ui}\\
		= & - 4 \gamma^{i j} \beta_j \left(\partial_u F_{uz} \right) F_{ui}\,.
	\end{split}
\end{equation}
The second term in Eq. (\ref{hkk3thirdthird}) is
\begin{equation}
	\begin{split}
		& - 4 \gamma^{ij} \left(\partial_u F_{uz} \right) \Gamma^{h}_{\ ij} F_{uh}\\
		= & - 4 \gamma^{ij} \left(\partial_u F_{uz} \right) \Gamma^{u}_{\ ij} F_{uu} - 4 \gamma^{ij} \left(\partial_u F_{uz} \right) \Gamma^{z}_{\ ij} F_{uz} - 4 \gamma^{ij} \left(\partial_u F_{uz} \right) \Gamma^{k}_{\ ij} F_{uk}\\
		= & - 4 \gamma^{ij} \left(\partial_u F_{uz} \right) \Gamma^{z}_{\ ij} F_{uz} - 4 \gamma^{ij} \left(\partial_u F_{uz} \right) \Gamma^{k}_{\ ij} F_{uk}\\
		= & 2 \gamma^{ij} \left(\partial_u F_{uz} \right) \left(\partial_u \gamma_{ij} \right) F_{uz} - 4 \gamma^{ij} \left(\partial_u F_{uz} \right) \hat{\Gamma}^{k}_{\ ij} F_{uk}\,.
	\end{split}
\end{equation}
Therefore, the third term in Eq. (\ref{hkk3third}) is obtained as 
\begin{equation}
	\begin{split}
		& - 4 k^a k^b g^{ce} g^{df} \left(\partial_c F_{ae} \right) \Gamma^{h}_{\ df} F_{bh}\\
		= & - 4 \gamma^{i j} \beta_j \left(\partial_u F_{uz} \right) F_{ui} + 2 \gamma^{ij} \left(\partial_u F_{uz} \right) \left(\partial_u \gamma_{ij} \right) F_{uz}\\
		& - 4 \gamma^{ij} \left(\partial_u F_{uz} \right) \hat{\Gamma}^{k}_{\ ij} F_{uk}\\
		= & - 4 \gamma^{i j} \beta_j \left(\partial_u F_{uz} \right) F_{ui} + 4 \gamma^{ij} \left(\partial_u F_{uz} \right) K_{ij} F_{uz}\\
		& - 4 \gamma^{ij} \left(\partial_u F_{uz} \right) \hat{\Gamma}^{k}_{\ ij} F_{uk}\\
		= & 0\,.
	\end{split}
\end{equation}

The fourth term in Eq. (\ref{hkk3third}) is
\begin{equation}\label{hkk3thirdfourth}
	\begin{split}
		& - 4 k^a k^b g^{ce} g^{df} \Gamma^{g}_{\ ca} F_{ge} \partial_d F_{bf} = - 4 g^{ce} g^{df} \Gamma^{g}_{\ cu} F_{ge} \partial_d F_{uf}\\
		= & - 4 \Gamma^{g}_{\ uu} F_{gz} \partial_u F_{uz} - 4 \gamma^{ij} \Gamma^{g}_{\ uu} F_{gz} \partial_i F_{uj} - 4 \Gamma^{g}_{\ zu} F_{gu} \partial_u F_{uz}\\
		& - 4 \gamma^{ij} \Gamma^{g}_{\ zu} F_{gu} \partial_i F_{uj} - 4 \gamma^{ij} \Gamma^{g}_{\ iu} F_{gj} \partial_u F_{uz} - 4 \gamma^{ij} \gamma^{kl} \Gamma^{g}_{\ iu} F_{gj} \partial_k F_{ul}\\
		= & - 4 \Gamma^{g}_{\ zu} F_{gu} \partial_u F_{uz} - 4 \gamma^{ij} \Gamma^{g}_{\ iu} F_{gj} \partial_u F_{uz}\,.
	\end{split}
\end{equation}
The repeated index $g$ should be further expanded. The first term of Eq. (\ref{hkk3thirdfourth}) is
\begin{equation}
	\begin{split}
		& - 4 \Gamma^{g}_{\ zu} F_{gu} \partial_u F_{uz}\\
		= & - 4 \Gamma^{u}_{\ zu} F_{uu} \partial_u F_{uz} - 4 \Gamma^{z}_{\ zu} F_{zu} \partial_u F_{uz} - 4 \Gamma^{i}_{\ zu} F_{iu} \partial_u F_{uz}\\
		= & - 4 \Gamma^{z}_{\ zu} F_{zu} \partial_u F_{uz} - 4 \Gamma^{i}_{\ zu} F_{iu} \partial_u F_{uz}\\
		= & - 2 \gamma^{ij} \beta_j F_{iu} \partial_u F_{uz}\,.
	\end{split}
\end{equation}
The second term of Eq. (\ref{hkk3thirdfourth}) is
\begin{equation}
	\begin{split}
		& - 4 \gamma^{ij} \Gamma^{g}_{\ iu} F_{gj} \partial_u F_{uz}\\
		= & - 4 \gamma^{ij} \Gamma^{u}_{\ iu} F_{uj} \partial_u F_{uz} - 4 \gamma^{ij} \Gamma^{z}_{\ iu} F_{zj} \partial_u F_{uz} - 4 \gamma^{ij} \Gamma^{k}_{\ iu} F_{kj} \partial_u F_{uz}\\
		= & 2 \gamma^{ij} \beta_i F_{uj} \partial_u F_{uz} - 2 \gamma^{ij} \gamma^{kl} \left(\partial_u \gamma_{il} \right) F_{kj} \partial_u F_{uz}\,.
	\end{split}
\end{equation}
Therefore, the fourth term in Eq. (\ref{hkk3third}) is obtained as 
\begin{equation}
	\begin{split}
		& - 4 k^a k^b g^{ce} g^{df} \Gamma^{g}_{\ ca} F_{ge} \partial_d F_{bf}\\
		= & - 2 \gamma^{ij} \beta_j F_{iu} \partial_u F_{uz} + 2 \gamma^{ij} \beta_i F_{uj} \partial_u F_{uz}\\
		& - 2 \gamma^{ij} \gamma^{kl} \left(\partial_u \gamma_{il} \right) F_{kj} \partial_u F_{uz}\\
		= & - 2 \gamma^{ij} \beta_j F_{iu} \partial_u F_{uz} + 2 \gamma^{ij} \beta_i F_{uj} \partial_u F_{uz}\\
		& - 4 \gamma^{ij} \gamma^{kl} K_{il} F_{kj} \partial_u F_{uz}\\
		= & 0\,.
	\end{split}
\end{equation}

The fifth term in Eq. (\ref{hkk3third}) is 
\begin{equation}
	\begin{split}
		& 4 k^a k^b g^{ce} g^{df} \Gamma^{g}_{\ ca} F_{ge} \Gamma^{h}_{\ db} F_{hf} = 4 g^{ce} g^{df} \Gamma^{g}_{\ cu} F_{ge} \Gamma^{h}_{\ du} F_{hf}\\
		= & 4 \Gamma^{g}_{\ uu} F_{gz} \Gamma^{h}_{\ uu} F_{hz} + 4 \Gamma^{g}_{\ uu} F_{gz} \Gamma^{h}_{\ zu} F_{hu} + 4 \gamma^{ij} \Gamma^{g}_{\ uu} F_{gz} \Gamma^{h}_{\ iu} F_{hj}\\
		& + 4 \Gamma^{g}_{\ zu} F_{gu} \Gamma^{h}_{\ uu} F_{hz} + 4 \Gamma^{g}_{\ zu} F_{gu} \Gamma^{h}_{\ zu} F_{hu} + 4 \gamma^{ij} \Gamma^{g}_{\ zu} F_{gu} \Gamma^{h}_{\ iu} F_{hj}\\
		& + 4 \gamma^{ij} \Gamma^{g}_{\ iu} F_{gj} \Gamma^{h}_{\ uu} F_{hz} + 4 \gamma^{ij} \Gamma^{g}_{\ iu} F_{gj} \Gamma^{h}_{\ zu} F_{hu} + 4 \gamma^{ij} \gamma^{kl} \Gamma^{g}_{\ iu} F_{gj} \Gamma^{h}_{\ ku} F_{hl}\\
		= & 4 \Gamma^{g}_{\ zu} F_{gu} \Gamma^{h}_{\ zu} F_{hu} + 4 \gamma^{ij} \Gamma^{g}_{\ zu} F_{gu} \Gamma^{h}_{\ iu} F_{hj} + 4 \gamma^{ij} \Gamma^{g}_{\ iu} F_{gj} \Gamma^{h}_{\ zu} F_{hu}\\
		& + 4 \gamma^{ij} \gamma^{kl} \Gamma^{g}_{\ iu} F_{gj} \Gamma^{h}_{\ ku} F_{hl}\,.
	\end{split}
\end{equation}
The index $g$ should be further expanded.
\begin{equation}\label{hkk3thirdfifth}
	\begin{split}
		& 4 \Gamma^{g}_{\ zu} F_{gu} \Gamma^{h}_{\ zu} F_{hu} + 4 \gamma^{ij} \Gamma^{g}_{\ zu} F_{gu} \Gamma^{h}_{\ iu} F_{hj} + 4 \gamma^{ij} \Gamma^{g}_{\ iu} F_{gj} \Gamma^{h}_{\ zu} F_{hu}\\
		& + 4 \gamma^{ij} \gamma^{kl} \Gamma^{g}_{\ iu} F_{gj} \Gamma^{h}_{\ ku} F_{hl}\\
		= & 4 \Gamma^{u}_{\ zu} F_{uu} \Gamma^{h}_{\ zu} F_{hu} + 4 \Gamma^{z}_{\ zu} F_{zu} \Gamma^{h}_{\ zu} F_{hu} + 4 \Gamma^{i}_{\ zu} F_{iu} \Gamma^{h}_{\ zu} F_{hu}\\
		& + 4 \gamma^{ij} \Gamma^{u}_{\ zu} F_{uu} \Gamma^{h}_{\ iu} F_{hj} + 4 \gamma^{ij} \Gamma^{z}_{\ zu} F_{zu} \Gamma^{h}_{\ iu} F_{hj} + 4 \gamma^{ij} \Gamma^{k}_{\ zu} F_{ku} \Gamma^{h}_{\ iu} F_{hj}\\
		& + 4 \gamma^{ij} \Gamma^{u}_{\ iu} F_{uj} \Gamma^{h}_{\ zu} F_{hu} + 4 \gamma^{ij} \Gamma^{z}_{\ iu} F_{zj} \Gamma^{h}_{\ zu} F_{hu} + 4 \gamma^{ij} \Gamma^{k}_{\ iu} F_{kj} \Gamma^{h}_{\ zu} F_{hu}\\
		& + 4 \gamma^{ij} \gamma^{kl} \Gamma^{u}_{\ iu} F_{uj} \Gamma^{h}_{\ ku} F_{hl} + 4 \gamma^{ij} \gamma^{kl} \Gamma^{z}_{\ iu} F_{zj} \Gamma^{h}_{\ ku} F_{hl} + 4 \gamma^{ij} \gamma^{kl} \Gamma^{m}_{\ iu} F_{mj} \Gamma^{h}_{\ ku} F_{hl}\\
		= & 4 \Gamma^{z}_{\ zu} F_{zu} \Gamma^{h}_{\ zu} F_{hu} + 4 \gamma^{ij} \Gamma^{z}_{\ zu} F_{zu} \Gamma^{h}_{\ iu} F_{hj} + 4 \gamma^{ij} \Gamma^{z}_{\ iu} F_{zj} \Gamma^{h}_{\ zu} F_{hu}\\
		& + 4 \gamma^{ij} \Gamma^{k}_{\ iu} F_{kj} \Gamma^{h}_{\ zu} F_{hu} + 4 \gamma^{ij} \gamma^{kl} \Gamma^{z}_{\ iu} F_{zj} \Gamma^{h}_{\ ku} F_{hl} + 4 \gamma^{ij} \gamma^{kl} \Gamma^{m}_{\ iu} F_{mj} \Gamma^{h}_{\ ku} F_{hl}\,.
	\end{split}
\end{equation}
The repeated index $h$ should be further expanded. The first term of Eq. (\ref{hkk3thirdfifth}) is
\begin{equation}
	\begin{split}
		4 \Gamma^{z}_{\ zu} F_{zu} \Gamma^{h}_{\ zu} F_{hu} = 0\,.
	\end{split}
\end{equation}
The second term of Eq. (\ref{hkk3thirdfifth}) is
\begin{equation}
	\begin{split}
		4 \gamma^{ij} \Gamma^{z}_{\ zu} F_{zu} \Gamma^{h}_{\ iu} F_{hj} = 0\,.
	\end{split}
\end{equation}
The third term of Eq. (\ref{hkk3thirdfifth}) is
\begin{equation}
	\begin{split}
		4 \gamma^{ij} \Gamma^{z}_{\ iu} F_{zj} \Gamma^{h}_{\ zu} F_{hu} = 0\,.
	\end{split}
\end{equation}
The fourth term of Eq. (\ref{hkk3thirdfifth}) is
\begin{equation}
	\begin{split}
		& 4 \gamma^{ij} \Gamma^{k}_{\ iu} F_{kj} \Gamma^{h}_{\ zu} F_{hu}\\
		= & 4 \gamma^{ij} \Gamma^{k}_{\ iu} F_{kj} \Gamma^{u}_{\ zu} F_{uu} + 4 \gamma^{ij} \Gamma^{k}_{\ iu} F_{kj} \Gamma^{z}_{\ zu} F_{zu} + 4 \gamma^{ij} \Gamma^{k}_{\ iu} F_{kj} \Gamma^{l}_{\ zu} F_{lu}\\
		= & 4 \gamma^{ij} \Gamma^{k}_{\ iu} F_{kj} \Gamma^{z}_{\ zu} F_{zu} + 4 \gamma^{ij} \Gamma^{k}_{\ iu} F_{kj} \Gamma^{l}_{\ zu} F_{lu}\\
		= & \gamma^{ij} \gamma^{km} \gamma^{ln} \beta_n \left(\partial_u \gamma_{im} \right) F_{kj} F_{lu}\,.
	\end{split}
\end{equation}
The fifth term of Eq. (\ref{hkk3thirdfifth}) is
\begin{equation}
	\begin{split}
		4 \gamma^{ij} \gamma^{kl} \Gamma^{z}_{\ iu} F_{zj} \Gamma^{h}_{\ ku} F_{hl} = 0\,.
	\end{split}
\end{equation}
The sixth term of Eq. (\ref{hkk3thirdfifth}) is
\begin{equation}
	\begin{split}
		& 4 \gamma^{ij} \gamma^{kl} \Gamma^{m}_{\ iu} F_{mj} \Gamma^{h}_{\ ku} F_{hl}\\
		= & 4 \gamma^{ij} \gamma^{kl} \Gamma^{m}_{\ iu} F_{mj} \Gamma^{u}_{\ ku} F_{ul} + 4 \gamma^{ij} \gamma^{kl} \Gamma^{m}_{\ iu} F_{mj} \Gamma^{z}_{\ ku} F_{zl} + 4 \gamma^{ij} \gamma^{kl} \Gamma^{m}_{\ iu} F_{mj} \Gamma^{n}_{\ ku} F_{nl}\\
		= & - \gamma^{ij} \gamma^{kl} \gamma^{mn} \beta_k \left(\partial_u \gamma_{in} \right) F_{mj} F_{ul} + \gamma^{ij} \gamma^{kl} \gamma^{mo} \gamma^{np} \left(\partial_u \gamma_{io} \right) F_{mj} \left(\partial_u \gamma_{kp} \right) F_{nl}\,.
	\end{split}
\end{equation}
Therefore, the fifth term in Eq. (\ref{hkk3third}) is obtained as 
\begin{equation}
	\begin{split}
		& 4 k^a k^b g^{ce} g^{df} \Gamma^{g}_{\ ca} F_{ge} \Gamma^{h}_{\ db} F_{hf}\\
		= & \gamma^{ij} \gamma^{km} \gamma^{ln} \beta_n \left(\partial_u \gamma_{im} \right) F_{kj} F_{lu} - \gamma^{ij} \gamma^{kl} \gamma^{mn} \beta_k \left(\partial_u \gamma_{in} \right) F_{mj} F_{ul}\\
		& + \gamma^{ij} \gamma^{kl} \gamma^{mo} \gamma^{np} \left(\partial_u \gamma_{io} \right) F_{mj} \left(\partial_u \gamma_{kp} \right) F_{nl}\\
		= & 2 \gamma^{ij} \gamma^{km} \gamma^{ln} \beta_n K_{im} F_{kj} F_{lu} - 2 \gamma^{ij} \gamma^{kl} \gamma^{mn} \beta_k K_{in} F_{mj} F_{ul}\\
		& + 4 \gamma^{ij} \gamma^{kl} \gamma^{mo} \gamma^{np} K_{io} F_{mj} K_{kp} F_{nl}\\
		= & 0\,.
	\end{split}
\end{equation}

The sixth term of Eq. (\ref{hkk3third}) is
\begin{equation}
	\begin{split}
		& 4 k^a k^b g^{ce} g^{df} \Gamma^{g}_{\ ca} F_{ge} \Gamma^{h}_{\ df} F_{bh} = 4 g^{ce} g^{df} \Gamma^{g}_{\ cu} F_{ge} \Gamma^{h}_{\ df} F_{uh}\\
		= & 4 \Gamma^{g}_{\ uu} F_{gz} \Gamma^{h}_{\ uz} F_{uh} + 4 \Gamma^{g}_{\ uu} F_{gz} \Gamma^{h}_{\ zu} F_{uh} + 4 \gamma^{ij} \Gamma^{g}_{\ uu} F_{gz} \Gamma^{h}_{\ ij} F_{uh}\\
		& + 4 \Gamma^{g}_{\ zu} F_{gu} \Gamma^{h}_{\ uz} F_{uh} + 4 \Gamma^{g}_{\ zu} F_{gu} \Gamma^{h}_{\ zu} F_{uh} + 4 \gamma^{ij} \Gamma^{g}_{\ zu} F_{gu} \Gamma^{h}_{\ ij} F_{uh}\\
		& + 4 \gamma^{ij} \Gamma^{g}_{\ iu} F_{gj} \Gamma^{h}_{\ uz} F_{uh} + 4 \gamma^{ij} \Gamma^{g}_{\ iu} F_{gj} \Gamma^{h}_{\ zu} F_{uh} + 4 \gamma^{ij} \gamma^{kl} \Gamma^{g}_{\ iu} F_{gj} \Gamma^{h}_{\ kl} F_{uh}\\
		= & 4 \Gamma^{g}_{\ zu} F_{gu} \Gamma^{h}_{\ uz} F_{uh} + 4 \Gamma^{g}_{\ zu} F_{gu} \Gamma^{h}_{\ zu} F_{uh} + 4 \gamma^{ij} \Gamma^{g}_{\ zu} F_{gu} \Gamma^{h}_{\ ij} F_{uh}\\
		& + 4 \gamma^{ij} \Gamma^{g}_{\ iu} F_{gj} \Gamma^{h}_{\ uz} F_{uh} + 4 \gamma^{ij} \Gamma^{g}_{\ iu} F_{gj} \Gamma^{h}_{\ zu} F_{uh} + 4 \gamma^{ij} \gamma^{kl} \Gamma^{g}_{\ iu} F_{gj} \Gamma^{h}_{\ kl} F_{uh}\,.
	\end{split}
\end{equation}
The index $g$ should be further expanded.
\begin{equation}\label{hkk3thirdsixth}
	\begin{split}
		& 4 \Gamma^{g}_{\ zu} F_{gu} \Gamma^{h}_{\ uz} F_{uh} + 4 \Gamma^{g}_{\ zu} F_{gu} \Gamma^{h}_{\ zu} F_{uh} + 4 \gamma^{ij} \Gamma^{g}_{\ zu} F_{gu} \Gamma^{h}_{\ ij} F_{uh}\\
		& + 4 \gamma^{ij} \Gamma^{g}_{\ iu} F_{gj} \Gamma^{h}_{\ uz} F_{uh} + 4 \gamma^{ij} \Gamma^{g}_{\ iu} F_{gj} \Gamma^{h}_{\ zu} F_{uh} + 4 \gamma^{ij} \gamma^{kl} \Gamma^{g}_{\ iu} F_{gj} \Gamma^{h}_{\ kl} F_{uh}\\
		= & 4 \Gamma^{u}_{\ zu} F_{uu} \Gamma^{h}_{\ uz} F_{uh} + 4 \Gamma^{z}_{\ zu} F_{zu} \Gamma^{h}_{\ uz} F_{uh} + 4 \Gamma^{i}_{\ zu} F_{iu} \Gamma^{h}_{\ uz} F_{uh}\\
		& + 4 \Gamma^{u}_{\ zu} F_{uu} \Gamma^{h}_{\ zu} F_{uh} + 4 \Gamma^{z}_{\ zu} F_{zu} \Gamma^{h}_{\ zu} F_{uh} + 4 \Gamma^{i}_{\ zu} F_{iu} \Gamma^{h}_{\ zu} F_{uh}\\
		& + 4 \gamma^{ij} \Gamma^{u}_{\ zu} F_{uu} \Gamma^{h}_{\ ij} F_{uh} + 4 \gamma^{ij} \Gamma^{z}_{\ zu} F_{zu} \Gamma^{h}_{\ ij} F_{uh} + 4 \gamma^{ij} \Gamma^{k}_{\ zu} F_{ku} \Gamma^{h}_{\ ij} F_{uh}\\
		& + 4 \gamma^{ij} \Gamma^{u}_{\ iu} F_{uj} \Gamma^{h}_{\ uz} F_{uh} + 4 \gamma^{ij} \Gamma^{z}_{\ iu} F_{zj} \Gamma^{h}_{\ uz} F_{uh} + 4 \gamma^{ij} \Gamma^{k}_{\ iu} F_{kj} \Gamma^{h}_{\ uz} F_{uh}\\
		& + 4 \gamma^{ij} \Gamma^{u}_{\ iu} F_{uj} \Gamma^{h}_{\ zu} F_{uh} + 4 \gamma^{ij} \Gamma^{z}_{\ iu} F_{zj} \Gamma^{h}_{\ zu} F_{uh} + 4 \gamma^{ij} \Gamma^{k}_{\ iu} F_{kj} \Gamma^{h}_{\ zu} F_{uh}\\
		& + 4 \gamma^{ij} \gamma^{kl} \Gamma^{u}_{\ iu} F_{uj} \Gamma^{h}_{\ kl} F_{uh} + 4 \gamma^{ij} \gamma^{kl} \Gamma^{z}_{\ iu} F_{zj} \Gamma^{h}_{\ kl} F_{uh} + 4 \gamma^{ij} \gamma^{kl} \Gamma^{m}_{\ iu} F_{mj} \Gamma^{h}_{\ kl} F_{uh}\\
		= & 4 \Gamma^{z}_{\ zu} F_{zu} \Gamma^{h}_{\ uz} F_{uh} + 4 \Gamma^{z}_{\ zu} F_{zu} \Gamma^{h}_{\ zu} F_{uh} + 4 \gamma^{ij} \Gamma^{z}_{\ zu} F_{zu} \Gamma^{h}_{\ ij} F_{uh}\\
		& + 4 \gamma^{ij} \Gamma^{z}_{\ iu} F_{zj} \Gamma^{h}_{\ uz} F_{uh} + 4 \gamma^{ij} \Gamma^{k}_{\ iu} F_{kj} \Gamma^{h}_{\ uz} F_{uh} + 4 \gamma^{ij} \Gamma^{z}_{\ iu} F_{zj} \Gamma^{h}_{\ zu} F_{uh}\\
		& + 4 \gamma^{ij} \Gamma^{k}_{\ iu} F_{kj} \Gamma^{h}_{\ zu} F_{uh} + 4 \gamma^{ij} \gamma^{kl} \Gamma^{z}_{\ iu} F_{zj} \Gamma^{h}_{\ kl} F_{uh} + 4 \gamma^{ij} \gamma^{kl} \Gamma^{m}_{\ iu} F_{mj} \Gamma^{h}_{\ kl} F_{uh}\,.
	\end{split}
\end{equation}
The repeated index $h$ should be further expanded. The first term of Eq. (\ref{hkk3thirdsixth}) is
\begin{equation}
	\begin{split}
		4 \Gamma^{z}_{\ zu} F_{zu} \Gamma^{h}_{\ uz} F_{uh} = 0\,.
	\end{split}
\end{equation}
The second term of Eq. (\ref{hkk3thirdsixth}) is
\begin{equation}
	\begin{split}
		4 \Gamma^{z}_{\ zu} F_{zu} \Gamma^{h}_{\ zu} F_{uh} = 0\,.
	\end{split}
\end{equation}
The third term of Eq. (\ref{hkk3thirdsixth}) is
\begin{equation}
	\begin{split}
		4 \gamma^{ij} \Gamma^{z}_{\ zu} F_{zu} \Gamma^{h}_{\ ij} F_{uh} = 0\,.
	\end{split}
\end{equation}
The fourth term of Eq. (\ref{hkk3thirdsixth}) is
\begin{equation}
	\begin{split}
		4 \gamma^{ij} \Gamma^{z}_{\ iu} F_{zj} \Gamma^{h}_{\ uz} F_{uh} = 0\,.
	\end{split}
\end{equation}
The fifth term of Eq. (\ref{hkk3thirdsixth}) is
\begin{equation}
	\begin{split}
		& 4 \gamma^{ij} \Gamma^{k}_{\ iu} F_{kj} \Gamma^{h}_{\ uz} F_{uh}\\
		= & 4 \gamma^{ij} \Gamma^{k}_{\ iu} F_{kj} \Gamma^{u}_{\ uz} F_{uu} + 4 \gamma^{ij} \Gamma^{k}_{\ iu} F_{kj} \Gamma^{z}_{\ uz} F_{uz} + 4 \gamma^{ij} \Gamma^{k}_{\ iu} F_{kj} \Gamma^{l}_{\ uz} F_{ul}\\
		= & 4 \gamma^{ij} \Gamma^{k}_{\ iu} F_{kj} \Gamma^{z}_{\ uz} F_{uz} + 4 \gamma^{ij} \Gamma^{k}_{\ iu} F_{kj} \Gamma^{l}_{\ uz} F_{ul}\\
		= & \gamma^{ij} \gamma^{km} \gamma^{ln} \beta_n \left(\partial_u \gamma_{im} \right) F_{kj} F_{ul}\,.
	\end{split}
\end{equation}
The sixth term of Eq. (\ref{hkk3thirdsixth}) is
\begin{equation}
	\begin{split}
		4 \gamma^{ij} \Gamma^{z}_{\ iu} F_{zj} \Gamma^{h}_{\ zu} F_{uh} = 0\,.
	\end{split}
\end{equation}
The seventh term of Eq. (\ref{hkk3thirdsixth}) is
\begin{equation}
	\begin{split}
		& 4 \gamma^{ij} \Gamma^{k}_{\ iu} F_{kj} \Gamma^{h}_{\ zu} F_{uh}\\
		= & 4 \gamma^{ij} \Gamma^{k}_{\ iu} F_{kj} \Gamma^{u}_{\ zu} F_{uu} + 4 \gamma^{ij} \Gamma^{k}_{\ iu} F_{kj} \Gamma^{z}_{\ zu} F_{uz} + 4 \gamma^{ij} \Gamma^{k}_{\ iu} F_{kj} \Gamma^{l}_{\ zu} F_{ul}\\
		= & 4 \gamma^{ij} \Gamma^{k}_{\ iu} F_{kj} \Gamma^{z}_{\ zu} F_{uz} + 4 \gamma^{ij} \Gamma^{k}_{\ iu} F_{kj} \Gamma^{l}_{\ zu} F_{ul}\\
		= & \gamma^{ij} \gamma^{km} \gamma^{ln} \beta_n \left(\partial_u \gamma_{im} \right) F_{kj} F_{ul}\,.
	\end{split}
\end{equation}
The eighth term of Eq. (\ref{hkk3thirdsixth}) is
\begin{equation}
	\begin{split}
		4 \gamma^{ij} \gamma^{kl} \Gamma^{z}_{\ iu} F_{zj} \Gamma^{h}_{\ kl} F_{uh} = 0\,.
	\end{split}
\end{equation}
The ninth term of Eq. (\ref{hkk3thirdsixth}) is
\begin{equation}
	\begin{split}
		& 4 \gamma^{ij} \gamma^{kl} \Gamma^{m}_{\ iu} F_{mj} \Gamma^{h}_{\ kl} F_{uh}\\
		= & 4 \gamma^{ij} \gamma^{kl} \Gamma^{m}_{\ iu} F_{mj} \Gamma^{u}_{\ kl} F_{uu} + 4 \gamma^{ij} \gamma^{kl} \Gamma^{m}_{\ iu} F_{mj} \Gamma^{z}_{\ kl} F_{uz} + 4 \gamma^{ij} \gamma^{kl} \Gamma^{m}_{\ iu} F_{mj} \Gamma^{n}_{\ kl} F_{un}\\
		= & 4 \gamma^{ij} \gamma^{kl} \Gamma^{m}_{\ iu} F_{mj} \Gamma^{z}_{\ kl} F_{uz} + 4 \gamma^{ij} \gamma^{kl} \Gamma^{m}_{\ iu} F_{mj} \Gamma^{n}_{\ kl} F_{un}\\
		= & - \gamma^{ij} \gamma^{kl} \gamma^{mn} \left(\partial_u \gamma_{in} \right) F_{mj} \left(\partial_u \gamma_{kl} \right) F_{uz} + 2 \gamma^{ij} \gamma^{kl} \gamma^{mo} \left(\partial_u \gamma_{io} \right) F_{mj} \hat{\Gamma}^{n}_{\ kl} F_{un}\,.
	\end{split}
\end{equation}
Therefore, the sixth term of Eq. (\ref{hkk3third}) is obtained as 
\begin{equation}
	\begin{split}
		& 4 k^a k^b g^{ce} g^{df} \Gamma^{g}_{\ ca} F_{ge} \Gamma^{h}_{\ df} F_{bh}\\
		= & \gamma^{ij} \gamma^{km} \gamma^{ln} \beta_n \left(\partial_u \gamma_{im} \right) F_{kj} F_{ul} + \gamma^{ij} \gamma^{km} \gamma^{ln} \beta_n \left(\partial_u \gamma_{im} \right) F_{kj} F_{ul}\\
		& - \gamma^{ij} \gamma^{kl} \gamma^{mn} \left(\partial_u \gamma_{in} \right) F_{mj} \left(\partial_u \gamma_{kl} \right) F_{uz} + 2 \gamma^{ij} \gamma^{kl} \gamma^{mo} \left(\partial_u \gamma_{io} \right) F_{mj} \hat{\Gamma}^{n}_{\ kl} F_{un}\\
		= & 2 \gamma^{ij} \gamma^{km} \gamma^{ln} \beta_n K_{im} F_{kj} F_{ul} + 2 \gamma^{ij} \gamma^{km} \gamma^{ln} \beta_n K_{im} F_{kj} F_{ul}\\
		& - 4 \gamma^{ij} \gamma^{kl} \gamma^{mn} K_{in} F_{mj} K_{kl} F_{uz} + 4 \gamma^{ij} \gamma^{kl} \gamma^{mo} K_{io} F_{mj} \hat{\Gamma}^{n}_{\ kl} F_{un}\\
		= & 0\,.
	\end{split}
\end{equation}

The seventh term in Eq. (\ref{hkk3third}) is
\begin{equation}
	\begin{split}
		& - 4 k^a k^b g^{ce} g^{df} \Gamma^{g}_{\ ce} F_{ag} \partial_d F_{bf} = - 4 g^{ce} g^{df} \Gamma^{g}_{\ ce} F_{ug} \partial_d F_{uf}\\
		= & - 8 \Gamma^{g}_{\ uz} F_{ug} \partial_u F_{uz} - 8 \gamma^{ij} \Gamma^{g}_{\ uz} F_{ug} \partial_i F_{uj} - 4 \gamma^{ij} \Gamma^{g}_{\ ij} F_{ug} \partial_u F_{uz}\\
		& - 4 \gamma^{ij} \gamma^{kl} \Gamma^{g}_{\ ij} F_{ug} \partial_k F_{ul}\\
		= & - 8 \Gamma^{g}_{\ uz} F_{ug} \partial_u F_{uz} - 4 \gamma^{ij} \Gamma^{g}_{\ ij} F_{ug} \partial_u F_{uz}\,.
	\end{split}
\end{equation}
The index $g$ should be expanded.
\begin{equation}\label{hkk3thirdseventh}
	\begin{split}
		& - 8 \Gamma^{g}_{\ uz} F_{ug} \partial_u F_{uz} - 4 \gamma^{ij} \Gamma^{g}_{\ ij} F_{ug} \partial_u F_{uz}\\
		= & - 8 \Gamma^{u}_{\ uz} F_{uu} \partial_u F_{uz} - 8 \Gamma^{z}_{\ uz} F_{uz} \partial_u F_{uz} - 8 \Gamma^{i}_{\ uz} F_{ui} \partial_u F_{uz}\\
		& - 4 \gamma^{ij} \Gamma^{u}_{\ ij} F_{uu} \partial_u F_{uz} - 4 \gamma^{ij} \Gamma^{z}_{\ ij} F_{uz} \partial_u F_{uz} - 4 \gamma^{ij} \Gamma^{k}_{\ ij} F_{uk} \partial_u F_{uz}\\
		= & - 8 \Gamma^{z}_{\ uz} F_{uz} \partial_u F_{uz} - 4 \gamma^{ij} \Gamma^{z}_{\ ij} F_{uz} \partial_u F_{uz}\,.
	\end{split}
\end{equation}
Therefore, the seventh term of Eq. (\ref{hkk3third}) is obtained as
\begin{equation}
	\begin{split}
		& - 4 k^a k^b g^{ce} g^{df} \Gamma^{g}_{\ ce} F_{ag} \partial_d F_{bf} = 2 \gamma^{ij} \left(\partial_u \gamma_{ij} \right) F_{uz} \partial_u F_{uz}\\
		= & 4 \gamma^{ij} K_{ij} F_{uz} \partial_u F_{uz}\\
		= & 0\,.
	\end{split}
\end{equation}

The eighth term in Eq. (\ref{hkk3third}) is
\begin{equation}
	\begin{split}
		& 4 k^a k^b g^{ce} g^{df} \Gamma^{g}_{\ ce} F_{ag} \Gamma^{h}_{\ db} F_{hf} = 4 g^{ce} g^{df} \Gamma^{g}_{\ ce} F_{ug} \Gamma^{h}_{\ du} F_{hf}\\
		= & 8 \Gamma^{g}_{\ uz} F_{ug} \Gamma^{h}_{\ uu} F_{hz} + 8 \Gamma^{g}_{\ uz} F_{ug} \Gamma^{h}_{\ zu} F_{hu} + 8 \gamma^{ij} \Gamma^{g}_{\ uz} F_{ug} \Gamma^{h}_{\ iu} F_{hj}\\
		& + 4 \gamma^{ij} \Gamma^{g}_{\ ij} F_{ug} \Gamma^{h}_{\ uu} F_{hz} + 4 \gamma^{ij} \Gamma^{g}_{\ ij} F_{ug} \Gamma^{h}_{\ zu} F_{hu} + 4 \gamma^{ij} \gamma^{kl} \Gamma^{g}_{\ ij} F_{ug} \Gamma^{h}_{\ ku} F_{hl}\\
		= & 8 \Gamma^{g}_{\ uz} F_{ug} \Gamma^{h}_{\ zu} F_{hu} + 8 \gamma^{ij} \Gamma^{g}_{\ uz} F_{ug} \Gamma^{h}_{\ iu} F_{hj} + 4 \gamma^{ij} \Gamma^{g}_{\ ij} F_{ug} \Gamma^{h}_{\ zu} F_{hu}\\
		& + 4 \gamma^{ij} \gamma^{kl} \Gamma^{g}_{\ ij} F_{ug} \Gamma^{h}_{\ ku} F_{hl}\,.
	\end{split}
\end{equation}
The index $g$ should be expanded.
\begin{equation}\label{hkk3thirdeighth}
	\begin{split}
		& 8 \Gamma^{g}_{\ uz} F_{ug} \Gamma^{h}_{\ zu} F_{hu} + 8 \gamma^{ij} \Gamma^{g}_{\ uz} F_{ug} \Gamma^{h}_{\ iu} F_{hj} + 4 \gamma^{ij} \Gamma^{g}_{\ ij} F_{ug} \Gamma^{h}_{\ zu} F_{hu}\\
		& + 4 \gamma^{ij} \gamma^{kl} \Gamma^{g}_{\ ij} F_{ug} \Gamma^{h}_{\ ku} F_{hl}\\
		= & 8 \Gamma^{u}_{\ uz} F_{uu} \Gamma^{h}_{\ zu} F_{hu} + 8 \Gamma^{z}_{\ uz} F_{uz} \Gamma^{h}_{\ zu} F_{hu} + 8 \Gamma^{i}_{\ uz} F_{ui} \Gamma^{h}_{\ zu} F_{hu}\\
		& + 8 \gamma^{ij} \Gamma^{u}_{\ uz} F_{uu} \Gamma^{h}_{\ iu} F_{hj} + 8 \gamma^{ij} \Gamma^{z}_{\ uz} F_{uz} \Gamma^{h}_{\ iu} F_{hj} + 8 \gamma^{ij} \Gamma^{k}_{\ uz} F_{uk} \Gamma^{h}_{\ iu} F_{hj}\\
		& + 4 \gamma^{ij} \Gamma^{u}_{\ ij} F_{uu} \Gamma^{h}_{\ zu} F_{hu} + 4 \gamma^{ij} \Gamma^{z}_{\ ij} F_{uz} \Gamma^{h}_{\ zu} F_{hu} + 4 \gamma^{ij} \Gamma^{k}_{\ ij} F_{uk} \Gamma^{h}_{\ zu} F_{hu}\\
		& + 4 \gamma^{ij} \gamma^{kl} \Gamma^{u}_{\ ij} F_{uu} \Gamma^{h}_{\ ku} F_{hl} + 4 \gamma^{ij} \gamma^{kl} \Gamma^{z}_{\ ij} F_{uz} \Gamma^{h}_{\ ku} F_{hl} + 4 \gamma^{ij} \gamma^{kl} \Gamma^{m}_{\ ij} F_{um} \Gamma^{h}_{\ ku} F_{hl}\\
		= & 8 \Gamma^{z}_{\ uz} F_{uz} \Gamma^{h}_{\ zu} F_{hu} + 8 \gamma^{ij} \Gamma^{z}_{\ uz} F_{uz} \Gamma^{h}_{\ iu} F_{hj} + 4 \gamma^{ij} \Gamma^{z}_{\ ij} F_{uz} \Gamma^{h}_{\ zu} F_{hu}\\
		& + 4 \gamma^{ij} \gamma^{kl} \Gamma^{z}_{\ ij} F_{uz} \Gamma^{h}_{\ ku} F_{hl}\,.
	\end{split}
\end{equation}
The repeated index $h$ should be further expanded. The third term of Eq. (\ref{hkk3thirdeighth}) is
\begin{equation}
	\begin{split}
		8 \Gamma^{z}_{\ uz} F_{uz} \Gamma^{h}_{\ zu} F_{hu} = 0\,.
	\end{split}
\end{equation}
The second term of Eq. (\ref{hkk3thirdeighth}) is
\begin{equation}
	\begin{split}
		8 \gamma^{ij} \Gamma^{z}_{\ uz} F_{uz} \Gamma^{h}_{\ iu} F_{hj} = 0\,.
	\end{split}
\end{equation}
The third term of Eq. (\ref{hkk3thirdeighth}) is
\begin{equation}
	\begin{split}
		& 4 \gamma^{ij} \Gamma^{z}_{\ ij} F_{uz} \Gamma^{h}_{\ zu} F_{hu}\\
		= & 4 \gamma^{ij} \Gamma^{z}_{\ ij} F_{uz} \Gamma^{u}_{\ zu} F_{uu} + 4 \gamma^{ij} \Gamma^{z}_{\ ij} F_{uz} \Gamma^{z}_{\ zu} F_{zu} + 4 \gamma^{ij} \Gamma^{z}_{\ ij} F_{uz} \Gamma^{k}_{\ zu} F_{ku}\\ 
		= & 4 \gamma^{ij} \Gamma^{z}_{\ ij} F_{uz} \Gamma^{z}_{\ zu} F_{zu} + 4 \gamma^{ij} \Gamma^{z}_{\ ij} F_{uz} \Gamma^{k}_{\ zu} F_{ku}\\
		= & - \gamma^{ij} \gamma^{km} \beta_m \left(\partial_u \gamma_{ij} \right) F_{uz} F_{ku}\,. 
	\end{split}
\end{equation}
The fourth term of Eq. (\ref{hkk3thirdeighth}) is
\begin{equation}
	\begin{split}
		& 4 \gamma^{ij} \gamma^{kl} \Gamma^{z}_{\ ij} F_{uz} \Gamma^{h}_{\ ku} F_{hl}\\
		= & 4 \gamma^{ij} \gamma^{kl} \Gamma^{z}_{\ ij} F_{uz} \Gamma^{u}_{\ ku} F_{ul} + 4 \gamma^{ij} \gamma^{kl} \Gamma^{z}_{\ ij} F_{uz} \Gamma^{z}_{\ ku} F_{zl} + 4 \gamma^{ij} \gamma^{kl} \Gamma^{z}_{\ ij} F_{uz} \Gamma^{m}_{\ ku} F_{ml}\\
		= & \gamma^{ij} \gamma^{kl} \beta_k \left(\partial_u \gamma_{ij} \right) F_{uz} F_{ul} - \gamma^{ij} \gamma^{kl} \gamma^{mo} \left(\partial_u \gamma_{ij} \right) F_{uz} \left(\partial_u \gamma_{ko} \right) F_{ml}\,.
	\end{split}
\end{equation}
Therefore, the eighth term in Eq. (\ref{hkk3third}) is obtained as 
\begin{equation}
	\begin{split}
		& 4 k^a k^b g^{ce} g^{df} \Gamma^{g}_{\ ce} F_{ag} \Gamma^{h}_{\ db} F_{hf}\\
		= & - \gamma^{ij} \gamma^{km} \beta_m \left(\partial_u \gamma_{ij} \right) F_{uz} F_{ku} + \gamma^{ij} \gamma^{kl} \beta_k \left(\partial_u \gamma_{ij} \right) F_{uz} F_{ul}\\
		& - \gamma^{ij} \gamma^{kl} \gamma^{mo} \left(\partial_u \gamma_{ij} \right) F_{uz} \left(\partial_u \gamma_{ko} \right) F_{ml}\\
		= & - 2 \gamma^{ij} \gamma^{km} \beta_m K_{ij} F_{uz} F_{ku} + 2 \gamma^{ij} \gamma^{kl} \beta_k K_{ij} F_{uz} F_{ul}\\
		& - 4 \gamma^{ij} \gamma^{kl} \gamma^{mo} K_{ij} F_{uz} K_{ko} F_{ml}\\
		= & 0\,.
	\end{split}
\end{equation}

The ninth term in Eq. (\ref{hkk3third}) is 
\begin{equation}
	\begin{split}
		& 4 k^a k^b g^{ce} g^{df} \Gamma^{g}_{\ ce} F_{ag} \Gamma^{h}_{\ df} F_{bh} = 4 g^{ce} g^{df} \Gamma^{g}_{\ ce} F_{ug} \Gamma^{h}_{\ df} F_{uh}\\
		= & 16 \Gamma^{g}_{\ uz} F_{ug} \Gamma^{h}_{\ uz} F_{uh} + 8 \gamma^{ij} \Gamma^{g}_{\ uz} F_{ug} \Gamma^{h}_{\ ij} F_{uh} + 8 \gamma^{ij} \Gamma^{g}_{\ ij} F_{ug} \Gamma^{h}_{\ uz} F_{uh}\\
		& + 4 \gamma^{ij} \gamma^{kl} \Gamma^{g}_{\ ij} F_{ug} \Gamma^{h}_{\ kl} F_{uh}\,.
	\end{split}
\end{equation}
The index $g$ should be further expanded.
\begin{equation}\label{hkk3thirdninth}
	\begin{split}
		& 16 \Gamma^{g}_{\ uz} F_{ug} \Gamma^{h}_{\ uz} F_{uh} + 8 \gamma^{ij} \Gamma^{g}_{\ uz} F_{ug} \Gamma^{h}_{\ ij} F_{uh} + 8 \gamma^{ij} \Gamma^{g}_{\ ij} F_{ug} \Gamma^{h}_{\ uz} F_{uh}\\
		& + 4 \gamma^{ij} \gamma^{kl} \Gamma^{g}_{\ ij} F_{ug} \Gamma^{h}_{\ kl} F_{uh}\\
		= & 16 \Gamma^{u}_{\ uz} F_{uu} \Gamma^{h}_{\ uz} F_{uh} + 16 \Gamma^{z}_{\ uz} F_{uz} \Gamma^{h}_{\ uz} F_{uh} + 16 \Gamma^{i}_{\ uz} F_{ui} \Gamma^{h}_{\ uz} F_{uh}\\
		& + 8 \gamma^{ij} \Gamma^{u}_{\ uz} F_{uu} \Gamma^{h}_{\ ij} F_{uh} + 8 \gamma^{ij} \Gamma^{z}_{\ uz} F_{uz} \Gamma^{h}_{\ ij} F_{uh} + 8 \gamma^{ij} \Gamma^{k}_{\ uz} F_{uk} \Gamma^{h}_{\ ij} F_{uh}\\
		& + 8 \gamma^{ij} \Gamma^{u}_{\ ij} F_{uu} \Gamma^{h}_{\ uz} F_{uh} + 8 \gamma^{ij} \Gamma^{z}_{\ ij} F_{uz} \Gamma^{h}_{\ uz} F_{uh} + 8 \gamma^{ij} \Gamma^{k}_{\ ij} F_{uk} \Gamma^{h}_{\ uz} F_{uh}\\
		& + 4 \gamma^{ij} \gamma^{kl} \Gamma^{u}_{\ ij} F_{uu} \Gamma^{h}_{\ kl} F_{uh} + 4 \gamma^{ij} \gamma^{kl} \Gamma^{z}_{\ ij} F_{uz} \Gamma^{h}_{\ kl} F_{uh} + 4 \gamma^{ij} \gamma^{kl} \Gamma^{m}_{\ ij} F_{um} \Gamma^{h}_{\ kl} F_{uh}\\
		= & 16 \Gamma^{z}_{\ uz} F_{uz} \Gamma^{h}_{\ uz} F_{uh} + 8 \gamma^{ij} \Gamma^{z}_{\ uz} F_{uz} \Gamma^{h}_{\ ij} F_{uh} + 8 \gamma^{ij} \Gamma^{z}_{\ ij} F_{uz} \Gamma^{h}_{\ uz} F_{uh}\\
		& + 4 \gamma^{ij} \gamma^{kl} \Gamma^{z}_{\ ij} F_{uz} \Gamma^{h}_{\ kl} F_{uh}\,.
	\end{split}
\end{equation}
The repeated index $h$ should be further expanded. The first term of Eq. (\ref{hkk3thirdninth}) is 
\begin{equation}
	\begin{split}
		16 \Gamma^{z}_{\ uz} F_{uz} \Gamma^{h}_{\ uz} F_{uh} = 0\,.
	\end{split}
\end{equation}
The second term of Eq. (\ref{hkk3thirdninth}) is
\begin{equation}
	\begin{split}
		8 \gamma^{ij} \Gamma^{z}_{\ uz} F_{uz} \Gamma^{h}_{\ ij} F_{uh} = 0\,.
	\end{split}
\end{equation}
The third term of Eq. (\ref{hkk3thirdninth}) is
\begin{equation}
	\begin{split}
		& 8 \gamma^{ij} \Gamma^{z}_{\ ij} F_{uz} \Gamma^{h}_{\ uz} F_{uh}\\
		= & 8 \gamma^{ij} \Gamma^{z}_{\ ij} F_{uz} \Gamma^{u}_{\ uz} F_{uu} + 8 \gamma^{ij} \Gamma^{z}_{\ ij} F_{uz} \Gamma^{z}_{\ uz} F_{uz} + 8 \gamma^{ij} \Gamma^{z}_{\ ij} F_{uz} \Gamma^{k}_{\ uz} F_{uk}\\
		= & 8 \gamma^{ij} \Gamma^{z}_{\ ij} F_{uz} \Gamma^{z}_{\ uz} F_{uz} + 8 \gamma^{ij} \Gamma^{z}_{\ ij} F_{uz} \Gamma^{k}_{\ uz} F_{uk}\\
		= & - 2 \gamma^{ij} \gamma^{km} \beta_m \left(\partial_u \gamma_{ij} \right) F_{uz} F_{uk}\,.
	\end{split}
\end{equation}
The fourth term of Eq. (\ref{hkk3thirdninth}) is
\begin{equation}
	\begin{split}
		& 4 \gamma^{ij} \gamma^{kl} \Gamma^{z}_{\ ij} F_{uz} \Gamma^{h}_{\ kl} F_{uh}\\
		= & 4 \gamma^{ij} \gamma^{kl} \Gamma^{z}_{\ ij} F_{uz} \Gamma^{u}_{\ kl} F_{uu} + 4 \gamma^{ij} \gamma^{kl} \Gamma^{z}_{\ ij} F_{uz} \Gamma^{z}_{\ kl} F_{uz} + 4 \gamma^{ij} \gamma^{kl} \Gamma^{z}_{\ ij} F_{uz} \Gamma^{m}_{\ kl} F_{um}\\
		= & 4 \gamma^{ij} \gamma^{kl} \Gamma^{z}_{\ ij} F_{uz} \Gamma^{z}_{\ kl} F_{uz} + 4 \gamma^{ij} \gamma^{kl} \Gamma^{z}_{\ ij} F_{uz} \Gamma^{m}_{\ kl} F_{um}\\
		= & \gamma^{ij} \gamma^{kl} \left(\partial_u \gamma_{ij} \right) F_{uz} \left(\partial_u \gamma_{kl} \right) F_{uz} - 2 \gamma^{ij} \gamma^{kl} \left(\partial_u \gamma_{ij} \right) F_{uz} \hat{\Gamma}^{m}_{\ kl} F_{um}\,.
	\end{split}
\end{equation}
Therefore, the ninth term in Eq. (\ref{hkk3third}) is obtained as 
\begin{equation}
	\begin{split}
		& 4 k^a k^b g^{ce} g^{df} \Gamma^{g}_{\ ce} F_{ag} \Gamma^{h}_{\ df} F_{bh}\\
		= & - 2 \gamma^{ij} \gamma^{km} \beta_m \left(\partial_u \gamma_{ij} \right) F_{uz} F_{uk} + \gamma^{ij} \gamma^{kl} \left(\partial_u \gamma_{ij} \right) F_{uz} \left(\partial_u \gamma_{kl} \right) F_{uz}\\
		& - 2 \gamma^{ij} \gamma^{kl} \left(\partial_u \gamma_{ij} \right) F_{uz} \hat{\Gamma}^{m}_{\ kl} F_{um}\\
		= & - 4 \gamma^{ij} \gamma^{km} \beta_m K_{ij} F_{uz} F_{uk} + 4 \gamma^{ij} \gamma^{kl} K_{ij} F_{uz} K_{kl} F_{uz}\\
		& - 4 \gamma^{ij} \gamma^{kl} K_{ij} F_{uz} \hat{\Gamma}^{m}_{\ kl} F_{um}\\
		= & 0\,.
	\end{split}
\end{equation}

Finally, the third term of Eq. (\ref{rehkk3}) is 
\begin{equation}
	\begin{split}
		4 k^a k^b \nabla_c F_{a}^{\ c} \nabla_d F_{b}^{\ d} = 4 \left(\partial_u F_{uz} \right) \partial_u F_{uz}\,.
	\end{split}
\end{equation}

The fourth term of Eq. (\ref{rehkk3}) is 
\begin{equation}\label{hkk3fourth}
	\begin{split}
		& 4 k^a k^b F_{a}^{\ c} R_{ce} F_{b}^{\ e} = 4 k^a k^b g^{ce} g^{df} F_{ae} R_{cd} F_{bf} = 4 g^{ce} g^{df} F_{ue} R_{cd} F_{uf}\\
		= & 4 F_{uz} R_{uu} F_{uz} + 4 \gamma^{ij} F_{uz} R_{ui} F_{uj} + 4 \gamma^{ij} F_{uj} R_{iu} F_{uz} + 4 \gamma^{ij} \gamma^{kl} F_{uj} R_{ik} F_{ul}\\
		= & 4 F_{uz} R_{uu} F_{uz}\,.
	\end{split}
\end{equation}
Therefore, the fourth term of Eq. (\ref{rehkk3}) is obtained as 
\begin{equation}
	\begin{split}
		& 4 k^a k^b F_{a}^{\ c} R_{ce} F_{b}^{\ e} = 4 F_{uz} R_{uu} F_{uz}\\
		= & 4 F_{uz} \left[- \frac{1}{2} \left(\partial_u \gamma^{ij} \right) \left(\partial_u \gamma_{ij} \right) - \frac{1}{2} \gamma^{ij} \partial_u^2 \gamma_{ij} - \frac{1}{4} \gamma^{ij} \gamma^{kl} \left(\partial_u \gamma_{ik} \right) \left(\partial_u \gamma_{jl} \right) \right] F_{uz}\\
		= & - 8 F_{uz} K^{ij} K_{ij} F_{uz} - 2 \gamma^{ij} F_{uz} \left(\partial_u^2 \gamma_{ij} \right) F_{uz} - 4 \gamma^{ij} \gamma^{kl} F_{uz} K_{ik} K_{jl} F_{uz}\\
		= & - 2 \gamma^{ij} F_{uz} \left(\partial_u^2 \gamma_{ij} \right) F_{uz}\,.
	\end{split}
\end{equation}

The fifth term of Eq. (\ref{rehkk3}) is 
\begin{equation}\label{hkk3fifth}
	\begin{split}
		& - 4 k^a k^b F_{b}^{\ c} R_{cda}^{\ \ \ e} F_{e}^{\ d}\\
		= & - 4 k^a k^b g^{cf} g^{dg} g^{eh} F_{bf} R_{cdah} F_{eg} = - 4 g^{cf} g^{dg} g^{eh} F_{uf} R_{cduh} F_{eg}\\
		= & - 4 \gamma^{ij} F_{uz} R_{uzuj} F_{iu} - 4 \gamma^{ij} F_{uz} R_{uiuz} F_{uj} - 4 \gamma^{ij} \gamma^{kl} F_{uz} R_{uiul} F_{kj}\\
		& - 4 \gamma^{ij} F_{uj} R_{iuuz} F_{uz} - 4 \gamma^{ij} \gamma^{kl} F_{uj} R_{iuul} F_{kz} - 4 \gamma^{ij} \gamma^{kl} F_{uj} R_{izul} F_{ku}\\
		& - 4 \gamma^{ij} \gamma^{kl} F_{uj} R_{ikuz} F_{ul} - 4 \gamma^{ij} \gamma^{kl} \gamma^{mn} F_{uj} R_{ikun} F_{ml}\\
		= & - 4 \gamma^{ij} \gamma^{kl} F_{uz} R_{uiul} F_{kj}\,.
	\end{split}
\end{equation}
Therefore, the fifth term of Eq. (\ref{rehkk3}) is obtained as 
\begin{equation}
	\begin{split}
		& - 4 k^a k^b F_{b}^{\ c} R_{cda}^{\ \ \ e} F_{e}^{\ d} = - 4 \gamma^{ij} \gamma^{kl} F_{uz} R_{uiul} F_{kj}\\
		= & - 4 \gamma^{ij} \gamma^{kl} F_{uz} \left[- \frac{1}{2} \gamma_{lm} \left(\partial_u \gamma^{mn} \right) \left(\partial_u \gamma_{in} \right) - \frac{1}{2} \partial_u^2 \gamma_{il} - \frac{1}{4} \gamma^{mn} \left(\partial_u \gamma_{in} \right) \left(\partial_u \gamma_{ml} \right) \right] F_{kj}\\
		= & 2 \gamma^{ij} F_{uz} \left(\partial_u \gamma^{kl} \right) \left(\partial_u \gamma_{il} \right) F_{kj} + 2 \gamma^{ij} \gamma^{kl} F_{uz} \left(\partial_u^2 \gamma_{il} \right) F_{kj}\\
		& + \gamma^{ij} \gamma^{kl} \gamma^{mn} F_{uz} \left(\partial_u \gamma_{in} \right) \left(\partial_u \gamma_{ml} \right) F_{kj}\\
		= & 8 \gamma^{ij} F_{uz} K^{kl} K_{il} F_{kj} + 2 \gamma^{ij} \gamma^{kl} F_{uz} \left(\partial_u^2 \gamma_{il} \right) F_{kj} + 4 \gamma^{ij} \gamma^{kl} \gamma^{mn} F_{uz} K_{in} K_{ml} F_{kj}\\
		= & 2 \gamma^{i(j} \gamma^{k)l} F_{uz} \left(\partial_u^2 \gamma_{il} \right) F_{[kj]}\\
		= & 0\,.
	\end{split}
\end{equation}

The sixth term of Eq. (\ref{rehkk3}) is obtained as 
\begin{equation}
	\begin{split}
		& - 2 k^a k^b R_{acde} F_{b}^{\ c} F^{de}\\
		= & - 2 k^a k^b g^{cf} g^{dg} g^{eh} R_{acde} F_{bf} F_{gh} = - 2 g^{cf} g^{dg} g^{eh} R_{ucde} F_{uf} F_{gh}\\
		= & - 2 \gamma^{ij} R_{uiuu} F_{uj} F_{zz} - 2 \gamma^{ij} R_{uiuz} F_{uj} F_{zu} - 2 \gamma^{ij} \gamma^{kl} R_{uiuk} F_{uj} F_{zl}\\
		& - 2 \gamma^{ij} R_{uizu} F_{uj} F_{uz} - 2 \gamma^{ij} R_{uizz} F_{uj} F_{uu} - 2 \gamma^{ij} \gamma^{kl} R_{uizk} F_{uj} F_{ul}\\
		& - 2 \gamma^{ij} \gamma^{kl} R_{uiku} F_{uj} F_{lz} - 2 \gamma^{ij} \gamma^{kl} R_{uikz} F_{uj} F_{lu} - 2 \gamma^{ij} \gamma^{kl} \gamma^{mn} R_{uikm} F_{uj} F_{ln}\\
		= & 0\,.
	\end{split}
\end{equation}

Based on the results derived from the preceding calculations, the expression of $H_{uu}^{(3)}$ under the linear-order constraints of the quantim corrections can be expressed as 
\begin{equation}\label{finalhkk3beforesimplify}
	\begin{split}
		H_{uu}^{(3)} = & 8 F_{uz} \partial_u^2 F_{uz} + 8 \gamma^{ij} F_{uz} \partial_u \partial_i F_{uj} - 8 \gamma^{ij} \gamma^{kl} F_{uz} \left(\partial_u K_{il} \right) F_{kj}\\
		& + 4 \gamma^{ij} F_{uz} \left(\partial_u^2 \gamma_{ij} \right) F_{uz} - 8 \gamma^{ij} F_{uz} \hat{\Gamma}^{k}_{\ ij} \partial_u F_{uk} + 4 \left(\partial_u F_{uz} \right) \partial_u F_{uz}\\
		& + 8 \gamma^{ij} \left(\partial_u F_{uj} \right) \partial_i F_{uz} + 4 \left(\partial_u F_{uz} \right) \partial_u F_{uz} - 2 \gamma^{ij} F_{uz} \left(\partial_u^2 \gamma_{ij} \right) F_{uz}\,.
	\end{split}
\end{equation}
The fourth and ninth terms in Eq. (\ref{finalhkk3beforesimplify}) are simplified as 
\begin{equation}
	\begin{split}
		4 \gamma^{ij} F_{uz} \left(\partial_u^2 \gamma_{ij} \right) F_{uz} - 2 \gamma^{ij} F_{uz} \left(\partial_u^2 \gamma_{ij} \right) F_{uz} = 2 \gamma^{ij} F_{uz} \left(\partial_u^2 \gamma_{ij} \right) F_{uz}\,.
	\end{split}
\end{equation}
The sixth and eighth terms in Eq. (\ref{finalhkk3beforesimplify}) are simplified as 
\begin{equation}
	\begin{split}
		4 \left(\partial_u F_{uz} \right) \partial_u F_{uz} + 4 \left(\partial_u F_{uz} \right) \partial_u F_{uz} = 8 \left(\partial_u F_{uz} \right) \partial_u F_{uz}\,.
	\end{split}
\end{equation}
Therefore, the expression of $H_{uu}^{(3)}$ in Eq. (\ref{finalhkk3beforesimplify}) can be further simplified as
\begin{equation}\label{finalhkk3aftersimplify}
	\begin{split}
		H_{uu}^{(3)} = & 8 F_{uz} \partial_u^2 F_{uz} + 8 \gamma^{ij} F_{uz} \partial_u \partial_i F_{uj} - 8 \gamma^{ij} \gamma^{kl} F_{uz} \left(\partial_u K_{il} \right) F_{kj}\\
		& + 2 \gamma^{ij} F_{uz} \left(\partial_u^2 \gamma_{ij} \right) F_{uz} - 8 \gamma^{ij} F_{uz} \hat{\Gamma}^{k}_{\ ij} \partial_u F_{uk} + 8 \left(\partial_u F_{uz} \right) \partial_u F_{uz}\\
		& + 8 \gamma^{ij} \left(\partial_u F_{uj} \right) \partial_i F_{uz}\,.
	\end{split}
\end{equation}
The first and sixth terms in Eq. (\ref{finalhkk3aftersimplify}) are simplified as 
\begin{equation}\label{finalhkk3firstsixthterms}
	\begin{split}
		& 8 F_{uz} \partial_u^2 F_{uz} + 8 \left(\partial_u F_{uz} \right) \partial_u F_{uz}\\
		= & 4 \left[2 F_{uz} \partial_u^2 F_{uz} + 2 \left(\partial_u F_{uz} \right) \partial_u F_{uz} \right]\\
		= & 4 \left[2 \partial_u \left(F_{uz} \partial_u F_{uz} \right)\right]\\
		= & 4 \left[\partial_u \left[F_{uz} \partial_u F_{uz} + \left(\partial_u F_{uz}\right) F_{uz} \right] \right]\\
		= & 4 \partial^2_u \left(F_{uz} F_{uz} \right)\,.
	\end{split}
\end{equation}
The second, fifth and seventh terms in Eq. (\ref{finalhkk3aftersimplify}) are simplified as 
\begin{equation}
	\begin{split}
		& 8 \gamma^{ij} F_{uz} \partial_u \partial_i F_{uj} - 8 \gamma^{ij} F_{uz} \hat{\Gamma}^{k}_{\ ij} \partial_u F_{uk} + 8 \gamma^{ij} \left(\partial_u F_{uj} \right) \partial_i F_{uz}\\
		= & 8 \gamma^{ij} F_{uz} \partial_i \partial_u F_{uj} - 8 \gamma^{ij} F_{uz} \hat{\Gamma}^{k}_{\ ij} \partial_u F_{uk} + 8 \gamma^{ij} \left(\partial_u F_{uj} \right) D_i F_{uz}\\
		= & 8 \gamma^{ij} F_{uz} \left(\partial_i \partial_u F_{uj} - \hat{\Gamma}^{k}_{\ ij} \partial_u F_{uk} \right) + 8 \gamma^{ij} \left(\partial_u F_{uj} \right) D_i F_{uz}\\
		= & 8 \gamma^{ij} F_{uz} D_i \left(\partial_u F_{uj} \right) + 8 \gamma^{ij} \left(\partial_u F_{uj} \right) D_i F_{uz}\\
		= & 8 \gamma^{ij} D_i \left(F_{uz} \partial_u F_{uj} \right)\,.
	\end{split}
\end{equation}
This result can be neglected due to the compactness of the cross-section on the null hypersurface $L$. Utilizing the symmetric property of the extrinsic curvature and the antisymmetric property of the electromagnetic field tensor, the third term in Eq. (\ref{finalhkk3aftersimplify}) is 
\begin{equation}
	\begin{split}
		- 8 \gamma^{ij} \gamma^{kl} F_{uz} \left(\partial_u K_{il} \right) F_{kj} = - 8 \gamma^{i(j} \gamma^{k)l} F_{uz} \left(\partial_u K_{il} \right) F_{[kj]} = 0\,.
	\end{split}
\end{equation}
On the other hand, the second-order derivative of the term $\sqrt{\gamma} F_{uz} F_{uz}$ with respect $u$ can be expressed as 
\begin{equation}\label{partial2ugammafuzfuz}
	\begin{split}
		& \partial_u^2 \left(\sqrt{\gamma} F_{uz} F_{uz} \right)\\
		= & \left(\partial_u^2 \sqrt{\gamma} \right) F_{uz} F_{uz} + 2 \left(\partial_u \sqrt{\gamma} \right) \partial_u \left(F_{uz} F_{uz} \right) + \sqrt{\gamma} \partial_u^2 \left(F_{uz} F_{uz} \right)\\
		= & \left(\partial_u^2 \sqrt{\gamma} \right) F_{uz} F_{uz} + 2 \left(\frac{1}{2 \sqrt{\gamma}} \gamma \gamma^{ij} \partial_u \gamma_{ij} \right) \partial_u \left(F_{uz} F_{uz} \right) + \sqrt{\gamma} \partial_u^2 \left(F_{uz} F_{uz} \right)\\
		= & \left(\partial_u^2 \sqrt{\gamma} \right) F_{uz} F_{uz} + 2 \left(\frac{1}{\sqrt{\gamma}} \gamma \gamma^{ij} K_{ij} \right) \partial_u \left(F_{uz} F_{uz} \right) + \sqrt{\gamma} \partial_u^2 \left(F_{uz} F_{uz} \right)\\
		= & \left(\partial_u^2 \sqrt{\gamma} \right) F_{uz} F_{uz} + \sqrt{\gamma} \partial_u^2 \left(F_{uz} F_{uz} \right)\\
		= & \frac{1}{2} \sqrt{\gamma} \gamma^{ij} \left(\partial_u^2 \gamma_{ij} \right) F_{uz} F_{uz} + \sqrt{\gamma} \partial_u^2 \left(F_{uz} F_{uz} \right)\,.
	\end{split}
\end{equation}
Therefore, combining the result of Eq. (\ref{finalhkk3firstsixthterms}) with the fourth term in Eq. (\ref{finalhkk3aftersimplify}) and multiplying by the square root of the induced metric, the final expression for $H_{uu}^{(3)}$ can be written as
\begin{equation}\label{hkk3finalresultwithgamma}
	\begin{split}
		\sqrt{\gamma} H_{uu}^{(3)} = & 4 \sqrt{\gamma} \partial^2_u \left(F_{uz} F_{uz} \right) + 2 \sqrt{\gamma} \gamma^{ij} F_{uz} \left(\partial_u^2 \gamma_{ij} \right) F_{uz}\\
		= & 4 \left[\sqrt{\gamma} \partial^2_u \left(F_{uz} F_{uz} \right) + \frac{1}{2} \sqrt{\gamma} \gamma^{ij} F_{uz} \left(\partial_u^2 \gamma_{ij} \right) F_{uz} \right]\\
		= & 4 \partial^2_u \left(\sqrt{\gamma} F_{uz} F_{uz} \right)\,,
	\end{split}
\end{equation}
where the result in Eq. (\ref{partial2ugammafuzfuz}) has been used in the last step. Based on the results in Eqs. (\ref{rhkk1resultfinalwithgamma}), (\ref{hkk2finalresultwithgamma}), and (\ref{hkk3finalresultwithgamma}), the final expression of the third components of the function $\mathcal{F} \left(u\,, x \right)$ in Eq. (\ref{apformalexpfunctionf}), incorporating the square root of the induced metric $\sqrt{\gamma}$ and constrained by the linear-order quantum corrections, is derived as
\begin{equation}\label{finalresultthirdcomponentoff}
	\begin{split}
		& \frac{1}{2} \sum_{i = 1}^{3} \sqrt{\gamma} a_i H_{uu}^{(i)}\\
		= & \partial_u^2 \left[\sqrt{\gamma} \left(a_1 \left(2 F_{uz} F_{uz} - 4 \gamma^{ij} F_{ui} F_{zj} - \gamma^{ij} \gamma^{kl} F_{ik} F_{jl} \right) + a_2 \left(F_{uz} F_{uz} - \gamma^{ij} F_{ui} F_{zj} \right) \right. \right.\\
		& \left. \left. + a_3 \left(2 F_{uz} F_{uz} \right) \right) \right]\,.
	\end{split}
\end{equation}

The fourth component of the function $\mathcal{F} \left(u, x \right)$ in Eq. (\ref{apformalexpfunctionf}), representing the variation of the generalized expansion along the light sheet $L$, requires further calculation. Based on the definition of the generalized expansion in Eq. (\ref{apdefgeneralexpansion}), the variation of the generalized expansion, incorporating the square root of the determinant of the induced metric $\sqrt{\gamma}$, is expressed as
\begin{equation}
	\begin{split}
		& \sqrt{\gamma} \partial_u \Theta \left(u\,, x \right) = \frac{1}{2} \sqrt{\gamma} \gamma^{ij} \left(\partial_u^2 \gamma_{ij} \right) s_{bh} \left(u\,, x \right) + \sqrt{\gamma} \partial_u^2 s_{bh} \left(u\,, x \right)\\
		= & \left(\partial_u^2 \sqrt{\gamma} \right) s_{bh} \left(u\,, x \right) + \sqrt{\gamma} \partial_u^2 s_{bh} \left(u\,, x \right)\\
		= & \partial_u^2 \left[\sqrt{\gamma} s_{bh} \left(u\,, x \right) \right]\,.
	\end{split}
\end{equation}
Since the generalized expansion $\Theta \left(u\,, x \right)$ in the fourth component of $\mathcal{F} \left(u\,, x \right)$ is exclusively associated with the terms of Wald entropy that involves the coupling constants, the explicit expression of the fourth component in the function $\mathcal{F} \left(u\,, x \right)$ can be derived using the density of black hole entropy $s_{bh} \left(u\,, x \right) $ from Eq. (\ref{aptotalentropydensity}) and the density of the Wald entropy from Eq. (\ref{apwaldentropydensity}). The expression of the fourth component is represented as
\begin{equation}\label{finalresultfourthcomponentoff}
	\begin{split}
		& \sqrt{\gamma} \partial_u \Theta \left(u\,, x\,, a \right) = \partial_u^2 \left[\sqrt{\gamma} s_{bh} \left(u\,, x\,, a  \right) \right] = \partial_u^2 \left[\sqrt{\gamma} \left(s_{\text{W}} - 1 + s_{\text{dyna}} \left(u\,, x \right) \right) \right]\\
		= & \partial_u^2 \left[\sqrt{\gamma} \left(a_1 \left(- 2 F_{uz} F_{uz} - 4 \gamma^{ij} F_{ui} F_{zj} + \gamma^{ij} \gamma^{kl} F_{ik} F_{jl} \right) \right. \right.\\
		& \left. \left. - a_2 \left(F_{uz} F_{uz} + \gamma^{ij} F_{ui} F_{zj} \right) - a_3 \left(2 F_{uz} F_{uz} \right) + s_{\text{dyna}} \left(u\,, x \right) \right) \right]\,,
	\end{split}
\end{equation}

Substituting the results of the third and fourth components of the function $\mathcal{F} \left(u\,, x \right)$ from Eqs. (\ref{finalresultthirdcomponentoff}) and (\ref{finalresultfourthcomponentoff}) into the expression for the function in Eq. (\ref{apformalexpfunctionf}), the final form of $\mathcal{F} \left(u\,, x \right)$, incorporating the square root of the induced metric $\sqrt{\gamma}$, is obtained as
\begin{equation}
	\begin{split}
		\sqrt{\gamma} \mathcal{F} \left(u\,, x \right) = & \sqrt{\gamma} \left(- K^{ij} K_{ij} - 2 F_{ui} F_{u}^{\ i} \right) + \frac{1}{2} \sum_{i = 1}^{3} \sqrt{\gamma} a_i H_{uu}^{(i)} + \sqrt{\gamma} \partial_u \Theta \left(u\,, x\,, a \right)\\
		= & \sqrt{\gamma} \left(- K^{ij} K_{ij} - 2 F_{ui} F_{u}^{\ i} \right)\\
		& + \partial_u^2 \left[\sqrt{\gamma} \left(a_1 \left(2 F_{uz} F_{uz} - 4 \gamma^{ij} F_{ui} F_{zj} - \gamma^{ij} \gamma^{kl} F_{ik} F_{jl} \right) \right. \right.\\
		& \left. \left. + a_2 \left(F_{uz} F_{uz} - \gamma^{ij} F_{ui} F_{zj} \right) + a_3 \left(2 F_{uz} F_{uz} \right) \right) \right]\\
		& + \partial_u^2 \left[\sqrt{\gamma} \left(a_1 \left(- 2 F_{uz} F_{uz} - 4 \gamma^{ij} F_{ui} F_{zj} + \gamma^{ij} \gamma^{kl} F_{ik} F_{jl} \right) \right. \right.\\
		& \left. \left. + a_2 \left(- F_{uz} F_{uz} - \gamma^{ij} F_{ui} F_{zj} \right) - a_3 \left(2 F_{uz} F_{uz} \right) + s_{\text{dyna}} \left(u\,, x \right) \right) \right]\\
		= & \sqrt{\gamma} \left(- K^{ij} K_{ij} - 2 F_{ui} F_{u}^{\ i} \right)\\
		& + \partial_u^2 \left[\sqrt{\gamma} \left(- 8 a_1 \gamma^{ij} F_{ui} F_{zj} - 2 a_2 \gamma^{ij} F_{ui} F_{zj} + s_{\text{dyna}} \left(u\,, x \right) \right) \right]\,.
	\end{split}
\end{equation}
Furthermore, the integral of the complete expression for the function $\mathcal{F} \left(u\,, x \right)$ on the cross-section $B(u)$ can be written as
\begin{equation}\label{finalfunctionfwithgamma}
	\begin{split}
		\sqrt{\gamma} \mathcal{F} \left(u\,, x \right) = & \int_{B(u)} d^{D-2} x \sqrt{\gamma} \left(- K^{ij} K_{ij} - 2 F_{ui} F_{u}^{\ i} \right)\\
		& + \partial_u^2 \int_{B(u)} d^{D-2} x \left[\sqrt{\gamma} \left(- 8 a_1 \gamma^{ij} F_{ui} F_{zj} - 2 a_2 \gamma^{ij} F_{ui} F_{zj} + s_{\text{dyna}} \left(u\,, x \right) \right) \right]\\
		& + \mathcal{O} \left(a_l K^{ij} K_{ij}\,, a_l F_{ui} F_{u}^{\ i}\,, a_l K_{ij} F_{uk} \right)\,,
	\end{split}
\end{equation}
where the third term encompasses all contributions to $\mathcal{F} \left(u\,, x \right)$ that are negligible during the derivation process. Since these terms have minimal weight and do not significantly influence the final properties of $\mathcal{F} \left(u\,, x \right)$, the explicit expressions of these terms need not be included in the final representation of $\mathcal{F} \left(u\,, x \right)$. Therefore, the final expression for the function $\mathcal{F} \left(u, x \right)$, incorporating the square root of the induced metric $\sqrt{\gamma}$ and subject to the linear-order constraints of quantum corrections, is provided in Eq. (\ref{finalfunctionfwithgamma}).

\end{document}